\documentclass[12pt]{book}
\usepackage{amsmath}
\usepackage{amsfonts}
\usepackage{amssymb}
\usepackage{epsfig}
\usepackage{times}
\usepackage{graphicx}
\usepackage{version}
\usepackage[french]{babel}
\usepackage{calc}  % to print time
\newcounter{hours}\newcounter{minutes}

\def\tw{\textwidth}
\def\beq{\begin{equation}}
\def\eeq{\end{equation}}
\def\beeq{\begin{eqnarray}}
\def\eeeq{\end{eqnarray}}
\numberwithin{equation}{section}

\begin{document}

\null

\noindent

\vskip 1.5cm
% \enlargethispage{3cm}
\thispagestyle{empty}
\vspace*{-1.3cm}
\centerline{\scshape 
\large Laboratoire de Physique Th\'eorique et Hautes \'Energies}
\vskip 1.5cm
\centerline{\Large \bfseries TH{\`E}SE DE DOCTORAT DE L'UNIVERSIT{\'E} PARIS VII}
\vskip .8cm
\centerline{\large Sp{\'e}cialit{\'e} : \bfseries\scshape PHYSIQUE TH{\'E}ORIQUE}
\vskip 2.0cm
\centerline{pr{\'e}sent{\'e}e par}
\vskip .6cm
\centerline{\Large \bf M. Redamy PEREZ RAMOS}
\vskip 1cm
\centerline{pour obtenir le grade de}
\vskip .6cm
\centerline{\Large \bf Docteur de l'Universit{\'e} Paris VII}
\vskip 2.0cm
\centerline{Sujet :}
\vskip 1cm
\centerline{\Large \bfseries \itshape DISTRIBUTIONS ET CORR\'ELATIONS HADRONIQUES}

\bigskip

\centerline{\Large \bfseries \itshape EN CHROMODYNAMIQUE QUANTIQUE}

\bigskip

\centerline{\Large \bfseries \itshape DANS L'APPROXIMATION DES ``PETITS X''}
\vskip .3cm
\centerline{\LARGE \bfseries \itshape }
\vskip .2cm
\centerline{\LARGE \bfseries \itshape }
\vskip 2cm
\noindent
Soutenue le 19 septembre 2006 devant le jury compos{\'e} de~:
\vskip 0.5cm
\begin{tabular}{cll}
MM. & Yuri DOKSHITZER, &directeur de th{\`e}se,\\
    & Val\'ery KHOZE, &rapporteur, \\
    & Bruno MACHET, &\\
    & Giuseppe MARCHESINI, &rapporteur,\\
    & Philippe SCHWEMLING,\\
\&  & Jean-Bernard ZUBER.
\end{tabular}
\cleardoublepage

\pagestyle{plain}

%%%%%%%%%%%%%%%%%%%%%%%%%%%%%%%%%%%%%%%%%%%%%%%%%%%%%%%%%%%%%%%%%%%%%%%%%%
%\newpage
\null

\tableofcontents

%%%%%%%%%%%%%%%%%%%%%%%%%%%%%%%%%%%%%%%%%%%%%%%%%%%%%%%%%%%%%%%%%%%%%%%%%%
\newpage
\null

%\chapter{Remerciements}
\chapter{Remerciements}

\section{\em Remerciements}
Je ne saurais terminer l'\'ecriture de cette th\`ese sans remercier
tous ceux qui,
d'une mani\`ere ou d'une autre, m'ont aid\'e \`a la r\'ealiser.

Je commence par adresser mes remerciements ainsi que toute ma reconnaissance
aux premiers enseignants que j'ai eus en physique quand je suis arriv\'e en
France, il y a six ans maintenant.

Je voudrais remercier plus sp\'ecialement
Alain Laverne pour ses corrections et discussions sur le premier chap\^itre
de cette th\`ese, Galliano Valent pour ses cours de licence et de ma\^itrise,
Jose Ocariz pour tous ses encouragements pendant mon stage de DEA au LPNHE,
et pour m'avoir fortement pouss\'e vers la physique
th\'eorique, Pierre Gazeau pour son soutien, Maurice Courbage
ainsi que Philippe Schwemling pour m'avoir donn\'e
les premi\`eres notions en physique des particules pendant les ann\'ees
de licence et de ma\^itrise.

Je remercie l'ensemble des enseignants du DEA de physique th\'eorique
de l'Ecole Normale Sup\'erieure,
en particulier, Costas Bachas, Pierre Fayet, Christophe Schweigert
et sa tr\`es gentille secr\'etaire Nicole Ribet.

Il y a d'autres personnes comme Jacques Chauveau, Lydia Ross et l'\'equipe
exp\'erimentale {\em BaBar} du LPNHE qui, d'une mani\`ere indirecte,
ont \'egalement jou\'e un r\^ole dans ma carri\`ere.

Je remercie bien s\^ur Yuri Dokshitzer qui m'a accept\'e comme \'etudiant
sans aucune contrainte. Je lui suis reconnaissant pour son savoir,
pour son intuition physique stup\'efiante ainsi que pour m'avoir donn\'e
un sujet int\'eressant o\`u il reste encore \'enorm\'ement de progr\`es \`a
faire. Je le remercie \'egalement
pour sa disponibilit\'e, pour ses exigences qui approchent la perfection,
ainsi que pour avoir toujours r\'epondu \`a mes questions.

Je remercie Bruno Machet car il a \'et\'e mon plus proche collaborateur.
Gr\^ace \`a lui et \`a l'aide de Yuri, j'ai publi\'e mes premiers articles
en Chromodynamique Quantique. C'est par cette collaboration et ses conseils
que j'ai appris \`a organiser mes id\'ees, \`a travailler plus efficacement,
\`a \'ecrire mes r\'esultats et \`a r\'eussir enfin \`a nager
 dans cette mer qu'est la physique th\'eorique. Je lui suis aussi tr\`es
reconnaissant car c'est lui qui m'a aid\'e \`a mener \`a bien l'ensemble
des d\'emarches administratives qui ont pr\'ec\'ed\'e la soutenance
de cette th\`ese.
Il a de m\^eme jou\'e un r\^ole tr\`es important pendant que je r\'edigeais
ce m\'emoire, par ses conseils concernant la mise en place des id\'ees et
par ses corrections de mes fautes de fran\c{c}ais.

Je remercie Gavin Salam pour son soutien pr\'ecieux en informatique.
Sans son aide je serais difficilement arriv\'e \`a finir \`a temps tous
les programmes que j'ai d\^u r\'ediger en Fortran-90. Je le remercie pour
sa disponibilit\'e ainsi que pour nos discussions en QCD.

Je remercie de m\^eme Fran\c{c}ois Arl\'eo pour ses encouragements
et nos discussions tout au long de cette th\`ese.

Je remercie l'ensemble du LPTHE, plus sp\'ecialement, ses deux directeurs
Laurent Baulieu et Olivier Babelon. C'est en effet un laboratoire
 dans lequel je me suis senti tr\`es bien accueilli depuis mon arriv\'ee.
J'adresse l'expression de toute ma reconnaissance aux secr\'etaires
Marie-Christine L\'evy, Annie Richard, Sylvie Dalla Foglia
et Val\'erie Sabourot. J'ai toujours b\'en\'efici\'e d'excellentes
conditions de travail et d'un mat\'eriel informatique de qualit\'e.
\`A ce sujet, j'adresse mes remerciements les plus
chaleureux \`a Marco Picco, Elys\'ee Macagny et \`a Damien Br\'emont.
Je dois de m\^eme toute ma reconnaissance \`a Denis Bernia qui m'avait
toujours encourag\'e et aid\'e quand il fallait relier des brochures
scientifiques.

Il y a aussi bien s\^ur tous mes coll\`egues et amis th\'esards du LPTHE,
LPT-Orsay, LPT-ENS.
J'adresse mes remerciements \`a Kyril Kazymirenko, Bruno Durin,
Nicolas Couchoud, Pedro Bordalo, Guillaume Bossard, Alexis Martin,
Yacine Dolivet, Jean Savinien, Alexey Lokhov,
Quentin Duret, Beno\^it Estienne, Emmanuel S\'erie, Hamed Ben Yahia et
au postdoc Francesco Bigazzi.

Je crois qu'un petit mot est
d\^u \`a Michel Bo\^iteux qui m'a tr\`es bien int\'egr\'e dans l'enseignement
sup\'erieur au Centre Universitaire des Saint-P\`eres, et qui m'a fait le
plaisir et l'honneur de venir assister \`a ma soutenance.

Je remercie bien s\^ur les membres du jury pour leur soutien, leur
encouragement, et la confiance en moi m\^eme qu'ils m'ont donn\'e pour la
soutenance. Je dois une gratitude particuli\`ere \`a mes deux rapporteurs
qui se sont d\'eplac\'es d'Italie et d'Angleterre.

Et il y a enfin tous ceux vers qui je tourne mon affection,
amiti\'e et confiance; que chacun s'y reconnaisse car c'est
\`a eux que j'adresse ces quelques ann\'ees de travail et d'effort.

%%%%%%%%%%%%%%%%%%%%%%%%%%%%%%%%%%%%%%%%%%%%%%%%%%%%%%%%%%%%%%%%%%%%%%%%%%%
\newpage\null

\section{\em Agradecimientos}

\vskip 1cm

En Cuba, quiero en especial agradecer
a mis padres Pedro y Susana que siempre me apoyaron desde peque\~no, y sobre todo,
mientras cursaba mis estudios en el Instituto Superior de Ciencias y Tecnolog\'{\i}a
Nucleares de La Habana; a mis abuelos Ramiro y Maximina que siempre estuvieron
a mi lado, tambi\'en recuerdo a mi familia, a t\'{\i}os como L\'azaro y Rosa;
es a ellos a quienes en este p\'arrafo, con mas amor y afecto, dedico este trabajo de
doctorado.

Vale la pena mencionar y recordar a mi maestra Margarita Oliva, a mi gran maestro de matem\'aticas de noveno grado Nelson Z\'u\~niga y a mi exelente profesor de f\'{\i}sica en el IPVCE F\'elix Mac\'{\i}as, fueron ellos inculcaron en mi la pasi\'on por estas
dos disciplinas. Tambi\'en, tuve exelentes profesores a quienes quiero
agredecer: Adriano, de matem\'aticas, Adonis, de f\'{\i}sica, Roberto, de qu\'imica, Mercedes, de ingl\'es, as\'i como al exelente director del centro Elpidio Morales.
Mientras cursaba estudios secundarios, tambi\'en se destacaron Regla, de historia,
Luc\'ia, de espa\~nol-literatura, Bararita, de educaci\'on laboral, Maira, de ingl\'es, entre otros.

En La Habana, recuerdo con mucha admiraci\'on a mis profesores de f\'{\i}sica general
y de matem\'aticas superiores Juan de Dios Garrido,
Valentina Bad\'{\i}a, Roberto Cruz y Mario Piris, quienes se inspiraron
del exelente modelo sovi\'etico para impartir sus clases.

En Par\'{\i}s cuento con numerosos amigos cubanos como L\'azaro e Ismel que
siempre me apoyaron y quienes adem\'as asistieron a la lectura de la tesis el d\'{\i}a
19 de septiembre del 2006. Fernando siempre me apoy\'o, adem\'as de haberme hecho
sobrepasar los momentos m\'as dif\'{\i}ciles de esta etapa, mientras pasaba horas en su
casa mirando los cap\'{\i}tulos de la comedia espa\~nola "Aqu\'{\i} no hay quien viva". Ese d\'{\i}a
tan especial cont\'e con la presencia de mi amiga espa\~nola Beatriz, a quien quiero
agradecer por ello y por haberme dado su mano durante los preparativos del brindis.
 Me hicieron adem\'as el honor de estar presentes mis amigos Antony (le Petit) y John,
 a todos ellos
expreso el sentimiento de mis m\'as sinceros reconocimientos.

En el Laboratorio de F\'{\i}sica Nuclear y Altas Energ\'{\i}as (LPNHE seg\'un las siglas en franc\'es) de Par\'{\i}s-Jussieu, quiero
agradecer en especial a Jos\'e Ocariz por su ayuda y apoyo mientras era estudiante en el Master de F\'{\i}sica Te\'orica y por haberme lanzado hacia la f\'{\i}sica te\'orica de
altas energ\'{\i}as. Tambi\'en recuerdo ah\'{\i} a Jacques Chauveau, Lidia Ross, a Julie
y a Florent Fayette que estuvieron presentes en la lectura de esta tesis.

Siempre estuve sostenido por Jean Marie, su madre Henri\`ette y su esposa Caruquita.
Merece la pena mencionar el apoyo incondicional que recib\'{\i} de mi amiga Th\'er\`ese Obrecht, gran periodista suiza, mientras cursaba el Master de F\'{\i}sica Te\'orica en
Par\'{\i}s. Jean Savinien estuvo dentro de mis mejores amigos durante esta etapa, y
tambi\'en Bruno Durin.

En La Habana, a pesar de la distancia, cont\'e siempre con el apoyo de amigos como
Alexander, Etian, Henry, Lidia, Luis y Normita; siempre tuve, por otra parte, el
de mi gran amiga Yadira de Trinidad.

En Madrid, quiero agredecer a las hermanas Sara e In\'es Rodr\'{\i}guez-Arguelles, mis
dos amigas europeas m\'as cercanas, al igual que a mis amigos cubanos y compan\~eros
de beca (12 y Malec\'on) Armando y el Chino.

%\centerline{\large\bf\em Remerciements}

%%%%%%%%%%%%%%%%%%%%%%%%%%%%%%%%%%%%%%%%%%%%%%%%%%%%%%%%%%%%%%%%%%%%%%%%%%

%\newpage\null
%\newpage\null

\part{TEXTE DE LA TH\`ESE}

\chapter{Introduction}
%\section{Introduction}

Dans le cadre de la Chromodynamique Quantique (CDQ, th\'eorie de jauge
de Yang-Mills) perturbative,
nous menons \`a bien une \'etude de quelques aspects
physiques, ainsi que des techniques math\'ematiques qui ont permis
de simplifier et de d\'ecrire l'\'evolution de la mati\`ere hadronique
dans les collisions leptoniques ou hadroniques
(``Deep Inelatisc Scattering'' (DIS), annihilation $e^+e^-$,
collisions $pp$ $\cdots$) \`a tr\`es haute \'energie.
D'apr\`es la CDQ, les hadrons (proton (p), neutron (n), m\'esons
($\pi^{\pm}$)
$\cdots$) sont des particules compos\'ees de quarks ($q$), anti-quarks
($\bar q$) et gluons ($g$) (partons); cette th\'eorie permet, en particulier,
d'\'etudier les interactions quark-gluon, anti-quark-gluon et gluon-gluon
au sein de ces particules.
Elle permet ainsi de quantifier le comportement
des interactions partoniques \`a tr\`es ``courtes distances''
(par rapport \`a la taille caract\'erisque des hadrons
$\sim10^{-13}\text{cm}$), o\`u, grâce \`a la libert\'e asymptotique,
on peut appliquer la th\'eorie des perturbations en s\'erie de puissances
de la constante de couplage
$\alpha_s$. Autrement dit, la CDQ perturbative d\'ecrit
avec succ\`es les processus dans lesquelles les effets \`a ``petite distance''
sont essentiels et, o\`u la valeur de la constante de couplage est faible
($\alpha_s\ll1$).

Prenons comme exemple le cas de l'annihilation $e^+e^-$ en une paire
quark-anti-quark,
c'est \`a dire la r\'eaction $e^+e^-\to q \bar q$. La production de
cette paire de particules charg\'ees est suivie par l'\'emission d'un
ensemble de gluons ($q(\bar q)\to gq(\bar q)$) qui, \`a leur tour,
donnent naissance
\`a d'autres gluons ($g\to gg$) et/ou \`a d'autres paires quark-anti-quark 
($g\to q\bar q$);
tous les partons sont soumis aux forces
de confinement de couleur et, par cons\'equent, ne se
d\'etectent pas s\'epar\'ement comme les leptons et les photons
en Electrodynamique
Quantique (EDQ). On appelle ``jet'' l'ensemble du syst\`eme partonique
produit qui a \'et\'e initi\'e par le quark ou l'anti-quark.

Dans les th\'eories de jauge comme l'EDQ, ou la CDQ, o\`u le m\'ediateur des
interactions est de masse nulle, la probabilit\'e de production d'un quantum
(un photon mou dans le cas de l'EDQ
ou d'un gluon mou en CDQ) de basse \'energie  (par rapport \`a l'\'energie
$E$ de la charge initiant le jet, $E\gg\omega$) est tr\`es grande
$\propto d\omega/\omega$ \cite{Peskin}\cite{IZuber}.
En effet, une
particule charg\'ee rayonne lorsque son champ coulombien stationnaire
est bris\'e sous l'action d'une perturbation externe.

\`A ce sujet, nous avons 
consacr\'e le premier chap\^itre de cette th\`ese; nous allons en particulier,
d\'emontrer le caract\`ere classique du rayonnement mou en EDQ,
il sera ainsi obtenu en \'Electrodynamique Classique (EDC) dans l'objectif de
mieux comprendre la nature, origine
et universalit\'e physiques de ce ph\'enom\`ene. Le cas o\`u l'impact subi
par la charge est instantan\'e (acc\'el\'eration infinie) sera
distingu\'e de celui o\`u l'on consid\`ere un transfert d'impulsion sur
un intervalle de temps fini (acc\'el\'eration finie) \cite{Niedermayer}.
Ceci permet de r\'egulariser la ``catastrophé ultra-violette''
\`a l'aide d'un
``cut-off'' physique naturel dont on g\'en\'eralise l'universalit\'e
\`a une trajectoire arbitraire suivie par la charge.
Nous allons de m\^eme d\'ecrire l'origine physique de la coh\'erence en
EDQ en consid\'erant le processus physique le plus simple
($e^-\to e^-\gamma$, \'emission d'un photon de bremsstrahlung);
une contrainte (``Angular Ordering'' en anglais) sur les angles
d'\'emission des photons mous (de bremsstrahlung)
et l'angle de diffusion d\'ecoule du calcul de la
section efficace de ce processus, lorsque l'on prend la moyenne
azimutale sur l'angle
d'ouverture du c\^one de rayonnement (c\^one de bremsstrahlung);
ici, on rencontre pour la premi\`ere fois le caract\`ere doublement
logarithmique de la distribution
($\propto\frac{d\omega}{\omega}\frac{d\Theta}{\Theta}$) des photons \'emis,
\`a savoir que le processus n'est pas domin\'e que par les \'emissions
de photons mous, mais aussi colin\'eaires. Cette \'etude se g\'en\'eralise
au cas de la CDQ, o\`u l'on
doit en outre tenir compte du nouveau degr\'e de libert\'e de la particule
charg\'ee, la couleur \cite{pQCDforbeginners}.
Nous ne nous int\'eressons qu'aux
gluons mous rayonn\'es qui n'emportent qu'une petite
fraction de l'\'energie totale du parton initial ($x=\omega/E\ll1$)
car ils sont \`a l'origine du plus grand nombre de particules produites
(les m\'esons l\'egers $\pi^{\pm}$, $K^{\pm}$,...) dans les jets hadroniques.

Dans le deuxi\`eme chap\^itre, pour des raisons p\'edagogiques,
nous donnons les \'etapes qui ont conduit au calcul des jets
\cite{Veneziano},
depuis le choix de la jauge axiale dans les diagrammes de Feynman
jusqu'\`a la construction de l'\'equation ma\^itresse satisfaite
par la fonctionnelle g\'en\'eratrice des grandeurs inclusives dans
les jets (voir \cite{EvEq}\cite{KO} et r\'ef\'erences incluses).
Dans la jauge axiale, on peut traiter les jets comme des objets
distincts qui \'emettent des gluons; on introduit ainsi le sch\'ema
de resommation ``DLA'' (``Double Logarithmic Approximation'' en anglais)
\cite{Fadin1} qui utilise les contraintes \'energ\'etiques ($E\gg\omega_i\gg\omega_{i+1}\dots$) et angulaires
($\Theta_{i}\gg\Theta_{i+1}$) rigoureuses sur les \'emissions successives
des gluons mous; ceci constitue l'ingr\'edient
principal (car $\alpha_s\log^2\sim1$) dans l'estimation d'observables
inclusives dans les jets hadroniques, telles que les
multiplicit\'es, le spectre, les corr\'elations. Mais il
n\'eglige le principe de conservation
de l'\'energie (recul du parton \'emetteur)
et l'\'evolution de la constante de couplage $\alpha_s$, et il s'av\`ere
 insuffisant pour faire des pr\'edictions que l'on puisse comparer avec
les donn\'ees exp\'erimentales. Par contre, l'approximation ``DLA''
constitue le point de d\'epart dans
la construction du sch\'ema probabiliste (consulter \cite{EvEq})
du calcul des jets. Elle
permet en plus de pr\'edire la {\em forme} des distributions inclusives et,
en particulier, de d\'ecrire  les ph\'enom\`enes
de {\em coh\'erence} des gluons mous en CDQ perturbative 
(voir \cite{EvEq} et r\'ef\'erences incluses). Le cas o\`u la
constante de couplage est fix\'ee \`a la duret\'e totale
($\alpha_s$ constant)
du jet et celui o\`u l'on consid\`ere son \'evolution dans le ``temps''
($t=d\Theta/\Theta$) sont distingu\'es dans l'\'evaluation
du spectre inclusif des gluon mous ainsi que dans le calcul des corr\'elations
entre deux particules. Nous donnons, en ``DLA'', les techniques utilis\'ees
dans \cite{DLA}, pour l'\'evaluation du spectre par la m\'ethode du col.
Elles sont d'importante utilit\'e pour la compr\'ehension de
l'article \ref{sub:article3}. De m\^eme, dans le calcul
des corr\'elations \`a deux particules, nous pr\'esentons les techniques
employ\'ees dans \ref{sub:article2}; ici, nous avons effectu\'e une
\'etude am\'elior\'ee de cette observable en CDQ perturbative.

Pourquoi ``am\'elior\'ee''? Vues les limitations du sch\'ema ``DLA'',
un traitement des corrections en ``Logarithmes Simples''
(``Single Logs'' en anglais) est n\'ecessaire.
Ceci est l'objectif du chap\^itre 3 o\`u nous discutons les sources
physiques qui sont \`a l'origine de ces corrections: 

\medskip

$\bullet$ la variation de la constante de couplage $\alpha_s(k_\perp^2)$
dans le ``temps'' d'\'evolution caract\'eristique ($t=d\Theta/\Theta$) du jet;

\medskip

$\bullet$ les d\'esint\'egrations d'un parton en deux partons
d'\'energies comparables $z\sim1$ (les corrections dites ``hard''
qui restaurent la conservation de
l'\'energie en utilisant l'expression exacte des fonctions de
fragmentation partonique de
Dokshitzer-Gribov-Lipatov-Altarelli-Parisi (DGLAP) \cite{DGLAP});

\medskip

$\bullet$ les r\'egions cin\'ematiques o\`u les angles successifs
d'\'emission  sont du m\^eme ordre de grandeur $\Theta_{i}\sim\Theta_{i+1}$.
Dans la solution de ce probl\`eme, la contrainte angulaire ``rigoureuse''
sur les angles d'\'emission
$\Theta_i\gg\Theta_{i+1}$ qui d\'ecoule de la coh\'erence des gluons
mous en DLA est
remplac\'ee par la contrainte angulaire ``stricte'' $\Theta_i\geq\Theta_{i+1}$
(voir \cite{EvEq} et r\'ef\'erences incluses).

\medskip

L'approximation correspondante est connue comme ``MLLA''
(``Modified Leading Logarithmic Approximation'' \cite{DKT} en anglais).
Elle tient compte des corrections sous-dominantes en ``SL''
(``Single Logs'' en anglais)
dans le ``Hamiltonien'' d'\'evolution partonique; elles sont de l'ordre
de $\gamma_0^2$,
o\`u $\gamma_0\propto\sqrt{\alpha_s}$ constitue la dimension anormale
des multiplicit\'es en ``DLA'' \cite{DLA}. Nous donnons l'\'equation
ma\^itresse que satisfait la fonctionnelle g\'en\'eratrice dans le cadre MLLA
\cite{EvEq}\cite{KO}. Elle permet d'obtenir les \'equations
d'\'evolution des distributions partoniques inclusives dans le domaine
des ``petits $x$''.  Elles sont utilis\'ees dans les articles
\ref{sub:article1}, \ref{sub:article2} et
\ref{sub:article3} qui sont l'objet principal de cette th\`ese.
Nous faisons de m\^eme le lien avec les \'equations DGLAP \cite{DGLAP}
dont on utilise les solution dans l'espace de Mellin
dans \ref{sub:article1}.

Dans l'article \ref{sub:article1}, nous effectuons le premier calcul
MLLA analytique des distributions
inclusives en fonction de l'impulsion transverse $k_{\perp}$ dans le
domaine des ``petits $x$''.
Nous utilisons l'approximation du ``limiting spectrum'' \cite{DKT}
(le cut-off colin\'eaire $Q_0$ est alors \'egal \`a l'\'echelle de
masse $\Lambda_{QCD}$, $Q_0=\Lambda_{QCD}$),
qui permet de bien
d\'ecrire le spectre inclusif d'une particule en fonction de l'\'energie.
Nous faisons des pr\'edictions aux \'energies du LEP, du Tevatron et
du future LHC. On d\'emontre comment tenir compte de l'\'evolution du
jet permet de restaurer la positivit\'e des
distributions, ce qui n'est pas le cas   en ``DLA'' \cite{EvEq},
o\`u elle est n\'eglig\'ee. L'intervalle de validit\'e  de
notre calcul en MLLA est donn\'e. Nous d\'emontrons qu'il est d'autant
plus grand que l'\'energie du jet est importante.
Les interf\'erences des gluons mous,
(ph\'enom\`enes de coh\'erence en CDQ)
\'etant \'ecrant\'ees par la divergence de la constante de
couplage dans le domaine des petits $k_\perp$, nous les rendons
visibles en prenant
une valeur non-r\'ealiste de l'\'energie totale du jet qui diminue
la valeur de $\alpha_s$
et qui permet de comparer la forme de cette distribution avec celle
qui a \'et\'e pr\'edite par ``DLA''.
Dans le paragraphe \ref{subsub:comparaisonkt} nous comparons nos
pr\'edictions pour la section efficace diff\'erentielle inclusive
en fonction de $k_\perp$
avec les donn\'ees exp\'erimentales pr\'eliminaires de CDF.
L'accord entre th\'eorie
et exp\'erience, via le param\`etre ph\'enom\'enologique ${\cal K}^{ch}$
(qui normalise le nombre de partons au nombre de hadrons charg\'es),
est excellent dans l'intervalle de validit\'e ``MLLA''; ceci permet
une fois de plus, de confirmer l'hypoth\`ese ``LPHD''
(``Local Hadron Parton Duality'' en anglais) dans le cas des grandeurs
inclusives.
Nous rappelons que l'hypoth\`ese ``LPHD'' suppose que les hadrons se
comportent ``comme des partons'' et que l'on peut donc leur attribuer
les m\^emes propri\'et\'es \cite{DKT}\cite{DKT}.

Dans l'article \ref{sub:article2}  \'etudions
les corr\'elations entre deux particules dans un jet
 en fonction de leur \'energie, dans le cadre ``MLLA''.
Le premier calcul des corr\'elations a \'et\'e effectu\'e par
Fong et Webber en 1991 \cite{FW}. Ils ont obtenu une expression
analytique simple seulement dans le cas o\`u
l'\'energie des particules est proche du maximum de leur distribution
inclusive (``distorted Gaussian'' \cite{FW1} en anglais).
Elle croit lin\'eairement en fonction de la somme
$[\ln(1/x_1)+\ln(1/x_2)]$ et est quadratique en fonction de la
diff\'erence $[\ln(1/x_1)-\ln(1/x_2)]$. Elle a de plus seulement
  \'et\'e \'evalu\'ee dans la limite $Q_0=\Lambda_{QCD}$.
Dans cette th\`ese, nous r\'esolvons au contraire de fa\c con exacte
les \'equations d'\'evolution,
en utilisant la logique qui a bien r\'eussi dans la description du spectre 
inclusif; \`a savoir, nous calculons la solution exacte
d'une \'equation ``MLLA'' (donc approch\'ee) dans le domaine des
``petits $x$''.
Cette expression est \'ecrite dans \ref{sub:article2} pour les jets
de quarks et de gluons, en termes des deriv\'ees logarithmiques du
spectre, puis nous la calculons num\'eriquement dans
l'approximation du ``limiting spectrum''.
Il est ainsi d\'emontr\'e que les corr\'elations, au lieu de cro\^itre
ind\'efiniment en fonction de cette somme, comme dans \cite{FW},
s'aplatissent, puis d\'ecroissent jusqu'\`a leur valeur minimale
$1$ (particules d\'ecorr\'el\'ees);
ceci est li\'e \`a la coh\'erence des gluons mous lorsque leurs
impulsions deviennent n\'egligeables.
Nous pr\'edisons aussi  des corr\'elations plus faibles
par rapport \`a l'analyse de Fong et Webber;
les explications sont pr\'esent\'ees.
Contrairement au cas du spectre, on s'attend \`a ce que les corr\'elations
fournissent un test plus r\'ealiste de la dynamique
hadronique par rapport \`a la dynamique partonique.
Le param\`etre ${\cal K}^{ch}$ se simplifie, en effet,
dans la d\'efinition de cette observable et l'hypoth\`ese ``LPHD''
peut \^etre sujette \`a caution.
Nous comparons nos pr\'edictions avec celles de Fong-Webber, ainsi
qu'avec les donn\'ees exp\'erimentales du LEP-I \cite{OPAL1}.
Dans le paragraphe compl\'ement
\ref{subsub:corrpaper}, nous donnons des explications suppl\'ementaires.

Dans  l'article \ref{sub:article3}, nous nous 
int\'eressons \`a l'\'evaluation du spectre par la m\'ethode du col,
dans le cas g\'en\'eral $Q_0\ne\Lambda_{QCD}$. Bien que s'agissant d'une
approximation, elle se r\'ev\`ele extr\^emenent performante tout en \'etant
beaucoup plus simple et \'economique \`a mettre en {\oe}uvre.
Elle permet de faire de tr\`es bonnes pr\'edictions
sur la forme, la position du pic de la distribution et sur
les effets de coh\'erence en CDQ.
Lorsque l'on prend les limites $Q_0\to\Lambda_{QCD}$ et
$Y+\lambda\to\infty$,
on obtient un tr\`es bon accord entre cette m\'ethode et la m\'ethode
exacte  du travail pr\'ec\'edent.
Puisque les corr\'elations \`a 2 particules y ont \'et\'e exprim\'ees
en fonction des d\'eriv\'ees logarithmiques du
spectre, ce sont elles que nous  attachons ensuite
\`a \'evaluer, afin d'obtenir, ce qui n'avait pas \'et\'e possible
auparavant, des expressions \`a $\lambda$ quelconque pour ces
corr\'elations. Nous pouvons ainsi \'etudier
leur d\'ependance en $\lambda$. Le ``limiting spectrum'' $\lambda =0$
semble \^etre le plus susceptible d'un accord avec l'exp\'erience.
L'analyse de Fong et Webber a donc pu  \^etre g\'en\'eralis\'ee, et on la
retrouve bien dans les limites appropri\'ees.

\subsection{Comparaison avec les travaux pr\'ec\'edents}

Les  premiers calculs effectu\'es dans l'approximation MLLA ont concern\'e 
les multiplicit\'es hadroniques des jets en CDQ et
le spectre inclusif d'une particule, \'evalu\'es en  fonction de l'\'energie
($\ell=\ln(1/x)$) (``hump-backed plateau''). On citera ainsi \cite{EvEq},
\cite{FW1} qui traite ``l'approximation gaussienne'', \cite{DKT}, \cite{DKT}
et \cite{KO}.
En ce qui concerne le spectre, l'accord entre pr\'edictions th\'eoriques et
les donn\'ees exp\'erimentales (LEP par exemple \cite{OPAL}\cite{TASSO}) est
remarquable. L'hypoth\`ese de dualit\'e locale parton hadron
(``LPHD'' en anglais) \cite{DKT} se trouve parfaitement confort\'ee par les
donn\'ees. 

Par contre, le seul calcul concernant les distributions inclusives
en fonction de l'impulsion transverse n'a \'et\'e effectu\'e qu'en DLA.
Il est expliqu\'e en d\'etails dans le paragraphe
\ref{subsection:distdoubdiff}. C'est dans l'article \ref{sub:article1}
que ce calcul a \'et\'e pour la premi\`ere fois g\'en\'eralis\'e au cadre
MLLA.

\medskip

Pour les corr\'elations \`a deux particules dans un jet, des
pr\'edictions ont \'et\'e obtenues en DLA \cite{DLA}, et en MLLA par
Fong \& Webber \cite{FW}. Le cas DLA sera discut\'e dans le paragraphe
\ref{sec:correlations}, qui ne d\'ecrit que certains traits de cette
observable. Les calculs MLLA de Fong \& Webber \cite{FW} \cite {FW1}
 ont \'et\'e obtenus dans le cadre restreint o\`u l'\'energie des deux
particules se trouve au voisinage du maximum de leur distribution
inclusive, en cons\'equence de quoi ni l'aplatissement attendu ni la
d\'ecroissance  en fonction de la somme $[\ln(1/x_1)+\ln(1/x_2)]$.
Cel\`a sera discut\'e dans le paragraphe \ref{subsub:FWApprox}.
Dans l'article \ref{sub:article2}, par la r\'esolution exacte des
\'equations d'\'evolution MLLA, via le formalisme de la fonctionnelle
g\'en\'eratrice, nous avons pu nous affranchir de cette
restriction et donner une solution valable pour tout $x$.
Le calcul a pu \^etre men\'e \`a bien analytiquement jusqu'au bout pour les
petits $x$ et dans le cas (limiting spectrum $\lambda=0$)
o\`u le cut-off colin\'eaire est \'egal \`a $\Lambda_{QCD}$. 
Nous avons \'egalement g\'en\'eralis\'e  au cadre MLLA la repr\'esentation
int\'egrale (\ref{eq:DLAalphasrun}) \`a $\lambda\ne0$.

\medskip

L'utilisation de la m\'ethode du col a permis, 
dans l'article \ref{sub:article3}, des progr\`es suppl\'ementaires. En
effet, si cette m\'ethode constitue une approximation, elle s'est
r\'ev\`el\'ee redoutablement efficace et pr\'ecise pour le calcul des
corr\'elations. Ainsi, on a pu s'affranchir du ``limiting spectrum'' et
donner des formules analytiques pour $\lambda \not = 0$, toujours dans le
cadre MLLA, ce qui n'avait jamais \'et\'e possible auparavant.
col du paragraphe \ref{subsection:col} \`a $\lambda\ne0$.
Les r\'esultats de Fong \& Webber sont reproduits dans les limites
appropri\'ees.

Si les r\'esultats obtenus pour les corr\'elations sont en bien meilleur
accord avec les r\'esultats exp\'erimentaux existants  de LEP
\cite{OPAL}\cite{TASSO} que ceux de Fong \& Webber ou ceux obtenus aussi
par la m\'etode du col mais en DLA (\ref{eq:Specalphasrun}),
il n'en subsiste pas moins un d\'esaccord avec les donn\'ees.

Les r\'esultats \`a venir (Tevatron, LHC)  sur les corr\'elations
constitueront donc un test important des pr\'edictions de la CDQ
perturbative, et de l'hypoth\`ese de dualit\'e locale parton hadron; cette
derni\`ere peut en effet \^etre plus sujette \`a caution en ce qui concerne
cette observable moins inclusive que les distributions \'etudi\'ees dans le
premier travail de cette th\`ese. L\'eventualit\'e d'un r\^ole non
n\'egligeable des corrections next-to-MLLA n'est pas non plus \`a \'ecarter
arbitrairement.

\medskip

Nous avons \'egalement discut\'e en d\'etail,  pour toutes les observables
\'etudi\'ees, les ph\'enom\`enes de coh\'erence des gluons mous \`a petit
$k_\perp$.

%%%%%%%%%%%%%%%%%%%%%%%%%%%%%%%%%%%%%%%%%%%%%%%%%%%%%%%%%%%%%%%%%%%%%%%%%%%%

%\section{Rayonnement mou en \'electrodynamique quantique (classique). 
%Extension \`a la chromodynamique quantique}

\chapter{Rayonnement mou en \'electrodynamique quantique (classique). 
Extension \`a la chromodynamique quantique}

Ce chap\^itre a pour but de rappeler les aspects essentiels du
rayonnement mou en Electrodynamique Quantique (EDQ) ainsi
qu'en Chromodynamique Quantique (CDQ). Nous allons obtenir le
courant d'accompagnement mou (bremsstrahlung) d'une particule
charg\'ee (souvent appel\'e ``rayonnement de freinage'' dans la litt\'erature) 
dans le cadre quantique \`a partir des diagrammes de Feynman
\`a l'ordre des arbres. Apr\`es avoir d\'emontr\'e son 
universalit\'e et sa nature classique,
nous l'obtiendrons en utilisant la th\'eorie classique du
rayonnement o\`u une nouvelle m\'ethode sera expos\'ee. Son interpr\'etation 
physique permettra de comprendre la forme de la distribution trouv\'ee
pour les photons \'emis \`a l'int\'erieur d'un certain cr\'eneau de rapidit\'e.
La g\'en\'eralisation au rayonnement des gluons mous en CDQ sera
automatique en ajoutant le nouvel ingr\'edient de la th\'eorie, la couleur.
Nous comparerons les ph\'enom\`enes de coh\'erence
dans les deux th\'eories.
Le chap\^itre sera conclu par une application des r\'esultats obtenus
aux deux  canaux de production possible du boson de Higgs dans les
futurs collisionneurs.

\section{Courant d'accompagnement mou d'une 
particule charg\'ee; m\'ethode quantique}
\label{subsection:CAM}

Nous consid\'erons un photon de bremsstrahlung (photon mou)
\'emis par une particule charg\'ee
(\'electron) sous l'action d'un champ externe 
(exemple: un champ \'electrostatique).

\begin{figure}[h]
\begin{center}
\includegraphics[height=7truecm,width=1.2\tw]{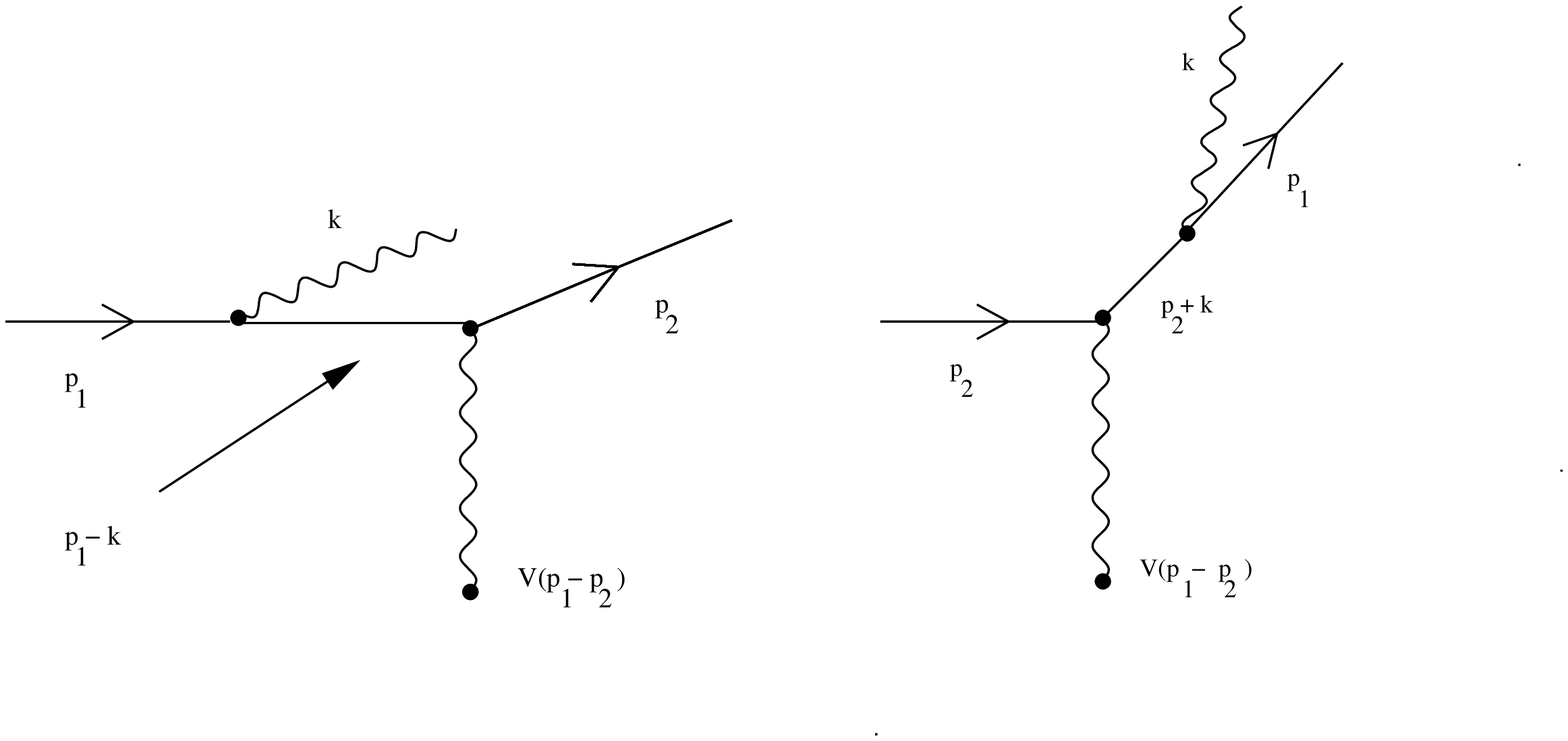}
\caption{\label{fig:bremsst} 
Diagrammes de Feynman d'un photon de bremsstrahlung \'emis sous l'action 
d'un champ externe.}
\end{center}
\end{figure}

Les diagrammes de Feynman \`a l'ordre des arbres sont donn\'ees dans la 
Fig.\ref{fig:bremsst} \cite{{pQCDforbeginners}}.
 $p_1\!=\!(p_1^0,\vec{p_1})$,
$p_2\!=\!(p_2^0,\vec{p_2})$ repr\'esentent les 
quadri-impulsions de l'\'electron entrant et sortant 
respectivement, et $k\!=\!(k^0,\vec{k})$ celle du photon r\'eel 
\footnote{Il s'agit d'un photon qui aurait d\^u \^etre r\'eabsorb\'e 
par l'\'electron
(photon virtuel) si celui-ci n'avait pas \'et\'e d\'evi\'e de sa 
trajectoire initiale.}
\'emis lors du processus.  En appliquant les r\`egles de
Feynman, les amplitudes correspondantes s'\'ecrivent sous la forme
\cite{Peskin} \cite{IZuber}:

\beq
M_i^{\mu}=e\,\bar{u}(p_2,s_2)V(p_2+k-p_1)\frac{m+p\!\!\!/_1-k\!\!\!/}
{m^2-(p_1-k)^2}\gamma^{\mu}u(p_1,s_1)
\eeq
lorsque l'\'electron est \'emis avant la diffusion (diagramme \`a gauche), et
\beq
M_f^{\mu}=e\,\bar{u}(p_2,s_2)\gamma^{\mu}\frac{m+p\!\!\!/_2+k\!\!\!/}
{m^2-(p_2+k)^2}V(p_2+k-p_1)u(p_1,s_1)
\eeq
lorsqu'il est a \'et\'e \'emis apr\`es (diagramme \`a droite).
$s_{1,2}$ \'etiqu\`etent l'\'etat de spin de l'\'electron,
$V$ l'amplitude d'interaction qui, en g\'en\'eral, d\'epend de
l'impulsion transf\'er\'ee 
(dans le cas de la diffusion sur un champ e.m. $V=\gamma_0$).
L'amplitude totale est donn\'ee par la somme des amplitudes

\beq
\nonumber
M^{\mu}=M_i^{\mu}+M_f^{\mu}.
\eeq

Nous utilisons l'approximation du photon mou \cite{Peskin}\cite{IZuber}
(on prend son \'energie tr\`es inf\'erieure \`a celle de
l'\'electron qui l'a \'emis et on n\'eglige le recul de celui-ci), 
$\omega\ll p_1^0,p_2^0$, on n\'eglige les termes en 
$k\!\!\!/$ au num\'erateur,
on utilise l'astuce de Dirac $p\!\!\!/\gamma^{\mu}\!=
\!-\gamma^{\mu}p\!\!\!/\!+\!2p^{\mu}$, 
ainsi que les \'equations suivantes pour des fermions 
(\'electrons) sur couche de masse:
\beeq\label{eq:mi}
(m+p\!\!\!/_1)\gamma^{\mu}u(p_1)=(2p_1^{\mu}+\gamma^{\mu}[m-p\!\!\!/_1])u(p_1)
=2p_1^{\mu}u(p_1),\\
\label{eq:mf}
\bar{u}(p_2)\gamma^{\nu}(m+p\!\!\!/_2)=\bar{u}(p_2)([m-p\!\!\!/_2]
\gamma^{\nu}+2p_2^{\nu})=\bar{u}(p_2)2p_2^{\nu}.
\eeeq
(\ref{eq:mi}) (\ref{eq:mf}) seront utilis\'ees pour simplifier $M_i^{\mu}$ et 
$M_f^{\mu}$ respectivement. Nous simplifions les d\'enominateurs en utilisant 
l'approximation ($p_i^2\approx m^2$) pour des \'electrons 
de tr\`es faible virtualit\'e et $k^2=0$ pour des photons r\'eels. Nous avons
\beeq\nonumber
m^2-(p_1-k)^2\approx 2(p_1.k),\\
\nonumber
m^2-(p_2+k)^2\approx -2(p_2.k).
\eeeq

L'amplitude totale $M^{\mu}$ est donc donn\'ee par l'expression suivante
\cite{pQCDforbeginners}

\beq\label{eq:Ma}
M^{\mu}=e\,j^{\mu}\times M_{el}.
\eeq

Ici $M_{el}$ repr\'esente l'\'el\'ement de matrice de Born de la partie 
non-radiative (\'elastique) de la diffusion, 

$$
M_{el}=\bar{u}(p_2,s_2)V(p_2-p_1)u(p_1,s_1)
$$

dans lequel le recul de l'\'electron a \'et\'e n\'eglig\'e
\footnote{Ceci entra\^\i ne une mauvaise estimation des observables que
l'on peut mesurer dans certains processus (voir chap\^itre \ref{sec:ADL}).}
, $q=p_2+k-p_1\simeq p_2-p_1$,
donc le principe de conservation de l'\'energie  viol\'e. 
Le courant de rayonnement mou associ\'e \`a une particule 
charg\'ee $j^{\mu}$ est donc donn\'e par l'expression

\beq\label{eq:courant}
j^{\mu}=\frac{p_1^{\mu}}{(p_1\cdot k)}-\frac{p_2^{\mu}}{(p_2\cdot k)}.
\eeq 

La factorisation de l'amplitude de diffusion 
(\ref{eq:Ma}) est tout \`a fait naturelle. 
Le quadri-courant $j^{\mu}$ ne d\'epend pas de la 
nature des particules charg\'ees, en particulier, 
du spin. Elle d\'epend des impulsions 
($p_1, p_2, k$) et de la charge (e) des particules 
entrante et sortante. Ceci a \'et\'e d\'emontr\'e 
dans les travaux de Low \cite{Low} pour des bosons charg\'es, et
g\'en\'eralis\'ee par 
Burnett et Kroll \cite{BandK} pour le cas des fermions.

Le courant (\ref{eq:courant}) est essentiellement classique. 
Il peut d'ailleurs \^etre obtenu
dans le cadre de la th\'eorie classique du rayonnement
\cite{Jackson}\cite{alain} en 
consid\'erant le potentiel induit
par une particule charg\'ee lorsqu'elle est soudainement 
d\'evi\'ee de sa trajectoire 
(acc\'el\'er\'ee) sous l'action d'un champ externe, comme je le montre
maintenant.

\section{Consid\'erations classiques sur le rayonnement;
acc\'el\'eration instantan\'ee infinie}

\begin{figure}[h]
\begin{center}
\includegraphics[height=6truecm,width=0.8\tw]{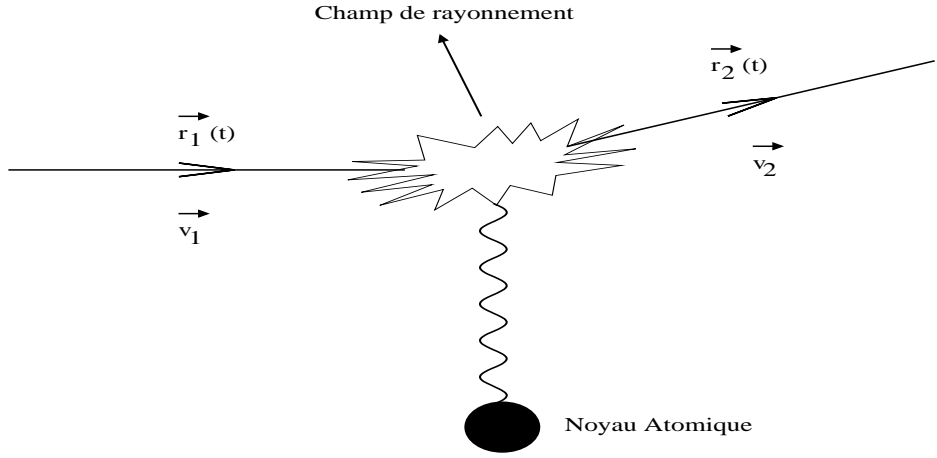}
\caption{\label{fig:rayonnement} 
Champ de rayonnement induit par une charge 
soudainement acc\'el\'er\'ee.}
\end{center}
\end{figure}

Dans la th\'eorie classique du rayonnement il est connu
qu'une charge acc\'el\'er\'ee
cr\'ee un champ comportant, outre une contribution de
type coulombien ($\propto 1/R^2$), une contribution de type rayonnement
($\propto 1/R$) \cite{Jackson}\cite{alain}.
Nous allons estimer le champ
induit par une particule charg\'ee lorsqu'elle subit une
d\'eviation instantan\'ee (choc)
\`a  un instant $t_0$. On prend $e=1$ pour simplifier.
Le courant \'electromagn\'etique
dans ce cas est donn\'e par les deux termes suivants

\begin{equation}\label{eq:acint}
\begin{array}{c}
 \vec{j}=\vec{j_1}+\vec{j_2} \end{array} \left\{
\begin{array}{c}
        \vec{j_1}=\vec{v}_1\,\delta^3(\vec{r}-\vec{v}_1t)\vartheta(t_0-t), \cr
        \vec{j_2}=\vec{v}_2\,\delta^3(\vec{r}-\vec{v}_2t)\vartheta(t-t_0),
 \end{array}
\right.
\end{equation}

o\`u $\vec{v}_{1,(2)}$ est la vitesse initiale (finale) de la particule
 lorsqu'elle se 
d\'eplace le long de sa trajectoire classique $\vec{r}=\vec{v}_it$ 
(en m\'ecanique classique nous pouvons parler de trajectoire,
voir Fig.\,\ref{fig:rayonnement}). 
$\vartheta$ repr\'esente
ici la fonction de Heaviside. Nous restaurons la covariance
de Lorentz en ajoutant \`a
(\ref{eq:acint}) la composante $j^0$
qui correspond \`a la densit\'e de charge du quadri-vecteur, puis
on d\'efinit $j^{\mu}$ d'apr\`es

\beq\label{eq:courant2}
j_i^{\mu}(t,\vec{r})=\left(j^0_i(t,\vec{r}),
\vec{j}_i(t,\vec{r})\right)\equiv v_i^{\mu}j^0_i,
\eeq

o\`u la quadri-vitesse s'exprime $v_i^{\mu}=(1,\vec{v}_i)$.

L'amplitude d'\'emission du champ de quadri-impulsion 
$(k_0=\omega,\vec{k})$ est
proportionnelle \`a la transform\'ee de Fourier du
courant \'electrique total:

\beq\nonumber
j_i^{\mu}(k)=\int_{-\infty}^{+\infty}dt\int d^3\vec{r}\,e^{ix^{\nu}k_{\nu}}\,j_i^{\mu}(t,\vec{r}),
\eeq
dont on d\'eterminera les deux termes correspondants
\`a la trajectoire de la Fig.\,\ref{fig:rayonnement}:

\beeq\label{eq:j1}
j_1^{\mu}(k)&=&\int_{-\infty}^{+\infty}dt\int d^3\vec{r}\,e^{ix^{\nu}k_{\nu}}\,j_1^{\mu}(t,\vec{r})
=v_1^{\mu}\int_{-\infty}^0d\tau\,e^{ik_0(t_0+\tau)-i(\vec{k}\cdot\vec{v}_1)\tau}\\
&=&\frac{-iv_1^{\mu}e^{ik^0t_0}}{k^0-(\vec{k}\cdot\vec{v}_1)}\label{eq:j1bis},
\eeeq

\beeq\label{eq:j2}
j_2^{\mu}(k)&=&\int_{-\infty}^{+\infty}dt\int d^3\vec{r}\,e^{ix^{\nu}k_{\nu}}\,j_2^{\mu}(t,\vec{r})
=v_2^{\mu}\int_0^{+\infty}d\tau\,e^{ik_0(t_0+\tau)-i(\vec{k}\cdot\vec{v}_2)\tau}\\
&=&\frac{iv_2^{\mu}e^{ik^0t_0}}{k^0-(\vec{k}\cdot\vec{v}_2)}\label{eq:j2bis}.
\eeeq

% \subsubsection{Calcul d\'etaill\'e de (\ref{eq:j1}) et (\ref{eq:j2})}

\begin{figure}[h]
\begin{center}
\includegraphics[height=5truecm,width=0.41\tw]{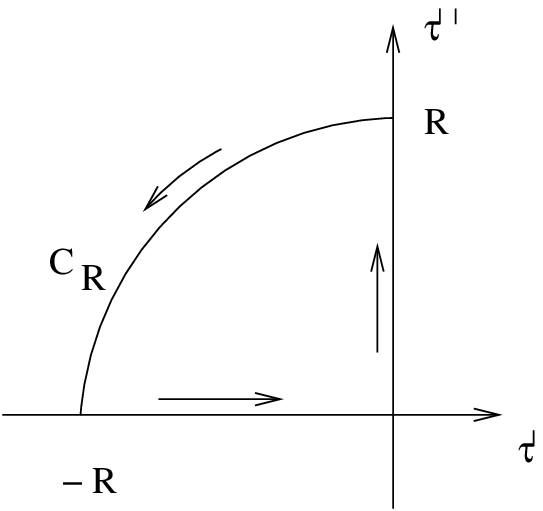}
\hfill
\includegraphics[height=5truecm,width=0.41\tw]{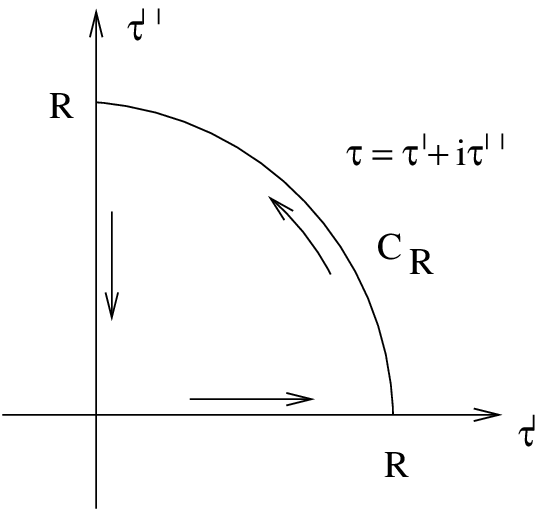}
\caption{\label{fig:contours} Contour d'int\'egration choisi pour \'evaluer 
(\ref{eq:j1}) (figure gauche) et (\ref{eq:j2}) (figure droite).}
\end{center}
\end{figure}

\subsection{Exemple: calcul de (\ref{eq:j1})}

\beq
I_1=\int_{-\infty}^{0}d\tau\,e^{i\left[k^0-(\vec{k}\cdot\vec{v}_1)\right]\tau}.
\eeq

On choisit le contour de la Fig.\,\ref{fig:contours} \`a gauche, 
on effectue le prolongement analytique $\tau=\tau'+i\tau''$
et on applique le th\'eor\`eme de Cauchy \cite{Shabat}

\beq\nonumber
\int_Cd\tau\,e^{i\left[k^0-(\vec{k}\cdot\vec{v}_1)\right]\tau}=\int_{-R}^{0}d\tau'
e^{i\left[k^0-(\vec{k}\cdot\vec{v}_1)\right]\tau'}+i\int_{0}^{R}d\tau''
e^{-\left[k^0-(\vec{k}\cdot\vec{v}_1)\right]\tau''}+
\int_{C_R}d\tau\,e^{i\left[k^0-(\vec{k}\cdot\vec{v}_1)\right]\tau}=0.
\eeq
On passe maintenant \`a la limite $R\!\rightarrow\!\infty$.
En vertu du lemme de Jordan \cite{Shabat}
l'int\'egrale sur le contour du cercle s'annule et on obtient:

\beq
I_1=\lim_{R\rightarrow\infty}\int_{-R}^{0}d\tau'
e^{i\left[k^0-(\vec{k}\cdot\vec{v}_1)\right]\tau'}=
\lim_{R\rightarrow\infty}-i\int_{0}^{R}d\tau''
e^{-\left[k^0-(\vec{k}\cdot\vec{v}_1)\right]\tau''}
=-\frac{i}{k^0-(\vec{k}\cdot\vec{v}_1)},
\eeq

puis $j^{\mu}_1(k)=v^{\mu}_1e^{i\omega t_0}I_1(k)$, d'o\`u 
on d\'eduit le r\'esultat
(\ref{eq:j1bis}). Le calcul pour (\ref{eq:j2}) se fait 
de fa\c con analogue en choisissant le contour 
\`a droite de la Fig.\,\ref{fig:contours} pour obtenir (\ref{eq:j2bis}).

La solution de l'\'equation de Maxwell \cite{Jackson} pour le quadri-potentiel 
induit par le quadri-courant
(\ref{eq:courant2}) s'\'ecrit sous la forme \cite{Jackson}:

\beeq\nonumber
A^{\mu}(x)&=&\int\frac{d^4k}{(2\pi)^4}\,
e^{-ix^{\mu}k_{\mu}}[-2\pi i\delta(k^2)] 
j^{\mu}(k)\\
&=&\int\frac{d^3k}{2\omega(2\pi)^3}\,
e^{-i\omega x^0+i(\vec{k}\cdot\vec{x})} A^{\mu}(k),
\eeeq
o\`u
% \beq\nonumber
$$
A^{\mu}(k)=A_2^{\mu}(k)-A_1^{\mu}(k);\\
$$
\beq\label{eq:champ}
A_i^{\mu}(k)=\frac{v_i^{\mu}e^{i\omega t_0}}
{\omega(1-v_i\cos\Theta_i)};\quad
\omega=\mid\!\vec{k}\!\mid, \quad 
(\vec{k}\cdot\vec{v}_i)\equiv\omega v_i\cos\Theta_i.
\eeq
Ici $\Theta_i$ repr\'esente l'angle d'\'emission 
entre l'impulsion du photon $\vec{k}$ et la direction 
du mouvement de la charge. Si l'on r\'ecrit (\ref{eq:champ}) 
sous la forme covariante \cite{pQCDforbeginners}
$$
\frac{v_i^{\mu}}{\omega-(\vec{k}\cdot\vec{v}_i)}=\frac{E_iv_i^{\mu}}
{E_i(\omega-(\vec{k}\cdot\vec{v}_i))}=\frac{p_i^{\mu}}{(p_i\cdot k)},
$$ 
$$
A_i^{\mu}(k)=\exp{(i\omega t_0)}\left(\frac{p_2^{\mu}}{(p_2\cdot k)}
-\frac{p_1^{\mu}}{(p_1\cdot k)}\right),
$$
on remarque que le quadri-potentiel classique est identique \`a 
celui qui a \'et\'e obtenu de fa\c con quantique
(\ref{eq:courant}) \`a une phase pr\`es $\exp{(i\omega t_0)}$.
Cette derni\`ere n'intervient pas
dans le calcul des sections efficaces lorsque 
l'on prend le module de l'amplitude au carr\'e.

L'analyse classique conduit au r\'esultat (\ref{eq:courant}) qui
a \'et\'e obtenu
au paragraphe pr\'ec\'edent \ref{subsection:CAM} dans la limite du photon mou.
Ce r\'esultat est naturel car, dans
cette approximation (recul de l'\'electron n\'egligeable),
il est l\'egitime de consid\'erer
que les charges se d\'eplacent sur des trajectoires classiques.

L'origine physique du courant (\ref{eq:courant}) peut \^etre 
interpret\'ee autrement.  Le champ d'une charge
ponctuelle qui se d\'eplace, par exemple, \`a vitesse constante,
est de nature convective 
(il accompagne la particule). Lorsque la charge est soudainement 
acc\'el\'er\'ee, le nouveau champ ne peut pas la suivre instantan\'ement.
La nouvelle configuration non-stationnaire
rayonne cet exc\`es d'\'energie ($\propto j^{\mu}j_{\mu}$) 
jusqu'\`a ce que la nouvelle configuration stationnaire du champ soit atteinte.

\subsection{Densit\'e ${\cal N}$ du nombre de photons rayonn\'es}
\label{subsubsection:Sray}

\begin{figure}[h]
\begin{center}
\includegraphics[height=4truecm,width=0.4\tw]{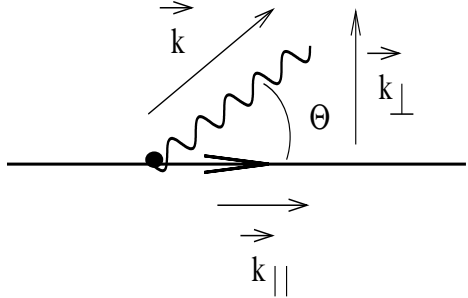}
\caption{\label{fig:photon} \'Emission d'un photon 
de quadri-impulsion $k=(\omega,\vec{k})$
par une particule charg\'ee, 
$\vec{k}_{\parallel}$ est l'impulsion longitudinale du photon et 
$\vec{k}_{\perp}$ son impulsion transverse .}
\end{center}
\end{figure}

\begin{figure}[h]
\begin{center}
\includegraphics[height=5truecm,width=0.55\tw]{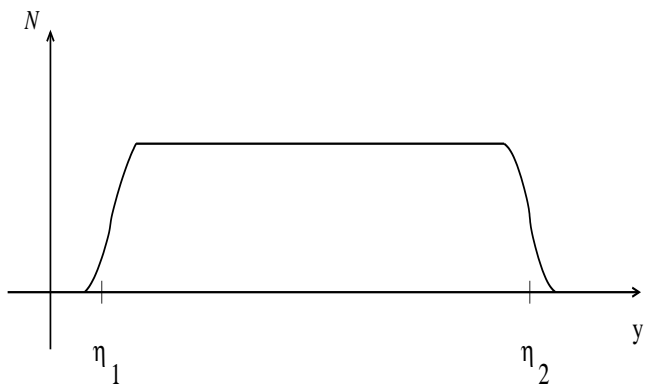}
\caption{\label{fig:plateau1}
Densit\'e du nombre de photons rayonn\'es \`a $k_{\perp}$
fix\'e en fonction de la rapidit\'e $y$ du photon.}
\end{center}
\end{figure}

On s'int\'eresse maintenant \`a l'\'evaluation de la densit\'e du nombre de 
photons rayonn\'es dans le cas d'une l'acc\'el\'eration 
instantan\'ee choisie le long de l'axe $x\equiv x_1$.
L'\'energie totale rayonn\'ee
dans le volume diff\'erentiel de l'espace de phase $d^3k$ 
est donn\'ee par l'expression\cite{IZuber}
\beq\label{eq:spectre}
dE\!=\!\frac{d^3k}{2(2\pi)^3}[-j^{\mu}(k)j^*_{\mu}(k)], 
\quad \omega\!=\!\mid\!\vec{k}\!\mid,
\quad d^3k\!=\!k_{\perp}\cosh y\,dy\,d^2k_{\perp}\!=
\!k_{\perp}^2\cosh y\,dy\,dk_{\perp}
d\phi,
\eeq
o\`u l'angle azimutal $0\leq\phi\leq2\pi$.
Nous avons introduit l'impulsion 
transverse $k_{\perp}$ du photon ainsi que sa rapidit\'e $y$
$$
k^{\mu}=(\omega=k_{\perp}\cosh y, k_{\perp}\sinh y, \vec{k}_{\perp} ).
$$
$y$ est li\'ee \`a l'angle z\'enithal $\Theta$ par l'expression suivante
(voir Fig.\,\ref{fig:photon}):
$$
\tan\Theta=\frac1{\sinh y}\Leftrightarrow 
y=\ln\frac1{\tan{\Theta/2}}.
$$
Elle montre qu'aux grands angles d'\'emission des photons correspondent
des petites rapidit\'es et {\em vice versa}. En introduisant la variable de
rapidit\'e de la charge $(\eta)$ selon
$$
v^{\mu}=(\cosh\eta,\sinh\eta,\vec{0}),
$$
on a
$$
v_i.k=k_{\perp}\cosh(\eta_i-y), \qquad j^{\mu}j^*_{\mu}\!=\!- \frac1{k^2_{\perp}}\left[\tanh(\eta_2-y)-\tanh(\eta_1-y)\right]^2.
$$
On d\'efinit la densit\'e (${\cal N }$) du nombre de photons rayonn\'es
($dE/\omega$) par unit\'e de volume ($d^3k/\omega$)
en fonction de $k_{\perp}$ et $y$ \cite{Niedermayer}:
\beq\label{eq:spectre1}
{\cal {N}} (y,k_{\perp})\!\equiv\displaystyle{\frac{\frac{dE}{\omega}}{\frac{d^3k}{\omega}}}\!=
\!\frac{dE}{k_{\perp}d^2k_{\perp}dy\cosh y}=\frac1{2(2\pi)^3}\left\{\frac1
{k^2_{\perp}}\left[\tanh(\eta_2-y)-\tanh(\eta_1-y)\right]^2\right\}.
\eeq
A $k_{\perp}$ fix\'e l'allure de (\ref{eq:spectre1}) est donn\'ee
par un plateau qui s'\'etend entre la rapidit\'e initiale ($\eta_1$) et
finale ($\eta_2$) de la charge acc\'el\'er\'ee, puis
s'annule exponentiellement au-del\`a (voir Fig.\ref{fig:plateau1}). 
Ce plateau est bien connu en \'electrodynamique classique sous le nom de
c\^one de rayonnement en $\Theta$ \cite{Jackson}. 
La d\'ependance de $\cal {N}$ en $1/k_{\perp}^2$ est 
remarquable (voir chap\^itres \ref{sec:ADL}, \ref{sec:MLLA},
les articles \ref{sub:article1}, \ref{sub:article2} et \ref{sub:article3}
qui font l'objet de cette th\`ese).
La densit\'e du nombre total de photons rayonn\'es \`a rapidit\'e $y$ fix\'ee 
\beq
\widetilde{{\cal {N}}}(y)\equiv
\int\frac{dE}{k_{\perp}d^2k_{\perp}dy\cosh y}d^2k_{\perp}
=\int {\cal {N}}(y,k_{\perp})d^2k_{\perp}\propto\ln [k_{\perp}]
\eeq

s'av\`ere logarithmiquement divergente en $0$ (catastrophe infrarouge)
et \`a l'infini (catastrophe ultraviolette)
pour le cas de l'acc\'el\'eration instantan\'ee. 
Cette derni\`ere se justifie puisque l'acc\'el\'eration est infinie
\`a $t=t_0$.  Nous allons donc r\'egulariser cette divergence
ultraviolette en consid\'erant le cas d'une 
acc\'el\'eration finie sur un intervalle de temps fini.

\medskip

\textbf{Remarque:} $\widetilde{\cal {N}}$ est bien
logarithmiquement divergente car on
int\`egre sur $\frac{dk_{\perp}}{k_{\perp}}$. 
On trouve cette divergence logarithmique dans
les th\'eories de jauge o\`u le m\'ediateur des interactions
(photons en EDQ, gluons en CDQ) est de masse nulle \cite{Peskin}\cite{IZuber}.

\section{Acc\'el\'eration finie: deux cas simples}

Dans ce paragraphe nous poursuivons les raisonnements
pr\'ec\'edents dans les cas les plus simples
d'une acc\'el\'eration finie \cite{Niedermayer}.
En particulier, nous consid\'erons le mouvement rectiligne
\`a acc\'el\'eration propre constante de la charge.

%\subsubsection{$\boldsymbol{1^{\text{er}}}$ cas: trajectoire rectiligne}
\subsection{Un premier cas de trajectoire rectiligne: plateau de largeur
infinie}
\label{subsubsection:1emeaccpfinie}

\begin{figure}[h]
\begin{center}
\includegraphics[height=5truecm,width=0.55\tw]{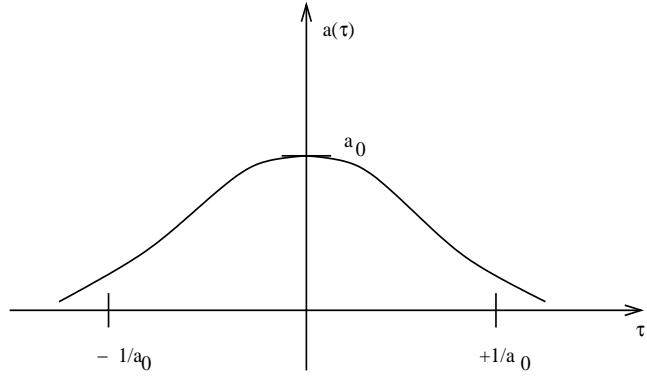}
\caption{\label{fig:accfini} Acc\'el\'eration finie 
transmise dans un intervalle de temps fini.}
\end{center}
\end{figure}

Soit la ligne d'univers donn\'ee par \cite{Niedermayer}

\beq\label{eq:lu1}
x^{\mu}(\tau)=\frac1{a_0}(\sinh a_0\tau,\cosh a_0\tau,\vec{0}), \quad 
-\infty<\tau<+\infty.
\eeq
o\`u $\tau$ est le temps propre, $a_0$ l'acc\'el\'eration propre, d\'efinis 
dans le r\'ef\'erentiel de la charge. (\ref{eq:lu1}) satisfait
$$
t^2-x^2=-\frac1{a_0^2},\qquad d\tau^2=dx_0^2-d\vec{x}^2
$$
(\ref{eq:lu1}) correspond au cas
d'une acc\'el\'eration finie transmise \`a une charge ultra-relativiste
($v=c=1$) dans
un intervalle infini de rapidit\'e, $\eta_1=-\infty$, $\eta_2=+\infty$. 
La vitesse de la charge s'\'ecrit sous la forme
$$
v(\tau)=\frac{dx^1(\tau)}{dx^0(\tau)}=\tanh a_0\tau,
$$
et son acc\'el\'eration
$$
a(\tau)=\frac{dv(\tau)}{dx^0(\tau)}=\frac{a_0}{\cosh^3(a_0\tau)}
$$
dont on trace l'allure dans la Fig.\ref{fig:accfini}.
On en d\'eduit que la charge subit, dans le r\'ef\'erentiel du laboratoire,
une acc\'el\'eration variable pendant l'intervalle de
temps $\simeq 1/a_0$ dont la valeur maximale est donn\'ee par
son acc\'el\'eration propre $a_0$ (constante dans le
ref\'erentiel de la charge) \`a $\tau=0$. On s'int\'eresse \`a
la transform\'ee de Fourier du courant rayonn\'e
\cite{Jackson}\cite{Niedermayer}

\beq\label{eq:courant3}
j^{\mu}(k)=\int d\tau\frac{dx^{\mu}}{d\tau}\,e^{ikx(\tau)}.
\eeq

En effectuant le produit scalaire $k^{\mu}x_{\mu}$,
l'exposant dans (\ref{eq:courant3}) prend la forme

$$
kx(\tau)= \frac{k_{\perp}}{a_0}\sinh(a_0\tau-y).
$$

Il s'av\`ere int\'eressant d'effectuer un boost sur le 
r\'ef\'erentiel du photon rayonn\'e ($K_y$)
qui se d\'eplace avec la rapidit\'e $y$. Dans $K_y$, 
le photon consid\'er\'e ne poss\`ede pas 
d'impulsion longitudinale ($k_{\parallel}=0$). 
Les nouvelles coordonn\'ees s'\'ecrivent en 
fonction des anciennes selon:
\beeq
\widetilde{x}^0(\tau)=
\left[\cosh y\,x^0(\tau)-\sinh y\,x^1(\tau)\right]=
\frac1{a_0}\sinh(a_0\tau-y),\cr
\widetilde{x}^1(\tau)=\left[-\sinh y\,x^0(\tau)+
\cosh y\,x^1(\tau)\right]=\frac1{a_0}\cosh(a_0\tau-y).\label{eq:boost}
\eeeq

Le courant (\ref{eq:courant3}) prend donc la forme simple
\beeq\nonumber
\widetilde{j}^0(k)&=&\int_{-\infty}^{+\infty}  
d\tau\cosh(a_0\tau-y)\exp\left[i\frac{k_{\perp}}{a_0}
\sinh(a_0\tau-y)\right]\\
&=&\frac1{k_{\perp}}\int_{-\infty}^{+\infty}d\tau
\frac{d}{d\tau}\left\{\exp\left[i\frac{k_{\perp}}{a_0}\sinh(a_0\tau-y)
\right]\right\}=0,\nonumber
\eeeq

\beq
\widetilde{j}^1(k)=\int_{-\infty}^{+\infty}d\tau\sinh(a_0\tau-y)
\exp\left[i\frac{k_{\perp}}{a_0}\sinh(a_0\tau-y)\right]=
i\frac2{k_{\perp}}[uK_1(u)],
\eeq

o\`u $u=\displaystyle{\frac{k_{\perp}}{a_0}}$ et 
$K_1(u)$ est ici la fonction de Macdonald (Bessel modifi\'ee)
d'ordre $1$. Son comportement asymptotique est donn\'e par

\begin{equation*}
 \left[uK_1\left(u\right)\right]^2=
\left\{ \begin{array}{cll}
                             1 & \mbox{pour} & u\ll1; \\
         u\pi\frac{e^{-2u}}{2} & \mbox{pour} & u\gg1.
        \end{array}\right.
\end{equation*}

Nous rappelons la repr\'esentation int\'egrale 
de la fonction de Macdonald sachant que

$$
\cosh\xi=-i\sinh(\xi+i\frac{\pi}2),
$$

$$
K_{\nu}(x)=\frac12\int_{-\infty}^{+\infty}
e^{-x\cosh\xi-\nu\xi}d\xi, \quad
K_0(x)=\frac12\int_{-\infty}^{+\infty}e^{-x\cosh\xi}d\xi,
$$
puis on prend la d\'eriv\'e de $K_0$ par rapport \`a $x$
$$
K_1(x)=-\frac{d}{dx}K_0(x)=\frac12
\int_{-\infty}^{+\infty}\cosh\xi\, e^{-x\cosh\xi}d\xi.
$$
Pour $u\ll1;\,\, k_{\perp}\ll a_0$, la divergence
que l'on trouvait dans le cas d'une l'acc\'el\'eration
infinie au paragraphe pr\'ec\'edent \ref{subsubsection:1emeaccpfinie}
est remplac\'ee par un
``cut-off'' sup\'erieur sur les fr\'equences, $k_{\perp}\approx a_0$.
A $y$ et $k_{\perp}$ fix\'es 
on obtient un plateau de hauteur $\ln[a_0]$ sur un intervalle
infini de rapidit\'e.
Dans ce cas, la densit\'e du nombre total de photons rayonn\'es \`a
rapidit\'e $y$ fix\'ee se calcule sans difficult\'e
(voir cas du mouvement rectiligne pour une \'equation
horaire quelconque en \cite{Niedermayer})
$$
\widetilde{{\cal {N}}}(y)=\int_{}^{a_0}{\cal {N}}
(y,k_{\perp})d^2k_{\perp}\propto \ln[a_0].
$$
Le spectre est ainsi domin\'e par la contribution logarithmique $\ln[a_0]$.
La nature de ce ``cut-off'' peut \^etre \'elucid\'ee
\`a partir des champs intervenant dans le
probl\`eme. Le champ total est donn\'e par la
superposition du champ coulombien entra\^\i n\'e
et du champ de rayonnement cr\'e\'e par l'acc\'el\'eration
de la charge. Son expression
exacte, n'\'etant pas importante dans ce probl\`eme,
nous nous contentons d'en donner le
comportement en fonction de la distance \`a laquelle
on observe le processus
$$
\vec{E}=\vec{E}_{\text{coul}}\left(\propto\frac1{R^2}\right)+
\vec{E}_{\text{ray}}
\left(\propto\frac1{R}\right).
$$ 

Le champ coulombien domine \`a petites distances (petit $R$),
tandis que que le champ
de rayonnement l'emporte \`a grandes distances (grand $R$).
Or, la seule distance
qui intervient dans le probl\`eme est donn\'ee par la quantit\'e ``$a_0^{-1}$''.
Ainsi, pour
$R> a_0^{-1}$, le champ de rayonnement l'emporte sur le champ coulombien.
Ceci \'equivaut \`a supposer que, dans cette r\'egion, on trouve des photons
 dont la longueur d'onde transverse $\lambda_{\perp}$ est sup\'erieure
\`a $a_0^{-1}$, $R\sim\lambda_{\perp}
\sim 1/k_{\perp}>a_0^{-1}\Rightarrow k_{\perp}<a_0$.

On peut \'egalement induire que les composantes de
 Fourier de plus hautes fr\'equences ($k_{\perp}>a_0$) suivent le
mouvement de la charge, tandis que celles de plus basses fr\'equences
 $k_{\perp}<a_0$
se sont \'ecart\'ees de celle-ci et continuent leur mouvement
le long de sa trajectoire.  Ceci entra\^\i ne l'apparition d'un
 c\^one de rayonnement (c\^one de bremsstrahlung) le long de la
 trajectoire rectiligne.

%\subsubsection{$\boldsymbol{2^{\text{\`eme}}}$ cas: trajectoire rectiligne}
\subsection{Un deuxi\`eme cas de trajectoire rectiligne: plateau de
largeur finie}
\label{subsubsection:4emeaccpfinie}

Nous rempla\c cons la ligne d'univers (\ref{eq:lu1}) par

\begin{equation}
x^\mu\left(\tau'\right) = \frac{1}{a_0}\left[\sinh a_0\tau', \frac{1}{2}\left(v_2e^{a_0\tau'}+v_1e^{-a_0\tau'}\right),\vec 0\right]
\label{eq:red21},\quad -\infty<\tau'<+\infty,
\end{equation}

o\`u l'on a introduit les param\`etres
suppl\'ementaires $v_1$ et $v_2$ qui la diff\'erencient de (\ref{eq:lu1}).
On peut v\'erifier par analyse dimensionnelle que $v_1$ et $v_2$ ont
la dimension d'une vitesse. Ici, $\tau'$ n'a plus le sens physique d'un temps
 propre car $dx_{\mu}dx^{\mu}\neq d\tau'^2$.

% Cette modification de la ligne d'univers
(\ref{eq:red21}) permet de r\'etr\'ecir le plateau de la densit\'e spectrale
des photons \'emis \`a  un intervalle fini de rapidit\'e, dont les bornes sont
donn\'ees par
$$
\eta_1=-\frac12\ln\left(\frac{1+v_1}{1-v_1}\right), \quad  \eta_2=+\frac12\ln\left(\frac{1+v_2}{1-v_2}\right); \quad
$$
si l'on pose $v_1=v_2=1$, on retrouve, comme pour (\ref{eq:lu1}),
un intervalle infini de rapidit\'e $\Rightarrow$
$\eta_1\rightarrow-\infty,\quad\eta_2\rightarrow+\infty$.

On s'int\'eresse \'egalement au boost (\ref{eq:boost}) sur le r\'ef\'erentiel
o\`u la composante longitudinale de l'impulsion $k$ de la particule est nulle 
($k_{\parallel}=0$); l'expression
de la composante $j^0$ du courant s'\'ecrit alors, dans le r\'ef\'erentiel
$K_y$, sous la forme
\begin{eqnarray*}
&&\widetilde{j}^0\left(k\right)
=\int\,d\tau'\left[\cosh{y}\cosh{a_0\tau'}
-\frac{1}{2}\sinh{y}\left(v_2
e^{a_0\tau'}-v_1e^{-a_0\tau'}\right)\right]\cr
\label{eq:red22}
&&\hskip2cm\exp\left\{i\frac{k_{\perp}}{a_0}
\left[\cosh{y}\sinh{a_0\tau'}
-\frac{1}{2}\sinh{y}
\left(v_2e^{a_0\tau'}+v_1e^{-a_0\tau'}\right)\right]\right\}
\label{eq:red23}
\end{eqnarray*}
qui peut se r\'ecrire plus simplement comme
\begin{equation*}
\widetilde{j}^0\left(k\right)=\frac1{k_{\perp}}
\int\,d\tau'\frac{d}{d\tau'}
\exp\left\{i\frac{k_{\perp}}{a_0}\left[\cosh{y}
\sinh{a_0\tau'}-\frac{1}{2}\sinh{y}\left(v_2e^{a_0\tau'}
+v_1e^{-a_0\tau'}\right)\right]\right\}=0;\\
\end{equation*}
nous donnons de m\^eme l'expression de sa composante spatiale
\begin{eqnarray}
&&\widetilde{j}^1\left(k\right)=\int\,d\tau'\left[\frac{1}{2}
\cosh{y}\left(v_2e^{a_0\tau'}-v_1e^{-a_0\tau'}
\right)-\sinh{y}\cosh{a_0\tau'}\right]
\label{eq:courant1}
\cr&&\hskip1cm\exp\left\{i\frac{k_{\perp}}{a_0}\left[\cosh{y}
\sinh{a_0\tau'}-\frac{1}{2}\sinh{y}\left(v_2e^{a_0\tau'}
+v_1e^{-a_0\tau'}\right)\right]\right\}.
\end{eqnarray}
Le calcul de (\ref{eq:courant1}) est d\'etaill\'e dans l'appendice 
\ref{subsection:4emeacceppropefinie}
\begin{equation*}
\begin{split}
\widetilde{j}^1\left(k\right)&=\frac{v_1+v_2}{D}\int\,
d\tau'\sinh\left(a_0\tau'+\chi\right)\exp\left[i
\frac{k_{\perp}}{a_0}D\sinh\left(a_0\tau'+\chi\right)\right]\\
&=i\frac{2}{k_{\perp}}\times\frac{v_1+v_2}{D^2}\,
\left[u_1K_1\left(u_1\right)\right]\\
&=i\frac{2}{k_{\perp}}\times\frac{1}{2}
[\tanh\left(\eta_2-y\right)-\tanh\left(\eta_1-y\right)]
\,[u_1K_1\left(u_1\right)],\\
\end{split}
\end{equation*}

avec $u_1=\displaystyle{\frac{k_{\perp}} {a_0}}D(y,\eta_1,\eta_2)$. $\chi$
est donn\'e dans le m\^eme appendice et $D$ a comme expression

$$
D^2(y,\eta_1,\eta_2)=
\displaystyle{\frac{\tanh\eta_2-\tanh\eta_1}
{\tanh\left(\eta_2-y\right)
-\tanh\left(\eta_1-y\right)}}.
$$

\begin{figure}[h]
\begin{center}
\includegraphics[height=5truecm,width=10truecm]{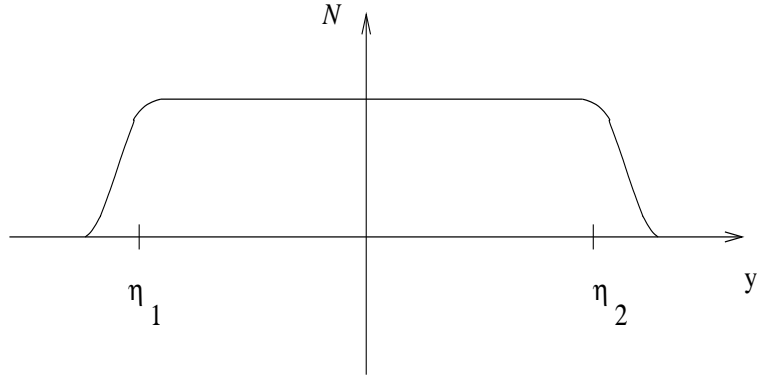}
\vskip .5cm
\caption{\label{fig:plateau2} Densit\'e du nombre de photons 
rayonn\'es \`a $k_{\perp}$ fix\'e en fonction de 
la rapidit\'e $y$}
\end{center}
\end{figure}

La densit\'e du nombre de photons rayonn\'es en fonction de $y$ et $k_{\perp}$
s'estime par analogie avec (\ref{eq:spectre1})
\begin{equation}
{\cal {N}}\left(y,k_{\perp}\right)=\frac1{2\left(2\pi\right)^3}
\left\{\frac{e^2}{k_{\perp}^2}[\tanh\left(\eta_2-y\right)-
\tanh\left(\eta_1-y\right)]^2\right\}
\end{equation}
dans la limite asymptotique qui donne le ``cut-off''
$k_{\perp}\approx a_0D(y,\eta_1,\eta_2)$.
Le plateau pour la distribution de la densit\'e du nombre de photons 
pr\'esente l'allure de la Fig.\ref{fig:plateau2} dont l'expression
est \'ecrite ci-dessous,
\begin{equation*}
 {\cal {N}} \left(y,k_{\perp}\right)\propto\left\{ \begin{array}{cll}
                      1 & \mbox{pour} & \eta_1<y<\eta_2; \\
    \exp\left[4(\mid\!\!\eta_2\!\!\mid-\mid\!\!y\!\!\mid)\right] & \mbox{pour} 
    & \mid\!\!y\!\!\mid\gg\mid\!\!\eta_2\!\!\mid; \\
    \exp\left[-4(\mid\!\!\eta_1\!\!\mid+\mid\!\!y\!\!\mid)\right] & \mbox{pour} & 
    \mid\!\!y\!\!\mid\gg\mid\!\!\eta_1\!\!\mid.
                              \end{array}\right.
\label{eq:red20}
\end{equation*}

Le cas (\ref{eq:red21}) g\'en\'eralise (\ref{eq:lu1})
et permet de retrouver, en particulier,
la distribution de la densit\'e spectrale des photons en fonction de $y$ \`a 
$k_{\perp}$ dans un intervalle $\eta_1\leq y\leq\eta_2$
(plateau de la Fig.\ref{fig:plateau1}).
On retrouve \'egalement le ``cut-off'' du cas (\ref{eq:lu1}) au terme
$D(y,\eta_1,\eta_2)$ pr\`es.  Ce ``cut-off'' d\'epend, bien s\^ur,
 de la vitesse (ou de la rapidit\'e) initiale et finale de la charge
consid\'er\'ee.  La densit\'e du nombre de photons
rayonn\'es en fonction de $y$ est finie et
sa valeur s'estime sans difficult\'e
$$
\widetilde{{\cal {N}}}(y,\eta_1,\eta_2)=
\int_{}^{a_0D(y,\eta_1,\eta_2)}{\cal {N}}
(y,k_{\perp})d^2k_{\perp}
\propto \ln\left[a_0D(y,\eta_1,\eta_2)\right].
$$
La hauteur du plateau est donc modifi\'ee par le
facteur $D(y,\eta_1,\eta_2)$.

Dans tous les cas discut\'es jusqu'\`a pr\'esent, nous avons
 rencontr\'e la d\'ependance ${\cal {N}}(k_\perp^2)
\propto1/k_{\perp}^2$. 
Nous avons de m\^eme pr\'edit la d\'ecroissance exponentielle du plateau
au-del\`a d'une certaine limite en rapidit\'e pour les 
photons \'emis \`a ``haute fr\'equence'' $k_{\perp}>a_0D$
(photons virtuels ou r\'eabsorb\'es par la particule).

\vskip 0.5cm

$\ast$ Soit le cas particulier o\`u l'on prend $v_1=v_2=v$.
Dans ce cas l'allure du plateau 
pour la densit\'e des photon rayonn\'es est la m\^eme,
par contre, il est sym\'etrique par rapport \`a l'axe
$\cal {N}$ et s'\'etend dans l'intervalle de rapidit\'e
$-\frac12\ln\left(\frac{1+v}{1-v}\right)\!<\!y\!<\!+
\frac12\ln\left(\frac{1+v}{1-v}\right)$. La valeur du
``cut-off'' est $k_{\perp}\approx a_0D(y,\eta)$.

\vskip 0.5cm

\begin{figure}[h]
\begin{center}
\includegraphics[height=4truecm,width=7truecm]{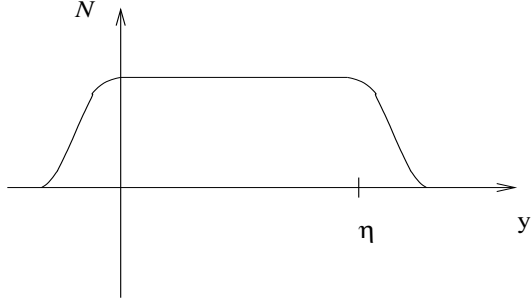}
\vskip .5cm
\caption{\label{fig:plateau3}
Densit\'e du nombre de photons rayonn\'es 
\`a $k_{\perp}$ fix\'e en fonction de la rapidit\'e $y$ }
\end{center}
\end{figure}

$\ast$ Si l'on ne s'int\'eresse qu'aux rapidit\'es
 positives, on peut poser $v_1=0$ et
$v_2=v$ dans (\ref{eq:red21}). Dans ce cas-ci, le plateau pour la
 densit\'e du nombre de photons rayonn\'es
 ne s'\'etend que dans l'intervalle des rapidit\'es positives
$0\!<\!y\!<\!\frac12\ln\left(\frac{1+v}{1-v}\right)$.
Son allure est repr\'esent\'ee par le plateau de la Fig.\,\ref{fig:plateau3}

\begin{equation*}
 {\cal {N}}\left(y,k_{\perp}\right)=\left\{ \begin{array}{cll}
                   1 & \mbox{pour} & \eta\gg y; \\
                  \exp\left\{4y\right\} & \mbox{pour} & y<0; \\
                  \exp\left\{4(\eta-y)\right\} & \mbox{pour} & y\gg\eta.
                              \end{array}\right.
\label{eq:red36}
\end{equation*}

La valeur du ``cut-off'' est ici
$k_{\perp}\approx a_0\Delta({y,\eta})$, o\`u

$$
\Delta^2(y,\eta)=\frac{\tanh\eta}{\tanh\left(\eta-y\right)+\tanh{y}}.
$$

\vskip 0.5cm

\textbf{Remarque:} Si l'on fait $y\approx\eta$ on
obtient $\Delta(\bar{v})\rightarrow1$ et 
$k_{\perp}\approx a_0$. Les photons \'emis \`a
$y\approx\eta$ (aux fr\'equences 
$k_{\perp}\gtrapprox a_0$) ne sont jamais
r\'eabsorb\'es par la charge et sont \'emis sous 
forme de rayonnement le long de la direction choisie:
le c\^one de bremsstrahlung appara\^it.

%\begin{center}
\section{Trajectoire arbitraire}
\label{subsection:Traj.Arb}
%\end{center}
Nous consid\'erons enfin une trajectoire quelconque pour le mouvement
de la charge. Nous rappelons l'expression du quadri-courant
\begin{equation}
\widetilde{j}^\mu\left(k\right)=\int\,dt\,\
 \frac{dx^\mu}{dt}e^{ikx\left(t\right)}.
\label{eq:red37}
\end{equation} 
\`A ce propos, on \'ecrit la quadri-impulsion du photon,
pour un mouvement quelconque, comme
\begin{center}
$k^\mu=\left(k^0=\omega,k^1 ,k^2, k^3\right)$.
\end{center}
Puisque $k^{\mu}k_{\mu}=0$ (photon r\'eel) on a $k_0=|\vec{k}|\mid$.
Nous r\'ecrivons (\ref{eq:red37}) sous la forme
\begin{equation}
\widetilde{j}^\mu\left(k\right)
=\int_{-\infty}^{+\infty}\,dt\,\ v^\mu \left(t\right) 
e^{ikx\left(t\right)}.
\end{equation}
On sectionne la trajectoire de la particule charg\'ee en trois parties.
D'abord, nous consid\'erons que sa vitesse itiale $v_1$
est constante pendant l'intervalle de temps
$\left(-\infty ,t_1\right)$, puis qu'elle subit l'action d'un champ
externe pendant $\left(t_1, t_2\right)$, pour \^etre enfin
acc\'el\'er\'ee jusqu'\`a la vitesse finale $v_2$. Pendant l'intervalle
$\left(t_2,+\infty \right)$ elle continue son mouvement inertiel jusqu`\`a
l'infini.  Le courant rayonn\'e prend la forme:
\begin{equation}
\begin{split}
\widetilde{j}^\mu\left(k\right)
 &= \widetilde{j}^\mu\left(k\right)_{-}
+\widetilde{j}^\mu
\left(k\right)_{c}+
\widetilde{j}^\mu\left(k\right)_{+} 
= \int_{-\infty}^{t_1}\,dt\,\ v^\mu
 \left(t\right)e^{ikx\left(t\right)}   \\
&+\int_{t_1}^{t_2}\,dt\,\ v^\mu
 \left(t\right)e^{ikx\left(t\right)}+
\int_{t_2}^{+\infty}\,dt\,\ v^\mu
 \left(t\right)e^{ikx\left(t\right)}\text{,}\\
\end{split}
\end{equation}
Il est connu des calculs pr\'ec\'edents sur le bremsstrahlung que:

\begin{equation*}
\widetilde{j}^\mu\left(k\right)_{-}=
 -i\displaystyle{\frac{v_1^\mu}{k\cdot v_1}}e^{ikv_1 t_1},
\end{equation*}
\begin{equation*}
\widetilde{j}^\mu\left(k\right)_{+}
=+i\displaystyle{\frac{v_2^\mu}{k\cdot v_2}}e^{ikv_2t_2},
\end{equation*}
et que
\begin{equation}\label{eq:jc}
\widetilde{j}^\mu\left(k\right)_{c}
=-i\int_{t_1}^{t_2}\,dt\,\
\left\{\displaystyle{\frac{v^\mu
 \left(t\right)}{k\cdot v\left(t\right)}}\right\}\,\
\displaystyle{\frac{d}{dt}}\,\ e^{ikx\left(t\right)}. 
\end{equation}
En int\'egrant (\ref{eq:jc}) par parties, on obtient
\begin{equation}
\begin{split}
\widetilde{j}^\mu\left(k\right)_c&=
 -i\left\{e^{ikx\left(t\right)}\,\
\displaystyle{\frac{v^\mu\left(t\right)}
{k\cdot v\left(t\right)}}\right\}^{t_2}_{t_1}
+ i\int_{t_1}^{t_2}\,dt\,\ 
\displaystyle{\frac{d}{dt}}
\left\{\displaystyle{\frac{v^\mu\left(t\right)}
{k\cdot v\left(t\right)}}\right\}e^{ikx\left(t\right)} \\
&= -i\displaystyle{\frac{v_2^\mu}{k\cdot v_2}}e^{ikv_2 t_2}+i
\displaystyle{\frac{v_1^\mu}{k\cdot v_1}}
e^{ikv_1 t_1}+i\int_{t_1}^{t_2}\,dt\,\ 
\displaystyle{\frac{d}{dt}}
\left\{\displaystyle{\frac{v^\mu\left(t\right)}
{k\cdot v\left(t\right)}}\right\}e^{ikx\left(t\right)},  \\
\end{split}
\end{equation}
et, finalement
\begin{equation}
\widetilde{j}^\mu\left(k\right)=
\int_{-\infty}^{+\infty}\,dt\,\ v^\mu \left(t\right)
e^{ikx\left(t\right)}
= i\int_{t_1}^{t_2}\,dt\,\
\displaystyle{\frac{d}{dt}}
\left\{\displaystyle{\frac{v^\mu\left(t\right)}{k\cdot v\left(t\right)}}\right\}
\,\ e^{ikx\left(t\right)}.
\label{eq:red37.5}
\end{equation}

L'identit\'e (\ref{eq:red37.5}) sera utile pour la g\'en\'eralisation
de notre analyse \`a une ligne d'univers arbitraire; elle permet
 notamment de ne s'int\'eresser qu'\`a l'intervalle de temps pendant lequel
la particule rayonne (temps de l'interaction avec un champ externe).

On d\'emontre, en particulier, que l'existence
d'un ``cut-off'' pour $k_{\perp}$, comme celui qui
a \'et\'e trouv\'e dans le cas d'une
trajectoire rectiligne est universel. Il sera donc
g\'en\'eralis\'e au cas d'une trajectoire quelconque.
D'abord, on consid\`ere que la phase 
$\phi\left(t\right)=kx\left(t\right)$
 varie peu dans l'intervalle de temps
$\left(t_1, t_2\right)$.
\begin{equation}
\phi \left(t_2 \right)-\phi \left(t_1 \right)
 \simeq \displaystyle{\frac{d\phi}
{dt}}\Delta t=
\left(\omega-\vec{k}\cdot\vec{v}\left(t\right)\right)\Delta t \ll 1 
\end{equation}
Si $\vec{k}$ est perpendiculaire \`a la vitesse
 $\vec{v}$, $\vec{k}.\vec{v}=0$,
$\omega\Delta t\approx 0$, on est dans le
 cas consid\'er\'e en \ref{subsubsection:Sray}
o\`u l'acc\'el\'eration est infinie.

On pose
$k=\left(\omega ,k_{\parallel}, \vec{k}_{\perp}\right)$
\footnote{l'impulsion longitudinale $k_{\parallel}$
 et l'impulsion transverse $k_{\perp}$ sont d\'efinies par rapport
 \`a la direction du mouvement de l'\'electron sortant},
$k^2=0={\omega}^2-{k_{\parallel}}^2-{k_{\perp}^2}$
(puisque les photons sont r\'eels) $\Rightarrow$ 
${k_{\perp}^2}={\omega}^2-{k_{\parallel}}^2$.
Puisque $\vec{v}\perp \vec{k_{\perp}}$:
\begin{equation*}
\left[\omega -k_{\parallel}\left(v+1-1\right)\right]\Delta t
= \left[\omega -k_{\parallel}+
k_{\parallel}\left(1-v\right)\right]\Delta t \\
=\left(\displaystyle{\frac{{\omega}^2-{k_{\parallel}^2}}
{\omega+k_{\parallel}}}+k_{\parallel}
\displaystyle{\frac{1-v^2}{1+v}}\right)\Delta t\ll 1,
\end{equation*}
que l'on peut r\'ecrire
\beq\label{eq:condition1}
\left(\displaystyle{\frac{k_{\perp}^2}
{\omega+k_{\parallel}}}+\omega \displaystyle{\frac{1-
v^2}{2}}\right)\Delta t\ll 1.
\eeq
Nous pouvons consid\'erer que
$\omega\approx k_{\parallel}\gg k_{\perp}$, ceci
impose

1) si $1-v=\cal{O}$$(1)$, le deuxi\`eme terme dans 
(\ref{eq:condition1}) est dominant $\Rightarrow$ 
$\displaystyle{\frac1{\lambda_{\parallel}}}=
\omega\ll \displaystyle{\frac1{\Delta t}}$
dans le cas relativiste;

2) si $1-v\ll 1, \text{le premier terme dans 
(\ref{eq:condition1}) est dominant} \Rightarrow 
\displaystyle{\frac1{t_f}}
\simeq\displaystyle{\frac{k_{\perp}^2}{2\omega}}
\ll\displaystyle{\frac1{\Delta t}}$ dans le cas
 ultra-relativiste qui nous int\'eresse. $t_f$
est le temps de formation de l'\'etat virtuel
 associ\'e \`a cette \'emission.

2) $\Rightarrow$ $k_{\perp}\ll 
\sqrt{\displaystyle{\frac{\omega}{\Delta t}}}\ll 
\displaystyle{\frac{\gamma}{\Delta t}}$.
 L'exponentielle dans l'expression du courant
(\ref{eq:red37.5}) peut
\^etre n\'eglig\'ee et on l'estime selon
\begin{equation*}
\widetilde{j}^\mu\left(k\right)=i\int_{t_1}^{t_2}\,dt\,\
\displaystyle{\frac{d}{dt}}\left\{\displaystyle{\frac{v^
\mu\left(t\right)}{k.v\left(t\right)}}\right\}
\,\ e^{ikx\left(t\right)}
\approx i\int_{t_1}^{t_2}\,dt\,\displaystyle{\frac{d}{dt}}
\left\{\displaystyle{\frac{v^\mu\left(t\right)}
{k.v\left(t\right)}}\right\}\\
=i\left[\frac{v^\mu\left(t_2\right)}{k.v\left(t_2\right)}-
\frac{v^\mu\left(t_1\right)}{k.v\left(t_1\right)}
\right],
\end{equation*}
tandis que les hautes fr\'equences sont supprim\'ees par le ``cut-off''
$k_{\perp}\approx \displaystyle{\frac{\gamma}{\Delta t}}$ dont le second
membre a la dimension d'une acc\'el\'eration propre.

\subsection{Courant dans le r\'ef\'erentiel
 o\`u $\boldsymbol{k'_{\parallel}=0}$, direction arbitraire}

\label{sub:dir.arb.}

Nous nous int\'eressons, comme dans les cas pr\'ec\'edents
\ref{subsubsection:1emeaccpfinie}
\ref{subsubsection:4emeaccpfinie}, \`a l'expression du courant dans le 
r\'ef\'erentiel o\`u l'impulsion longitudinale du photon 
est nulle $k'_{\parallel}=0$, pour une direction maintenant
 arbitraire du mouvement.
Nous effectuons un boost sur le r\'ef\'erentiel $K_y$ o\`u la rapidit\'e du
photon est \'egale \`a $y$.  On d\'efinit la vitesse du boost selon
$\vec{\beta}=\tanh{y}\,\vec{n}_{\vec{\beta}}$.
 Les transformations de Lorentz
de l'espace-temps ($x^{\mu}$) et de la quadri-impulsion ($k^{\mu}$) 
le long d'une direction arbitraire $\vec{n}_{\vec{\beta}}$ sont
donn\'ees respectivement par \cite{alain}
$$
t'=\gamma(\beta)\left(t-\vec{\beta}\cdot\vec{r}\right),\quad \omega'=\gamma(\beta)\left(\omega-\vec{\beta}\cdot\vec{k}\right),\
$$
$$
\vec{r}\,'=\vec{r} + \frac{\gamma(\beta)-1}{\beta^2}
(\vec{\beta}\cdot\vec{r})\vec{\beta}
-\gamma(\beta)\vec{\beta}\,t,\quad \vec{k}\,'=\vec{k}
\frac{\gamma(\beta)-1}{\beta^2}(\vec{\beta}\cdot\vec{k})\vec{\beta} -
\gamma(\beta)\vec{\beta}\,\omega,
$$
o\`u $\gamma(\beta)=1/\sqrt{1-\beta^2}
=\cosh{y}$ est le facteur de Lorentz. Le quadri-vecteur
$k^{\mu}=(\omega=k^0,k_{\parallel},
\vec{k}_{\perp})$ ob\'eit \`a la m\^eme loi
de transformation que $x^{\mu}=(t, \vec{r})$.
Il est de m\^eme utile d'obtenir le courant dans le r\'ef\'erentiel
o\`u la composante longitudinale de sa quadri-impulsion est nulle
$k'\,^{\mu}=(\omega',0,\vec{k}\,'_{\perp})$
\footnote{$\vec{k}\,'=(0, \vec{k}\,'_{\perp})$.}, ainsi,
$\vec{k}\,'\cdot\vec{r}\,'\!=\!=\!\vec{k}\,'_{\perp}
\cdot\vec{r}\,'\!=\!0$ $(\vec{k}\,'\cdot\vec{v}\,'\!=\!\vec{k}\,'_{\perp}\cdot\vec{v}\,'\!=\!0)$ 
o\`u $\vec{v}\,'$ est la vitesse de la charge dans le nouveau rep\`ere.
 L'invariance de Lorentz entra\^\i ne
$\omega t-\vec{k}\cdot\vec{r}=\omega 't'-\vec{k}\,'\cdot\vec{r}\,'=\omega 't'$.
De plus, $k'.v'(t')=\omega'$ car $\vec{k}\,'_{\perp}\perp\vec{v}\,'$
 dans le nouveau rep\`ere, et $v'^{\mu}=(1,\vec{v}\,')$.
De nouveau (pour une trajectoire arbitraire)
\beq
j'\,^0=\int_{-\infty}^{+\infty}dt'e^{i\omega't'}=0,
\eeq
et l'expression pour la composante spatiale \`a trois
dimensions peut \^etre obtenue \`a partir de (\ref{eq:red37.5})
\beq
\vec{j}\,'= \frac{ie}{\omega'}\int_{t_1'}^{t_2'}dt'
\,\frac{d}{dt'}[\vec{v}\,'(t')]\,e^{i\omega't'}=
 \frac{ie}{\omega'}\int_{t_1'}^{t_2'}dt'
\,\vec{a}\,'(t')\,e^{i\omega't'}=
\frac{ie}{\omega'}\vec{\widetilde{a}}\,'(\omega').
\eeq
Si maintenant on consid\`ere que
 la phase $\phi'(t)=\omega't'$ varie peu dans l'intervalle de temps 
$[t_1, t_2]$, $\omega'(t_2'-t_1')\ll 1$, nous retrouvons le ``cut-off''
en $\omega'=k_{\perp}'\ll1/(t_2'-t_1')=[a_0']$ qui est homog\`ene
 \`a une acc\'el\'eration propre. Par cons\'equent,

\beq\label{eq:cour.arb}
\vec{j}\,'= \frac{ie}{k_{\perp}'}\left[\vec{v}\,'(t_2')-\vec{v}\,'(t_1')\right].
\eeq

% o\`u la vitesse dans le nouveau rep\`ere s'obtient
%  \`a partir des transformation de Lorentz
% 
% \beeq
% \vec{v}\,'(t')&=&\frac{d\vec{r}\,'}{dt'}=
% \frac{\frac{d\vec{r}\,'}{dt}}{\frac{dt'}{dt}}=
% \frac{\vec{v}+\frac{\gamma(\beta)-1}{\beta^2}
% (\vec{\beta}\cdot\vec{v})\vec{\beta}
%  - \gamma(\beta)\vec{\beta}}{\gamma(\beta)
% (1-\vec{\beta}\cdot\vec{v})},\nonumber\\
% &=& \frac{\vec{v}-\frac{\gamma(\beta)}
% {\gamma(\beta)+1}\vec{\beta}}
% {\gamma(\beta)(1-\vec{\beta}\cdot\vec{v})}
% -\frac{\gamma(\beta)}{\gamma(\beta)+1}\vec{\beta}.
% \nonumber
% \eeeq
% 
% \vskip 0.5cm

Nous retrouvons la condition 2) du paragraphe pr\'ec\'edent
\ref{subsection:Traj.Arb} en utilisant la relation
entre $\Delta t=t_2-t_1$ et $\Delta t'=t_2'-t_1'$ (dilatation du temps)
qui d\'ecoule des transformations de Lorentz, soit $\Delta t=\gamma\Delta t'$.
 Cette derni\`ere entra\^\i ne 
$k_{\perp}=k_{\perp}'\approx\displaystyle{\frac{\gamma}{\Delta t}}$.
 Ce r\'esultat n'est pas surprenant car $\vec{k}_{\perp}$
 est perpendiculaire \`a la vitesse de la charge.

On remarque dans (\ref{eq:cour.arb}) l'apparition
 d'un premier c\^one de bremsstrahlung le long
de la direction de la charge entrante, et d'un deuxi\`eme
 le long de la charge sortante, car celle-ci
est d\'evi\'ee de sa trajectoire initiale.
 En particulier, nous venons de prouver que
la nature du courant est ind\'ependante de la ligne
 d'univers suivie par la charge entre les
\'ev\'enements $t_1$$(t_1')$ et $t_2$$(t_2')$.

L'universalit\'e du ``cut-off'' est donc d\'emontr\'ee
dans le cas g\'en\'eral (direction arbitraire du mouvement).
Sa nature est classique et se g\'en\'eralise \`a toute trajectoire.
Ici, nous avons d\'ecrit les aspects g\'en\'eraux des distributions
uniformes en fonction de la rapidit\'e ($\log\Theta$).
Comme il sera d\'emontr\'e prochainement, pour que
la particule charg\'ee rayonne, il faut que l'angle d'\'emission
soit inf\'erieur \`a l'angle de diffusion de la charge
($\Theta_\gamma<\Theta_{\text{d}}$) par rapport \`a sa direction initiale.

En CDQ, la charge change de direction de mouvement ainsi que d'\'etat de
couleur; or, du fait que le courant de couleur se conserve, ce dernier
entra\^\i ne l'apparition d'un c\^one de rayonnement aux angles
sup\'erieurs \`a l'angle de diffusion du parton \'emetteur
$(\Theta_g>\Theta_{\text{d}})$. La contrainte
($\Theta_\gamma<\Theta_{\text{d}}$) en EDQ est remplac\'ee
par une contrainte angulaire (``Angular Ordering'' en anglais)
 sur les angles des \'emissions successives des gluons mous
en CDQ.

\section{Section efficace du rayonnement mou}
\label{subsection:SERM}

Dans le but de calculer la section efficace du rayonnement mou 
en EDQ \cite{pQCDforbeginners}\cite{Peskin}
(voir Fig.\,\ref{fig:bremsst}), on prend le module
au carr\'e de l'amplitude (\ref{eq:Ma})
que l'on projette sur les \'etats de polarisation ($\lambda$)
du photon, on effectue la somme
 sur ($\lambda$) en tenant de m\^eme en consid\'eration l'espace de phase du
photon mou
\beq\label{eq:amplitude}
dW=e^2\sum_{\lambda=1,2}|\epsilon^{\lambda}_{\mu}\,j^{\mu}|^2
\frac{\omega^2\,d\omega\,d\Omega_{\gamma}}{2\,\omega\,(2\pi)^3}\,dW_{el}.
\eeq
o\`u $dW_{el}=\mid\!\!M_{el}\!\!\mid^2$. La somme s'effectue
sur les deux \'etats de polarisation physiques
du photon r\'eel, qui sont d\'ecrits par des vecteurs
normalis\'es orthogonaux \`a  la quadri-impulsion $k$ du photon
et entre eux, soit
$$
\epsilon_{\lambda}^{\mu}(k)\cdot\epsilon_{\mu,\lambda'}^{*}(k)
=-\delta_{\lambda\lambda'},
\qquad \epsilon_{\lambda}^{\mu}(k)\cdot k_{\mu}=0;
\qquad \lambda,\lambda'=1,2.
$$
Dans ces conditions, les vecteurs de polarisation peuvent \^etre choisis
de plusieurs mani\`eres. N\'eanmoins, cette incertitude n'affecte pas
le calcul des observables physiques, gr\^ace \`a l'invariance de jauge.
C'est pour cel\`a que le tenseur de polarisation peut s'\'ecrire sous la forme
\beq\label{eq:sumlambda}
\sum_{\lambda=1,2}\epsilon_{\lambda}^{\mu}
\epsilon_{\lambda}^{*\nu}=-g^{\mu\nu}+
\text{tenseur proportionnel \`a } k^{\mu}k^{\nu}.
\eeq
Puisque le courant se conserve $j^{\mu}k_{\mu}=0$,
on peut n\'egliger le dernier terme de (\ref{eq:sumlambda}).
Dans le but de calculer la production des photons mous, au lieu
d'utiliser les polarisations physiques, nous pouvons simplement,
en vertu de l'invariance de jauge, choisir la plus simple (jauge de Feynman)
$$
\sum_{\lambda=1,2}\epsilon_{\lambda}^{\mu}
\epsilon_{\lambda}^{*\nu}=-g^{\mu\nu}.
$$
On d\'efinit alors le nombre de photons
 de bremsstrahlung produits en divisant (\ref{eq:amplitude}) \`a gauche par
le module au carr\'e du terme de Born
\beeq\label{eq:DL}
dN\equiv\frac{dW}{dW_{el}}&=&-\frac{\alpha}{4\pi^2}\,(j^{\mu})^2
\,\omega\, d\omega\, d\Omega_{\gamma}\\
&\simeq&\frac{\alpha}{\pi}\,\frac{d\omega}{\omega}\,\frac{d\Omega_{\gamma}}{2\pi}
\frac{1-\cos\Theta_d}{(1-\cos\Theta_1)(1-\cos\Theta_2)},\nonumber
\eeeq
o\`u $\alpha=e^2/4\pi$ repr\'esente la constante de couplage des interactions
 \'electromagn\'etiques.

Dans (\ref{eq:DL}) nous avons pris la limite relativiste $1-v_1,\,1-v_2\ll 1$:
\beq\label{eq:courcar}
-(j^{\mu})^2=\frac{2\,(p_1\cdot p_2)}{(p_1\cdot k)\,(p_2\cdot k)}
\left(1+{\cal {O}}\left(\frac{m^2}{p_0^2}\right)\right)\simeq
\frac2{\omega^2}\frac{(1-\vec{n}_1\cdot\vec{n}_2)}
{(1-\vec{n}_1\cdot\vec{n})((1-\vec{n}_2\cdot\vec{n}))}
\eeq
o\`u $\vec{n}$, $\vec{n}_1$ et $\vec{n}_2$ sont des
 vecteurs unitaires qui repr\'esentent
les directions du photon \'emis, de l'\'electron
 entrant et de l'\'electron sortant respectivement par rapport \`a l'axe Ox.
 (\ref{eq:courcar}) ne tient pas compte des contributions dans les r\'egions
\`a tr\`es petit angle $\Theta^2_i\lesssim(1-v_i^2)
=m^2/p^2_{0i}\ll 1$ car le rayonnement des photons
 mous s'annule alors (``c\^one d'extinction''). Si le
photon est \'emis \`a petit angle par rapport \`a la particule entrante
 $\Theta_1\ll\Theta_2\simeq \Theta_d$, le spectre du rayonnement (\ref{eq:DL})
prend la forme
\beq
\label{eq:doublelog}
dN\simeq\frac{\alpha}{\pi}\,\frac{\sin\Theta_1d\Theta_1}{(1-\cos\Theta_1)}\,
\frac{d\omega}{\omega}\simeq\frac{\alpha}{\pi}\,\frac{d\Theta_1^2}{\Theta_1^2}
\frac{d\omega}{\omega}.
\eeq
Deux c\^ones de bremsstrahlung apparaissent
 le long des directions de la particule entrante 
et sortante. A l'int\'erieur des c\^ones
 le rayonnement poss\`ede une double d\'ependance 
logarithmique qui montre une divergence
 molle en $d\omega/\omega$ (car le photon est mou) et une autre colin\'eaire
en $d\Theta^2/\Theta^2$ (car le photon est \'emis \`a petit angle).
Cette double contribution logarithmique (\ref{eq:doublelog})
est un ingr\'edient dans la construction des \'equations d'\'evolution
partoniques dans les jets hadroniques (voir les articles \ref{sub:article1},
\ref{sub:article2} et \ref{sub:article3}). Si l'on choisit les variables
$$
\ell=\ln[\omega],\qquad y=\ln[\Theta]
$$
on r\'ecrit (\ref{eq:doublelog}) sous la forme
\beq\label{eq:elly}
dN\simeq\frac{2\alpha}{\pi}\,d\ell\, dy
\eeq
que l'on rencontre dans les \'equations d'\'evolution partoniques du spectre
et des corr\'elations des particules (3.12,3.13) et (3.35,3.36)
dans l'article \ref{sub:article2}.

\section{Introduction \`a la coh\'erence}
\label{subsection:coher}

Dans la jauge de Feynman, le terme d'interf\'erence entre les deux
\'emetteurs en (\ref{eq:DL}) est dominant, \`a savoir
\beq
dN\propto-\left[\frac{p_1^{\mu}}{(p_1\cdot k)}
-\frac{p_2^{\mu}}{(p_1\cdot k)}\right]^2
\approx\frac{2(p_1.p_2)}{(p_1\cdot k)(p_2\cdot k)}.
\eeq
On ne peut donc pas d\'eceler quelle
 partie du rayonnement est associ\'ee \`a la particule
charg\'ee entrante ou sortante.
 Nous consid\'erons alors le calcul d\'etaill\'e de (\ref{eq:amplitude})
en fonction des polarisations physiques du photon. Dans cet objectif,
on se place dans la jauge de radiation (hors sources); dans celle-ci,
le tri-vecteur $\vec{A}$ satisfait $\vec{\nabla}\cdot\vec{A}=0$, alors que
la composante temporelle est nulle, c'est \`a dire
$A_0\equiv 0$. Dans cette jauge le photon est donc d\'ecrit par
 deux tri-vecteurs orthogonaux entre eux et \`a son impulsion, on a
\beq
(\vec{\epsilon}_{\lambda}\cdot\vec{\epsilon}_{\lambda'})
=\delta_{\lambda\lambda'},\,\,\,
\,\,\,\,(\vec{\epsilon_{\lambda}}\cdot\vec{k}=0).
\eeq
Ainsi, il ne reste que deux \'etats de polarisation physiques.
On effectue ensuite la somme sur ($\lambda$) et on trouve
\beq
dN\propto\sum_{\lambda=1,2}\mid\vec{j}(k)
\cdot\vec{e}_{\lambda}\mid^2=
\sum_{\alpha,\beta=1\dots 3}\vec{j}^{\alpha}(k)
\cdot[\delta _{\alpha\beta}-
\vec{n}_{\alpha}\cdot\vec{n}_{\beta}]\,
\vec{j}^{\beta}(k),
\eeq
% 
% % 
o\`u $\alpha$, $\beta$ repr\'esentent les trois coordonn\'ees spatiales.
On r\'ecrit le courant mou (\ref{eq:courant}) sous la forme d'un tri-vecteur
$p_i^{\mu}\rightarrow\vec{v}_ip_{0i}$, on utilise les expressions suivantes
\beq\label{eq:proj1}
(\vec{v}_i)_{\alpha}\left[\delta_{\alpha\beta}
-\frac{k_{\alpha}k_{\beta}}
{\vec{k}^2}\right](\vec{v}_i)_{\beta}
=v_i^2\sin^2\Theta_i,
\eeq
\beq\label{eq:proj2}
(\vec{v}_1)_{\alpha}\left[\delta_{\alpha\beta}
-\frac{k_{\alpha}k_{\beta}}
{\vec{k}^2}\right](\vec{v}_2)_{\beta}=
v_1v_2(\cos\Theta_{12}-\cos\Theta_1\cos\Theta_2),
\eeq
et on obtient 
\beq\label{eq:DL1}
dN=\frac{\alpha}{\pi}\left\{{\cal {R}}_1 +
 {\cal {R}}_2 - 2 {\cal {J}} \right\}
\frac{d\omega}{\omega}\frac{d\Omega}{4\pi}.
\eeq
Ici,
\beq\label{eq:Ri}
{\cal {R}}_i=\frac{v_i^2\,\sin^2\Theta_i}
{(1-v_i\cos\Theta_i)^2},\,\,\,\,\,\,i=1,2,
\eeq
\beq\label{eq:JJ}
{\cal {J}}\equiv \frac{v_1v_2(\cos\Theta_{12}
-\cos\Theta_1\cos\Theta_2)}
{(1-v_1\cos\Theta_1)(1-v_2\cos\Theta_2)}.
\eeq
Les contributions ${\cal {R}}_1,\,{\cal {R}}_2$ (\ref{eq:Ri}) sont
ind\'ependantes; elle sont  attach\'ees
au rayonnement de la charge initiale et diffus\'ee respectivement.
L'expression du terme d'interf\'erence est 
donn\'e par ${\cal {J}}$ (\ref{eq:JJ}). La somme des trois contributions
d\'ecrit alors l'\'emission coh\'erente des photons.
On peut \'egalement v\'erifier que (\ref{eq:DL1})
est \'equivalente \`a (\ref{eq:DL}) dans la jauge
de Feynman, soit
$$
{\cal {R}}_{\text{coh\'er}}\equiv{\cal {R}}_{\text{ind\'ep}}-2{\cal {J}}=-\omega^2(j^{\mu})^2, \qquad
{\cal {R}}_{\text{ind\'ep}}\equiv{\cal {R}}_1 + {\cal {R}}_2.
$$
\textbf{Remarque:} la physique dans ce probl\`eme est la m\^eme
que celle qui explique l'interf\'erence de la lumi\`ere quand celle-ci
traverse les trous d'Young:  l'intensit\'e de la lumi\`ere
sur un \'ecran situ\'e \`a une distance finie des
orifices (cette distance \'etant tr\`es 
sup\'erieure \`a celle qui s\'epare les trous)
 est proportionnelle \`a la somme ind\'ependante
des modules au carr\'e du champ \'electrique et
du terme d'interf\'erence.

\subsection{Le r\^ole de l'interf\'erence;
contrainte sur les angles d'\'emission}
\label{subsection:roleinterf}

Dans la limite ultra relativiste ($v_i\rightarrow 1$) on a
\beq
{\cal {R}}_1\simeq\frac{\sin^2\Theta_1}{(1-\cos\Theta_1)^2}=\frac2{a_1}-1,
\eeq
\beq\label{eq:interf}
{\cal {J}}\simeq\frac{\cos\Theta_{12}-\cos\Theta_1\cos\Theta_2}{(1-\cos\Theta_1)
(1-\cos\Theta_2)}=\frac{a_1+a_1-a_{12}}{a_1.a_2}-1
\eeq
o\`u l'on a introduit les notations $\vec{n}=\displaystyle{\frac{\vec{k}}{\omega}}$,
$\vec{n}_i=\displaystyle{\frac{\vec{p}_i}{\mid\!\!\vec{p}_i\!\!\mid}}$ et
$$
a_1=1-\vec{n}\cdot\vec{n}_1=1-\cos\Theta_1, \quad a_2=1-\cos\Theta_2,
$$
$$
a_{12}=1-\vec{n}_1\cdot\vec{n}_2=1-\cos\Theta_\text{d}.
$$
Les termes $a_i$ sont n\'egligeables quand les angles deviennent petits
$a\simeq\frac12\Theta^2$. La contribution apport\'ee par ${\cal R}_i$
au rayonnement prend une d\'ependance logarithmique qui s'\'etend jusqu'aux
limites des grands angles $a\lesssim 1$
\beq
dN_1\propto {\cal {R}}_1\sin\Theta d\Theta\propto\frac{da_1}{a_1}.
\eeq
N\'eanmoins, le terme d'interf\'erence supprime le rayonnement lorsque l'angle
d'\'emission devient sup\'erieur \`a l'angle de diffusion de la charge
\beq
dN\propto{\cal {R}}_{\text{coh\'er}.}\sin\Theta
d\Theta=2a_{12}\frac{da}{a_1.a_2}
\propto\frac{da}{a^2}\propto\frac{d\Theta^2}{\Theta^2},\quad a_1\simeq a_2\gg a_{12}.
\eeq
Dans le but de quantifier ces effets de coh\'erence on \'ecrit
les contributions
$$
V_1={\cal {R}}_1-{\cal {J}}=\frac2{a_1}-\frac{a_1+a_2-a_{12}}{a_1.a_2}=
\frac{a_{12}+a_2-a_1}{a_1.a_2};
$$
$$
V_2={\cal {R}}_2-{\cal {J}}=\frac2{a_2}-\frac{a_1+a_2-a_{12}}{a_1.a_2}=
\frac{a_{12}+a_1-a_2}{a_1.a_2};
$$
\beq\label{eq:Rcoher}
{\cal {R}}_{\text{coh\'er.}}=V_1+V_2.
\eeq
L'amplitude d'\'emission $V_i$ peut \^etre encore consid\'er\'ee comme
\'etant associ\'ee \`a la charge $i$ ($V_1$ est singulier si
$a_1\rightarrow0$ et {\em vice versa}).
Puisque $V_1$ d\'epend de la direction de son partenaire $2$, on ne peut pas
consid\'erer des probabilit\'es ind\'ependantes mais conditionnelles.
Puisque l'\'emission des photons mous
pr\'esente une sym\'etrie azimutale,
on prend la moyenne de $V$ sur l'angle d'\'emission du quantum rayonn\'e
par rapport \`a l'angle $(\vec{n},\vec{n}_1)$ et on en d\'eduit
que la probabilit\'e $V_1(\vec{n},\vec{n}_1;\vec{n}_2)$ s'annule
\`a l'ext\'erieur du
c\^one d'angle d'ouverture $\Theta_{\text{d}}$ (angle de diffusion), soit
\footnote{voir le calcul de cette int\'egrale dans l'appendice
\ref{sub:Vangulaire}.}
\beq\label{eq:moyV}
<V_1>_{\text{azimut}}\equiv\int_0^{2\pi}
\frac{d\phi_{\vec{n},\vec{n_1}}}{2\pi}V_1(\vec{n},\vec{n}_1;\vec{n}_2)
=\frac2{a_1}\vartheta(a_{12}-a_1).
\eeq
Or, $a_2$ change sous l'int\'egrale (\ref{eq:moyV}), tandis que $a_1$ et
$a_{12}$ sont fix\'es. Le r\'esultat d\'ecoule de l'int\'egrale angulaire
\beq\label{eq:a2}
\int_{0}^{2\pi}\frac{d\phi_{n,n_1}}{2\pi}\frac{1}{a_2}=\frac1{\mid\!\cos\Theta_1-\cos
\Theta_{\text{d}}\!\mid}=\frac1{\mid\!a_{12}-a_1\!\mid}.
\eeq 
Nous pouvons conclure que le r\'esultat s'exprime comme la somme
des probabilit\'es associ\'ees
\`a l'apparition de deux c\^ones de bremsstrahlung ind\'ependants d'angle
d'ouverture $\Theta_{\text{d}}/2$ centr\'es
sur les directions des vecteurs $\vec{n}_1$ et $\vec{n}_2$.
Au d\'el\`a de cette r\'egion, la section efficace s'annule,
 autrement dit, le rayonnement des photons mous est supprim\'e.

Cette propri\'et\'e est connu en anglais comme {\em Angular Ordering} (AO),
il s'agit d'une contrainte d'apr\`es laquelle les angles sont forc\'ement
ordonn\'es (dans le sens $\Theta_{\text{d}}\geq\Theta_{\gamma}$);
si ce n'\'etait pas le cas, il n'existerait pas d'\'emission. Cette condition
constitue l'un des ingr\'edients dont on doit tenir compte
dans la g\'en\'eralisation des \'equations de 
Dokshitzer-Gribov-Lipatov-Altarelli-Parisi, DGLAP
\cite{DGLAP} aux domaines de petits $x$ (fraction de l'\'energie
emport\'ee par un parton dans un jets, ex. collisions 
$e^+e^-$), \`a savoir, dans l'approximation MLLA
(Modified Leading Logarithmic Approximations) qui d\'ecrit
la structure interne des jets partoniques en CDQ.

\section{Interpr\'etation de la contrainte sur les angles
d'\'emission de photons en m\'ecanique quantique relativiste}
\label{sub:Interp}

Pour quelle raison le rayonnement est-il supprim\'e lorsque l'angle
d'\'emission
d\'epasse l'angle de diffusion? Pour r\'epondre \`a cette question nous
utilisons quelques aspects essentiels
de la m\'ecanique quantique. Un \'electron physique est une charge
entour\'ee de son propre champ coulombien.
Du point de vue quantique, le champ coulombien
associ\'e \`a la charge peut \^etre interpr\'et\'e comme un ensemble de
photons virtuels \'emis et aussit\^ot r\'eabsorb\'es par celle-ci
au bout d'un intervalle de temps fini. Les processus d'\'emission
et de r\'eabsortion virtuels forment un \'etat que l'on appelle
``\'electron physique''.

Cette coh\'erence est partiellement perturb\'ee lorsque
la charge subit une perturbation externe. Par
cons\'equent, une partie des fluctuations intrins\`eques du champ
stationnaire est rayonn\'ee sous forme de photons r\'eels et, ainsi,
un c\^one de bremsstrahlung appara\^it le long
de la direction du mouvement initial de la particule. Finalement,
dans le processus de 
reg\'en\'eration du nouveau champ coulombien qui suit la direction
finale du mouvement, un deuxi\`eme c\^one de bremsstrahlung
appara\^it le long de la charge diffus\'ee.

L'intervalle de temps qui s'\'ecoule
entre l'\'emission et la r\'eabsortion du photon de quadri-impulsion $k$
par l'\'electron de quadri-impulsion $p_1$
est proportionnel \`a l'inverse du propagateur
de l'\'etat virtuel de quadri-impulsion
$(p_1-k)$ (voir Fig.\,\ref{fig:bremsst} gauche), soit
\beq
t_{\text{fluct}}\sim\frac{E_1}{\mid\!\!m^2-(p_1-k)^2\!\!\mid}=\frac{E_1}{2p_1.k}\sim\frac1{\omega\Theta^2}
\simeq\frac{\omega}{k_{\perp}^2},
\eeq
o\`u l'on a effectu\'e l'approximation des angles colin\'eaires:
$k_{\perp}\approx\omega\Theta\ll k_{\parallel}\approx\omega$.
Le temps de la fluctuation peut \^etre important pour de faibles
\'energies $\omega$ et devenir le param\`etre fondamental
de certains processus en EDQ. On consid\`ere un \'electron qui a diffus\'e
sur un champ \'electrique externe sous
un angle $\Theta_d$ par rapport \`a la direction de sa trajectoire initiale.
Que le photon \'emis soit r\'eabsorb\'e d\'epend, en particulier,
de la position finale de la particule par
rapport \`a la coordonn\'ee attendue par celle-ci.
Nous allons donc comparer le d\'eplacement de la 
charge $\Delta\vec{r}$ par rapport \`a la largeur du champ du photon,
$\lambda_{\parallel}\sim\omega^{-1}$,
$\lambda_{\perp}\sim k_{\perp}^{-1}$ o\`u $\lambda$ est sa longueur
d'onde. On a
\beq
\Delta r_{\parallel}\sim\mid\!\!v_{2\parallel}-v_{1\parallel}\!\!\mid t_{\text{fluct}}\sim\Theta_d^2
\frac1{\omega\Theta^2}=\left(\frac{\Theta_d}{\Theta}\right)^2\lambda_{\parallel}
\Leftrightarrow\lambda_{\parallel};
\eeq
\beq
\Delta r_{\perp}\sim c\Theta_d
t_{\text{fluct}}\sim\Theta_d\frac1{\omega\Theta^2}=\left(\frac{\Theta_d}{\Theta}\right)
\lambda_{\perp}\Leftrightarrow\lambda_{\perp}.
\eeq
Si l'angle de diffusion est grand, $\Theta_d\sim 1$, le d\'eplacement
de la charge est sup\'erieur \`a sa longueur d'onde pour toute valeur
de $\Theta$; les deux c\^ones de bremsstrahlung apparaissent.
Si l'angle de diffusion est petit $\Theta_d\ll 1$, il n'y a que
les photons \'emis \`a $\Theta\lesssim\Theta_d$ qui peuvent {\em observer}
la d\'eviation de la charge et,
par cons\'equent, r\'ealiser la perturbation de l'\'etat coh\'erent. C'est 
pour ces raisons que le rayonnement de bremsstrahlung n'a lieu qu'aux
angles qui sont inf\'erieurs \`a l'angle de diffusion;
ceci est \'equivalent au ``cut-off'' que l'on trouve pour
$k_{\perp}$ dans \ref{sub:dir.arb.}.
Les autres composantes ont une longueur d'onde importante et peuvent \^etre
facilement r\'easorb\'ees par la particule.  Le temps de formation
du rayonnement est court aux grands angles d'\'emission et les photons
qui sont \'emis \`a $\Theta>\Theta_\text{d}$
ne r\'ealisent pas la d\'eviation de la charge.
L'origine physique du ``cut-off'' en $k_{\perp}$ est
associ\'ee \`a l'apparition de cette {\em contrainte} pour les angles
$\Theta\lesssim\Theta_\text{d}$ (AO) et sera utilis\'ee dans
l'int\'egration des \'equations 
d'\'evolution des observables physiques dans les travaux
\ref{sub:article1}, \ref{sub:article2} et \ref{sub:article3}.

Pour conclure ce paragraphe, on remarque que dans le cas o\`u
$\Theta\gg\Theta_\text{d}$ (le photon ne r\'ealise pas
la d\'eviation de la charge), $p_1\approx p_2\approx p$ et, par cons\'equent,
le courant (\ref{eq:courant}) s'annule. 
\section{Rayonnement de bremsstrahlung en CDQ}
\label{eq:BremouCDQ}

Nous g\'en\'eralisons le courant de bremsstrahlung (\ref{eq:courant})
en CDQ pour l'\'emission des gluons mous en injectant le nouvel
ingr\'edient de la th\'eorie, c'est \`a dire la couleur.
Ainsi, les facteurs de Lorentz $(p_i^{\mu}/p_ik)$ doivent \^etre
multipli\'es par les facteurs de couleur \cite{Cheng}
associ\'es \`a l'interaction quark-gluon,
c'est \`a dire les matrices de Gell-Mann \cite{pQCDforbeginners} (voir Fig.8)

\beq\label{eq:coulcour}
j^{\mu}=\left[t^bt^a\left(\frac{p_1^{\mu}}{(p_1\cdot k)}\right)-
t^at^b\left(\frac{p_2^{\mu}}{(p_2\cdot k)}\right)\right].
\eeq 

On d\'efinit $A_i=\displaystyle\frac{p_i^{\mu}}{(p_i\cdot k)}$ et on
utilise la d\'ecomposition
$$
t^at^b=\frac1{2N_c}\delta_{ab}+\frac12(d_{abc}+if_{abc})t^c,
$$
que l'on peut r\'ecrire sous la forme
\beeq
j^{\mu}&=&\frac12(A_1-A_2)\left\{t^b,t^a\right\}+(A_1+A_2)
\left[t^b,t^a\right]\\
&=&\frac12(A_1-A_2)\left(\frac1{N_c}\delta^{ab}+d^{abc}t^c\right)
-\frac12(A_1+A_2)if^{abc}t^c.
\eeeq
Dans l'objectif d'en d\'eduire la probabilit\'e de l'\'emission,
il faut d\'eterminer
$j^{\mu}j^*_{\mu}$ et effectuer la somme sur les \'etats de couleur.
Nous avons
$$
\sum_{a,b}\left(\frac1{2N_\text{c}}\delta_{ab}\right)^2=\left(\frac1{2N_\text{c}}\right)^2
(N_\text{c}^2-1)=\frac1{2N_\text{c}} C_\text{F};
$$
$$
\sum_{a,b}\left(\frac12d_{abc}t^c\right)^2=\frac14\frac{N_c^2-4}{N_c}(t^c)^2=
\frac{N_c^2-4}{4N_c} C_F;
$$
$$
\sum_{a,b}\left(\frac12if_{abc}t^c\right)^2=\frac14N_c(t^c)^2 C_F.
$$
Le facteur commun $C_\text{F}$ appartient au terme de Born de la section
efficace (non-radiative). Le spectre du rayonnement prend la forme
\beeq
dN&\propto&\frac1{C_F}\sum_{\text{couleur}}j^{\mu}\cdot j^*_{\mu}=
\left(\frac1{2N_c}+\frac{N_c^2-4}{4N_c}\right)(A_1-A_2)\cdot(A_1-A_2)\\
&+&\frac{N_c}4(A_1+A_2)\cdot(A_1+A_2),\nonumber
\eeeq
que l'on peut simplifier en
\beq
dN\propto C_\text{F}(A_1-A_2)\cdot (A_1-A_2) + N_\text{c}A_1\cdot A_2.
\eeq
Les points symbolisent la somme que l'on doit effectuer sur les
\'etats de polarisation des gluons.
Nous utiliserons les projecteurs (\ref{eq:proj1}) et (\ref{eq:proj2})
pour d\'eterminer cette somme
\beq
A_1\cdot A_2\equiv \sum_{\lambda=1,2}\left(A_1e^{(\lambda)}\right)
\left(A_2e^{(\lambda)}\right)^*={\cal {J}}\quad \left\{\neq-(A_1A_2)\right\}.
\eeq

La section efficace se simplifie finalement \`a l'expression suivante

\beq\label{eq:colsecteff}
dN\propto C_F\,{\cal {R}}_{\text{coh\'er}}+N_c\,{\cal {J}}.
\eeq

Le terme proportionnel \`a $C_F$ en (\ref{eq:colsecteff}) repr\'esente
les deux c\^ones de bremsstrahlung que l'on a d\'ej\`a trouv\'es
dans le cas abelien en EDQ (\ref{eq:Rcoher}), ils sont
respectivement centr\'es sur les direction du mouvement du quark
entrant et diffus\'e
$\Theta_1, \Theta_2\leq\Theta_\text{d}$ (contrainte sur les angles ``AO'').
Le deuxi\`eme terme,
d'origine non-abelienne, est proportionnel \`a la charge de couleur du gluon.
Il a (\ref{eq:interf}) comme expression, soit
\beq\label{eq:interf1}
{\cal {J}}=\frac{a_1+a_1-a_{12}}{a_1.a_2}-1,
\eeq
qui reste non-singulier dans la r\'egion d'\'emission aux petits angles
$\Theta_1\ll\Theta_\text{d}$, $\Theta_2\ll\Theta_\text{d}$. 
En m\^eme temps, il remplit
la r\'egion qui correspond aux angles d'\'emission sup\'erieur \`a l'angle
de diffusion du quark $\Theta=\Theta_1\approx\Theta_2\gg\Theta_\text{d}$ o\`u

\beq\label{eq:comeqcd}
dN\propto d\Omega\,{\cal {J}}\propto\sin\Theta\,d\Theta
\left(\frac2{a}-1\right)\propto\frac{d\Theta^2}{\Theta^2}.
\eeq

Si l'on int\`egre sur l'angle azimutal autour de la direction de mouvement
du quark entrant, on obtient

\beq
\int\frac{d\phi_1}{2\pi}{\cal {J}}=\frac1{a_1}\left(1+\frac{a_1-a_{12}}
{\mid\!\! a_1-a_{12}\!\!\mid}\right)=\frac2{a_1}\vartheta(\Theta_1-\Theta_\text{d})-1.
\eeq

\begin{figure}[h]
\begin{center}
\includegraphics[height=4truecm,width=13truecm]{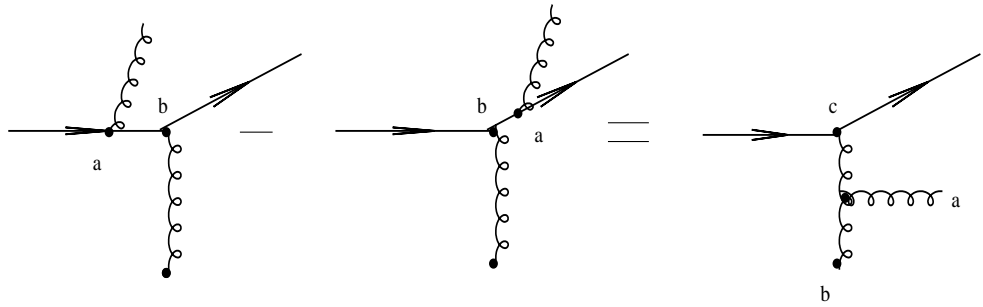}
\vskip .5cm
\caption{\label{fig:gluonbrem} Repr\'esentation diagrammatique de la r\'elation de commutation}
\end{center}
\end{figure}

C'est ainsi qu'un troisi\`eme c\^one de bremsstrahlung, d'origine
non-abelienne appara\^it.
On peut aussi intuiter ce r\'esultat sans effectuer
les calculs pr\'ec\'edents. Pour toute repr\'esentation $R$
de $SU(N_c)$, on peut r\'ecrire (\ref{eq:coulcour}) sous la forme
\beq
j^{\mu}=T^bT^a\frac{p_1^{\mu}}{p_1\cdot k}-T^aT^b\frac{p_2^{\mu}}{p_1\cdot k}
\approx\left(T^bT^a-T^aT^b\right)\frac{p^{\mu}}{p_1\cdot k}.
\eeq
On rappelle la forme g\'en\'erale de la relation de commutation de $SU(N_c)$
$$
\left[T^a(R),T^b(R)\right]=i\sum_cf_{abc}T^c(R),
$$
qui peut se repr\'esenter diagrammatiquement sous la forme de
la Fig.\ref{fig:gluonbrem}

tel que l'on obtient
$$
j^{\mu}(j_{\mu})^*\propto(if_{abc})^2\propto N_c,
$$
qui est le facteur de couleur associ\'e au module au carr\'e du diagramme 
\`a droite dans la Fig.\ref{fig:gluonbrem}.
Ainsi, le rayonnement \`a $\Theta>\Theta_\text{d}$ ne d\'epend pas
de la nature (\'etat de couleur) du projectile. 

Les gluons de bremsstrahlung se transforment,
lorsque les distances entre partons croissent,
en hadrons observables. Dans le prochain
paragraphe on donne un exemple d'application
de ces propri\'et\'es \`a l'identification du canal de production
du boson de Higgs.

\section{Application \`a l'identification des canaux possibles dans la
production du boson de Higgs \cite{higgs}}
\label{subsection:Higgs}

Si on utilise la rapidit\'e comme variable, elle vaut $\eta=\ln\Theta^{-1}$
pour les \'emissions de gluons colin\'eaires. Ainsi,
la distribution logarithmique angulaire
(\ref{eq:comeqcd}) pr\'esente un plateau de rapidit\'e $dN\propto d\eta$.

\begin{figure}[h]
\begin{center}
\includegraphics[height=5truecm,width=10truecm]{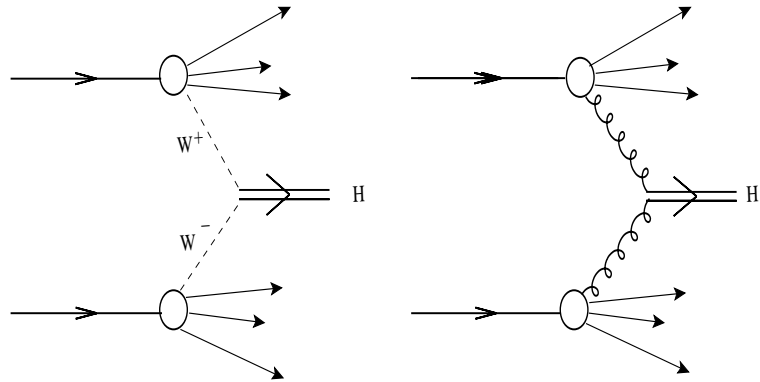}
\vskip .5cm
\caption{\label{fig:higgs} 
Diffusion $W^+W^-$ (gauche) et $gg$ (droite) dans la production du boson de Higgs }
\end{center}
\end{figure}

Or, \`a tr\`es haute \'energie, on conna\^it deux m\'ecanismes possibles
(ou canaux de production) qui interviennent dans la production
du boson de Higgs,
il s'agit de
$W^+W^-\rightarrow H$ et de la fusion gluon-gluon $gg\rightarrow H$
(voir Fig.\,\ref{fig:higgs}).
Puisque l'impulsion transf\'er\'ee est de l'ordre de la masse du 
boson dans le canal $t$, $(-t)\sim M^2_H$, sa production
est un processus qui fait intervenir les interaction fortes
(quark-gluon, anti-quark-gluon, gluon-gluon).
Par cons\'equent, les quarks diffus\'es, \'etant soudainement 
acc\'el\'er\'es, rayonnent des gluons de bremsstrahlung.
L'angle de diffusion dans ce processus est donn\'e par
$\Theta_\text{d}^2\simeq\mid\!\!t\!\!\mid\!\!\!/s\sim M^2_H/s$ 
(voir appendice \ref{subsection:higgs}).
Dans le cas du canal $WW$, les bosons $W$ n'\'echangent pas des \'etats de
couleur, \`a savoir les quarks entrants n'\'echangent que leur 
\'etat de saveur; ainsi,
le courant (\ref{eq:courant}) est suffisant pour pr\'edire l'allure de la 
distribution en fonction de la rapidit\'e des hadrons
qui accompagnent le Higgs, \`a un facteur de
couleur pr\`es. Dans le cas de la fusion $gg$, les quarks sortants prennent
un autre degr\'e de couleur sous l'\'echange des gluons.
Le courant mou rayonn\'e est donc donn\'e par (\ref{eq:coulcour}).
L'allure de la distribution hadronique finale est donn\'ee par
un plateau pratiquement uniforme \cite{higgs} dont la hauteur au centre
($\eta$ petit ou $\Theta$ grand), d'apr\`es (\ref{eq:colsecteff}), est
proportionnelle \`a $N_c$. En effet, il n'y a que le terme
d'interf\'erence (${\cal J}$), qui l'emporte dans l'approximation 
des grands angles (c\^one de bremsstrahlung d'origine non-abelienne).
Si l'on s'\'eloigne de cette r\'egion, soit vers les grandes rapidit\'es
(petits angles), c'est \`a dire vers les r\'egions de 
fragmentation ($\eta_{\text{max}}>\mid\eta\mid>\ln\Theta_\text{d}^{-1}$, 
o\`u $\Theta<\Theta_\text{d}$), les deux c\^ones de bremsstrahlung 
(comme en EDQ) donnent une densit\'e hadronique proportionnelle \`a 
$2\times C_F\approx N_c$.

Dans le cas $WW$, il n'y a qu'une contribution, c'est \`a dire le
terme $\propto C_F$ dans (\ref{eq:colsecteff}). Un ``gap'' \cite{higgs}
de rapidit\'e qui s'\'etend sur l'intervalle
 $\mid\eta\mid<\ln(\sqrt{s}/M_H)$ appara\^it dans la r\'egion 
$\Theta_\text{d}<\Theta$ alors que la densit\'e hadronique dans la 
partie $\Theta<\Theta_\text{d}$, de largeur 
$\Delta\eta=\eta_{max}-\ln\Theta_\text{d}\simeq\ln M_H$, est
donn\'ee par deux pics de hauteur $\propto C_F$ 
(soit une densit\'e $\propto 2\times C_F$). Cet exemple montre comment la CDQ
aide \`a l'identification des canaux de production du Higgs dans le
futur LHC (Large Hadron Collider).

%%%%%%%%%%%%%%%%%%%%%%%%%%%%%%%%%%%%%%%%%%%%%%%%%%%%%%%%%%%%%%%%%%%%%%%%%%%%
%\newpage

%\section{Approximation Doublement Logarithmique (DLA)} 
%\label{sec:ADL}

\chapter{Approximation Doublement Logarithmique (DLA)} 
\label{sec:ADL}

Dans ce chap\^itre nous commen\c cons l'\'etude du sch\'ema probabiliste
qui a men\'e au calcul des observables physiques dans
les jets hadroniques.
Nous donnons, en particulier, les d\'etails techniques expos\'es par Fadin
dans son article \cite{Fadin1} sur l'estimation de la section efficace de 
production de $N$ gluons, o\`u il a d\'evelopp\'e le sch\'ema
de resommation DLA.

%Dans cette partie de la th\`ese, on d\'etaille
%les calculs et raisonnement
%de l'article \cite{Fadin1}.
Nous consid\'erons en premier le cas le plus simple, \`a savoir
l'annihilation $e^+e^-$ en une paire quark-anti-quark, suivie
de l'\'emission de deux gluons de bremsstrahlung
($e^+e^-\!\rightarrow\!q\bar{q}g_1g_2$).
Nous nous limiterons, dans  l'\'emission de gluons mous, \`a l'approximation
qui n\'eglige le recul de la particule charg\'ee \'emettrice.
Elle ne fait intervenir que les contributions doublement
(pour infrarouge et colin\'eaire) logarithmiques dont
on a d\'ej\`a rencontr\'e la forme au
paragraphe \ref{subsection:SERM}, eq.\ref{eq:elly}. Elle constitue
le principale ingr\'edient dans l'estimation des observables physiques que
 l'on peut mesurer dans les grands collisionneurs des particules
 (LEP, Tevatron, LHC), sa contribution, \'etant dominante

\beq\label{eq:LLADLA}
\frac{\alpha_s}{\pi}\ll 1, \qquad \underbrace{\frac{\alpha_s}{\pi}
\log \frac{Q^2}{Q_0^2}\ll1}_{LLA},
\qquad \underbrace{\frac{\alpha_s}{\pi}
\log^2\frac{Q^2}{Q_0^2}\sim1}_{DLA}.
\footnote{\text{$Q$ est la duret\'e du processus et $Q_0$, le
``cut-off'' (masse du gluon) ou l'impulsion transverse minimale du gluon}}
\eeq

Nous d\'ecrivons aussi la structure interne des jets hadroniques
en imposant des contraintes sur les angles d'\'emissions des gluons mous,
compatibles avec (\ref{eq:LLADLA}), de sorte qu'il ne se
produise pas d'interf\'erences destructives entre eux 
({\em coh\'erence} des gluons mous en CDQ,
dont on a d\'ej\`a expliqu\'e l'origine physique au chap\^itre
pr\'ec\'edent).

Une bonne compr\'ehension de ces questions nous permettra de construire
l'amplitude ($M_N$)  associ\'ee \`a la production de $N$ gluons de
bremsstrahlung;
on \'etendra le cas de la production de deux gluons au cas de l'\'emission
de $N$ gluons $e^+e^-\rightarrow q\bar{q}+Ng$. Nous utiliserons les techniques
expos\'ees au chap\^itre pr\'ec\'edent pour simplifier
les expressions obtenues. On remarquera \`a nouveau 
la factorisation des amplitudes molles (ph\'enom\`ene classique)
dans le terme de Born.
Nous tiendrons aussi compte des corrections virtuelles \`a une boucle; ce 
r\'esultat sera toutefois donn\'e sans d\'emonstration.

Nous ferons un r\'esum\'e du formalisme du calcul des jets
\cite{Veneziano} \`a l'aide de la Fonctionelle G\'en\'eratrice
 (FG) qui satisfait une \'Equation Ma\^itresse (EM)
(voir \cite{EvEq} et r\'ef\'erences incluses).
Celle-ci constitue le point de d\'epart dans la construction
des \'equations d'\'evolution que satisfont
les observables inclusives dans les jets hadroniques.
Nous nous int\'eresserons au cas du spectre inclusif d'une particule
\`a l'int\'erieur d'un jet dans le cadre DLA \cite{DLA}\cite{DFK}
et donnerons les id\'ees, ainsi que les techniques qui facilitent la
compr\'ehension des travaux \ref{sub:article1}, \ref{sub:article2}
et \ref{sub:article3}. Nous effectuons, en particulier, le
calcul d\'etaill\'e du spectre par la m\'ethode du col
(voir \ref{sub:article3}).
Nous donnerons, dans le paragraphe \ref{sec:correlations},
les techniques utilis\'ees dans l'article \ref{sub:article2}
pour le calcul exacte des corr\'elations \`a deux particules \`a petit $x$,
dans un jet.

\section{Amplitudes multi-gluoniques \`a l'ordre des arbres pour la
collision \newline $\boldsymbol{e^+e^-\rightarrow q\bar{q}+Ng}$  ($\boldsymbol{N=}$
nombre de gluons mous  colin\'eaires rayonn\'es)}

Nous consid\'erons le processus le plus simple en CDQ, soit la collision 
leptonique $e^+e^-\!\rightarrow\!q\bar{q}$\, \`a tr\`es haute \'energie.

\vskip 0.5cm

\subsection{Notations et variables utilis\'ees}

\beeq
p_+&=&(E_+,\vec{p}_+)--------\text{quadri-impulsion du quark $q$},\nonumber\\
p_-&=&(E_-,\vec{p}_-)------\text{quadri-impulsion de l'anti-quark $\bar{q}$},\nonumber\\
k_i&=&(\omega_i,\vec{k}_i)-------\text{quadri-impulsion du gluon mou $g_i$},\nonumber\\
e^{\lambda}_{i\mu}&\!-\!&\text{quadri-vecteur de polarisation du gluon mou ($\lambda=0,\dots,3$)}.\nonumber,\\
a_i&-& \text{indice de couleur du gluon $i$}.
\eeeq

\subsection{Contrainte \'energ\'etique}

\vskip 0.5cm

Dans l'approximation DLA on cherche la contribution logarithmique 
la plus importante au calcul de la section efficace de production
de $N$ gluons de bremsstrahlung \cite{EvEq}:

\beq\label{eq:sec.eff}
d\sigma_N\propto\int\mid\!\!M_N\!\!\mid^2\prod_{i=1}^N\frac{d^3k_i}{2\omega_i}
\propto (g_s^2\log^2)^N
\eeq

Dans l'objectif d'isoler une telle contribution, on d\'emontrera que
chaque gluon mou et colin\'eaire apporte deux contributions
logarithmiques dont le produit est dominant (voir \ref{eq:LLADLA}).
C'est pour cel\`a que les \'emissions des gluons que l'on consid\`ere 
doivent se succ\'eder de sorte que leur \'energie d\'ecroisse
consid\'erablement pendant l'\'evolution du jet

\beq\label{eq:SEO}
E_{\pm}\gg\omega_1\gg\omega_2\gg\dots\gg\omega_N.
\eeq

En effet, si $\omega_i\sim\omega_k$, il se produirait, au moins,
la perte d'une contribution logarithmique molle.

\vskip 0.5cm

\subsection{Choix de jauge}

\vskip 0.5cm

Pour simplifier l'analyse des logarithmes colin\'eaires,
nous utiliserons une jauge dite ``physique'' o\`u le vertex associ\'e
au gluon dans les cas $q\rightarrow qg$ et 
$g\rightarrow gg$ s'annule pour les \'emissions colin\'eaires.
Il s'agit de la jauge planaire qui a \'et\'e utilis\'ee dans le cas de
la Diffusion Profond\'ement In\'elastique \cite{DDT} (``DIS'' en anglais).
Dans cette jauge, le propagateur du gluon s'\'ecrit sous la forme
\beq\label{jauge}
G^{ab}_{\mu\nu}(k)=\delta^{ab}\frac{d_{\mu\nu}(k)}{k^2+i\epsilon},
\eeq
o\`u
\beq\label{dmunu}
d_{\mu\nu}(k)=g_{\mu\nu}-\frac{k_{\mu}c_{\nu}+c_{\mu}k_{\nu}}{(k\cdot c)}.
\eeq
Nous consid\'erons un vecteur de jauge $c_{\mu}$ proportionnel
\`a la quadri-impulsion
totale de la paire $e^+e^-$, $c_{\mu}=\alpha(p_++p_-)$:
\beeq\nonumber
c_{\mu}&=&(1,\vec{c});\qquad \vec{c}=0\quad \text{dans le r\'ef\'erentiel du centre de masse},\\
\vec{c}&=&\frac{\vec{p}_++\vec{p}_-}{E_++E_-}\quad \text{dans tout autre r\'ef\'erentiel,}\nonumber
\eeeq
c'est \`a dire $\alpha=1/(E_++E_-)$. Maintenant on s'int\'eresse
\`a l'\'evaluation du produit scalaire  $p_+^{\mu}d_{\mu\nu}(k)p_-^{\nu}$.
Nous avons,
\beq\nonumber
p_+^{\mu}d_{\mu\nu}(k)p_-^{\nu}=p_+\cdot p_--\frac{(p_+\cdot k)(c\cdot p_-)+(p_+\cdot c)(k\cdot p_-)}{(k\cdot c)},
\eeq
or
$$
k\cdot c=\frac{k\cdot p_++k\cdot p_-}{E_++E_-},\qquad c\cdot p_+=c\cdot p_-
=\frac{p_+\cdot p_-}{E_++E_-},
$$
donc $p_+^{\mu}d_{\mu\nu}(k)p_-^{\nu}=0$.
Cette int\'eressante propri\'et\'e de la jauge 
planaire sugg\`ere que les quarks $q$ et $\bar{q}$ \'emettent
des gluons ind\'ependamment.
Nous pouvons donc traiter les jets comme des objets qui
n'interf\`erent pas. Dans le cas du ``DIS'', cette propri\'et\'e
a permis de passer des diagrammes de Feynman aux
diagrammes en \'echelle \cite{DDT}.

Dans le calcul exact de la section efficace du processus
(\ref{eq:sec.eff}), on doit
effectuer une somme sur les \'etats de polarisation ``physiques'' 
$e^{(1)}_{\mu}$ et $e^{(2)}_{\mu}$ de $N$ gluons ``r\'eels'':
$$
\left(e^{(1,2)}_i(k_i)\cdot k_i\right)=0, \qquad \left(e^{(1,2)}_i\cdot c\right)=0,
\qquad (e)^2=-1.
$$
Dans l'analyse des diagrammes de Feynman on a des gluons virtuels
portant les quatre \'etats de polarisations possibles
(tranverses et longitudinales).
Cependant, on va d\'emontrer que l'importance des polarisations
longitudinales est n\'egligeable par rapport aux polarisations 
transverses, autrement dit,
la projection des amplitudes sur les \'etats de polarisations transverses
(physiques)
est dominante par rapport \`a celle que l'on obtient lorsqu'on les projette
sur les \'etats de polarisations longitudinales (non-physiques)
$e^{(0)}_{\mu}$ et $e^{(3)}_{\mu}$ (voir \cite{EvEq} et r\'ef\'erences
incluses)

\beq\label{eq:dmunu}
d_{\mu\nu}(k)=-\sum_{\lambda=0}^{3}\,e^{(\lambda)}_{\mu}(k)e^{(\lambda)}_{\nu}(k),
\eeq

\beq\label{eq:pol.longit}
e^{(0,3)}_{\mu}(k)=\frac{k_{\mu}\pm\sqrt{k^2}c_{\mu}}
{\left[2\omega(\omega\pm\sqrt{k^2})\right]^{1/2}}.
\eeq

\section{Amplitude du processus
$\boldsymbol{e^+e^-\!\rightarrow\!q\bar{q}g_1g_2}$}

\begin{figure}[h]
\begin{center}
\includegraphics[height=5.5truecm,width=0.45\tw]{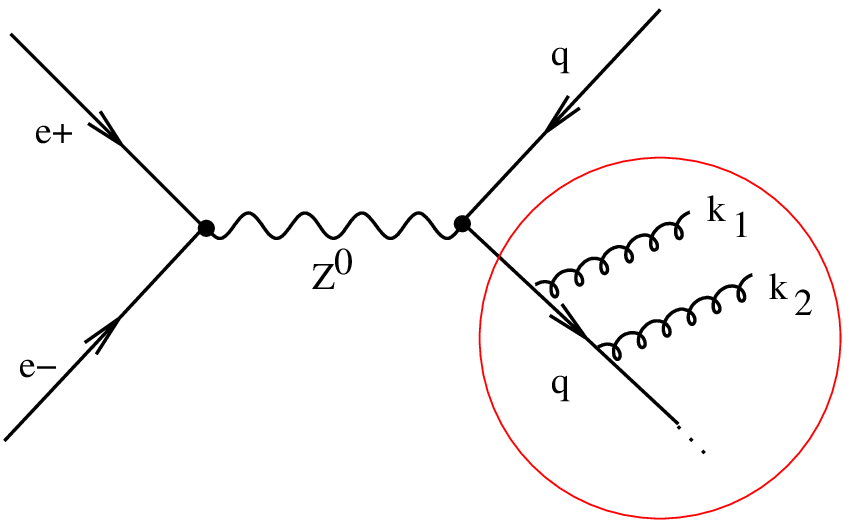}
\hfill
\includegraphics[height=5.5truecm,width=0.45\tw]{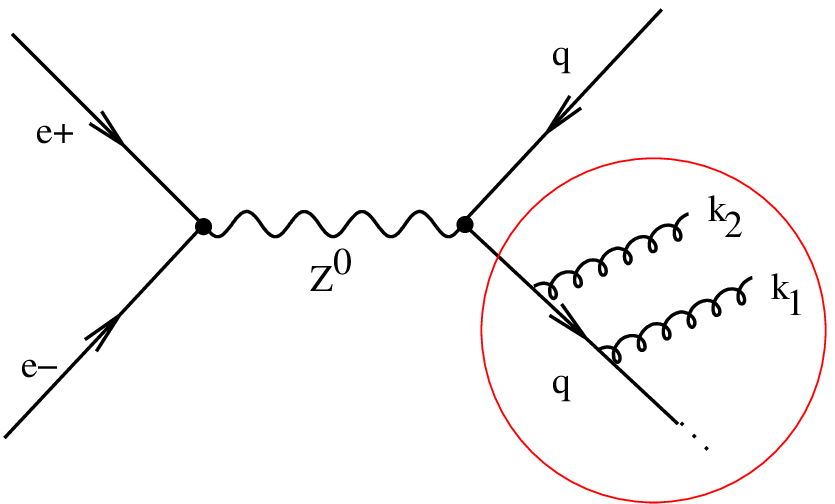}
\caption{\label{fig:bremsst1} Processus consid\'er\'e, 
$e^+e^-\rightarrow q\bar{q}g_1g_2$ \`a gauche et $e^+e^-\rightarrow q\bar{q}g_2g_1$
\`a droite.}
\end{center}
\end{figure}

Les diagrammes de la Fig.\ref{fig:bremsst1} sont du m\^eme type
qu'en EDQ pour le processus que l'on consid\`ere.
L'expression des amplitudes est obtenue \`a partir  des r\`egles de Feynman 
\cite{Peskin}\cite{IZuber}:
\begin{subequations}
\beeq\nonumber
\widetilde{M}^a&=&M_0({e^+e^-\!\!\rightarrow\!q\bar{q}})\frac{m+p\!\!\!/_+-k\!\!\!/_1-k\!\!\!/_2}
{m^2-(p_+-k_1-k_2)^2}\,g_s\,t^{a_2}\,\gamma^{\mu}\,e_{2,\mu}(k_2)\\
&&\frac{m+p\!\!\!/_+-k\!\!\!/_1}
{m^2-(p_+-k_1)^2}\,g_s\,t^{a_1}\,\gamma^{\nu}\,e_{1,\nu}(k_1)\,u(p_+,s_+),\label{eq:Maa}
\eeeq
\beeq\nonumber
\widetilde{M}^b&=&M_0({e^+e^-\!\!\rightarrow\!q\bar{q}})\frac{m+p\!\!\!/_+-k\!\!\!/_2-k\!\!\!/_1}
{m^2-(p_+-k_2-k_1)^2}\,g_s\,t^{a_1}\,\gamma^{\mu}\,e_{1,\mu}(k_1)\\
&&\frac{m+p\!\!\!/_+-k\!\!\!/_2}
{m^2-(p_+-k_2)^2}\,g_s\,t^{a_2}\,\gamma^{\nu}\,e_{2,\nu}(k_2)\,u(p_+,s_+)
,\label{eq:Mb}
\eeeq
\begin{figure}[h]
\begin{center}
\includegraphics[height=6truecm,width=0.5\tw]{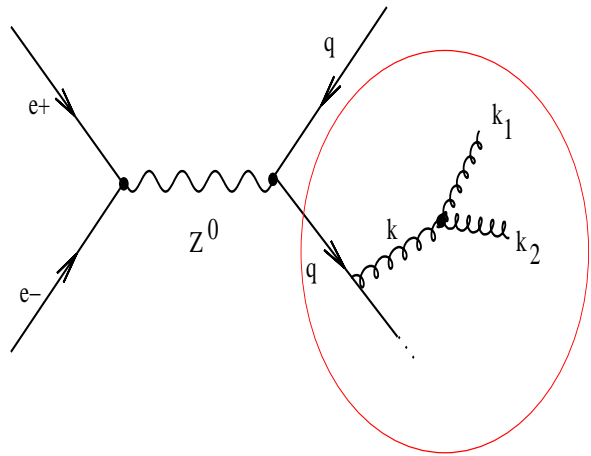}
\caption{\label{fig:bremsst2} Processus consid\'er\'e, 
$e^+e^-\rightarrow q\bar{q}g_1g_2$ avec un vertex \`a trois gluons, d\'ej\`a
\'evoqu\'e au chap\^itre pr\'ec\'edent}\label{fig:Mc}
\end{center}
\end{figure}
En CDQ on doit ajouter une amplitude suppl\'ementaire, Fig.\ref{fig:Mc},
o\`u l'on inclut le couplage non-abelien \`a trois gluons,
\beeq\nonumber
\widetilde{M}^c&=&M_0({e^+e^-\!\!\rightarrow\!q\bar{q}})\frac{m+(p\!\!\!/_+-k\!\!\!/)}{m^2-(p_+-k)^2}\,g_s\,\gamma^{\sigma}
\,\frac{d_{\rho\sigma}(k)}{k^2}\\
&&if_{a_1a_2c}\,t^c\,g_s\,e_1^{\mu}(k_1)\,e_2^{\nu}(k_2)\,
\gamma_{\mu\nu\rho}(k_1,k_2,-k)\,u(p_+,s_+),
\label{eq:Mc}
\eeeq
\end{subequations}
$\gamma_{\mu\nu\rho}$ est le facteur de Lorentz associ\'e au
vertex non-abelien \cite{Cheng}
\beeq\label{eq:gamma3gluons}
\gamma_{\mu\nu\rho}(k_1,k_2,k_3)&=&g_{\mu\nu}(k_2-k_1)_{\rho}+g_{\rho\mu}(k_1-k_3)_{\nu}+g_{\nu\rho}
(k_3-k_2)_{\mu}\\
&=&g_{\mu\nu}(k_2-k_1)_{\rho}+g_{\rho\mu}(2k_1+k_2)_{\nu}-g_{\nu\rho}
(k_1+2k_2)_{\mu}.\nonumber
\eeeq
\begin{figure}[h]
\begin{center}
\includegraphics[height=5truecm,width=0.4\tw]{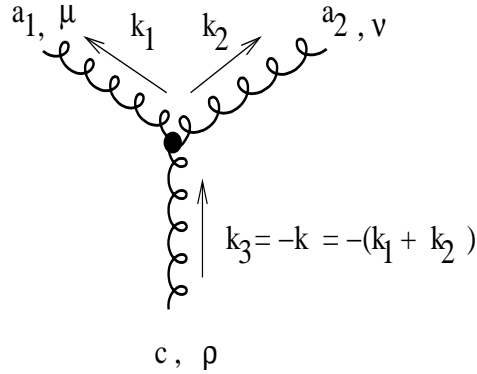}
\caption{\label{fig:threegvertex} Vertex \`a trois gluons.}
\end{center}
\end{figure}
Nous simplifions (\ref{eq:Maa}) en utilisant les astuces du paragraphe
(\ref{subsection:CAM}) pour l'\'emission de gluons
de bremsstrahlung ($\omega_i\ll E_{\pm}$).
Dans ce cas on effectue une suite de transformations
qui n\'egligent le recul des partons 
$$
m+p\!\!\!/_+-k\!\!\!/_1-k\!\!\!/_2\approx m+p\!\!\!/_+-k\!\!\!/_1
\approx m+p\!\!\!/_+,
$$
$$
m^2-(p_+-k_1)^2=m^2-p_+^2+2k_1.p_+-k_1^2\approx2(k_1.p_+),
$$
\beeq\nonumber
(m+p\!\!\!/_+)\gamma^{\nu}e_{1,\nu}(k_1)u(p_+,s_+)\!&\!\!=\!\!&\!\left[2p_+^{\nu}e_{1,\nu}(k_1)+
\gamma^{\nu}e_{1,\nu}(k_1)(m-p\!\!\!/_+)\right]u(p_+,s_+)\\
\!&\!\!=\!\!&\!2(e_1p_+)u(p_+,s_+),\nonumber
\eeeq
$$
m^2-(p_+-k_1-k_2)^2=m^2-p_+^2+2(k_1+k_2).p_+-k_1^2-k_2^2\approx2(k_1+k_2).p_+,
$$
et finalement
$$
(m+p\!\!\!/_+)\gamma^{\mu}e_{2,\mu}(k_2)u(p_+,s_+)=2(e_2\cdot p_+)u(p_+,s_+).
$$
L'amplitude de la Fig.\,\ref{fig:bremsst1} \`a gauche prend la forme 
$\widetilde{M}^a\!=\!M_0u(p_+,s_+)M^a$ o\`u
\beq\label{eq:Mabis}
M^a=g_s^2\,\frac{e_2\cdot p_+}{k_2\cdot p_+}\,
\frac{e_1\cdot p_+}{(k_1+k_2)\cdot p_+}\,t^{a_2}t^{a_1}
\eeq
est l'amplitude associ\'ee au processus mou, tandis que $M_0u_+$
est le terme de Born.
L'amplitude de la 
Fig.\,\ref{fig:bremsst1} \`a droite s'obtient par une simple
permutation des indices $1$ et $2$, soit $\widetilde{M}^b=M_0u(p_+,s_+)M^b$, o\`u
\beq\label{eq:Mbbis}
M^b=g_s^2\,\frac{e_1\cdot p_+}{k_1\cdot p_+}\,
\frac{e_2\cdot p_+}{(k_2+k_1)\cdot p_+}\,t^{a_1}t^{a_2}.
\eeq
Nous constatons encore une fois la factorisation du terme de bremsstrahlung
de la partie non-radiative. On trouve de plus, dans (\ref{eq:Mabis}) et
(\ref{eq:Mbbis}), le produit de deux facteurs de Lorentz
de la forme $\frac{p_i^{\mu}}{(kp_i)}$ qui correspondent
aux \'emissions ind\'ependantes des deux gluons. En effectuant les
m\^emes simplifications on arrive \`a l'expression
$\widetilde{M}^c=M_0u(p_+,s_+)M^c$, o\`u
\beq\label{eq:Mcbis}
M^c=g_s^2e_1^{\mu}e_2^{\nu}\gamma_{\mu\nu\rho}(k_1,k_2,-k)\frac{d^{\,\rho\sigma}(k)}{k^2}
\frac{p_{+\sigma}}{k\cdot p_+}if_{a_1a_2c}t^c.
\eeq
L'int\'egration sur les directions des \'emissions des gluons $\vec{n}_i$ 
(voir \ref{subsection:roleinterf}) n'appara\^it que dans les 
d\'enominateurs des propagateurs de Feynman (quarks, gluon virtuel) car 
$d^{\,{\rho\sigma}}(k)$ ne pr\'esente
pas de d\'ependance angulaire dans le d\'enominateur. 
Par cons\'equent, les  r\'egions cin\'ematiques o\`u les int\'egrations
sur les 
directions $\vec{n}_1$ et $\vec{n}_2$ donnent les contributions logarithmiques 
d'origine molle et colin\'eaire les plus importantes sont les suivantes:
\begin{subequations}
\beeq
k_1\cdot p_+\approx \frac{E_+}2\omega_1\Theta_1^2\gg k_2\cdot p_+\approx \frac{E_+}2\omega_2\Theta_2^2\Longleftrightarrow
\Theta_1^2&\gg&\frac{\omega_2}{\omega_1}\Theta_2^2,\label{eq:cont1}\\
k_2\cdot p_+\approx \frac{E_+}2\omega_2\Theta_2^2\gg k_1\cdot p_+\approx \frac{E_+}2\omega_1\Theta_1^2\Longleftrightarrow
\Theta_2^2&\gg&\frac{\omega_1}{\omega_2}\Theta_1^2,\label{eq:cont2}\\
k\cdot p_+\approx\frac{E_+}2(\omega_1\Theta^2_1+\omega_2\Theta^2_2),
\quad k^2\approx\omega_1\omega_2\Theta^2_{12};\quad
\frac{\omega_1}{\omega_2}\Theta_1^2&\gg&\Theta_2^2\gtrsim\Theta_{1}^2,
\label{eq:cont3}
\eeeq 
\end{subequations}
pour les amplitudes (\ref{eq:Mabis}), (\ref{eq:Mbbis}) et (\ref{eq:Mcbis}) respectivement,
avec, $\Theta_i=(\vec{k}_i,\vec{p}_+)$, $\Theta_{12}=(\vec{k}_1,\vec{k}_2)$. 

\bigskip

\textbf{Exemple}: Avec (\ref{eq:cont1}), le d\'enominateur de
(\ref{eq:Mabis}) s'estime sans difficult\'e
$$
\frac1{(k_2\cdot p_+)}\times\frac1{(k_1+k_2)\cdot p_+}\simeq
\frac1{\omega_2\Theta_2^2(\omega_1\Theta_1^2+\omega_2\Theta_2^2)}\sim
\frac1{\omega_1\Theta_1^2}\times\frac1{\omega_2\Theta_2^2}.
$$
Apr\`es avoir pris le module au carr\'e et avoir int\'egr\'e sur
l'espace de phase on peut \'ecrire l'expression de la section efficace
de fa\c con symbolique
$$
d\sigma_2\propto\left(\frac{\alpha_s}{\pi}\right)^2\int
\left(\frac{d\omega_1}{\omega_1}\frac{d\Theta_1^2}{\Theta_1^2}\right)
\int\left(\frac{d\omega_2}{\omega_2}\frac{d\Theta_2^2}{\Theta_2^2}\right)
\sim\left(\frac{\alpha_s}{\pi}\log^2\right)^2.
$$
Ainsi, pour $N=2$, on a $(\log^2)^2$, soit, deux
contributions molles et colin\'eaires, soit $(\log^2)^2$ 
dans le calcul de la section efficace $d\sigma_2$
du processus, c'est \`a dire la plus grande contribution
logarithmique. Si l'on prend $\omega_1\Theta_1^2\sim \omega_2\Theta_2^2$
o\`u $\omega_1\Theta_1^2\ll \omega_2\Theta_2^2$
(on fait cro\^itre l'angle $\Theta_2$ consid\'erablement 
dans les deux cas car $\omega_1\gg\omega_2$) on obtiendrait
$$
\frac1{(k_2\cdot p_+)}\times\frac1{(k_1+k_2)\cdot p_+}\sim\left(\frac1{\omega_2\Theta_2^2}\right)^2,
$$
dans lequel se produit la perte d'une contribution logarithmique.
C'est pour cel\`a que l'on doit ajouter l'amplitude (\ref{eq:Mbbis})
qui est compatible avec la contrainte (\ref{eq:cont2}).
Avec la condition (\ref{eq:cont3}) pour l'amplitude (\ref{eq:Mcbis}),
on obtient une double contribution logarithmique \`a condition
que $\Theta_2$ soit sup\'erieur \`a $\Theta_1$, c'est \`a dire
$\Theta_2\gtrsim\Theta_{1}$. Dans le cas contraire, si on fait cro\^itre
$\Theta_{2}$ $(\Theta_2\gg\Theta_1)$ jusqu'\`a la limite
$\omega_1\Theta_1^2\sim\omega_2\Theta_2^2$ 
on aurait $\Theta_2\sim\Theta_{12}$, et on perdrait
un $\log$ d'origine colin\'eaire.

On doit donc simplifier (\ref{eq:Mcbis}) et d\'emontrer que le r\'esultat
peut se mettre sous la forme (\ref{eq:Mabis}) et (\ref{eq:Mbbis}). 
On utilise la condition (\ref{eq:cont3}) pour estimer les \'etats
de polarisation du gluon virtuel $k$. On tient compte de
la contrainte $\omega_1\gg\omega_2\Longleftrightarrow k_1\gg k_2,
\quad k=k_1+k_2\approx k_1$. On oriente le vecteur $\vec{k}$ le long
de l'axe $\vec{u}_z$, ainsi $\vec{k}_1$, $\vec{k}_2$ et $\vec{k}$
se trouvent sur le m\^eme plan
($\vec{u}_x, \vec{u}_z$). Avec ceci, les \'etats de polarisations
transverses du gluon virtuel sont sur le plan ($\vec{u}_x, \vec{u}_y$).
Avant de passer au calcul, on donne les outils suivants
$$
\vec{k}=\vec{k}_1+\vec{k}_2,\, \text{on proj\`ete sur}\quad \vec{k}_2\,
\Rightarrow \vec{k}\cdot\vec{k}_2=\mid\!\!\vec{k}\!\!\mid\omega_2\cos(\vec{k},\vec{k}_2)=
\vec{k}_1\cdot\vec{k}_2+\omega_2^2\approx\omega_1\omega_2\cos\Theta_{12},
$$
puis, si l'on n\'eglige la virtualit\'e du gluon, on obtient
$(\vec{k},\vec{k}_2)\approx\Theta_{12}$.
De plus, les polarisations transverses des gluons quasi-r\'eels se localisent
sur un plan perpendiculaire aux vecteurs $\vec{k}_i$.

On projette d'abord le quadri-vecteur
$e_1^{\mu}e_2^{\nu}\gamma_{\mu\nu\rho}$ sur les \'etats de polarisations 
transverses du gluon virtuel 
\footnote{\text{dans (\ref{eq:projgluon}) on 
d\'emontre que le terme du milieu dans (\ref{eq:gamma3gluons})
est dominant.}}
\beeq\nonumber
e_1^{\mu}e_2^{\nu}\gamma_{\mu\nu\rho}(k_1,k_2,-k)e^{(1,2)}_{\rho}(k)&\approx&
-(e_1\cdot e_2)\left[e^{(1,2)}(k)\cdot k_1\right]
+2[e_1\cdot e^{(1,2)}(k)][e_2\cdot k_1]\\ \nonumber\\
&-&[e_2\cdot e^{(1,2)}(k)](e_1\cdot k_1)
=2[e_1^{(1,2)}\cdot e^{(1,2)}(k)](e_2\cdot k_1)\label{eq:projgluon}\nonumber\\ \nonumber \\
&\approx&2(e_2\cdot k_1)=2(e_2^{(1,2)}\cdot k_1)\sim\omega_1
\cos(\frac{\pi}2-\Theta_{12})\nonumber\\ \nonumber\\
&\sim&\omega_1\Theta_{12}.
\eeeq
Finalement (voir (\ref{eq:pol.longit}))
\beq
e_1^{\mu}e_2^{\nu}\gamma_{\mu\nu\rho}(k_1,k_2,-k)e^{(1,2)}_{\rho}(k)\sim\omega_1\Theta_{12}
\gg \sqrt{\omega_1\omega_2\Theta_{12}^2}\approx\sqrt{k^2}\sim
e_1^{\mu}e_2^{\nu}\gamma_{\mu\nu\rho}e^{(0,3)}_{\rho}(k).\label{eq:ineq1}
\eeq
On s'int\'eresse maintenant \`a la projection de l'impulsion du quark
\'emetteur sur les \'etats de polarisations transverses du gluon
interm\'ediaire ($k$):
\beq
e^{(1,2)}_{\sigma}(k)p_+^{\sigma}=E_+\cos(\frac{\pi}2-\Theta)\approx E_+\Theta\sim E_+\Theta_1,
\eeq
on utilise la contrainte (\ref{eq:cont3}), et l'on peut
\'ecrire ($\Theta\sim\Theta_1$)
\footnote{dans tous les cas $\Theta$, $\Theta_1$,
$\Theta_2$ et $\Theta_{12}\ll1$.}
\beeq
e^{(1,2)}_{\sigma}(k)p_+^{\sigma}&\sim& E_+\Theta_1\sim\frac{E_+}{\omega_1}
(\omega_1\Theta_1+\omega_2\Theta_2)\gg\frac{E_+}{\omega_1}
(\omega_1\Theta_1^2+\omega_2\Theta_2^2
\pm\sqrt{\omega_1\omega_2\Theta_{12}^2})\label{eq:ineq2}\nonumber\\ \\
&\sim&\frac{(p_+\cdot k)\pm\sqrt{k^2}(p_+\cdot c)}{\omega}\sim
e^{(0,3)}_{\sigma}(k)p_+^{\sigma}.\nonumber
\eeeq
Les in\'egalit\'es (\ref{eq:ineq1}) et (\ref{eq:ineq2}) permettent
de n\'egliger les polarisations longitudinales par rapport
aux polarisations transverses du gluon virtuel. 
Par cons\'equent, le propagateur $d_{\rho\sigma}$ dans l'amplitude
(\ref{eq:Mcbis}) peut \^etre remplac\'e par le tenseur transverse
\beq\label{eq:tentra}
g^{\perp}_{\rho\sigma}\equiv-\sum_{\lambda=1,2}e^{(\lambda)}_{\rho}(k)e^{(\lambda)}_{\sigma}(k),
\eeq
$$
g_{\rho0}=g_{0\sigma}=g_{00}=0,
$$
$$
g_{ij}=\delta_{ij}-\frac{k_ik_j}{\vec{k}^2}\qquad (i,j=1,2,3),
$$
et le vertex du gluon est domin\'e par le terme
$\propto g_{\rho\mu}$ tel qu'il a
\'et\'e d\'emontr\'e dans (\ref{eq:projgluon}) 
$$
\gamma_{\mu\nu\sigma}(k_1,k_2,-k)\approx 2g_{\mu\rho}k_{1\nu},
$$
puis
\beeq
e^{\mu}_1e^{\nu}_2\gamma_{\mu,\nu\sigma}(k_1,k_2,-k)d_{\rho\sigma}p_+^{\sigma}&\approx&2
(e_2\cdot k)e^{\mu}_1g^{\perp}_{\rho\sigma}p_+^{\sigma}\nonumber\\
&=&2(e_2\cdot k)\left[-(\vec{e}_1\cdot\vec{p}_+)
+\frac{(\vec{e}_1\cdot\vec{k})(\vec{k}\cdot\vec{p}_+)}{\vec{k}^2}\right]
\label{eq:sproduct1}.
\eeeq
Maintenant on utilise (\ref{eq:cont3}),
$\vec{e}_1\cdot\vec{k}=\vec{e}_1\cdot\vec{k}_2\approx\omega_2\Theta_{12}$,
$\vec{k}\cdot\vec{p}_+\approx E_+\omega_1$, $\vec{k}^2\approx\omega_1^2$,
$$
(\vec{e}_1\cdot\vec{p}_+)=E_+\cos(\frac{\pi}{2}-\Theta_1)\approx E_+\Theta_1
\gg \frac{(\vec{e}_1\cdot\vec{k})(\vec{k}\cdot\vec{p}_+)}{\vec{k}^2}\sim E_+
\frac{\omega_2}{\omega_1}\Theta_{12},
$$
on peut donc n\'egliger le second terme \`a l'int\'erieur des crochets dans
(\ref{eq:sproduct1}) et r\'ecrire
$$
e^{\mu}_1e^{\nu}_2\gamma_{\mu,\nu\sigma}(k_1,k_2,-k)d_{\rho\sigma}p_+^{\sigma}
\approx-2(e_2\cdot k_1)(\vec{e}_1\cdot\vec{p}_+)=2(e_2\cdot k_1)
(e_1^{\mu}g_{\mu\sigma}p_+^{\sigma})=2(e_2\cdot k_1)(e_1\cdot p_+).
$$
Avec ceci, $k^2\approx 2k_1\cdot k_2$, $k\cdot p_+=(k_1+k_2)\cdot 
p_+$, et on obtient
\beq\label{eq:Mcter}
M^c=g_s^2\frac{e_2\cdot k_1}{k_1\cdot k_2}\,
\frac{e_1\cdot p_+}{(k_1+k_2)\cdot p_+}\,if_{a_1a_2c}t^c.
\eeq
\section{Contrainte angulaire pour $\boldsymbol{N=2}$ (nombre de gluons
rayonn\'es)}

Les amplitudes (\ref{eq:Mabis}), (\ref{eq:Mbbis}) et (\ref{eq:Mcter})
peuvent se simplifier davantage si l'on utilise les contraintes sur
les angles d'\'emission li\'ees \`a la coh\'erence.
Par exemple, les r\'egions d'int\'egration (\ref{eq:cont1}) et 
(\ref{eq:cont3}) se chevauchent et les amplitudes correspondantes
(\ref{eq:Mabis}) et (\ref{eq:Mcter}) interf\`erent.
On d\'emontre qu'en utilisant les contraintes angulaires ind\'ependantes
suivantes, on \'elimine ces interf\'erences (on appelle
$\Theta_{q\bar{q}}\equiv\Theta_{\pm}$):
\begin{subequations}
\beq\label{eq:contr1}
I. \qquad \Theta_{\pm}\quad>\quad\Theta_1\quad\gg\quad\Theta_2,
\eeq
\beq\label{eq:contr2}
II. \qquad \Theta_{\pm}\quad>\quad\Theta_2\quad\gg\quad\Theta_1,
\eeq
\beq\label{eq:contr3}
III. \qquad \Theta_{12}\quad\ll\quad\Theta_1\approx\Theta_2\quad<\quad\Theta_{\pm}.
\eeq
\end{subequations}
Dans la r\'egion $I$, la seule contribution est donn\'ee par
l'amplitude $M^a$; cette r\'egion angulaire est bien compatible avec
la contrainte (\ref{eq:cont1}).
Dans cette configuration, les deux gluons mous n'interf\`erent pas et
les deux \'emissions sont ind\'ependantes. En effectuant
l'approximation $k_1p_+\gg k_2p_+$, on peut simplifier $M^a$
$$
M_I=g_s^2\frac{e_2\cdot p_+}{k_2\cdot p_+}\,\frac{e_1\cdotp_+}
{k_1\cdot p_+}t^{a_2}t^{a_1}.
$$
La contrainte $II$ se divise en deux sous-r\'egions cin\'ematiques
possibles:
$$
\Theta_2^2\quad\gg\quad\frac{\omega_1}{\omega_2}\Theta_1^2,\quad 
\text{$\Theta_1$ doit \^etre tr\`es petit car $\frac{\omega_1}{\omega_2}
\!\gg\!1$}
$$
et
$$
\frac{\omega_1}{\omega_2}\Theta_1^2\quad\gg\quad\Theta_2^2\quad\gg\quad\Theta_{1}^2,\quad
\text{pourvu que $\omega_1\gg\omega_2$}.
$$
Dans le premier cas (le m\^eme que \ref{eq:cont2}), il n'y a que
$M^b$ qui domine, avec $k_2\cdot p_+\gg k_1\cdot p_+$, et on peut \'ecrire
\beq
M_{II}=g_s^2\frac{e_1\cdot p_+}
{k_1\cdot p_+}\,\frac{e_2\cdot p_+}{k_2\cdot p_+}t^{a_1}t^{a_2}\label{eq:Mbter}.
\eeq
Dans la deuxi\`eme sous-r\'egion on doit tenir compte des amplitudes $M^a$ et
$M^c$ puisqu'elles sont compatibles avec (\ref{eq:cont1}) et
(\ref{eq:cont2}) respectivement.
Cependant, $\Theta_{12}\approx\Theta_2$ car $\vec{k}_1$ et 
$\vec{p}_+$ sont quasi-colin\'eaires. Ceci permet d'estimer
$$
\frac{e_2\cdot k_1}{k_2\cdot k_1}\approx
\frac{e_2\cdot p_+}{k_2\cdot p_+},\quad k_1\cdot p_+\gg k_2\cdot p_+.
$$
On \'evalue la somme $M^a+M^c$ dans cette approximation, soit
\beeq\nonumber
M^a+M^c&\approx& g_s^2\frac{e_2\cdot p_+}
{k_2\cdot p_+}\,\frac{e_1\cdot p_+}{k_1\cdot p_+}\left[t^{a_2}t^{a_1}
+if_{a_1a_2c}t^c\right]\\
&=&g_s^2\frac{e_2\cdot p_+}{k_2\cdot p_+}\,\frac{e_1\cdot p_+}
{k_1\cdot p_+}t^{a_1}t^{a_2}\equiv M_{II}.\nonumber
\eeeq
Donc (\ref{eq:Mbter}) a lieu dans toute la r\'egion cin\'ematique que
l'on consid\`ere
\footnote{il s'agit du m\^eme cas qui s'applique \`a la production
du boson de Higgs, voir \ref{subsection:Higgs}.}. 
Finalement, l'amplitude $M^c$ domine dans la r\'egion $III$ o\`u la paire
$g_1g_2$ est consid\'er\'ee quasi-colin\'eaire.
Ici $k_1\cdot p_+\gg k_2\cdot p_+$ et on a 
$$
M_{III}=g_s^2\frac{e_2\cdot k_1}{k_2\cdot k_1}\,\frac{e_1\cdot p_+}
{k_1\cdot p_+}if_{a_1a_2c}t^c.
$$
En effet, l'amplitude totale associ\'ee \`a l'\'emission des
gluons mous dans ce processus est donn\'ee par la somme des 3 amplitudes
\beeq\nonumber
M_2^{\text{tot}}\!&\!=\!&\!
M_I+M_{II}+M_{III}=g_s^2\left[\frac{e_2\cdot p_+}
{k_2\cdot p_+}\,
\frac{e_1\cdot p_+}{k_1\cdot p_+}t^{a_2}t^{a_1}
+\frac{e_1\cdot p_+}{k_1\cdot p_+}\,\frac{e_2\cdot p_+}
{k_2\cdot p_+}
t^{a_1}t^{a_2}\right.\\
&&\hskip 4.2cm\left.+\frac{e_2\cdot k_1}{k_2\cdot k_1}\,\frac{e_1\cdot p_+}{k_1\cdot p_+}if_{a_1a_2c}t^c\right],
\eeeq
mais elle se r\'eduit \`a $M_2^{\text{tot}}=M_I$ dans la r\'egion $I$, \`a
$M_2^{\text{tot}}=M_{II}$ dans la r\'egion $II$ et/ou \`a
 $M_2^{\text{tot}}=M_{III}$ dans la r\'egion $III$. Ces aspects ont \'egalement
\'et\'e trait\'es dans \cite{EvEq}.

L'id\'ee de l'approximation DLA consiste \`a consid\'erer tous les 
diagrammes dont la contribution au calcul de la section efficace
est doublement logarithmique (gluons mous et colin\'eaires), dans une
configuration bien pr\'ecise pour les angles d'\'emission.
On peut en faire le bilan en pr\'ecisant les \'etapes de la chaine de
transformations qui conduisent aux r\'esultats.

\bigskip

$\ast$ Fig.\ref{fig:bremsst1}a(b): le quark de quadri-impulsion $p_+$
\'emet d'abord un gluon $k_1(k_2)$, puis $k_2(k_1)$;
on attribue la contrainte angulaire $I(II)$ 
\`a ce diagramme; en outre, $E_+\gg\omega_1\gg\omega_2$;

\bigskip

$\ast$ Fig.\ref{fig:bremsst2}c: le quark \'emet un gluon $k_1$ qui,
\`a son tour, \'emet le gluon de plus faible impulsion $k_2$;
on lui attribue la contrainte angulaire $III$;

\bigskip

$\ast$ on associe un facteur de Lorentz classique de la forme 
$\displaystyle{\frac{e\cdot p}{k\cdot p}}$ \`a chaque 
\'emission molle;

\bigskip

$\ast$ les angles d'\'emission sont contraints de diminuer le long de
la cha\^\i ne des \'emissions (contrainte angulaire);

\bigskip

$\ast$ on associe un facteur de couleur \`a chaque diagramme.

\bigskip

Ceci se g\'en\'eralise aux \'emissions des gluons par l'anti-quark.

\section{Contrainte angulaire \`a tous les ordres}

\begin{figure}[h]
\begin{center}
\includegraphics[height=6truecm,width=0.8\tw]{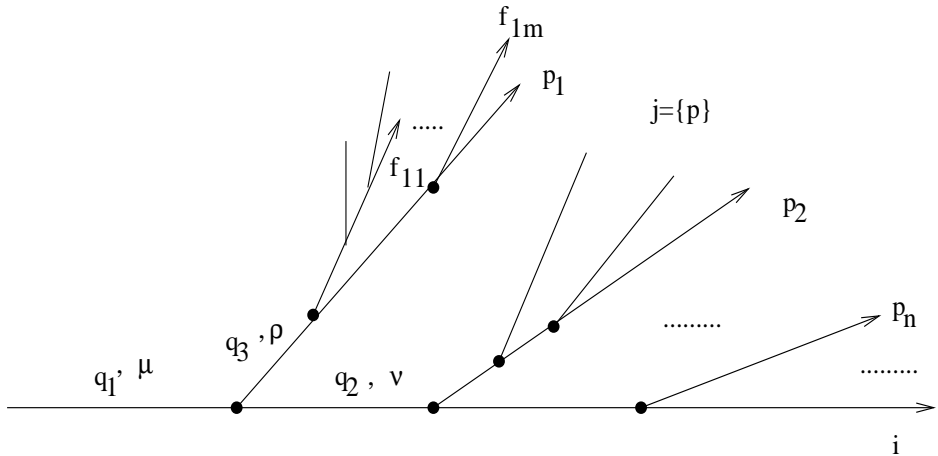}
\caption{\label{fig:Ngluon} Sch\'ema de multiplication d'un gluon
\'emis par $q$ ou $\bar{q}$}
\end{center}
\end{figure}

Dans la Fig.\ref{fig:Ngluon} on consid\`ere l'\'emission de 
$N$ gluons de bremsstrahlung dans un jet initi\'e par un parton $i$ qui a
\'et\'e \'emis par le quark $q$ ou l'anti-quark $\bar{q}$.
Chaque point le long de $i$ repr\'esente un vertex \`a trois 
partons o\`u l'on appelle $i$ le parent et $j$ l'enfant, tel que 
$i\rightarrow i+j$, on a $j=p_1,\,p_2,\dots$ ($p$ labelle le p\`ere).
L'\'energie de chaque gluon virtuel le long de chaque 
branche est consid\'er\'ee du m\^eme ordre de grandeur, ce qui est en accord 
avec la condition (\ref{eq:SEO}), en d'autres mots, on n\'eglige
le recul des gluons \'emetteurs (c'est pour cel\`a que l'on dessine des
lignes droites continues). De plus, on associe la r\'egion
$\Gamma_{{\cal {V}}_{ba}}$ \`a l'espace des angles d'\'emission
qui correspond \`a ${\cal {V}}_{ba}$: le vertex $a\rightarrow a+b$,
tel que si $a=i$, on a $i\rightarrow i+j$ le long de $i$ et si $a=j$,
alors $j\rightarrow j+f$ le long de $j$ ($f$ le parton enfant).
Les angles des gluons rayonn\'es doivent d\'ecro\^itre
le long de la branche choisie en vertu de la coh\'erence des gluons mous.
Cette contrainte doit \^etre satisfaite
\`a compter du vertex $\gamma^*/Z^0\rightarrow q\bar{q}$.
On a, d'apr\`es la contrainte angulaire, et pour chaque vertex,
le long de la m\^eme branche:
\beeq
&&\hskip -0.7cm\Gamma_{{\cal {V}}_{p_1\pm}}(k_{p_1}, \Theta_{p_1\pm}):
\left\{k^0_{p_1}\equiv\omega_{p_1}\!\ll\! E_{\pm};\; 
\Theta_{{p_1}\pm}\!\ll\!\Theta_{\pm};
\; k_{\perp {p_1}}\!\approx\!\omega_{p_1}
\Theta_{p_1\pm}\!>\!Q_0\right\};\quad\text{si $i=q(\bar{q})$},\cr\cr
\label{eq:inequal1}
&&\hskip -0.7cm\Gamma_{{\cal {V}}_{f_{11}p_1}}(k_{f_{11}}, \Theta_{f_{11}p_1}):
\left\{k^0_{f_{11}}\equiv\omega_{f_{11}}\!\ll\! \omega_{p_1};\; \Theta_{f_{11}p_1}\!
\ll\!\Theta_{p_1i};
\; k_{\perp f_{11}}\!\approx\!\omega_{f_{11}}\Theta_{f_{11}p_1}\!>\!Q_0\right\},\\
&&\hskip 7cm\cdots\cr
&&\hskip -0.7cm\Gamma_{{\cal {V}}_{f_{1m}p_1}}(k_{f_{1m}}, \Theta_{f_{1m}p_1})\!:\!
\left\{k^0_{{f_{1m}}}\!\equiv\!\omega_{f_{1m}}\!\ll\! \omega_{p_1};\;
\Theta_{f_{1m}p_1}\!\ll\!\Theta_{f_{1m\!-\!1}p_1};
\; k_{\perp f_{1m}}\!\approx\!\omega_{f_{1m}}\Theta_{f_{1m}p_1}\!>\!Q_0\right\},
\notag
\eeeq
$\Theta_{ba}$ est l'angle d'\'emission du gluon $b$ par rapport
\`a son parton parent $a$ d'impulsion $k_a$, 
$\Theta_{\pm}$ est l'angle entre le quark et l'anti-quark.
(\ref{eq:inequal1}) se g\'en\'eralise \`a chaque \'emission
molle $j=p_1,p_2,\dots$
On associe \`a toute la branche ${\cal {B}}_{ji}$ 
l'espace des angles $\Gamma_{{\cal {B}}_{ji}}$ qui est donn\'e
par l'union de toutes les r\'egions angulaires attribu\'ees \`a chaque vertex:
\beq\label{eq:unionv}
\Gamma_{{\cal {B}}_{p_1i}}(k_{p_1i}, \Theta_{p_1i})=
\Gamma_{{\cal {V}}_{p_1i}}(k_{p_1}, \Theta_{p_1i})\cup
\Gamma_{{\cal {V}}_{f_{11}p_1}}(k_{f_{11}}, \Theta_{f_{11}p_1})\cup\dots\cup
\Gamma_{{\cal {V}}_{f_{1m}p_1}(k_{f_{1m}}}, \Theta_{f_{1m}p_1}),
\eeq
tandis que
\beq\label{eq:interv}
\Gamma_{{\cal {V}}_{p_1i}}(k_{p_1}, \Theta_{p_1i})\!\cap\!
\Gamma_{{\cal {V}}_{f_{11}p_1}}(k_{f_{11}}, \Theta_{f_{11}p_1})\!\cap\!\dots\!\cap\!
\Gamma_{{\cal {V}}_{f_{1m}p_1}}(k_{f_{1m}}, \Theta_{f_{1m}p_1})\!=\!0
% \Gamma_{{\cal {V}}_{f_{1m}p_1}}(k_{f_{1m}}, \Theta_{f_{1m}p_1}).
\eeq
traduit leur ind\'ependance, c'est \`a dire que
les r\'egions ne se chevauchent pas. (\ref{eq:unionv}) 
et (\ref{eq:interv}) se g\'en\'eralisent $\forall j$.
Cette branche constitue un jet.
L'angle $\Theta_{ji}$ est tr\`es sup\'erieur \`a la dimension
angulaire du jet et \'egalement aux angles d'\'emission des gluons
mous qui le constituent.
On peut de m\^eme \'ecrire la condition suivante pour les angles, qui
d\'ecoule de (\ref{eq:unionv}) et (\ref{eq:interv})
\beq\label{eq:inequal1.1}
(\vec{k}_j,\vec{k}_i)\gg(\vec{k}_{f_{j1}},\vec{k}_j)\gg(\vec{k}_{f_{j2}},\vec{k}_j)
\gg\dots(\vec{k}_{f_{jm}},\vec{k}_j).
\eeq
On appelle 
$\Gamma_{{\cal {D}}}$ la r\'egion enti\`ere donn\'ee par l'union de toutes
les branches $p_1,\,p_2,\dots,p_n$ et,
par cons\'equent, associ\'ee \`a la production de $N$ gluons
(pour tout le diagramme $\cal {D}$):
\beq\nonumber
\Gamma_{{\cal {D}}}(k_{p_1i}, \Theta_{p_1i})
=\Gamma_{{\cal {B}}_{p_1i}}(k_{p_1}, \Theta_{p_1i})\cup
\Gamma_{{\cal {B}}_{p_2i}}(k_{p_2}, \Theta_{p_2i})\cup\dots\cup
\Gamma_{{\cal {B}}_{p_ni}}(k_{p_n}, \Theta_{p_ni}),
\eeq
tel que
\beq\nonumber
\Gamma_{{\cal {B}}_{p_1i}}(k_{p_1}, \Theta_{p_1i})\cap
\Gamma_{{\cal {B}}_{p_2i}}(k_{p_2}, \Theta_{p_2i})\cap\dots
\Gamma_{{\cal {B}}_{p_ni}}(k_{p_n}, \Theta_{p_ni})=0,
% \Gamma_{{\cal {B}}_{p_ni}}(k_{p_n}, \Theta_{p_ni}),
\eeq
ou encore,
\beq\label{eq:inequal2}
1\gg(k_{p_1},\vec{p}_{\pm})\gg\dots\gg(k_{p_n},\vec{p}_{\pm})\quad\text{si}\quad i=q(\bar{q}),
\eeq
sinon
\beq\label{eq:inequal3}
1\gg(k_{p_1},\vec{k}_i)\gg\dots\gg(k_{p_n},\vec{k}_i)\quad\text{pour}\quad i\ne q(\bar{q}).
\eeq
On multiplie maintenant par le
facteur de couleur $\cal {G}$ associ\'e \`a chaque vertex: $t^a$ pour 
$q(\bar{q})\rightarrow q(\bar{q})+g$ et $if_{a_1a_2c}$
pour $g\rightarrow gg$, o\`u
$a_1(a_2)$ repr\'esente le gluon le moins (plus) \'energ\'etique.
Finalement, on attribue au diagramme de la Fig.\ref{fig:Ngluon} 
l'\'el\'ement de matrice suivant dans la r\'egion 
$\Gamma_{{\cal {D}}}(k\equiv{\cal {P}},
\Theta\gtrsim\Theta_{p_1i})$, c'est \`a dire dans
la r\'egion $\Gamma_{\cal {D}}({\cal {P}},\Theta)$:
\beq\label{eq:MNgluons}
M_N=g_s^N(-1)^m\prod_{i=1}^{N}\frac{(e_i{\cal {P}}_i)}{(k_i{\cal {P}}_i)}
{\cal {G}},
\eeq
o\`u $m$ est le nombre de gluons \'emis par $\bar{q}$.
Nous rappelons qu'il n'y a pas d'interf\'erence entre les \'emissions
issues du quark et de l'anti-quark. Nous avons consid\'er\'e l'union
de toutes les branches comme un seul jet d'angle d'ouverture
l\'eg\`erement sup\'erieur \`a l'angle de la premi\`ere \'emission,
soit $p_1$.

\vskip 0.5cm

\textbf{Remarque:} ${\cal {P}}_i$ est l'impulsion de l'un des partons 
``r\'eels'' finaux et non pas celle d'un \'etat virtuel.
L'\'equation \ref{eq:MNgluons} repr\'esente ce que l'on appelle
``soft insertion rules'' en anglais \cite{BCM}.

\vskip 0.5cm

Pour le cas $e^+e^-\!\rightarrow\!q\bar{q}+Ng$ nous allons suivre la
m\^eme d\'emarche que pour le cas $N=2$. Dans un premier temps
nous allons simplifier le mieux possible les diagrammes de Feynman
dans la r\'egion qui donne une contribution doublement
logarithmique, en utilisant la jauge planaire et les polarisations
transverses des gluons finaux. Dans la deuxi\`eme 
partie nous d\'emontrerons que la somme des contributions des
diagrammes de Feynman dans chacune des r\'egions de
(\ref{eq:inequal1})-(\ref{eq:inequal3}) donne (\ref{eq:MNgluons}).

\vskip 0.5cm

\subsection{Premi\`ere partie de la d\'emonstration}

\vskip 0.5cm

\begin{figure}[h]
\begin{center}
\includegraphics[height=5truecm,width=0.57\tw]{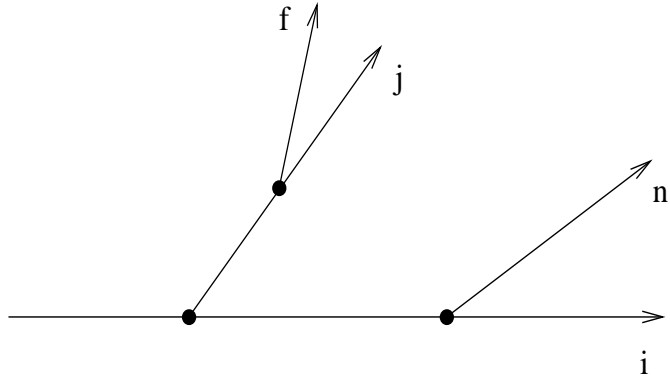}
\caption{\label{fig:ijgluons} Diagramme de la Fig.\ref{fig:Ngluon} simplifi\'e.}
\end{center}
\end{figure}

On consid\`ere les propagateurs des \'etats virtuels dans la ligne $i$
(voir Fig.\ref{fig:Ngluon}) car ce sont les seuls qui donnent une
d\'ependance logarithmique angulaire. On a $i=0,1,\dots,N$, o\`u $0$
correspond au quark ou l'anti-quark. Les 
propagateurs ont la forme $q_i^2=\left(\sum_t k_t\right)^2$.
Puisque la particule $i$ a l'\'energie la plus importante,
d'apr\`es la condition (\ref{eq:SEO})
$\omega_i\gg\omega_{p_1},\,\omega_{p_2}$, on a
$$
k_i\cdot(k_{p_1}+k_{p_2})\sim\omega_i(\omega_{p_1}\Theta_{p_1i}^2+
\omega_{p_2}\Theta_{p_2i}^2)\gg \omega_{p_1}\omega_{p_2}(\Theta_{p_1i}^2+\Theta_{p_2i}^2)
\gtrsim\omega_{p_1}\omega_{p_2}\Theta_{p_1p_2}^2\sim k_{p_1}k_{p_2},
$$
qui se g\'en\'eralise \`a toute \'emission arbitraire $j$. 
Dans le cas g\'en\'eral nous pouvons donc \'ecrire
\beq\label{eq:propag}
q_i^2=\left(\sum_t k_t\right)^2\approx 2k_i\sum_{t\ne i}k_t.
\eeq
On consid\`ere maintenant $i\rightarrow i+j$, puis $j\rightarrow j+f$.
On doit alors imposer pour ce vertex la condition
(voir Fig.\ref{fig:ijgluons})
\beq\label{eq:contg1}
\omega_{j}\Theta_{ji}^2\gg\omega_{f}\Theta_{fj}^2.
\eeq
Lorsque d'autres gluons sont \'emis apr\`es $j$, pour le $n^{\text{\`eme}}$
gluon, soit $i\rightarrow i+n$, nous avons
\beq\label{eq:contg2}
\omega_j\Theta_{ji}^2\gg\omega_n\Theta_{ni}^2.
\eeq
Nous allons d\'emontrer que si (\ref{eq:contg1}) et (\ref{eq:contg2})
sont satisfaites, on peut n\'egliger la d\'ependance en $k_f$ (ou $k_n$)
dans tous les propagateurs qui s\'eparent ce vertex de celui o\`u
la paire $q\bar{q}$ a \'et\'e produite par le
boson $\gamma^*/Z^0$.
Ces \'etats virtuel appartiennent soit \`a la ligne $i$, soit \`a un
certain parton $h$, tel que $h\rightarrow h+i$. Dans le premier cas,
en prenant (\ref{eq:propag}) en consid\'eration, $k_f$ et $k_n$
apparaissent sous la forme
\beq\label{eq:propag2}
2k_i\cdot(k_j+k_f+k_n+\dots)\approx\omega_i(\omega_j\Theta_{ji}^2+\omega_f\Theta_{fi}^2+
\omega_n\Theta_{ni}^2+\dots)
\eeq
avec $j=p_1$, $n\geqslant2$.
Dans le deuxi\`eme cas on a
\beq\label{eq:propag3}
2k_h\cdot(k_i+k_j+k_f+k_n+\dots)\approx\omega_h(\omega_i\Theta_{ih}^2+\omega_j\Theta_{jh}^2
+\omega_f\Theta_{fh}^2+
\omega_n\Theta_{nh}^2+\dots).
\eeq
Que la d\'ependance en $k_n$ puisse \^etre n\'eglig\'ee dans (\ref{eq:propag2})
d\'ecoule de la condition (\ref{eq:contg2}), et dans le deuxi\`eme cas, de 
(\ref{eq:contg2}) et des relations suivantes:
$$
\omega_n\ll\omega_i,\quad
\Theta_{nh}^2\lesssim\Theta_{ni}^2+\Theta_{ih}^2,
$$
\beq\label{eq:contg3}
\omega_i\gg\omega_j,\quad\omega_i\Theta_{ih}^2+\omega_j\Theta_{jh}^2
\sim\omega_i\Theta_{ih}^2+\omega_j(\Theta_{jh}^2+\Theta_{ih}^2)\gtrsim
\omega_i\Theta_{ih}^2+\omega_j\Theta_{ji}^2;
\eeq
de plus, 
$$
\omega_i\Theta_{ih}^2+\omega_j\Theta_{ji}^2
\gg\omega_i\Theta_{ih}^2+\omega_n\Theta_{ni}^2\gg
\omega_n(\Theta_{ih}^2+\Theta_{ni}^2)\gtrsim\omega_n\Theta_{nh}^2.
$$ 
\textbf{Conclusion:}
$k_n$ peut \^etre n\'eglig\'e par rapport \`a $k_i+k_j$ car 
$\omega_i\Theta_{ih}^2+\omega_j\Theta_{jh}^2\gg
\omega_n\Theta_{nh}^2$.
Quant \`a $k_f$, il peut se n\'egliger dans le premier cas si l'on tient
compte de (\ref{eq:contg1}) et de la relation
\beq\label{eq:contg4}
\omega_f\ll\omega_j,\quad 
\Theta_{fi}^2\lesssim\Theta_{fj}^2+\Theta_{ji}^2;
\eeq
en effet, 
$$\omega_j\Theta_{ji}^2\sim\omega_j\Theta_{ji}^2+\omega_f\Theta_{fj}^2\gg
\omega_f(\Theta_{fj}^2+\Theta_{ji}^2)\gtrsim\omega_f\Theta_{fi}^2\Rightarrow
\omega_j\Theta_{ji}^2\gg\omega_f\Theta_{fi}^2.
$$
Dans le deuxi\`eme cas, on utilise (\ref{eq:contg1}), (\ref{eq:contg3}), 
$\Theta_{ih}>\Theta_{jh}$ et
$$
\omega_{f}\ll\omega_{j},\quad\Theta_{fh}^2\lesssim\Theta_{fj}^2+\Theta_{jh}^2
$$
o\`u l'on a remplac\'e $h$ par $i$ dans (\ref{eq:contg4}).
En effet, \`a partir de (\ref{eq:contg3}) 
$$\omega_i\Theta_{ih}^2
+\omega_j\Theta_{jh}^2\gtrsim\omega_i\Theta_{ih}^2+\omega_j\Theta_{ji}^2\gg
\omega_i\Theta_{jh}^2+\omega_f\Theta_{fj}^2\gg\omega_f(\Theta_{jh}^2+\Theta_{fj}^2)
\gtrsim\omega_f\Theta_{fh}^2.
$$
\textbf{Conclusion:} $k_n$ et $k_f$ peuvent \^etre n\'eglig\'es par rapport
\`a $k_i+k_j$.  La contribution logarithmique dominante est donc
donn\'ee par l'\'etat le plus virtuel, soit l'\'etat de plus faible temps
de fluctuation.

Nous allons d\'emontrer maintenant que si l'une des contraintes
(\ref{eq:contg1}) ou (\ref{eq:contg2}) n'est pas satisfaite,
rien que sur un vertex on perd alors une contribution logarithmique.
Pour cel\`a, on va consid\'erer le vertex de la 
Fig.\ref{fig:Ngluon} o\`u l'on a $q_1=q_2+q_3$.
D'apr\`es ce qu'on vient de d\'emontrer
$$
q_2^2=2k_i\cdot k_n\simeq\omega_i\omega_n\Theta_{ni}^2,\quad q_3^2=2k_j\cdot k_f\simeq
\omega_j\omega_f\Theta_{fi}^2,
$$
$$
q_1^2=2k_i\cdot(k_j+k_n+k_f)\simeq\omega_i(\omega_j\Theta_{ji}^2+\omega_n\Theta_{ni}^2+\omega_f
\Theta_{fi}^2).
$$
En effet,
$$
M_N\propto\frac1{q_1^2}\times\frac1{q_2^2}\times\frac1{q_3^2}\dots
$$
o\`u on n'a pas tenu compte des num\'erateurs. Si les conditions
(\ref{eq:contg1}) et (\ref{eq:contg2}) sont satisfaites, alors
$$
M_N\propto\frac1{\omega_j\Theta_{ji}^2}\times\frac1{\omega_n\Theta_{ni}^2}\times
\frac1{\omega_f\Theta_{fj}^2}
$$
et ainsi, on gagne une contribution doublement logarithmique molle
et colin\'eaire dans le calcul de la section efficace pour chaque
\'emission de ce type.  Par contre, la violation de la condition
(\ref{eq:contg1}) entra\^\i ne
$\omega_f\Theta_{fj}^2\gtrsim\omega_j\Theta_{ji}^2$;
or, si l'on tient compte de $\omega_j\gg\omega_f$, on doit avoir
$\Theta_{fj}\gg\Theta_{ji}$.
Puisque
$\Theta_{fi}+\Theta_{ji}\geqslant\Theta_{fj}
\geqslant\mid\!\!\Theta_{fi}-\Theta_{ji} \!\!\mid$,
on a
$$
\Theta_{fi}\simeq\Theta_{fj}\quad \text{et} 
\quad q_1^2\simeq\omega_i(\omega_j\Theta_{ji}^2
+\omega_n\Theta_{ni}^2+\omega_f\Theta_{fj}^2)
\sim\omega_i(\omega_n\Theta_{ni}^2+\omega_f\Theta_{fj}^2)
$$
ind\'ependamment de la contrainte (\ref{eq:contg2}). Avec ceci on aurait
$$
M_N\propto\frac1{\omega_n\Theta_{ni}^2+\omega_f\Theta_{fj}^2}
\times\frac1{\omega_n\Theta_{ni}^2}
\times\frac1{\omega_f\Theta_{fj}^2}
$$
et on perdrait en $\log\Theta$.
Remarquons que \textbf{dans l'approximation doublement 
logarithmique, chacun des d\'enominateurs des propagateurs doit
donner une contribution logarithmique angulaire lorsque l'on int\`egre
sur les angles d'\'emission}; puisque $q_1^2$ est d\'etermin\'e
par les m\^emes angles que $q_2^2$ et $q_3^2$,
ceci entra\^\i ne la perte d'un $\log$.  Le m\^eme argument est
vrai si l'on consid\`ere que (\ref{eq:contg2}) n'est pas
satisfaite.

Nous avons ainsi d\'emontr\'e que les contraintes 
(\ref{eq:contg1}) et (\ref{eq:contg2}) d\'eterminent la r\'egion
qui donne une contribution doublement logarithmique.
Nous concluons que le d\'enominateur du propagateur
de la ligne $i$ peut s'\'ecrire sous la forme simplifi\'ee $2k_ik_j$.

\begin{figure}[h]
\begin{center}
\includegraphics[height=4truecm,width=0.4\tw]{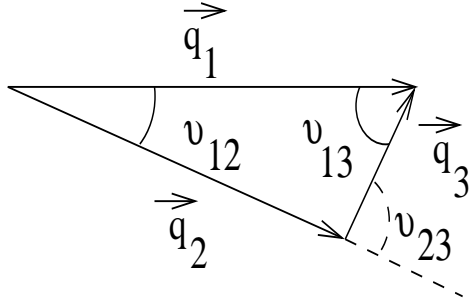}
\caption{\label{fig:triangle} Triangle $\vec{q}_1=\vec{q}_2+\vec{q}_3$.}
\end{center}
\end{figure}

On \'etudie maintenant la simplification des num\'erateurs. On
d\'emontre que comme dans le cas $N=2$, $\forall\, N$, seulement 
les polarisations physiques ($\lambda=1,2$) contribuent au calcul de
la section efficace.
Prenons le m\^eme vertex  et consid\'erons la quantit\'e
\beq\label{eq:Vlambda}
V_{\lambda_1\lambda_2\lambda_3}=
e^{\lambda_1}_{\mu}(q_1)e^{\lambda_2}_{\nu}(q_2)e^{\lambda_3}_{\rho}(q_3)
\gamma^{\mu\nu\rho}(-q_1,q_2,q_3),
\eeq
o\`u $\gamma^{\mu\nu\rho}$ est le facteur de Lorentz associ\'e au
vertex \`a trois gluons. On va d\'eterminer
$V_{\lambda_1\lambda_2\lambda_3}$ pour les diff\'erentes valeurs de $\lambda$.
D'abord, on exprime les quantit\'es dans (\ref{eq:Vlambda})
en fonction de $\omega_n$, $\omega_f$, $\omega_j$, $\omega_i$,
et des angles $\Theta_{ji}$, $\Theta_{ni}$ et $\Theta_{fj}$.
Dans le r\'ef\'erentiel du centre de masse de la paire $q\bar{q}$,
le quadri-vecteur de jauge devient $c=(1,\vec{0})$, or
\beeq\label{eq:gaugecm}
(q_1\cdot c)\!&\!\!=\!\!&\!q_1^0\approx(q_2\cdot c)=q_2^0\simeq\omega_i,
\qquad (q_3\cdot c)=q_3^0\simeq\omega_j,\\ \nonumber \\
q_1^2\!&\!\!\simeq\!\!&\!\omega_i\omega_j\Theta_{ji}^2,\qquad q_2^2\simeq\omega_i\omega_n\Theta_{ni}^2,\qquad
q_3^2\simeq\omega_j\omega_f\Theta_{fj}^2.\nonumber
\eeeq
On estime les angles $(\widehat{\vec{q}_1,\vec{q}_2})=\vartheta_{12}$, 
$(\widehat{\vec{q}_1,\vec{q}_3})=\vartheta_{13}$ et
$(\widehat{\vec{q}_2,\vec{q}_3})=\vartheta_{23}$.
G\'eom\'etriquement, la condition $\vec{q}_1=\vec{q}_2+\vec{q}_3$
se repr\'esente sous la forme donn\'ee  par le triangle de la
Fig.\ref{fig:triangle} et on a
\beq\label{eq:triangle1}
\frac{\mid\!\!\vec{q}_1\!\!\mid}{\sin(\pi-\vartheta_{23})}=
\frac{\mid\!\!\vec{q}_2\!\!\mid}{\sin\vartheta_{13}}=
\frac{\mid\!\!\vec{q}_3\!\!\mid}{\sin\vartheta_{12}}.
\eeq
Pour $\vartheta_{12}\!\ll\!1$ ($\vec{q_1}$ et $\vec{q_2}$
quasi-colin\'eaires), on a de m\^eme  $\vartheta_{13}\!\ll\!1$
et  $\vartheta_{23}\!\ll\!1$. Dans ce cas on simplifie (\ref{eq:triangle1})
et on obtient
$$
\frac{\vartheta_{12}}{\mid\!\!\vec{q}_3\!\!\mid}=
\frac{\vartheta_{13}}{\mid\!\!\vec{q}_2\!\!\mid}=
\frac{\vartheta_{23}}{\mid\!\!\vec{q}_1\!\!\mid},
$$
puis selon (\ref{eq:contg1}), (\ref{eq:contg2}), et (\ref{eq:gaugecm})
on obtient
$$
\omega_i\omega_j\Theta_{ji}^2\simeq q_1^2=q_2^2+q_3^2+2q_2q_3\simeq q_2^0q_3^0
\vartheta_{23}^2\simeq\omega_i\omega_j\vartheta^2_{23},
$$
et on a 
$$
\vartheta_{23}\simeq\Theta_{ji},\qquad \vartheta_{13}\simeq\Theta_{ji},
\qquad \vartheta_{12}\simeq\left(\frac{\omega_j}{\omega_i}\right)\Theta_{ji}\ll\Theta_{ji}.
$$
On d\'etermine (\ref{eq:Vlambda}) pour les \'etats de polarisations
transverses en prenant explicitement l'expression (\ref{eq:gamma3gluons})
de $\gamma_{\mu\nu\rho}$, soit
$$
\gamma_{\mu\nu\rho}(-q_1,q_2,q_3)=g_{\mu\nu}(q_1+q_2)_{\rho}+g_{\rho\mu}(-q_1-q_3)_{\nu}
+g_{\nu\rho}(q_3-q_2)_{\mu};
$$
mais $q_1\simeq q_2\ll q_3$, donc
$$
\gamma_{\mu\nu\rho}(-q_1,q_2,q_3)\simeq g_{\mu\nu}2q_{2\rho}-g_{\rho\mu}q_{2\nu}
-g_{\nu\rho}q_{1\mu};
$$
\beeq\nonumber
V_{t_1t_2t_3}(q_1,q_2,q_3)\!\!&\!\simeq\!&\!\!(e^{(t_1)}(q_1)\cdot e^{(t_2)}(q_2))
2(e^{(t_3)}(q_3)\cdot q_2)\!-\!(e^{(t_1)}(q_1)\cdot e^{(t_3)}(q_3))(e^{(t_2)}(q_2)\cdot q_2)
\\ \nonumber \\
\!\!&\!\!-\!\!&\!\!(e^{(t_2)}(q_2)\cdot e^{(t_3)}(q_3))(e^{(t_1)}(q_1)\cdot q_1)
=(e^{(t_1)}(q_1)\cdot e^{(t_2)}(q_2))
2(e^{(t_3)}(q_3)\cdot q_2)\nonumber
\\ \nonumber\\
\!\!&\!\sim\!&\!\! \omega_i\Theta_{ji},\qquad t_{1,2,3}=1,2,\label{eq:Vttt}
\eeeq
o\`u l'on a tenu compte de $(e^{(t_i)}(q_i)\cdot q_i)=0$ pour $i=1,2$.
Il est de m\^eme n\'ecessaire de remarquer que
\beq\label{eq:poltrans}
(e^{(t_3)}(q_3)\cdot q_2)\simeq(e^{(t_3)}(q_3)\cdot k_i).
\eeq
En effet, si l'on consid\`ere que, dans la cr\'eation de l'\'etat virtuel
d'impulsion $q_2$, en plus du gluon $i$, des gluons suppl\'ementaires
d'impulsions $k_s$ sont \'emis aux angles
d'\'emission l\'eg\`erement inf\'erieurs \`a $\Theta_{ji}$,
on a $q_2=k_i+\sum_{s}k_s$.
Soit $\vartheta_s=(\widehat{\vec{q}_3,\vec{k}_s})$. Alors, \`a partir de 
$\omega_j\omega_s\vartheta_s^2\lesssim
2q_3k_s\ll q_1^2\simeq\omega_i\omega_j\Theta_{ji}^2$, on conclut
$[e^{(t_3)}(q_3)\cdot k_s]\sim\omega_s\vartheta_s\ll\omega_i\Theta_{ji}$,
et, avec ceci, on en d\'eduit (\ref{eq:poltrans}).
On consid\`ere maintenant le cas o\`u l'un des vecteurs $e^{(\lambda)}(q)$
est non-physique (polarisations longitudinales,
$\lambda\equiv\ell=0,3$). Par exemple, dans
\beeq\nonumber
V_{\ell_1t_2t_3}&\!\!\simeq\!\!& (e^{(\ell_1)}(q_1)\cdot e^{(t_2)}(q_2))
(2q_2\cdot e^{(t_3)}(q_3)) - (e^{(\ell_1)}(q_1)\cdot e^{(t_3)}(q_3))(q_2\cdot e^{(t_2)}(q_2))\\
\nonumber\\
&\!\!-\!\!&(e^{(t_2)}(q_2)\cdot e^{(t_3)}(q_3))(q_1\cdot e^{(\ell_1)}(q_1)),
\eeeq
les deux premiers termes sont nuls et on a
\beq
V_{\ell_1t_2t_3}\simeq-(e^{(t_2)}(q_2)\cdot e^{(t_3)}(q_3))(q_1\cdot e^{(\ell_1)}(q_1)),
\eeq
or $e^{(t_2)}(q_2)\cdot e^{(t_3)}(q_3)\sim\cos(\pi-\Theta_{ji})=-\cos(\Theta_{ji})
\approx-1+{\cal {O}}(\Theta_{ji}^2)$, puis 
$$
q_1\cdot e^{(\ell_1)}(q_1)=\frac{q_1^2\pm\sqrt{q_1^2}\omega_i}{\left[2\omega_i
(\omega_i\pm\sqrt{q_1^2})\right]^{1/2}}\approx\frac{\omega_i\left(\omega_j
\Theta_{ji}^2\pm\sqrt{\omega_i\omega_j\Theta_{ji}^2}\right)}{\left[2\omega_i^2\left(1\pm
\sqrt{\frac{\omega_j}{\omega_i}\Theta_{ji}^2}\right)\right]^{1/2}},
$$
mais $\omega_j\Theta_{ji}^2\ll \sqrt{\omega_i\omega_j\Theta_{ji}^2}$ et
$1\gg\sqrt{\frac{\omega_j}{\omega_i}\Theta_{ji}^2}$, donc 
$q_1\cdot e^{(\ell_1)}(q_1)\sim \sqrt{\omega_i\omega_j\Theta_{ji}^2}$, et finalement
\beq\label{eq:Vltt}
V_{\ell_1t_2t_3}\sim\sqrt{\omega_i\omega_j\Theta_{ji}^2}.
\eeq
Si maintenant on compare (\ref{eq:Vttt}) et (\ref{eq:Vltt}) on voit que
$$
V_{t_1t_2t_3}\sim\omega_i\Theta_{ji}\gg V_{\ell_1t_2t_3}\sim\sqrt{\omega_i\omega_j\Theta_{ji}^2}.
$$

Ce r\'esultat se g\'en\'eralise \`a toute autre contraction sur l'une
ou plus des polarisations longitudinales. Les expressions peuvent
\^etre trouv\'ees dans \cite{Fadin1}.

Cette comparaison d\'emontre que la projection des amplitudes (\ref{eq:Maa}),
(\ref{eq:Mb}), (\ref{eq:Mc}) sur les polarisations transverses des gluons
donne la contribution la plus importante.
La m\^eme analyse se g\'en\'eralise au cas
du vertex $q\rightarrow q+g$. D'ailleurs, le d\'enominateur du 
propagateur peut se mettre sous la forme $2k_j(k_0\equiv E_{\pm})$
dans le vertex pour l'\'emission du gluon $j$. Les projections sur les
polarisations transverses l'emportent sur les projections sur les
polarisations longitudinales, ainsi que le d\'emontrent
 les estimations suivantes (elles peuvent \^etre obtenues facilement):
$$
(e^{(1,2)}(q_3)\cdot k_0)\sim\omega_0\Theta_{j0},\qquad 
(e^{(0,3)}(q_3)\cdot k_0)\sim\omega_0\Theta_{j0}^2+(\omega_0/\omega_j)
\sqrt{\omega_j\omega_f\Theta_{fj}^2}
$$
o\`u l'on tient compte de (\ref{eq:contg1}).
Par cons\'equent, cette analyse d\'emontre que, dans l'expansion
(\ref{eq:dmunu}), on  peut ne garder que les polarisations physiques,
de sorte que $d_{\rho\sigma}(k,c)\simeq g_{\rho\sigma}^{\perp}(k)$,
dont l'expression est \'ecrite dans (\ref{eq:tentra}).
De (\ref{eq:Vttt}) et (\ref{eq:poltrans}) on en d\'eduit que
seulement $g_{\mu\nu}2k_i^{\sigma}$ donne une contribution
non nulle dans le vertex $i\rightarrow i+j$, suite \`a la projection
sur les \'etats de polarisations transverses. La m\^eme d\'emarche
peut \^etre men\'ee \`a bien comme dans le cas de l'\'emission de deux
gluons, \`a savoir, nous allons d\'emontrer qu'apr\`es cette 
simplification des vertex, $g_{\rho\sigma}^{\perp}$ peut \^etre remplac\'ee
par $g_{\rho\sigma}$. Nous allons consid\'erer les propagateurs
de la ligne $j$. Leurs num\'erateurs interviennent dans l'\'el\'ement
de matrice sous la forme suivante
$$
2k_i^{\rho}g_{\rho\rho_3}^{\perp}(q_3)g_{\rho_3\rho_4}^{\perp}(q_4)\dots
g_{\rho_{n-1}\rho_{n}}^{\perp}(q_n)e^{\rho_n}_j,
$$
o\`u $q_3,q_4,\dots,q_n$ sont les impulsions des propagateurs dans la
ligne $j$, $g^{\perp}_{\mu\nu}(q_a)=-\delta_{\mu\nu}+q^{\mu}_aq^{\nu}_a/q^2_a$,
 $e_j$ est le vecteur de polarisation du gluon $j$. Les indices de
Lorentz prennent les valeurs des trois coordonn\'ees spatiales $1,2,3$.
Il suffit de d\'emontrer, par analogie avec le cas $N=2$, que le terme
en $\delta$ donne la contribution dominante. La multiplication du
deuxi\`eme terme dans $g^{\perp}$ par $e_j$ donne
la quantit\'e $2\omega_i(q_ae_j)/\omega_j$, o\`u $q_a$ est l'impulsion
d'un certain \'etat virtuel de la ligne $j$ qui se d\'esint\`egre en un
autre \'etat du m\^eme type, puis d'autres gluons d'impulsion $k_t$
sont \'emis, tel que $q_a=k_j+\sum_{t}k_t$. Puisque $e_jk_j=0$, nous avons
$$
\omega_i(q_a\cdot e_j)/\omega_j\simeq\sum_t(\omega_t\omega_i/\omega_j)\vartheta_t,
$$
o\`u $\vartheta_t=(\widehat{\vec{k}_j,\vec{k}_t})$.
Le terme donn\'e par $\delta$ est $\sim\omega_i\Theta_{ji}$. Or
$$
\omega_j\omega_t\vartheta^2_t\simeq2k_j\cdot k_t\lesssim q_3^2\simeq
\omega_j\omega_f\Theta_{fj}^2\ll\omega_j^2\Theta_{ji}^2,
$$
On divise par $\omega_j$ et on obtient 
$(\omega_t\omega_i/\omega_j)\vartheta_t\ll\omega_i\Theta_{ji}$.
La premi\`ere partie de la d\'emonstration est ainsi termin\'ee.

\vskip 0.5cm

\subsection{Deuxi\`eme partie de la d\'emonstration}

\vskip 0.5cm

On doit maintenant se convaincre que les conditions (\ref{eq:inequal1.1}) 
et (\ref{eq:inequal3}) couvrent bien l'espace des angles d'\'emission
et que celles-ci interviennent dans le cadre de l'approximation doublement
logarithmique, \`a savoir, que l'on doit attribuer le diagramme
$\cal {D}$ \`a l'amplitude (\ref{eq:MNgluons}) dans la r\'egion
$\Gamma_{\cal {D}}$. Le diagramme $\cal {D}$,
respectant les contraintes sur les angles d'\'emission,
a \'et\'e construit dans la Fig.\ref{fig:Ngluon}.
Nous allons effectuer cette d\'emonstration par induction.
Pour l'\'emission d'un gluon (N=1) cet \'enonc\'e se confirme
sans difficult\'e car il n'existe qu'un seul diagramme
(on consid\`ere l'\'emission d'un seul gluon par 
$q(\bar{q})$) et la r\'egion qui lui correspond est identique
\`a celle qui donne une contribution doublement logarithmique
$(\Theta_{g,q(\bar{q})}\ll\Theta_{\pm})$. 
Par la suite, on assume que cet \'enonc\'e se g\'en\'eralise
\`a l'\'emission de $N-1$ gluons de bremsstrahlung
et on d\'emontrera qu'il est vrai dans le cas de l'\'emission de $N$ gluons.
\begin{figure}[h]
\begin{center}
\includegraphics[height=5truecm,width=0.5\tw]{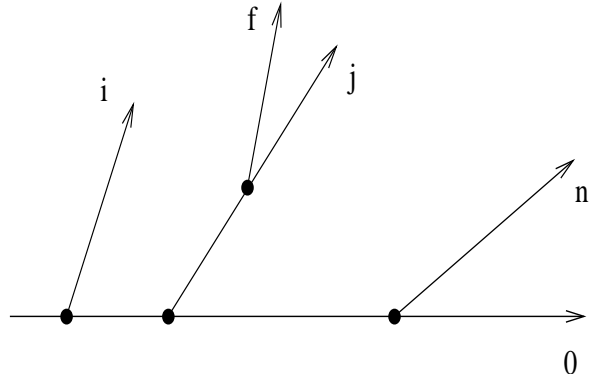}
\caption{\label{fig:ijgluons1} Diagramme correspondant \`a la production de $N$ gluons.}
\end{center}
\end{figure}

On consid\`ere les diff\'erentes mani\`eres d'ins\'erer un
gluon $i$ dans un diagramme \`a $N\!-\!1$ gluons.
Nous allons construire ces diagrammes de sorte que les contraintes angulaires
(\ref{eq:contg1}) et (\ref{eq:contg2}) soient satisfaites
dans la r\'egion $\Gamma_{\cal {D}}$, dans l'objectif d'obtenir une
 contribution doublement logarithmique 
La pr\'esence du terme $g_s(e_i\cdot p_+)/(k_i\cdot p_+)$
(sans facteur de couleur) se comprend facilement:
si le gluon $i$ est \'emis par la particule $j$, alors, 
$g_s(e_i\cdot k_j)/(k_i\cdot k_j)$ appara\^it
dans l'\'el\'ement de matrice, mais, vue l'importance de l'angle
$\Theta_{i0}$ qui entra\^\i ne $\Theta_{ij}\simeq\Theta_{i0}\gg\Theta_{j0}$,
on a $(e_i\cdot k_j)/(k_i\cdot k_j)\simeq
(e_i\cdot k_0)/(k_i\cdot k_0)=(e_i\cdot p_+)/(k_i\cdot p_+)$.
Par cons\'equent, le probl\`eme qui se pose est donn\'e par le facteur
de couleur. En effet, si $\omega_i$ est la fr\'equence 
maximale, le facteur de couleur $t^{a_i}$ est correct car la
seule possibilit\'e d'avoir une contribution doublement logarithmique
est celle qui est donn\'ee par l'insertion du gluon $i$
sur le quark juste avant l'\'emission du gluon $j$, 
soit celle qui est donn\'ee par le diagramme de 
la Fig.\ref{fig:ijgluons1}. La contrainte donnant cette contribution
s'\'ecrit sous la forme $\omega_i\Theta_{i0}^2\gg\omega_j\Theta_{j0}^2$.
\begin{figure}[h]
\begin{center}
\includegraphics[height=3.5truecm,width=6.0truecm]{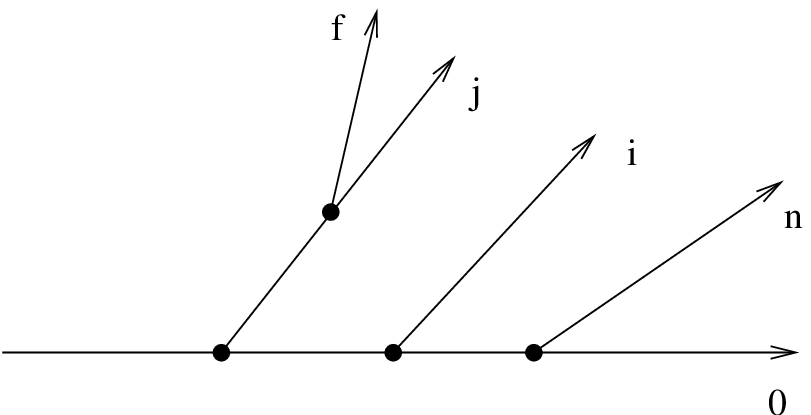}
\hskip 2.5cm
\includegraphics[height=3.5truecm,width=6.0truecm]{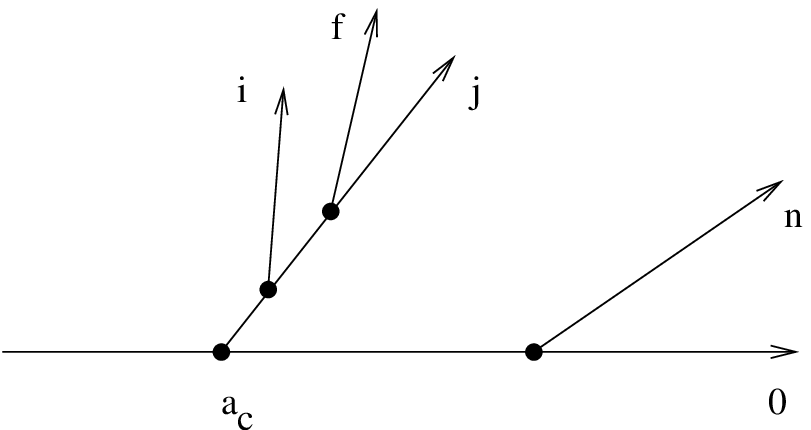}
\caption{\label{fig:ijgluons23} Diagrammes \`a prendre en consid\'eration
lorsque l'on fait d\'ecro\^itre la fr\'equence $\omega_i$.}
\end{center}
\end{figure}
Par contre, si on fait d\'ecro\^itre $\omega_i$ ind\'efiniment de sorte que 
l'on arrive \`a la r\'egion $\omega_i\Theta_{i0}^2\ll\omega_j\Theta_{j0}^2$,
($\omega_i\Theta_{i0}^2
\gg\omega_j\Theta_{j0}^2$ ne donne aucune contribution) mais si
encore $\omega_i\Theta_{i0}^2\gg\omega_n\Theta_{n0}^2$ et 
$\omega_i\Theta_{i0}^2\gg\omega_f\Theta_{f0}^2$, ce diagramme n'apporterait
aucune contribution;
or, dans ce cas on aurait deux diagrammes qui en donnerait une,
celui o\`u l'on ins\`ere le gluon $i$ entre
$j$ et $n$, Fig.\ref{fig:ijgluons23} (gauche) et le deuxi\`eme, o\`u
l'on ins\`ere $i$ sur la ligne $j$ avant l'\'emission de $f$,
Fig.\ref{fig:ijgluons23} (droite).  Cependant, on r\'ealise
que la somme des diagrammes de la Fig.\ref{fig:ijgluons23} donne,
en utilisant la relation de commutation du groupe $SU(3)$
\cite{Peskin}\cite{Cheng} $[t^{a_i},t^{a_j}]=if_{a_ia_ja_c}t^{a_c}$, 
la m\^eme contribution $g_s(e_ip_+)/(k_ip_+)t^{a_i}$, 
qui correspond au diagramme de la Fig.\ref{fig:ijgluons1}.
En fait, la somme
\beeq\nonumber
g_s((e_ip_+)/(k_ip_+)t^{a_j}t^{a_i}+(e_ik_j)/(k_ik_j)if_{a_ia_ja_c}t^{a_c})
\!&\!\simeq\!&\! g_s(e_i.p_+)/(k_i.p_+)(t^{a_j}t^{a_i}+if_{a_ia_ja_c}t^{a_c})\\
\!&\!=\!&\!\left[g_s(e_i.p_+)/(k_i.p_+)t^{a_i}\right]t^{a_j}\nonumber
\eeeq
dans l'approximation $\Theta_{ij}\simeq\Theta_{i0}\gg\Theta_{j0}$
mention\'ee ci-dessus. 
Si l'on continue de faire d\'ecro\^itre $\omega_i$ de sorte
que l'on franchit la r\'egion 
$\omega_i\Theta_{i0}^2\sim\omega_n\Theta_{n0}^2$ et / ou 
$\omega_i\Theta_{i0}^2\sim\omega_f\Theta_{f0}^2$,
il faudrait faire le raisonnement
pr\'ec\'edent:  on prendrait deux diagrammes suppl\'ementaires
pour chaque cas tels que le r\'esultat de la somme donne les diagrammes
de la Fig.\ref{fig:ijgluons23}.
La deuxi\`eme partie de la d\'emonstration est ainsi faite.
On a donc d\'emontr\'e que l'amplitude (\ref{eq:MNgluons}) correspond
ainsi \`a l'\'emission de $N$ gluons dans la r\'egion $\Gamma_{\cal {D}}$.

\section{Section efficace du processus et introduction du facteur de
forme de Sudakov}

Apr\`es avoir \'evalu\'e
le carr\'e de la somme des amplitudes (\ref{eq:MNgluons}) pour tous les
diagrammes $\cal {D}$, en plus d'avoir pond\'er\'e par l'espace de phase
des \'emissions et d'avoir
effectu\'e la somme sur
 les polarisations transverses $\lambda=1,2$ des gluons \'emis, on
obtient
\beq\label{eq:SENG}
d\sigma_N=d\sigma_0\prod_{i=1}^N\frac{g_s^2}{(2\pi)^3}C_A
\frac{d^3k_i}{2\omega_i}
\frac{{\cal {P}}_i^2\sin^2\Theta_{k_i}}{(k_i.{\cal {P}}_i)^2},
\eeq
o\`u $d\sigma_0$ est la section efficace de Born qui correspond
\`a l'annihilation $e^+e^-\rightarrow q\bar{q}$. On remarque
que la factorisation de l'amplitude associ\'ee
\`a l'emission de $N$ gluons, sur laquelle on a dej\`a insist\'e
depuis le premier chap\^itre, entra\^\i ne \'egalement la factorisation
du facteur de rayonnement dans le calcul de la section efficace.
$C_A$ est le facteur de couleur qui correspond \`a l'\'emission
d'un gluon par un quark $A\equiv F, \quad C_F=(N_c^2-1)/2N_c$
et $A\equiv G,\quad C_G=N_c$, \`a l'\'emission d'un gluon par un gluon
\footnote{$C_F$ et $N_c$ sont respectivement les Casimirs de la
repr\'esentation fondamentale et de la repr\'esentation adjointe
du groupe $SU(N_c)$, pour $N_c=3$, $C_F=4/3$.}.

Puisque la section efficace est domin\'ee par l'\'emission des gluons
colin\'eaires ($\Theta$ petit), le facteur de rayonnement se simplifie comme
$$
\frac{{\cal {P}}_i^2\sin^2\Theta_{k_i}}{(k_i.{\cal {P}}_i)^2}=\frac1{\omega^2}
\frac{\sin^2\Theta_{ki}}{(1-\cos\Theta_{ki})^2}\approx\frac4{\omega^2\Theta_{ki}^2}
\approx\frac4{k^2_{i\perp}}.
$$
On peut de plus introduire la constante de couplage des interactions fortes
$\alpha_s=g_s^2/4\pi$ et r\'ecrire (\ref{eq:SENG}) de fa\c con \`a ce que l'on
puisse mieux appr\'ecier le caract\`ere Doublement Logarithmique (DL) du
rayonnement
\beq\label{eq:SER}
d{\cal {K}}(\vec{k})\equiv\frac{d\omega}{\omega}\frac{d^2k_{\perp}}{2\pi k_{\perp}^2}
\frac{2C_A}{\pi}.
\eeq
L'insertion des corrections virtuelles \`a une boucle
dans le calcul de la section efficace consiste \`a multiplier
l'\'el\'ement de matrice (\ref{eq:MNgluons}) par le facteur
suivant \cite{Fadin1}\cite{EvEq}
(facteur de forme de Sudakov, sans d\'emonstration), soit
\beq
{\cal {F}}=\exp\left\{-\frac12\left[\omega_{F}(p_+,1)+\omega_F(p_-,1)+
\sum_{i=1}^N\omega_{G}(k_i,\Theta_i)\right]\right\}
\eeq
o\`u $\Theta_i$ est l'angle entre le gluon (enfant) et son parent
$$
\vec{k}_i.\vec{\cal {P}}_i=\mid\!\!\vec{k}_i\!\!\mid\mid\!\!\vec{\cal {P}}_i\!\!\mid
\cos\Theta_i.
$$
Ce facteur, d\'ependant de la topologie du diagramme $\cal {D}$,
est \'etroitement li\'e \`a la section efficace du rayonnement (\ref{eq:SER}).
 A savoir, la fonction $\omega_{G(F)}$ d\'ecrit la probabilit\'e
totale de l'\'emission d'un gluon mou par un gluon (ou quark) \`a 
l'int\'erieur d'un c\^one de demi-angle d'ouverture $\Theta$:
\beq\label{eq:VC}
\omega_{G}(p,\Theta)=\int_{\Gamma(p,\Theta)}\frac{g_s^2}{(2\pi)^3}\,N_c\,
\frac{d^3k}{2\omega}\frac{\vec{p}^{\,\,2}\,\sin^2\Theta_k}{(k.p)^2}\approx
\int_{\Gamma(p,\Theta)}d{\cal {K}}(\vec{k}),
\eeq
puis dans le cadre de l'approximation DL
$$
\omega_F(p,\Theta)\approx\frac{C_F}{N_c}\omega_G(p,\Theta).
$$
L'int\'egration sur $\Theta_k$ a comme limite sup\'erieure $\Theta$
et celle sur $k$, $p^0$.
Finalement, en tenant compte des corrections virtuelles la section
efficace du rayonnement s'\'ecrit sous la forme
\beq\label{eq:SERCV}
d\sigma=d\sigma_0\sum_{\cal {D}}{\cal {F}}^2\prod_id{\cal {K}}(\vec{k}_i).
\eeq
On rappelle que pour chaque diagramme $\cal {D}$, les gluons \'emis 
$k_1,k_2,\dots k_N$ sont ordonn\'es par rapport \`a leurs angles d'\'emission
d'apr\`es la contrainte $\Gamma_{\cal {D}}$ que l'on r\'ecrit ci-dessous
$$
\Gamma_{\cal {D}}({\cal {P}}, \Theta):\left\{k^0\equiv\omega<{\cal {P}}^0;
\qquad \Theta_{\widehat{\vec{k}\vec{\cal {P}}}}\equiv\Theta_k<\Theta;\qquad
k_{\perp}\approx\omega\Theta_{k}>Q_0\right\},
$$
o\`u $\cal {P}$ est l'impulsion $k$ du parent \'emetteur, $\Theta$
est l'angle de l'\'emission qui a donn\'e $\cal {P}$.

\vskip 0.5cm

\subsection{D\'efinition de l'angle d'ouverture du jet}:

$\Theta_i$ dans (\ref{eq:VC}) est le demi-angle d'ouverture total du jet,
soit l'angle \`a l'int\'erieur duquel seront \'emis tous les partons.
Dans la production de deux quarks, ce param\`etre d'\'evolution est
le plus important et a \'et\'e pris $\sim 1$ car une valeur
un peu plus large serait au-del\`a de l'approximation DLA.
Tous les gluons dans le jet sont \'emis aux angles inf\'erieurs
\`a $\Theta_i$.

\section{M\'ethode de la Fonctionnelle G\'en\'eratrice 
\cite{Veneziano}}
\label{sub:MFG}

La notion de Fonctionnelle G\'en\'eratrice (FG) est depuis longtemps
exploit\'ee en physique et en math\'ematiques. Par exemple,
si l'on d\'eveloppe la fonction $G(u)=\exp(u)$ en s\'erie de Taylor
au voisinage du point $u=0$, on peut dire que
celle-ci g\'en\`ere les coefficients $a_n$ d'apr\`es l'expression suivante:

$$
a_n\equiv\left[\left(\frac{d}{du}\right)^nG(u)\right]_{\left\{u=0\right\}}.
$$

Exemples:

$$
G(u)=u\exp{(u)}\Rightarrow0,1,2\dots n\quad \text{nombres naturels,}
$$

$$
G(u)=u/(e^u-1)\Rightarrow B_n\quad \text{s\'eries de Bernoulli,}
$$

$$
G(u)=\exp{(2xu-u^2)}\Rightarrow H_n(x)\quad \text{polynomes d'Hermite, etc}.
$$

Nous suivons ici cette m\^eme logique. On peut
consid\'erer que notre section efficace du rayonnement de $N$ gluons
$d\sigma_N$ est le $N^{\text{\`eme}}$ coefficient du d\'eveloppement
de Taylor d'un certain objet ``g\'en\'erateur'' qui retient l'information
du processus consid\'er\'e en CDQ.
Cet objet ne peut \^etre une fonction mais une fonctionnelle
car les s\'eries qu'il doit g\'en\'erer sont des fonctions
(par exemple, elles d\'ependent des tri-impulsions des N-
gluons) et non pas des nombres.

%L'id\'ee principale de cette m\'ethode est expliqu\'ee
%dans l'appendice \ref{sub:IFG} \cite{EvEq}.

On est ainsi amen\'e \`a remplacer  la section efficace
``exclusive'' de production de $N$ gluons de bremsstrahlung $d\sigma_N$ 
(\ref{eq:SERCV}) par la fonctionnelle g\'en\'eratrice
$d\sigma\left\{u\right\}$ des coefficients
$d\sigma_N$ de l'expansion de Taylor
de $d\sigma\left\{u\right\}$ par rapport \`a une fonction dite de ``sondage''
$u(k)$ %{\bf COMPL\`ETEMENT INCOMPR\'EHENSIBLE}:
\beq\label{eq:Ex.CS}
d\sigma^{\text{excl}}_N=\left[\left(\prod_{i=1}^{N}d^3k_i\frac{\delta}{\delta u(k_i)}\right)
d\sigma\left\{u\right\}\right]_{\left\{u=0\right\}}.
\eeq
Pour obtenir $d\sigma\left\{u\right\}$ on utilise (\ref{eq:SERCV})
et on multiplie les deux membres de (\ref{eq:Ex.CS}) par la fonction
de sondage $u(k)$, on int\`egre sur l'espace des angles $\Gamma$, on utilise
$$
\left(\frac{\delta}{\delta u(k_i)}\right)u(k)\equiv\delta^3(\vec{k}_i-\vec{k}),\qquad
\prod_{i=1}^{N}\int_{\Gamma}d^3k_i\,\delta^3(\vec{k}_i-\vec{k})=1
$$
et on obtient
\beq\label{eq:SEFG}
d\sigma\left\{u\right\}=d\sigma_0\sum_{N=0,1,\dots\infty}{\cal {F}}^2\prod
\int_{\Gamma_{{\cal{P}}(k),\Theta({\cal {P}})}}d{\cal {K}}(\vec{k})\,u(k).
\eeq
La notion de FG est tr\`es utile dans l'\'etude des grandeurs inclusives.
Pour obtenir la section efficace inclusive partonique d'ordre $N$
on doit faire agir l'op\'erateur suivant sur $d\left\{u\right\}$
et \'evaluer le r\'esultat \`a $u=0$
\beeq\nonumber
d\sigma_N^{\text{incl}}\!&\!=\!&\!\left[\left(\prod_{i=1}^{N}d^3k_i\,\frac{\delta}{\delta u(k_i)}\right)
\sum_{m=0}^{\infty}\frac1{m!}\left(\prod_{j=1}^{m}d^3k_j\,\frac{\delta}{\delta u(k_j)}
\right)d\sigma\left\{u\right\}\right]_{\left\{u=0\right\}}\\ \nonumber\\
\!&\!=\!&\!\left[\left(\prod_{i=1}^{N}d^3k_i\,\frac{\delta}{\delta u(k_i)}\right)
\exp\left\{\int d^3k\,\frac{\delta}{\delta u(k)}\right\}{d\sigma \left\{u\right\}}\right]_{\left\{u=0\right\}}\nonumber\\ \nonumber\\
\!&\!=\!&\!\left[\left(\prod_{i=1}^{N}d^3k_i\,\frac{\delta}{\delta u(k_i)}\right)
d\sigma\left\{u\right\}\right]_{\left\{u=1\right\}},\label{eq:inclCS}
\eeeq
ce qui est \'equivalent au d\'eveloppement en s\'erie de Taylor de la 
fonctionnelle au voisinage de $u(k)=1$.

Les formules pr\'ec\'edentes s'appliquent aux deux cas de processus
exclusifs et inclusifs.
%le sens des quantit\'es exclusives et inclusives {\bf CA VEUT DIRE QUOI
%????}.
Dans le premier cas, (\ref{eq:Ex.CS}), 
pour obtenir la section efficace de production exclusive de $N$ particules, 
o\`u l'on tient compte de toutes les particules qui sont produites
\`a l'\'etat final,  on prend les d\'eriv\'ees de la fonctionnelle
g\'en\'eratrice au point $u=0$.
Dans le deuxi\`eme cas (\ref{eq:inclCS}), on obtient la section efficace
inclusive de production de $N$ particules en effectuant la somme sur l'espace
de phase d'un ensemble de $m$ particules dont on ne tient pas compte
dans l'\'etat final; ceci revient \`a prendre les d\'eriv\'ees de la
fonctionnelle g\'en\'eratrice au voisinage de $u=1$.

\vskip 0.5cm

\textbf{Remarque:} dans le cas de l'\'evolution ind\'ependante des
jets initiaux ($q$ et $\bar{q}$) on peut r\'epresenter (\ref{eq:SEFG})
comme le produit de deux FG's
$$
d\sigma^{e^+e^-}\left\{u\right\}=d\sigma_0Z_F(p_+,1;\left\{u\right\})
Z_F(p_-,1;\left\{u\right\}).
$$
\subsection{Fonctionnelle G\'en\'eratrice pour les jets de quarks et
de gluons}

On prend un seul gluon d'impulsion $l$ qui ne rayonne pas et qui
appartient \`a une certaine branche. L'expression de sa FG d\'ecrit
un sous-jet constitu\'e par une seule particule et s'\'ecrit sous la forme
\beq\label{eq:FGGl}
Z(l,\Theta_l;\left\{u\right\})=u(l)e^{-\omega_G(l,\Theta_l)}=u(l)\exp
\left\{\int_{\Gamma_{(l,\Theta_l)}}d{\cal {K}}\cdot[-1]\right\}.
\eeq
$\Theta_l$ est l'angle d'\'emission du gluon d'impulsion $l$ qui joue
le r\^ole d'angle d'ouverture du sous-jet $l$. $u(l)$ est supprim\'e
par l'exponentielle du facteur de forme qui d\'ecoule de la fonction
${\cal {F}}^2=e^{-\omega_G(l,\Theta_l)}$ car il n'\'emet que des gluons
virtuels.  Cette suppression diminue alors consid\'erablement
la probabilit\'e pour que ce gluon rayonne.
L'expression (\ref{eq:FGGl}) est \'ecrite \`a partir de (\ref{eq:SEFG}).

\begin{figure}[h]
\begin{center}
\includegraphics[height=4.5truecm,width=0.60\tw]{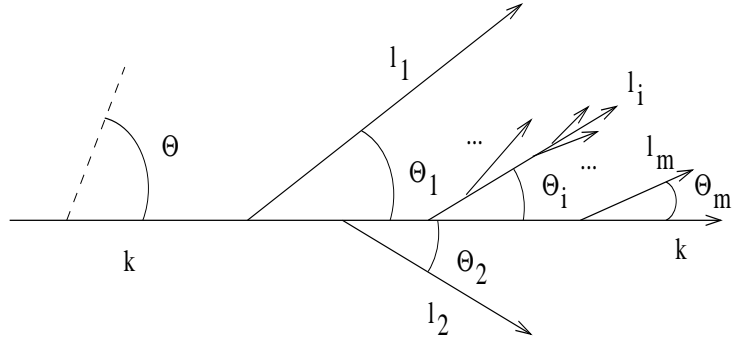}
\caption{\label{fig:inclCS}\'Emission d'un ensemble de gluons mous
d'impulsion $l_i$ par un parton d'impulsion $k$.}
\end{center}
\end{figure}

\begin{figure}[h]
\begin{center}
\includegraphics[height=4.5truecm,width=0.62\tw]{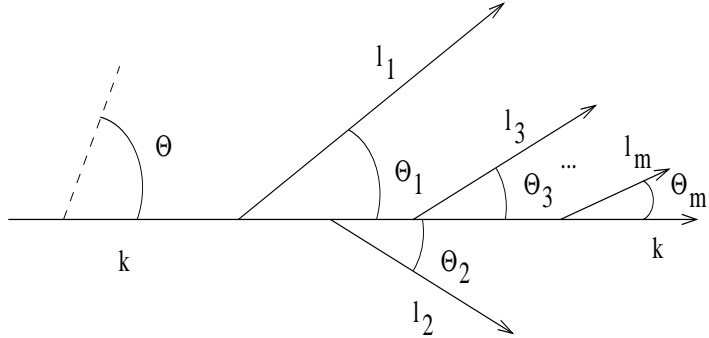}
\caption{\label{fig:InclCSi}\'Emission d'un ensemble de gluons mous
d'impulsion $l_i$ par un parton d'impulsion $k$.}
\end{center}
\end{figure}

Consid\'erons maintenant que ce gluon rayonne quelques gluons
``\'el\'ementaires'' (des gluons qui ne rayonnent pas).
Nous appelons $k$ son impulsion, $\Theta$ son angle d'\'emission
et $l_1,l_2\dots l_m$ les impulsions de ces enfants respectivement,
tels que $\Theta>\Theta_1>\Theta_2>\dots>\Theta_m$ (voir 
Fig.\ref{fig:inclCS}).
\beeq\nonumber
Z(k,\Theta;\left\{u\right\})\!\!&\!\!=\!\!&\!\!u(k)e^{-\omega_G(k,\Theta)}\int_{\Gamma(k,\Theta)}
d{\cal {K}}(\vec{l}_1)\,u(l_1)\,e^{-\omega_G(l_1,\Theta_1)}\times\\ \nonumber\\
&&\hskip-2.5cm\times\int_{\Gamma(k,\Theta_1)}
d{\cal {K}}(\vec{l}_2)\,u(l_2)\,e^{-\omega_G(l_2,\Theta_2)}\dots\int_{\Gamma(k,\Theta_{m-1})}
d{\cal {K}}(\vec{l}_m)\,u(l_m)\,e^{-\omega_G(l_m,\Theta_m)}.
\eeeq
Si l'on prend maintenant l'expression (\ref{eq:FGGl}) et qu'on  la remplace
dans chaque terme de l'\'equation pr\'ec\'edente,
en consid\'erant que $l_1\approx l_2\approx\dots\approx l_m$
(les fr\'equences ne sont plus strictement ordonn\'ees, bien que la
contrainte sur les angles doit \^etre toujours satisfaite,
on a $m!$ mani\`eres indiscernables de les placer) on a
\beq\label{eq:mgluons}
Z(k,\Theta;\left\{u\right\})=u(k)e^{-\omega_G(k,\Theta)}\frac1{m!}
\left[\int_{\Theta>\Theta_l}d{\cal {K}}(\,\vec{l}\,)\,Z(l,\Theta_l;\left\{u\right\})\right]^m.
\eeq
Pour un grand nombre d'\'emissions, 
on effectue la somme sur $m$ jusqu'\`a l'infini, on utilise (\ref{eq:FGGl})
et on r\'ecrit (\ref{eq:mgluons}) sous la forme
\beeq\nonumber
Z(k,\Theta;\left\{u\right\})\!\!&\!\!=\!\!&\!\!u(k)\exp\left\{\int_{\Gamma(k,\Theta)}
d{\cal {K}}(\,\vec{l}\,)\,\left[Z(l,\Theta_l;\left\{u\right\})-1\right]\right\}\\ \nonumber\\
\!\!&=&\!\!u(k)\exp\left\{\int_{\Gamma(k,\Theta)}
\frac{dl}{l}\,\frac{d^2\vec{l}_{\perp}}{2\pi l_{\perp}^2}\,\frac{2N_c\alpha_s}{\pi}\,
\left[Z(l,\Theta_l;
\left\{u\right\})-1\right]\right\}.\label{eq:PME}
\eeeq
Nous devons maintenant d\'emontrer que (\ref{eq:PME}) se g\'en\'eralise,
que les \'emissions soient \'el\'ementaires ou non, on aura ainsi obtenu
l'\'Equation Ma\^itresse satisfaite par la FG dans l'approximation DL.
La d\'emonstration est simple. Il suffit en effet 
de consid\'erer un gluon d'impulsion $l_i$ qui \'emet \`a son tour
 d'autres gluons (Fig.\ref{fig:InclCSi}). On effectue
les m\^emes d\'emarches et on obtient, en respectant la contrainte angulaire
\`a l'int\'erieur du sous-jet, un terme dans (\ref{eq:mgluons}) qui
s'\'ecrit sous la forme
$$
\int_{\Gamma_{(l_i,\Theta_i)}}d{\cal{K}}(\vec{l}_i)\,Z(l_i,\Theta_{i};
\left\{u\right\}),
$$ 
telle que:
\beq\nonumber
Z(l_i,\Theta_i;\left\{u\right\})=u(l_i)\exp\left\{\int_{\Gamma(l_i,\Theta_i)}
d{\cal {K}}(\,\vec{l}_i\,)\,\left[Z({l_i},\Theta_{i};\left\{u\right\})-1\right]\right\}.\nonumber
\eeq
Ceci peut se g\'en\'eraliser $\forall\, i$ et, ensuite, on finit la
d\'emonstration en r\'ep\`etant les \'etapes pr\'ec\'edentes:
$m$ devient le nombre total de sous-jets \`a l'int\'erieur du jet.
On d\'emontre ainsi que pour d\'ecrire le jet, il suffit de d\'ecrire
la premi\`ere \'emission, ce qui est une cons\'equence des contraintes
sur les angles d'\'emission.

%\vskip 0.5cm

\subsection{L'\'Equation Ma\^itresse (EM)}

%\vskip 0.5cm

On donne (voir \cite{EvEq} et les r\'ef\'erences ci-incluses)

\beeq\label{eq:EMFF}
Z_A(k,\Theta;\left\{u\right\})\!\!&\!\!=\!\!&\!\!u(k)e^{-\omega_G(k,\Theta)}+\\ \nonumber\\
&&\hskip -2.5cm\int_{\Gamma(k,\Theta)}\frac{d\omega'}{\omega'}\frac{d^2{k'}\!\!_{\perp}}
{{k}'\!\!_{\perp}^{\,\,\,2}}
\frac{2C_A\alpha_s}{\pi}e^{-\omega_A(k,\Theta)+\omega_A(k,\Theta')}Z_A(k,\Theta';
\left\{u\right\})Z_G(k',\Theta';\left\{u\right\})\nonumber
\eeeq
pour l'interpr\`eter physiquement, puis on d\'emontrera qu'elle est
\'equivalente \`a (\ref{eq:PME}).

\begin{figure}[h]
\begin{center}
\includegraphics[height=5truecm,width=0.60\tw]{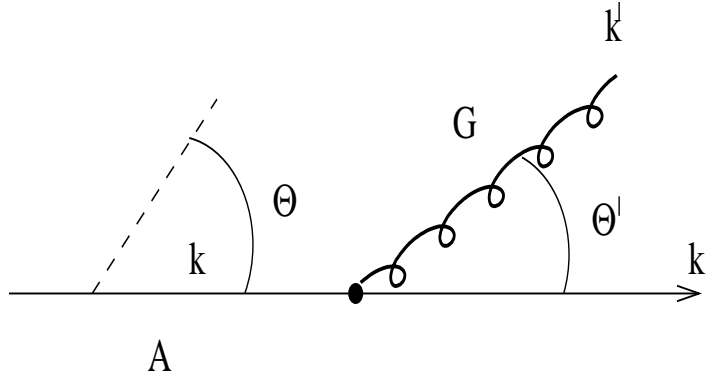}
\caption{\label{fig:agem}\'Emission d'un gluon mou d'impulsion
$k'\approx\omega'\Theta'$ par un parton $A$ d'impulsion $k$.}
\end{center}
\end{figure}
\begin{itemize}
\item{La fonction $Z_A$ repr\'esente la Fonctionelle G\'en\'eratrice (FG)
des grandeurs inclusives pour les jets de quark ou de gluon 
($A$=$G$, $Q$, $\bar{Q}$;\quad $C_G=N_c=3$ et $C_{Q,\bar{Q}}=C_F=4/3$)};

\item{$Z_A=Z_G^{C_F/N_c}$} dans l'approximation DL;

\item{le premier terme de l'\'equation d\'ecrit le jet sans \'emission
d'autres gluons};

\item{dans l'int\'egrand on reconna\^it la structure doublement 
logarithmique, $k'=\left(\omega',\vec{k}'\right)$ est la quadri-impulsion
de la premi\`ere \'emission molle \`a partir du ``parent'' A
(dont l'amplitude est proportionnelle au facteur de couleur correspondant
$C_A$). Cette \'emission est suivie par l'\'evolution de deux jets
G et A de quadri-impulsions k' et k (recul de $A$ n\'eglig\'e) 
respectivement, et $\Theta'=(\widehat{\vec{k}',\vec{k}})$
(voir Fig.\ref{fig:agem}}), la contrainte angulaire impose que les
angles d'\'emission des partons \'emis par la suite
doivent \^etre tr\`es inf\'erieurs \`a $\Theta'$;

\item{l'int\'egration sur $\Gamma$ impose 
 
\begin{center}
$\omega'$\,\,\,$<$\,\,\,
$k^0=|\vec{k}|$,\,\,\,\,\,\,     $\Theta'< \Theta$\\
\end{center}
}
o\`u $\Theta$ est l'angle d'ouverture du jet (param\`etre externe), 
$\Theta\sim 1$ pour \^etre conforme \`a DLA.
La condition $\Theta'<\Theta$ est
une contrainte cin\'ematique. $\Theta$ joue de m\^eme
le r\^ole de param\`etre d'\'evolution;
\item{l'exponentielle dans l'int\'egrand garantit que la premi\`ere
\'emission est celle du gluon de quadri-impulsion $k'$ et que
$A$ n'\'emet que des gluons virtuels dans l'intervalle
angulaire $(\Theta,\Theta')$}.
\end{itemize}
Le param\`etre d'\'evolution $\Theta$ n'intervient que dans le facteur
de forme du parton $A$, soit
$$
u(k)\exp{\left[-\omega_G(k, \Theta)\right]}
$$ 
et sert de limite sup\'erieure d'int\'egration dans les \'equations
d'\'evolution
$$
\int_{\Gamma\left(k,\Theta\right)}\displaystyle{\frac{d\omega'}
{\omega'}}\displaystyle{\frac{d^2k_{\perp}'}{2\pi k_{\perp}'^2}}=
\int_{Q_0/k^0}^{\Theta}\displaystyle{\frac{d\Theta'}
{\Theta'}}\int_{Q_0/\Theta'}^{k^0}\displaystyle{\frac{d\omega'}{\omega'}}
\int_{0}^{2\pi}\displaystyle{\frac{d\phi'}{2\pi}}.
$$
Ici, dans la deuxi\`eme int\'egration, on fixe l'angle et on int\`egre
sur l'\'energie, dans la premi\`ere l'\'energie est fix\'ee
\`a $k^0$. La forme suivante est aussi valable:
$$
\int_{\Gamma\left(k,\Theta\right)}\displaystyle{\frac{d\omega'}
{\omega'}}\displaystyle{\frac{d^2k_{\perp}'}{2\pi k_{\perp}'^2}}=
\int_{Q_0/\Theta}^{k^0}\displaystyle{\frac{d\omega'}{\omega'}}
\int_{Q_0/\omega'}^{\Theta}\displaystyle{\frac{d\Theta'}
{\Theta'}}
\int_{0}^{2\pi}\displaystyle{\frac{d\phi'}{2\pi}}.
$$
Discutons d'abord ces limites d'int\'egration dans le dernier cas:
\begin{itemize}

\item{$\Theta\geq\Theta'$ est une contrainte cin\'ematique, de plus 
$k'_{\perp}\approx\omega'\Theta'\geqslant Q_0$, d'o\`u $\Theta'\geqslant Q_0/\omega'$,
finalemet $Q_0/\omega'\leqslant\Theta'\leqslant\Theta$}
\item{l'\'energie de l'\'emission \'etant inf\'erieure \`a celle du
parton qui l'a \'emis, $\omega'\ll k^0$ (approximation molle), et 
avec $\omega'\Theta\geqslant Q_0$}, nous avons bien
$Q_0/\Theta\leqslant\omega'\leqslant k^0$; ici $k^0$ est une
\'energie. $Q_0$ est le ``cut-off'' colin\'eaire ou l'impulsion
transverse minimale des gluon \'emis.

\end{itemize}
On prend la d\'eriv\'ee de l'\'equation (\ref{eq:EMFF}) par rapport
\`a $\Theta$ sur le produit $Z\exp{(-\omega)}$:
$$
\frac{\partial}{\partial\ln\Theta}\left[e^{-\omega_A(k,\Theta)}Z_A(k,\Theta)\right]=
e^{-\omega_A(k,\Theta)}\int_{Q_0/\Theta}^{k^0}\frac{d\omega'}{\omega'}\int_{0}^{2\pi}
\frac{d\phi'}{2\pi}\frac{2C_A\alpha_s}{\pi}Z_A(k,\Theta)Z_G(k',\Theta),
$$
on utilise 
$$
\frac{\partial}{\partial\ln\Theta}\left[\omega_G(k,\Theta)\right]=
\int_{Q_0/\Theta}^{k^0}\frac{d\omega'}{\omega'}\int_{0}^{2\pi}
\frac{d\phi'}{2\pi}\frac{2C_A\alpha_s}{\pi}\,[\,1\,]
$$ 
et on obtient
\beeq\nonumber
\frac{\partial}{\partial\ln\Theta}Z_A(k,\Theta)\!&\!=\!&\!\int_{Q_0/\Theta}^{k^0}
\frac{d\omega'}{\omega'}\int_{0}^{2\pi}
\frac{d\phi'}{2\pi}\frac{2C_A\alpha_s}{\pi}
[Z_A(k,\Theta)Z_G(k',\Theta)-Z_A(k,\Theta)]\\ \nonumber\\
\!&\!=\!&\!Z_A(k,\Theta)\int_{Q_0/\Theta}^{k^0}
\frac{d\omega'}{\omega'}\int_{0}^{2\pi}
\frac{d\phi'}{2\pi}\frac{2C_A\alpha_s}{\pi}
[Z_G(k',\Theta)-1].\label{eq:DEM}
\eeeq
L'int\'egration de cette \'equation avec la condition initiale 
(pour l'\'emission de plus petit angle)
$$
Z_A(k,\Theta;\left\{u\right\})\vert_{k^0\Theta=Q_0}=u(k),
$$
entra\^\i ne
\begin{eqnarray}
Z_A\left(k,\Theta;\left\{u\right\}\right)=u\left(k\right)
\exp\left(\int_{\Gamma\left(k,\Theta\right)}\frac{d\omega'}
{\omega'}\frac{d^2k{'}\!\!_{\perp}}{2\pi k{'}\!\!_{\perp}^{\,\,\,2}}\frac{C_A}{N_c}\,
\gamma_0^2(k{'}\!\!_{\perp}^{\,\,\,2})\,
\left[Z_G\left(k',\Theta';\left\{u\right\}\right)-1\right]\right).
\label{eq:red_1}
\end{eqnarray}
Nous avons introduit l'expression de la constante anormale $\gamma_0^2$ en DLA
(voir \cite{EvEq} et r\'ef\'erences incluses):
\beq\label{eq:gammadefTH}
\gamma_0^2(k_{\perp}^2)=\displaystyle{\frac{2N_c\,\alpha_s\left(k_{\perp}^2\right)}
{\pi}}=\frac2{\beta\ln(k_{\perp}^2/\Lambda^2)};
\eeq
elle d\'etermine le taux de croissance des multiplicit\'es dans les jets
en fonction de l'\'energie (voir \ref{subsection:muldimanorm}).
Dans (\ref{eq:gammadefTH})
$$
\beta=\frac1{4N_c}\left(\frac{11}3N_c-\frac43 T_R\right)
,\quad T_R=\frac12n_f
$$
o\`u $n_f$ est le nombre de fermions,
$n_f=3$ est le nombre de fermions l\'egers.
La condition de normalisation
\beq\label{eq:CN}
Z_A\left(k,\Theta;\left\{u\right\}\right)\mid_{u=1}\equiv 1.
\eeq
garantit que la section efficace totale est ind\'ependante de 
$Q_0$. On a donc d\'emontr\'e que (\ref{eq:EMFF}) est bien \'equivalente
\`a (\ref{eq:PME}).

Avec ceci on peut obtenir les \'equations d'\'evolution pour les
grandeurs inclusives, telles que le spectre des particules
(voir \ref{sub:article1}), les corr\'elations \'a deux particules
dans un jet (voir \ref{sub:article2} et \ref{sub:article3}).
En g\'en\'eral, on obtient ces observables en
prenant $n$ fois la d\'eriv\'ee fonctionnelle
de la FG par rapport aux fonctions de sondage:
$$
D^{(n)}(k_1,\dots,k_n)=\delta^nZ(\left\{u\right\})/\delta u(k_1)\dots\delta u(k_n)\vert_{u=1},
$$
et les corr\'elations:
$$
\Gamma^{(n)}(k_1,\dots,k_n)=\delta^{n}\ln Z(\left\{u\right\})/\delta u(k_1)\dots u(k_n)\vert_{u=1}.
$$
\`A la place de $\Gamma$ on a utilis\'e ${\cal C}$ dans les articles
\ref{sub:article2} et \ref{sub:article3} pour noter le
corr\'elateur \`a deux particules, soit pour $n=2$ on a:
$$
{\cal C}\equiv\Gamma^{(2)}=\frac{D^{(2)}}{D_1D_2}.
$$

\section{Spectre inclusif d'une particule $\boldsymbol p$ dans un jet}
\label{subsection:HBP}

Pour obtenir la distribution inclusive 
$D^{(1)}\equiv D_A^p$ ou le spectre inclusif d'une particule $p$ (de
quadri-impulsion $k_p$) 
dans le jet $A$ on prend la d\'eriv\'ee fonctionnelle de l'EM
(\ref{eq:red_1}) par rapport \`a $u(k_p)$ au voisinage de $u=1$
\cite{EvEq}:
\begin{equation}
D_A^p(k_p,\Theta)\equiv \frac{\delta}{\delta u\left(k_p\right)}
Z_A\left(k,\Theta;\left\{u\right\}\right)\Big\vert_{u=1},
\end{equation}
$k$ \'etant la quadri-impulsion du parton initiant le jet $A$.
Dans la suite, on consid\`ere
$k^2\approx0$, $k_p^2\approx0$, ce qui permet d'\'ecrire $E=k^0\approx
\mid\!\!\vec{k}\!\!\mid$, $E_p=k_p^0\approx\mid\!\!\vec{k}_p\!\!\mid$.
On a par d\'efinition
$$
xD_A^p(x,\Theta)=E_p\frac{\delta}{\delta u\left(k_p\right)}
Z_A\left(k,\Theta;\left\{u\right\}\right)\Big\vert_{u=1}
$$
o\`u l'on appelle $x=E_p/E$, la fraction de 
l'\'energie $E$ du jet emport\'ee par la particule $p$
d'\'energie $E_p$, d\'efinie dans le centre de masse de la particule de 
quadri-moment $k$.

On fait agir $\delta/\delta{u(k_p)}$ sur l'\'equation (\ref{eq:DEM})
en utilisant la condition de normalisation (\ref{eq:CN}) et on obtient:
\begin{equation}\label{eq:partialspec}
\frac{\partial}{\partial\ln{\Theta}}D_A^p\left(k_p,\Theta\right)=\frac{C_A}{N_c}
\int_{Q_0/(\Theta)_{min}}^{E}\frac{d\omega'}{\omega'}\,
\gamma_0^2\,D_A^p\left(\omega',\Theta\right),
\end{equation}
que l'on int\`egre sur $\Theta$:
\begin{equation}
D_A^p\left(k_p,\Theta\right)=\delta_A^p\delta\left(1-\frac{E_p}{E}\right)+\frac{C_A}{N_c}
\int_{Q_0/E_p}^{\Theta}\frac{d\Theta'}{\Theta'}\int_{Q_0/(\Theta)_{min}}^
{E}\frac{d\omega'}{\omega'}\,\gamma_0^2\,D_A^p\left(\omega',\Theta'\right);
\label{eq:integ}
\end{equation}
le premier terme de (\ref{eq:integ}) r\'epr\'esente le parton $A$
sans d'autres \'emissions, et $\Theta_{min}\!\!=\!\!Q_0/E_p$.
On effectue le changement de variables suivant
(le m\^eme que dans \ref{sub:article2} en ``MLLA'')
\begin{subequations}
\beeq
y&=&\ln\left(E_p\Theta/Q_0\right),\quad
\ell=\ln\left(E/E_p\right)\equiv\ln{\frac1x}\\ \nonumber\\
y'&=&\ln\left(E_p\Theta'/Q_0\right),\quad
\ell'=\ln\left(\omega'/E_p\right)\equiv\ln\frac{z}{x},
\eeeq
\end{subequations}
o\`u $z=\omega'/E$.  Avec ceci,
$Q_0/(\Theta)_{min}\leq\omega'\leq E$ $\Leftrightarrow$ $0\leq\ell'\leq\ell$
et $Q_0/E_p\leq\Theta'\leq\Theta$ $\Leftrightarrow$ $0\leq y'\leq y$.
L'\'equation pour le spectre inclusif d'une particule dans un jet
$A$ se r\'ecrit alors sous la forme simple:
\begin{equation}\label{eq:SpectreDLA}
D_A^p\left(\ell,y\right)=\delta_A^p\delta\left(\ell\right)+
\frac{C_A}{N_c}\int_{0}^{\ell}d\ell'\,\int_{0}^{y}dy'\,
\gamma_0^2(\ell'+y')\,D_A^p\left(\ell',y'\right)
\end{equation}
o\`u l'on a \'ecrit (\ref{eq:gammadefTH}) comme
($k_\perp\approx\omega'\Theta'$)
$$
\gamma_0^2(\omega'\Theta')=\frac{1}{\beta\ln\left(\displaystyle{\frac{\omega'\Theta'}
{\Lambda_{QCD}}}\right)}=
\frac1{\beta\left(\ln\displaystyle{\frac{\omega'}{E_p}}+\ln
\displaystyle{\frac{E_p\Theta'}{Q_0}}+
\lambda\right)}\equiv\gamma_0^2(\ell'+y')=
\frac{1}{\beta(\ell'+y'+\lambda)}.
$$
avec $\lambda=\ln(Q_0/\Lambda_{QCD})$. (\ref{eq:SpectreDLA}) est
g\'en\'eralis\'ee au cadre ``MLLA'' par les \'equations (3.12) et
(3.13) dans l'article \ref{sub:article2}.
Dans le cadre DLA, on fixe $\gamma_0^2$ \`a la valeur
de la virtualit\'e du jet $Q=2E\sin\Theta/2\stackrel{\Theta\ll1}{=}E\Theta$
(duret\'e du processus), soit, pour $\omega'=E$ et $\Theta'=\Theta$,
nous avons
\begin{equation}
\gamma_0^2 = \frac1{\beta(\ell+y+\lambda)}=\frac{1}{\beta(Y_{\Theta}+\lambda)},\qquad 
\ell+y=Y_{\Theta}.
\label{eq:gammabetaTH}
\end{equation}
La relation entre le spectre inclusif d'une particule dans un jet initi\'e
par un quark (anti-quark) et celui d'une particule dans un jet initi\'e
par un gluon en DLA est la suivante:
\beq\label{eq:DQDG}
D_Q^p=\frac{C_F}{N_c}D_G^p.
\eeq
Si l'on veut obtenir le spectre des particules dans un jet $A$
d'\'energie totale $E$ et angle d'ouverture $\Theta$ on doit fixer la somme
$$
Y_{\Theta}\equiv \ell+y=\ln\displaystyle{\frac{E\Theta}{Q_0}}.
$$

\subsection{Solution de l'\'equation (\ref{eq:SpectreDLA}) pour
 $\boldsymbol{\alpha_s}$ constant. 
Transform\'ee de Mellin et ``Hump-Backed plateau''}

\begin{figure}[h]
\begin{center}
\includegraphics[height=7truecm,width=0.60\tw]{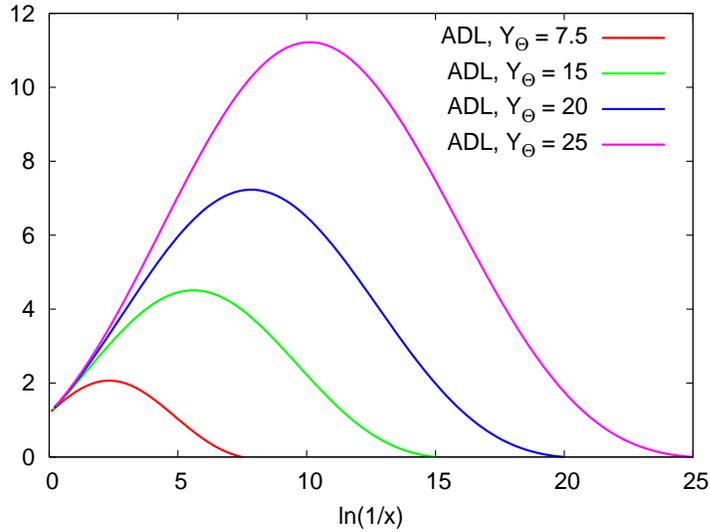}
\caption{\label{fig:SDLA} Spectre inclusif des particules dans deux jets de quark.}
\end{center}
\end{figure}

Dans cette approximation (DLA) on ne tient pas compte de l'\'evolution de la
constante de couplage $\alpha_s$ (ou de la constante anormale $\gamma_0$).
Sa valeur est par cons\'equent fix\'ee \`a celle qui
correspond \`a la premi\`ere \'emission (gluon $G$ \'emis par $A$
\`a l'angle $\Theta'\approx\Theta$)
$$
\gamma_0^2\left(Y_{\Theta}+\lambda\right)=4N_c\frac{\alpha_s}{2\pi}=\frac1{\beta (Y_{\Theta}+\lambda)},\quad \text{o\`u}\quad 
\lambda=\ln\frac{Q_0}{\Lambda}.
$$ 
On peut alors r\'esoudre  (\ref{eq:SpectreDLA}) en effectuant une
transformation de Mellin
\begin{equation}
D_A^p\left(\ell,y\right)=\iint_{C}\frac{d\omega\, d\nu}
{\left(2\pi i\right)^2}\,e^{\omega \ell}
\,e^{\nu y}\,{\cal D}_A^p\left(\omega,\nu\right)
\end{equation}
o\`u le contour $C$ pour chaque int\'egration 
se trouve \`a droite de toute singularit\'e sur le plan complexe.
Nous pouvons \'etendre les limites inf\'erieures d'int\'egration sur
$\ell$ et $y$ \`a $-\infty$ car les int\'egrales correspondantes
dans le plan complexe sur $\omega$ et $\nu$ sont nulles.
% En effet, pour que les int\'egrales converges, si $\ell,y<0$, 
% on doit fermer le contour \`a droite, du coup l'int\'egrale sur le cercle s'annule \`a 
% l'infini et comme il n'y a aucun p\^ole \`a l'int\'erieur le r\'esultat est nul. 
Nous obtenons, en extrayant $\gamma_0^2$ du symbole
de l'int\'egrale, le ``propagateur'' dans l'espace de Mellin:
\begin{equation}\label{eq:DSMellin}
{\cal {D}}_A^p\left(\omega,\nu\right)=\displaystyle{\frac{1}{\nu-\gamma_0^2/\omega}}
\end{equation}
qui pr\'esente un p\^ole en $\omega_0\nu_0=\gamma_0^2$. 
Nous avons utilise' la repr\'esentation suivante pour $\delta(\ell)$
$$
\delta(\ell)=\iint\frac{d\omega\,d\nu}{(2\pi i)^2}e^{\omega\ell}\,\frac{e^{\nu y}}{\nu}.
$$
On effectue la premi\`ere int\'egration sur $\nu$, puisque
$y>0$ on ferme le contour \`a gauche de sorte que l'on inclut le p\^ole
 $\nu_0=\gamma_0^2/\omega$. On prend maintenant l'int\'egrale sur $\omega$:
\begin{equation}
D_A^p\left(\ell,y\right)=\int_{C'}\frac{d\omega}{2\pi i}\exp{\left(\omega\ell+\left(\gamma_0^2/\omega\right)y\right)}
\end{equation}
qui est, en effet, la repr\'esentation int\'egrale de la fonction de Bessel
de premi\`ere esp\`ece.  La solution de (\ref{eq:SpectreDLA}) est 
\begin{equation}
D_A^p\left(\ell,y\right)=\delta_A^p\delta\left(\ell\right)+\frac{C_A}{N_c}
\gamma_0\sqrt{\frac{y}{\ell}}I_1\left(2\gamma_0\sqrt{\ell\,y}\right).
\label{eq:red_2}
\end{equation}
Pour en d\'eduire le spectre inclusif des particules
dans le jet en fonction de $\ell=\ln(1/x)$ par exemple,
on doit fixer $Y_{\Theta}$ et remplacer $y$
dans (\ref{eq:red_2}) par $y=Y_{\Theta}-\ell$. Dans le cas de l'annihilation 
$e^+e^-$ (deux jets de quarks avec $\Theta\sim1$),
le premier terme est nul car $p=G\ne Q (\bar{Q})$, et nous avons 
\begin{equation}
D_{Q,\bar{Q}}^G(\ln(1/x))\equiv\left(\frac{dN}{d\ln(1/x)}\right)_{Q,\bar{Q}}^G=\frac{2C_F}{N_c}
\gamma_0\sqrt{\frac{\ln(Ex/Q_0)}{\ln(1/x)}}\,I_1\left(2\gamma_0\sqrt{\ln\frac{Ex}{Q_0}\,\ln
\frac1x}\right).
\label{eq:Spectreln1/x}
\end{equation}
qui est repr\'esent\'e sur la Fig.\ref{fig:SDLA} pour
$Y_{\Theta}=7.5, 15, 20, 25$. 
C'est ce que l'on connait dans la litt\'erature comme
``hump-backed plateau''\cite{DLA}\cite{DFK}\cite{FW1}.
On pose $D_G^p\equiv D$ et on r\'ecrit:
\beq\label{eq:DSpec}
D(\ell,y)=\delta(\ell)+\gamma_0\sqrt{\frac{y}{\ell}}
\,I_1\left(2\gamma_0\sqrt{y\,\ell}\right).
\eeq
Dans la limite asymptotique ($E$ tr\`es grand), cette expression, avec
$y=Y_{\Theta}\!-\!\ell$, peut \^etre remplac\'ee par (voir \cite{Shabat})
\beq\label{eq:specconst}
D(\ell)=\frac{dN}{d\ln(1/x)}\simeq
\frac12\sqrt{\frac{\gamma_0(Y_{\Theta}-\ell)^{1/2}}{{\pi\,\ell^{3/2}}}}
\exp{\left(2\gamma_0\sqrt{\ell{(Y_{\Theta}\!-\!\ell)}}\right)}
\eeq
Le spectre (\ref{eq:Spectreln1/x}) a une propri\'et\'e int\'eressante,
il pr\'esente un maximum asymptotique pour la valeur $\ell_{max}=
Y/2$; c'est ce que l'on observe sur la figure Fig.\ref{fig:SDLA}.
L'origine de cette bosse est li\'ee \`a la
structure doublement logarithmique des \'emissions
(molles et colin\'eaires). En particulier, on remarque
que la distribution d\'ecro\^it lorsque
$\ell$ augmente, soit lorsque $x$ diminue. On aurait pu s'attendre
\`a ce qu'elle augmente car l'espace de phase, \'etant de plus en plus large,
augmenterait la probablit\'e pour la production de ces particules.
N\'eanmoins, puisque les particules sont de plus en plus 
molles, l'angle $\Theta$ doit augmenter de sorte que la condition $k_{\perp}
\approx\omega\Theta\geqslant Q_0$ soit toujours satisfaite;
dans cette r\'egion, les particules sont \'emises
\`a plus grand angle que pr\'evu et par cons\'equent, les gluons mous
interf\`erent destructivement. C'est le ph\'enom\`ene qu'on connait comme
{\em coh\'erence des gluons mous} en CDQ\cite{DLA}\cite{DFK}.
Il a lieu dans la r\'egion de l'espace de phase o\`u l'impulsion 
transverse est tr\`es faible, ici, pour $y\!\rightarrow\!0$.
Ceci se voit dans la d\'ecroissance de la distribution (Fig.\ref{fig:SDLA}
dans la limite $\ell\!\rightarrow\!Y_{\Theta}$, soit lorsque $x$ d\'ecro\^it
(voir aussi \cite{EvEq} et r\'ef\'erences incluses).

\subsection{D\'eriv\'ees logarithmiques
(utiles pour l'article \ref{sub:article2} et \ref{sub:article3})}
\label{sub:derlog}

Il est tr\`es utile pour le chap\^itre suivant
de donner la d\'efinition et l'ordre de grandeur
des d\'eriv\'ees logarithmique du spectre. On d\'efinit
\beq\label{eq:LogD}
\psi(\ell,y)=\ln [D(\ell,y)].
\eeq
On prend les d\'eriv\'ees par rapport \`a $\ell$ et $y$
\beq
\psi_{\ell}(\ell,y)=\frac{D_{\ell}(\ell,y)}{D(\ell,y)}=\gamma_0\sqrt{\frac{y}{\ell}}, \qquad
\psi_y(\ell,y)=\frac{D_y(\ell,y)}{D(\ell,y)}=\gamma_0\sqrt{\frac{\ell}{y}}.
\eeq
Dans les deux cas on constate que $\psi_{\ell}={\cal {O}}(\gamma_0)$ et 
$\psi_{y}={\cal {O}}(\gamma_0)$. Si on passe \`a la seconde
d\'eriv\'ee on voit que
$\psi_{\ell\ell}=\psi_{yy}=\psi_{\ell y}={\cal {O}}(\gamma_0^3)$ sachant que
$1/\ell$ ou $1/y$ $\propto{\cal {O}}(\gamma_0^2)$.

\section{Distribution doublement diff\'erentielle inclusive et
distribution angulaire inclusive de la particule 
d\'etect\'ee}
\label{subsection:distdoubdiff}

Nous pr\'esentons le seul calcul qui a \'et\'e effectu\'e dans le cas des
distributions inclusives en fonction de l'impulsion transverse des partons
\'emis (voir \cite{EvEq} et r\'ef\'erences incluses).
Nous l'avons g\'en\'eralis\'e au cadre MLLA dans l'article
\ref{sub:article1}. 
La distribution inclusive doublement diff\'erentielle est obtenue
en diff\'erentiant (\ref{eq:red_2}) par rapport \`a l'angle $\Theta$
de la particule d\'etect\'ee, soit par rapport \`a $y$
\beq\label{eq:dddist}
\frac{dD}{dy}\equiv\frac{d^2N}{d\ell\,dy}=\frac{d^2N}
{d\ell\, d\ln\Theta}=\gamma_0^2\,I_0\left(2\gamma_0\sqrt{\ell\, y}\right).
\eeq

\begin{figure}[h]
\begin{center}
\includegraphics[height=5truecm,width=0.48\tw]{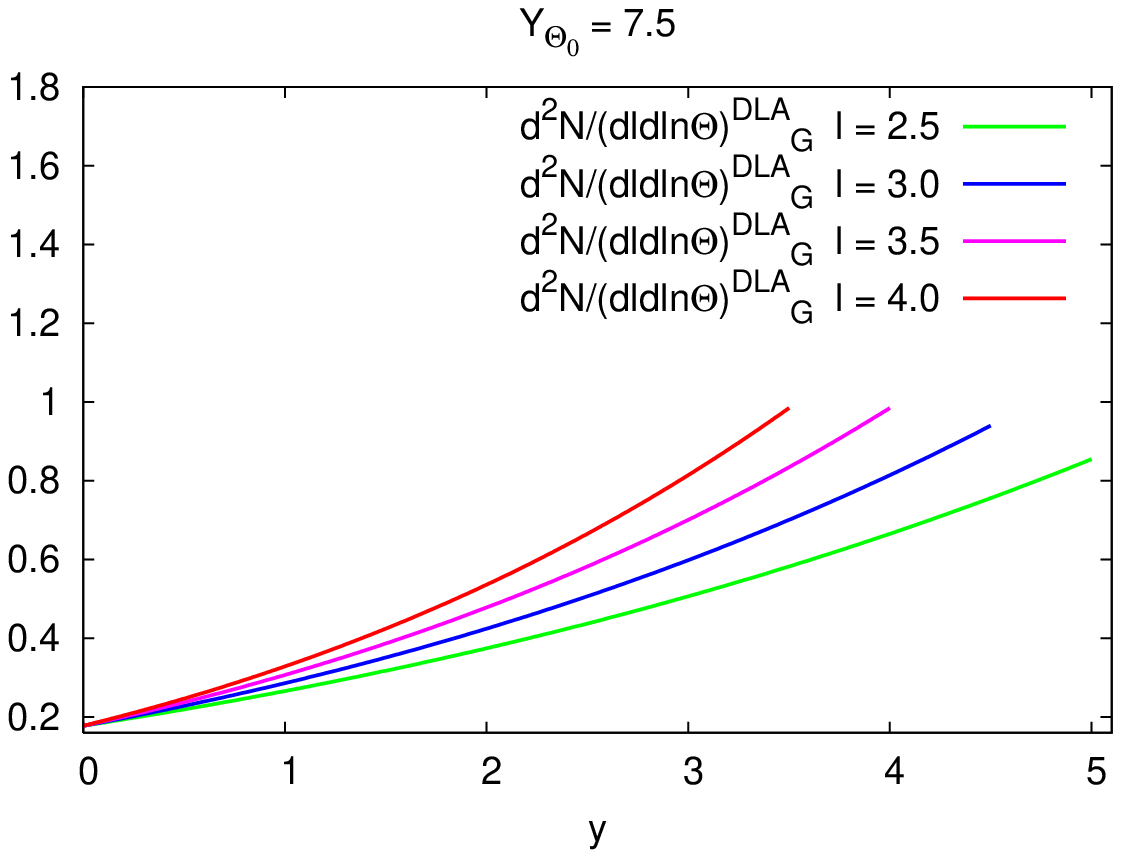}
\hfill
\includegraphics[height=5truecm,width=0.48\tw]{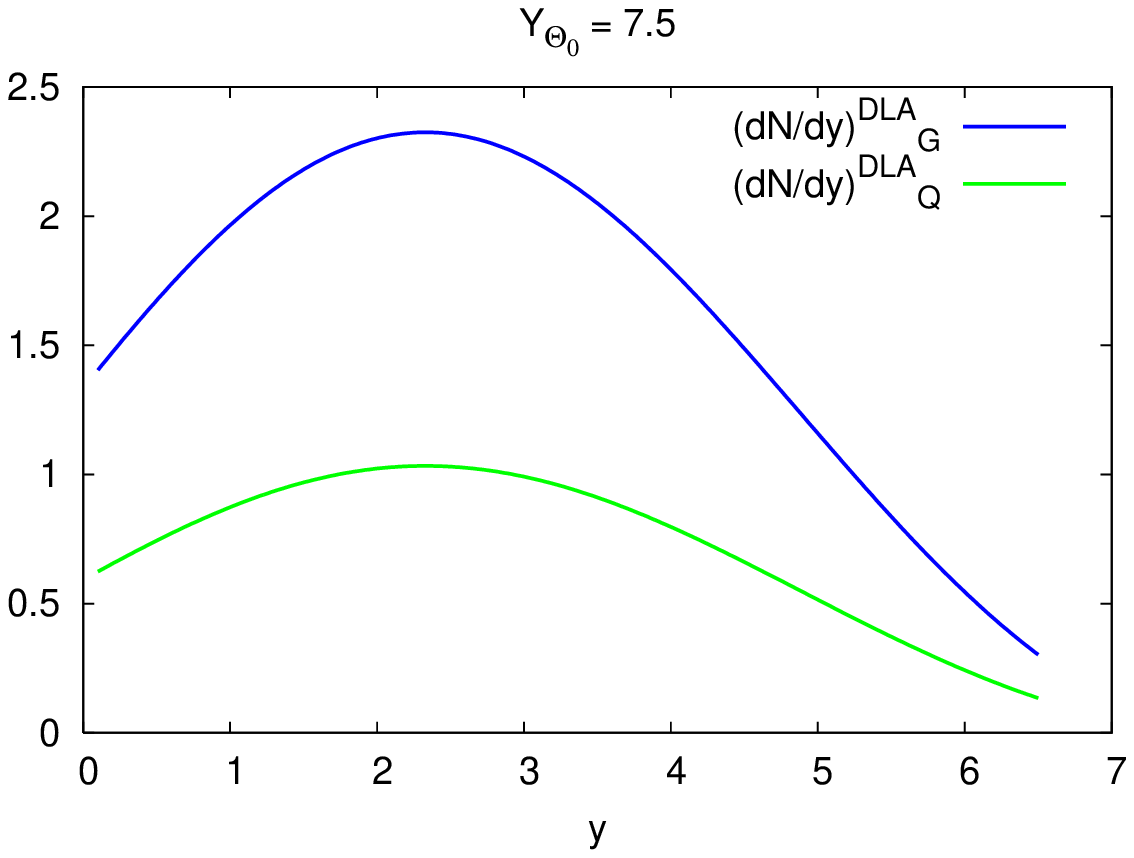}
\caption{\label{fig:DDLAy7.5} Distribution doublement
inclusive (gauche) pour $\ell$ fix\'e en fonction de $y$ et distribution
angulaire inclusive (droite).}
\end{center}
\end{figure}

Pour en d\'eduire la distribution angulaire , ou la distribution en fonction
de $y=\ln k_{\perp}$ on int\`egre (\ref{eq:dddist}) dans l'intervalle
$0\!\leqslant\!\ell\!\leqslant\!Y_{\Theta_0}\!-\!y$, c'est \`a dire
 sur l'\'energie des particules qui se trouvent
dans le c\^one d'angle d'ouverture $\Theta$ (qui est
inf\'erieur \`a l'ouverture totale $\Theta_0$ du jet).
L'expression pour la distribution angulaire inclusive est obtenue de
la fa\c con suivante
\beq\label{eq:daincl}
\frac{dN}{d\ln k_{\perp}}\equiv\int_{0}^{Y_{\Theta_0}-y}d\ell\,\frac{d^2N}{d\ell\,dy}=
\gamma_0^2\int_{0}^{Y_{\Theta_0}-y}d\ell\,I_0\left(2\gamma_0\sqrt{\ell\, y}\right).
\eeq
Nous pouvons repr\'esenter (\ref{eq:dddist}) en fonction de $y$ pour plusieurs 
valeurs de $\ell$, c'est \`a dire qu'on fixe l'\'energie
de la particule d\'etect\'ee et on trace sa d\'ependance en fonction de
$k_\perp$ (voir Fig.\ref{fig:DDLAy7.5} gauche). De m\^eme, on peut
repr\'esenter (\ref{eq:daincl}) num\'eriquement pour
plusieurs valeurs de $Y_{\Theta_0}$ (voir Fig.\ref{fig:DDLAy7.5} droite).

Dans cette approximation (DLA), on n\'eglige l'\'evolution du jet
entre $\Theta_0$ et $\Theta<\Theta_0$.
Il sera d\'emontr\'e dans l'article \ref{sub:article1},
que l'\'evolution du jet entre les angles
$\Theta$ et $<\Theta$ entra\^\i ne des corrections qui ne sont pas
n\'egligeables.  On peut en d\'eduire les courbes pour les distributions
\`a l'int\'erieur d'un jet de quark en multipliant par le facteur
$C_F/N_c$. Une fois de plus, on constate, dans la d\'ecroissance de
la Fig.\ref{fig:DDLAy7.5} lorsque $y\rightarrow0$ les effets 
de coh\'enrece des gluons mous qui sont \'emis dans cette r\'egion.

\vskip 0.5cm

\textbf{Remarque:} La forme des distributions de la Fig.\ref{fig:DDLAy7.5}
\'et\'e compar\'ee avec celle que nous avons obtenue dans le cadre MLLA
dans l'appendice \textbf{E.3} de l'article \ref{sub:article1}.

\section{Multiplicit\'es des jets et interpr\'etation de la dimension
anormale}
\label{subsection:muldimanorm}

Dans ce paragraphe, on applique les techniques de la FG au calcul des 
multiplicit\'es dans les jets. Dans cet objectif,
on remplace $u(k)$ par une constante dans (\ref{eq:red_1})
car cette observable, contrairement au cas du spectre inclusive,
ne d\'epend ni de l'\'energie $E$ ni de l'angle d'ouverture du jet $\Theta$
s\'epar\'ement, mais de leur produit $Q=E\Theta$, soit de l'impulsion
charact\'eristique du jet (ou de la virtualit\'e totale du jet). On pose
$$
Z(k,\Theta;u(k))=Z(y;u),\quad y=\ln\frac{k\Theta}{Q_0}\equiv\ln\frac{Q}{Q_0}
$$
et on \'ecrit l'\'equation ma\^itresse qui permet d'obtenir les \'equations
des multiplicit\'es et des fluctuations des multiplicit\'es 
(corr\'elateurs de multiplicit\'es) dans les jets:
\begin{equation}\label{eq:Zenfyu}
Z(y;u)=u\exp\left(\int_0^ydy'(y-y')\gamma_0^2(y')[Z(y';u)-1]\right),\quad 
y'\equiv\ln\frac{k_\perp}{Q_0}.
\end{equation}
Les conditions initiales et de normalisation s'\'ecrivent maintenant sous la
forme:
\beq\label{eq:ZZu}
Z(0;u)=u,\qquad Z(y;1)=1.
\eeq
Pour obtenir (\ref{eq:Zenfyu}) nous avons effectu\'e la cha\^\i ne
de transformations suivante:
$$
\int_{\Gamma(k,\Theta)}\frac{d\omega'}{\omega'}\frac{d^2k'_\perp}{2\pi k_{\perp}^{'2}}=
\int_{Q_0/\Theta}^{k^0}\frac{d\omega'}{\omega'}\int_{Q_0}^{\omega'\Theta}\frac{
d(\omega'\Theta')}{\omega'\Theta'}=
\int_{\ln Q_0}^{\ln(k^0\Theta)}d\ln(\omega'\Theta)
\int_{\ln Q_0}^{\ln(\omega'\Theta)}
d\ln k_\perp.
$$
Il y a deux mani\`eres de traiter les fluctuations des multiplicit\'es:
soit on \'etudie la distribution $P_n=\sigma^n/\sigma$ des \'ev\`enements
sur le nombre total des particules produites,  soit on mesure les
corr\'elateurs des multiplicit\'es inclusives
$n_k=\big<\!n(n-1)\dots(n-k+1)\!\big>$.
D'apr\`es (\ref{eq:Ex.CS}) qui d\'efinit les propri\'et\'es exclusives des
distributions partoniques en termes de la FG, la probabilit\'e de production
de $n$ particules s'\'ecrit alors sous la forme:
$$
P_n=\frac1{n!}\prod_{i=1}^{n}\left(\int d^3k_i\frac{\delta}{\delta(u(k_i))}\right)
Z(\left\{u\right\})\bigg|_{\left\{u=0\right\}}.
$$
La deriv\'ee varitionnelle sur les fonctions $u(k)$, suivie de
l'int\'egration sur l'espace de phase des quadri-impulsions des partons,
 est \'equivalente \`a la diff\'erentiation de la fonction
$Z(u),\;u(k)\equiv u$ que l'on a d\'ej\`a \'evoqu\'ee:
$$
P_n(y)=\frac1{n!}\left(\frac{d}{du}\right)^nZ(y;u)\bigg|_{\left\{u=0\right\}};\qquad
Z(y;u)=\sum_{n=0}^{\infty}P_n(y)u^n.
$$
Il est utile de remarquer que la distribution des multiplicit\'es est
normalis\'ee \`a l'unit\'e:
$$
\sum_{n=0}^{\infty}P_n=\sum_{n=0}^{\infty}\frac1{n!}\left(\frac{d}{du}\right)^n
Z(u)\bigg|_{\left\{u=0\right\}}=e^{d/du}Z(u)\bigg|_{\left\{u=0\right\}}=
Z(u)\bigg|_{\left\{u=1\right\}}=1.
$$

%\vskip 0.5cm

\subsection{Corr\'elateurs des multiplicit\'es}

Par d\'efinition
$$
n_k(y)\equiv\sum_{n=k}^{\infty}n(n-1)\dots(n-k+1)P_n(y).
$$
Cette proc\'edure \'equivaut au calcul du $k^{\text{\`eme}}$ coefficient
de l'expansion de Taylor de $Z$ au voisinage de $u=1$
\beq\label{eq:Taylorexp}
n_k(y)=\left[\left(\frac{d}{du}\right)^k\sum_{n=0}^{\infty}u^nP_n(y)\right]
\Bigg|_{\left\{u=1\right\}}=\left(\frac{d}{du}\right)^kZ(y;u)
\Bigg|_{\left\{u=1\right\}}.
\eeq
Par cons\'equent, l'expansion des FG's au voisinage de $u=1$ peut
s'\'ecrire en termes des corr\'elateurs des multiplicit\'es comme
$$
Z(y;u)=\sum_{k=0}^{\infty}\frac{(u-1)^k}{k!}n_k(y),\quad (n_0\equiv0).
$$

%\vskip 0.5cm

\subsection{Multiplicit\'e moyenne des partons}

Elle  correspond \`a la valeur $k=1$ dans l'expression
(\ref{eq:Taylorexp}): $<n>=n_1=\bar n$. On diff\'erentie (\ref{eq:Zenfyu})
deux fois par rapport \`a $y$, puis par rapport \`a $u$,
on pose $u=1,\;Z=1$ et on obtient l'\'equation pour $\bar n(y)$ dans le
cadre DLA:
\beq\label{eq:eqbarn}
\bar n''(y)=\gamma_0^2(y)\bar n(y)
\eeq
avec les conditions initiales $\bar n(0)=1$, $\bar n'(0)=0$. Ici,
$$
\gamma_0^2(k_\perp)=\frac1{\beta\ln\displaystyle{\frac{k_\perp}{\Lambda_{QCD}}}}=
\frac1{\beta\left(\displaystyle{\ln\frac{k_\perp}{Q_0}+\ln\frac{Q_0}{\Lambda_{QCD}}}
\right)}\equiv\gamma_0^2(y)=\frac1{\beta(y+\lambda)},\quad\lambda=\ln\frac{Q_0}
{\Lambda_{QCD}}.
$$

%\vskip 0.5cm

\subsection{Solution de l'\'equation (\ref{eq:eqbarn}) pour
$\boldsymbol{y+\lambda,\lambda\gg1}$}

%\vskip 0.5cm

On r\'esout (\ref{eq:eqbarn}) en effectuant une transformation de 
Mellin-Laplace que l'on \'ecrit sous la forme:
\beq\label{eq:nbarmellin}
\bar n(y)=\int_C\frac{d\omega}{2\pi i}\,e^{\omega y}\,\bar n(\omega),
\eeq
o\`u $C$ est le contour d'int\'egration que l'on choisit \`a droite de toute
singularit\'e. On substitue (\ref{eq:nbarmellin}) dans (\ref{eq:eqbarn}) et
on obtient l'\'equation diff\'erentielle satisfaite par $\bar n(\omega)$:
\beq
\frac{d\bar n(\omega)}{d\omega}
=\left(\lambda-\frac2{\omega}-\frac1{\beta\omega^2}\right)\bar n(\omega),
\eeq
que l'on int\`egre, puis l'on remplace dans (\ref{eq:nbarmellin}) pour obtenir
\beq\label{eq:barno}
\bar n(y)=const\times\int_C\frac{d\omega}{2\pi i}
\frac{\exp\left[{\displaystyle{(y+\lambda)\omega+
1/{\beta\omega}}}\right]}{\omega^2},\quad 
const=1\Bigg/\!\!\!\int_C\frac{d\omega}{2\pi i}
\frac{\exp\left[{\displaystyle{\lambda\omega+
1/{\beta\omega}}}\right]}{\omega^2}
\eeq
o\`u la constante d\'ecoule de la condition $\bar n(0)=1$.
Or, dans la limite asymptotique $y+\lambda,\lambda\gg1$ qui garantit, en
particulier, la convergence de la s\'erie perturbative $\alpha_s/\pi\ll1$,
les repr\'esentations dans (\ref{eq:barno}) peuvent \^etre estim\'ees
par la m\'ethode du col. On n'\'ecrit
pas la constante qui se simplifient dans le produit de:
$$
\bar n(y)\propto(y+\lambda)^{1/4}\exp\left(\frac2{\sqrt{\beta}}
\sqrt{y+\lambda}\right),
\qquad
const\propto\lambda^{-1/4}\exp\left(-\frac2{\sqrt\beta}\sqrt\lambda\right)
$$
et on obtient
\begin{equation}\label{eq:multDLA}
\bar n(y)\stackrel{y+\lambda,\lambda\gg1}{\simeq}\left(\frac{y+\lambda}{\lambda}
\right)^{1/4}\exp\left[\frac2{\sqrt\beta}\left(\sqrt{y+\lambda}-\sqrt{\lambda}
\right)\right].
\end{equation}
Cette expression s'allie \`a celle de l'approximation WKB \cite{EvEq} par
la repr\'esentation
$$
\bar n(y)=\exp\left\{\int^ydy'\gamma_0(y')\right\},
$$
ainsi, la quantit\'e $\gamma_0$ s'identifie \`a la dimension anormale
qui d\'etermine le taux de croissance des multiplicit\'es en fonction
de l'\'energie dans les jets hadroniques.

\section{Repr\'esentation int\'egrale du spectre dans l'espace de Mellin;
cas $\boldsymbol{\alpha_s(k_{\perp})}$ variable}

On consid\`ere maintenant l'effet de la variation de la constante de
couplage sur le spectre. On ne peut donc pas extraire $\gamma_0^2$
de l'int\'egrale. Nous allons exposer les techniques n\'ecessaires
pour estimer le spectre, pour la premi\`ere fois, dans le 
cadre de l'approximation o\`u on introduit les corrections en logs simples
(MLLA, voir \ref{sub:article3}). On effectue le changement de variable suivant
\begin{equation}
F\left(\ell,y\right)=\gamma_0^2\,D\left(\ell,y\right)=
\frac1{\beta}\left(\ell+y+\lambda\right)^{-1}D\left(\ell,y\right),
\end{equation}
o\`u 
\beq\label{eq:gamma02}
\gamma_0^2\equiv\gamma_0^2(\ell+y)=\frac1{\beta(\ell+y+\lambda)},
\eeq
on r\'ecrit l'\'equation (\ref{eq:SpectreDLA}) sous la forme:
\begin{equation}\label{eq:eqalphasrun}
\beta\left(\ell+y+\lambda\right)F
\left(\ell,y\right)=\delta\left(\ell\right)+
\int_{0}^{\ell}d\ell'\int_{0}^{y}dy'\,F\left(\ell',y'\right)
\end{equation}
et on effectue une transformation de Mellin sur la fonction $F(\ell,y)$, soit
\begin{equation*}
F\left(\ell,y\right)=\iint\frac{d\omega d\nu}
{\left(2\pi i\right)^2}\,e^{\omega\ell+\nu y}{\cal F}\left(\omega,\nu\right),
\end{equation*}
le membre de gauche de l'\'equation (\ref{eq:eqalphasrun}) devient,
au facteur $\beta$ pr\`es, apr\`es une int\'egration par parties 
\beq\nonumber
\iint\frac{d\omega\, d\nu}{\left(2\pi i\right)^2}
\left[\left(\frac{\partial}{\partial\omega}
\!+\!\frac{\partial}{\partial\nu}\!+\!\lambda\right)
e^{\omega\ell+\nu y}\right]{\cal F}\left(\omega,\nu\right)
=\iint\frac{d\omega\, d\nu}
{\left(2\pi i\right)^2}\left(\lambda{\cal F}\!-\!\frac{\partial{\cal F}}
{\partial\omega}\!-\!\frac{\partial{\cal F}}
{\partial\nu}\right)\,e^{\omega\ell+\nu y}.
\eeq
L'\'equation dans l'espace $(\omega$,$\nu)$ devient:
\begin{equation}\label{eq:Fequat}
\beta\left(\lambda{\cal F}-\frac{\partial{\cal F}}{\partial\omega}-\frac{\partial {\cal F}}{\partial\nu}\right)=\frac{1}{\nu}+\frac{{\cal F}}{\omega\nu}.
\end{equation}
On d\'ecouple cette \'equation en effectuant le changement de variables
suivant:
$$
\omega'=\frac{\omega+\nu}{2}\quad \text{et}\quad \omega'=\frac{\omega-\nu}{2}
$$
et on r\'ecrit (\ref{eq:Fequat}) sous la forme
\begin{equation*}
\beta\left(\lambda{\cal F}-\frac{d{\cal F}}{d\omega'}\right)=
\frac{1}{\omega'-\nu'}+\frac{{\cal F}}{\omega'^2-\nu'^2}.
\end{equation*}
Il s'agit d'une \'equation lin\'eaire non-homog\`ene que l'on peut
r\'esoudre facilement.
La solution de l'\'equation homog\`ene est la suivante:
\begin{equation*}
{\cal F}\left(\omega,\nu\right)=\frac1{\beta}\int_{0}^{\infty}
\frac{ds}{\nu+s}\left(\frac{\omega\left(\nu+s\right)}
{\left(\omega+s\right)\nu}\right)^{1/\beta\left(\omega-\nu\right)};
\end{equation*}
nous l'avons r\'ecrite en fonction des variables pr\'ec\'edentes. On obtient
une repr\'esentation int\'egrale pour le spectre inclusif d'une particule avec
$\alpha_s$ variable
\begin{equation}\label{eq:DLAalphasrun}
D\left(\ell,y\right)=\left(\ell\!+\!y\!+\!\lambda\right)\iint\frac{d\omega\, d\nu}
{\left(2\pi i\right)^2}e^{\omega\ell+\nu y}
\int_{0}^{\infty}\frac{ds}{\nu+s}
\left(\frac{\omega\left(\nu+s\right)}
{\left(\omega+s\right)\nu}\right)^{1/\beta\left(\omega-\nu\right)}\,e^{-\lambda s}.
\end{equation}
Cette repr\'esentation a \'et\'e donn\'ee, sans d\'emonstration, dans
\cite{DLA} et \cite{DFK}. Elle est g\'en\'eralis\'ee dans l'article
\ref{sub:article3} au cadre MLLA.

\vskip 0.5cm

\textbf{Remarque:} Dans la limite $\alpha_s$-constante, on doit trouver
(\ref{eq:DSpec}).  Ceci revient \`a poser $\ell=y=0$ dans (\ref{eq:gamma02}),
soit
\footnote{cette limite n'est bien s\^ur pas physique vu que $Q\ne Q_0$.}
$$
\gamma_0^2(\ell+y)=\frac1{\beta(\ell\!+\!y\!+\!\lambda)}
\stackrel{\lambda\gg\ell+y}{=}
\gamma_0^2=\frac1{\beta\lambda};\quad\text{avec}\quad\lambda\gg1\quad
\Rightarrow\quad\gamma_0^2\ll1.
$$
Ceci signifie que l'on ne tient pas compte de l'\'evolution de la constante
de couplage.  Tout se passe comme si on donnait \`a la premi\`ere \'emission
la valeur de l'impulsion transverse minimale $Q_0$.
Nous engageons le lecteur \`a \'etudier le paragraphe 3.4 de l'article
\ref{sub:article3}.  Pour $\lambda\gg1$, on doit avoir $s\ll1$ 
de sorte que l'on \'evalue l'int\'egrale dans le domaine de $s$ o\`u
le r\'esultat est non-n\'egligeable. On effectue l'expansion
dans l'int\'egrand (voir aussi le paragraphe 3.4 de \ref{sub:article3}):
\beeq\nonumber
\frac1{\nu+s}\left(\frac{\omega\left(\nu+s\right)}
{\left(\omega+s\right)\nu}\right)^{1/\beta\left(\omega-\nu\right)}&\approx&
\frac1{\nu}\left(1+\frac{\omega-\nu}{\omega\nu}s\right)^{1/\beta\left(\omega-\nu\right)}
\\ \nonumber\\
&&\hskip -4cm
\approx\frac1{\nu}\left[1+\frac1{\beta\omega\nu}s+\frac1{2!}
\left(\frac{s}{\beta\omega\nu}\right)^2+\frac1{3!}
\left(\frac{s}{\beta\omega\nu}\right)^3+\dots\right],\nonumber
\eeeq
quand on int\`egre sur $s$, on utilise 
$\int_0^{\infty}s^n\,e^{-\lambda s}=\frac{n!}{\lambda^n}$ et on obtient
\beeq\nonumber
{\cal {D}}(\omega,\nu)
&\approx&\frac1{\nu}\left[1+\frac1{\omega\nu}\frac1{\beta\lambda}+
\left(\frac1{\omega\nu}\right)^2\left(\frac1{\beta\lambda}\right)^2+
\left(\frac1{\omega\nu}\right)^3\left(\frac1{\beta\lambda}\right)^3+\dots\right]\\ \nonumber\\
&=&\frac1{\nu-\gamma_0^2/\omega}\nonumber
\eeeq
en accord avec l'expression du ''propagateur'' (\ref{eq:DSMellin}).

\vskip 0.5cm

\textbf{Remarque:} La limite $\lambda=0$ ne peut pas \^etre prise car la 
repr\'esentaion (\ref{eq:DLAalphasrun}) diverge.

\section{Estimation du spectre par la m\'ethode du col \cite{Shabat}}
\label{subsection:col}

Nous donnons ici  l'estimation du spectre (\ref{eq:DLAalphasrun})
en pr\'esentant la m\'ethode du col de mani\`ere d\'etaill\'ee
(car elle sera utilis\'ee dans \ref{sub:article3}). On utilise
les techniques et notations de \cite{DLA} que l'on g\'en\'eralisera au
cadre MLLA dans l'approximation $Q_0\ne\Lambda_{QCD}$
(voir article \ref{sub:article3}).  On d\'efinit la fonction \cite{DLA}
\begin{equation}
\sigma\left(s\right)=\frac1{\beta(\omega-\nu)}
\ln{\frac{\omega\left(\nu+s\right)}{\left(\omega+s\right)\nu}}-\lambda s
\end{equation}
dont on prend la d\'eriv\'ee pour obtenir la valeur $s_0$ qui l'annule,
$\sigma'(s_0)=0$
\begin{equation*}
\sigma'\left(s_0\right)=\frac1{\beta\left(\nu+s\right)
\left(\omega+s\right)}-\lambda=0,
\end{equation*}
et on obtient $s_0(\omega,\nu,\lambda)=\frac{1}{2}\left(\sqrt{\frac{4}{\beta\lambda}+
\left(\omega-\nu\right)^2}-\left(\omega+\nu\right)\right)$.

Pour v\'erifier qu'on peut bien appliquer cette m\'ethode,
il faut d\'eterminer le signe de $\sigma''(s_0)$; cel\`a donne
$$
\sigma''(s_0)=-\beta\lambda^2\sqrt{\frac{4}{\beta\lambda}+(\omega-\nu)^2}<0,
$$
donc on peut effectuer l'expansion de Taylor:
$$
\sigma(s)=\sigma(s_0)+\frac12\sigma''(s_0)(s-s_0)^2+{\cal {O}}((s-s_0)^2).
$$
On obtient ainsi une int\'egrale gaussienne dont le r\'esultat
sera d'autant meilleur que $\lambda\gg1$,
\beeq\nonumber
\int_{0}^{\infty}\frac{ds}{\nu+s}
\left(\frac{\omega\left(\nu+s\right)}
{\left(\omega+s\right)\nu}\right)^{1/\beta\left(\omega-\nu\right)}\,e^{-\lambda s}
\!&\!\approx\!&\!\frac{e^{\sigma(s_0)}}{\nu+s_0}\int_0^{\infty}ds\,e^{
-\frac12\mid\!\sigma''(s_0)\!\mid(s-s_0)^2}\\ \nonumber\\
&&\hskip -7.0cm=\frac{e^{\sigma(s_0)}}{\nu+s_0}
\frac1{\sqrt{\mid\!\sigma''(s_0)\!\mid}}\int_{-\xi_0}^{\infty}d\xi\,e^{-\xi^2}
=\frac{e^{\sigma(s_0)}}{\nu+s_0}
\sqrt{\frac{\pi}{2\mid\!\sigma''(s_0)\!\mid}}\left[1-\text{erf}(-\xi_0)\right]
\nonumber\\\nonumber\\
&&\hskip -7.0cm=\sqrt{\frac{\pi}2}\displaystyle{\frac{e^{\sigma(s_0)}}{(\nu+s_0)
\sqrt{\mid\!\sigma''(s_0)\!\mid}}}G(\xi_0)\stackrel{\lambda\gg1}{\approx}
2\sqrt{\frac{\pi}2}\displaystyle{\frac{e^{\sigma(s_0)}}{(\nu+s_0)
\sqrt{\mid\!\sigma''(s_0)\!\mid}}}\label{eq:ints}.
\eeeq
En effet, $\xi_0(\omega,\nu,\lambda)\!\!=\!\!s_0
\sqrt{\mid\!\sigma''(s_0)\!\mid}$ et 
$G(\xi_0)\!\!=\!\![1-\text{erf}(-\xi_0)]$, avec 
$\xi_0\simeq\lambda^{1/4}\Rightarrow G\rightarrow2$
\footnote{la fonction $\text{erf}(x)$ tombe tr\`es vite vers $-1$,
($1$) lorsque $x<0$, ($x>0$).}. 
Nous avons alors pour $D(\ell,y)$ l'estimation suivante 
\begin{equation}
D\left(\ell,y\right)\approx2\sqrt{\frac{\pi}2}(\ell+y+\lambda)
\iint\frac{d\omega\, d\nu}{\left(2\pi i\right)^2}\,
\displaystyle{\frac{e^{\phi\left(\omega,\nu,\ell,y\right)}}{(\nu+s_0)
\sqrt{\mid\!\sigma''(s_0)\!\mid}}},
\end{equation}
o\`u
\beq\label{eq:phiexpTH}
\phi\left(\omega,\nu,\ell,y\right)=\omega\ell+\nu y+
\frac1{\beta\left(\omega-\nu\right)}
\ln{\frac{\omega\left(\nu+s_0\right)}{\left(\omega+s_0\right)\nu}}-
\lambda s_0(\omega,\nu,\lambda).
\eeq
On utilise la m\'ethode du col une deuxi\`eme fois, maintenant
sur la double int\'egration pour ainsi d\'eterminer le point de col
$(\omega_0,\nu_0)$ qui satisfait
$$
\frac{\partial\phi}{\partial\omega}=
\frac{\partial\phi}{\partial\nu}=0,
$$
et on obtient les \'equations
\begin{subequations}
\begin{equation}\label{eq:deromegaTH}
\frac{\partial\phi}{\partial\omega}=\ell-\frac1{\beta\left(\omega-\nu\right)^2}
\ln{\frac{\omega\left(\nu+s_0\right)}{\left(\omega+s_0\right)\nu}}+\frac1
{\beta\omega\left(\omega-\nu\right)}-\lambda\frac{\left(\nu+s_0\right)}
{\left(\omega-\nu\right)}=0,
\end{equation}
\begin{equation}\label{eq:dernuTH}
\frac{\partial\phi}{\partial\nu}=y+\frac1{\beta\left(\omega-\nu\right)^2}
\ln{\frac{\omega\left(\nu+s_0\right)}{\left(\omega+s_0\right)\nu}}-
\frac1{\beta\nu\left(\omega-\nu\right)}+\lambda
\frac{\left(\omega+s_0\right)}{\left(\omega-\nu\right)}=0.
\end{equation}
\end{subequations}
On additionne et on soustrait (\ref{eq:deromegaTH}) et (\ref{eq:dernuTH})
terme \`a terme et on obtient respectivement les relations
\begin{subequations}
\begin{equation}
\omega_0\nu_0=\frac1{\beta\left(\ell+y+\lambda\right)},
\end{equation}
\begin{equation}\label{eq:ymlTH}
\begin{split}
y-\ell&=\frac1{\beta\left(\omega_0-\nu_0\right)}
\left(1/\omega_0+1/\nu_0\right)-\frac2
{\beta\left(\omega_0-\nu_0\right)^2}\ln{\frac{\omega_0
\left(\nu_0+s_0\right)}{\left(\omega_0+s_0\right)\nu_0}}\\
&-\lambda\frac{\omega_0+\nu_0+2s_0}{\omega_0-\nu_0},
\end{split}
\end{equation}
\end{subequations}
satisfaites par $(\omega_0,\nu_0)$ avec:
\begin{equation}
\left(\omega_0+s_0\right)\left(\nu_0+s_0\right)=\frac1{\beta\lambda}.
\end{equation}
Si on utilise (\ref{eq:deromegaTH}) et (\ref{eq:dernuTH}) pour remplacer 
$\ell$ et $y$ dans (\ref{eq:phiexpTH}), on obtient
\beq\label{eq:phiexpbisTH}\
\phi(\omega_0,\nu_0,\ell,y)=\frac{2}
{\beta\left(\omega_0-\nu_0\right)}\ln{\frac{\omega_0\left(\nu_0+s_0\right)}
{\left(\omega_0+s_0\right)\nu_0}}.
\eeq
On introduit les variables auxiliaires $\mu,\upsilon$ pour param\'etriser
le point de col:
\begin{equation}\label{eq:muupsilonTH}
\omega_0\left(\nu_0\right)=\frac1{\sqrt{\beta(\ell\!+\!y\!+\!\lambda)}}
e^{\pm\mu(\ell,y)},\qquad \left(\omega_0+s_0\right)
\left(\nu_0+s_0\right)=\frac1{\sqrt{\beta\lambda}}e^{\pm\upsilon(\ell,y)}.
\end{equation}
Avec celles-ci on a pour (\ref{eq:ymlTH})
\begin{subequations}
\begin{equation}\label{eq:ratiomunuTH}
\frac{y-\ell}{y+\ell}=\frac{\left(\sinh 2\mu-2\mu\right)-\left(\sinh 2\upsilon-2\upsilon\right)}{2\left(\sinh^2\mu-\sinh^2\upsilon\right)}
\end{equation}
et puisque $\omega_0-\nu_0=(\omega_0-s_0)-(\nu_0-s_0)$
\begin{equation}\label{eq:relmunuTH}
\frac{\sinh\upsilon}{\sqrt{\lambda}}=\frac{\sinh\mu}{\sqrt{\ell+y+\lambda}}.
\end{equation}
\end{subequations}
Finalement (\ref{eq:phiexpTH}) devient
$$
\phi(\mu,\upsilon)=\frac2{\sqrt{\beta}}\left(\sqrt{\ell+y+\lambda}-\sqrt{\lambda}\right)
\,\frac{\mu-\upsilon}{\sinh\mu-\sinh\upsilon}.
$$
On peut maintenant effectuer le d\'eveloppement limit\'e suivant
pour estimer l'int\'egrale au voisinage du point de col:
\beeq\nonumber
\phi(\omega,\nu,\ell,y)\!\!&\!\!=\!\!&\!\!\phi(\omega_0,\nu_0,\ell,y)+
\frac12\,\frac{\partial^2\phi}{\partial\omega^2}\,(\omega_0,\nu_0)(\omega-\omega_0)^2+
\frac12\,\frac{\partial^2\phi}{\partial\nu^2}\,(\omega_0,\nu_0)(\nu-\nu_0)^2\\\nonumber\\
&&\hskip -3cm+\frac{\partial^2\phi}{\partial\omega\partial\nu}
(\omega_0,\nu_0)\,(\omega-\omega_0)(\nu-\nu_0)
+{\cal {O}}\left[(\omega-\omega_0)^2,(\omega-\omega_0)(\nu-\nu_0),(\nu-\nu_0)^2\right].
\eeeq
Nous aurons besoin d'introduire le d\'eterminant pour estimer
cette double int\'egration par la m\'ethode du col:
$$
DetA\equiv\frac{\partial^2\phi}{\partial\omega^2}\frac{\partial^2\phi}{\partial\nu^2}-
\left(\frac{\partial^2\phi}{\partial\omega\partial\nu}\right)^2.
$$
Les expressions des d\'eriv\'ees partielles  et celle du d\'eterminant sont 
donn\'ees dans l'appendice \ref{subsection:DSPHID}.
Avec celles-ci, l'expression pour le d\'eterminant
\footnote{Cette expression n'avait jamais \'et\'e calcul\'ee auparavant;
en particulier, son expression sera utile pour l'article \ref{sub:article3}.
Elle permet d'obtenir les corrections en ``single logs'' qui sont li\'ees
\`a la variation de la constante de couplage $\alpha_s$.}
s'\'ecrit en fonction de (\ref{eq:muupsilonTH}) sous la forme:
\begin{eqnarray}\label{eq:determinantTH}
DetA(\mu,\nu)&=&\beta\,(\ell\!+\!y\!+\!\lambda)^3
\left[\frac{(\mu\!-\!\upsilon)\cosh\mu\cosh\upsilon\!+\!\cosh\mu\sinh\upsilon
\!-\!\sinh\mu\cosh\upsilon}
{\sinh^3\mu\cosh\upsilon}\right].\cr
&&
\end{eqnarray}
\begin{figure}[h]
\begin{center}
\includegraphics[height=5truecm,width=0.45\tw]{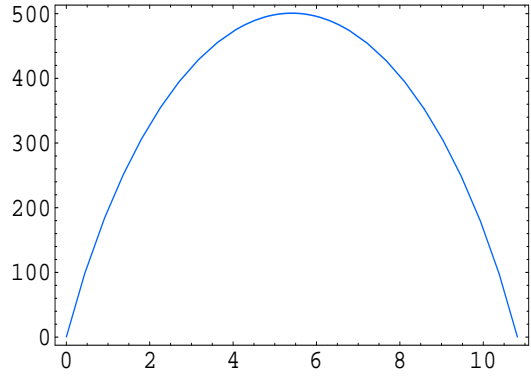}
\caption{\label{fig:DetA} D\'eterminant $DetA(\mu(\ell,Y),\nu(\ell,Y))$ 
en fonction de $\ell=\ln(1/x)$ pour $Y=10\gg\lambda=2.0$ .}
\end{center}
\end{figure}
On choisit le contour de sorte que l'on obtienne le r\'esultat en
int\'egrant le long de l'axe  imaginaire; on obtient
une double int\'egration gaussienne dont
on \'ecrit la r\'eponse dans la limite $\ell+y+\lambda\gg1$
(voir \ref{subsection:DSPHID}): 
\begin{equation}
\label{eq:Dexp}
\ln[{D\left(\ell,y\right)}]\approx \frac2{\sqrt{\beta}}\left(\sqrt{\ell+y+\lambda}-\sqrt{\lambda}\right)
\frac{\mu-\upsilon}{\sinh\mu-\sinh\upsilon}+\upsilon+\ln[{\cal N}(\mu,\upsilon,\lambda)],
\end{equation}
o\`u
$$
{\cal N}(\mu,\upsilon,\lambda)=\frac12(\ell\!+\!y\!+\!\lambda)
\frac{\left(\frac{\beta}{\lambda}\right)^{1/4}}{\sqrt{\pi\cosh\upsilon\,
DetA(\mu,\upsilon)}}.
$$
Nous obtenons dans cette approximation   une estimation du
spectre $D(\ell,Y)$ en DLA qui inclut la normalisation (avec
l'expression du d\'eterminant ici obtenu); on fixe la somme $\ell\!+\!y=Y$:
\beq\label{eq:Specalphasrun}
D(\ell,Y)\approx{\cal N}(\mu,\upsilon,\lambda)\exp\left[
\frac2{\sqrt{\beta}}\left(\sqrt{Y+\lambda}-\sqrt{\lambda}\right)
\frac{\mu-\upsilon}{\sinh\mu-\sinh\upsilon}+\upsilon\right].
\eeq
Or, dans cette approximation, puisqu'on ne s'int\'eresse qu'\`a la forme
de la distribution, on n\'eglige les contributions qui apportent
des effets sous-dominants.  Dans ce cas, on \'ecrit simplement le r\'esultat
de \cite{DLA} qui n\'eglige ${\cal N}$ et ne donne que l'allure de
la distribution:
$$
D(\ell,Y)\simeq\exp\left[\frac2{\sqrt{\beta}}\left(\sqrt{Y+\lambda}-\sqrt{\lambda}\right)
\frac{\mu-\upsilon}{\sinh\mu-\sinh\upsilon}\right]
$$
qui a permis de pr\'edire l'existence du ``hump-backed plateau''.
Nous remarquons que le facteur pre-exponentiel ${\cal N}$ de la distribution
(\ref{eq:Specalphasrun}) est instable dans la limite infrarouge 
$\lambda\rightarrow0$; cependant, la forme de la distribution
donn\'ee par la fonction
dans l'exponentielle est, elle, stable. Or, on sait que pour garantir 
la convergence de la s\'erie perturbative, il faut
$\alpha_s/\pi\ll1\Leftrightarrow\lambda\gg1$, ce qui est en accord avec
la condition d'application de la m\'ethode du col ayant permis 
d'estimer l'int\'egration sur $s$ (\ref{eq:ints}) dans (\ref{eq:DLAalphasrun}).
Nous pouvons utiliser l'expression de la multiplicit\'e moyenne
(\ref{eq:multDLA}) dans la m\^eme limite et renormaliser
(\ref{eq:Specalphasrun}) par
$$
\bar{n}(Y)\approx\frac12\left(\frac{Y+\lambda}{\lambda}\right)^{1/4}
\exp\left[\frac2{\sqrt{\beta}}\left(\sqrt{Y+\lambda}-\sqrt{\lambda}\right)\right]
$$
pour obtenir
\beq\label{eq:SpecNorm}
\frac{D(\ell,Y)}{\bar{n}(Y)}\approx \sqrt{\frac{\beta^{1/2}(Y+\lambda)^{3/2}}
{\pi\cosh\upsilon DetA(\mu,\upsilon)}}\!\exp\!\left[
\frac2{\sqrt{\beta}}\left(\sqrt{Y+\lambda}\!-\!\sqrt{\lambda}\right)
\!\left(\frac{\mu-\upsilon}{\sinh\mu-\sinh\upsilon}\!-\!1\right)+\upsilon\right].
\eeq
Par contre, (\ref{eq:SpecNorm}) est bien stable pour
$\lambda\rightarrow0$; nous en donnons l'allure dans la
Fig.\ref{fig:SpecNormDLA} pour $Y=10$ et $\lambda=1.0,\,1.5,\,2.0,\,2.5$. 
Nous rencontrons la m\^eme forme (``hump-backed plateau'') que dans
la Fig.\ref{fig:SDLA}.
A partir de l'expression dans l'exponentielle de (\ref{eq:SpecNorm}), on peut 
v\'erifier que la position du maximum correspond \`a $\ell_{max}=Y/2$.
\begin{figure}[h]
\begin{center}
\includegraphics[height=7truecm,width=0.6\tw]{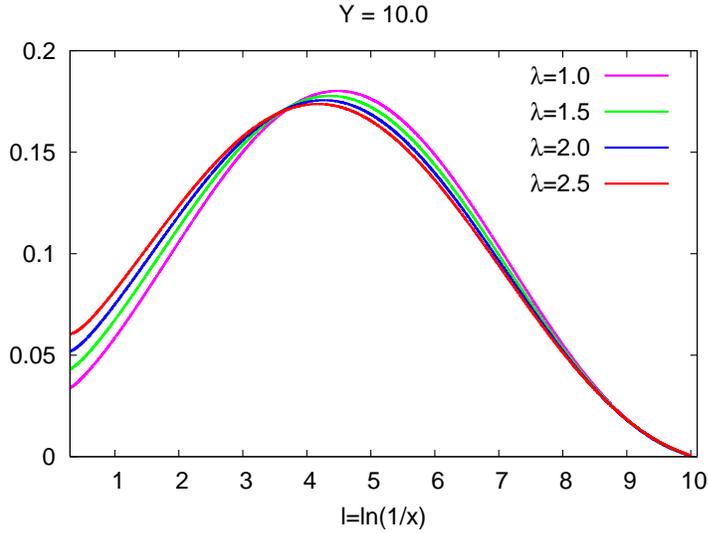}
\caption{\label{fig:SpecDLArun} Spectre normalis\'e $D(\ell,Y)/\bar{n}(Y)$ en fonction de $\ell=\ln(1/x)$ pour $Y\approx10$ et $\lambda=1.0,\,1.5,\,2.0,\,2.5$.}
\label{fig:SpecNormDLA}
\end{center}
\end{figure}
En effet, en regardant le membre de gauche de (\ref{eq:ratiomunuTH}),
on en d\'eduit que $\mu$ et $\upsilon$ doivent \^etre petits au
voisinage de ce point; dans ce cas, (\ref{eq:ratiomunuTH}) et
(\ref{eq:relmunuTH}) deviennent (utile pour la suite)
\beq
Y-2\ell\stackrel{\mu,\upsilon\sim0}{\simeq}\frac23\,\frac{\left(Y+\lambda\right)^{3/2}-\lambda^{3/2}}
{\left(Y+\lambda\right)^{1/2}}\,\mu,
\eeq
o\`u on a d\'ej\`a remplac\'e $\mu$ par
\beq
\upsilon\stackrel{\mu,\upsilon\sim0}{\simeq}\left(\frac{\lambda}{Y+\lambda}\right)^{1/2}\mu.
\eeq
Le d\'eveloppement limit\'e au voisinage de ce point de la fonction 
dans l'exponentielle
\footnote{On ne d\'erive pas sur le dernier terme ($\upsilon$) dans
l'exponentielle de (\ref{eq:SpecNorm}) car ceci donnerait des
corrections au-del\`a du cadre DLA.}
de (\ref{eq:SpecNorm}) que l'on appelle $f(\mu,\upsilon)$
s'\'ecrit sous la forme
$$
f(\mu,\upsilon)\simeq-\frac2{\sqrt{\beta}}
\frac3{(Y+\lambda)^{3/2}-\lambda^{3/2}}\frac{(Y/2-\ell)^2}2+{\cal O}\Big((Y/2-\ell)^2\Big),
$$
o\`u nous avons utilis\'e
\beq\label{eq:dermullTH}
\frac{\partial\mu}{\partial\ell}\simeq-3\frac{(Y+\lambda)^{1/2}}
{(Y+\lambda)^{3/2}-\lambda^{3/2}}.
\eeq
On s'int\'eresse maintenant \`a l'expression du d\'eterminant dans cette
limite.  On l'\'ecrit d'abord sous la forme suivante que l'on \'evalue
au voisinage de $\ell_{max}\approx Y/2$:
\beeq\nonumber
DetA\!&\!\stackrel{\mu,\upsilon\sim0}=
\!&\!\beta(Y\!+\!\lambda)^3\frac{\left(\mu\!-\!\upsilon\right)
\left(1\!+\!\frac12
\mu^2\right)\left(1\!+\!\frac12\upsilon^2\right)\!+\!(1\!+\!\frac12
\mu^2)\left(\upsilon\!+\!\frac16\upsilon^3\right)\!-\!\left(\mu\!+\!\frac16\mu^3\right)
\left(1\!+\!\frac12\upsilon^2\right)}{\mu^3\left(1\!+\!\frac12\upsilon^2\right)}\\\notag\\
\!&\!=\!&\!\frac13\beta(Y\!+\!\lambda)^3\left(1\!-\!\frac{\upsilon^3}{\mu^3}\right)=
\frac13\beta(Y+\lambda)^3\left[1-\left(\frac{\lambda}{Y+\lambda}\right)^{3/2}\right].
\eeeq
Finalement, 
$$
\lim_{\mu,\upsilon\rightarrow0}
\sqrt{\frac{\beta^{1/2}(Y+\lambda)^{3/2}}{\pi DetA(\mu,\upsilon)}}\!=\!
\left(\frac{3}{\pi\sqrt{\beta}\left[(Y+\lambda)^{3/2}
\!-\!\lambda^{3/2}\right]}\right)^{1/2},
$$
qui permet de donner la forme de la distribution au
voisinage de ce point  \cite{DLA}\cite{DFK}
\beq\nonumber
\frac{D(\ell,Y)}{\bar{n}(Y)}\approx\left(\frac{3}{\pi\sqrt{\beta}\left[(Y+\lambda)^{3/2}
-\lambda^{3/2}\right]}\right)^{1/2}\!\!\exp\left(-\frac2{\sqrt{\beta}}\,\frac3{(Y+\lambda)^{3/2}
-\lambda^{3/2}}\frac{\left(Y/2-\ell\right)^2}2\right).
\eeq
La forme correspond \`a une distribution gaussienne.
Pour $\ell\!\gg\! Y/2$ nous trouvons \`a nouveau la
d\'ecroissance rapide de la distribution qui est associ\'ee 
aux ph\'enom\`enes de coh\'erence en CDQ; elle constitue
 l'une des pr\'edictions 
les plus importantes de  CDQ perturbative.

\subsection{Deux limites utiles}

\label{subsub:limitesut}

Deux limites sont int\'eressantes pour les formules
(\ref{eq:ratiomunuTH}) et (\ref{eq:relmunuTH}):

\begin{itemize}
\item{$\ell+y\gg\lambda$ (bien que $\lambda\gg1$) entra\^\i ne $\upsilon\ll\mu$
et par cons\'equent
$$
\frac{y-\ell}{y+\ell}\approx
\frac{\sinh 2\mu-2\mu}
{2\sinh^2\mu},
$$
au voisinage du maximum $\ell\approx Y/2$ nous avons
\beq\label{eq:muY/2}
\frac{Y-2\ell}Y\approx\frac23\,\mu;
\eeq
dans cette limite le spectre devient:
$$
\ln[D(\ell,y)]\propto2\sqrt{\frac{\ell+y}{\beta}}\,
\frac{\mu}{\sinh\mu}.
$$
}
\item{$\lambda\gg\ell+y$ ($\alpha_s$-constant) donne $\mu\approx\upsilon$
(voir paragraphe 3.4 de \ref{sub:article3})
\beq\label{eq:mualpconst}
\frac{y-\ell}{y+\ell}\approx\tanh\mu\qquad\Rightarrow\qquad\mu\approx\frac12\ln\frac{y}{\ell}
\eeq
et
$$
D(\ell,y)\simeq\frac12\sqrt{\frac{\gamma_0y^{1/2}}{\pi\ell^{3/2}}}
\exp\left(2\gamma_0\sqrt{\ell\,y}\right)
$$
}
\end{itemize}
qui est (\ref{eq:specconst}).

\subsection{Remarque concernant les articles \ref{sub:article2}
et \ref{sub:article3}}

Il est int\'eressant de remarquer que bien que le spectre normalis\'e
(\ref{eq:Specalphasrun})  par ${\cal N}$ soit divergent dans la
limite $\lambda\rightarrow0$, ce n'est pas le cas des d\'eriv\'es
logarithmiques o\`u l'on fait dispara\^itre la d\'ependance en
$\lambda$ du terme $\ln{\cal N}(\upsilon,\lambda)$.

% \subsubsection{Spectre pour $\boldsymbol{\lambda=0}$}
% 
% Cette limite peut \^etre prise dans l'\'equation diff\'erentielle (\ref{eq:Fequat}), dans 
% ce cas, on doit en \'echange r\'esoudre
% 
% \begin{equation}
% \frac{\partial F}{\partial\omega}+\frac{\partial F}{\partial\nu}=-\frac1{\beta}\left(\frac{1}{\nu}+\frac{F}{\omega\nu}\right)
% \end{equation}
% dont on \'ecrit directement la repr\'esentation de Mellin pour le spectre
% 
% \beq
% D(\ell,y)=(\ell+y)\iint\frac{d\omega\,d\nu}{(2\pi i)^2}\,e^{\omega\ell+\nu y}
% \left(\frac{\omega}{\nu}\right)^{1/\beta(\omega-\nu)}.
% \eeq
% 
% Son estimation par la m\'ethode du col donne
% 
% \beq
% D(\ell,y)\stackrel{\lambda=0}{\approx}\frac1{2\pi}\sqrt{\frac{\sinh^3\mu}{\beta(\ell+y)
% (\mu\cosh\mu-\sinh\mu)}}\exp{\left(\frac2{\sqrt{\beta}}\,\sqrt{\ell+y}\,\frac{\mu}
% {\sinh\mu}\right)},
% \eeq
% o\`u
% 
% $$
% \frac{y-\ell}{y+\ell}\approx
% \frac{\sinh 2\mu-2\mu}
% {2\sinh^2\mu}.
% $$
% 
% %%%%%%%%%%%%%%%%%%%%%%%%%%%%%%%%%%%%%%%%%%%%%%%%%%%%%%%%%%%%%%%%%%%%%%%%%%%%%%%%%%%%%%%%%%%%%%%%

% \newpage

%%%%%%%%%%%%%%%%%%%%%%%%%%%%%%%%%%%%%%%%%%%%%%%%%%%%%%%%%%%%%%%%%%%%%%%%%%%%%%%%%%%%%%%%%%%%%%%%

%\section{Corr\'elations en \'energie entre deux gluons mous produits dans
%un jet initi\'e par un gluon ou un quark dans le cadre DLA}
%\label{sec:correlations}

\chapter{Corr\'elations en \'energie entre deux gluons mous produits dans
un jet initi\'e par un gluon ou un quark dans le cadre DLA}
\label{sec:correlations}

\begin{figure}[h]
\begin{center}
\includegraphics[height=5truecm,width=0.45\tw]{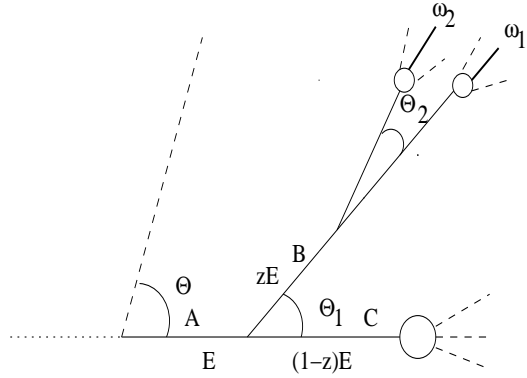}
\caption{\label{fig:correlations} Corr\'elations entre deux particules d'\'energies
$\omega_1$ et $\omega_2$.}
\end{center}
\end{figure}

On prend deux fois la d\'eriv\'ee fonctionnelle de l'\'equation
ma\^itresse (\ref{eq:EMFF}) par rapport aux fonctions de sondage
$u\left(k_1\right)$ et $u\left(k_2\right)$ pour obtenir l'\'equation
satisfaite par la distribution inclusive doublement diff\'erentielle
de deux particules dans un jet initi\'e par le parton $A$ \cite{DLA}:
\begin{equation}
\omega_1\,\omega_2\frac{1}{\sigma}\frac{d^2\sigma}
{d\omega_1\,d\omega_2}\equiv D^{\left(2\right)}_A
\left(E,\Theta;\omega_1,\omega_2\right);
\end{equation}

de plus
$$
D_A^{(2)}\left(k_1,k_2,\Theta\right)=\frac{\delta^2}{\delta u(k_1)\delta u(k_2)}
Z\big(k_1,k_2,\Theta;u(k_1),u(k_2)\big)\Big\vert_{u=1}.
$$
Il convient aussi de d\'efinir cette grandeur sous la forme suivante:
$$
x_1\,x_2D_A^{(2)}\big(x_1,x_2,\ln E\Theta\big)=\omega_1\,\omega_2
\frac{\delta^2}{\delta u(k_1)\delta u(k_2)}
Z\big(k_1,k_2,\Theta;u(k_1),u(k_2)\big)\Big\vert_{u=1}
$$
o\`u $x_{1,2}=\omega_{1,2}/E$ sont les fractions d'\'energie emport\'ees
par les deux particules dont on \'etudie la corr\'elation dans le jet. 
Nous \'ecrivons le r\'esultat de cette diff\'erentiation en utilisant
la condition (\ref{eq:CN}); de plus on prend $\omega_1>\omega_2$ de sorte
que l'on puisse fixer la limite inf\'erieure d'int\'egration sur la fraction
d'\'energie $z$ \`a $\omega_1/E$. Pour $z=1$, il n'y a pas d'\'emission.
Puis on rappelle que $Y_{\Theta}=\ln\frac{E\Theta}{Q_0}$.
Ceci permet d'\'ecrire le syst\`eme d'\'equation
suivant pour les jets de quark ($A\equiv Q$) et de gluon ($A\equiv G$)
\begin{eqnarray}\nonumber
\frac{\partial}{\partial\ln{\Theta}}x_1\,x_2D^{(2)}_A
\left(x_1,x_2, Y_{\Theta}\right)\!&\!=\!&\!\frac{C_A}{N_c}\int_{\omega_1/E}^{1}
\frac{dz}{z}\,\gamma_0^2\left[\frac{x_1}{z}\frac{x_2}{z}D^{(2)}_G
\left(\frac{x_1}{z},\frac{x_2}{z},Y_{\Theta}+\ln z\right)\right.\\\nonumber\\
\cr&&\hskip -4.0cm\left.+x_1D_A
\left(x_1,Y_{\Theta}\right)\frac{x_2}zD_G(\frac{x_2}z,Y_{\Theta}+\ln z)
+\frac{x_1}zD_G
\left(\frac{x_1}z,Y_{\Theta}+\ln z\right)x_2D_A
\left(x_2,Y_{\Theta}\right)\right]\nonumber\\\nonumber\\
\cr\!&\!=\!&\!\frac{C_A}{N_c}\int_{\omega_1/E}^{1}
\frac{dz}{z}\,\gamma_0^2\,\frac{x_1}{z}\frac{x_2}{z}D^{(2)}_G
\left(\frac{x_1}{z},\frac{x_2}{z},Y_{\Theta}+\ln z\right)\label{eq:CorrADLA}\\\nonumber\\
\cr&&\hskip-5.0cm+x_1D_A
\left(x_1,Y_{\Theta}\right)\left[\frac{\partial}{\partial\ln\Theta}
x_2D_A\left(x_2,Y_{\Theta}\right)\right]+\left[\frac{\partial}{\partial\ln\Theta}
x_1D_A\left(x_1,Y_{\Theta}\right)\right]x_2D_A
\left(x_2,Y_{\Theta}\right);\nonumber
\end{eqnarray}
nous avons utilis\'e l'\'equation int\'egro-diff\'erentielle
pour le spectre inclusive d'une particule (\ref{eq:partialspec})
$$
\frac{\partial}{\partial\ln\Theta}x\,D_A\left(x,Y_{\Theta}\right)=
\frac{C_A}{N_c}\int_{\omega_1/E}^1\frac{dz}{z}\,\gamma_0^2\,\frac{x}z\,
D_G(\frac{x}z,Y_{\Theta}+\ln z);
$$
nous allons r\'ecrire la derni\`ere ligne de (\ref{eq:CorrADLA}) sous
la forme compacte
$$
\frac{\partial}{\partial\ln\Theta}\left[x_1D_A(x_1,Y_{\Theta})x_2D_A(x_2,Y_{\Theta})\right]
$$
et la soustraire dans les deux membres de la m\^eme \'equation pour la mettre
sous la forme
\begin{eqnarray*}
\cr&&\hskip-0.5cm\frac{\partial}{\partial\ln{\Theta}}\left[x_1x_2D^{(2)}_A
\left(x_1,x_2,Y_{\Theta}\right)-x_1D_A
\left(x_1,Y_{\Theta}\right)x_2D_A
\left(x_2,Y_{\Theta}\right)\right]\\
\cr&&\hskip3.5cm=\frac{C_A}{N_c}\int_{\omega_1/E}^{1}\frac{dz}{z}
\,\gamma_0^2\,\frac{x_1}z\frac{x_2}zD^{(2)}_G\left(\frac{x_1}z,\frac{x_2}z,Y_{\Theta}
+\ln z\right).
\end{eqnarray*}
Nous int\'egrons cette \'equation sur $\Theta$. La contrainte angulaire
rigoureuse sur les angles des \'emissions successives en DLA impose
$\Theta\geq\Theta_1\gg\Theta_2$. Or, le plus
petit angle est celui de l'\'emission $\omega_2$. Toutefois, son impulsion
transverse ne peut pas \^etre inf\'erieure au cut-off colin\'eaire 
$k_{2,\perp}\approx\omega_2\Theta_2\geq\omega_2(\Theta_2)_{min}= Q_0$,
par cons\'equent, l'angle d'int\'egration
$\Theta_1$ a les bornes suivantes, voir Fig.\ref{fig:correlations}
$$
\Theta\!\geq\!\Theta_1\!\gg\!(\Theta_2)_{min}\!\approx\!Q_0/\omega_2;
\qquad \Theta_1\gg\Theta_2,
$$
% %
\beeq
&&x_1x_2D^{(2)}_A\left(x_1,x_2,Y_{\Theta}\right)-
x_1D_A\left(x_1,Y_{\Theta}\right)x_2
D_A\left(x_2,Y_{\Theta}\right)=\\\nonumber\\
&&\hskip4cm\frac{C_A}{N_c}\int_{\omega_1/E}^{1}\frac{dz}{z}
\int_{Q_0/\omega_2}^{\Theta}\,\frac{d\Theta_1}{\Theta_1}\gamma_0^2\,
\frac{x_1}z\frac{x_2}zD^{(2)}_G\left(\frac{x_1}z,\frac{x_2}z,Y_{\Theta'}+\ln z\right)\nonumber,
\eeeq
on introduit par d\'efinition la fonction de corr\'elation
\begin{center} $x_1x_2\widetilde{D}^{(2)}_A\left(x_1,x_2,Y_{\Theta}\right)
=x_1x_2D^{(2)}_A\left(x_1,x_2,Y_{\Theta}\right)
-x_1D_A\left(x_1,Y_{\Theta}\right)x_2
D_A\left(x_2,Y_{\Theta}\right)$,
\end{center}
dont on cherche l'\'equation que l'on met sous la forme
\begin{eqnarray*}
&&x_1x_2\widetilde{D}^{(2)}_A\left(x_1,x_2,Y_{\Theta}\right)=
\frac{C_A}{N_c}\int_{\omega_1/E}^{1}\frac{dz}{z}\int_{Q_0/\omega_2}^{\Theta}
\frac{d\Theta_1}{\Theta_1}
\,\gamma_0^2\,\left[\frac{x_1}z\frac{x_2}z\widetilde{D}^{(2)}_G
\left(\frac{x_1}z,\frac{x_2}z,Y_{\Theta'}+\ln z\right)\right.\\\nonumber\left.\right.\\
\cr&&\hskip4.5cm\left.+\frac{x_1}zD_G
\left(\frac{x_1}z,Y_{\Theta'}+\ln z\right)\frac{x_2}zD_G
\left(\frac{x_2}z,Y_{\Theta'}+\ln z\right)\right],
\end{eqnarray*}
on effectue le changement de variables
$$
\ell=\ln\frac{z}{x_1},\quad \omega_1/E\leq z\leq1
\Leftrightarrow0\leqslant\ell\leqslant\ell_1,\quad\ell_1=\ln\frac1{x_1}
$$ 
et 
$$
y=\ln\frac{\omega_2\Theta_1}{Q_0},\quad  (\Theta_2)_{min}\leq\Theta_1\leq\Theta
\Leftrightarrow 0\leqslant y\leqslant y_2,\quad 
y_2=\ln\frac{\omega_2\Theta}{Q_0},
$$
on d\'efinit $\eta=\ln\frac{\omega_1}{\omega_2}=\ln\frac{x_1}{x_2}>0$
de sorte que $\ell_2=\ell_1+\eta$ et $y_1=y_2+\eta$. Nous avons
$$
\int_{\omega_1/E}^{1}\frac{dz}z=\int_0^{\ell_1}d\ell,
\qquad\int_{Q_0/\omega_2}^{\Theta}\frac{d\Theta_1}{\Theta_1}=\int_0^{y_2}dy.
$$
Nous allons \'egalement utiliser le changement de variables suivant
$$
xD(x)\equiv D\big(\ell=\ln(1/x)\big).
$$
On arrive finalement \`a l'\'equation suivante pour les corr\'elations
\begin{eqnarray}
\label{eq:correq}
\cr&&\hskip-1cm\widetilde{D}^{\left(2\right)}_A
\left(\ell_1,y_2;\eta\right)\!=\!\frac{C_A}{N_c}\!\!\int_{0}^{\ell_1}
\!\!d\ell\int_{0}^{y_2}\!\!dy\gamma_0^2(\ell\!+\!y)\!\!
\left[\widetilde{D}^{\left(2\right)}_G
\left(\ell,y;\eta\right)
\!+\!D_G\left(\ell\!+\!\eta,y\right)
D_G\left(\ell,y\!+\!\eta\right)\right]\!\!.\cr
&&
\end{eqnarray}\smallskip
o\`u nous avons utilis\'e \cite{DLA}
\begin{equation}\label{eq:alpgasbisTH}
\gamma_0^2(zE\Theta_1)=
\frac{1}{\beta\left(\ln\displaystyle{\frac{z}{x_1}}+
\ln\displaystyle{\frac{x_2E\Theta_1}{Q_0}}+
\ln\displaystyle{\frac{x_1}{x_2}}+\lambda\right)}
\equiv\gamma_0^2(\ell+y)=\frac1{\beta(\ell+y+\eta+\lambda)}.
\end{equation}
Dans le cas $z=1$ et $\Theta_1=\Theta$ nous avons
$$
\gamma_0^2=\frac1{\beta(\ell_1+y_2+\eta+\lambda)}=\frac1{\beta(Y+\lambda)},\qquad
\ell_1+y_2+\eta=Y.
$$
Le syst\`eme d'\'equations coupl\'ees (\ref{eq:correq}) est g\'en\'eralis\'e
au cadre MLLA dans l'article \ref{sub:article2}
(voir les \'equations (3.35) et (3.36)), du \`a la n\'ecessit\'e d'inclure
les corrections qui garantissent la conservation de l'\'energie dans
les processus de branchement; elles ont \'et\'e \'egalement \'ecrites
 sous forme diff\'erentielle dans \cite{FW}.

\section{Solution de l'\'equation (\ref{eq:correq})}

On introduit les techniques que nous avons utilis\'ees dans \ref{sub:article2}
pour r\'esoudre (\ref{eq:correq}). Dans cet objectif, on utilise
une m\'ethode it\'erative de solution, soit
$$
D_A^{(2)}(\ell_1,y_2;\eta)
={\cal {C}}_A(\ell_1,y_2;\eta) D_{A1}(\ell_1,y_1)\,D_{A2}(\ell_2,y_2).
$$
On diff\'erencie (\ref{eq:correq}) par rapport \`a $\ell_1$ et $y_2$,
on obtient l'\'equation diff\'erentielle en DLA
$$
\widetilde{D}^{(2)}_{A\ell,y}
=\frac{C_A}{N_c}\gamma_0^2\left(\widetilde{D}_G^{(2)}+D_{G1}D_{G2}\right).
$$
Ici, on se place dans l'approximation o\`u l'on consid\`ere $\alpha_s$
constante dans l'objectif de retrouver le r\'esultat de \cite{DLA};
un traitement complet du probl\`eme qui tiendra compte des corrections
en logs simples sera fait dans \ref{sub:article2}. Nous allons par
cons\'equent n\'egliger les d\'eriv\'ees de la fonction ${\cal {C}}_A$
$$
\big({\cal {C}}_A-1\big) \big(D_{A1,\ell y}\,D_{A2}\!+\!D_{A1,\ell}\,D_{A2,y}
\!+\!D_{A1,y}\,D_{A2,\ell}\!+\!D_{A1}\,D_{A2,\ell y}\big)\!=\!\frac{C_A}{N_c}\gamma_0^2
\big({\cal {C}}_G D_{G1}D_{G2}\big),
$$
or 
$$
D_{Ak,\ell y}=\frac{C_A}{N_c}\gamma_0^2\,D_{Gk},\quad D_{Ak}=
\frac{C_A}{N_c}D_{Gk},\quad k=1,2; \quad\text{donc}
$$
\beeq\nonumber
\left(\frac{C_A}{N_c}\right)^2\big({\cal {C}}_A-1\big)\left(2\gamma_0^2D_{G1}D_{G2}\!+\!D_{G1,\ell}\,D_{G2,y}\!+\!D_{G1,y}\,D_{G2,\ell}\right)
\!\!&\!\!=\!\!&\!\!
\frac{C_A}{N_c}\gamma_0^2\left({\cal {C}}_G D_{G1}D_{G2}\right)\\
&&\hskip -4cm=\frac{C_A}{N_c}\gamma_0^2\left[({\cal {C}}_G-1) D_{G1}D_{G2}+D_{G1}D_{G2}\right],\nonumber
\eeeq
on utilise 
$$
{\cal {C}}_G-1=\frac{C_A}{N_c}({\cal {C}}_A-1),
$$
on divise par $\left(\frac{C_A}{N_c}\right)^2\gamma_0^2D_1D_2$ et on
obtient la solution suivante en DLA qui est \'ecrite en termes des
deriv\'ees logarithmiques du spectre $\psi=\ln[D(\ell,y)]$
(voir paragraphe \ref{sub:derlog}):
\beq\label{eq:soleqcorr}
{\cal {C}}_A-1=\frac{N_c}{C_A}\frac1{1+\displaystyle{\frac{\psi_{G1,\ell}\psi_{G2,y}
+\psi_{G1,y}\psi_{G2,\ell}}{\gamma_0^2}}},
\eeq
Dans le cas du processus $e^+e^-\rightarrow q\bar{q}$ nous avons
(voir \cite{FW})
$$
R\equiv{\cal {C}}_{e^+e^-\!\rightarrow q\bar{q}}=\frac12+\frac12{\cal {C}}_F.
$$
Pour $\alpha_s$ constante on utilise les d\'eriv\'ees logarithmiques
du paragraphe \ref{sub:derlog} et on \'ecrit
$$
{\cal {C}}_A-1=\frac{N_c}{C_A}\frac1{1+\sqrt{\displaystyle{\frac{y_1}{\ell_1}}}
\sqrt{\displaystyle{\frac{\ell_2}{y_2}}}+
\sqrt{\displaystyle{\frac{\ell_1}{y_1}}}\sqrt{\displaystyle{\frac{y_2}{\ell_2}}}},
$$
puis on remplace les quotients par $\mu$ selon (\ref{eq:mualpconst})
(voir paragraphe \ref{subsub:limitesut}) et on \'ecrit
$$
\sqrt{\displaystyle{\frac{y_1}{\ell_1}}}
\sqrt{\displaystyle{\frac{\ell_2}{y_2}}}+
\sqrt{\displaystyle{\frac{\ell_1}{y_1}}}
\sqrt{\displaystyle{\frac{y_2}{\ell_2}}}=e^{(\mu_1-\mu_2)}+e^{-(\mu_1-\mu_2)}=
2\cosh{(\mu_1-\mu_2)},
$$
puis
$$
1+2\cosh{(\mu_1-\mu_2)}=3+4\sinh^2\left(\frac{\mu_1-\mu_2}{2}\right)
$$
et finalement, comme dans \cite{DLA}, on \'ecrit la solution DLA des
corr\'elations \`a deux particules dans un jet
\beq\label{eq:solmueqcorr}
{\cal {C}}_A(\ell_1,y_2;\eta)=1+\frac{N_c}{3C_A}\frac1{1+\displaystyle{\frac43}\sinh^2
\left(\displaystyle{\frac{\mu_1(\ell_1,y_1)-\mu_2(\ell_2,y_2)}{2}}\right)}.
\eeq
La solution dominante de l'\'equation peut de m\^eme \^etre \'ecrite
pour $\alpha_s$ variable si on utilise la d\'efinition de $\mu$
donn\'ee par (\ref{eq:ratiomunuTH}) et (\ref{eq:relmunuTH}). Dans ce cas
on a $\psi_{\ell}=\omega_0+\dots=\gamma_0 e^{\mu}+\dots$ et
$\psi_y=\nu_0+\dots=\gamma_0 e^{-\mu}+\dots$, et on obtient la m\^eme
expression (\ref{eq:solmueqcorr}) avec la d\'efinition de $\mu$ trouv\'ee
dans le cas physique (dans l'article \ref{sub:article3} nous donnons
les expression des deriv\'ees logarithmiques obtenues pas la m\'ethode
du col, voir (30a) et (30b)). Cette expression d\'ecrit la d\'ecroissance
des corr\'elations lorsque $\mid\!\!\eta\!\!\mid$ augmente,
soit lorsque le gluon d'\'energie $\omega_2$ est de plus en plus mou.
Dans cette limite, les effets de coh\'erence dominent et
${\cal {C}}_A\rightarrow1$; ceci peut se lire directement sur
l'\'equation int\'egrale d'\'evolution (\ref{eq:correq}): lorsque 
$y_2\rightarrow0\Rightarrow\omega_2\Theta_2\rightarrow Q_0$
l'int\'egration est nulle et le corr\'elateur tombe vers l'unit\'e.
Pour $\eta=0\Rightarrow\omega_1=\omega_2$, elle 
coincide avec l'expression du corr\'elateur des multiplicit\'es en DLA,
soit \cite{DFK}
$$
\frac{<n(n-1)>_{q\bar{q}}}{\bar{n}^2_{q\bar{q}}}=1+\frac13 N_c/2C_F
$$
dans le cas de l'annihilation $e^+e^-$.

%\vskip 0.5cm

\subsection{Solution de l'\'equation  (\ref{eq:correq})
 avec $\boldsymbol{\alpha_s}$ variable}

%\vskip 0.5cm

Il est utile d'obtenir, dans la solution de l'\'equation (\ref{eq:correq}),
les termes correcteurs li\'es \`a l'\'evolution du jet.
Si (\ref{eq:soleqcorr}) est une fonction r\'eguli\`ere, on peut
diff\'erencier ${\cal {C}}_A$ par rapport \`a $\ell_1$ et $y_2$.
On obtient en diff\'erentiant les deux membres de (\ref{eq:correq})
(ceci permet d'introduire les techniques de calcul utilis\'ees dans
l'article \ref{sub:article2})
\beeq\nonumber
\left(\!\frac{C_A}{N_c}\!\right)^2\left[\gamma_0^2\big({\cal {C}}_A-1\big)\left(2\!+\!\displaystyle{\frac{\psi_{G1,\ell}\psi_{G2,y}
\!+\!\psi_{G1,y}\psi_{G2,\ell}}{\gamma_0^2}}\right)+
{\cal {C}}_A\big(\chi_{A,\ell y}\!+\!\chi_{A,\ell}\chi_{A,y}\big)\right.\\
\left.+{\cal {C}}_A\Big(\chi_{A,\ell}\big(\psi_{G1,y}+\psi_{G2,y}\big)+
\chi_{Ay}\big(\psi_{G1,\ell}+\psi_{G2,\ell}\big)\Big)\right]D_{G1}D_{G2}\label{eq:solequalvar}\\
&&\hskip-9cm =\left(\frac{C_A}{N_c}\right)^2\gamma_0^2\left[\big({\cal {C}}_A-1\big) D_{G1}D_{G2}+\frac{N_c}{C_A}D_{G1}D_{G2}\right],
\nonumber
\eeeq
on divise par $\left(\frac{C_A}{N_c}\right)^2\gamma_0^2D_1D_2$, on definit
$$
\chi_A=\ln\left[{\cal {C}}_A\right],
$$
o\`u ${\cal {C}}_A$ est donn\'e par (\ref{eq:soleqcorr}),
$$
\delta_{A1}=\frac1{\gamma_0^2}\left[\chi_{A,\ell}\big(\psi_{G1,y}+\psi_{G2,y}\big)+
\chi_{Ay}\big(\psi_{G1,\ell}+\psi_{G2,\ell}\big)\right],
$$
$$
\delta_{A2}=\frac1{\gamma_0^2}\left[\chi_{A,\ell y}\!+\!\chi_{A,\ell}\chi_{A,y}\right]
$$
et on r\'ecrit (\ref{eq:solequalvar}) comme
$$
\Big({\cal {C}}_A-1\Big)\left(1+\!\displaystyle{\frac{\psi_{G1,\ell}\psi_{G2,y}
\!+\!\psi_{G1,y}\psi_{G2,\ell}}{\gamma_0^2}}+\delta_{A1}+\delta_{A2}\right)=
\frac{N_c}{C_A}\left(1-\delta_{A1}-\delta_{A2}\right);
$$
finalement, la solution de l'\'equation pour $\alpha_s$ variable 
s'\'ecrit sous la forme
$$
{\cal {C}}_A(\ell_1,y_2;\eta)=1+\frac{N_c}{C_A}\frac{1-\delta_{A1}-\delta_{A2}}{1+
\displaystyle{\frac{\psi_{G1,\ell}\psi_{G2,y}
\!+\!\psi_{G1,y}\psi_{G2,\ell}}{\gamma_0^2}}+\delta_{A1}+\delta_{A2}}
$$
que l'on r\'ecrit de mani\`ere compacte ci-dessous:
\beq\label{eq:DLAcorr}
{\cal {C}}_A(\ell_1,y_2;\eta)=1+\frac{N_c}{C_A}\frac{1-\delta_c}{1+\Delta+\delta_c}
\eeq
avec
$$
\delta_c=\delta_1+\delta_2,\qquad \Delta=\displaystyle{\frac{\psi_{G1,\ell}\psi_{G2,y}
\!+\!\psi_{G1,y}\psi_{G2,\ell}}{\gamma_0^2}}.
$$

\subsection{Ordre de grandeur des corrections
(voir aussi le paragraphe 4.2 de \ref{sub:article2})}

On a
$$
\Delta={\cal {O}}(1),
$$
$$
\chi_{\ell}\propto\frac1{\gamma_0^2}\left[\psi_{\ell\ell}\psi_{y},
\,\,\,\psi_{\ell}\psi_{y\ell}\right]\Rightarrow\chi_{\ell}={\cal {O}}\big(\gamma_0^2\big),
$$
$$
\chi_{\ell y}\propto \frac1{\gamma_0^2}\psi_{\ell\ell}\psi_{yy}\Rightarrow
\chi_{\ell y}={\cal {O}}\big(\gamma_0^4\big).
$$
Cel\`a donne $\delta_1={\cal {O}}(\gamma_0)$ et
$\delta_2={\cal {O}}\big(\gamma_0^2\big)$; ils
repr\'esentent des corrections MLLA et NMLLA respectivement qui doivent
\^etre  n\'eglig\'ees dans le cadre DLA.
Nous pouvons en tenir compte pour d\'eceler le r\^ole de la variation de
la constante de couplage par rapport au r\'esultat (\ref{eq:solmueqcorr}).
Dans l'appendice F de \ref{sub:article2} on compare
(\ref{eq:DLAcorr}) avec le r\'esultat en MLLA qui tient compte de la
conservation de l'\'energie dans les cascades partoniques.

\section{Approximation de Fong \& Webber \cite{FW} en DLA
(voir aussi \cite{KO})}
\label{subsub:FWApprox}

Dans l'approximation de Fong \& Webber pour le cas des corr\'elations
\cite{FW}, l'\'energie des deux particules se trouve au voisinage du
maximum de leur distribution inclusive \cite{FW1}, c'est \`a dire au
voisinage de $Y/2$; $\ell_1\sim\ell_2\simeq Y/2$. Dans cette limite
$$
\left(\frac{\ell_1-\ell_2}{Y}\right)^2\ll1
$$
et on utilise l'expression de $\mu$ (\ref{eq:muY/2})
au voisinage du maximum du spectre inclusif; on a
$$
\mu_k\simeq\frac32-3\frac{\ell_k}Y;
$$
puis
$$
\frac{\mu_1-\mu_2}2\simeq-\frac32\left(\frac{\ell_1-\ell_2}Y\right)\ll1,
$$
on peut donc effectuer l'expansion de Taylor suivante
\beeq
\frac1{1+\displaystyle{\frac43}\sinh^2
\left(\displaystyle{\frac{\mu_1(\ell_1,y_1)-\mu_2(\ell_2,y_2)}{2}}\right)}
&\stackrel{\ell_1\sim\ell_2\simeq Y/2}{\simeq}&
\frac1{1+\frac43\frac94\left(\displaystyle{\frac{\ell_1-\ell_2}Y}\right)^2}\\
&\simeq&1-3\left(\frac{\ell_1-\ell_2}Y\right)^2;
\eeeq

finalement
\beq\label{eq:FWDLA}
{\cal{C}}_A(\ell_1,y_2;\eta)\stackrel{\ell_1\sim\ell_2\simeq
Y/2}{\simeq}1+\frac{N_c}{3C_A}-\frac{N_c}{C_A}\left(\frac{\ell_1-\ell_2}Y\right)^2,
\eeq
qui est en accord avec le r\'esultat obtenu dans \cite{FW} en DLA.
Pour $\alpha_s$ constant on utilise l'expression de
$\mu$ (\ref{eq:mualpconst})
$$
\mu=\frac12\ln\left(\frac{Y-\ell}{\ell}\right)\stackrel{\ell\simeq Y/2}{\simeq}-\frac2Y(\ell-Y/2),
$$
avec celle-ci
$$
\frac{\mu_1-\mu_2}2\simeq-\left(\frac{\ell_1-\ell_2}Y\right)\ll1,
$$
et on obtient une expression du corr\'elateur en DLA
\beq\label{eq:FWDLAac}
{\cal{C}}_A(\ell_1,y_2;\eta)\stackrel{\ell_1\sim\ell_2\simeq Y/2}{\simeq}1+\frac{N_c}{3C_A}-\frac{4N_c}{9C_A}\left(\frac{\ell_1-\ell_2}Y\right)^2
\eeq
diff\'erente de (\ref{eq:FWDLA}) pour $\alpha_s$ constante.
On reporte le lecteur au paragraphe 3.4 de l'article \ref{sub:article3};
dans celui-ci nous expliquons dans quelle
limite on peut consid\'erer $\alpha_s$ comme une constante.

\section{Conclusions et motivations}

Le sch\'ema de resommation DLA constitue, \`a petit $x$,
le point de d\'epart et le principal ingr\'edient
dans l'\'evaluation des grandeurs inclusives
en CDQ perturbative.
Cette approche permet notamment de d\'ecrire la dynamique et la production
multipartoniques. Elle a \'et\'e utilis\'ee pour pr\'edire la forme des
distributions inclusives, l'ordre de grandeur des multiplicit\'es
hadroniques dans les jets ainsi que le comportement des corr\'elations
\`a deux particules.
N\'eanmoins, que l'on consid\`ere $\alpha_s$ constante
ou variable, elle est insuffisante pour effectuer
des pr\'edictions raisonnables que l'on puisse comparer avec les donn\'ees
exp\'erimentales des grands collisionneurs de particules
(LEP, Tevatron et futur LHC).
En effet, l'approche DLA n\'eglige le principe
de conservation de l'\'energie par la sous estimation du recul
des particules charg\'ees \'emettrices.

%Apr\`es avoir calcul\'e, au premier paragraphe, le courant d'accompagnement
%mou en EDQ au premier chap\^itre (voir (\ref{eq:Ma})), nous l'avons ensuite
%utilis\'ee dans l'\'evaluation de la section efficace de la production
%de photons mous en EDQ dans le paragraphe \ref{subsection:SERM}. On a
%ainsi d\'emontr\'e la structure doublement logarithmique du rayonnement
%mou et on a donn\'ee sa forme dans (\ref{eq:elly}).
%Puis ceci a \'et\'e g\'en\'eralis\'e au cas des \'emissions de gluons
%mous en CDQ, o\`u l'on resomme sur l'ensemble des \'emissions molles et
%colin\'eaires (\ref{eq:MNgluons}).
Cette approximation, m\^eme dans la limite des hautes \'energies,
surestime consid\'erablement la production hadronique, les distribution
inclusives des particules et les corr\'elations,
du fait qu'elle viole la conservation de l'\'energie.
Par cons\'equent, dans l'objectif d'effectuer des pr\'edictions raisonnables,
un traitement consistent et rigoureux des corrections en logarithmes
simples (``Single Logs'' (SL) en anglais) s'av\`ere n\'ecessaire. Il s'agit
de l'approximation MLLA dont on donnera toutes les sources physiques au
prochain chap\^itre.
Ceci constitue l'objectif des travaux \ref{sub:article1} pour le cas des
distributions inclusives en fonction de l'impulsion transverse ($k_\perp$),
puis \ref{sub:article2} et \ref{sub:article3} pour le cas des corr\'elations.

\chapter{Approximation Logarithmique Dominante Modifi\'ee (MLLA)}
\label{sec:MLLA}

La lecture de ce chap\^itre aidera \`a la compr\'ehension des travaux
\ref{sub:article1}, \ref{sub:article2} et \ref{sub:article3}.
On y \'etudie  les techniques perturbatives qui permettent
de  d\'ecrire les propri\'et\'es des particules de petite fraction d'\'energie
($x=k/E_{jet}\ll1$) produites dans les collisions hadroniques.
N'emportant qu'une partie n\'egligeable de l'\'energie totale du jet, elles
 font partie de la majorit\'e des particules qui y sont produites.

\section{Corrections en logarithmes simples (SL) aux cascades DLA}

Nous avons d\'emontr\'e au chap\^itre pr\'ec\'edent que la 
contrainte angulaire rigoureuse fournit les bases de l'interpr\'etation 
probabiliste des processus de branchement des gluon mous rayonn\'es. Dans
le cadre DLA:
$$
k_f\ll k_j\ll k_i,\qquad \Theta_{fj}\ll\Theta_{ji},
$$
o\`u l'on rappelle que $i$ est le parton initiant le jet, $j$ correspond
\`a une premi\`ere \'emission et $f$ est son parton ``enfant''.

L'approche DLA est trop stricte pour donner des pr\'edictions
raisonnables. Quantitativement, elle surestime les grandeurs mesurables
(multiplicit\'es des gluons, spectre inclusive des gluon rayonn\'es,
corr\'elations) dans les cascades partoniques car elle ignore la
conservation de l'\'energie lors des \'emissions de gluons mous.
Elle surestime, en effet, l'\'energie des partons qui se multiplient le plus
facilement, soit les partons dont l'\'energie est proche du maximum de la
distribution inclusive ($\ell_{max}=Y/2$) (voir les paragraphes
\ref{subsection:HBP} et \ref{subsection:col}).
Cette approximation tient compte des termes $\propto\sqrt{\alpha_s}$
qui interviennent dans la constante anormale $\gamma$, mais
n\'eglige les contributions $\Delta\gamma\sim\alpha_s$,
donc celles d'ordre relatif $\sqrt{\alpha_s}$.
En raison de quoi on introduit dans ce chap\^itre les corrections
simplement logarithmiques (SL) qui tiennent compte des termes n\'eglig\'es
en DLA, pour ainsi viser \`a des pr\'edictions raisonnables qui puissent \^etre
compar\'ees avec les donn\'ees exp\'erimentales des grands collisionneurs.

\medskip

Lorsque l'on construit ainsi le sch\'ema probabiliste en tenant compte des
corrections doublement et simplement logarithmiques, on obtient une meilleure
pr\'ecision qui d\'ecoule des contributions associ\'ees \`a
l'accroissement du nombre d'interf\'erences dont on ne tient pas compte
en DLA. L'id\'ee fondamentale du sch\'ema probabiliste est fond\'ee sur
le principe selon lequel, dans une cascade, on ne tient compte que des
plus proches voisins de mani\`ere semblable au cas du mod\`ele 
d'Ising en th\'eorie statistique des champs.

Pour comprendre et \'evaluer les corrections en logs simples que l'on
doit ajouter aux contributions DLA, nous allons faire appel, de nouveau,
\`a la notion de la Fonctionnelle G\'en\'eratrice. Cette technique
est raisonnable pour d\'ecrire la structure interne des cascades partoniques.
Sa forme peut s'exprimer symboliquement comme \cite{EvEq}:

\begin{equation}\label{eq:Zgamma}
Z=C(\alpha_s(t))\exp\left[\int^t\gamma(\alpha_s(t'))dt'\right].
\end{equation}

Cette repr\'esentation tient compte du fait que les \'emissions successives
(par rapport au param\`etre $t$ ``temps d'\'evolution'')
\'el\'ementaires ind\'ependantes s'exponentient. Elle montre la propri\'et\'e
de localit\'e intrins\`eque du sch\'ema probabiliste.
A savoir, la deriv\'ee de (\ref{eq:Zgamma}) par rapport au ``temps $t$''
fait appa\^itre un pr\'efacteur $\gamma(\alpha_s(t))$ qui montre que
le taux de variation de $Z$ (et ainsi, du contenu partonique dans la cascade)
dans le temps $t$
est d\'etermin\'e par la quantit\'e $\gamma(t)$, dont la valeur ne d\'epend que
de l'\'echelle de temps ``t''; par cons\'equent, elle ne garde pas
la trace du pass\'e du syst\`eme.

Si l'on compare cette observation avec les notations introduites au
premier chap\^itre de \cite{EvEq}, o\`u le sch\'ema de resommation LLA 
(``Leading Logarithmic Approximation'') a \'et\'e  trait\'e, il
appara\^\i t naturel d'attribuer \`a $\gamma$ la d\'efinition de constante anormale
et \`a $C$, celle de fonction coefficient.

D'apr\`es la contrainte angulaire, le ``temps d'\'evolution''
$t$ dans (\ref{eq:Zgamma}) doit \^etre li\'e \`a l'angle d'ouverture du jet
$dt=d\Theta/\Theta$. Ceci signifie que toutes les contributions qui
sont singuli\`eres par rapport \`a l'angle entre les \'emissions
partoniques doivent \^etre incorpor\'ees dans l'exponentielle de
(\ref{eq:Zgamma}). C'est ainsi que l'int\'egrale sur la constante
anormale $\gamma$ contient les cha\^\i nes de Markov des d\'esint\'egrations
successives qui respectent la contrainte angulaire (AO). 
Par ailleurs, le pr\'efacteur $C$, \'etant sans divergence de masse
ou libre de toute singularit\'e colin\'eaire, d\'ecrit les configurations
partoniques \`a plus grand angle dans l'\'evolution du syst\`eme.

Les termes successifs de la s\'erie perturbative qui interviennent dans
le calcul de $\gamma(\alpha_s)$ peuvent s'\'ecrire sous la forme
symbolique (voir \cite{EvEq} et r\'ef\'erences ci-incluses)
\begin{equation}\label{eq:gammapow}
\gamma=\sqrt\alpha_s + \alpha_s + \alpha_s^{3/2} + \alpha_s^2 + \cdots;
\end{equation}
ceci  am\'eliore de la fiabilit\'e dans la description
des d\'esint\'egrations partoniques \`a petit angle $\Theta\ll1$ et ainsi, dans 
l'\'evolution du jet. De plus, tenir compte des corrections dans le coefficient
$$
C=1+\sqrt\alpha_s+\alpha_s+\cdots
$$
permet de consid\'erer l'ensemble du nombre croissant des jets dans lesquels
$\Theta_{ij}\sim1$.

\subsection{Estimation de $\boldsymbol{\gamma(\alpha_s)}$}

L'estimation de $\gamma(\alpha_s)$ d\'ecoule des \'equations d'\'evolution DLA
pour les multiplicit\'es des jets dont la d\'ependance s'exprime
en fonction de la duret\'e totale, c'est \`a dire le produit
de l'\'energie et de l'angle d'ouverture total du jet:
\begin{equation}\label{eq:nmultip}
N(p\Theta)\approx\int^{\Theta}\frac{d\Theta'}{\Theta'}
\left[\int^1_0\frac{dz}z4N_c\frac{\alpha_s}{2\pi}\right]
N(zp\Theta').
\end{equation}
Si l'on compare (\ref{eq:nmultip}) avec (\ref{eq:Zgamma}), on r\'ealise
que le terme entre crochets repr\'esente la dimension anormale.
Puisque les deux termes de (\ref{eq:nmultip}) contiennent des multiplicit\'es,
donc deux fonctions du m\^eme ordre de grandeur, les deux
int\'egrations logarithmiques doivent se compenser. Ceci revient
\`a effectuer l'analyse dimensionnelle de (\ref{eq:nmultip}), soit
\beq\label{eq:alphasell}
\int dt'\int \frac{dz}z\alpha_s\sim1 \Rightarrow \ell^2\alpha_s\sim1 \Rightarrow
\ell\sim\alpha_s^{-1/2},
\eeq
o\`u nous avons utilis\'e les notations 
$\int dt'\sim\int \frac{dz}{z}\sim \log\equiv \ell$.  Par d\'efinition,
\begin{equation}\label{eq:DLAgamma}
\gamma^{DLA}(\alpha_s)=\int\frac{dz}z\alpha_s=\alpha_s\ell\sim\sqrt{\alpha_s}.
\end{equation}
Lorsqu'on int\`egre sur $\gamma(\alpha_s)$, (estim\'e \`a partir de
(\ref{eq:Zgamma}) (\ref{eq:DLAgamma})), on arrive \`a l'exponentielle
charact\'eristique $\exp{(c\sqrt{\ln{E}})}$ qui d\'ecrit le taux
de croissance des multiplicit\'es des jets en fonction de l'\'energie.
Le terme sous-dominant, c'est \`a dire celui en $\Delta\gamma\sim\alpha_s$,
entra\^\i ne \'egalement une d\'ependance non-n\'egligeable dans
l'\'energie $\exp{(c_1\ln\ln{E/\Lambda})}\propto\alpha_s(E)^{-c_1}$
qui compense le terme croissant donn\'e par DLA.

Pour d\'ecrire ces effets de mani\`ere consistante on doit \'etudier
les sources physiques de ces corrections \`a partir de

\bigskip

$\bullet$ \textbf{la variation de} $\boldsymbol{\alpha_s(k_\perp)}$: elle
est due \`a l'\'eventuelle influence de la d\'ependance de la constante
de couplage en $\ln z$, soit $\alpha_s(\ln z)$, lors de l'\'emission
d'un gluon $g$ mou.
\begin{figure}[h]
\vbox{
\begin{center}
\includegraphics[height=3.5truecm,width=14truecm]{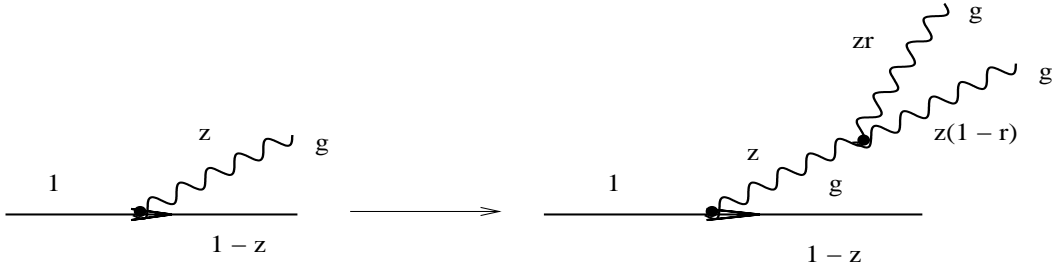}
\caption{\label{fig:runalphas}Effets de l'\'evolution de la constante de
couplage sur $\gamma(\alpha_s)$}
\end{center}
}
\end{figure}
Pour les diagrammes de la Fig.\ref{fig:runalphas} \`a gauche et \`adroite,
on obtient respectivement les contributions
$$
\gamma\!=\!\int\alpha_s\frac{dz}z\!=\!\alpha_s\ell\sim\sqrt
\alpha_s\!\equiv\!\gamma^{DLA},\quad 
\Delta\gamma\!=\!\int\alpha_s\frac{dz}z\alpha_s\frac{d(zr)}{(zr)}
\!\stackrel{zr=z'}{=}\!
\int\underbrace{\alpha_s^2\ell}_{\Delta\alpha_s}
\frac{dz'}{z'}=\alpha_s^2\ell^2\sim\alpha_s.
$$
Ainsi, contrairement au cas DLA o\`u l'on fixe la constante anormale,
il faut, dans l'objectif d'obtenir une estimation plus r\'ealiste
des observables, 
tenir compte de son \'evolution au cours de chaque \'emission. Ceci se 
g\'en\'eralise au cas d'un certain nombre d'\'emissions, pour trois
vertex on aurait $\alpha_s^3\ell^3\sim\alpha_s^{3/2}$, pour quatre 
$\alpha_s^4\ell^4\sim\alpha_s^2$, on trouve ainsi d'autres corrections d'ordre
sup\'erieur;

\bigskip

$\bullet$ \textbf{la production de partons dits ``\'energ\'etiques''}:
en DLA on n\'eglige le recul du parton \'emetteur et, par cons\'equent,
le principe de conservation de la quadri-impulsion.
Nous devons aussi consid\'erer les d\'esint\'egrations en deux
partons dont l'\'energie est du m\^eme ordre de grandeur $z\sim1$, et
resommer leurs contributions
$$
\Delta\gamma=\int\alpha_s dz\sim\alpha_s.
$$
Les diagrammes donnant ce type de corrections sont les suivants 
($g\to gg$, $g\to q\bar q$, $q\to gq$):
\vbox{
\begin{center}
\includegraphics[height=3.5truecm,width=14truecm]{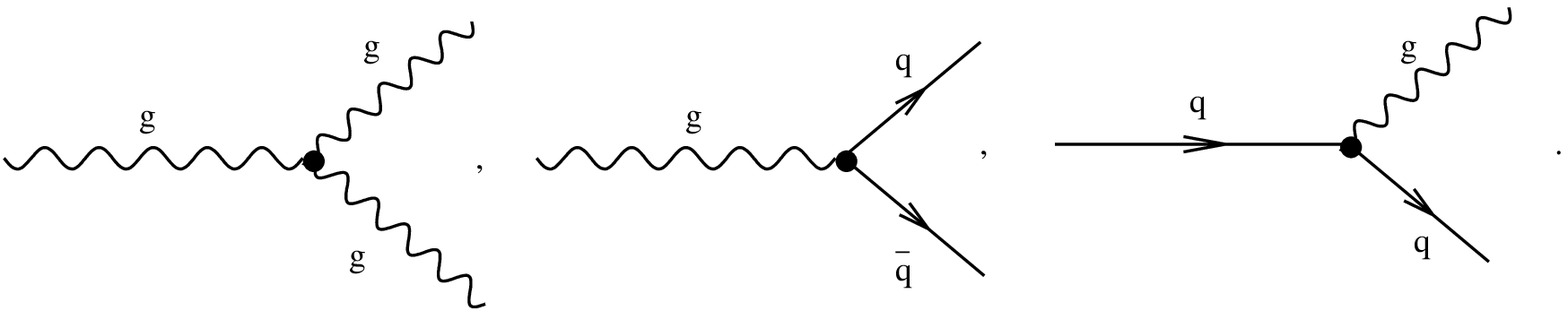}
% \epsfig{file=Figs/gqq.eps, height=3.5truecm,width=14truecm}
% \caption{Diagrammes o\`u l'on tient compte du recul des partons \'emetteurs}
\end{center}
}
En particulier, le diagramme du centre $g\to q\bar q$ n'est pas pr\'esent
en DLA.

\bigskip

$\bullet$ \textbf{l'int\'egration angulaire exacte}: dans la r\'egion
cin\'ematique o\`u les angles sont du m\^eme ordre de grandeur
$\Theta_{fj}\sim\Theta_{ji}\sim\Theta_{fi}$
lors des \'emissions doublement molles ($g\to ggg$ et $q\to qgg$)
d'une paire de gluons. Dans ce cas on a deux vertex, donc un $\alpha_s^2$
qui intervient dans le carr\'e de l'amplitude
$$
\Delta\gamma=\int\alpha_s^2\frac{dz_1}{z_1}\frac{dz_2}{z_2}=\alpha_s^2\ell^2\sim
\alpha_s.
$$
Lorsqu'on tient compte de ces effets, on obtient symboliquement en MLLA
$$
\gamma^{MLLA}=\sqrt\alpha_s+\alpha_s.
$$

\section{Probabilit\'e de d\'esint\'egration partonique dans le cadre MLLA}

La section efficace diff\'erentielle en MLLA s'\'ecrit sous la forme:
\begin{equation}
d\sigma_A^{BC}=\frac{\alpha_s(k_{\perp}^2)}{2\pi}\Phi_A^{BC}(z)dzV(\vec{n})
\frac{d\Omega}{8\pi},\label{eq:MLLACS}
\end{equation}
o\`u
\begin{equation}\label{eq:Vangulaire}
V_{j(i)}^f(\vec{n})=\frac{a_{fi}+a_{ji}-a_{fj}}{a_{fj}a_{fi}}
\end{equation}
et $\Phi_A^{BC}(z)$ sont les fonctions de d\'esint\'egration partoniques
d'Alterelli-Parisi \cite{DGLAP}, obtenues \`a 1-boucle en fonction de
$z$, la fraction d'\'energie emport\'ee par l'un des
enfants de $A$ ($B$ ou $C$):
\begin{eqnarray}
\label{eq:splitTH}
&& \Phi_q^{q[g]}(z) = C_F\, \frac{1+z^2}{1-z} , \quad
\Phi_q^{g[q]}(z)=C_F\, \frac{1+(1-z)^2}{z} , \\
\label{eq:cstTH}
&& \Phi_g^{q[\bar{q}]}(z) = T_R\left( z^2+(1-z)^2\right) , \quad
\Phi_g^{g[g]}(z)= 2C_A\left(\frac{1-z}{z}+ \frac{z}{1-z} +
z(1-z)\right),\cr
&& C_A=N_c,\quad C_F=(N_c^2-1)/2N_c,\quad T_R=1/2,
\end{eqnarray}
$N_c$ est le nombre de couleurs. De plus,
$$
a_{ik}=q^2\frac{p_i\cdot p_k}{(p_i\cdot q)(p_k\cdot q)}=1-\vec{n}_i\cdot \vec{n}_k=1-\cos\Theta_{ik}.
$$
Ainsi, int\'egrer sur l'expression exacte des fonctions (\ref{eq:splitTH})
revient \`a inclure les corrections en $\alpha_s$ qui restaurent
la conservation de la quadri-impulsion: 
$\Delta\gamma=\int\alpha_sz^ndz\sim\alpha_s$.  Les fonctions de
d\'esint\'egration (\ref{eq:splitTH}) ont \'et\'e en premier obtenues
dans le cadre de l'\'evolution partonique du genre espace
(virtualit\'e croissante des enfants) de la diffusion profond\'ement
in\'elastique  \cite{pQCDforbeginners}, mais elles interviennent, aussi,
dans le cas de l'\'evolution de genre temps (virtualit\'e d\'ecroissante
des enfants) qui d\'etermine l'\'evolution des jets jusqu'au stade de
l'hadronisation \cite{DKT}.
Cette propri\'et\'e est connue comme
``la relation de r\'eciprocit\'e de Gribov-Lipatov'' \cite{pQCDforbeginners}
et constitue l'une des plus
\'el\'egantes sym\'etries que les fonctions (\ref{eq:splitTH}) satisfont.

Puisque les gluons (quarks) sont rayonn\'es (\'emis) arbitrairement le long
du parton $j$ (sym\'etrie cylindrique), on int\`egre (\ref{eq:MLLACS})
en prenant la moyenne azimutale le long du c\^one d'angle d'ouverture
$\Theta_{ji}$ (voir appendice \ref{sub:Vangulaire})
\beq\label{eq:moyazit}
\Big<V_{j(i)}^f(\vec{n})\Big>_{moyenne\; azimutale}=\int_0^{2\pi}\frac{d\phi}{2\pi}V_{j(i)}^f(\vec{n})=\frac2{a_{fj}}
\vartheta(a_{ji}-a_{fj}),
\eeq
o\`u $\vartheta$ est la fonction de Heaviside.
C'est ainsi que (\ref{eq:MLLACS}) se r\'eduit \`a la section efficace
diff\'erentielle du processus $A\to B+C$ dans un jet:
\begin{equation}
d\sigma_A^{BC}=\frac{\alpha_s(k_{\perp}^2)}{2\pi}\Phi_A^{BC}(z)dz
\frac{d\Theta^2}{\Theta^2}=\frac{\alpha_s(k_{\perp}^2)}{\pi}\Phi_A^{BC}(z)dz\,dt,\quad
dt=\frac{d\Theta}{\Theta}
\end{equation}
\`a l'int\'erieur du c\^one $\Theta_{ji}\geq\Theta_{fj}$ et 
s'annule $d\sigma_A^{BC}=0$ \`a l'ext\'erieur $\Theta_{ji}<\Theta_{fj}$.
Par cons\'equent, la contrainte angulaire rigoureuse
$\Theta_{ji}\gg\Theta_{fj}$ en DLA 
est remplac\'ee par la contrainte angulaire stricte
$\Theta_{ji}\geq\Theta_{fj}$ en MLLA (voir \ref{sub:article2}).

\section{Equation Ma\^itresse dans le cadre de l'approximation MLLA}

Lorsque l'on \'etudie des observables telles que les multiplicit\'es, 
les fluctuations des multiplicit\'es, le spectre inclusif d'une particule
ou les corr\'elations dans les syst\`emes partoniques, on peut  remplacer le 
noyau $V(\vec{n})$ dans la section efficace (\ref{eq:MLLACS}) par la moyenne 
azimutale (\ref{eq:moyazit}). Ceci permet de construire des \'equations 
d'\'evolution simples pour les FG's qui satisfont la contrainte angulaire
stricte  d\'ecoulant de (\ref{eq:moyazit}).

\begin{figure}[h]
\begin{center}
\includegraphics[height=5truecm,width=0.48\tw]{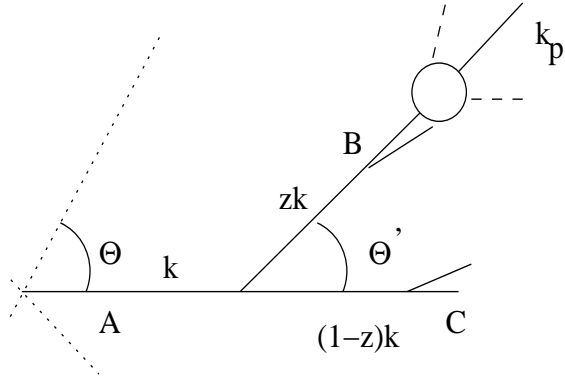}
\caption{\label{fig:vertex}Branchement $A\rightarrow B+C$.}
\end{center}
\end{figure}

Le syst\`eme d'\'equations, inspir\'e de (\ref{eq:red_1}) en DLA, pour les 
fonctionnelles $Z_G$, $Z_{F\equiv Q,\bar{Q}}$ qui d\'ecrit l'ensemble
partonique d'un jet initi\'e par un gluon ($G$) ou un quark, anti-quark 
($F\equiv Q\bar Q$) d'impulsion initiale $p$ et qui produit un jet d'angle
d'ouverture $\Theta$ est donn\'e par ($A,\,B,\,C=F, G$)
\beeq\notag
Z_A(p,\Theta;u(k))\!\!&\!\!=
\!\!&\!\!e^{-\omega_A(p\Theta)}u_A(k=p)+\frac12\sum_{B,C}\int_{Q_0/p}^{\Theta}
\frac{d\Theta'}{\Theta'}\int_0^1dz\,e^{-\omega_A(p\Theta)+\omega_A{p\Theta'}}\\\notag\\
&&\times\frac{\alpha_s(k^2_{\perp})}{2\pi}\Phi_A^{BC}(z)Z_B(zp,\Theta';u)Z_C((1-z)p,\Theta';u).\label{eq:MLLAGF}
\eeeq
Le premier terme dans le membre de  droite de (\ref{eq:MLLAGF})
correspond au cas o\`u le parton $A$ qui initie le jet ne rayonne pas. Cette 
probabilit\'e est, en particulier,
supprim\'ee par le facteur de forme de Sudakov dans l'exponentielle.
Le terme dans l'int\'grale d\'ecrit le premier branchement
$A\rightarrow B+C$ (voir Fig.\ref{fig:vertex})
et $\Theta'\leq\Theta$ (condition cin\'ematique) 
est l'angle entre $B$ et $C$. La multiplication par l'exponentielle 
dans l'int\'egrand garantit que cette \'emission est le premier
\'ev\`enement et que nul d'autre n'a lieu entre l'angle d'ouverture
du jet $\Theta$ et $\Theta'$; 
autrement dit, la probalit\'e pour qu'il y ait une \'emission
\`a l'int\'erieur du cr\'eneau angulaire $(\Theta,\Theta')$
est supprim\'ee par le facteur de Sudakov. $Z_B$ et $Z_C$ constituent
le point de d\'epart dans l'\'evolution des sous-jets $B$ et $C$;
leur \'energie est inf\'erieure \`a celle du parton $A$.
De nouveaux partons sont ainsi \'emis par $B$ et $C$, tels qu'on associe,
\`a chaque vertex, une nouvelle \'equation du type (\ref{eq:MLLAGF}).
Ceci met notamment
en \'evidence le caract\`ere it\'eratif des processus de branchement qui a
inspir\'e ce mod\`ele pour ainsi d\'ecrire la dynamique partonique des jets
(``{\em parton shower picture}'' en anglais).

\noindent

%\vskip 0.5cm

\subsection{Facteurs de forme de Sudakov en MLLA}

%\vskip 0.5cm

Les expressions des facteurs de forme de Sudakov sont les suivantes:
\beq\label{eq:omegaF}
\omega_F=\int_{Q_0/p}^{\Theta}\frac{d\Theta'}{\Theta'}\int_0^1dz
\frac{\alpha_s(k^2_{\perp})}{2\pi}\Phi_F^F(z);\qquad(F=Q,\bar Q),
\eeq
\beq\label{eq:omegaG}
\omega_G=\int_{Q_0/p}^{\Theta}\frac{d\Theta'}{\Theta'}\int_0^1dz
\frac{\alpha_s(k^2_{\perp})}{2\pi}
\left[\frac12\Phi_G^G(z)+n_f\Phi_G^F(z)\right].
\eeq
Les singularit\'es infrarouges et colin\'eaires, pr\'esentes dans
(\ref{eq:MLLAGF}-\ref{eq:omegaG}) doivent \^etre r\'egularis\'ees en 
imposant les contraintes perturbatives relatives \`a l'impulsion
transverse des partons \'emis. Nous l'\'ecrivons ici sous la forme
$$
k_{\perp}\approx pz(1-z)\Theta'>Q_0.
$$
On multiplie (\ref{eq:MLLAGF}) par $e^{\omega_A(p\Theta)}$, puis on d\'erive
les deux membres de l'\'equation par rapport \`a $\ln\Theta$ et on obtient
\beeq\notag
\frac{d}{d\ln\Theta}Z_A(p,\Theta;u(k))\!\!&\!\!+\!\!&\!\!Z_A(p,\Theta;u(k))
\frac{d\omega_A(p\Theta)}{d\ln\Theta}=
\frac12\sum_{B,C}\int_0^1dz\\
&&\times\frac{\alpha_s(k^2_{\perp})}{2\pi}\Phi_A^{BC}(z)Z_B(zp,\Theta;u)
Z_C((1-z)p,\Theta;u)\label{eq:MLLAGFbis},
\eeeq
or,
$$
\frac{d\omega_A(p\Theta)}{d\ln\Theta}=\frac12\sum_{B,C}\int_0^1dz
\frac{\alpha_s(k^2_{\perp})}{2\pi}\Phi_A^{BC}(z),
$$
ce qui permet de r\'ecrire (\ref{eq:MLLAGFbis}) sous la forme compacte
\begin{eqnarray}
\frac{d}{d\ln\Theta}Z_A\left(p,\Theta;\{u\}\right)
\!\!&\!\!=\!\!&\!\!\frac{1}{2}\sum_{B,C}\int_{0}^{1}dz\>
\Phi_A^{B[C]}(z)\ \frac{\alpha_s\left(k^2_{\perp}\right)}{\pi} \cr
&&\hskip -1cm
\Big(Z_B\big(zp,\Theta;\{u\}\big)\ Z_C\big((1-z)p,\Theta;\{u\}\big)
\!-\!Z_A\big(p,\Theta;\{u\}\big)\Big).
\label{eq:MLLAFGter}
\end{eqnarray}
L'\'equation (\ref{eq:MLLAFGter}) contient l'information des 
propri\'et\'es MLLA du jet moyenn\'ee sur l'angle azimutal. La
$n^{\text{\`eme}}$ deriv\'ee variationnelle de $Z_A$ par rapport
aux fonctions de sondage $u(k_i)$ au voisinage de $u=0$ permet d'obtenir
la $n^{\text{\`eme}}$ section efficace $d\sigma^n$ exclusive 
qui correspond \`a la production de $n$ partons. L'expansion de $Z_A$
autour de $u=1$ permet, \`a son tour, de g\'en\'erer les distributions
inclusives et les corr\'elations (voir le chap\^itre \ref{sec:ADL}).

%\medskip

\subsection{Condition initiale et normalisation}

\noindent

{\bf La condition initiale} pour r\'esoudre les syst\`eme d'\'equations
(\ref{eq:MLLAFGter}) est 
\begin{equation}\label{eq:condinit}
Z_A(p,\Theta;\left\{u\right\})|_{p\Theta=Q_0}=u_A(k=p).
\end{equation}
$Q_0$ est l'impulsion transverse minimale. Si l'impulsion de $A$ est
celle du cut-off, il ne se produit pas de branchement et le jet n'est 
constitu\'e que du parton qui l'a initi\'e, c'est ce que l'on appelle
``terme de Born''. Ceci se lit directement sur l'\'equation 
(\ref{eq:MLLAGF}), en effet, l'int\'egration devient nulle dans cette limite.

\medskip

\noindent

{\bf La normalisation} des fonctionnelles g\'en\'eratrices,
d\'ej\`a rencontr\'ee au chap\^itre \ref{sec:ADL}
\begin{equation}\label{eq:normcond}
Z_A(p,\Theta;\left\{u\right\})|_{u(k)\equiv1}=1
\end{equation}
peut se v\'erifier dans le cadre MLLA sans difficult\'e. 
En effet, si on pose $Z\equiv1$ et $u\equiv1$ dans (\ref{eq:MLLAGF})
on obtient
$$
e^{\omega_A(p\Theta)}=1+\frac12\sum_{B,C}\int_{Q_0/p}^{\Theta}
\frac{d\Theta'}{\Theta'}\int_0^1dz\frac{\alpha_s(k^2_{\perp})}{2\pi}
\Phi_A^{BC}(z)e^{\omega_A(p\Theta')};
$$
ceci entra\^\i ne directement (\ref{eq:omegaF}) et (\ref{eq:omegaG}) 
pour les probabilit\'es totales de d\'esint\'egration partoniques.

Le spectre inclusif d'une particule en MLLA s'obtient
\`a partir de (\ref{eq:MLLAFGter}) en prenant la deriv\'ee
${\delta}/{\delta u(k_a)}$ de la fonctionnelle g\'en\'eratrice 
$Z_A$ et en tenant compte de (\ref{eq:condinit}) et (\ref{eq:normcond})
\begin{equation}
\frac{d}{dY} \,
x{D}^a_A(x,Y) = \int_0^1 dz
\sum_B \Phi_A^B(z)\,\frac{\alpha_s(k_{\perp}^2)}{\pi}
 \left[
\frac{x}{z}  {D}^a_B\left(\frac{x}{z},Y+\ln z\right)
- \frac1{2} x{D}^a_A(x,Y) \right]
\label{eq:SIeveqTH}
\end{equation}
o\`u
$$
Y=\ln\frac{E\Theta}{Q_0},\quad x=E_p/E.
$$
% \subsubsection{Cin\'ematique, param\`etre de Feynman}
% 
% Nous nous proposons d'expliquer l'origine de la vraie variable de
% Feynman $x_F$ qui intervient dans le l'ancien mod\`ele des partons.
% 

\section{Lien entre (\ref{eq:SIeveqTH}) et 
les \'equations de Dokshitzer-Gribov-Lipatov-Alterelli-Parisi (DGLAP)
des fonctions de fragmentation partonique}

Les \'equations d'\'evolution MLLA (\ref{eq:MLLAFGter}) sont identiques aux 
\'equations de DGLAP \cite{DGLAP}
\`a un d\'etail pr\`es: la translation par
$\ln z$ de la variable $Y$ qui caract\'erise l'\'evolution du jet de duret\'e
$Q$. \'Etant une cons\'equence directe de la contrainte angulaire (AO), cette
modification est n\'egligeable dans le cadre de l'approximation logarithmique
dominante (LLA) en $(\alpha_sY)$ pour des partons suffisamment
\'energ\'etiques:
$|\ln z|<|\ln x|={\cal O}(1)$. C'est le cas des partons dans la diffusion
profond\'ement in\'elastique (DIS).

\subsection{Cin\'ematique ``DIS'', variable de Bjorken}

\begin{figure}[h]
\begin{center}
\includegraphics[height=6truecm,width=0.48\tw]{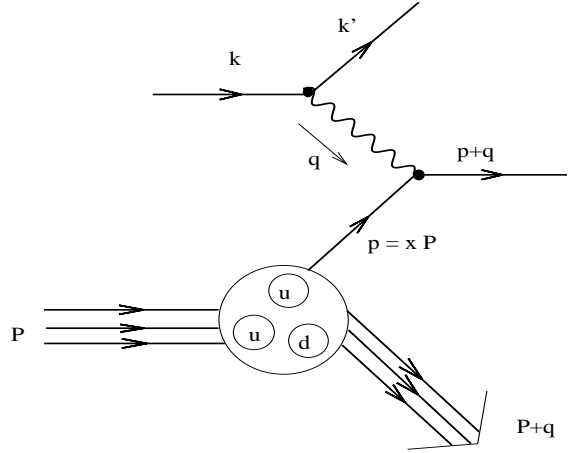}
\caption{\label{fig:DIS1} Diffusion profond\'ement in\'elastique.}
\end{center}
\end{figure}

Dans la diffusion profond\'ement in\'elastique, on consid\`ere le processus
dans lequel un lepton (\'electron, positron etc) ultra-relativiste
diffuse sur une cible (proton, neutron, noyau etc) (voir Fig.\ref{fig:DIS1}).
L'\'energie du lepton est suffisamment importante pour qu'il se 
produise une interaction du type \'electrofaible 
(\'echange d'un $\gamma^*$, $Z^0$, $W^{\pm}$) avec l'un des constituants de
la cible (quarks etc). Dans ce processus, l'impulsion $q$ du boson,
du genre espace $Q^2=-q^2\gg M^2_p$,
est transf\'er\'ee du lepton incident vers la cible. Ceci
provoque sa brisure en un \'etat multi-partonique$\to$syst\`eme
multi-hadronique.

Soit $M_p$ la masse invariante du proton, $P$ ($P^2=M_p^2$)
sa quadri-impulsion,
$k$ et $k'$ ($k^2\approx k'^2\approx0$) les quadri-impulsions de l'\'electron 
incident et sortant respectivement, et $P+q$ celle du syst\`eme
multi-hadronique.
On appelle $x$ la fraction de la quadri-impulsion $P$ emport\'ee par
le constituant frapp\'e; on peut donc l'\'ecrire comme $p=xP$.
La quadri-impulsion transf\'er\'ee est $q=k-k'$ et on s'int\'eresse
\`a l'\'evaluation de son carr\'e:
\begin{equation*}
q^2=(k-k')^2=-2k.k'=-2(k_0k'_0-\vec{k}.\vec{k'})=-4k_0k'_0
\sin^2\left(\frac12(\widehat{\vec{k},\vec{k'}})\right)<0
\end{equation*}
est bien du genre espace, donc $Q^2=-q^2>0$.
Soit $W$ la quantit\'e (invariant de masse) qui mesure l'in\'elasticit\'e du
processus dans le syst\`eme hadronique produit:
\begin{equation}\label{eq:bjorken}
W^2=(P+q)^2-M_p^2=q^2+2P.q=s(1-x)\quad\text{avec}\quad x=\frac{Q^2}{2P.q}\leq1,
\end{equation}
o\`u $x$ est la variable de Bjorken. Si $x=1$,
le processus est dit \'elastique. ``x'' est la variable qui intervient
dans les \'equations d'\'evolution partoniques de DGLAP, alors que dans
les cas DLA et MLLA, il s'agit du param\`etre de Feynman, soit de
la fraction d'\'energie totale du jet emport\'ee par le parton.

\subsection{Equations d'\'evolution}

L'espace de phase associ\'e au processus du genre espace 
$A\to B+C$ qui d\'etermine l'\'evolution des fonctions de structure
dans la diffusion profond\'ement in\'elastique (``DIS'') a la forme
suivante \cite{pQCDforbeginners}:
\begin{equation}
d\sigma_A^{BC}=\frac{\alpha_s(k_{\perp}^2)}{2\pi}\frac{dz}{z}
\frac{dk_{\perp}^2}{k_{\perp}^2}\Phi_A^{BC}(z),
\end{equation}
$z$ est la fraction de l'impulsion longitudinale emport\'ee par le parton
$B$. Les fonctions $\Phi$ d\'efinies dans (\ref{eq:splitTH}) jouent le r\^ole de 
``Hamiltonien'' des observables partoniques dans le cadre LLA 
(``Leading Logarithmic Approximation'' ou resommation de logarithmes
s colin\'eaires).
\begin{figure}[h]
\begin{center}
\includegraphics[height=6truecm,width=0.48\tw]{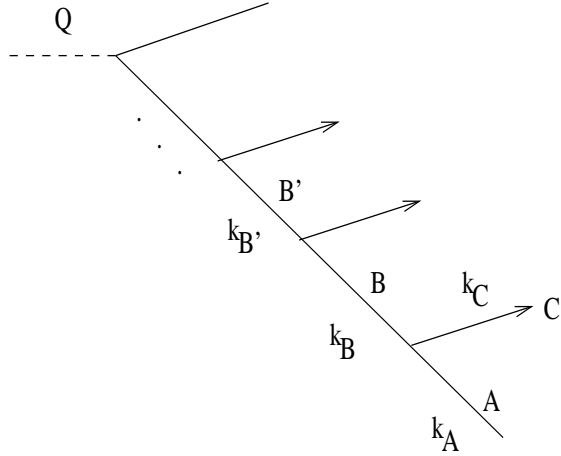}
\caption{\label{fig:DIS} Diffusion profond\'ement in\'elastique.}
\end{center}
\end{figure}
Dans l'environnement ``DIS'', le parton initial $A$ de virtualit\'e n\'egative
(genre espace) produit $B[z]$ de plus grande impulsion transverse
$|k_B^2|\gg|k_A^2|$, ainsi que $C[1-z]$ de virtualit\'e n\'egative
(genre temps).
Le parton C g\'en\`ere un sous-jet de partons ($\to$ hadrons) secondaires
\`a la fin du processus. Puisqu'on ne s'int\'eresse pas \`a la structure
des \'etats finaux dans les processus inclusifs, mais aux distribution
des \'etats initiaux, on int\`egre sur la masse invariante du sous-jet $C$.
L'int\'egration est domin\'ee par la r\'egion $k_C^2\ll |k_B^2|$
o\`u l'\'etat $C$ est pris sur couche de masse si l'on compare avec $B$
(le m\^eme argument a lieu pour $A$); cette r\'egion donne, en effet, la
contribution logarithmique dominante (``LLA'') au calcul de
la section efficace.
Les d\'esint\'egrations partoniques en ``LLA'' fournissent un
transfert d'impulsion important d'un parton cible $A$ (``r\'eel'')
vers un \'etat $C$ (``r\'eel'') par l'interm\'ediaire d'un \'etat $B$
de plus grande virtualit\'e.  \`A son tour, $B\equiv A'$, 
``r\'eel'' par rapport \`a $B'$ ($|k_{B'}^2|\gg|k_A^2|$),
devient la nouvelle cible \`a atteindre par le lepton et ainsi de suite.
La contrainte angulaire stricte sur les angles d'\'emission que l'on
a expliqu\'ee en MLLA est remplac\'ee par la contrainte sur les
impulsions des \'etats partoniques successifs en LLA, c'est \`a dire
$$
(|k_0|^2\equiv)|k_A^2|\ll|k_B^2|\ll|k_{B'}^2|\cdots\ll Q^2,
$$
o\`u $Q$ est la duret\'e du processus. Contrairement au cas MLLA,
le temps de formations des \'etats, de plus en plus virtuels,
est de moins en moins grand.
Dans ce cas, un syst\`eme d'\'equations identique \`a (\ref{eq:SIeveqTH}) 
est obtenu:
\begin{equation}\label{eq:DGLAP1}
\frac{\partial}{\partial Q^2}D_A^B(x,Q^2,k_0^2)=\frac{\alpha_s(Q^2)}{4\pi}
\sum_C\int_0^1\frac{dz}{z}\Phi_C^{B}(z)\left[D_A^C(\frac{x}{z},Q^2,k_0^2)
-z^2D_A^B(x,Q^2,k_0^2)\right]
\end{equation}
lorsque l'on it\`ere les cellules de $Q^2$ \`a $k_0^2$;
inversement, on obtient
\begin{equation}\label{eq:DGLAP2}
\frac{\partial}{\partial k_0^2}D_A^B(x,Q^2,k_0^2)=-\frac{\alpha_s(Q^2)}{4\pi}
\sum_C\int_0^1\frac{dz}{z}\Phi_A^{C}(z)\left[D_C^B(\frac{x}{z},Q^2,k_0^2)
-z^2D_A^B(x,Q^2,k_0^2)\right].
\end{equation}
Les fonctions de structure partoniques 
$D_A^B(x,Q^2,k_0^2)$ d\'ecrivent la probabilit\'e de trouver
\`a l'int\'erieur du nuage form\'epar la particule $A$
un parton du type $B$ de fraction d'impulsion longitudinale $x\leq1$;
le signe ``$=$'' correspond \`a la valeur de la virtualit\'e maximale $Q^2$ 
que $B$ peut atteindre. La somme sur les polarisations et les \'etats
de couleur est prise en compte. Nous consid\'erons, au m\^eme t\^itre,
les distributions de quarks, anti-quarks et gluons ``habill\'es''
\`a l'int\'erieur des quarks, anti-quarks et gluons,
tels que $A,B=q,\bar q,g$ ou $A,B=F,G$.
Ces distribution partoniques LLA ne d\'ependent pas de $k_0^2$ et $Q^2$
mais de la combinaison $\xi$ des deux quantit\'es d\'efinies par
\begin{equation}
d\xi(k^2)\equiv\frac{\alpha_s(k^2)}{4\pi}\frac{dk^2}{k^2},
\qquad\xi(Q^2)=\int_{\mu^2}^{Q^2}\frac{dk^2}{k^2}\frac{\alpha_s(k^2)}{4\pi},
\end{equation}
on remarque ainsi que les fonctions $D_A^B$ d\'ependent de la diff\'erence
$\Delta\xi$, par cons\'equent
\begin{eqnarray}
D_A^B(x,Q^2,k_0^2)&=&D_A^B(x,\Delta\xi),\cr\cr
\Delta\xi=\xi(Q^2)-\xi(k_0^2)&\approx&\frac1{4N_c\beta}
\ln\frac{\alpha_s(k_0^2)}{\alpha_s(Q^2)}=\frac1{4N_c\beta}
\ln\frac{\ln(Q^2/\Lambda^2)}{\ln(k_0^2/\Lambda^2)}\label{eq:deltaxi}
\end{eqnarray}
o\`u l'on a utilis\'e l'expression de la constante de couplage \`a
une boucle pour d\'eterminer l'expression analytique de $\Delta\xi$
\cite{EvEq}.  La variable $\xi$ peut \^etre
trait\'ee comme un ``temps d'\'evolution'' et la matrice $\Phi$ comme le
``Hamiltonien'' du syst\`eme. Nous rappelons que, dans le cas MLLA,
l'\'evolution du syst\`eme est du genre temps; le ``temps d'\'evolution''
dans ce cas est donn\'e par la variable $t=d\Theta/\Theta$
qui s'impose naturellement comme une cons\'equence in\'evitable 
de la contrainte angulaire (AO).
Le syst\`eme d'\'equations d'\'evolution qui s'\'ecrit sous forme
int\'egrale, qui satisfait les conditions initiales \'ecrites ci-dessous et
qui unifie (\ref{eq:DGLAP1}) et (\ref{eq:DGLAP2}),
s'\'ecrit sous la m\^eme forme que
l'\'equation originale de Bethe-Salpeter \cite{EvEq}
\begin{eqnarray}\label{eq:dglap}
D_A^B(x,\xi)&=&\delta_A^B\delta(1-x)+\sum_C\int_0^{\xi}d\xi'\int_0^1\frac{dz}z
\Phi_C^B(z)\left[D_C^B(\frac{x}z,\xi')-z^2D_A^B(x,\xi')\right]\cr\cr
&=&\sum_C\int_0^{\xi}d\xi'\int_0^1\frac{dz}z\Phi_A^C(z)\left[D_C^B(\frac{x}z,\xi')
-z^2D_A^B(x,\xi')\right].
\end{eqnarray}
La nature des processus de branchement partoniques s'av\`ere utile
pour exprimer les distributions en termes de la transformation
de Laplace-Mellin; ceci permet par ailleurs
d'\'ecrire leur produit de convolution en un produit simple des
distributions partoniques ind\'ependantes qui correspondent \`a chaque
\'emission dans l'espace conjugu\'e.

On introduit les distributions partoniques $D_A^B(j)$ dans l'espace de
Laplace-Mellin sous la forme
\begin{equation}
D_A^B(j,\xi)\equiv\int_0^1dx\,x^{j-1}D_A^B(x,\xi)
\end{equation}
que l'on ins\`ere dans (\ref{eq:dglap}) pour obtenir:
\begin{equation}\label{eq:mellineq}
\frac{\partial}{\partial\xi}D_A^B(j,\xi)=\sum_C\left(\Phi_A^C(j)-
\delta_A^C\phi_C\right)D_C^B(j,\xi)=\sum_C\left(\Phi_C^B(j)-
\delta_C^B\phi_C\right)D_A^C(j,\xi)
\end{equation}
o\`u les notations suivantes ont \'et\'e introduites:
\begin{subequations}
\begin{eqnarray}
\Phi_A^C(j)&\equiv&\int_0^1dz\,z^{j-1}\Phi_A^C(z),\\
\phi_F&\equiv&\int_0^1dz\,z[\Phi_F^F(z)+\Phi_F^G(z)]=\int_0^1dz
\Phi_F^F(z),\\
\phi_G&\equiv&\int_0^1dz\,z[\Phi_G^G(z)+2n_f\Phi_G^F(z)]=
\int_0^1dz\,[z\Phi_G^G(z)+n_f\Phi_G^F(z)].
\end{eqnarray}
\end{subequations}

La solution de cette \'equation avec la condition initiale
\begin{equation}\label{eq:condinit11}
D_A^B(j,\xi=0)=\delta_A^B
\end{equation}
donne les distributions en fonction de $x$ lorsqu'on inverse la transform\'ee
de Mellin
\begin{equation}
D_A^B(x,\xi)=\int_{(\Gamma)}\frac{dj}{2\pi i}\,x^{-j}\,D_A^B(j,\xi)
\end{equation}
dont l'int\'egration s'effectue le long de l'axe imaginaire. Le contour 
d'int\'egration $\Gamma$ est choisi de sorte qu'il inclut toutes les 
singularit\'es de $D(j)$ dans le plan complexe ($\Re j>1$). Il est commode de
repr\'esenter les distributions partoniques $D_A^B$ sous une forme matricielle
comme,
\begin{equation}
D_A^B(j,\xi)=\left(e^{\widehat{H}\xi}\right)_A^B,
\end{equation}
que l'on \'ecrit en fonction du ``Hamiltonien'' du syst\`eme:
\begin{equation}
\widehat{H}_A^B(j)=\Phi_A^B(j)-\delta_A^B\phi_A.
\end{equation}
Dans cette repr\'esentation l'\'etat d'\'evolution partonique forme un vecteur
qui s'\'ecrit sous la forme
\begin{equation}
(F_{NS},F_S,G).
\end{equation}
La premi\`ere composante d\'ecrit le quark de valence
(distribution non-singlet par rapport \`a la saveur du groupe),
le deuxi\`eme d\'ecrit la combinaison singlet des saveurs
(quarks de la ``mer'' et anti-quarks), et la troisi\`eme correspond
\`a la propagation des gluons. Dans cette base, le ``Hamiltonien''
prend la forme \cite{DGLAP}
\begin{equation}\label{eq:hamiltonien}
\widehat{H} =\left(\!\!
\begin{array}{ccc}
        \nu_F(j) & 0 & 0 \cr
\cr
        0 & \nu_F(j) & 2n_f\Phi_G^F(j) \cr
\cr
        0 & \Phi_F^G(j) & \nu_G(j)
 \end{array}
\!\!\right)
\end{equation}
o\`u les trajectoires r\'egularis\'ees des quarks et des gluons ont \'et\'e 
introduites:
\begin{eqnarray}\label{eq:nuF}
\nu_F(j)&\equiv&\int_0^1dz(z^{j-1}-1)\Phi_F^F(z)\\
\nu_G(j)&\equiv&\int_0^1dz\left[(z^{j-1}-z)\Phi_G^G(z)-n_f\Phi_G^F(z)\right].
\label{eq:nuG}
\end{eqnarray}
Les expression analytiques de (\ref{eq:nuF}), (\ref{eq:nuG}) peuvent \^etre 
\'ecrites en fonction de la fonction standard $\psi$:
\begin{equation}
\psi(j)=\frac{d}{dj}\ln\Gamma(j);\qquad\psi(j+1)=\psi(j)+1,\qquad\psi(1)=-\gamma_E
\end{equation}
o\`u $\gamma_E\approx0.5772$ est la constante d'Euler.
Finalement, (\ref{eq:nuF}), (\ref{eq:nuG}) et les transform\'ees de
Mellin des fonctions $\Phi_F^G$, $\Phi_G^F$ satisfont
\begin{subequations}
\begin{eqnarray}
\nu_F(j)&=&=-C_F\left[4\psi(j+1)+4\gamma_E-3-\frac2{j(j+1)}\right],\\
\nu_G(j)&=&-4N_c[\psi(j+1)+\gamma_E]+\frac{11N_c}3-\frac{2n_f}3
+\frac{8N_c(j^2+j+1)}{j(j^2-1)(j+2)},\\
\Phi_F^G(j)&=&2C_F\frac{j^2+j+2}{j(j^2-1)},\\
\Phi_G^F(j)&=&\frac{j^2+j+2}{j(j+1)(j+2)}.
\end{eqnarray}
\end{subequations}
D'apr\`es (\ref{eq:hamiltonien}), les quarks de valence se propagent
librement le long de la trajectoire $\nu_F(j)$, tandis que les quarks
de la mer se m\'elangent avec les \'etats gluoniques.
La diagonalisation de (\ref{eq:hamiltonien})
donne les ``fr\'equences propres'':
\begin{equation}
\nu_{\pm}=\frac12\left\{\nu_F(j)+\nu_G(j)\pm
\sqrt{[\nu_F(j)-\nu_G(j)]^2+8n_f\Phi_F^G(j)\Phi_G^F(j)}\right\}.
\end{equation}
On donne sans d\'emonstration les solutions de (\ref{eq:mellineq})
dans l'espace de Mellin \cite{EvEq}\cite{DDT}:

\vskip 0.5cm
\begin{subequations}
1. Distribution des quarks de valence (non-singlet)
\begin{equation}\label{eq:Dval}
D^{val}(j,\xi)=e^{\nu_F\xi}.
\end{equation}

2. Quark de la mer + anti-quarks dans le quark:
\begin{equation}
D^{sea}_F(j,\xi)=\frac{\nu_F(j)-\nu_-(j)}{\nu_+(j)-\nu_-(j)}e^{\nu_+\xi}
+\frac{\nu_+(j)-\nu_F(j)}{\nu_+(j)-\nu_-(j)}e^{\nu_-\xi}-e^{\nu_F\xi}.
\end{equation}
3. Distribution d'un gluon dans un quark:
\begin{equation}
D_F^G(j,\xi)=\frac{\Phi_F^G(j)}{\nu_+(j)-\nu_-(j)}\left(e^{\nu_+\xi}-
e^{\nu_-\xi}\right).
\end{equation}
4. Distribution des quarks+anti-quark dans un gluon:
\begin{equation}
D_G^F(j,\xi)=\frac{2n_f\Phi_G^F(j)}{\nu_+(j)-\nu_-(j)}\left(e^{\nu_+\xi}-
e^{\nu_-\xi}\right).
\end{equation}
5. Distribution d'un gluon dans un gluon:
\begin{equation}\label{eq:DGG}
D^G_G(j,\xi)=\frac{\nu_+(j)-\nu_F(j)}{\nu_+(j)-\nu_-(j)}e^{\nu_+\xi}
+\frac{\nu_F(j)-\nu_-(j)}{\nu_+(j)-\nu_-(j)}e^{\nu_-\xi}.
\end{equation}
\end{subequations}
% 

%\section{Compl\'ements des articles}

\chapter{Compl\'ements des articles}

Des d\'etails techniques seront donn\'es dans ce chap\^itre afin de
faciliter la compr\'ehension de certains aspects techniques des articles
\ref{sub:article1}, \ref{sub:article2} et \ref{sub:article3}.

\section{Inclusive hadronique distributions inside one jet at high
energy colliders at ``Modified Leading Approximation'' of Quantum 
Chromodynamics \ref{sub:article1}}

\subsection{Corr\'elation entres deux particules produites dans
l'annihilation $\boldsymbol{e^+e^-}$ \cite{DDT}\cite{DD}}

On consid\`ere la section efficace semi-inclusive de 
l'annihilation $e^+e^-$ en deux particules $h_1$ et $h_2$
($e^+e^-\to q\bar q h_1h_2$) de fractions
d'\'energie $x_1$ et $x_2$ et d'angle relatif de s\'eparation 
$\Theta$ dans un jet (l'angle $\Theta$ dans le r\'ef\'erentiel du 
laboratoire co\"\i ncide avec celui de la paire $q\bar q$ dans le centre 
de masse). Dans la jauge planaire, le processus ressemble \`a une cascade
de branchements de d\'esint\'egrations partoniques ind\'ependantes.
Consid\'erons la d\'esint\'egration du parton $A$ en $B$ et $C$, soit
$A\to B+C$, qu'\`a leur tour s'hadronisent pour produire les particules
$h_1$ et $h_2$, voir la Fig.1 de \ref{sub:article1}.

\begin{figure}[h]
\vbox{
\begin{center}
\includegraphics[height=12truecm,width=10truecm]{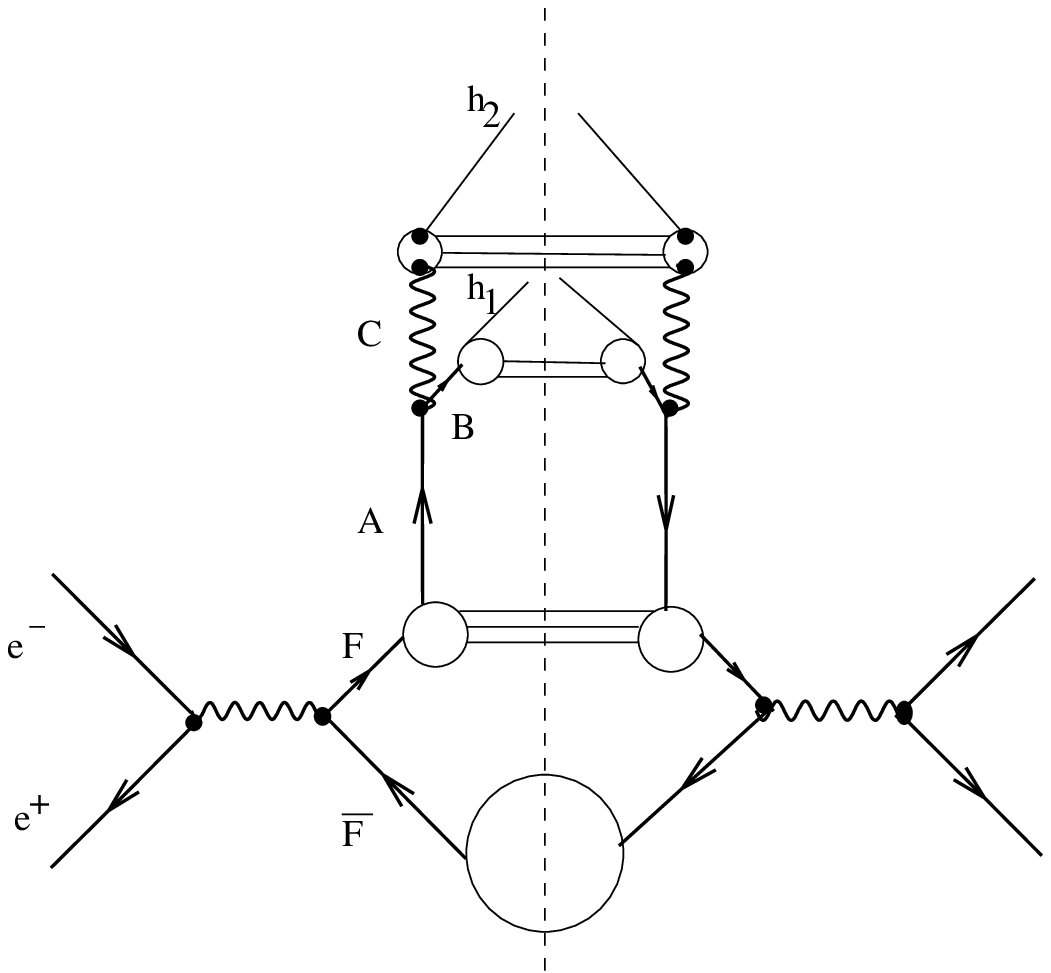}
\caption{\label{fig:amplitudecarr}Carr\'e de l'amplitude de la Fig.1 de l'article}
\end{center}
}
\end{figure}

La cin\'ematique est la suivante. Soit $u$ la fraction de l'\'energie du 
quark (ou de l'anti-quark) emport\'ee par le parton $A$ et $uz$, $u(1-z)$
celles des partons $B$ et $C$ respectivement; nous allons int\'egrer sur les
variables $u$ et $z$. Avoir d\'etect\'e les particules $h_1$ et $h_2$
d'angle de s\'eparation r\'elatif $\Theta$ permet de d\'eterminer,
en particulier, le point de branchement o\`u l'\'emission du parton
$A$ a eu lieu dans la cascade partonique. Or, la contrainte angulaire (AO),
permet de justifier l'approximation d'apr\`es laquelle l'angle entre
les partons $B$ et $C$ est tr\`es proche de celui
 entre les hadrons $h_1$ et $h_2$. De plus, les masses
invariantes des jets hadroniques issus de $B$ et $C$ sont tr\`es
inf\'erieures \`a leur impulsion transverse relative, de sorte que la
virtualit\'e du parton $C$ peut s'exprimer simplement en fonction
de l'angle d'ouverture $\Theta$ entre les hadrons d\'etect\'es comme
$$
k_A^2=(k_B+k_C)^2\approx2k_Bk_C=2\left(\frac{q}2uz\right)\left(\frac{q}2u(1-z)\right)
(1-\cos\Theta).
$$
Les premiers stades du processus, illustr\'es dans la Fig.1 de
\ref{sub:article1} et la Fig.\ref{fig:amplitudecarr}, 
\`a savoir la production du parton virtuel $A$, ainsi que les
produits de sa d\'esint\'egration, sont d\'ecrits par les fonctions
de structure d'annihilation dont la d\'ependance est douce en $k_A^2$.
Par cons\'equent, le pr\'efacteur d'ordre 1 qui est li\'e \`a
la redistribution de l'\'energie dans le
processus $A\to B+C$ peut (dans le cadre DLA) \^etre consid\'er\'e \`a part:
$$
k_A^2\sim q^2\sin^2\Theta/2\stackrel{\Theta\ll1}{\sim}q^2\Theta^2.
$$
Ainsi, fixer l'angle $\Theta$ entre les deux hadrons d\'etermine le temps de
production (``date de naissance'') du parton p\`ere
$A$, $\xi=\xi(q^2\Theta^2)$ (voir \ref{eq:deltaxi}).

Dans le but d'obtenir l'\'equation diff\'erentielle qui nous int\'eresse,
on doit multiplier l'amplitude de la Fig.1 par sa complexe conjugu\'ee
et int\'egrer sur l'impulsion des particules non d\'etect\'ees dans
le processus (voir Fig.\ref{fig:amplitudecarr}); nous devons
exprimer l'amplitude en \'echelle en termes des fonctions de
structure correspondantes.
Nous donnons la d\'efinition des fonctions de structure d'annihilation.
Si \`a l'\'etat final, $A$ (hadron ou parton avec des degr\'es de couleur)
est l'unique particule de fraction d'\'energie $z_A$ qui a
\'et\'e d\'etect\'ee,
la section efficace diff\'erentielle est donn\'ee par l'expression
\begin{equation}
\frac{d\sigma}{dz_Ad\Omega_{jet}}=\left(\frac{d\sigma}{d\Omega_{jet}}
\right)_0\sum_{F}^{2n_f}e^2_F\bar D_F^A\Big(z_A,\xi(q^2)-\xi(k_A^2)\Big),
\end{equation}
o\`u $(d\sigma/d\Omega_{jet})_0$ est la section efficace de Born correspondante
\`a l'annihilation $e^+e^-$ en une paire de quarks $q\bar q$:
\begin{equation}
\left(\frac{d\sigma}{d\Omega_{jet}}\right)_0=3\frac{\pi\alpha^2}
{2q^2}\frac{1+\cos^2\Psi}{2\pi},
\end{equation}
$\Psi$ est l'angle entre l'axe de la collision et celui de production de la
paire. $k^2$ repr\'esente la virtualit\'e de la particule d\'etect\'ee.
La fonction 
de structure $\bar D$ est li\'ee \`a l'amplitude invariante $\bar M_F^C$

\vbox{
\begin{center}
\includegraphics[height=3.5truecm,width=0.35\tw]{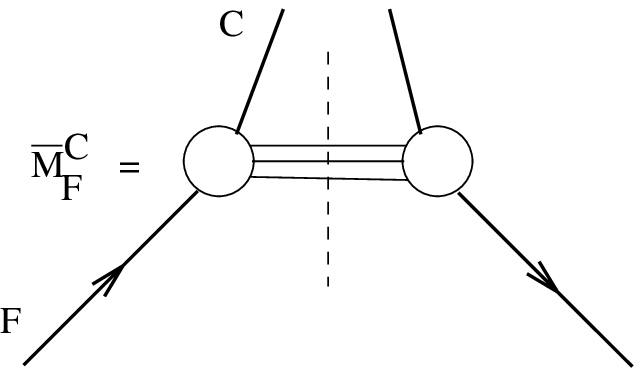}
\end{center}
}

moyenn\'ee sur les \'etats de polarisation et de couleur de la particule
C(F) et int\'egr\'ee sur la vitualit\'e du parton sortant $F$ jusqu'\`a la
limite sup\'erieure $q^2$ du ``partonom\`etre'', par la formule
\begin{equation}
z_A\bar D_F^A(z_A,\xi(q^2)-\xi(k_A^2))=\bar M_F^A.
\end{equation}
Le parton $A$ de virtualit\'e $k_A^2\sim q^2\sin^2\Theta/2$ joue,
au stade final du processus, le m\^eme r\^ole que le photon virtuel
$q^2$ \`a son stade initial.
Par cons\'equent, la production inclusive des hadrons $h_1$ et $h_2$ dans les 
sous-jets $B$ et $C$ est respectivement d\'ecrite par les fonctions
de structure:
$$
\bar D_B^{h_1}\Big(\frac{x_1}{z_B},\xi(k^2_A)-\xi(\mu^2)\Big),\quad
\bar D_C^{h_2}\Big(\frac{x_2}{z_C},\xi(k^2_A)-\xi(\mu^2)\Big).
$$
La section efficace diff\'erentielle s'\'ecrit enfin sous la forme:
\begin{eqnarray}
\frac{d\sigma}{d\Omega_{jet}dx_1dx_2d\ln\left(\sin^2\frac{\Theta}2\right)
\frac{d\phi}{2\pi}}\!\!&\!\!=\!\!&\!\!\left(\frac{d\sigma}{d\Omega_{jet}}
\right)_0\sum_{F,A,B,C}\!\!e_F^2\int\frac{du}{u^2}\int\frac{dz}{z(1-z)}
\bar D_F^A(u,\xi_q-\xi_0)\cr\cr
&&\hskip -2cm\Phi_A^{BC}(z)\frac{\alpha_s(k_A^2)}{4\pi}
\bar D_B^{h_1}\bigg(\frac{x_1}{zu},\xi_A\bigg)
\bar D_C^{h_2}\bigg(\frac{x_2}{(1-z)u},\xi_A\bigg)\label{eq:secteff12}
\end{eqnarray}
qui devient 2.1 (on utilise la d\'efinition $\xi(\mu^2)=0$, $\xi(k^2_A)=\xi_A$).

%\vskip 0.5cm

\subsection{Espace de phase dans (\ref{eq:secteff12})}

%\vskip 0.5cm

\`A l'origine, l'int\'egration dans (\ref{eq:secteff12}) 
(voir Fig.\ref{fig:amplitudecarr}) doit avoir lieu
sur les quatre composantes de la quadri-impulsion $k$, tandis que
le r\'esultat ne pr\'esente que les int\'egrations sur les fractions
d'\'energie de $h_1$ et $h_2$. Comment se fait-il que les autres
composantes, disons celles de $k_A^2$ soient fix\'ees?
Nous avons d\'ej\`a mentionn\'e l'analogie entre la d\'esint\'egration
$A\to h_1+h_2+X$ et l'annihilation. Si l'on se pla\c cait dans
le ref\'erentiel de la particule $A$, les partons $B$ et $C$ se
d\'eplaceraient dans le sens oppos\'e.
Par analogie avec l'annihilation semi-inclusive $e^+e^-$ en une paire de deux
particules, on peut \'ecrire la probabilit\'e du processus $A\to h_1+h_2$
sous la forme:
\beq\label{eq:DDrepres}
\frac{d\sigma^{A\to h_1+h_2+X}}{dk_{\perp}^2}\propto\frac1{k_{\perp}^2}
\frac{\partial}{\partial\ln k_{\perp}^2}\left\{\bar D_C^{h_1}(x_1, k_{\perp}^2)
\bar D_B^{h_2}(x_2,k_{\perp}^2)\right\},
\eeq
o\`u
$$
x_1\equiv\frac{2p_1k}{k^2},\qquad x_2\equiv\frac{2p_2k}{k^2},
$$
et $k_{\perp}$ est la composante du quadri-moment $k_A$ du ``photon''
virtuel $A$ projet\'e sur l'hyperplan form\'e par $(p_1, p_2)$,
tel que l'on effectue la d\'ecomposition de Sudakov suivante:
$$
k_A=\frac1{x_1}p_1+\frac1{x_2}p_2+k_{\perp}.
$$
Avec la repr\'esentation (\ref{eq:DDrepres}) pour la probabilit\'e
de d\'esint\'egration, on peut int\'egrer sur le quadri-moment $k_A$
que l'on \'ecrit en termes des variables de Sudakov:
$$
dk_A^2=d\left(\frac1{x_1}\right)d\left(\frac1{x_2}\right)d^2k_{\perp}
\frac{(2p_1p_2)}2=\frac{k_A^2}2\frac{dx_1}{x_1}\frac{dx_2}{x_2}d^2k_{\perp}.
$$
Puisqu'en LLA, la contribution essentielle est uniquement domin\'ee
par la r\'egion $k_{\perp}^2\ll k_{A}^2$ nous prenons l'int\'egrale
de (\ref{eq:DDrepres}) sur la deriv\'ee totale de la composante
perpendiculaire $k_{\perp}$:
$$
dk_A^2\frac{d\sigma^{A\to h_1+h_2}}{dk_{\perp}^2}\propto
\frac{dx_1}{x_1}\frac{dx_2}{x_2}\bar D_C^{h_1}(x_1, k_{\perp}^2)
\bar D_B^{h_2}(x_2,k_{\perp}^2).
$$
Par cons\'equent, pour compenser la d\'ependance douce
de (\ref{eq:DDrepres})  en fonction de $\ln k_{\perp}^2$
en LLA, la quadri-impulsion du parton p\`ere $A$ doit, en moyenne,
se trouver sur le plan des quadri-impulsions des hadron $h_1$ et $h_2$:
$$
k_A\approx\frac1{x_1}p_1+\frac1{x_2}p_2,\qquad k_A^2\approx q^2\sin^2\frac{\Theta}2.
$$
Pour obtenir la section efficace du processus (\ref{eq:secteff12}), 
on int\`egre sur l'\'energie de la particule $A$ et le transfert d'\'energie
relatif $z$ dans $A\to B+C$:
$$
\frac{dx_1}{x_1}\frac{dx_2}{x_2}=\frac{dz}{z(1-z)}\frac{du}u
$$
 et on ins\`ere les facteurs de normalisation qui lient
l'amplitude invariante aux fonctions de structure.

\subsection{Calcul des termes intervenant dans l'expression du courant
de couleur}

Nous avons, en particulier, utilis\'e les solutions des \'equations 
d'\'evolution de DGLAP (\ref{eq:dglap}) dans le calcul
de $<\!u\!>_{A_0}^A$ et $\delta\!<\!u\!>_{A_0}^A$ qui sont d\'efinis
en termes des fonctions de fragmentation
partoniques $D_{A_0}^A$ dans les formules (4.7) et (4.8) respectivement:
\begin{equation}\label{eq:udeltau}
<\!u\!>_{A_0}^A\approx\int_0^1duuD_{A_0}^A(u,E\Theta_0,E\Theta),\quad
<\!u\!>_{A_0}^A\approx\int_0^1du(u\ln u)D_{A_0}^A(u,E\Theta_0,E\Theta).
\end{equation}
On utilise (\ref{eq:Dval})-(\ref{eq:DGG}) pour calculer (\ref{eq:udeltau}) dans
\ref{sub:article1}, soit
$$
<\!u\!>_{A_0}^A={\cal D}_{A_0}^A(2,\xi(E\Theta_0)-\xi(E\Theta)),\quad
\delta<\!u\!>_{A_0}^A=\frac{d}{dj}{\cal D}_{A_0}^A(2,\xi(E\Theta_0)-\xi(E\Theta))
{\bigg|_{j=2}}
$$
dont les solutions ont \'et\'e explicitement donn\'ees dans l'appendice
C du m\^eme article.

\subsection{Comparaison des pr\'edictions avec les r\'esultats
pr\'eliminaires de CDF}
\label{subsub:comparaisonkt}
\begin{figure}
\vbox{
\begin{center}
\includegraphics[height=6.5truecm,width=0.49\tw]{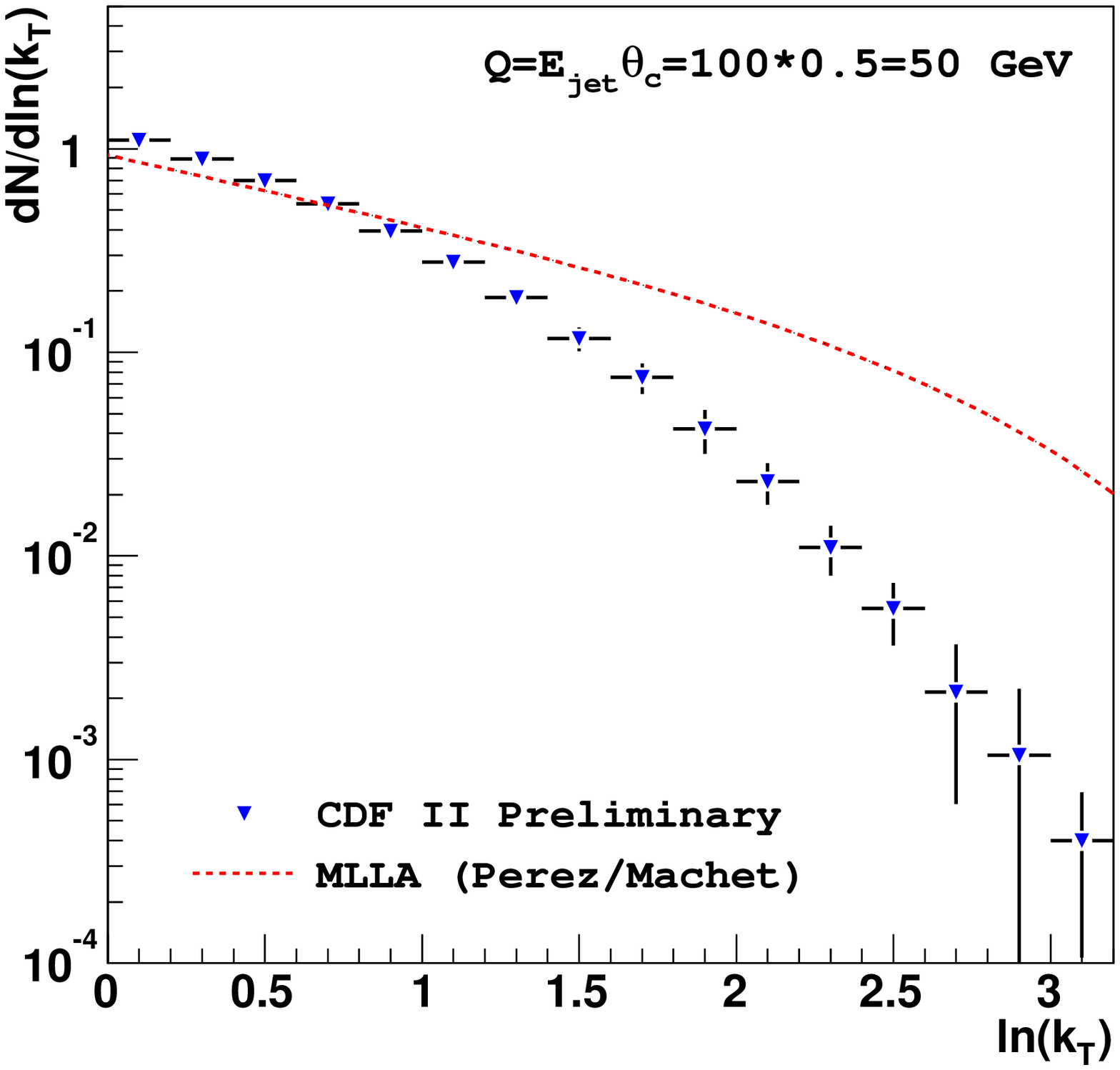}
\includegraphics[height=6.5truecm,width=0.49\tw]{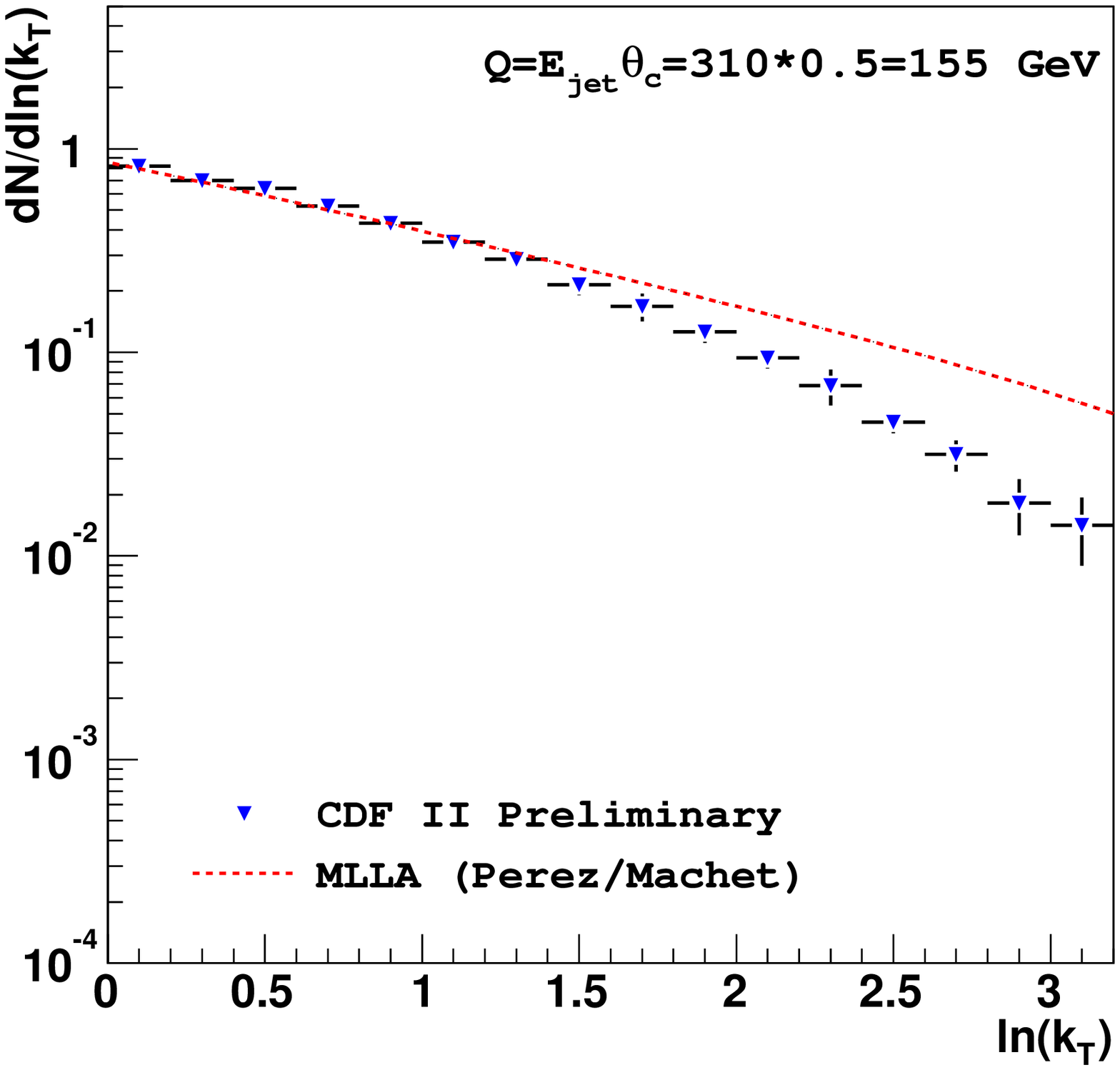}
\caption{\label{fig:sergo} Comparaison des pr\'edictions MLLA avec les 
donn\'ees de CDF pour $Q=55$ GeV ($Y=5.2$, gauche) et $Q=155$ GeV ($Y=6.4$, droite).}
\end{center}
}
\end{figure}

Nous comparons nos pr\'edictions pour les distributions inclusives
en fonction de l'impulsion transverse $k_{\perp}$ avec les r\'esultats
pr\'eliminaires de CDF dans le cas du m\'elange des jets de quarks et
de gluons que nous avons \'evoqu\'e au paragraphe 5.4.1.
Le nombre total de particules charg\'ees repr\'esente 60\% du nombre
total de particules produites. Puisque $N^{ch}\approx
N^{MLLA}$ \footnote{le nombre de particules charg\'ees est
approximativement \'egal au nombre de particules tel qu'il est
pr\'edit par le sch\'ema MLLA} (r\'esultat exp\'erimetal)
o\`u $N^{MLLA}$ est la pr\'ediction MLLA des multiplicit\'es
dans les jets hadroniques \cite{MW}, ${\cal K}^{ch}\approx0.56\pm0.10$.
Ainsi, la distribution des particules charg\'ees en fonction de $k_\perp$
est donn\'ee par la combinaison lin\'eaire au facteur ${\cal K}^{ch}$
pr\`es, c'est \`a dire
\begin{equation*}
\left(\frac{dN}{d\ln k_{\perp}}\right)^{ch}={\cal K}^{ch}
\left[\omega\left(\frac{dN}{d\ln k_{\perp}}\right)_g+(1-\omega)
\left(\frac{dN}{d\ln k_{\perp}}\right)_q\right]
\end{equation*}
o\`u $\omega=0.44$ est le param\`etre de m\'elange.
La Fig.\ref{fig:sergo} confirme, en particulier,
l'accord remarquable entre nos pr\'edictions et les donn\'ees
\`a petit $k_{\perp}=xE\Theta$ (car petit ``$x$'' et ``$\Theta$'')
dans l'intervalle de validit\'e MLLA. Pour 
$Y=5.2$ ($Q=55$ GeV) nous avons pr\'edit
$1\leq\ln(k_{\perp}/Q_0)\leq2.7$ $\Rightarrow$
$0 \leq \ln (k_{\perp}/1\,GeV) \leq 1.5$ et pour $Y=6.4$ ($Q=155$ GeV),
$1\leq\ln(k_{\perp}/Q_0)\leq3.9$ $\Rightarrow$ $0 \leq \ln (k_{\perp}/1\,GeV) \leq 2.6$.
A savoir, l'intervalle de confiance est d'autant plus grand
 que l'\'echelle d'\'energie est 
importante. Voir les explications d\'etaill\'ees dans \ref{sub:article1}.

Cette accord
remarquable constitue \'egalement une nouvelle confirmation de l'hypoth\`ese de dualit\'e
locale parton hadron (LPHD).
\section{Two-particle correlations inside one jet at
``Modified Leading Logarithmic Approximation'' of Quantum Chromodynamics
I: Exact solution of the evolution equations at ``small
$\boldsymbol x$'', \ref{sub:article2}}

\label{subsub:corrpaper}

L'unique calcul des corr\'elations qui avait \'et\'e effectu\'e
jusqu'alors dans le cadre MLLA est celui par Fong et Webber \cite{FW} en 1991.
Ils ont \'ecrit les \'equations d'\'evolution MLLA sous forme
diff\'erentielle et les ont r\'esolues dans l'approximation o\`u les 
\'energies des deux partons sont
quasiment identiques, en plus de les situer au voisinage
du maximum de leur distribution inclusive \cite{EvEq}:
$$
\ell_1\approx\ell_2\approx\frac{Y}2+a\gamma_0.
$$
Nous avons effectu\'e un calcul de cette observable, en
CDQ perturbative, qui nous a permis
d'\'etendre le r\'esultat \`a toutes les valeurs possibles de $x$.
Nous avons trouv\'e, pour l'intervalle de validit\'e de l'approximation des
petits ``$x$'', la limite inf\'erieure $x_{min}\approx0.08$
($\ell_{max}\approx2.5$) au seuil d'\'energie des pr\'esents 
acc\'el\'erateurs, celle-ci est en accord avec celle que l'on
trouve dans l'article \ref{sub:article1} pour le cas des
 distributions inclusives en fonction de $k_{\perp}$.

%\vskip 0.5cm

\subsection{Comparaison entre les corr\'elations en DLA et MLLA,
appendice F}

%\vskip 0.5cm

Nous avons compar\'e la formule DLA des corr\'elations (\ref{eq:DLAcorr})
avec la (5.2) de \ref{sub:article2}. Ceci permet d'observer le r\^ole
de la variation de la constante de couplage ainsi que l'importance de
la conservation de l'\'energie dans l'\'etude de cette observable.

\section{Two-particle correlations inside one jet at "Modified Leading Lo-
garithmic Approximation" of Quantum Chromodynamics ;
II : Steepest descent evaluation of the single inclusive distribution
at small $\boldsymbol{x}$, \ref{sub:article3}}

\subsection{Spectre inclusif d'une particule en MLLA; m\'ethode du col}

Les \'equations d'\'evolution qui d\'ecoulent de (\ref{eq:MLLAFGter}), 
ont \'et\'e obtenues dans la partie I de \ref{sub:article3},
dans l'approximation des petits $x$:
 \begin{equation}
G(\ell,y) = \delta(\ell)
+\int_0^{\ell} d\ell'\int_0^{y} dy' \gamma_0^2(\ell'+y')\Big(
 1  -a\delta(\ell'-\ell) \Big) G(\ell',y'),
\label{eq:solgg}
\end{equation}
\begin{equation}
Q(\ell,y)= \delta(\ell) + \frac{C_F}{N_c}\int_0^\ell d\ell'\int_0^y dy'
\gamma_0^2(\ell'+y')\Big( G(\ell',y')
-\frac34\delta(\ell'-\ell) \Big) G(\ell',y').
\label{eq:solqq}
\end{equation}
Si on fait l'analyse dimensionnelle de (\ref{eq:solgg}) et (\ref{eq:solqq})
en utilisant $\ell\sim\alpha_s^{-1/2}$ (voir \ref{eq:alphasell}),
on constate que le terme $\propto1$ est bien
${\cal O}(1)$ et que ceux qui sont proportionnels \`a $a$ et $\frac34$
incluent les corrections simplement logarithmiques
${\cal O}(\sqrt{\alpha_s})$. Elles d\'ecoulent
de l'int\'egration exacte sur les fonctions de d\'esint\'egrations
partoniques (\ref{eq:splitTH}). La d\'ependance de
la dimension anormale en $(\ell+y)$, ainsi que l'int\'egration
exacte sur $y(\Theta)$ donnent des corrections du m\^eme ordre de grandeur.

\begin{figure}[h]
\begin{center}
\includegraphics[height=6truecm,width=0.5\tw]{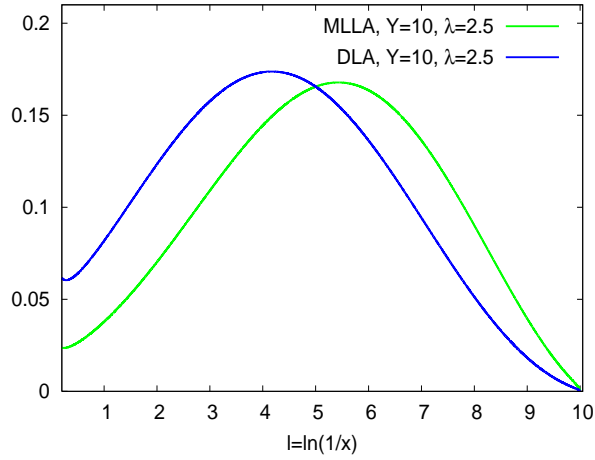}
\caption{\label{fig:MLLAlambdaTH}Spectre inclusif en DLA 
(\ref{eq:SpecNorm}, bleu), en MLLA (\ref{eq:SpecNormMLLATH}, vert) pour 
$Y=10.0$, $\lambda=2.5$; $\ell_{max}^{DLA}=5.0$, $\ell_{max}^{MLLA}\approx6.3$.}
\end{center}
\end{figure}

L'expression de $G$, solution exacte de (\ref{eq:solgg}) dans l'espace
de Mellin, qui g\'en\'eralise (\ref{eq:DLAalphasrun})
\footnote{Ici on change de notation, en effet, il s'agit de la m\^eme
fonction, soit $G\equiv D$} au cadre MLLA est d\'emontr\'ee dans
l'appendice D de \ref{sub:article2}.
\begin{equation}\label{eq:MLLAalphasrunTH}
G\left(\ell,y\right)=\left(\ell\!+\!y\!+\!\lambda\right)\!\!\iint\frac{d\omega\, d\nu}
{\left(2\pi i\right)^2}e^{\omega\ell+\nu y}
\!\!\int_{0}^{\infty}\frac{ds}{\nu+s}\!\!
\left(\frac{\omega\left(\nu+s\right)}
{\left(\omega+s\right)\nu}\right)^{1/\beta\left(\omega-\nu\right)}\!\!
\left(\frac{\nu}{\nu+s}\right)^{a/\beta}\,e^{-\lambda s}.
\end{equation}
Le terme en puissance de $a/\beta$ fait la diff\'erence par rapport \`a
(\ref{eq:DLAalphasrun}) et peut \^etre consid\'er\'e comme une faible
perturbation du terme dominant en DLA. L'objectif est d'estimer 
(\ref{eq:MLLAalphasrunTH}) par la m\'ethode du col;
il est ainsi suffisant de remplacer le point de col $(\omega_0,\nu_0)$
obtenu en DLA (\ref{eq:muupsilonTH}) dans cette correction
(voir l'appendice \ref{subsec:metcol}); le r\'esultat s'\'ecrit alors
sous la forme
\beeq\notag
G(\ell,Y)\approx\widehat{{\cal N}}(\mu,\upsilon,\lambda)\exp\Big[
\frac2{\sqrt{\beta}}\left(\sqrt{Y+\lambda}-\sqrt{\lambda}\right)
\frac{\mu-\upsilon}{\sinh\mu-\sinh\upsilon}+\upsilon-\frac{a}{\beta}
(\mu-\upsilon)\Big]
\label{eq:SpecalphasrunmllaTH}
\eeeq
pour $Y+\lambda\gg1$ et $\lambda\gg1$, o\`u 
$$
\widehat{{\cal N}}={\cal N}\times\left(\frac{Y+\lambda}{\lambda}\right)^
{\displaystyle{-\frac12\frac{a}{\beta}}}.
$$
La fonction ${\cal N}(\mu,\upsilon)$
est donn\'ee dans (\ref{eq:Dexp}). Nous pouvons de m\^eme normaliser 
(\ref{eq:SpecalphasrunmllaTH}) par l'expression en MLLA de la
multiplicit\'e d'un jet, c'est \`a dire
$$
\bar{n}(Y)\approx\frac12\left(\frac{Y+\lambda}{\lambda}\right)^
{\displaystyle{-\frac12\frac{a}{\beta}+\frac14}}
\exp\left[\frac2{\sqrt{\beta}}\left(\sqrt{Y+\lambda}-\sqrt{\lambda}\right)\right]
$$
pour r\'ecrire (\ref{eq:SpecalphasrunmllaTH}) sous la forme qui g\'en\'eralise 
(\ref{eq:SpecNorm}) au cadre MLLA
\beeq\notag
\frac{G(\ell,Y)}{\bar{n}(Y)}\!\!&\!\!\approx\!\!&\!\!\sqrt{\frac{\beta^{1/2}(Y+\lambda)^{3/2}}
{\pi\cosh\upsilon DetA(\mu,\upsilon)}}\!\exp\!\Big[
\frac2{\sqrt{\beta}}\left(\sqrt{Y+\lambda}\!-\!\sqrt{\lambda}\right)
\!\left(\frac{\mu-\upsilon}{\sinh\mu-\sinh\upsilon}\!-\!1\right)+\upsilon\Big.\\
&&\hskip 4.5cm\Big.-\frac{a}{\beta}
(\mu-\upsilon)\Big]
\label{eq:SpecNormMLLATH}.
\eeeq
Le maximum de la distribution peut \^etre d\'etermin\'e \`a partir de 
(\ref{eq:SpecNormMLLATH}) en utilisant (\ref{eq:dermullTH}). Ceci entra\^\i ne
\beq\label{eq:ellmaxmllaTH}
\ell_{max}=\underbrace{\frac{Y}2+\frac12\frac{a}{\beta}
\left(\sqrt{Y+\lambda}-\sqrt{\lambda}\right)}_{MLLA}>\underbrace{\frac{Y}2}_{DLA}.
\eeq
(\ref{eq:SpecNormMLLATH}) reproduit la forme gaussienne de la distribution 
au voisinage de (\ref{eq:ellmaxmllaTH})
\beq\nonumber
\frac{G(\ell,Y)}{\bar{n}(Y)}\approx\left(\frac{3}{\pi\sqrt{\beta}\left[(Y+\lambda)^{3/2}
-\lambda^{3/2}\right]}\right)^{1/2}\!\!\exp\left(-\frac2{\sqrt{\beta}}\,\frac3{(Y+\lambda)^{3/2}
-\lambda^{3/2}}\frac{\left(\ell_{max}-\ell\right)^2}2\right).
\eeq
La position du maximum du spectre inclusif d'une particule est donc
d\'ecal\'ee vers les plus grandes valeurs de $\ell$, soit vers les
plus petits $x$; ce d\'ecalage est une cons\'equence de la conservation
de l'\'energie dans les processus de branchements partoniques. Dans la 
Fig.\ref{fig:MLLAlambdaTH}, nous comparons le spectre normalis\'e
en DLA (\ref{eq:SpecNorm}) et celui en MLLA (\ref{eq:SpecNormMLLATH}).
\begin{figure}[h]
\begin{center}
\includegraphics[height=6truecm,width=0.48\tw]{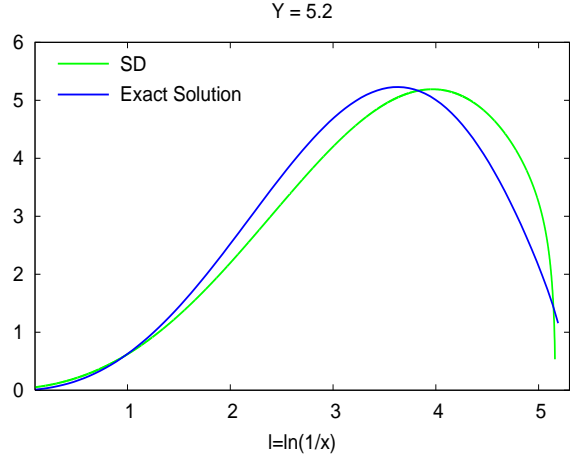}
\caption{\label{fig:MLLASTESpecTH} Forme du spectre inclusive donn\'ee par la 
m\'ethode du col et la m\'ethode exacte: formule 7.55 de \cite{EvEq}.}
\end{center}
\end{figure}
Bien que (\ref{eq:SpecalphasrunmllaTH}) n'est pas stable dans la
limite infrarouge $\lambda\to0$, car il faut que $\lambda\gg1$
pour garantir la convergence des m\'ethodes
perturbatives, on s'int\'eresse \`a cette limite.
Cela nous a permis de comparer, dans la Fig.\ref{fig:MLLASTESpecTH},
la forme du spectre obtenue par la m\'ethode du col et celle qui a
\'et\'e obtenue \`a partir de (\ref{eq:MLLAalphasrunTH}) pour
$\lambda=0$ \cite{EvEq}\cite{DKT} (``limiting spectrum'' en anglais).
En effet, on obtient une bonne allure pour la distribution car celle-ci
ne d\'epend pas du param\`etre $\lambda$, seulement sa normalisation
en d\'epend.

La m\'ethode du col permet par cons\'equent de donner la vraie forme
du spectre, la position du pic et reproduit de m\^eme son allure
\`a $\lambda\ne0$ ($Q_0\ne\Lambda_{QCD}$). Cette derni\`ere
a \'et\'e donn\'ee en effectuant une int\'egration
num\'erique de la formule (7.12) de \cite{EvEq} dans le plan complexe
\cite{DKT}.

%\vskip 0.5cm

\subsection{Deriv\'ees logarithmiques obtenues par la m\'ethode du col (utile pour la paragraphe 2.4 de \ref{sub:article3})}

%\vskip 0.5cm

\begin{figure}[h]
\begin{center}
\includegraphics[height=5truecm,width=0.48\tw]{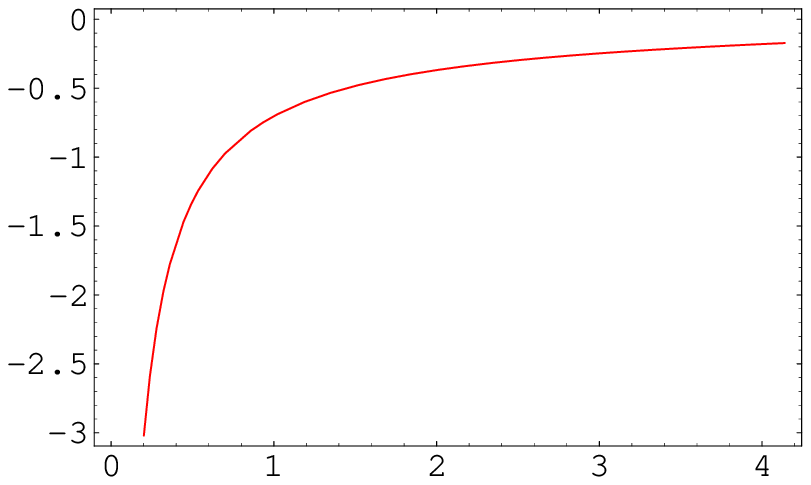}
\hfill
\includegraphics[height=5truecm,width=0.48\tw]{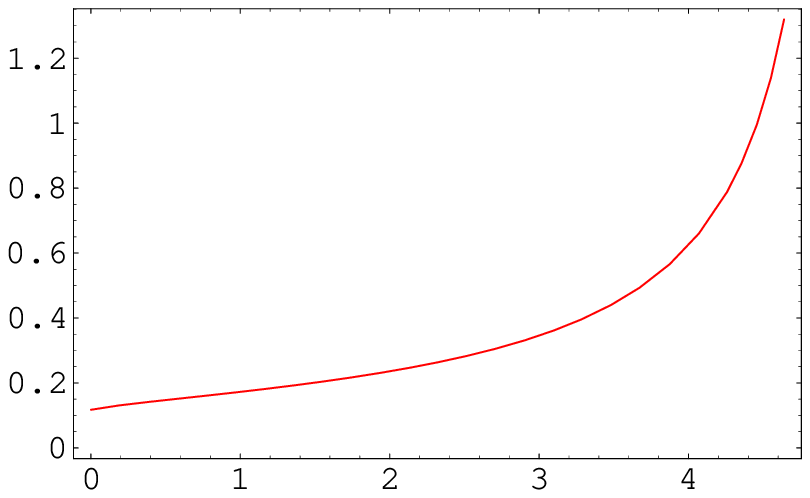}
\caption{\label{fig:dermunuelly} D\'eriv\'ees $\frac{\partial\mu}{\partial\ell}
$ et $\frac{\partial\mu}{\partial y}$ pour $Y\approx5.2$, $\lambda=0$ en fonction
de $\ln(1/x)$.}
\end{center}
\end{figure}

Certains d\'etails des calculs effectu\'es dans \ref{sub:article3}
sont donn\'es ici afin de mieux comprendre l'origine des corrections MLLA;
apr\`es avoir exponenti\'e la d\'ependance en $(\ell,y)$ du terme 
$\widehat{{\cal N}}$ et avoir obtenu 
la fonction dans l'exponentielle de (\ref{eq:SpecalphasrunmllaTH})
que l'on a \'ecrite sous la forme:
\beq\label{eq:phipsi}
\psi=\phi+\delta\psi,
\eeq
o\`u 
\beq\label{eq:phibar}
\phi=\frac2{\sqrt{\beta}}\left(\sqrt{\ell+y+\lambda}-\sqrt{\lambda}\right)
\frac{\mu-\upsilon}{\sinh\mu-\sinh\upsilon}
\eeq
est le terme dominant qui a \'et\'e trouv\'e en DLA et
\beq\label{eq:phiprimeTH}
\delta\psi=-\frac12\left(1+\frac{a}{\beta}\right)\ln(\ell+y+\lambda)-\frac{a}{\beta}\mu
+\left(1+\frac{a}{\beta}\right)\upsilon+\frac12\ln[Q(\mu,\upsilon)],
\eeq
est le terme sous-dominant au sens o\`u ses d\'eriv\'ees
donnent les corrections MLLA. On d\'efinit
$$
Q(\mu,\upsilon)=\frac{\sinh^3\mu}
{(\mu\!-\!\upsilon)\cosh\mu\cosh\upsilon\!+\!\cosh\mu\sinh\upsilon\!-\!\sinh\mu
\cosh\upsilon}.
$$
Maintenant on s'int\'eresse aux deriv\'ees de (\ref{eq:phibar}) et
(\ref{eq:phiprimeTH}) dont le r\'esultat a \'et\'e donn\'e sans d\'emonstration
dans l'article. Par d\'efinition du point de col:
$$
\bar{\phi}_{\ell}=\omega_0=\gamma_0e^{\mu},\quad 
\bar{\phi}_{y}=\nu_0=\gamma_0e^{-\mu},
$$
ceci peut \^etre de m\^eme v\'erifi\'e si l'on prend explicitement
la deriv\'ee de (\ref{eq:phibar}).
Les expression pour les d\'eriv\'ees sont en effet \'ecrites en
fonction de $\frac{\partial\mu}{\partial\ell}$, $\frac{\partial\upsilon}{\partial\ell}$,
$\frac{\partial\mu}{\partial y}$ et $\frac{\partial\upsilon}{\partial y}$
comme
\beq\label{eq:derphiprimell}
\delta\psi_{\ell}\equiv\frac{\partial\delta\psi}{\partial\ell}=-\frac12\left(1+\frac{a}{\beta}\right)\beta\gamma_0^2+
{\cal {L}}(\mu,\upsilon)\frac{\partial\mu}{\partial\ell}+{\cal {K}}(\mu,\upsilon)\frac{\partial\upsilon}{\partial\ell},
\eeq
\beq\label{eq:derphiprimey}
\delta\psi_y\equiv\frac{\partial\delta\psi}{\partial y}=-\frac12\left(1+\frac{a}{\beta}\right)\beta\gamma_0^2+
{\cal {L}}(\mu,\upsilon)\frac{\partial\mu}{\partial y}+{\cal {K}}(\mu,\upsilon)\frac{\partial\upsilon}{\partial y},
\eeq
o\`u on a d\'efini:
$$
{\cal {L}}(\mu,\upsilon)=-\frac{a}{\beta}+L(\mu,\upsilon),\quad
L(\mu,\upsilon)=\frac12\frac{\partial}{\partial\mu}
\ln[Q(\mu,\upsilon)],
$$
et
$$
{\cal {K}}(\mu,\upsilon)=1+\frac{a}{\beta}+K(\mu,\upsilon),\quad
K(\mu,\upsilon)=\frac12\frac{\partial}{\partial\upsilon}
\ln[Q(\mu,\upsilon)].
$$
Un calcul explicite permet de d\'eterminer
$$
L(\mu,\upsilon)=\frac32\frac{\cosh\mu}{\sinh\mu}-
\frac12\frac{(\mu-\upsilon)\cosh\upsilon\sinh\mu+\sinh\upsilon\sinh\mu}
{(\mu-\upsilon)\cosh\mu\cosh\upsilon+\cosh\mu\sinh\upsilon-\sinh\mu\cosh\upsilon},
$$
et
$$
K(\mu,\upsilon)=-\frac12\sinh\upsilon\frac{(\mu-\upsilon)\cosh\mu-\sinh\mu}
{(\mu-\upsilon)\cosh\mu\cosh\upsilon+
\cosh\mu\sinh\upsilon-\sinh\mu\cosh\upsilon}.
$$
\`A partir de (\ref{eq:relmunuTH}) on en d\'eduit:
$$
\frac{\partial\upsilon}{\partial\ell}=\frac{\sinh\upsilon}{\cosh\upsilon}
\left(\frac{\cosh\mu}{\sinh\mu}\frac{\partial\mu}{\partial\ell}-
\frac12\beta\gamma_0^2\right)
$$
qui permet de r\'ecrire (\ref{eq:derphiprimell})(\ref{eq:derphiprimey})
sous la forme
$$
\delta\psi_{\ell}=-\frac12\left(1+\frac{a}{\beta}
+\tanh\upsilon\,{\cal {K}}(\mu,\upsilon)\right)\beta\gamma_0^2+
\bigg({\cal {L}}(\mu,\upsilon)+\tanh{\upsilon}
\coth{\mu}\,{\cal {K}}(\mu,\upsilon)\bigg)\frac{\partial\mu}{\partial\ell},
$$
$$
\delta\psi_{y}=-\frac12\left(1+\frac{a}{\beta}
+\tanh\upsilon\,{\cal {K}}(\mu,\upsilon)\right)\beta\gamma_0^2+
\bigg({\cal {L}}(\mu,\upsilon)+\tanh{\upsilon}
\coth{\mu}\,{\cal {K}}(\mu,\upsilon)\bigg)\frac{\partial\mu}{\partial y}.
$$
Un calcul explicite o\`u on utilise (\ref{eq:ratiomunuTH}) donne
\beq\label{eq:dermul}
\frac{\partial\mu}{\partial\ell}=-\frac12\beta\gamma_0^2
\left(\frac{2y+\lambda}{\ell+y+\lambda}+
\frac{\sinh^3\upsilon}{\sinh^2\mu\cosh\upsilon}\right)Q(\mu,\upsilon)\cosh\upsilon,
\eeq
\beq\label{eq:dermuy}
\frac{\partial\mu}{\partial y}=\frac12\beta\gamma_0^2
\left(\frac{2\ell+\lambda}{\ell+y+\lambda}-
\frac{\sinh^3\upsilon}{\sinh^2\mu\cosh\upsilon}\right)Q(\mu,\upsilon)\cosh\upsilon.
\eeq
On peut maintenant r\'eexprimer (\ref{eq:dermul}) et (\ref{eq:dermuy})
sous la forme
$$
\frac{2y+\lambda}{\ell+y+\lambda}=1+\frac{y-\ell}{\ell+y+\lambda}=
1+\frac{(\sinh2\mu-2\mu)-(\sinh2\upsilon-2\upsilon)}{2\sinh^2\mu},
$$
$$
\frac{2\ell+\lambda}{\ell+y+\lambda}=1-\frac{y-\ell}{\ell+y+\lambda}=
1-\frac{(\sinh2\mu-2\mu)-(\sinh2\upsilon-2\upsilon)}{2\sinh^2\mu},
$$
o\`u on a utilis\'e (\ref{eq:ratiomunuTH}); on obtient ainsi
\beq
\frac{\partial\mu}{\partial\ell}=-\frac12\beta\gamma_0^2\,g_+(\mu,\upsilon)
Q(\mu,\upsilon)\cosh\upsilon,\quad \frac{\partial\mu}{\partial y}=\frac12\beta\gamma_0^2\,g_-(\mu,\upsilon)Q(\mu,\upsilon)\cosh\upsilon,
\eeq
avec
$$
g_+(\mu,\upsilon)=1\!+\!\frac{(\sinh2\mu\!-\!2\mu)\!-\!(\sinh2\upsilon-2\upsilon)}
{2\sinh^2\mu}\!+\!\frac{\sinh^3\upsilon}{\sinh^2\mu\cosh\upsilon}
$$
et
$$
g_-(\mu,\upsilon)=1\!-\!\frac{(\sinh2\mu\!-\!2\mu)\!-
\!(\sinh2\upsilon\!-\!2\upsilon)}{2\sinh^2\mu}
\!-\!\frac{\sinh^3\upsilon}{\sinh^2\mu\cosh\upsilon}.
$$
On d\'eveloppe les deux derni\`eres expressions et on a respectivement
$$
g_+(\mu,\upsilon)=\frac{\sinh\mu\cosh\upsilon(\sinh\mu+\cosh\mu)-
(\mu-\upsilon)\cosh\upsilon-\sinh\upsilon}{\sinh^2\mu\cosh\upsilon},
$$
$$
g_-(\mu,\upsilon)=\frac{\sinh\mu\cosh\upsilon(\sinh\mu-\cosh\mu)+
(\mu-\upsilon)\cosh\upsilon+\sinh\upsilon}{\sinh^2\mu\cosh\upsilon}
$$
puis, un peu d'alg\`ebre permet de les mettre sous la forme commode
$$
g_+(\mu,\upsilon)Q(\mu,\upsilon)\cosh\upsilon=1+e^{\mu}\widetilde{Q}(\mu,\upsilon),
\quad
g_-(\mu,\upsilon)Q(\mu,\upsilon)\cosh\upsilon=1+e^{-\mu}\widetilde{Q}(\mu,\upsilon)
$$
avec
$$
\widetilde{Q}(\mu,\upsilon)=\frac{\cosh\mu\sinh\mu\cosh\upsilon-
(\mu-\upsilon)\cosh\upsilon-\sinh\upsilon}
{(\mu-\upsilon)\cosh\mu\cosh\upsilon+\cosh\mu\sinh\upsilon-\sinh\mu\cosh\upsilon}.
$$.
On peut donc r\'eexprimer (\ref{eq:dermul}) et (\ref{eq:dermuy})
sous la forme simple
\beq\label{eq:dermul1TH}
\frac{\partial\mu}{\partial\ell}=-\frac12\beta\gamma_0^2
\left[1+e^{\mu}\widetilde{Q}(\mu,\upsilon)\right],\qquad
\frac{\partial\mu}{\partial y}=\frac12\beta\gamma_0^2
\left[1+e^{-\mu}\widetilde{Q}(\mu,\upsilon)\right]
\eeq
dont on donne l'allure des d\'eriv\'ees dans la Fig.\ref{fig:dermunuelly}.
Les expression MLLA pour les d\'eriv\'ees se simplifient aux expressions
\beeq\nonumber
\psi_{\ell}(\mu,\upsilon)\!\!&\!\!=\!\!&\!\!\gamma_0e^{\mu}\!+\!
\frac12a\gamma_0^2\Big[e^{\mu}\widetilde{Q}(\mu,\upsilon)\!-\!\tanh\upsilon
\!-\!\tanh\upsilon\coth\mu\Big(1\!+\!e^{\mu}\widetilde{Q}(\mu,\upsilon)\Big)\Big]\\
&&-\frac12\beta\gamma_0^2\Big[1\!+\!\tanh\upsilon
\Big(1\!+\!K(\mu,\upsilon)\Big)+C(\mu,\upsilon)
\Big(1\!+\!e^{\mu}\widetilde{Q}(\mu,\upsilon)\Big)\Big]\label{eq:derpsi'lTH},
\eeeq
\beeq\nonumber
\psi_{y}(\mu,\upsilon)\!\!&\!\!=\!\!&\!\!\gamma_0e^{-\mu}\!-\!\frac12a\gamma_0^2
\Big[2\!+\!e^{-\mu}\widetilde{Q}(\mu,\upsilon)\!+\!\tanh\upsilon
\!-\!\tanh\upsilon\coth\mu\Big(1\!+\!e^{-\mu}\widetilde{Q}(\mu,\upsilon)\Big)\Big]\\
&&-\frac12\beta\gamma_0^2\Big[1\!+\!\tanh\upsilon
\Big(1\!+\!K(\mu,\upsilon)\Big)-C(\mu,\upsilon)
\Big(1\!+\!e^{-\mu}\widetilde{Q}(\mu,\upsilon)\Big)\Big]\label{eq:derpsi'yTH}
\eeeq
o\`u nous avons d\'efini
$$
C(\mu,\upsilon)=L(\mu,\upsilon)+
\tanh\upsilon\coth\mu\Big(1 + K(\mu,\upsilon)\Big).
$$
En effect, bien que $L(\mu,\upsilon)$ diverge lorsque
$\mu,\upsilon\rightarrow0$, nous
trouvons une complicit\'e au sein de cette expression. C'est pour cette
raison que nous l'avons ainsi d\'efinie. On a en effet
$$
\lim_{\mu,\upsilon\rightarrow0}\left[L(\mu,\upsilon)+
\tanh\upsilon\coth\mu K(\mu,\upsilon)\right]=\lim_{\mu,\upsilon\rightarrow0}
\frac{2-3\frac{\upsilon^2}{\mu^2}-\frac{\upsilon^3}{\mu^3}}{4\left(1-\frac{\upsilon^3}
{\mu^3}\right)}\mu=0;
$$
de m\^eme
$$
\lim_{\mu,\upsilon\rightarrow0}
\tanh\upsilon\coth\mu\left(1+e^{\pm\mu}\widetilde{Q}(\mu,\upsilon)\right)=
\lim_{\mu,\upsilon\rightarrow0}=\frac{3\frac{\upsilon}{\mu}}{1-\frac{\upsilon^3}
{\mu^3}}=\frac{3\sqrt{\frac{\lambda}{Y+\lambda}}}{1-
\left(\frac{\lambda}{Y+\lambda}\right)^{3/2}}.
$$
Nous avons alors d\'emontr\'e
que (\ref{eq:derpsi'lTH}) et (\ref{eq:derpsi'yTH}), en plus d'\^etre
stables dans la limite $\lambda\to0$, sont 
parfaitement r\'eguli\`eres et peuvent \^etre utilis\'ees,
par exemple, dans l'estimation des corr\'elations entre deux
particules dans un jet.

\section{V\'erification des \'equations (\ref{eq:SpectreDLA}) et
 (\ref{eq:solgg}) par la solution du col}

Nous allons simultan\'ement v\'erifier que les \'equations d'\'evolution
DLA (\ref{eq:SpectreDLA}) et MLLA (\ref{eq:solgg}) sont satisfaites
par (\ref{eq:Specalphasrun}) et (\ref{eq:SpecalphasrunmllaTH}) respectivement.
Nous mettons l'\'equation (\ref{eq:solgg}) sous la forme diff\'erentielle
$$
G_{\ell y} = \gamma_0^2 \left(G - a G_\ell\right) \!+\! {\cal {O}}{\big(\gamma_0^4G\big)}
$$
que l'on peut r\'ecrire en fonction des d\'eriv\'ees logarithmiques
\beq\label{eq:MLLApsiTH}
 \psi_\ell\psi_y +\psi_{\ell y} \>=\>
 \gamma_0^2\left(1-a\psi_\ell\right)  \>+\> {\cal {O}}\big({\gamma_0^4}\big),
\eeq
nous avons n\'eglig\'e les corrections next-to-MLLA (NMLLA)
${\cal {O}}{\big(\gamma_0^4\big)}$ (d'ordre relatif $\gamma_0^2$) qui
d\'ecoulent de la diff\'erentiation de la constante anormale $\gamma_0^2$ 
dans le terme sous-dominant $\propto a$.

Nous devons par contre nous assurer
que \eqref{eq:MLLApsiTH} est bien v\'erifi\'ee en incluant les termes
${\cal {O}}\big({\gamma_0^3}\big)$.

Dans les termes sous-dominants nous pouvons poser
$\psi\to \phi$ (voir eqs.\ref{eq:phipsi}
et \ref{eq:phibar}):
\beq
  (\phi_\ell + \delta\psi_\ell)(\phi_y + \delta\psi_y) + \phi_{\ell
  y} = \gamma_0^2(1-a\phi_\ell).
\eeq
On s\'electionne les termes correctifs et on les met \`a droite de
l'\'equation
\beq\label{eq:collectTH}
  a\gamma_0^2\phi_\ell 
+ \left[\,\phi_\ell \delta\psi_y + \phi_y\delta\psi_\ell\,\right] + \phi_{\ell
  y} \>=\>  \gamma_0^2 -\phi_\ell\phi_y.
\eeq
Puis, par d\'efinition du point de col
$$
  \phi_\ell = \omega_0=\gamma_0e^\mu,\quad \phi_y = \nu_0=\gamma_0e^{-\mu}\,,
$$
on en d\'eduit que le membre droit de l'\'equation est nul
et que l'on doit avoir
\beq\label{eq:hastobeTH}
   \omega_0 a\gamma_0^2 + \left[\,\omega_0 \delta\psi_y +
     \nu_0\delta\psi_\ell\,\right] + \frac{d\omega_0}{dy} \>=\> 0\,,
\eeq
ou de fa\c con \'equivalente,
\beq
   \omega_0 \left(a\gamma_0^2 + \delta\psi_y\right) +
     \nu_0\delta\psi_\ell + \frac{d\omega_0}{dy} \>=\> 0\,.
\eeq
On collecte d'abord les termes $\propto a$:
\begin{eqnarray*}
&&a\gamma_0^3\left[e^{\mu}-e^{\mu}-\frac12\widetilde{Q}-\frac12\tanh\upsilon\, e^{\mu}
+\frac12\tanh\upsilon\coth\mu\, e^{\mu} + \frac12\tanh\upsilon\coth\mu \,
\widetilde{Q}\right.\\
&&\left.+\frac12\widetilde{Q}-\frac12\tanh\upsilon\,e^{-\mu}-\frac12
\tanh\upsilon\coth\mu\, e^{-\mu}-\frac12\tanh\upsilon\coth\mu \,
\widetilde{Q}\right]\\
&&=a\gamma_0^3\left[-\tanh\upsilon\cosh\mu+\tanh\upsilon\coth\mu\sinh\mu\right]
\equiv0
\end{eqnarray*}
il reste alors de faire la m\^eme chose pour les termes $\propto\beta$
$$
\frac{d\omega_0}{dy}=\frac12\beta\gamma_0^3\widetilde{Q},
$$
\begin{eqnarray*}
&&-\beta\gamma_0^3\left[\frac12e^{\mu}+\frac12\tanh\upsilon
\Big(1\!+\!K\Big)e^{\mu}-\frac12C\,e^{\mu}-\frac12C\widetilde{Q}+
\frac12e^{-\mu}+\frac12\tanh\upsilon
\Big(1\!+\!K\Big)e^{-\mu}\right.\\
&&\left.+\frac12C\,e^{-\mu}+\frac12C\widetilde{Q}\right]
=-\beta\gamma_0^3\left[\cosh\mu+\tanh\upsilon\cosh\mu\Big(1\!+\!K\Big)
-C\sinh\mu-\frac12\widetilde{Q}\right]
\end{eqnarray*}
et on obtient
\begin{eqnarray*}
&&-\beta\gamma_0^3\left[\cosh\mu-\sinh\mu\,L-\frac12\widetilde{Q}+
\tanh\upsilon\cosh\mu\Big(1\!+\!K\Big)-\tanh\upsilon\cosh\mu\Big(1\!+\!K\Big)\right],
\end{eqnarray*}
les deux derniers termes se simplifient. On construit maintenant
\begin{eqnarray*}
\widetilde{Q}(\mu,\upsilon)-2\cosh\mu\!\!&\!\!=\!\!&\!\!
-3\cosh\mu+\sinh\mu\frac{(\mu-\upsilon)\cosh\upsilon\sinh\mu+\sinh\upsilon\sinh\mu}
{(\mu-\upsilon)\cosh\mu\cosh\upsilon+\cosh\mu\sinh\upsilon-\sinh\mu\cosh\upsilon}\\
\!\!&\!\!=\!\!&\!\!-2\sinh\mu L(\mu,\upsilon)
\end{eqnarray*}
et finalement
$$
-\beta\gamma_0^3\left[\cosh\mu-\sinh\mu\,L-\frac12\widetilde{Q}\right]\equiv0.
$$
On a ainsi simultan\'ement v\'erifi\'e que nos solutions DLA
(\ref{eq:SpecNorm}) ($a=0$, $\beta$) et MLLA (\ref{eq:SpecNormMLLATH})
($a\ne0$, $\beta$) trouve\'es \`a partir de la m\'ethode du col 
satisfont l'\'equation d'\'evolution dans l'approximation o\`u
on n\'eglige les puissance de $\gamma_0$ sup\'erieure \`a $3$, 
${\cal {O}}(\gamma_0^4)$.

%\vskip 0.5cm

\subsection{Application au cas des corr\'elations entre deux particules}

%\vskip 0.5cm

Nous donnons le calcul d\'etaill\'e de la fonction $\Delta'$
(voir eq.(37) de \ref{sub:article3})
pr\`es du maximum de la distribution inclusive. Nous avons
$$
\lim_{\mu,\upsilon\rightarrow0}C(\mu,\upsilon)=\left(\frac{\lambda}{Y+\lambda}\right)^{1/2}
,\qquad\lim_{\mu,\upsilon\rightarrow0}K_i=\frac32\frac{\upsilon^2_i}{\mu^3_i-\upsilon^3_i};
$$
on ne garde que les termes lin\'eaires en $\mu$
\begin{eqnarray*}
\Delta'&\stackrel{\ell_1\sim\ell_2\simeq Y/2}{\simeq}&-a\gamma_0\left[2+\mu_1+\mu_2+
\left(\frac{\lambda}{Y+\lambda}\right)^{1/2}\!\!\!(\mu_1+\mu_2)-
\left(\frac{\lambda}{Y+\lambda}\right)^{1/2}\!\!\!(\mu_1+\mu_2)\right]\\
&&-\beta\gamma_0\left[2-
\left(\frac{\lambda}{Y+\lambda}\right)^{1/2}\!\!\!(\mu_1+\mu_2)+
\left(\frac{\lambda}{Y+\lambda}\right)^{1/2}\!\!\!(\mu_1+\mu_2)
+3\frac{\left(\frac{\lambda}{Y+\lambda}\right)^{3/2}}{1-
\left(\frac{\lambda}{Y+\lambda}\right)^{3/2}}\right].
\end{eqnarray*}
Nous obtenons finalement:
\beeq
\Delta'&\stackrel{\ell_1\sim\ell_2\simeq Y/2}{\simeq}&-a\gamma_0\left[2+\mu_1+\mu_2\right]-\beta\gamma_0\left[2
+3\frac{\left(\frac{\lambda}{Y+\lambda}\right)^{3/2}}{1-
\left(\frac{\lambda}{Y+\lambda}\right)^{3/2}}\right]\nonumber\\
&&=-a\gamma_0\left[2+\mu_1+\mu_2\right]-\beta\gamma_0\left[2
+3\frac{\lambda^{3/2}}{(Y+\lambda)^{3/2}-\lambda^{3/2}}\right].
\eeeq
Nous avons en effet n\'eglig\'e les termes quadratiques.

%\vskip 0.5cm

\subsection{R\'esultats de la m\'ethode du col}

%\vskip 0.5cm

Cette m\'ethode nous a donc permis de donner l'expression analytique du 
spectre et ainsi, celles des corr\'elations,
\'etendues au cas $Q_0\ne\Lambda_{QCD}$.
Nous avons de m\^eme g\'en\'eralis\'e l'approche des corr\'elations
propos\'ee par Fong et Webber au cas g\'en\'eral $\lambda\ne0$.
La limite $\lambda=0$ nous a permis de v\'erifier la compatibilit\'e
de nos calculs avec ceux qui les ont
pr\'ec\'ed\'es \cite{DLA}\cite{FW}.

%\vskip 0.5cm

\subsection{R\'esultats de \ref{sub:article3} pour les corr\'elations,}
$\boldsymbol{\lambda=0}$

%\vskip 0.5cm

\begin{figure}
\vbox{
\begin{center}
\includegraphics[height=6truecm,width=0.48\tw]{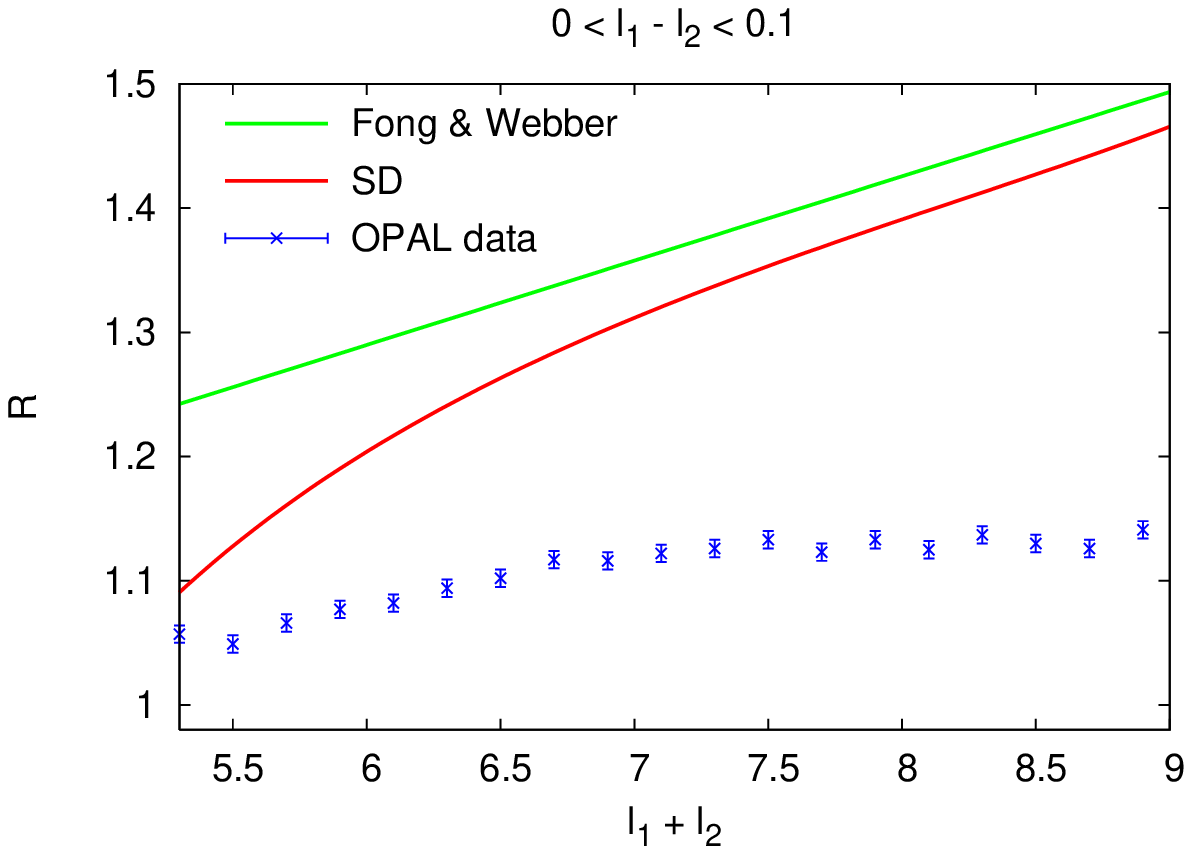}
\hfill
\includegraphics[height=6truecm,width=0.48\tw]{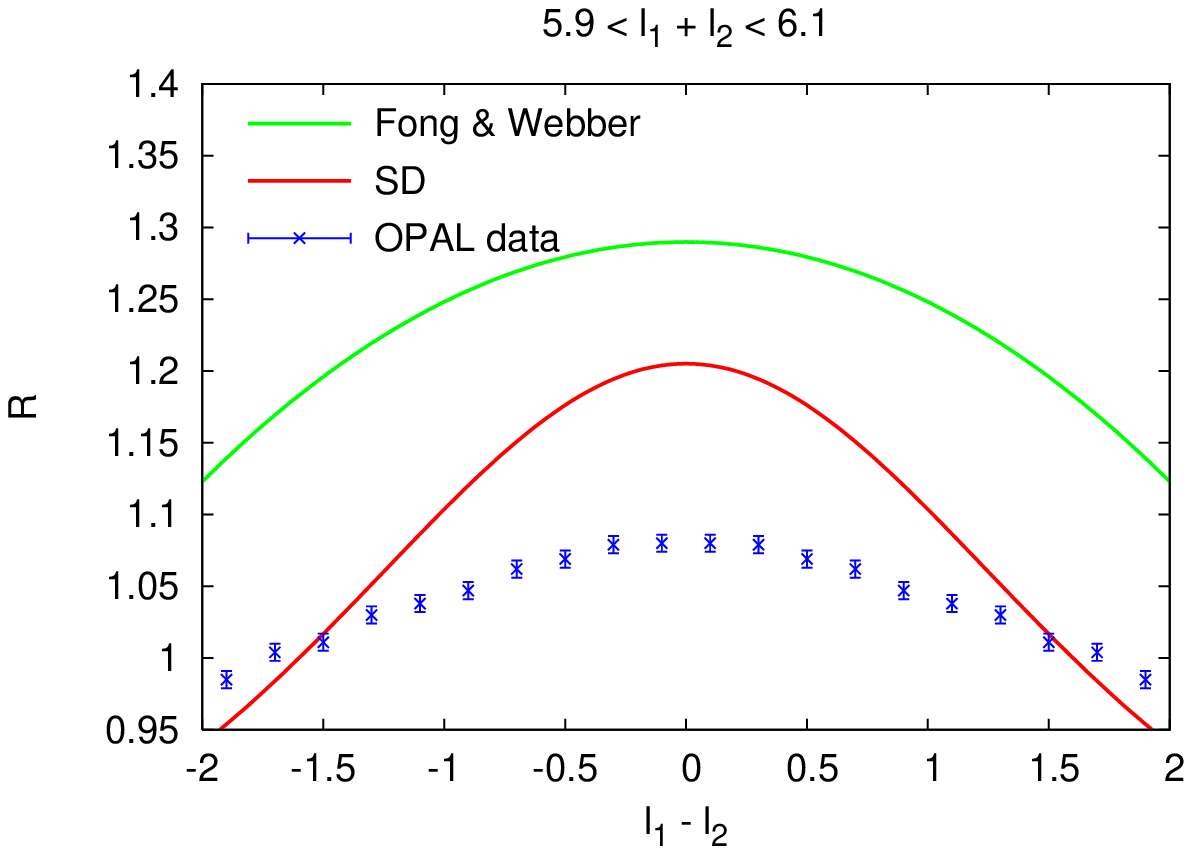}
\vskip .5cm
\includegraphics[height=6truecm,width=0.48\tw]{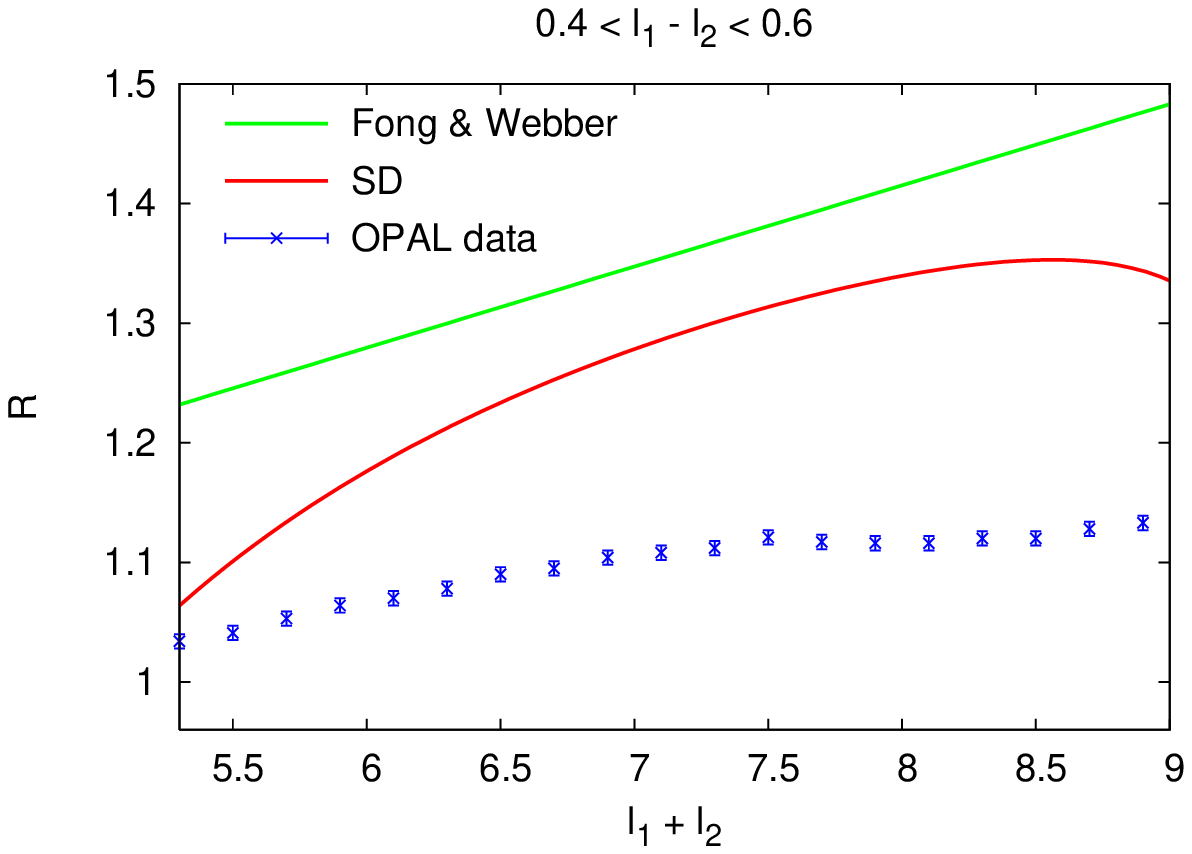}
\hfill
\includegraphics[height=6truecm,width=0.48\tw]{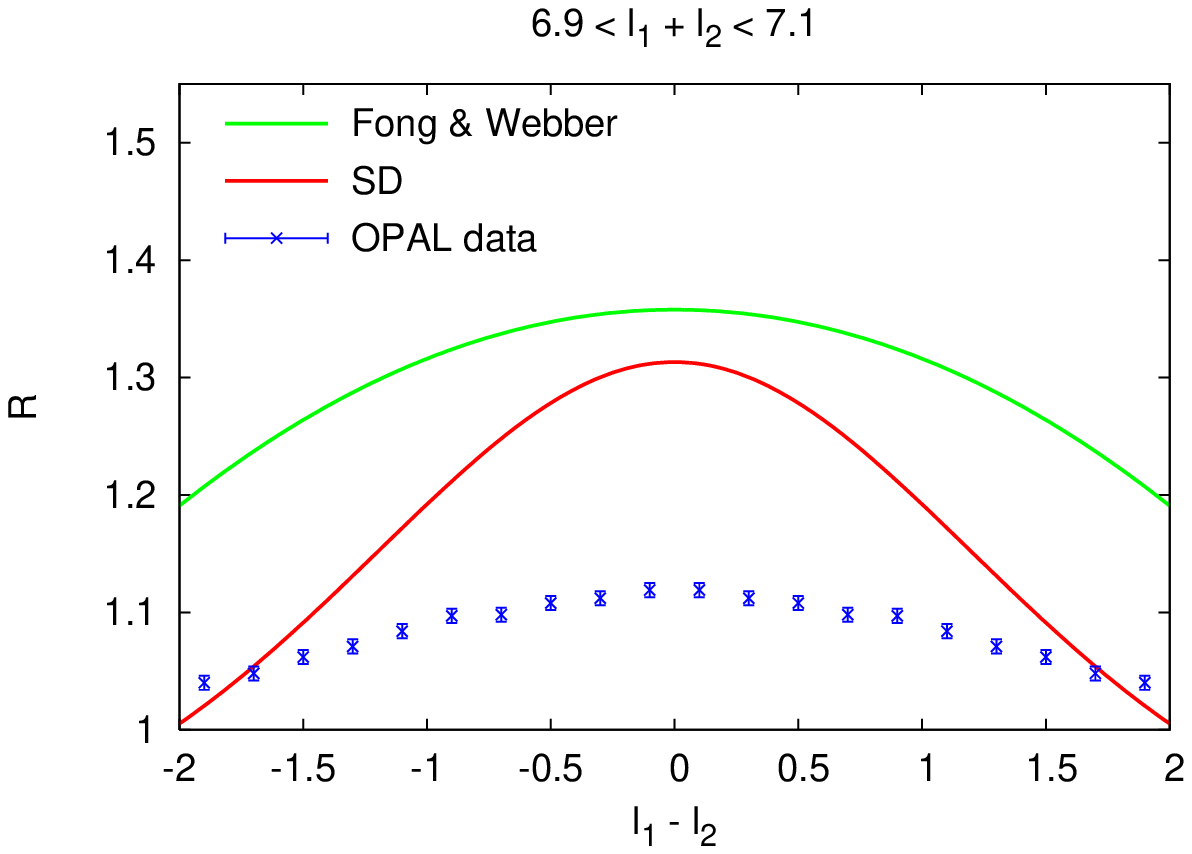}
\vskip .5cm
\includegraphics[height=6truecm,width=0.48\tw]{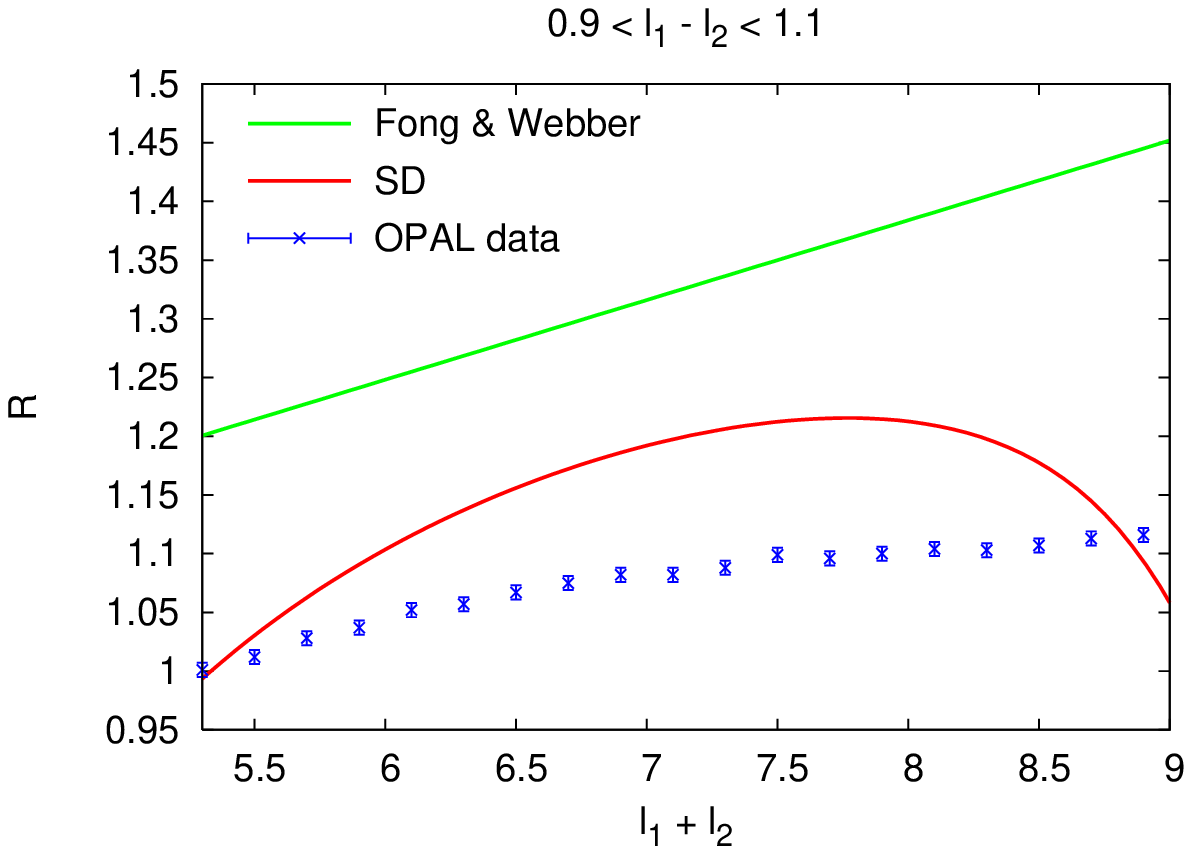}
\hfill
\includegraphics[height=6truecm,width=0.48\tw]{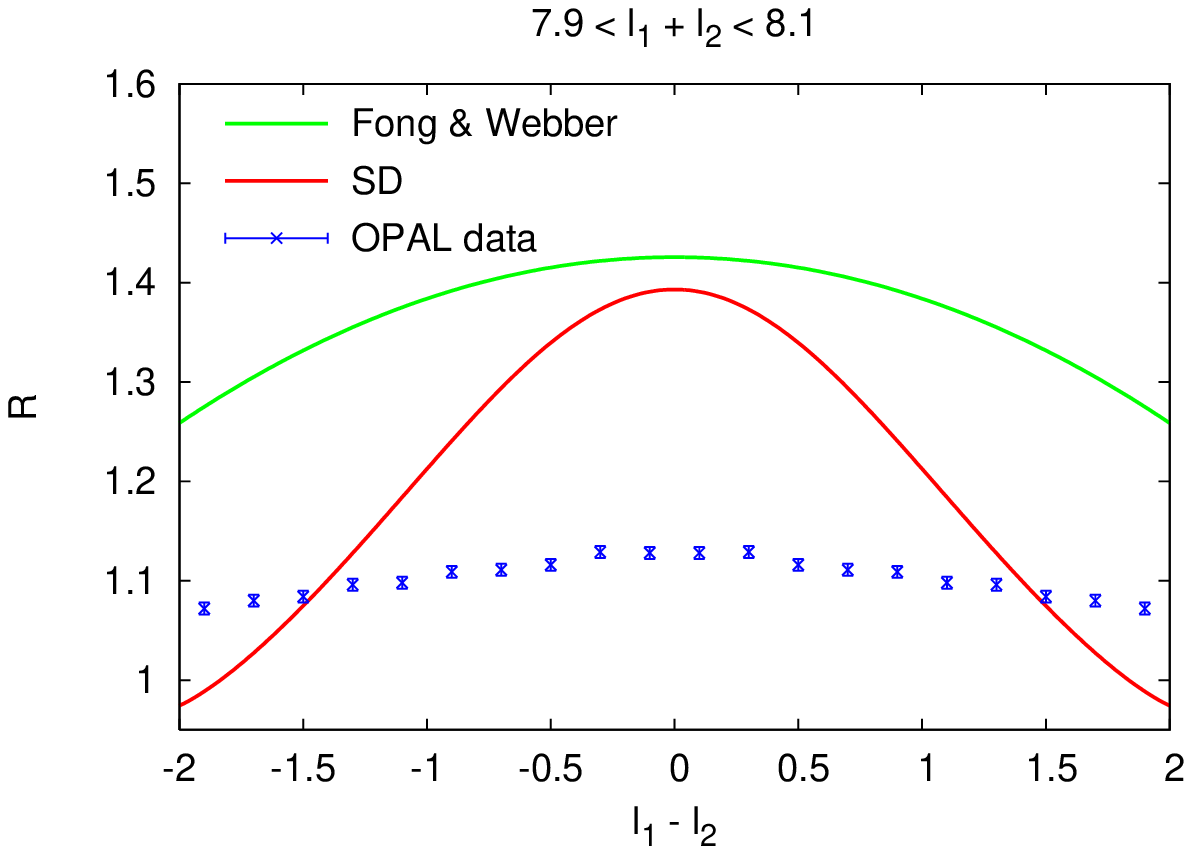}
\vskip .5cm
\caption{Corr\'elation $R$ entre 2  particules produites dans
 $e^+e^-\rightarrow q\bar{q}$, compar\'ee avec les donn\'ees de OPAL et
avec l'approximation de Fong \& Webber ($\lambda=0$)}
\label{fig:corrqqbarTH}
\end{center}
}
\end{figure}

Dans la Fig.\ref{fig:corrqqbarTH}, nous donnons nos pr\'edictions
pour les corr\'elations $R$ (\ref{eq:RR}) \`a partir de la m\'ethode du col
(pour $\lambda=0$);
on les compare avec les r\'esultats de Fong-Webber et les donn\'ees
exp\'erimentales du LEP-I au pic du $Z^0$, $Y=5.2$
($E\Theta=91.2\,\text{GeV}$).
Pour deux jets de quark, nous donnons l'expression des corr\'elations
\beq\label{eq:RR}
R=\frac12+\frac12{\cal C}_q
\eeq
o\`u ${\cal C}_q$ est la fonction de corr\'elation entre deux
particules dans un jet de quark; elle a \'et\'e calcul\'ee dans
\ref{sub:article3}. Le r\'esultat de la comparaison donn\'ee par cette figure
est tr\`es similaire \`a celui de l'article \ref{sub:article2}.
Donc, si les r\'esultats num\'eriques sont tr\`es similaires,
cette fa\c con d'estimer les corr\'elations s'av\`ere beaucoup plus
\'economique et rapide que celle de l'article \ref{sub:article2}.

%%%%%%%%%%%%%%%%%%%%%%%%%%%%%%%%%%%%%%%%%%%%%%%%%%%%%%%%%%%%%%%%%%%%%%%%%%%%
%\newpage

%\section{Conclusions}

\chapter{Conclusions}

Apr\`es avoir donn\'e les explications de base qui facilitent la
compr\'ehension des trois articles  \ref{sub:article1}, \ref{sub:article2} et
\ref{sub:article3}, nous  concluons  l'ensemble de ce travail.

Dans l'article \ref{sub:article1}, j'ai obtenu l'expression de
la section efficace inclusive doublement diff\'erentielle
valable pour toute valeur de la fraction d'\'energie $x$ de la particule
observ\'ee. La solution exacte (pour $Q_0=\Lambda_{QCD}$) des \'equations
d'\'evolution MLLA a \'et\'e utilis\'ee pour calculer analytiquement
 cette observable ainsi que la section efficace inclusive diff\'erentielle
dans la limite des petits $x$. Les r\'esultats obtenus montrent
de grandes diff\'erences avec
ceux obtenus dans la cas (``na\"\i f'') o\`u l'on ne consid\`ere pas l'\'evolution
du jet entre son angle d'ouverture initiale $\Theta_0$ et l'angle
d'\'emission de la particule d\'etect\'ee. En particulier,
la positivit\'e de la distribution  est restaur\'ee dans l'intervalle
de validit\'e de l'approximation utilis\'ee.
Cette \'evolution est l'origine physique des corrections MLLA, que
l'on ne rencontre pas dans le cas DLA, elles croissent en fonction de $x$
et d\'ecroissent lorsque l'impulsion transverse du hadron sortant cro\^\i t.

Pour que l'approximation des petits $x$ reste valable, sa valeur minimale a
\'et\'e obtenue \`a partir de l'analyse des corrections,
au seuil d'\'energie des pr\'esents acc\'el\'erateurs:
$x\leq x_{max}\approx0.08$ ($\ell\geq\ell_{min}\approx2.5$)
qui est  en remarquable accord avec celle obtenue dans le cas des
corr\'elations (voir \ref{sub:article2}).
\`A $Y_{\Theta_0}=\ln(Q/Q_0)$ fix\'e, la borne inf\'erieure
sur $\ell$ est remplac\'ee par une borne sup\'erieure sur $y$ ou
$k_\perp$, soit $y\leq y_{max}\approx Y_{\Theta_0}-\ell_{min}$. Ainsi,
l'approximation sera valable dans le domaine des petits $k_\perp$. 

D'un autre cot\'e, pour  garantir la convergence de la s\'erie
perturbative, la constante de couplage ne doit pas devenir  sup\'erieure \`a
1.  La valeur minimale de $y$ a ainsi \'et\'e fix\'ee \`a $1$,
ce qui correspond \`a $y\geq y_{min}\approx1$, soit
${k_\perp}_{min}\gtrsim0.8$ GeV.

Ceci a \'et\'e ainsi confirm\'e par l'analyse (voir Fig.\ref{fig:sergo})
de CDF. L'accord est excellent avec nos pr\'edictions dans  l'intervalle
de validit\'e mentionn\'e, tandis que 
plus de particules que la nature en produit sont pr\'edites
 \`a grand $k_\perp$.
La distribution de gluons mous obtenue a \'et\'e multipli\'ee
par ${K}^{ch}\approx0.56$, ce qui l'a ainsi  adapt\'ee
\`a la distribution observ\'ee des particules charg\'ees,
qui ne repr\'esentent que 60\% du nombre total des particules produites.
Une fois de plus, l'hypoth\`ese ``LPHD'' n'est pas remise en question dans
ce cas d'une variable inclusive.

La coh\'erence des gluons mous se trouve parmi les ph\'enom\`enes les plus
r\'eguli\`erement mentionn\'es \`a propos des observables inclusives
des jets. Dans le cas des distributions inclusives qui ont \'et\'e
calcul\'ees en fonction de l'impulsion transverse ($k_\perp$) (article
\ref{sub:article1}), elle est \'ecrant\'ee par la divergence de la constante
de couplage \`a petit $k_\perp$. Par cons\'equent, la forme des
distributions MLLA est  diff\`erente de celle obtenue en DLA,
 o\`u l'on ne consid\`ere pas la variation de $\alpha_s$.
C'est aussi pour cel\`a que, en augmentant l'\'energie totale du jet
($\alpha_s$ diminue alors),  l'apparition d'un
maximum dans la distribution a \'et\'e montr\'ee,
suivie d'une d\'ecroissance \`a petit $k_\perp$ associ\'ee, elle,
\`a l'interf\'erence des gluons mous dans cette r\'egion de l'espace de phase.

Pour am\'eliorer ces pr\'edictions en CDQ perturbative,
un calcul exact de la distribution (3.11) dans \ref{sub:article1}, ainsi que
son extension au del\`a du ``limiting spectrum'' sont
n\'ecessaires. Ceci permettra d'\'etendre l'intervalle de validit\'e
de notre approximation ainsi que de pr\'edire la forme de la distribution
 \`a petit $k_\perp$.

Dans l'article \ref{sub:article2} j'ai calcul\'e les corr\'elations
entre deux particules dans un jet. Pour la premi\`ere
fois, la d\'emonstration permettant d'obtenir les \'equations d'\'evolution
MLLA aussi bien dans le cas du spectre que dans celui des corr\'elations a
\'et\'e donn\'ee.
Ont \'et\'e pris en consid\'eration: la contrainte angulaire dans
l'\'emission des gluons mous ({\em Angular Ordering}) et
la conservation de l'\'energie et les effets associ\'es
\`a l'\'evolution de la constante de couplage. Les \'equations
d'\'evolution MLLA des corr\'elations ont \'et\'e r\'esolues de fa\c con
it\'erative. Ceci  a permis, en particulier, de g\'en\'eraliser
les r\'esultats de Fong \& Webber \cite{FW}, qui n'\'etaient
valables qu'au voisinage du maximum de la distribution inclusive des partons.
L'intervalle (qui doit donc \^etre exclus)
 dans lequel les corr\'elations deviennent n\'egatives (${\cal C}-1<0$) a
\'et\'e donn\'e; ceci a lieu quand l'\'energie de l'un des partons devient
 tr\`es sup\'erieure \`a l'\'energie de l'autre, autrement dit, quand
l'espace de phase de la particule la plus molle est r\'eduit.
En m\^eme temps, les corr\'elations s'annulent
(${\cal C}\to1$) lorsque l'un des partons devient tr\`es mou
($\ell\equiv\ln(1/x)\to Y= \ln E\Theta/Q_0$). Ceci s'explique de fa\c con
analogue \`a ce qui se passe dans
le cas des distributions inclusives \`a petit $k_\perp$,
\`a savoir, \`a partir de la coh\'erence des gluons mous.
Du point de vue qualitatif, les pr\'edictions obtenues sont en meilleur accord
avec les donn\'ees du LEP-I qu'avec celles de
Fong \& Webber \cite{FW}. Cependant, il y reste encore une diff\'erence
notoire, d'autant plus marqu\'ee \`a petit $x$. Dans cette r\'egion,
les effets non-perturbatifs pourraient se  faire sentir de
fa\c con non n\'egligeable, et limiter l'application de l'hypoth\`ese
``LPHD'' au cas de cette observable moins inclusive que les distributions
\'etudi\'ees dans \ref{sub:article1}.
Les donn\'ees exp\'erimentales de CDF (\`a venir) seront  compar\'ees aux
r\'esultats de ce travail, et pourront \'eclairer cette question.

Dans l'article \ref{sub:article3} je me suis int\'eress\'e
\`a l'\'evaluation du spectre inclusif d'une particule par la m\'ethode du col.
Je me suis inspir\'e des travaux de Dokshitzer, Fadin, Khoze et Troyan
\cite{DLA}\cite{DFK} en DLA ({\em Leading Order}) et je les ai
g\'en\'eralis\'es au cas MLLA ({\em{Next-to-Leading Order}}).
Ceci m'a permis de donner, en particulier,  l'expression asymptotique
du spectre \`a $\lambda\ne0$ ($Q_0\ne\Lambda_{QCD}$) en MLLA, et de
retrouver \`a $\lambda=0$ la position de son maximum ainsi que 
sa forme gaussienne au voisinage de ce point.
Les deriv\'ees logarithmiques, importantes pour le calcul post\'erieur des
corr\'elations, \'etant finies dans la limite $\lambda\to0$, peuvent
\^etre compar\'ees avec leur expression exacte
(voir D.2 dans \ref{sub:article2}).
Bien que cette m\'ethode (approch\'ee) n'est valable que, strictement
parl\'e, dans la limite asymptotique $Y+\lambda,\lambda\gg1$,
l'accord obtenu entre  cette m\'ethode et l'expression exacte du travail
pr\'ec\'edent est remarquable, m\^eme \`a l'\'echelle d'\'energie du LEP-I
($Q=91.2$ GeV). C'est pour cel\`a que je l'ai utilis\'ee
 pour \'evaluer les corr\'elations d'une fa\c con plus rapide et
\'economique. Elle a permis, en outre, de faire des pr\'edictions
pour celles ci  \`a $\lambda\ne0$, ce qui \'etait irr\'ealisable
avec la m\'ethode exacte pr\'ec\'edente, pour des raisons techniques.
J'ai ainsi pu d\'emontrer que, si l'on augmente $Q_0$, on r\'eduit l'espace
de phase disponible pour les gluons mous et les corr\'elations augmentent.
Ceci a permis, en particulier, de confirmer quantitativement que
le ``limiting spectrum'' reste le meilleur candidat pour la
description des donn\'ees exp\'erimentales, celles ci demeurant pour le
moment au dessus des pr\'edictions.

S'il s'av\`ere que les donn\'ees exp\'erimentales \`a venir laissent
subsister un d\'esaccord avec nos pr\'edictions th\'eoriques MLLA,
des calculs NMLLA seraient \`a envisager, conjointement avec des
interrogations sur la validit\'e de l'hypoth\`ese LPHD.

Enfin, une continuation naturelle des m\^emes techniques
pr\'esent\'ees dans cette th\`ese concerne le probl\`eme ``KNO'' 
\cite{DLA}\cite{EvEq}
(Koba, Nielsen, Olesen) sur la loi de ``scaling''
des fluctuations des multiplicit\'es dans les jets. Jusqu'\`a pr\'esent,
aucun calcul th\'eorique n'a \'et\'e \`a m\^eme de tenir compte de
l'\'evolution du jet, c'est \`a dire des effets de variation de
$\alpha_s$, ni des ``corrections fortes'' qui apparaissent lorsque
l'on satisfait le principe de conservation de l'\'energie.

%%%%%%%%%%%%%%%%%%%%%%%%%%%%%%%%%%%%%%%%%%%%%%%%%%%%%%%%%%%%%%%%%%%%%%%%%%%%
%%%%%%%%%%%%%%%%%%%%%%%%%%%%%%%%%%%%%%%%%%%%%%%%%%%%%%%%%%%%%%%%%%%%%%%%%%%%

%\newpage

%\appendix

\chapter{Appendices}

%{\bf\Large APPENDICES}

%%%%%%%%%%%%%%%%%%%%%%%%%%%%%%%%%%%%%%%%%%%%%%%%%%%%%%%%%%%%%%%%%%%%%%%%%%%%

\section{Rayonnement en \'electrodynamique classique et en
chromodynamique quantique}
\label{section:app1}

\subsection{Calcul concernant \ref{subsubsection:4emeaccpfinie}}
\label{subsection:4emeacceppropefinie}

On pose:
\begin{equation*}
e^{a_0\tau'}=\sinh{a_0\tau'}+\cosh{a_0\tau'}
\end{equation*}
\begin{equation*}
e^{-a_0\tau'}=\cosh{a_0\tau'}-\sinh{a_0\tau'},
\end{equation*}
et on r\'ecrit
\begin{equation*}
\begin{split}
&\left[\cosh{y}+\frac{1}{2}\left(v_1-v_2\right)\sinh{y}\right]
\sinh{a_0\tau}-\frac{1}{2}\left(v_1+v_2\right)
\sinh{y}\cosh{a_0\tau'}\\
&=A\sinh{a_0\tau'}+B\cosh{a_0\tau'}=D\sinh\left(a_0\tau'+\chi\right),
\end{split}
\end{equation*}
o\`u:
$$
D^2=\displaystyle{\frac{\cosh\left(\eta_2-y\right)
\cosh\left(\eta_1-y\right)}{\cosh{\eta_2}\cosh{\eta_1}}}=
\displaystyle{\frac{v_1+v_2}{\tanh\left(\eta_2-y\right)
-\tanh\left(\eta_1-y\right)}}
$$

$$
\chi=\displaystyle{\tanh^{-1}\frac{v_1+v_2}{v_2-v_1-2\coth{y}}}.
$$
On d\'ecompose l'int\'egrand de la fa\c con suivante:
\\
\begin{equation*}
\begin{split}
&\left[\frac{1}{2}\left(v_2-v_1\right)\cosh{y}-\sinh{y}\right]
\cosh{a_0\tau'}+\frac{1}{2}\left(v_1+v_2\right)
\cosh{y}\sinh{a_0\tau'}=\\
&\widetilde{A}\left\{\left[\cosh{y}+\frac{1}{2}\left(v_1-v_2\right)
\sinh{y}\right]\sinh{a_0\tau'}-
\frac{1}{2}\left(v_1+v_2\right)\sinh{y}\cosh{a_0\tau'}\right\}+\\
&\frac{\widetilde{B}}{a_0}\frac{d}{d\tau'}\left\{\left[\cosh{y}+
\frac{1}{2}\left(v_1-v_2\right)\sinh{y}
\right]\sinh{a_0\tau'}-\frac{1}{2}\left(v_1+v_2\right)\sinh{y}
\cosh{a_0\tau'}\right\}.
\end{split}
\end{equation*}

Ici: $$
\widetilde{A}=\displaystyle\frac{{v_1+v_2}}{D^2}=
\tanh\left(\eta_2-y\right)-\tanh\left(\eta_1-y\right).$$
L'int\'egrale $\propto\widetilde{B}$ s'annule, celle qui est proportionnelle 
\`a $\widetilde{A}$ se calcule \`a partir de la r\'epr\'esentation int\'egrale
de la fonction de Bessel modifi\'ee de seconde esp\`ece.

\subsection{Calculs concernant \ref{subsection:coher} (Angular Ordering)}
\label{sub:Vangulaire}

$\ast$ On int\`egre (\ref{eq:Ri}) sur les angles d'\'emission pour trouver
la probabilit\'e totale ind\'ependante de rayonnement:

\beq
<{\cal {R}}_1>\equiv\int\frac{d\Omega}{4\pi}\frac{v_1^2\sin^2\Theta}{(1-v_1\cos\Theta)^2}=
\int_0^\pi\frac{v_1^2\sin^3\Theta\, d\Theta}{(1-v_1\cos\Theta)^2}\int_0^{2\pi}d\phi=
{\cal {I}}(v_1)
\eeq

qui peut s'int\'egrer facilement en effectuant le changement
de variables suivant

$$
u=\cos\Theta,\qquad du=-\sin\Theta d\Theta.
$$
On doit r\'esoudre

\beeq
{\cal {I}}(v_1)&=&\frac{v_1^2}2\left(\int_{-1}^1\frac{du}{(1-v_1u)^2}-\int_{-1}^1\frac{u^2\,du}
{(1-v_1u)^2}\right)=\frac1{v_1}\ln\frac{1+v_1}{1-v_1}-2\nonumber\\
&=&2\left(\frac{\eta}{\tanh\eta}-1\right);\quad \text{avec} \quad v=\tanh\eta\nonumber.
\eeeq

La deuxi\`eme int\'egrale dans l'expression pr\'ec\'edente se d\'ecompose en

\beq
\int_{-1}^1\frac{u^2\,du}
{(1-v_1u)^2}=\frac1{v_1^2}\left(2+\int_{-1}^1\frac{du}{(1-v_1u)^2}-2
\int_{-1}^1\frac{du}{(1-v_1u)}\right).\nonumber
\eeq

$\eta$ a le sens d'un ``angle hyperbolique'' en fonction duquel
on peut exprimer la quadri-impulsion de l'\'electron

$$
E_1=m_1\cosh\eta, \quad \mid\!\!\vec{p}_1\!\!\mid=m_1\sinh\eta.
$$

$\ast$ On peut v\'erifier l'invariance de Lorentz de la probabilit\'e
totale d'\'emission (\ref{eq:DL1})

$$
\left\{{\cal {R}}_1+{\cal {R}}_2-2{\cal {J}}\right\}\frac{d\omega}{\omega}d\Omega=
\left[-j^2_{\mu}\right]\frac{d^3k}{\omega}
$$

o\`u $\displaystyle{\frac{d^3k}{\omega}}=2\int d^4k\,\delta^4(k^2)$.

$\ast$ On calcule l'int\'egrale suivante dans la limite ultra-relativiste:

\beq\label{eq:moyJ}
\lim_{v_1,v_2\rightarrow1}<{\cal {J}}>=\int\frac{d\Omega_{\vec{n}}}{4\pi}\,\frac
{\vec{n}_1\vec{n}_2-(\vec{n}\vec{n}_1)(\vec{n}\vec{n}_2)}{(1-\vec{n}\vec{n}_1)(1-\vec{n}\vec{n}_2)}.
\eeq

$$
(\ref{eq:moyJ})=\int_{-1}^{1}\frac{d\cos\Theta_1}2\int_0^{2\pi}\frac{d\phi}{2\pi}\,\frac
{\vec{n}_1\vec{n}_2-(\vec{n}\vec{n}_1)(\vec{n}\vec{n}_2)}{(1-\vec{n}\vec{n}_1)(1-\vec{n}\vec{n}_2)}=
\frac12\int_{-1}^{1}dx<{\cal {J}}>_{\phi}\quad (\text{avec}\quad x=\cos\Theta_1).
$$

On effectue les d\'ecompositions

\beeq
\frac
{\vec{n}_1\vec{n}_2-(\vec{n}\vec{n}_1)(\vec{n}\vec{n}_2)}
{(1-\vec{n}\vec{n}_1)(1-\vec{n}\vec{n}_2)}&=&\frac{\vec{n}\vec{n}_1}{1-\vec{n}\vec{n}_1}+
\frac{\vec{n}_1\vec{n}_2-\vec{n}\vec{n}_1}{1-\vec{n}\vec{n}_1}\frac1{1-\vec{n}\vec{n}_2}
\nonumber\\
&=&\frac{x}{1-x}+\frac{\vec{n}_1\vec{n}_2-x}{1-x}\frac1{1-\vec{n}\vec{n}_2}\nonumber,
\eeeq

$$
\vec{n}\vec{n}_2=\cos\Theta_2=\vec{n}_z\cdot\vec{n}_{2z}+\vec{n}^{\perp}\cdot\vec{n}_2^{\perp}=
\cos\Theta_1\cos\Theta_{12}+\sin\Theta_1\sin\Theta_{12}\cos\phi.
$$

Or

\beeq
<{\cal {J}}>_{\phi}&=&\frac{x}{1-x}+\frac{\vec{n}_1\vec{n}_2-x}{1-x}
\int_0^{2\pi}\frac{d\phi}{a+b\cos\phi}\nonumber\\
&=&\frac1{1-x}\left\{1+(\vec{n}_1\vec{n}_2-x)\int_0^{2\pi}
\frac{d\phi_{\vec{n},\vec{n}_1}}{2\pi}\frac1{a_2}\right\}-1,
\eeeq

o\`u $a=1-\cos\Theta_1\cos\Theta_{12}$ et $b=1-\sin\Theta_1\sin\Theta_{12}$.
L'int\'egrale peut se calculer facilement

\beeq\nonumber
(\ref{eq:a2})&=&\int_0^{2\pi}\frac{d\phi_{\vec{n},\vec{n}_1}}{2\pi}\frac1{a_2}\equiv
\int_0^{2\pi}\frac{d\phi}{a+b\cos\phi}=\left[\frac2{\sqrt{a^2-b^2}}
\arctan\left(\sqrt{\frac{a-b}{a+b}}\tan\frac{\phi}2\right)\right]_0^{2\pi}\\
&=&\frac1{\mid\!\vec{n}_1\vec{n}_2-x\!\mid}\equiv\frac1{\mid\!a_{12}-a_1\!\mid}
\eeeq

de sorte que l'\'evaluation entra\^\i ne

$$
<{\cal {J}}>_{\phi}=\frac1{1-x}\left\{1+\frac{\vec{n}_1\vec{n}_2-x}
{\mid\!\vec{n}_1\vec{n}_2-x\!\mid}\right\}-1.
$$

En introduisant la fonction de Heaviside

$$
\vartheta({\vec{n}_1\vec{n}_2-x})=\frac12\left\{1+\frac{\vec{n}_1\vec{n}_2-x}
{\mid\!\vec{n}_1\vec{n}_2-x\!\mid}\right\},
$$
on r\'ecrit

$$
<{\cal {J}}>_{\phi}=2\frac{\vartheta({\vec{n}_1\vec{n}_2-x})}{1-x}-1
$$

et finalement

\beeq\nonumber
\lim_{v_1,v_2\rightarrow1}<{\cal {J}}>&=&\int_{-1}^{1}dx
\left[\frac{\vartheta(\vec{n}_1\vec{n}_2-x)}{1-x}-1/2\right]=\int_{-1}^{\vec{n}_1\vec{n}_2}
\frac{dx}{1-x}-1\nonumber\\
&=&\ln\frac2{1-\vec{n}_1\vec{n}_2}-1.
\eeeq

Puis on s'int\'eresse \`a la quantit\'e $<\!V_1\!>_{\text{azimuth}}=
<\!({\cal {R}}_1-{\cal {J}})\!>$
\footnote{$\vartheta(\vec{n}_1\vec{n}_2-x)=\vartheta[1-x-(1-\vec{n}_1\vec{n}_2)])=
\vartheta(a_1-a_{12})$, puis $1-\vartheta(a_1-a_{12})=\vartheta(a_{12}-a_1)$}:

$$
<\!V_1\!>_{\text{azimuth}}=\int_0^{2\pi}\frac{d\phi_{\vec{n},\vec{n}_1}}{2\pi}
V_1(\vec{n},\vec{n}_1;\vec{n}_2)=2\frac{1-\vartheta({\vec{n}_1\vec{n}_2-x})}{1-x}
\equiv\frac2{a_1}\vartheta(a_{12}-a_1),
$$

$$
<\!V_2\!>_{\text{azimuth}}=\frac2{a_2}\vartheta(a_{12}-a_2).
$$

\subsection{Production du boson de Higgs}
\label{subsection:higgs}

Soient $p_1=E(1,\vec{u})$, $p_2=E(1,-\vec{u})$ les quadri-impulsions
des quarks entrants,  $p_1'=E(1,\vec{v})$, $p_2'=E(1,-\vec{v})$ celles
des quarks sortants d\'efinies dans le centre
de masse. On d\'efinit $\Theta_\text{d}=\frac12(\vec{u},\vec{v})$.

\smallskip

Les variables de Mandelstam s'\'ecrivent

$$
s=(p_1+p_2)^2=(p_1'+p_2')^2=4E^2,
$$

$$
t=(p_1-p_1')^2=(p_2-p_2')^2=-2p_1\cdot p_1'=-2E^2(1-\vec{v}\cdot\vec{u})\approx-4E^2\Theta_\text{d}^2
\quad \text{pour $\Theta_\text{d}$ petit},
$$

d'o\`u

$$
\Theta_\text{d}^2\approx\frac{\mid t\mid}{s}\approx\frac{M_H^2}{s}.
$$

%%%%%%%%%%%%%%%%%%%%%%%%%%%%%%%%%%%%%%%%%%%%%%%%%%%%%%%%%%%%%%%%%%%%%%%%%%%%%
%\newpage

\section{Compl\'ements utiles pour le chap\^itre \ref{sec:ADL}}

%\subsection{Id\'ee de la Fonctionnelle G\'en\'eratrice (FG)}
%\label{sub:IFG}
%
%La notion de Fonctionnelle G\'en\'eratrice (FG) a \'et\'e longtemps exploit\'ee
%dans la physique et les math\'ematiques. Par exemple, si l'on d\'eveloppe la fonction
%$G(u)=\exp(u)$ en s\'erie de Taylor au voisinage du point $u=0$, on peut dire que
%celle-ci g\'en\`ere les coefficients $a_n$ d'apr\`es l'expression suivante:
%
%$$
%a_n\equiv\left[\left(\frac{d}{du}\right)^nG(u)\right]_{\left\{u=0\right\}}.
%$$
%
%Exemples:
%
%$$
%G(u)=u\exp{(u)}\Rightarrow0,1,2\dots n\quad \text{nombres naturels,}
%$$
%
%$$
%G(u)=u/(e^u-1)\Rightarrow B_n\quad \text{s\'eries de Bernoulli,}
%$$
%
%$$
%G(u)=\exp{(2xu-u^2)}\Rightarrow H_n(x)\quad \text{p\^olynomes d'Hermite, etc etc}.
%$$
%
%Nous suivons cette m\^eme logique dans \ref{sub:MFG}, o\`u l'on peut consid\'erer que
%notre section efficace du rayonnement de $N$ gluons $d\sigma_N$ est le 
%$N^{\text{\`eme}}$ coefficient du d\'eveloppement de Taylor d'un certain objet 
%``g\'en\'erateur'' qui retient l'information du processus consid\'er\'e en CDQ.
%Cet objet ne peut \^etre une fonction mais une fonctionnelle car les s\'eries qu'il doit
%g\'en\'erer sont des fonctions (par exemple, elles d\'ependent des tri-impulsions des N-
%gluons) et non pas des nombres.

\subsection{D\'eriv\'ees secondes
$\boldsymbol{\frac{\partial^2\phi}{\partial\omega^2}$, 
$\frac{\partial^2\phi}{\partial\nu^2}$,
$\frac{\partial^2\phi}{\partial\omega\partial\nu}}$
et expression du d\'eterminant $\boldsymbol{DetA}$
en fonction de $\boldsymbol{\omega}$, $\boldsymbol{\nu}$}
\label{subsection:DSPHID}

L'astuce pour calculer ces d\'eriv\'es secondes \`a partir de
(\ref{eq:deromegaTH}) et (\ref{eq:dernuTH}) consiste \`a les r\'ecrire
sous la forme

$$
\frac{\partial\phi}{\partial\omega}=\frac{2\omega-\nu}{\omega-\nu}\ell+
\frac{\nu}{\omega-\nu}-\frac{\phi}{\omega-\nu}-\lambda\frac{\nu+2s_0}{\omega-\nu}+
\frac1{\beta\omega(\omega-\nu)},
$$

$$
\frac{\partial\phi}{\partial\nu}=\frac{\omega-2\nu}{\omega-\nu}y-
\frac{\omega}{\omega-\nu}+\frac{\phi}{\omega-\nu}+\lambda\frac{\omega+2s_0}{\omega-\nu}-
\frac1{\beta\nu(\omega-\nu)},
$$

les expressions des d\'eriv\'ees secondes sont donn\'ees par

\beeq\nonumber
\frac{\partial^2\phi}{\partial\omega^2}&=&-\frac{\nu}{(\omega-\nu)^2}(\ell\!+\!y\!+\!\lambda)
+\frac{\phi}{(\omega-\nu)^2}-\frac{2\omega-\nu}{\beta\omega^2(\omega-\nu)^2}\\\nonumber\\
&+&\frac4{\beta(\omega-\nu)^2(2s_0+\omega+\nu)},\nonumber
\eeeq

\beeq\nonumber
\frac{\partial^2\phi}{\partial\nu^2}&=&-\frac{\omega}{(\omega-\nu)^2}(\ell\!+\!y\!+\!\lambda)
+\frac{\phi}{(\omega-\nu)^2}+\frac{\omega-2\nu}{\beta\nu^2(\omega-\nu)^2}\\\nonumber\\
&+&\frac4{\beta(\omega-\nu)^2(2s_0+\omega+\nu)},\nonumber
\eeeq

\beeq\nonumber
\frac{\partial^2\phi}{\partial\omega\partial\nu}&=&\frac{\omega}{(\omega-\nu)^2}(\ell\!+\!y\!+\!\lambda)
-\frac{\phi}{(\omega-\nu)^2}+\frac1{\beta\omega(\omega-\nu)^2}\\\nonumber\\
&-&\frac4{\beta(\omega-\nu)^2(2s_0+\omega+\nu)}.\nonumber
\eeeq
% % \begin{equation}\label{eq:acint}
% \begin{array}{c}
%  \vec{j}=\vec{j_1}+\vec{j_2} \end{array} \left\{
% \begin{array}{c}
%         \vec{j_1}=\vec{v}_1\,\delta^3(\vec{r}-\vec{v}_1t)\cdot\vartheta(t_0-t), \cr
%         \vec{j_2}=\vec{v}_2\,\delta^3(\vec{r}-\vec{v}_2t)\cdot\vartheta(t-t_0),
%  \end{array}
% \right.
% \end{equation}

\begin{equation}
\begin{array}{c}
 A =\end{array} \left(
\begin{array}{c}
        \frac{\partial^2\phi}{\partial\omega^2}\qquad
 \frac{\partial^2\phi}{\partial\omega\partial\nu} \cr
\cr
        \frac{\partial^2\phi}{\partial\nu\partial\omega}\qquad
 \frac{\partial^2\phi}{\partial\nu^2}
 \end{array}
\right)
\end{equation}

Finalement l'expression pour le d\'eterminant est
$$
DetA=(\ell+y+\lambda)^2\left[\frac{\beta(\omega+\nu)\phi-4}{(\omega-\nu)^2}+\frac{4(\omega+\nu)}
{(\omega-\nu)^2(2s_0+\omega+\nu)}\right].
$$

Nous utilisons de m\^eme le r\'esultat de l'int\'egration gaussienne suivante

$$
\int\prod_{i=1}^{N}dx_i\,e^{-\frac12x^T\,A\,x}=\frac{(2\pi)^{N/2}}{(det\,A)^{1/2}}.
$$

\subsection{Astuce pour la m\'ethode du col}

\label{subsec:metcol}

On veut estimer l'int\'egrale suivante par la m\'ethode du col:

\beq\label{eq:Icol}
I=\iint \frac{d\omega\,d\nu}{(2\pi)^2}\, e^{\Phi(\omega,\nu)}
\eeq
o\`u $\Phi(\omega,\nu)=\phi(\omega,\nu)+\phi'(\omega,\nu)$.
La fonction $\phi'(\omega,\nu)$
repr\'esente une correction face \`a la fonction dominante $\phi$,
soit $\phi\gg\phi'$. Soit $(\omega_0,\nu_0)$ le point de col
dans l'exponentielle de l'int\'egrale dominante

$$
I_0=\iint \frac{d\omega\,d\nu}{(2\pi)^2}\, e^{\phi(\omega,\nu)},
$$
tel que
$$
\frac{\partial\phi}{\partial\omega}(\omega_0,\nu_0)=
\frac{\partial\phi}{\partial\nu}(\omega_0,\nu_0)=0.
$$
L'estimation de l'int\'egrale $I_0$ au voisinage de $(\omega_0,\nu_0)$
est donn\'ee par l'expression suivante
$$
I_0\approx\frac{e^{\phi(\omega_0,\nu_0)}}{2\pi\sqrt{DetA(\omega_0,\nu_0)}}.
$$

Finalement, on remplace le point de col $(\omega_0,\nu_0)$ dans
le terme sous-dominant de (\ref{eq:Icol}) pour obtenir

$$
I\approx\frac{e^{\phi(\omega_0,\nu_0)+\phi'(\omega_0,\nu_0)}}
{2\pi\sqrt{DetA(\omega_0,\nu_0)}}.
$$

%%%%%%%%%%%%%%%%%%%%%%%%%%%%%%%%%%%%%%%%%%%%%%%%%%%%%%%%%%%%%%%%%%%%%%%%%%%%%
%%%%%%%%%%%%%%%%%%%%%%%%%%%%%%%%%%%%%%%%%%%%%%%%%%%%%%%%%%%%%%%%%%%%%%%%%%%%%

%\newpage\null

\part{ARTICLES}

\null

\noindent

\vskip 9 cm

\chapter{Inclusive hadronic distributions inside one jet at high energy
colliders at ``Modified Leading Logarithmic Approximation'' of Quantum
Chromodynamics}
\label{sub:article1}

%%%%%%%%%%%%%%%%%%%%%%%%%%%%%%%%%%%%%%%%%%%%%%%%%%%%%%%%%%%%%%%%%%%%%%%%%%%%%

%\newpage\null
% \include{Inclusive}

\begin{titlepage}

\setcounter{page}{137}

%%%%%%%%%%%%%%%%%%%%%%%%%%%%%%%%%%%%%%%%%%%%%%%%%%%%%%%%%%%%%%%%%%%%%%%%%%%%%%

January 2006 (revised March 2006)\hfill hep-ph/0512236, JHEP 04 (2006) 043

\vskip 5cm

\centerline{\bf INCLUSIVE HADRONIC DISTRIBUTIONS INSIDE ONE JET
AT HIGH ENERGY COLLIDERS}

\smallskip

\centerline{\bf AT ``MODIFIED LEADING LOGARITHMIC APPROXIMATION''}

\smallskip

\centerline{\bf OF QUANTUM CHROMODYNAMICS}

\vskip .75 cm

\centerline{R. Perez-Ramos
\footnote{E-mail: perez@lpthe.jussieu.fr}
\& B. Machet
\footnote{E-mail: machet@lpthe.jussieu.fr}
}

\baselineskip=15pt

\smallskip
\centerline{\em Laboratoire de Physique Th\'eorique et Hautes Energies
\footnote{LPTHE, tour 24-25, 5\raise 3pt \hbox{\tiny \`eme} \'etage,
Universit\'e P. et M. Curie, BP 126, 4 place Jussieu,
F-75252 Paris Cedex 05 (France)}}
\centerline{\em Unit\'e Mixte de Recherche UMR 7589}
\centerline{\em Universit\'e Pierre et Marie Curie-Paris6; CNRS;
Universit\'e Denis Diderot-Paris7}

\vskip 1.5cm

{\bf Abstract}: After demonstrating their general expressions valid at all $x$,
double differential 1-particle inclusive distributions 
inside a quark and a gluon jet produced in a hard process, together with the
inclusive $k_\perp$ distributions, are calculated at small $x$ in the
Modified Leading Logarithmic Approximation (MLLA), as functions of the
transverse momentum $k_\perp$ of the outgoing hadron.
Results are compared with
the Double Logarithmic Approximation (DLA) and a naive DLA-inspired
evaluation; sizable corrections are exhibited,  which, associated with the
requirement to stay in a perturbative regime, set the limits of
the interval where our calculations can be trusted.
We give predictions for the LHC and Tevatron colliders.

\vskip 1 cm

{\em Keywords: perturbative Quantum Chromodynamics,
jets, high-energy colliders}

\vfill

%\null\hfil\epsffile{LogoCNRS.ps}

%%%%%%%%%%%%%%%%%%%%%%%%%%%%%%%%%%%%%%%%%%%%%%%%%%%%%%%%%%%%%%%%%%%%%%%%%%%%%%

\end{titlepage}

%%%%%%%%%%%%%%%%%%%%%%%%%%%%%%%%%%%%%%%%%%%%%%%%%%%%%%%%%%%%%%%%%%%%%%%%%%%%%%

\baselineskip=15pt

%%%%%%%%%%%%%%%%%%%%%%%%%
\section{INTRODUCTION}
%%%%%%%%%%%%%%%%%%%%%%%%%

\vskip .4cm

In high energy collisions, perturbative Quantum Chromodynamics (pQCD)
successfully predicts inclusive energy spectra of particles in jets.
They have been determined within the
Modified Leading Logarithmic Approximation (MLLA) ~\cite{EvEqI}~\cite{FWI}
as  functions of the logarithm of the energy ($\ln(1/x)$) and the result
is in nice agreement with the data of -- $e^+e^-$ and hadronic --
 colliders and of deep inelastic scattering (DIS) (see for example
\cite{OPAL1I} \cite{CdF} \cite{DIS}).
Though theoretical predictions have been derived for small $x$
(energy fraction of one parton inside the jet, $x\ll 1$)
\footnote{as the exact solution of the MLLA evolution equations}
, the agreement turns out to hold even for $x\sim 1$. 
The shape of the inclusive spectrum can even be successfully described by
setting the infrared transverse momentum  cutoff $Q_0$ as low as the
intrinsic QCD scale $\Lambda_{QCD}$ (this is the so-called
``limiting spectrum'').

\bigskip

This work concerns the production of two hadrons inside  a high energy jet
(quark or gluon); they hadronize out of two partons 
at the end of a cascading process that we calculate in pQCD;
considering this transition as a ``soft'' process is
the essence of the ``Local Parton Hadron Duality'' (LPHD) hypothesis
\cite{EvEqI} \cite{DKTM} \cite{KOI}, that experimental data have,
up to now, not put in jeopardy.

\bigskip

More specifically, we study, in the MLLA scheme of resummation,
the double differential inclusive 1-particle distribution
and the inclusive $k_\perp$ distribution as functions of  the transverse
momentum of the emitted hadrons; they have up to now only been investigated
in DLA (Double Logarithmic Approximation) \cite{EvEqI}.
After giving  general expressions valid at all $x$, we are concerned in
the rest of the paper with the small $x$ region (the range of which is
extensively discussed) where explicit analytical formul{\ae} can be obtained;
  we furthermore consider the limit $Q_0
\approx \Lambda_{QCD}$, which leads to tractable results.
We deal with jets of small aperture; as far as hadronic colliders are
concerned, this has in particular the advantage to avoid interferences
between ingoing and outgoing states.

\bigskip

The paper is organized as follows:

\bigskip

$\bullet$\quad The description of the process, the
notations and conventions are presented in section \ref{section:descri}.
We  set there the general formula of the inclusive 2-particle
differential cross section for the production of two
hadrons $h_1$ and $h_2$  at angle $\Theta$ within a jet of opening angle
$\Theta_0$,  carrying respectively  the fractions $x_1$ and $x_2$ of
the jet energy $E$;  the axis of the jet is identified
with the direction of the energy flow.

\medskip

$\bullet$\quad In section \ref{section:EA}, we determine the 
 double differential inclusive 1-particle distribution
$\frac{d^2N}{d\ln\left(1/x_1\right)\,d\ln\Theta}$ for the hadron
$h_1$ emitted with the energy
fraction $x_1$ of the jet energy $E$, at an angle $\Theta$ with respect to
the jet axis.
This expression is valid for all $x$; it however only simplifies for $x \ll
1$, where an analytical expression can be obtained; this concerns the
rest of the paper. 

\medskip

$\bullet$\quad In section \ref{section:lowEA}, we go to the small $x$
region and determine
$\frac{d^2N}{d\ln\left(1/x_1\right)\,d\ln\Theta},\ x_1\ll 1$ 
both for a gluon jet and for a quark jet.
It is plotted as a function of $\ln k_\perp$ (or $\ln\Theta$)
for different values of $\ell_1 = \ln(1/x_1)$; the role of the
opening angle $\Theta_0$ of the jet is also considered; we compare in
particular the MLLA calculation  with a naive approach,
inspired by DLA calculations, in which  furthermore the evolution of the
starting jet from  $\Theta_0$, its initial aperture, to the angle $\Theta$
between the two outgoing hadrons  is not taken into account.

The MLLA expressions of the average gluon and quark color currents $<C>_g$ and $<C>_q$
involve potentially large corrections with respect to their expressions at
leading order; the larger the (small) $x$ domain extends, the larger they
are; keeping then under control  sets the bound
$\ell \equiv \ln\frac{1}{x}\geq 2.5$.

\medskip

$\bullet$\quad In section \ref{section:ktdist}, we study the inclusive 
$k_\perp$ distribution $\frac{dN}{d\ln k_\perp}$, which is
 the  integral of
$\frac{d^2N}{dx_1\,d\ln\Theta}$ with respect to $x_1$;
It is shown in particular how MLLA corrections ensure its positivity.
The domain of validity of our predictions is discussed; it is a $k_\perp$ interval,
 limited by the necessity of staying in the perturbative regime
 and the range of applicability of our small $x$ approximation;
it increases with the jet hardness.
The  case of mixed gluon and quark jets is evoked.

\medskip

$\bullet$\quad A conclusion briefly summarizes the results of this work and
comments on its extensions under preparation.

\medskip
\medskip

Five appendices complete this work;

\medskip

$\bullet$\quad  Appendix \ref{section:exactsol} is dedicated to the 
MLLA evolution equation for the partonic fragmentation functions
$D_g^{g\ or \ q}$ and their exact solutions \cite{Perez}\cite{Perez2}.
  They are
plotted, together with their derivatives with respect to $\ln(1/x)$
and $\ln k_\perp$.
This eases the understanding of the figures in the core of the paper and
shows the consistency of our calculations.

\medskip

$\bullet$ \quad Appendix \ref{section:leadingxF}
  presents the explicit expressions at leading order for
the average color currents of partons $<C>_{A_0}$.

\medskip

$\bullet$ \quad Appendix \ref{section:udeltau}
 completes section \ref{section:lowEA} and appendix  \ref{section:leadingxF}
 by providing explicit {formul\ae} necessary to evaluate
the MLLA corrections $\delta\!<C>_{A_0}$ to the average color currents;

\medskip

$\bullet$ \quad While the core of the paper mainly give results for LHC,
 Appendix \ref{section:LEP} provides an overview at LEP and Tevatron energies.
It is shown  how,
considering too large values of $x$ ($\ln\frac{1}{x} < 2$) endanger the
positivity of $\frac{d^2N}{d\ell\,d\ln k_\perp}$ at low $k_\perp$.
Curves are also given  for $\frac{dN}{d\ln k_\perp}$;
the range of applicability of our approximation is discussed in relation
with the core of the paper.

\medskip

$\bullet$ \quad in Appendix \ref{section:DLAI}, we compare the DLA and MLLA
approximations for the spectrum, the
double differential 1-particle inclusive distribution, and the inclusive
$k_\perp$ distribution.

\vskip .7 cm

%%%%%%%%%%%%%%%%%%%%%%%%%%%%%%%%%%%%%%%%%%%%%%%%%%%%%%%%%%%%%%%%%%%%%%%%%%%%%%
\section{THE PROCESS UNDER CONSIDERATION}
\label{section:descri}
%%%%%%%%%%%%%%%%%%%%%%%%%%%%%%%%%%%%%%%%%%%%%%%%%%%%%%%%%%%%%%%%%%%%%%%%%%%%%%

\vskip .4cm

It is depicted in Fig.~1 below.  In a hard collision, a parton $A_0$
is produced, which can be a quark or a gluon
\footnote{
in $p-p$ or $p-\bar p$ collisions, two partons collide which
can create $A_0$ either as a quark or as a gluon; in the deep inelastic
scattering (DIS) and in $e^+ e^-$ colliders,
a vector boson ($\gamma$ or $Z$) decays
into a quark-antiquark pair, and $A_0$ is a quark (or an antiquark);
}
.
$A_0$, by a succession of partonic emissions  (quarks, gluons), produces
a jet of opening angle $\Theta_0$, which, in particular, contains
the parton $A$; $A$ splits  into $B$ and $C$, which hadronize respectively
into the two hadrons $h_1$  and $h_2$ (and other hadrons).
$\Theta$ is the angle between $B$ and $C$.

Because the virtualities of $B$ and $C$ are much smaller than that
of $A$ \cite{DDTI},
$\Theta$ can be considered to be close to the angle between $h_1$ and $h_2$
\cite{DDTI}\cite{DDI}; angular ordering is also a necessary condition for
this property to hold.

\bigskip

\vbox{
\begin{center}
\epsfig{file=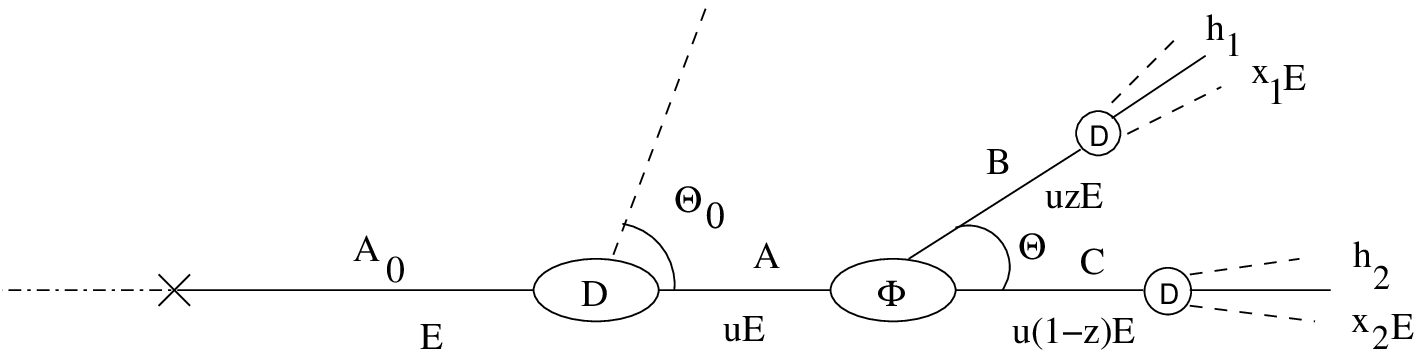, height=4.5truecm,width=11.9truecm}
\vskip .4cm
{\em Fig.~1: process under consideration: two hadrons $h_1$ and $h_2$ inside
one jet.}
\end{center}
}

\bigskip

$A_0$ carries the energy $E$.
 With a probability $D_{A_0}^A$, it
gives rise to the (virtual) parton $A$, which carries the fraction
$u$ of the energy $E$;
$\Phi_A^{BC}(z)$ is the splitting function of $A$ into $B$ and $C$,
carrying respectively the fractions $uz$ and $u(1-z)$ of $E$;
$h_1$ carries the fraction $x_1$ of $E$; $h_2$ carries the fraction $x_2$
of $E$; $D_B^{h_1}\left(\displaystyle\frac{x_1}{uz},uzE\Theta,Q_0\right)$
and $D_C^{h_2} \left(\displaystyle\frac{x_2}{u(1-z)},u(1-z)E\Theta,Q_0\right)$
are their respective energy distributions.

One has $\Theta \leq\Theta_0$.
On the other hand, since $k_{\perp} \geq Q_0$ ($Q_0$  is the collinear
cutoff), the emission angle
must satisfy $\Theta \geq \Theta_{min}=Q_0/(xE)$, $x$ being the fraction of the
energy $E$ carried away by this particle (see also subsection
\ref{subsection:notations} below).

The following expression for the inclusive double differential 2-particle
 cross section has been demonstrated in \cite{DDTI} \cite{DDI}:
\begin{eqnarray}
&&\frac{d\sigma}{d\Omega_{jet}\,
dx_1\,dx_2\,d\ln\left({\sin^2\displaystyle\frac{\Theta}{2}}\right)\,\displaystyle\frac{d\varphi}{2\pi}}
= 
\left(\frac{d\sigma}{d\Omega_{jet}}\right)_0\; \sum_{A,B,C} 
 \int \frac{du}{u^2} \int dz \Bigg[ \frac{1}{z(1-z)}
 \frac{\alpha_s(k_{\perp}^2)}{4\pi}\cr
 &&\Phi_A^{BC}\left(z\right)
 D_{A_0}^{A}\left(u,E\Theta_0,uE
 \Theta\right)
D_{B}^{h_1}\left(\frac{x_1}{uz},
uzE\Theta,Q_0\right)
 D_{C}^{h_2}\left(\frac{x_2}{u\left(1-z\right)},u
 (1-z)E\Theta,Q_0\right)\Bigg],\cr
&&
\label{eq:basic}
\end{eqnarray}
where $\left(\displaystyle\frac{d\sigma}{d\Omega_{jet}}\right)_0$ is the Born cross section
for the production of $A_0$, $\Omega_{jet}$ is the solid angle of the jet
 and $\varphi$ is the azimuthal angle between $B$ and $C$.

$\alpha_s(q^2)$ is the QCD running coupling constant:
\begin{equation}
\alpha_s(q^2) = \frac{4\pi}{4N_c\;\beta\ln\displaystyle
\frac{q^2}{\Lambda_{QCD}^2}},
\label{eq:alphas}
\end{equation}
where $\Lambda_{QCD} \approx$ a few hundred MeV is the intrinsic scale
of QCD and
\begin{equation}
\beta=\frac{1}{4N_c}\left(\frac{11}{3} N_c - \frac{4}{3}T_R\right) \\
\label{eq:beta}
\end{equation}
is the first term in the perturbative expansion of the $\beta$-function,
$N_c$ is the number of colors, $T_R = n_f/2$, where $n_f$ is the number
of light quark flavors ($n_f=3$); it is convenient to scale all relevant
parameters in units of $4N_c$.

In (\ref{eq:basic}), the integrations over $u$ and $z$ are performed
from $0$ to $1$;
the appropriate step functions ensuring $uz \geq x_1$, $u(1-z) \geq x_2$
(positivity of energy) are included in $D_B^{h_1}$ and $D_C^{h_2}$.

\vskip .7 cm

\subsection{Notations and variables}
\label{subsection:notations}
%%%%%%%%%%%%%%%%%%%%%%%%%%%%%%%%%%%%%%

\vskip .4cm

The notations and conventions, that are used above and throughout the paper
 are the following. For any given particle with 4-momentum $(k_0,\vec k)$,
transverse momentum $k_\perp \geq Q_0$ ($k_\perp$ is the modulus of the
trivector $\vec k_\perp$),
carrying the fraction $x=k_0/E$ of the jet energy $E$, one defines

\begin{equation}
   \ell = \ln\frac{E}{k_0}=\ln(1/x), \quad y = \ln \frac{k_{\perp}}{Q_0}.
\end{equation}
$Q_0$ is the infrared cutoff parameter (minimal transverse momentum).

If the radiated parton is emitted with an angle $\vartheta$ with
respect to the direction of the jet, one has
\begin{equation}
k_{\perp} = |\vec k|\sin{\vartheta}
\approx k_0\sin{\vartheta}.
\label{eq:appro}
\end{equation}
The r.h.s. of (\ref{eq:appro}) uses $|\vec k| \approx k_0$, resulting from
the property that the virtuality $k^2$ of the emitted parton is negligible in
the logarithmic approximation.
For collinear emissions ($\vartheta \ll 1$), $k_{\perp}\sim |\vec k|\vartheta
\approx k_0\vartheta$.

One also defines the variable $Y_\vartheta$
\begin{equation}
Y_\vartheta =\ell + y = \ln\left(E\frac{k_\perp}{k_0}\frac{1}{Q_0}\right)
\approx \ln \frac{E\vartheta}{Q_0};
\label{eq:defY}
\end{equation}
to the opening angle $\Theta_0$ of the jet corresponds
\begin{equation}
Y_{\Theta_0} = \ln\frac{E\Theta_0}{Q_0};
\label{eq:Y0}
\end{equation}
$E\Theta_0$ measures the ``hardness'' of the jet.
Since $\vartheta < \Theta_0$, one has the
condition, valid for any emitted soft parton off its ``parent''
\begin{equation}
Y_\vartheta < Y_{\Theta_0}.
\label{eq:condY}
\end{equation}

The partonic fragmentation  function $D_a^b(x_b,Q,q)$
represents the probability of finding
the parton $b$ having the fraction $x_b$ of the energy of $a$
inside the dressed parton $a$;  the virtuality (or transverse momentum)
$k_a^2$ of $a$ can go up to $|Q^2|$,  that of $b$ can go down to $|q^2|$.

\vskip .7 cm

\subsection{The jet axis}
\label{subsection:axis}
%%%%%%%%%%%%%%%%%%%%%%%%%%%

\vskip .4cm

The two quantities studied in the following paragraphs (double
differential 1-particle inclusive distribution and inclusive
 $k_\perp$ distribution)
refer to the direction (axis) of the jet, with respect to which the angles
are measured.
We identify it with the direction of the energy flow.

The double differential 1-particle inclusive distribution
$\frac{d^2N}{dx_1 d\ln\Theta}$ is accordingly defined
by summing the inclusive double differential 2-particle cross section
 over all $h_2$ hadrons and integrating it over
their energy fraction $x_2$ {\em with a weight which is the energy
($x_2$) itself};
it measures the angular distribution of
an outgoing hadron $h_1$ with energy fraction $x_1$ of the jet energy,
produced at an angle $\Theta$ with respect to the direction of
the energy flow.

Once the axis has been fixed, a second (unweighted) integration with
respect to the energy of the other hadron ($x_1$) leads to the inclusive
 $k_\perp$ distribution $\frac{dN}{d\ln k_\perp}$.

\vskip .7 cm

%%%%%%%%%%%%%%%%%%%%%%%%%%%%%%%%%%%%%%%%%%%%%%%%%%%%%%%%%%%%%%%%%%%%%%%%%%%%%%
\section{DOUBLE DIFFERENTIAL 1-PARTICLE INCLUSIVE DISTRIBUTION 
$\boldsymbol{\displaystyle\frac{d^2N}{dx_1\;d\ln\Theta}}$}
\label{section:EA}
%%%%%%%%%%%%%%%%%%%%%%%%%%%%%%%%%%%%%%%%%%%%%%%%%%%%%%%%%%%%%%%%%%%%%%%%%%%%%%

\vskip .4cm

After integrating trivially over the azimuthal angle
 (at this approximation the cross-section does not depend on it),
and going to small $\Theta$, the positive quantity $\frac{d^2N}{dx_1\,d\ln
\Theta}$ reads

\begin{equation}
\frac{d^2N}{dx_1\,
d\ln \Theta}
=
\sum_{h_2}\int _{0}^{1} dx_2\ {x_2}\;
\frac{d\sigma}
{d\Omega_{jet}\,
dx_1\,dx_2\,
d\ln\Theta}\,
\frac{1}{\left(\displaystyle\frac{d\sigma}{d\Omega_{jet}}\right)_0}.
\label{eq:eac}
\end{equation}

We use the energy conservation sum rule \cite{Politzer}
\begin{equation}
\sum_{h}\int _{0}^{1} dx\,x D_C^{h} (x,\ldots) =1
\label{eq:ensr}
\end{equation}
expressing that all partons $h_2$ within a dressed parton ($C$) carry
the total momentum of $C$,
then make the change of variable $v=\frac{x}{u(1-z)}$
where $u(1-z)$ is the upper kinematic limit for $x_2$, to get
\begin{equation}
\sum_{h_2}\int _{0}^{u(1-z)}
dx_2\,x_2 D_C^{h_2}
\left(\frac{x_2}{u(1-z)},u(1-z) E\Theta,Q_0\right)
=u^2(1-z)^2,
\label{eq:sumrule}
\end{equation}
and finally obtain the desired quantity;
\begin{equation}
  \frac{d^2N}{dx_1\,d\ln\Theta}=
\sum_{A,B}
\int du \int dz\ 
\frac{1-z}{z} \frac{\alpha_s\left(k_{\perp}^2\right)}
{2\pi}
  \Phi_A^B(z)
  D_{A_0}^A\left(u,E\Theta_0,uE\Theta\right)
  D_{B}^{h_1}\left(\frac{x_1}{uz},uzE \Theta,Q_0\right)
\label{eq:2};
\end{equation}
the summation index $C$ has been suppressed since knowing $A$ and $B$
fixes $C$.

We can transform (\ref{eq:2}) by using the following trick:
\begin{equation}
\int du\,\int \frac{dz}{z}(1-z)=\int du\,\int\frac{dz}{z}
-\int d(uz)\int\frac{du}{u},
\label{eq:trick}
\end{equation}
and (\ref{eq:2}) becomes
\begin{equation}
\begin{split}
  \frac{d^2N}{dx_1\,d\ln\Theta}
= 
&\sum_{A}\int
du\,
  D_{A_0}^{A}(u,E\Theta_0,uE\Theta)
\sum_{B}
  \int
\frac{dz}{z}
  \frac{\alpha_s\left(k_{\perp}^2\right)}
{2\pi}
\Phi_A^B (z)
  D_{B}^{h_1}\left(\frac{x_1}{uz},uzE\Theta,Q_0\right)\\
&- \sum_{B}\int
d(uz)D_{B}^{h_1}\left (\frac{x_1}{uz},uzE\Theta,Q_0\right)
  \sum_{A}\int
\frac{du}{u}\frac{\alpha_s (k_{\perp}^2)}{4\pi}
  \Phi_A^B\left(\frac{uz}u\right)
  D_{A_0}^A(u,E\Theta_0,uE\Theta).
\end{split}
\label{eq:inter}
\end{equation} 
We then make use of the two complementary DGLAP (see also the beginning of
section \ref{section:lowEA}) evolution equations
~\cite{DGLAPI} 
which contain the Sudakov form factors  $d_A$ and $d_B$ of the partons
$A$ and $B$ respectively:
\begin{equation}
d_A^{-1}(k_{A}^2)\frac{d}{d\ln k_A^2}
  \left[d_A(k_{A}^2)
    D_A^{h_1}\left(\frac{x_1}{u},uE\Theta,Q_0\right)\right]
    =
\frac{\alpha_s(k_{\perp}^2)}{4\pi}
\sum_{B}\int
  \frac{dz}{z}
  \Phi_A^{B}\left(z\right) D_{B}^{h_1}\left(\frac{x_1}{uz},uzE\Theta,Q_0\right),
\label{eq:dglap1}
\end{equation}
\begin{equation}
 d_B(k_{B}^2)\frac{d}{d\ln k_B^2}\left[d_B^{-1}
 (k_{B}^2)
 D_{A_0}^B(w,E\Theta_0,wE\Theta)\right]
=
- \frac{\alpha_s(k_{\perp}^2)}{4\pi}
\sum_{A} \int
 \frac{du}{u}
\Phi_A^B \left(\frac{w}{u}\right) D_{A_0}^A(u,E\Theta_0,uE\Theta);
\label{eq:dglap2}
\end{equation}
the variable $uz$  occurring in (\ref{eq:trick}) has been introduced;
in (\ref{eq:dglap1}) and (\ref{eq:dglap2}),
$(uE\Theta)^2$ refers respectively to the virtualities $k_A^2$ and $k_B^2$
of $A$ and $B$.
Using (\ref{eq:dglap1}) and (\ref{eq:dglap2}),
(\ref{eq:inter}) transforms into 
\begin{equation}
\begin{split}
  \frac{d^2N}{\displaystyle dx_1\,d\ln\Theta}
&=\sum_{A}\int
du
  \,D_{A_0}^A(u,E\Theta_0,uE\Theta)
d_A^{-1}(k_{A}^2)\frac{d}{d\ln k_A^2}\left[d_A
(k_{A}^2)
  D_A^{h_1}\left(\frac{x_1}{u},uE\Theta,Q_0\right)\right]\\
  &+\sum_B\int
dw\,D_B^{h_1}\left(\frac{x_1} {w},wE\Theta,Q_0\right)
  d_B(k_{B}^2)\frac{d}{d\ln k_B^2}\left[d_B^{-1}
  (k_{B}^2) D_{A_0}^B\left(w,E\Theta_0,wE\Theta\right)\right].
\end{split}
\label{eq:dist}  
\end{equation}
$D_A^{h_1}$ depends  on the virtuality of $A$ through the variable
\cite{EvEqI}
$\Delta \xi = \xi(k_A^2) -\xi(Q_0^2) =
\displaystyle\frac{1}{4N_c\beta}\ln\left(
\displaystyle\frac{\ln(k_A^2/\Lambda_{QCD}^2)}{\ln(Q_0^2/\Lambda_{QCD}^2)}\right)$
and elementary kinematic considerations \cite{DDTI} lead to
$k_A^2 \sim \left(uE \Theta\right)^2$.

By renaming $B\rightarrow A$ and  $w\rightarrow u$,
(\ref{eq:dist}) finally becomes
\begin{equation}
\begin{split}
 \frac{d^2N}{dx_1\,d\ln{\Theta}}
&=
\sum_A\int
du
 \left[D_{A_0}^A (u,E\Theta_0,uE\Theta)d_A^{-1}(k_{A}^2)
\frac{d} {d\ln \Theta}
\left[d_A(k_{A}^2)
D_A^{h_1} \left(\frac{x_1}{u},uE\Theta,Q_0\right)\right]\right.\\
&\left.+D_A^{h_1}\left(\frac{x_1}{u},uE\Theta,Q_0\right)
d_A(k_{A}^2)
\frac{d}{d\ln \Theta}
 \left[d_A^{-1}(k_{A}^2)D_{A_0}^A\left(u,E\Theta_0,uE\Theta\right)
 \right]\right]\\
&=\sum_A\frac{d}{d\ln \Theta}\left[\int
du
 D_{A_0}^{A}\left(u,E\Theta_0,uE\Theta\right)D_A^{h_1}
\left( \frac{x_1}{u},uE\Theta,Q_0\right)\right],
\end{split}
\label{eq:dist1}
\end{equation}
and one gets
\begin{equation}
\frac{d^2N}{dx_1\,d\ln{\Theta}}=
\frac{d}{d\ln\Theta}F_{A_0}^{h_1}\left(x_1,
\Theta,E,\Theta_0\right)
\label{eq:DDI}
\end{equation}
with
\begin{equation}
 F_{A_0}^{h_1}\left(x_1,\Theta,E,\Theta_0\right)
\equiv
\sum_{A}\int
du
 D_{A_0}^A\left(u,E\Theta_0,uE\Theta\right)D_A^{h_1}\left(\frac{x_1}
 {u},uE\Theta,Q_0\right);
%}
\label{eq:F}
\end{equation}

\medskip

$F$ defined in (\ref{eq:F}) is the inclusive double differential
distribution in $x_1$ and $\Theta$  with respect to the energy flux
(the energy fraction of the hadron $h_1$ within the registered
energy flux) and is represented by the convolution of the two functions
$D_{A_0}^A$ and $D_A^h$.

The general formula (\ref{eq:DDI}) is valid for all $x_1$; its 
analytical expression in the small $x_1$ region will be written in
the next section.

\vskip .7 cm

%%%%%%%%%%%%%%%%%%%%%%%%%%%%%%%%%%%%%%%%%%%%%%%%%%%%%%%%%%%%%
\section{SOFT APPROXIMATION (SMALL-$\boldsymbol x_1$) FOR
$\boldsymbol{\displaystyle\frac{d^2N}{d\ell_1\;d\ln k_\perp}}$}
\label{section:lowEA}
%%%%%%%%%%%%%%%%%%%%%%%%%%%%%%%%%%%%%%%%%%%%%%%%%%%%%%%%%%%%%

\vskip .4cm

At $\ell_1$ fixed, since  $y_1=\ln (k_{\perp}/Q_0)$ and
$Y = \ln(E\Theta/Q_0) = \ell_1 + y_1$, $dy_1 = d\ln k_\perp = d\ln\Theta$
and we write hereafter $\frac{d^2N}{d\ell_1\;d\ln k_\perp}$ or
 $\frac{d^2N}{d\ell_1\;d y_1}$ instead of
$\frac{d^2N}{d\ell_1\;d\ln\Theta}$.

Since the $u$-integral (\ref{eq:F}) is dominated by $u={\cal O}(1)$
\footnote{$D_A^{h_1}\left(\frac{x_1}{u},uE\Theta,Q_0\right) \approx
(u/x_1)\times$ (slowly varying function) -- see (\ref{eq:Dlowx2}) --
and the most singular possible behavior
of $D_{A_0}^A(u, E\Theta_0, uE\Theta,Q_0)$, which could enhance the
contribution of small $u$, is $\sim 1/u$; however, the integrand then
behaves like Const. $\times$ (slowly\ varying\ function) and the
contribution of small
$u$ to the integral is still negligible.}
, the DGLAP
\cite{EvEqI} partonic distributions  $D_{A_0}^A(u, \ldots)$  are to be used
and, since, on the other hand,  we restrict to small $x_1$, 
$x_1/u \ll 1$ and the MLLA inclusive $D_A^{h_1}((x_1/u),\ldots)$ are
requested. The latter will be taken as the exact solution (see
\cite{Perez}) of the (MLLA) evolution equations 
that we briefly sketch out, for the sake of completeness,
in appendix \ref{section:exactsol}.
MLLA evolution equations accounts for the constraints of angular
ordering (like DLA but unlike DGLAP equations) and of energy-momentum
conservation (unlike DLA).

For soft hadrons, the behavior of the function
$D_{A}^{h_1}(x_1,E\Theta,Q_0)$ at $x_1\ll 1$  is \cite{EvEqI}
\begin{equation}
 D_A^{h_1}(x_1,E\Theta,Q_0)\approx\frac{1}{x_1}\rho_A^{h_1}\left(\ln\frac{1}{x_1},
 \ln\frac{E\Theta}{Q_0}\equiv Y_\Theta\right),
\label{eq:Dlowx}
\end{equation}
where $\rho_A^{h_1}$ is a slowly varying function of two logarithmic
variables that describes the
``hump-backed'' plateau.

For $D_A^{h_1}\left(\displaystyle\frac{x_1}{u},uE\Theta,Q_0\right)$ occurring in
(\ref{eq:F}), this yields
\begin{equation}
D_A^{h_1}\left(\frac{x_1}{u},uE\Theta,Q_0\right) \approx
\frac{u}{x_1}\rho_A^{h_1}\left(\ln\frac{u}{x_1},\ln u + Y_\Theta\right).
\label{eq:Dlowx2}
\end{equation}
Because of (\ref{eq:defY}), one has
\begin{equation}
\rho_A^{h}(\ell, Y_\Theta) = \rho_A^{h_1}(\ell, \ell + y) = \tilde
D_A^{h}(\ell, y),
\label{eq:rhoD}
\end{equation}
and, in what follows, we shall always consider the functions
\begin{equation}
xD_A(x,E\Theta,Q_0)=\tilde D_A(\ell,y).
\label{eq:Dtilde}
\end{equation}

The expansion of
$\rho_A^{h_1}\left(\ln\displaystyle\frac{u}{x_1},\ln u + Y_\Theta\right)$
around $u=1$ ($\ln u=\ln 1$)  reads
\begin{eqnarray}
\frac{x_1}{u}D_A^{h_1}\left(\frac{x_1}{u},uE\Theta,Q_0\right)
&=& \rho_A^{h_1}(\ell_1 + \ln u,Y_\Theta + \ln u)
= \rho_A^{h_1}(\ell_1 + \ln u, y_1 + \ell_1 + \ln u)\cr
&=& \tilde D_A^{h_1}(\ell_1 + \ln u, y_1)
= \tilde D_A^{h_1}(\ell_1,y_1) + \ln u \; \frac{d}{d\ell_1}\tilde
D_A^{h_1}(\ell_1, y_1) + \ldots,\cr
&&
\label{eq:rhodev}
\end{eqnarray}
such that

\vbox{
\begin{eqnarray}
x_1 F_{A_0}^{h_1}(x_1,\Theta,E,\Theta_0)
&\approx&
\sum_A \int du\;
{u} D_{A_0}^A(u,E\Theta_0,uE\Theta)
\left(\tilde D_A^{h_1}(\ell_1,y_1) + \ln u\;
\frac{d\tilde D_A^{h_1}(\ell_1,y_1)}{d\ell_1}
\right)\cr
&=& \sum_A \left[\int du\, u D_{A_0}^A(u,E\Theta_0,uE\Theta)\right]
 \tilde D_A^{h_1}(\ell_1,y_1)\cr
&&  + \sum_A \left[\int du\, u\ln u D_{A_0}^A(u,E\Theta_0,uE\Theta)\right]
\frac{d\tilde D_A^{h_1}(\ell_1,y_1)}{d\ell_1};
\label{eq:Fdev}
\end{eqnarray}
}

the second line in (\ref{eq:Fdev}) is the ${\cal O}(1)$
 main contribution; the third line, which accounts for the derivatives,
including the variation of $\alpha_s$, makes up corrections of relative order
 ${\cal O}(\sqrt{\alpha_s})$ with respect to the leading terms
(see also (\ref{eq:correc})), which have never been considered before;
since, in the last
line of (\ref{eq:Fdev}), $u\leq 1 \Rightarrow \ln u \leq 0$ and
$\frac{d \tilde D_{A}^{h_1}}{d\ell_1}$ is positive
(see appendix \ref{subsection:deriv}), the corresponding correction is
negative.
A detailed discussion of all corrections  is made in subsections
\ref{subsection:colcur} and \ref{subsection:correcs}

It is important for further calculations that (\ref{eq:F}) has now factorized.

While (\ref{eq:F}) (\ref{eq:Fdev}) involve (inclusive)
{\em hadronic} fragmentation functions $\tilde D_A^{h_1} = \tilde
D_g^{h_1}$ or $\tilde D_q^{h_1}$,
the MLLA {\em partonic} functions $\tilde D_A^b(\ell, y)$ satisfy
the evolution equations (\ref{eq:eveqincl}) with
exact solution (\ref{eq:ifD}), demonstrated in
\cite{Perez} and recalled in appendix \ref{section:exactsol}.
The link between the latter
($\tilde D_g^g$, $\tilde D_q^g$,  $\tilde D_g^q$,  $\tilde D_q^q$) and the
former goes as follows.
 At small $x$, since quarks are secondary products of gluons, for a given
``parent'', the number
of emitted quarks is a universal function of the number of emitted gluons:
the upper indices of emitted partons are thus correlated, and
we can  replace  in (\ref{eq:Fdev}) the inclusive fragmentation functions by
the partonic ones,  go to the functions
$\tilde D_A(\ell, y)$, where the upper index (which we will omit)
 is indifferently $g$ or $q$,
and rewrite
\begin{equation}
x_1 F_{A_0}^{h_1}(x_1,\Theta,E,\Theta_0)
\approx
\sum_A \Big( <u>^A_{A_0} + \delta \! <u>^A_{A_0}
\psi_{A,\ell_1}(\ell_1, y_1)
\Big)
\tilde D_A(\ell_1,y_1),
\label{eq:xF2}
\end{equation}
with
\footnote{In (\ref{eq:udef}), $u$ is integrated form $0$ to $1$, while,
kinematically, it cannot get lower than $x_1$; since we are
working at small $x_1$, this approximation is reasonable.}
\begin{eqnarray}
<u>^A_{A_0}&=&\int_{0}^{1}du\,uD_{A_0}^A\left(u,E\Theta_0,uE\Theta\right)
\approx \int_{0}^{1}du\,uD_{A_0}^A\left(u,E\Theta_0,E\Theta\right),\cr
\delta \! <u>^A_{A_0}&=&
\int_{0}^{1}du\,(u\ln u)D_{A_0}^A\left(u,E\Theta_0,uE\Theta\right)
\approx \int_{0}^{1}du(\,u\ln u)D_{A_0}^A\left(u,E\Theta_0,E\Theta\right),\cr
&&
\label{eq:udef}
\end{eqnarray}
and
\begin{equation}
\psi_{A,\ell_1}(\ell_1,y_1) = \frac{1}{\tilde D_A(\ell_1, y_1)}
\frac{d \tilde D_A(\ell_1, y_1)}{d\ell_1}.
\label{eq:psidef}
\end{equation}

Thus, for a gluon jet
\begin{eqnarray}
x_1 F_{g}^{h_1}\left(x_1,\Theta,E,\Theta_0\right)
&\approx&
<u>^g_{g} \tilde D_g(\ell_1,y_1)
+ <u>^q_{g} \tilde D_q(\ell_1,y_1)\cr
 &+& \delta \! <u>^g_{g}\psi_{g,\ell_1}(\ell_1,y_1) 
\tilde D_g(\ell_1,y_1)
\cr
&+&  \delta \! <u>^q_{g}\psi_{q,\ell_1}(\ell_1,y_1)
\tilde D_q(\ell_1,y_1),\cr
&&
\label{eq:Fg}
\end{eqnarray}
and for a quark jet
\begin{eqnarray}
x_1 F_{q}^{h_1}\left(x_1,\Theta,E,\Theta_0\right)
&\approx&
<u>^g_{q}\tilde D_g(\ell_1,y_1)
+ <u>^q_{q} \tilde D_q(\ell_1,y_1)\cr
&+& \delta \! <u>^g_{q}\psi_{g,\ell_1}(\ell_1,y_1)
 \tilde D_g(\ell_1,y_1)\cr
&+& \delta \! <u>^q_{q}\psi_{q,\ell_1}(\ell_1,y_1)
\tilde D_q(\ell_1,y_1).\cr
&&
\label{eq:Fq}
\end{eqnarray}
It turns out (see \cite{EvEqI}) that the MLLA corrections to
the formul\ae
\begin{equation}
\tilde D_q^g \approx \frac{C_F}{N_c}\tilde D_g^g, \quad
\tilde D_q^q \approx \frac{C_F}{N_c} \tilde D_g^q,
\label{eq:DgDq}
\end{equation}
do not modify the results and we use (\ref{eq:DgDq}) in the
following.
We rewrite accordingly (\ref{eq:Fg}) and (\ref{eq:Fq})

\vbox{
\begin{eqnarray}
x_1 F_{g}^{h_1}\left(x_1,\Theta,E,\Theta_0\right)
&\approx&
\frac{<C>_g^0 + \delta \! <C>_g}{N_c}\;\tilde D_g(\ell_1,y_1)
\equiv \frac{<C>_g}{N_c}\;\tilde D_g(\ell_1,y_1),\cr
x_1 F_{q}^{h_1}\left(x_1,\Theta,E,\Theta_0\right)
&\approx&
\frac{<C>_q^0 + \delta \! <C>_q}{N_c}\;\tilde D_g(\ell_1,y_1)
\equiv \frac{<C>_q}{N_c}\;\tilde D_g(\ell_1,y_1),\cr
&&
\label{eq:Fgq}
\end{eqnarray}
}

with 
\begin{eqnarray}
<C>_{g}^0&=&<u>^g_{g} N_c+<u>^q_{g} C_F,\cr
&&\cr
<C>_{q}^0&=&<u>^g_{q} N_c+<u>^q_{q} C_F,
\label{eq:C}
\end{eqnarray}
and where we have called
\begin{eqnarray}
\delta \!<C>_{g}&=&N_c\;\delta \!<u>^g_{g}\psi_{g,\ell_1}(\ell_1,y_1)
 +C_F\;\delta \!<u>^q_{g}\psi_{q,\ell_1}(\ell_1,y_1),\cr
\delta \!<C>_{q}&=&N_c\;\delta \!<u>^g_{q}\psi_{g,\ell_1}(\ell_1,y_1)
+C_F\;\delta \!<u>^q_{q}\psi_{q,\ell_1}(\ell_1,y_1).
\label{eq:deltaC}
\end{eqnarray}
$<C>_{A_0}$ is the average color current of partons caught by the
calorimeter.

Plugging (\ref{eq:Fgq}) into (\ref{eq:DDI}) yields the general formula
\begin{equation}
\left(\frac{d^2N}{d\ell_1\, d\ln k_\perp}\right)_{q,g}=
\frac{d}{d y_1}\left[\frac{<C>_{q,g}}{N_c}\;\tilde
D_g(\ell_1,y_1)\right]
\label{eq:gen}
\end{equation}
\medskip

The first line of (\ref{eq:Fg}) and (\ref{eq:Fq}) are the leading terms,
the second and third lines are  corrections.
Their relative order  is easily determined by the following
relations (see (\ref{eq:gamma0}) for the definition of $\gamma_0$)
\begin{eqnarray}
&& \frac{d^2N}{d\ell_1\,d\ln k_\perp} = \frac{<C>_{q,g}}{N_c}
\frac{d}{dy_1}\tilde D_g(\ell_1,y_1) 
+ \frac{1}{N_c} \tilde D_g(\ell_1,y_1) \frac{d}{dy_1}<C>_{q,g},\cr
&& \frac{d}{dy_1}\tilde D_g(\ell_1,y_1) = {\cal O}(\gamma_0)
= {\cal O}(\sqrt{\alpha_s}),\cr
&& \frac{d}{dy_1}<C>_{q,g} = {\cal O}(\gamma_0^2) = {\cal
O}(\alpha_s);
\label{eq:correc}
\end{eqnarray}
 The different contributions are discussed in subsections
\ref{subsection:colcur} and \ref{subsection:correcs} below.

\medskip

$\bullet$\ $\frac{d\tilde D_g(\ell,y)}{d\ln k_\perp}
\equiv \frac{d\tilde D_g(\ell,y)}{dy}$ (see the beginning of this section)
occurring in (\ref{eq:gen}) is plotted
in Fig.~12 and 13 of appendix \ref{section:exactsol},
and $\frac{d \tilde D_g(\ell, y)}{d\ell}$ occurring in
(\ref{eq:xF2}) (\ref{eq:psidef}) is plotted in
Figs.~14 and 15.

\medskip

$\bullet$\ The expressions for the leading terms of
$x_1 F_{A_0}^{h_1}\left(x_1,\Theta,E,\Theta_0\right)$ 
together with the ones of  $<C>_{g}^0$ and $<C>_{q}^0$ are given in
appendix \ref{section:leadingxF}.

\medskip

$\bullet$\ The calculations of $\delta \! <C>_g$ and $\delta \! <C>_q$ are detailed
in appendix \ref{section:udeltau}, where the explicit analytical expressions
for the $<u>$'s and $\delta <u>$'s are also given.

\medskip

We call ``naive'' the approach'' in which
one disregards the evolution of the jet between $\Theta_0$ and
$\Theta$; this amounts to taking to zero the derivative of $<C>_{q,g}$ in
(\ref{eq:gen});
(\ref{eq:Cquark}), (\ref{eq:Cgluon}), (\ref{eq:c1c2c3}) then yield
\begin{equation}
<C>_g^{naive} = N_c,\quad <C>_q^{naive}= C_F.
\label{eq:Cnaive}
\end{equation}

\vskip .7cm

\subsection{The average color current $\boldsymbol{<C>_{A_0}}$}
\label{subsection:colcur}
%%%%%%%%%%%%%%%%%%%%%%%%%%%%%%%%%%%%%%%%%%%%%%%%%%%%%%%%%%%%%%%%

\vskip .4cm

On Fig.~2 below, we plot, for $Y_{\Theta_0}=7.5$,
  $<C>_q^0$, $<C>_q^0 + \delta\!<C>_q$, $<C>_g^0$, $<C>_g^0 + \delta\!<C>_g$
as functions of $y$, for $\ell = 2.5$ on the left and $\ell = 3.5$ on the
right. Since $\Theta \leq \Theta_{0}$, the curves
stop at $y$ such that $y + \ell = Y_{\Theta_{0}}$; they reach then their
respective asymptotic values $N_c$ for $<C>_g$ and $C_F$ for $<C>_q$, at
which $\delta\! <C>_q$ and $\delta\! <C>_g$ also vanish (see also
the naive approach (\ref{eq:Cnaive})).
These corrections also
vanish at $y=0$ because they are proportional to the logarithmic
derivative $(1/\tilde D(\ell,y))(d\tilde D(\ell,y)/d\ell)$
(see (\ref{eq:deltaC}))
which both vanish, for $q$ and $g$,
at $y=0$ (see appendix \ref{section:exactsol}, and Figs.~16-17); there, the
values of $<C>_g$ and $<C>_q$ can be determined from
(\ref{eq:Cquark})(\ref{eq:Cgluon}).

The curves corresponding to LEP and Tevatron working conditions,
$Y_{\Theta_0}=5.2$, are shown in appendix \ref{section:LEP}.

\bigskip

\vbox{
\begin{center}
\epsfig{file=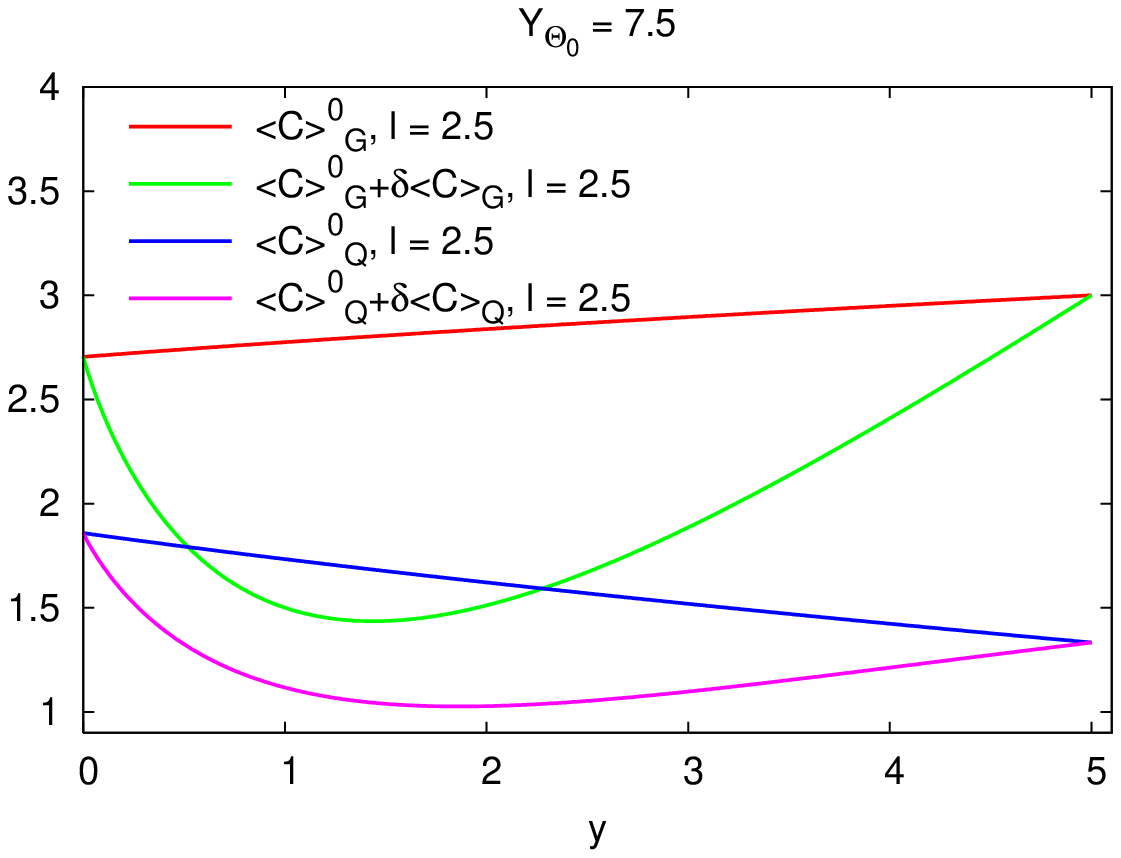, height=5truecm,width=7.5truecm}
\hfill
\epsfig{file=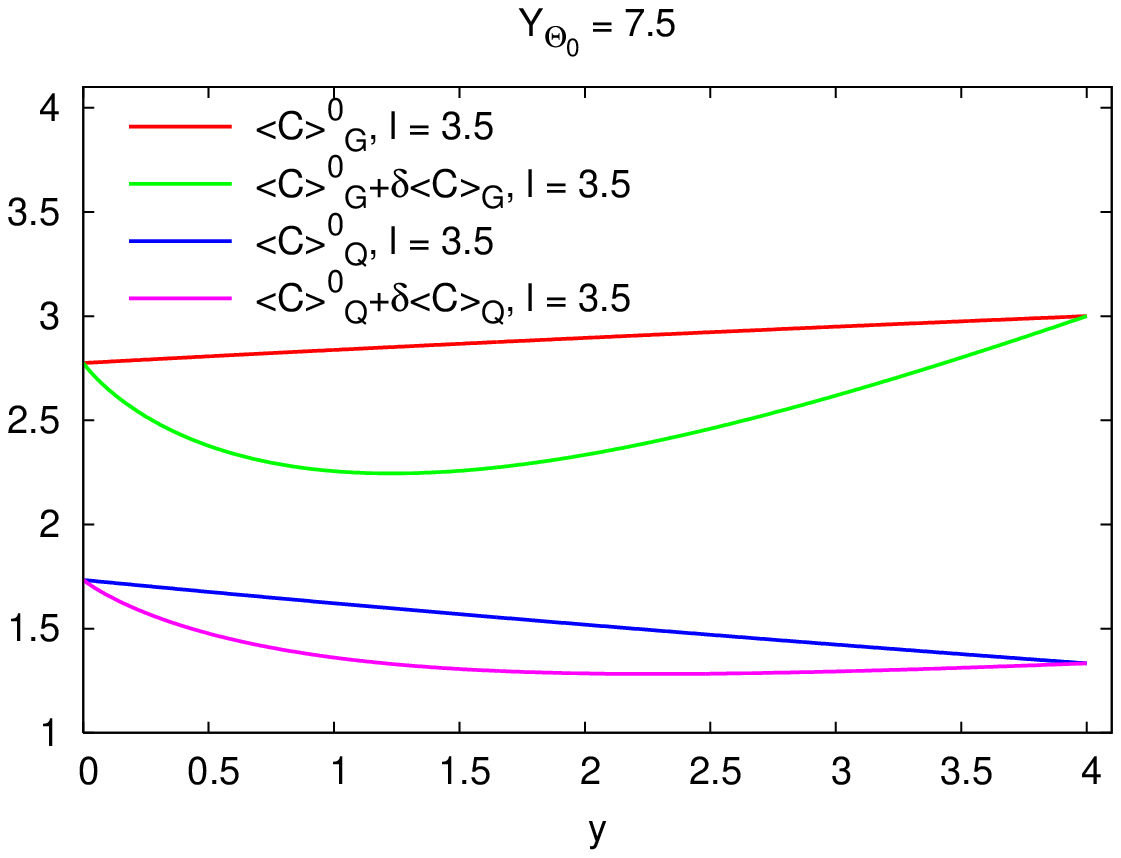, height=5truecm,width=7.5truecm}
\end{center}

\centerline{\em Fig.~2: $<C>_{A_0}^0$ and $<C>_{A_0}^0 + \delta\!<C>_{A_0}$ for quark and gluon
jets, as functions of $y$,}

\centerline{\em for $Y_{\Theta_0}=7.5$, $\ell=2.5$ on the left and
$\ell=3.5$ on the right.}
}

\bigskip

Two types of MLLA corrections arise in our calculation, which are
easily visualized on Fig.~2:

$\ast$\ through the expansion (\ref{eq:rhodev}) around $u=1$,
the average color current $<C>^0_{A_0}$ gets modified by $\delta\!<C>_{A_0}
\leq 0$
of relative order ${\cal O}(\sqrt{\alpha_s})$; it is represented on Fig.~2
by the vertical difference between the straight lines ($<C>_{A_0}^0$) and
the curved ones ($<C>^0_{A_0} +\delta\!<C>_{A_0}$);

$\ast$\ the derivative of $<C>^0_{A_0}$ with respect to $y$ is itself of
relative order ${\cal O}(\sqrt{\alpha_s})$ with respect to that of $\tilde
D_g$; it is the slopes of the straight lines in Fig.~2.

The $y$ derivatives of $<C>_{A_0}^0 +
\delta\!<C>_{A_0}$  differ from the ones of the leading $<C>^0_{A_0}$;
this effect combines the two types of MLLA corrections mentioned above:
the derivation of $<C>$ with respect to $y$ and the existence of
$\delta\!<C>$.

For  $Y_{\Theta_0}=7.5$, the $\delta\!<C>$ correction can represent
$50\%$ of $<C>_{g}$ at  $\ell=2.5$ and $y\approx 1.5$; for
higher values of $\ell$ (smaller $x$), as can be seen on the right figure,
its importance decreases; it is remarkable that, when $\delta\!<C>$ is
large, the corrections to $\frac{d<C>}{dy}$ with respect to
$\frac{d<C>^0}{dy}$ become small, {\em and vice-versa}: at both extremities
of the curves for the color current, the $\delta\!<C>$ corrections vanish,
but their slopes are very different from the ones of the
 straight lines corresponding to $<C>^0$.

So, all corrections that we have uncovered are potentially
large, even $\frac{d\delta\!<C>}{dy}$, which is the $y$ derivative of a
MLLA corrections. This raises the question of the validity of our
calculations. Several conditions need to be fulfilled at the same time:

$\ast$\ one must stay in the perturbative regime, which needs $y_1\geq 1$
($k_\perp > 2.72 \Lambda_{QCD} \approx .7\,$GeV;
this condition excludes in particular the zone of very large increase of
$\frac{d^2N}{d\ell_1\,d\ln k_\perp}$ when $y_1 \to 0$ (this property is
linked to the divergence of the running
coupling constant of QCD $\alpha_s(k_\perp^2) \to \infty$
when $k_\perp \to \Lambda_{QCD}$).

$\ast$\ $x$ must be small, that is $\ell$ large enough, since this is the
limit at which we have obtained analytical results; we see on Fig.~2 that
it cannot go reasonably below $\ell = 2.5$;
this lower threshold turns out to be of the same order magnitude as the one found
in the forthcoming study of 2-particle correlations
inside one jet in the MLLA approximation \cite{Perez2};

$\ast$\ (MLLA) corrections to the leading behavior must stay under control
(be small ``enough''); if one only looks at the size of the $\delta\!<C>$
corrections at $Y_{\Theta_0}=7.5$, it would be very tempting to
exclude $y\in [.5, 2.5]$; however this is
without taking into account the $y$ derivatives of $<C>$, which also play an
important role, as stressed above;
our attitude, which will be confirmed or not by experimental results, is to
only globally constrain the overall size of all corrections by setting $x$
small enough.

\bigskip

Would the corrections become
excessively large, the expansion (\ref{eq:rhodev}) should be pushed one
step further, which corresponds to next-to-MLLA (NMLLA) corrections; this
should then be associated with NMLLA evolution equations for the inclusive
spectrum, which lies out of the scope of the present work. 

\medskip

Though $\delta\!<C>$ can be large, specially at small
values of $\ell$, the positivity of $<C>^0 + \delta\! <C>$ is always
preserved on the whole allowed range of $y$.

\medskip

The difference between the naive and MLLA calculations lies in 
neglecting or not  the evolution of the jet between $\Theta_0$ and
$\Theta$, or, in practice, in considering or not the average
color current $<C>_{A_0}$ as a constant.

\vskip .5 cm

We present below our results for a gluon and for a quark jet.
We choose two values  $Y_{\Theta_0} = 7.5$,  which can be associated with
the LHC environment
\footnote{Sharing equally the $14$ TeV of available center of mass energy
between the six constituent partons of the two colliding nucleons yields
$E\approx 2.3$ TeV by colliding parton, one considers a jet
opening angle of $\Theta \approx .25$ and $Q_0\approx\Lambda_{QCD}
\approx 250$\,MeV; this gives $Y=\ln\frac{E\Theta}{Q_0}\approx 7.7$.
\label{footnote:LHC}}
, and the unrealistic $Y_{\Theta_0}= 10$ (see appendix \ref{section:LEP}
for  $Y_{\Theta_0}=5.2$ and $5.6$, corresponding to the LEP and Tevatron
 working conditions).
For each value of $Y_{\Theta_0}$ we  make the plots for two values
of $\ell_1$, and compare one of them with the naive approach.

In the rest of the paper we always consider the limiting case
$Q_0 \to \Lambda_{QCD} \Leftrightarrow \lambda \approx 0$,
\begin{equation}
\lambda = \ln\frac{Q_0}{\Lambda_{QCD}}.
\label{eq:lambda}
\end{equation}

The curves stop at their kinematic limit $y_{1\,max}$ such that
$y_{1\,max} + \ell_1 = Y_{\Theta_0}$.

\vskip .7 cm

\subsection{$\boldsymbol{\displaystyle\frac{d^2N}{d\ell_1\,d\ln{ k_\perp}}}$
at small $\boldsymbol x_1$: gluon jet}
%%%%%%%%%%%%%%%%%%%%%%%%%%%%%%%%%%%%%%%%%%%%%%%%%%%%%%%%%%%%%%%%%%%%%%%%%%%%%%

\vskip .4cm

On Fig.~3 below is plotted the double differential distribution
$\frac{d^2N}{d\ell_1\,d\ln{ k_\perp}}$ of a parton
inside a gluon jet as a function of $y_1$ for different values of $\ell_1$
(fixed).

\vbox{
\begin{center}
\epsfig{file=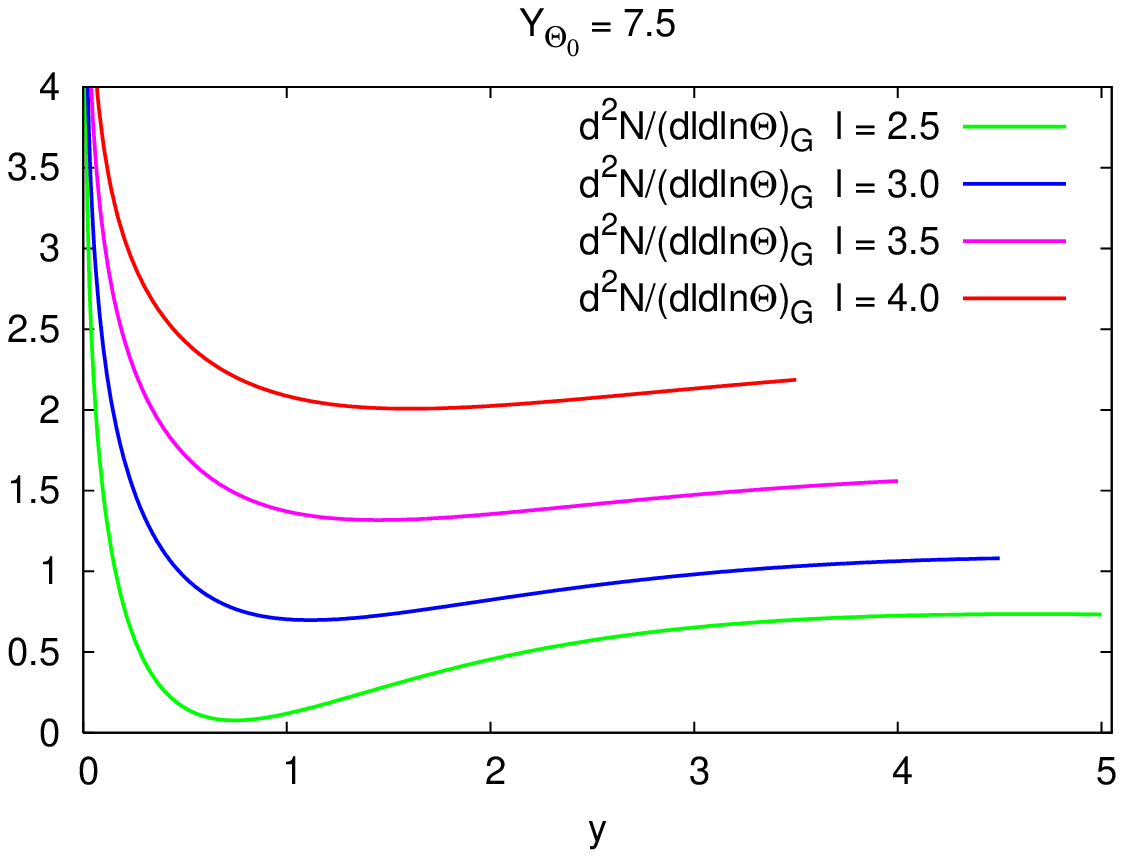, height=5truecm,width=7.5truecm}
\hfill
\epsfig{file=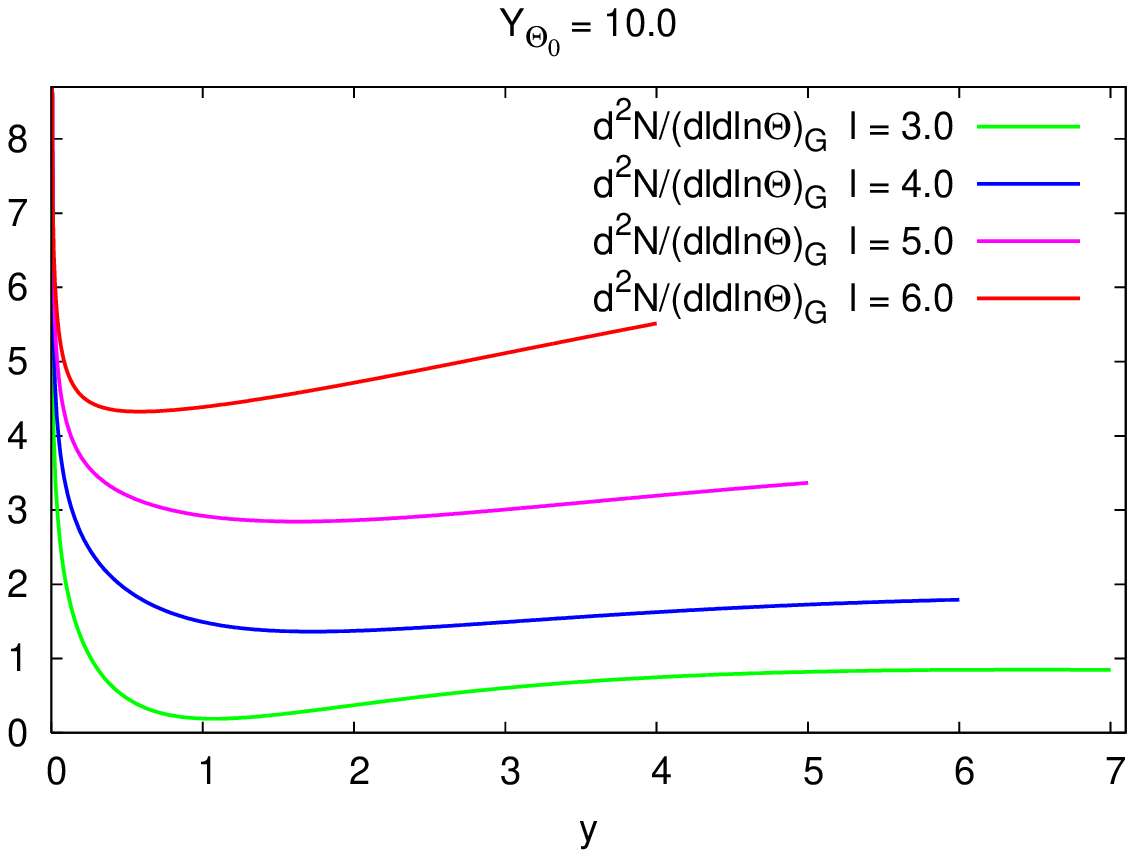, height=5truecm,width=7.5truecm}
\end{center}

\centerline{\em Fig.~3: $\frac{d^2N}{d\ell_1\;d\ln k_\perp}$ for a gluon jet.}
}

\bigskip
\bigskip

On Fig.~4 are compared, for a given value of  $\ell_1$,
the two following cases:

\medskip

$\ast$ the first corresponds to the full formul{\ae} (\ref{eq:Fgq})
(\ref{eq:gen});

$\ast$ the second corresponds to the naive approach (see the definition
above (\ref{eq:Cnaive}))
\begin{equation}
 \left(\frac{d^2N}{d\ell_1\,d\ln{ k_\perp}}\right)^{naive}_{g}
=\frac{d}{dy_1}\tilde D_g(\ell_1,y_1);
\label{eq:EAgnaive}
\end{equation}
$\displaystyle\frac{d\tilde D_g(\ell_1,y_1)}{dy_1}$ is given in (\ref{eq:derivy}).

\vbox{
\begin{center}
\epsfig{file=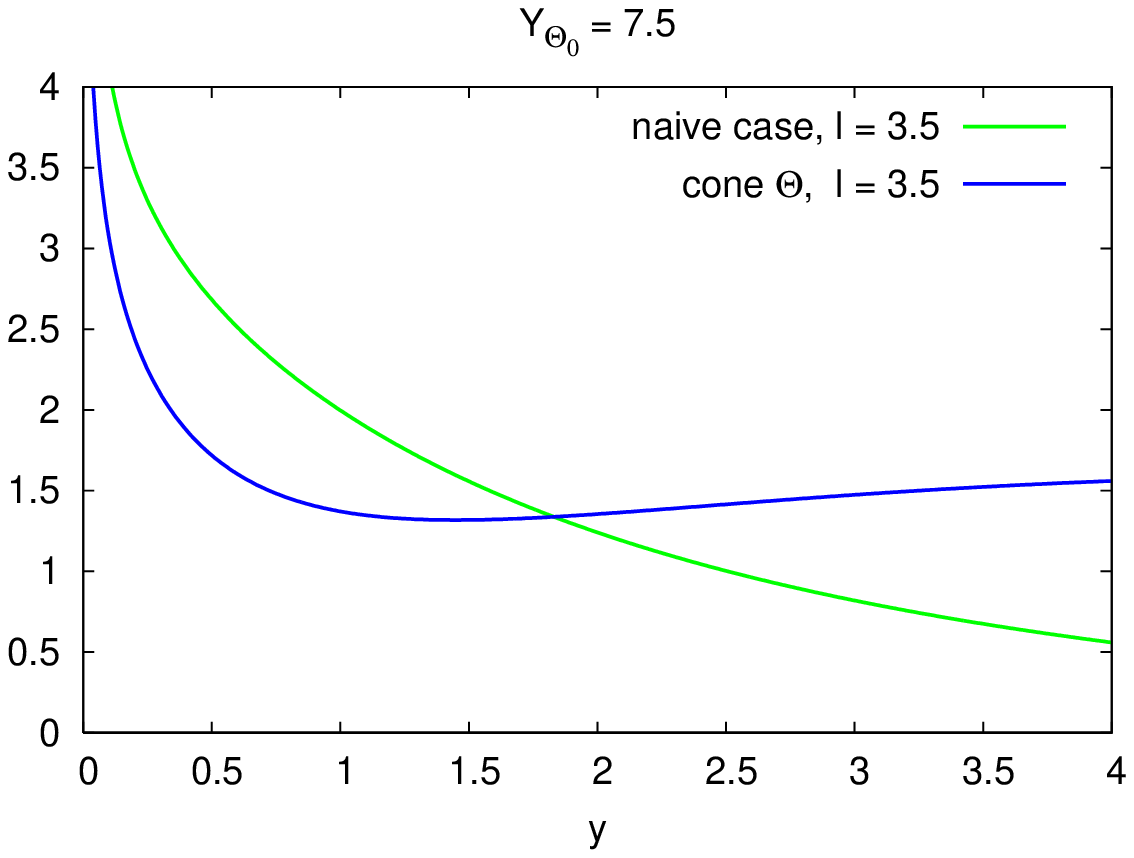, height=5truecm,width=7.5truecm}
\hfill
\epsfig{file=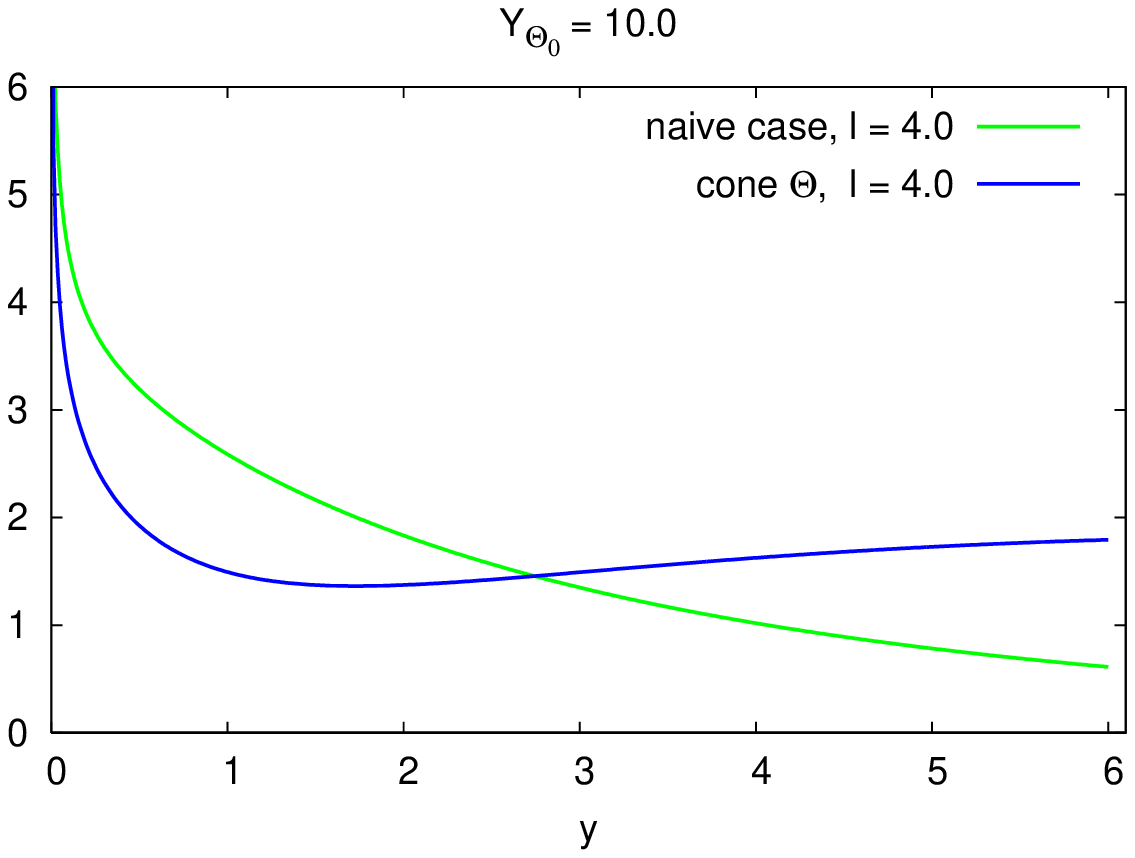, height=5truecm,width=7.5truecm}
\end{center}

\centerline{\em Fig.~4: $\frac{d^2N}{d\ell_1\;d\ln k_\perp}$ for a gluon jet
at fixed $\ell_1$,  MLLA and naive approach.}
}

\bigskip
\bigskip

The raise of the distribution at large $k_\perp$ is due to the positive
corrections already mentioned in the beginning of this section, which
arise from the evolution of the jet between $\Theta$ and $\Theta_0$.

\vskip .7 cm

\subsection{$\boldsymbol{\displaystyle\frac{d^2N}{d\ell_1\,d\ln{ k_\perp}}}$
at small $\boldsymbol x_1$: quark jet}
%%%%%%%%%%%%%%%%%%%%%%%%%%%%%%%%%%%%%%%%%%%%%%%%%%%%%%%%%%%%%%%%%%%%%%%%%%%%%

\vskip .4cm

On Fig.~5 is plotted
the double differential distribution
$\frac{d^2N}{d\ell_1\,d\ln{ k_\perp}}$ of a parton
inside a quark jet as a function of $y_1$ for different values of $\ell_1$
(fixed).

\vbox{
\begin{center}
\epsfig{file=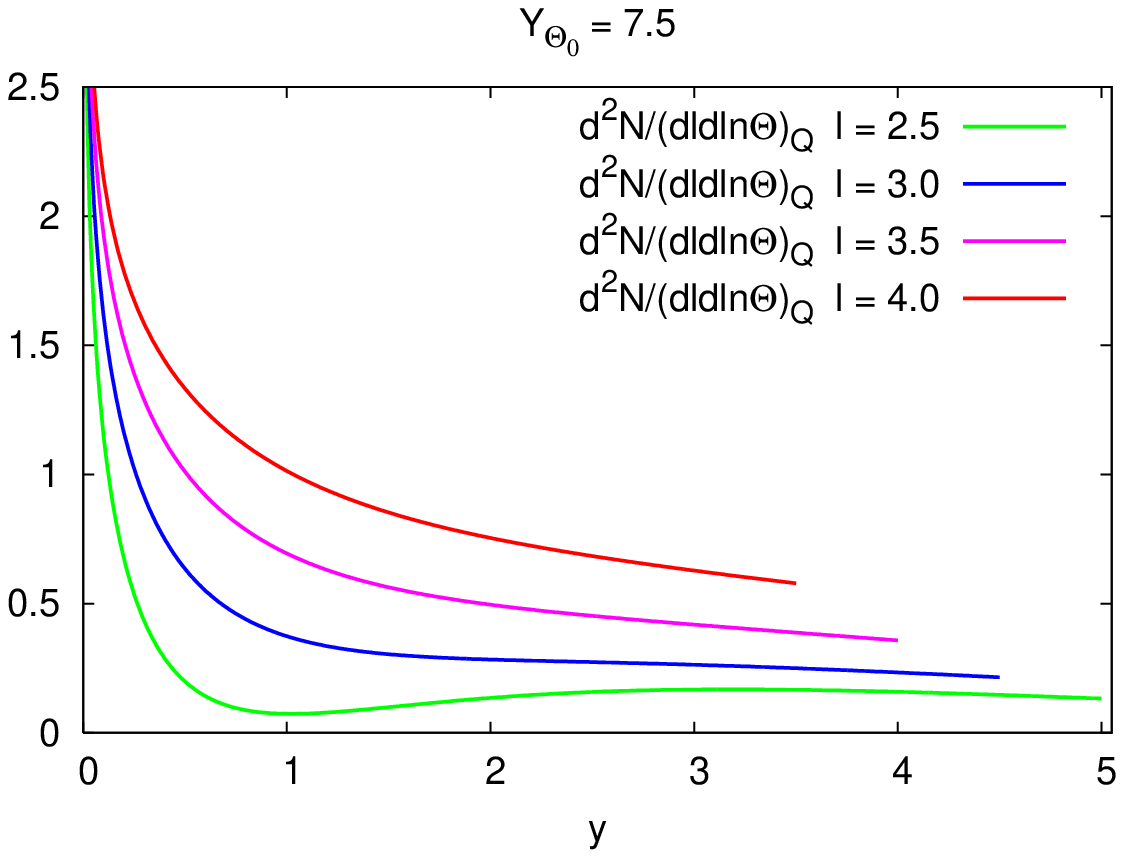, height=5truecm,width=7.5truecm}
\hfill
\epsfig{file=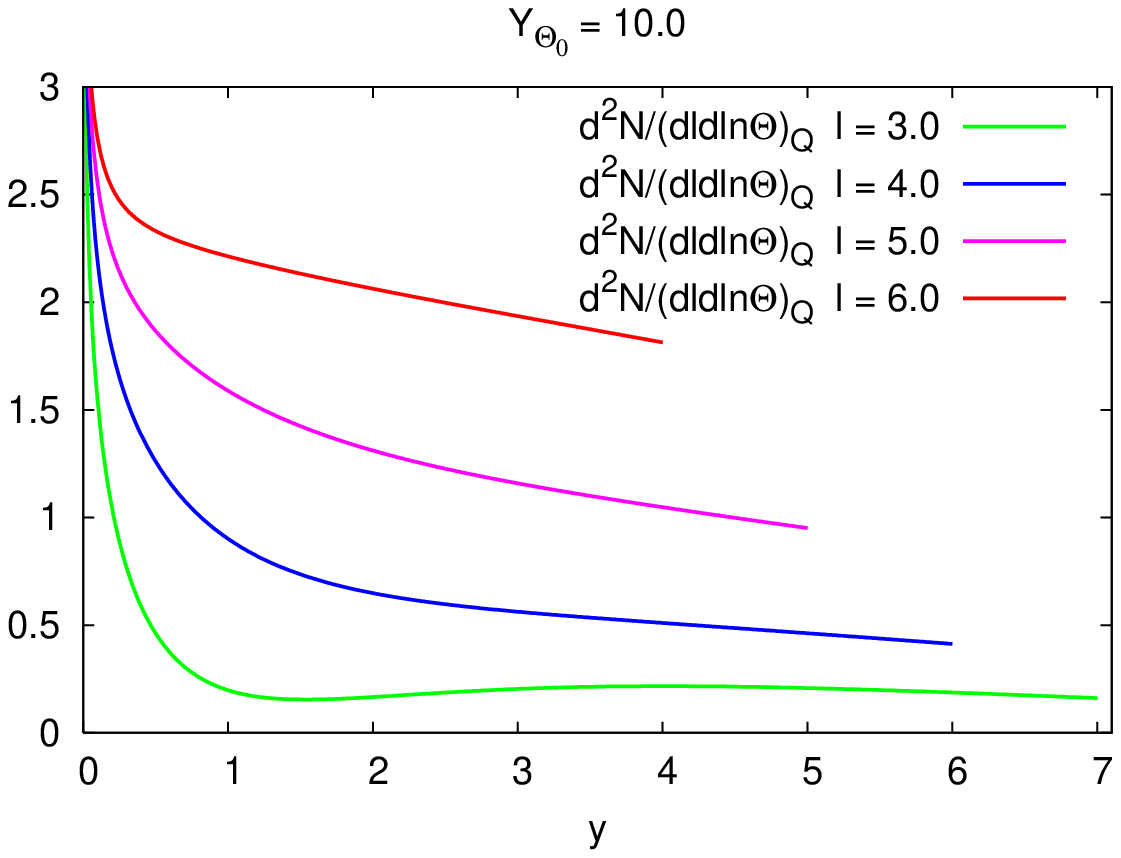, height=5truecm,width=7.5truecm}
\end{center}

\centerline{\em Fig.~5: $\frac{d^2N}{d\ell_1\;d\ln k_\perp}$ for a quark jet.}

}

\bigskip
\bigskip

On Fig.~6 are compared, for a given $\ell_1$ fixed,
the full formul{\ae} (\ref{eq:Fgq})
(\ref{eq:gen})  and the naive approach
\begin{equation}
 \left(\frac{d^2N}{d\ell_1\,d\ln k_\perp}\right)_{q}^{naive}
=\frac{C_F}{N_c}\frac{d}{dy_1} \tilde D_g(\ell_1,y_1).
\label{eq:EAqnaive}
\end{equation}

\vbox{
\begin{center}
\epsfig{file=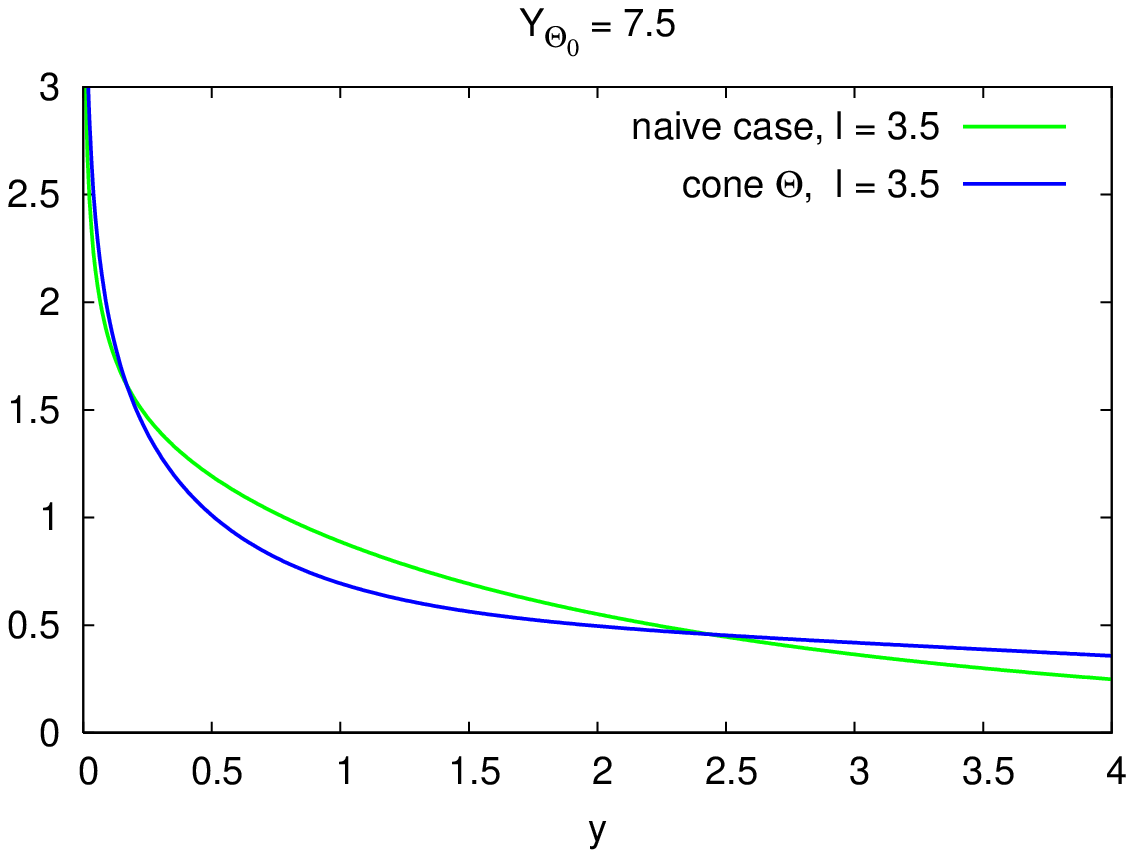, height=5truecm,width=7.5truecm}
\hfill
\epsfig{file=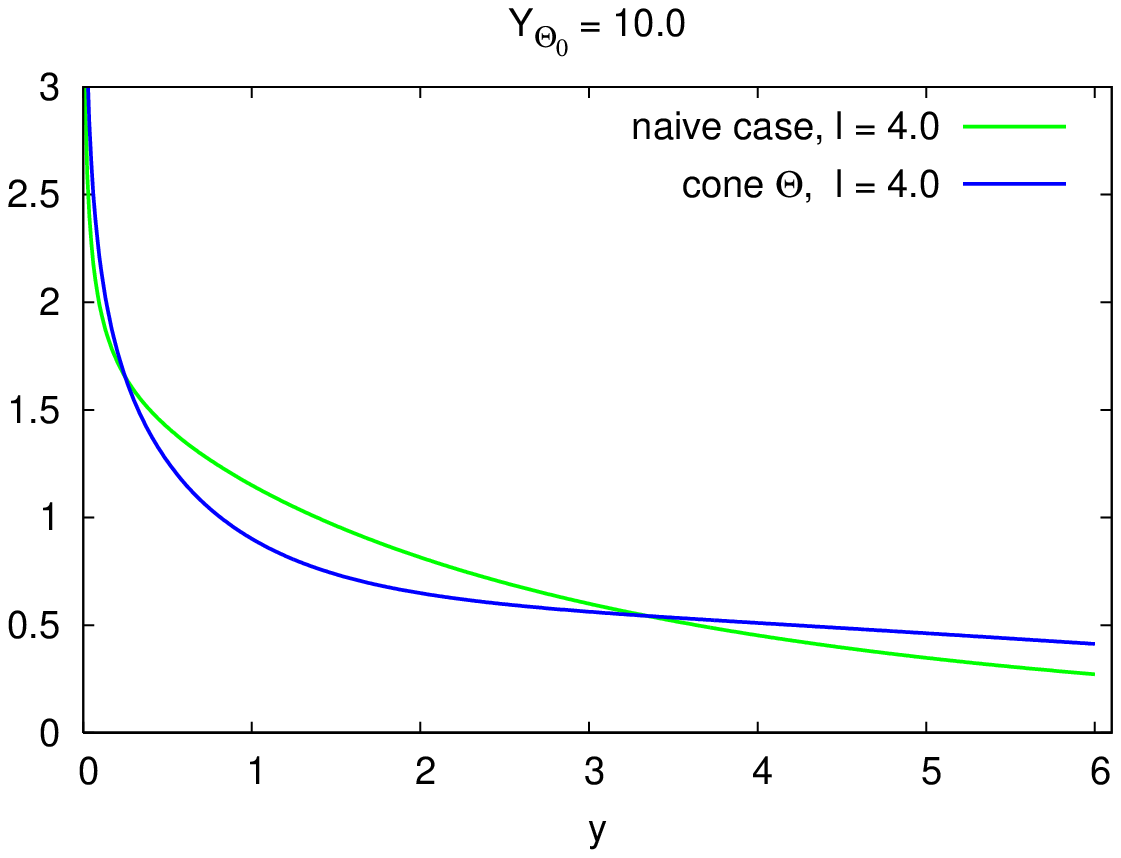, height=5truecm,width=7.5truecm}
\end{center}

\centerline{\em Fig.~6: $\frac{d^2N}{d\ell_1\;d\ln k_\perp}$ for a quark jet
at fixed $\ell_1$,  MLLA and naive approach.}

}

\bigskip
\bigskip

We note, like for gluon jets, at large $y$, a (smaller) increase of the
distribution,
due to taking into account the jet evolution between $\Theta$ and
$\Theta_0$.

\vskip .7 cm

\subsection{Comments}
\label{subsection:correcs}
%%%%%%%%%%%%%%%%%%%%%%%%%%%%%

\vskip .4cm

The gluon distribution is always larger than the quark distribution; this
can also be traced in Fig.~2  which
measures in particular the ratio of the color currents $<C>_g/<C>_q$.

\medskip

The curves for  $\frac{d^2N}{d\ell_1\,d\ln{ k_\perp}}$
have been drawn for $\ell_1 \equiv \ln (1/x_1) \geq 2.5$; going below this
threshold exposes to excessively large MLLA corrections.

\medskip

The signs of the two types of MLLA corrections pointed at in subsection
\ref{subsection:colcur} vary with $y$: $\delta\!<C>$
always brings a negative correction to $<C>^0$, and to
$\frac{d^2N}{d\ell_1\,d\ln{ k_\perp}}$;  for $y \geq 1.5$,
the slope of $<C>$ is always larger that the one of
$<C>^0$, while for $y \leq 1.5$ it is the opposite. 
 It is accordingly not surprising
that, on Figs.~4 and 6, the relative positions of the curves corresponding to
the MLLA calculation and  to a naive calculation change with the value of $y$.
At large $y$, one gets a growing behavior
 of $\frac{d^2N}{d\ell_1\,d\ln{ k_\perp}}$ for gluon jets (Fig.~4),
and a slowly decreasing one for quark jets (Fig.~6), which could not have
been anticipated {\em a priori}.

\medskip

We study in appendix \ref{subsection:doubleDLA}, how  MLLA results
compare with DLA \cite{DLAI} \cite{DLA1},
in which the running of $\alpha_s$ has been ``factored out''.

\vskip .7 cm

%%%%%%%%%%%%%%%%%%%%%%%%%%%%%%%%%%%%%%%%%%%%%%%%%%%%%%%%
\section{INCLUSIVE $\boldsymbol{k_\perp}$ DISTRIBUTION
$\boldsymbol{\displaystyle\frac{dN}{d\ln k_\perp}}$}
\label{section:ktdist}
%%%%%%%%%%%%%%%%%%%%%%%%%%%%%%%%%%%%%%%%%%%%%%%%%%%%%%%%

\vskip .4cm

Another quantity of interest is the  inclusive $k_\perp$ distribution which
is defined by
\begin{equation}
\left(\frac{dN}{d\ln k_\perp}\right)_{g\ or\ q} = \int dx_1
\left(\frac{d^2N}{dx_1\, d\ln k_\perp}\right)_{g\ or\ q}
\equiv \int_{\ell_{min}}^{Y_{\Theta_0}-y} d\ell_1
\left(\frac{d^2N}{d\ell_1\, d\ln k_\perp}\right)_{g\ or\ q};
\label{eq:ktdist}
\end{equation}
it measures the transverse momentum distribution of one particle with
respect to the direction of the energy flow (jet axis).

We have introduced in (\ref{eq:ktdist}) a lower bound of integration
$\ell_{min}$ because our calculations are valid for small $x_1$, that is for
large $\ell_1$. In a first step we take $\ell_{min}=0$, then vary it to
study the sensitivity of the calculation to the region of large $x_1$.

We plot below the inclusive $k_\perp$ distributions for gluon and quark
jets, for the same two values $Y_{\Theta_0}=7.5$ and
$Y_{\Theta_0}=10$ as above, and compare them, on the same graphs, with the
``naive calculations'' of the same quantity.

\vskip .7 cm

\subsection{Gluon jet; $\boldsymbol{\ell_{min}=0}$}
%%%%%%%%%%%%%%%%%%%%%%%%%%%%%%%%%%%%%%%%%%%%%%%%%%%

\vskip .4cm

\vbox{
\begin{center}
\epsfig{file=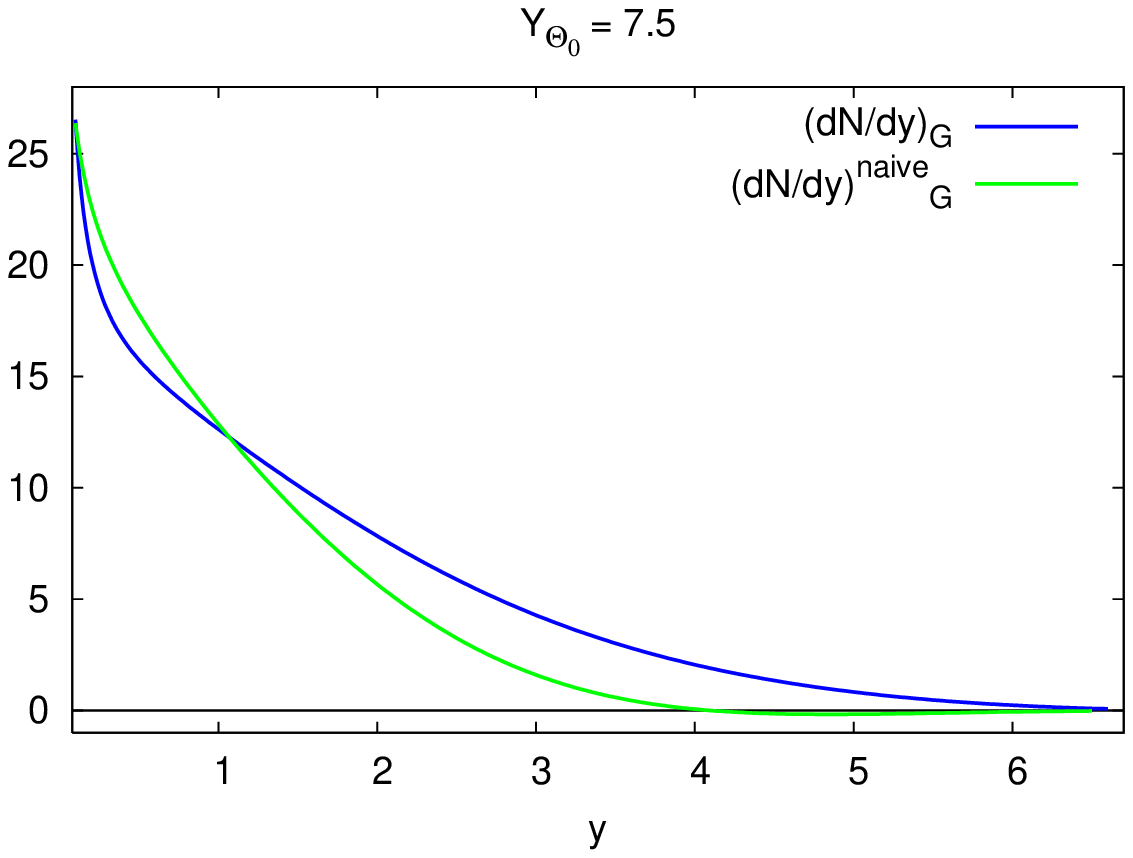, height=5truecm,width=7.5truecm}
\hfill
\epsfig{file=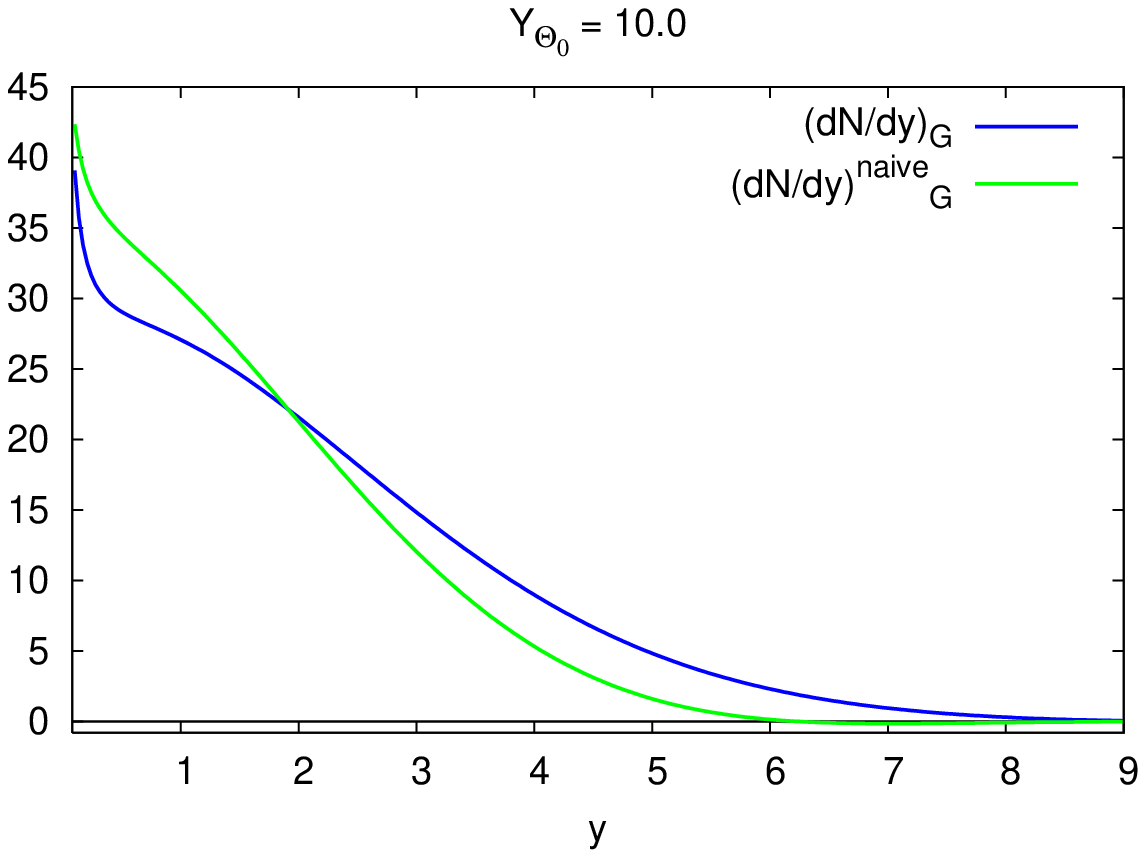, height=5truecm,width=7.5truecm}
\end{center}

\centerline{\em Fig.~7:  $\frac{d{N}} {d\ln k_\perp}$  for a gluon jet,
MLLA and naive approach,}

\centerline{\em for $\ell_{min=0}$,
$Y_{\Theta_0} =7.5$ and $Y_{\Theta_0}=10$.}
}

\vskip .3cm

\vbox{
\begin{center}
\epsfig{file=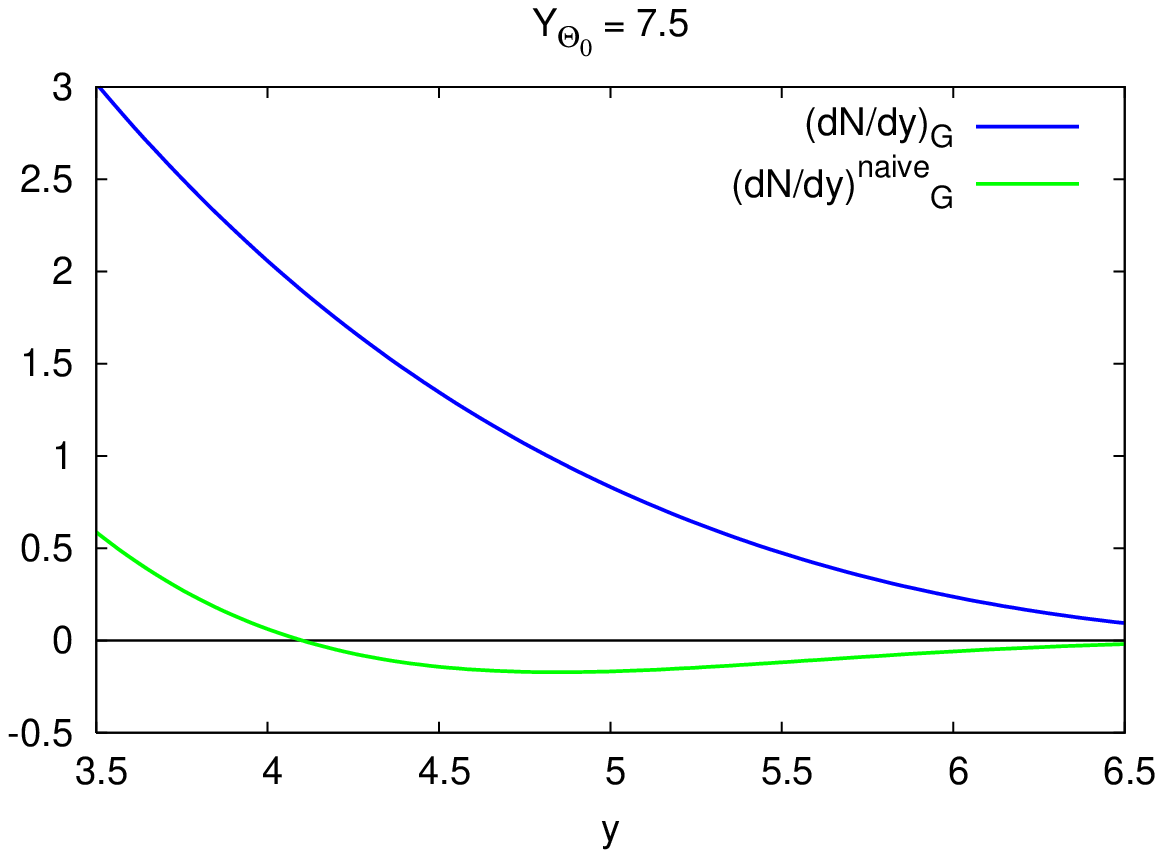, height=5truecm,width=7.5truecm}
\hfill
\epsfig{file=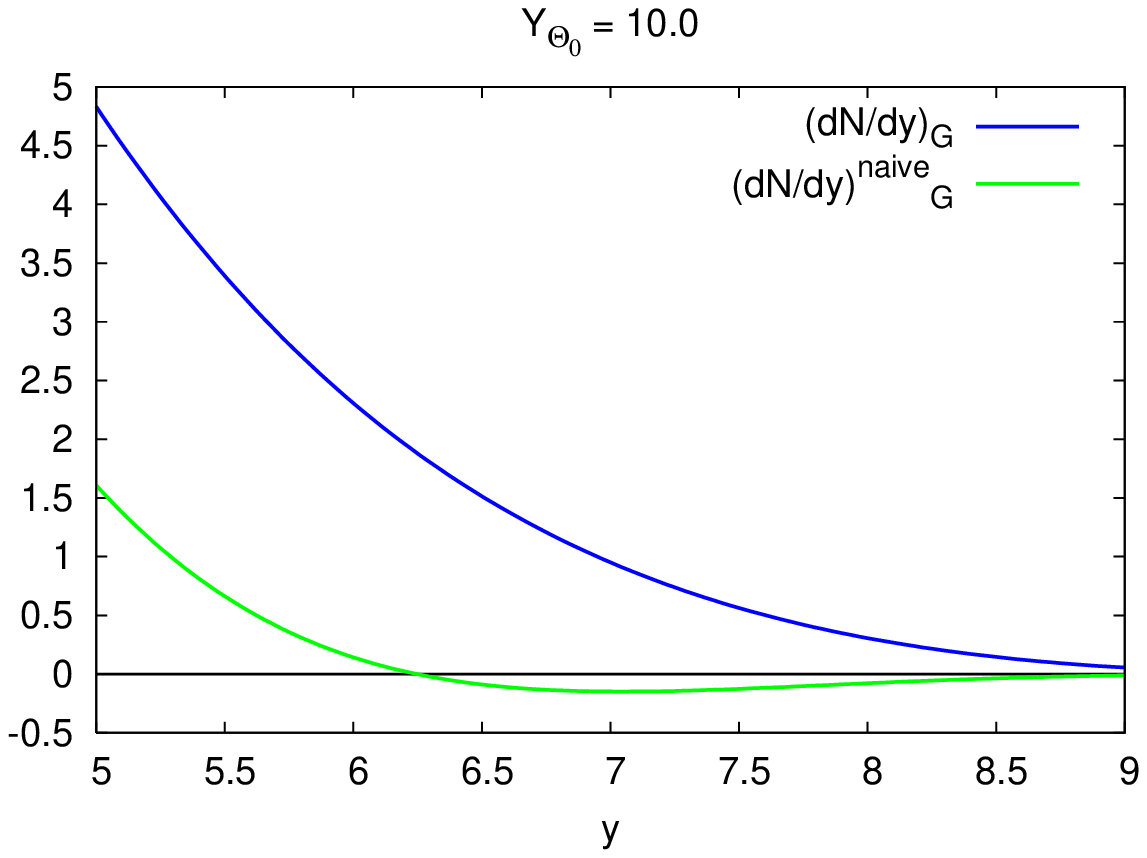, height=5truecm,width=7.5truecm}
\end{center}

\centerline{\em Fig.~8: enlargements of Fig.~7 at large $k_\perp$}
}

\vskip .7 cm

\subsection{Quark jet; $\boldsymbol{\ell_{min}=0}$}
%%%%%%%%%%%%%%%%%%%%%%%%%%%%%%%%%%%%%%%%%%%%%%%%%%%

\vskip .4cm

%\vbox{
\begin{center}
\epsfig{file=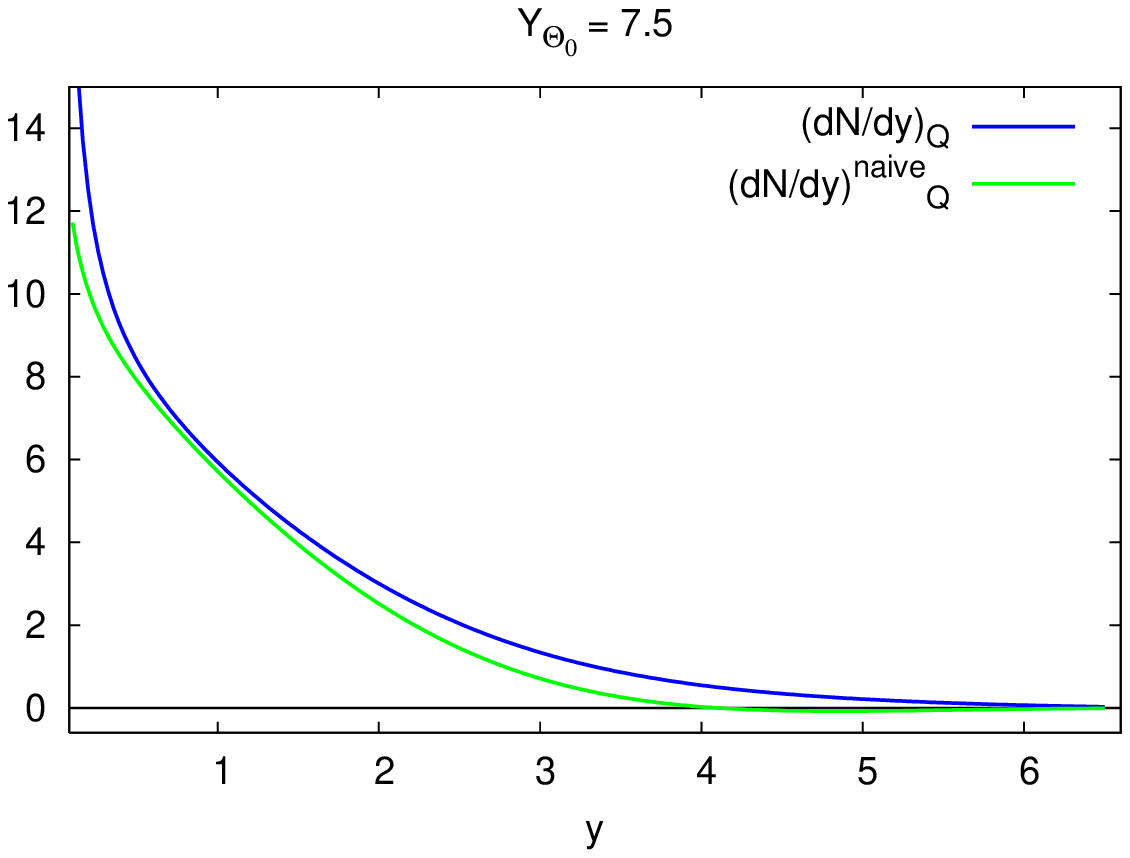, height=5truecm,width=7.5truecm}
\hfill
\epsfig{file=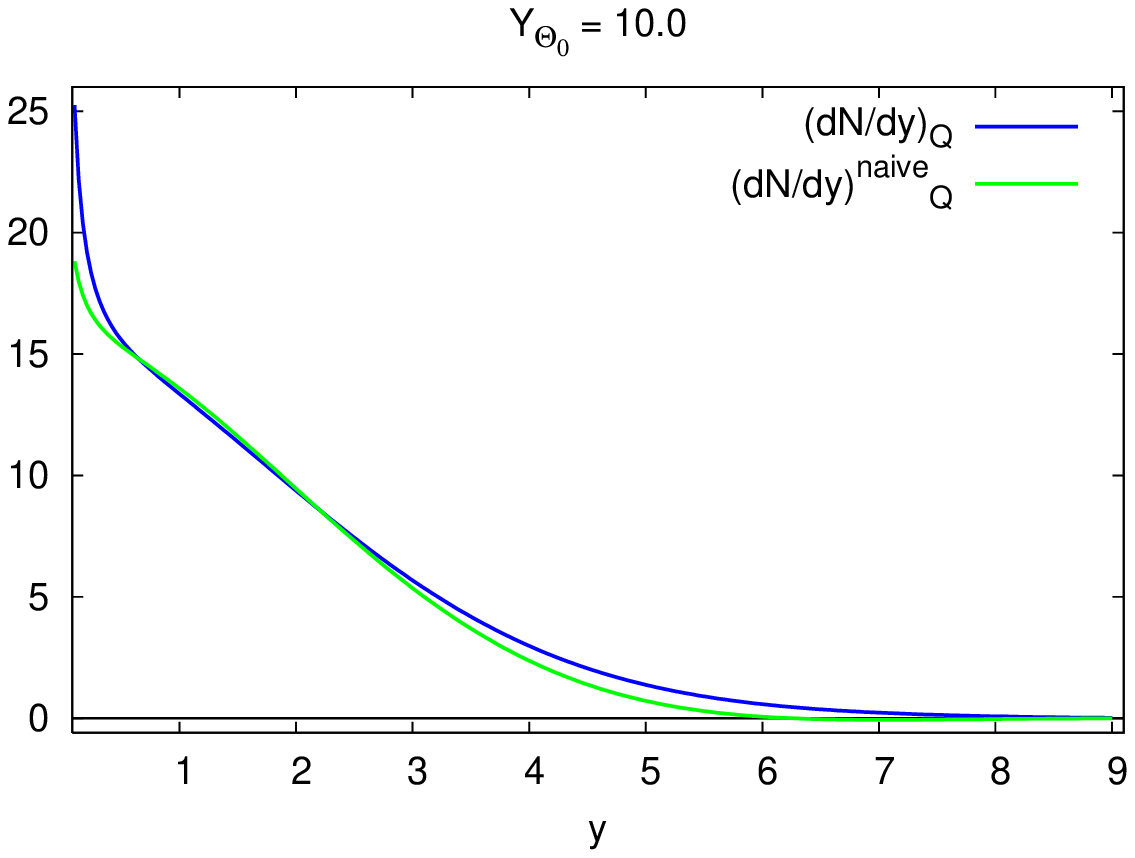, height=5truecm,width=7.5truecm}
\end{center}

\centerline{\em Fig.~9:  $\frac{d{N}} {d\ln k_\perp}$ for a quark jet,
MLLA and naive approach,}

\centerline{\em  for $\ell_{min=0}$,
$Y_{\Theta_0} =7.5$ and $Y_{\Theta_0}=10$.}
%}

\vskip .3cm

%\vbox{
\begin{center}
\epsfig{file=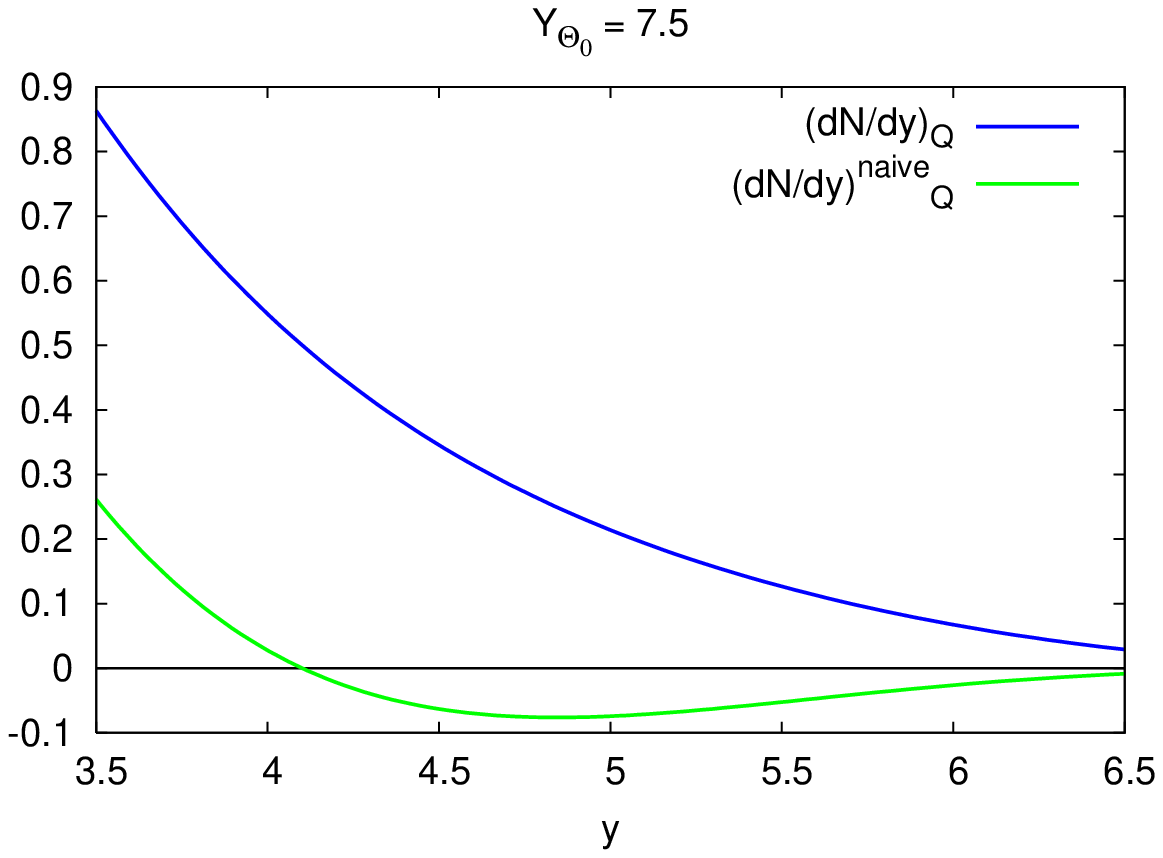, height=5truecm,width=7.5truecm}
\hfill
\epsfig{file=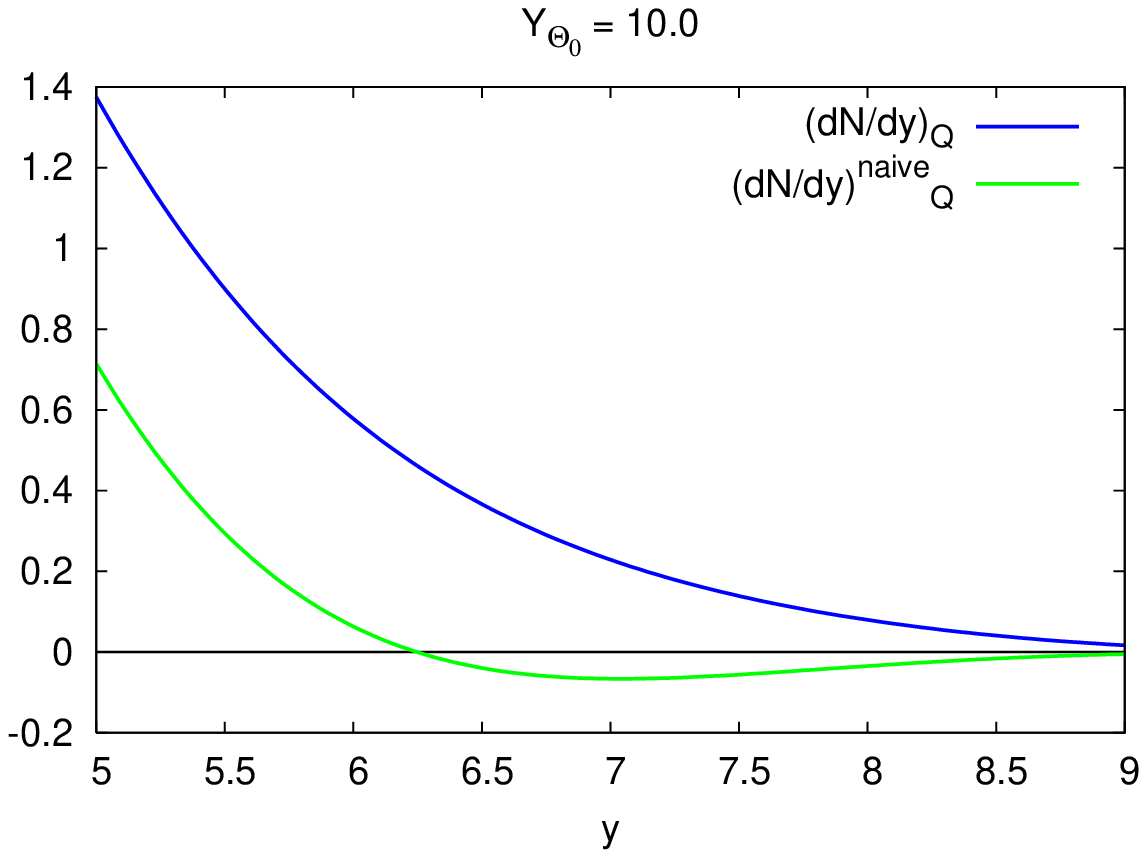, height=5truecm,width=7.5truecm}
\end{center}

\centerline{\em Fig.~10: enlargements of Fig.~9 at large $k_\perp$}
%}

\vskip .7 cm

\subsection{Role of the lower limit of integration $\boldsymbol{\ell_{min}}$}
%%%%%%%%%%%%%%%%%%%%%%%%%%%%%%%%%%%%%%%%%%%%%%%%%%%%%%%%%%%%%%%%%%%%%%%%%%%%%

\vskip .4cm

To get an estimate of the sensitivity of the calculation of
$\frac{dN}{d\ln\,k_\perp}$ to the lower bound of integration
in (\ref{eq:ktdist}), we plot in Fig.~11 below the two results obtained at
$Y_{\Theta_0}=7.5$ for $\ell_{min}=2$ and $\ell_{min}=0$, for a gluon jet
(left) and a quark jet (right).

\bigskip

\vbox{
\begin{center}
\epsfig{file=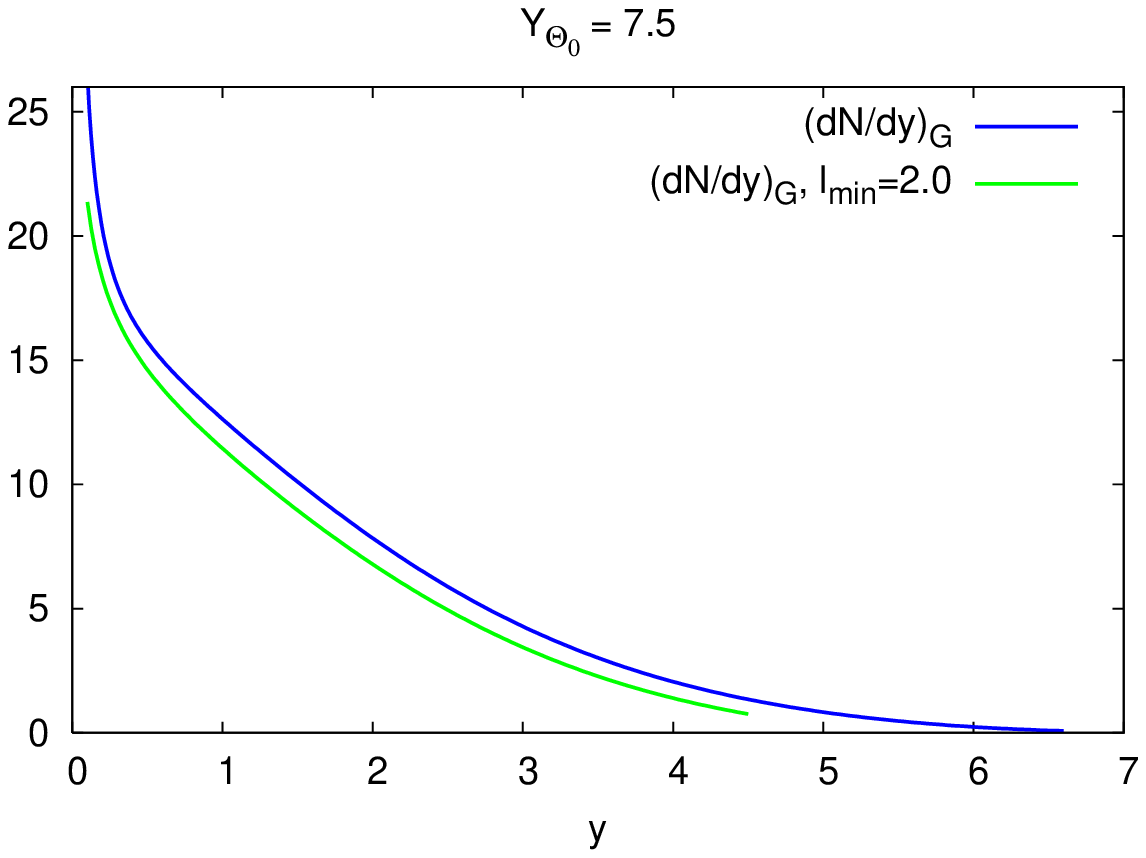, height=5truecm,width=7.5truecm}
\hfill
\epsfig{file=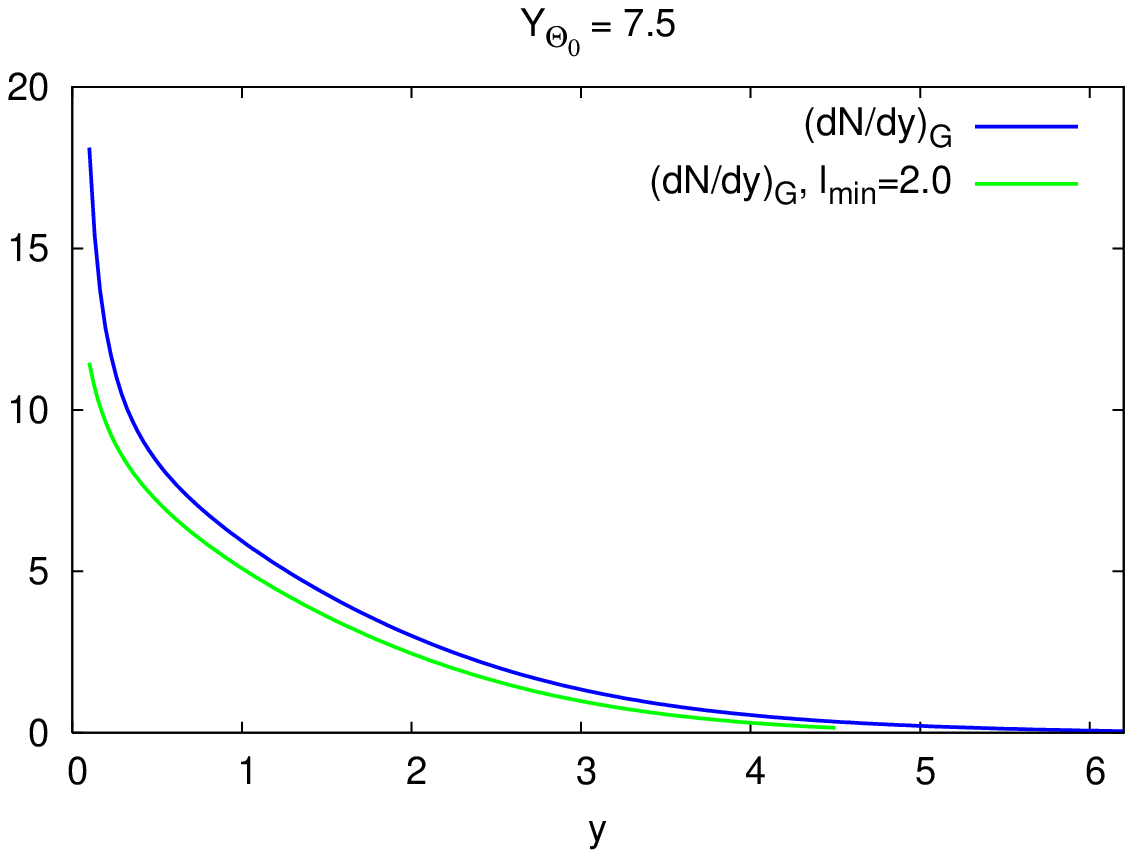, height=5truecm,width=7.5truecm}
\end{center}

\centerline{\em Fig.~11: $\frac{dN}{d\ln\,k_\perp}$ with $\ell_{min}=2$ and
$\ell_{min}=0$}
\centerline{\em for gluon (left) and quark (right) jet.}
}

\bigskip
\bigskip

The shapes of the corresponding distributions are identical; they only
differ by a vertical shift which is small in the perturbative region
$y \geq 1$ (restricting the domain of integration -- increasing
$\ell_{min}$ -- results as expected in a decrease of $\frac{dN}{d\ln
k_\perp}$).
This shows that, though our calculation is only valid at
small $x_1$, the sensitivity of the final result to this parameter is small.

\vskip .7 cm

\subsection{Discussion}
\label{subsection:discuss}
%%%%%%%%%%%%%%%%%%%%%%%%%%%%

\vskip .4cm

MLLA corrections are seen on Fig.~8 and Fig.~10 to cure the problems
of positivity which occur in the naive approach.

The range of $\ell_1$ integration in the definition (\ref{eq:ktdist}) of
$\frac{dN}{d\ln k_\perp}$ should be such that, at least,
 its upper bound corresponds to $x_1$ small enough; we have seen in the
discussion of MLLA corrections to
the color current in subsection \ref{subsection:colcur} that one should
reasonably consider $\ell_1 \geq 2.5$; at fixed $Y_{\Theta_0}$ this yields
the upper bound $y_1 \leq Y_{\Theta_0}-2.5$, that is, at LHC $y_1 \leq 5$.

On the other side, the perturbative regime we suppose to start at $y_1 \geq
1$. These  mark the limits of the interval where our
calculation can be trusted $1 \leq y_1 \leq 5$ at LHC.
For $y_1 <1$ non-perturbative corrections will dominate, and for
$y_1 > Y_{\Theta_0} - \ell_1^{min} \approx Y_{\Theta_0} -2.5$, the
integration defining $\frac{dN}{d\ln k_\perp}$ ranges over values of $x_1$
which lie outside our small $x$ approximation and for which the MLLA
corrections become accordingly out of control.

On the curves of Figs.~7 and 9 at $Y_{\Theta_0}=10$, the small $y$ region
exhibits a bump which comes from the competition between two phenomena: the
divergence of $\alpha_s(k_\perp^2)$ when $k_\perp \to Q_0$ and coherence
effects which deplete multiple production at very small momentum.
The separation of these two effects is still more visible at
$Y_{\Theta_0}=15$, which is studied in appendix \ref{subsection:ktDLA},
where a comparison with DLA calculations is performed.
At smaller $Y_{\Theta_0}$, the divergence of $\alpha_s$ wins
over coherence effects and the bump disappears.

The curves corresponding to the LEP and Tevatron working conditions are
given in appendix \ref{section:LEP}.

\vskip .7 cm

\subsubsection{Mixed quark and gluon jets}
\label{subsection:mixed}
%%%%%%%%%%%%%%%%%%%%%%%%%%%%%%%%%%%%%%%%%%%

\vskip .4cm

In many experiments, the nature of the jet (quark or gluon) is not
determined, and one simply detects outgoing hadrons, which can originate
from  either type; one then introduces a ``mixing'' parameter $\omega$,
which is to be determined experimentally, such that, for example if one
deals with the inclusive $k_\perp$ distribution
\begin{equation}
\left(\frac{dN}{d\ln k_{\perp}}\right)_{mixed}=
\omega\left(\frac{dN}{d\ln k_{\perp}}\right)_{g}+\left(1-\omega\right)
\left(\frac{dN}{d\ln k_{\perp}}\right)_{q}.
\end{equation}
It is in this framework that forthcoming data from the LHC will be
compared with our theoretical predictions; since outgoing
charged hadrons are detected, one introduces the phenomenological
parameter ${\cal K}^{ch}$ \cite{EvEqI}\cite{KOI}
 normalizing  partonic distributions to the ones
of charged hadrons
\begin{equation}
\Bigg(\frac {dN}{d\ln k_\perp}\Bigg)^{ch} = {\cal K}^{ch}\Bigg(\frac
{dN}{d\ln k_\perp}\Bigg)_{mixed}.
\label{eq:K}
\end{equation}

\vskip .7cm

%%%%%%%%%%%%%%%%%%%%%%%%
\section{CONCLUSION}
%%%%%%%%%%%%%%%%%%%%%%%%

\vskip .4cm

After deducing a general formula, valid for all $x$, for the double
differential 2-particle inclusive cross section for  jet production in a
hard collision process,
the exact solutions of the MLLA evolution equations 
\cite{Perez} have been used to perform a small $x$ calculation of the
double differential 1-particle inclusive distributions and of the inclusive
$k_\perp$ distributions for quark and gluon jets.

Sizable differences with the naive approach in which one forgets the
jet evolution between its  opening angle $\Theta_0$ and the emission angle
$\Theta$  have been found; their role is emphasized to recover, in
particular, the positivity of the distributions.

MLLA corrections increase with  $x$ and decrease when the transverse
momentum $k_\perp$ of the outgoing hadrons gets larger;
that they stay ``within control''
requires in practice that the small $x$ region should not be extended
beyond $\ell < 2.5$; it is remarkable that
similar bounds arise in the study of 2-particle correlations
\cite{Perez2}. At fixed $Y_{\Theta_0}$, the lower bound for $\ell$
translates into an upper bound for $y$; this fixes in particular the upper
limit of confidence for our calculation of $\frac{dN}{d\ln k_\perp}$; above
this threshold, though $k_\perp$ is larger (more
``perturbative''), the small $x$ approximation is no longer valid.

The ``divergent'' behavior of the MLLA distributions for $y \to 0$ forbids
extending the confidence
domain of MLLA  lower that $y \geq 1$, keeping away from the singularity
of $\alpha_s(k_\perp^2)$ when $k_\perp \to \Lambda_{QCD}$.

The two (competing) effects of coherence (damping of multiple production at
small momentum) and  divergence of
$\alpha_s(k_\perp^2)$ at small $k_\perp$ for the inclusive $k_\perp$
distribution have been exhibited.

MLLA and DLA calculations have been compared; in ``modified'' MLLA
calculations, we have furthermore factored out the $\alpha_s$ dependence to
ease the comparison with DLA.

While the goal of this work is a comparison of our theoretical predictions with
forthcoming data from LHC and  Tevatron, we have also given results for
LEP.  LHC energies will provide a larger trustable
 domain of comparison with theoretical predictions at small $x$.

Further developments of this work aim at getting rid of the limit $Q_0
\approx \Lambda_{QCD}$ and extending the calculations to a larger range of
values of $x$; then, because of the lack of analytical expressions,
the general formul{\ae}
(\ref{eq:DDI}) and (\ref{eq:F}) should be numerically investigated,
which will also  provide a deeper insight into the connection
between DGLAP and MLLA evolution equations \cite{Perez-Salam}.

\vskip .4 cm

{\em \underline{Acknowledgments}: It is a pleasure to thank M. Cacciari,
Yu.L. Dokshitzer and G.P. Salam for many stimulating discussions, and
for expert help in numerical calculations.
R. P-R. wants to specially  thank Y.L. Dokshitzer for his guidance and
encouragements.}

%}

%%%%%%%%%%%%%%%%%%%%%%%%%%%%%%%%%%%%%%%%%%%%%%%%%%%%%%%%%%%%%%%%%%%%%%%%%%%%
%%%%%%%%%%%%%%%%%%%%%%%%%%%%%%%%%%%%%%%%%%%%%%%%%%%%%%%%%%%%%%%%%%%%%%%%%%%%
\newpage

%\appendix

{\bf\Large APPENDIX}

\vskip .75 cm

%%%%%%%%%%%%%%%%%%%%%%%%%%%%%%%%%%%%%%%%%%%%%%%%%%%%%%%%%%%%%%%%%%%%%%%%%%%%
\section{EXACT SOLUTION OF THE  MLLA EVOLUTION EQUATION FOR THE
FRAGMENTATION FUNCTIONS; THE SPECTRUM AND ITS DERIVATIVES }
\label{section:exactsol}
%%%%%%%%%%%%%%%%%%%%%%%%%%%%%%%%%%%%%%%%%%%%%%%%%%%%%%%%%%%%%%%%%%%%%%%%%%%%

\vskip .5cm

\subsection{MLLA evolution equation for a gluon jet}
%%%%%%%%%%%%%%%%%%%%%%%%%%%%%%%%%%%%%%%%%%%%%%%%%%%%%%

\vskip .5cm

Because of (\ref{eq:DgDq}),
we will only write the evolution equations for  gluonic fragmentation
functions $D_g^b$.

The partonic structure functions $D_a^b$ satisfy an evolution equation which is
best written when expressed in terms of the variables $\ell$ and $y$ and
the functions $\tilde D_a^b$ defined by \cite{EvEqI}
(see also (\ref{eq:Dlowx}) (\ref{eq:rhoD})):
\begin{equation}
x_b D_a^b(x_b,k_a,q) = \tilde D_a^b(\ell_b,y_b).
\label{eq:D2}
\end{equation}

The parton content $\tilde D_g$ of a gluon is shown in \cite{Perez} to
satisfy the evolution equation ($Y$ and $y$ are linked by (\ref{eq:defY}))
\begin{equation}
\tilde D_g(\ell,y) = \delta(\ell)+ \int_0^{y} dy' \int_0^{\ell} d\ell'
\gamma_0^2
(\ell'+y')\left[1  -a\delta(\ell'\!-\!\ell) \right] \tilde D_g(\ell',y'),
\label{eq:eveqincl}
\end{equation}
where the anomalous dimension $\gamma_0(y)$ is given by ($\lambda$ is defined
in (\ref{eq:lambda}))
\begin{equation}
  \gamma_0^2(y)=4N_c\frac{\alpha_s(k_{\perp}^2)}{2\pi} \approx
  \frac1{\beta(y+ \lambda)}.
\label{eq:gamma0}
\end{equation}
(see the beginning of section \ref{section:descri}
for $\beta$, $T_R$, $C_F$, $\alpha_s$, $N_c$) and
\begin{equation}
a=\frac1{4N_c}\left[\frac{11}3 N_c + \frac{4}{3}T_R 
\left(1-\frac{2\,C_F}{N_c}\right) \right]; \quad  C_F= 4/3\ for\ SU(3)_c.
\label{eq:adef}
\end{equation}
The (single logarithmic) subtraction term proportional to $a$
in (\ref{eq:eveqincl}) accounts
for {\em gluon $\to$ quark}\/ transitions in parton cascades as well
as for energy conservation -- the so-called
``hard corrections'' to parton cascading --.

No superscript has been written in the structure
functions $D_g$ because the same equation is valid indifferently for
$D_g^g$ and $D_g^q$ (see section \ref{section:lowEA}).
One considers that the same evolution equations govern
the (inclusive) hadronic distributions $D_g^h$ (Local Hadron Parton Duality).

\vskip .75 cm

\subsection{Exact solution of the MLLA evolution equation for particle spectra}
%%%%%%%%%%%%%%%%%%%%%%%%%%%%%%%%%%%%%%%%%%%%%%%%%%%%%%%%%%%%%%%%%%%%%%%%%%%%%%%
 
\vskip .5cm

The exact solution of the evolution equation (\ref{eq:eveqincl}), which
includes constraints of energy conservation and
the running of $\alpha_s$, is
demonstrated in ~\cite{Perez} to be given by the following Mellin's
representation
\begin{equation}
\begin{split}
 \tilde D_g\left(\ell,y,\lambda\right)
&= \left(\ell + y +\lambda\right) 
\int\frac{d\omega}{2\pi i}\int\frac{d\nu}{2\pi i}\, e^{\omega\ell+\nu y}\\
 &\int_{0}^{\infty}\frac{ds}{\nu+s}\left(\frac{\omega\left(\nu+s\right)}
 {\left(\omega+s\right)\nu}\right)^{1/(\beta(\omega-\nu))}
 \left(\frac{\nu}{\nu+s}\right)^{a/\beta}\,e^{-\lambda s}.
\end{split}
\label{eq:red4}
\end{equation}
From (\ref{eq:red4}) and taking the 
high energy limit $\ell+y \equiv Y\gg\lambda$ 
\footnote{$Y\gg\lambda \Leftrightarrow E\Theta \gg Q_0^2/\Lambda_{QCD}$
is not strictly equivalent to $Q_0 \to \Lambda_{QCD}$ (limiting spectrum).
\label{footnote:cutoff}}
one gets ~\cite{EvEqI}\cite{KOI} the explicit formula  

\begin{equation}
 \tilde D_g(\ell,y) = \frac{\ell + y}{\beta B(B+1)} \int \frac{d\omega}{2\pi i}
 \> e^{-\omega y}\> \Phi\big(A+1,B+2,\omega(\ell+y)\big),
\label{eq:confrep}
\end{equation}
where $\Phi$ is the confluent hypergeometric function the integral 
representation of which reads ~\cite{GR} ~\cite{SDP} 
\begin{eqnarray}
 &&\Phi(A+1,B+2,\omega Y) = \Gamma(B+2)\,(\omega Y)^{-B-1}
 \int \frac{dt}{(2\pi i)} \frac{t^{-B}}{t(t-1)}
 \left(\frac{t}{t-1}\right)^A  e^{\omega Y t};\cr
&& \cr
&& \text{with}\quad A = \frac{1}{\beta \omega},\quad
B=\displaystyle{\frac{a}{\beta}},\quad \Gamma(n)=\int_0^\infty d\chi \,
\chi^{n-1}e^{-\chi}.
\label{eq:hyperg}
\end{eqnarray}
Exchanging the $t$ and $\omega$ integrations of
(\ref{eq:confrep}) (\ref{eq:hyperg}) and going from  $t$ to
the new variable $\displaystyle \alpha = \frac{1}{2} \ln\frac{t}{t-1}$,
(\ref{eq:confrep}) becomes
\begin{equation}
\tilde  D_g(\ell,y) = 2\frac{\Gamma(B)}{\beta}
\Re\left( \int_0^\frac{\pi}{2}
  \frac{d\tau}{\pi}\, e^{-B\alpha}\  {\cal F}_B(\tau,y,\ell)\right),
\label{eq:ifD}
\end{equation}
where the integration is performed with respect to $\tau$ defined by
$\displaystyle \alpha = \frac{1}{2}\ln\frac{y}{\ell}  + i\tau$,
\begin{eqnarray}
{\cal F}_B(\tau,y,\ell) &=& \left[ \frac{\cosh\alpha
-\displaystyle{\frac{y-\ell}{y+\ell}}
\sinh\alpha} 
 {\displaystyle \frac{\ell +
y}{\beta}\,\frac{\alpha}{\sinh\alpha}} \right]^{B/2}
  I_B(2\sqrt{Z(\tau,y,\ell)}), \cr
&& \cr
&& \cr
 Z(\tau,y,\ell) &=&
\frac{\ell + y}{\beta}\,
\frac{\alpha}{\sinh\alpha}\,
 \left(\cosh\alpha
%+ (1-2\zeta)
-\frac{y-\ell}{y+\ell}
\sinh\alpha\right); 
\label{eq:calFdef}
\end{eqnarray}
$I_B$ is the modified Bessel function of the first kind.

\vskip .75 cm

\subsection{The spectrum}
%%%%%%%%%%%%%%%%%%%%%%%%%

\vskip .5cm

On Fig.~12 below, we represent, on the left, the spectrum
as a function of the transverse momentum
(via $y$) for fixed $\ell$ and, on the right,  as a function
of the energy (via $\ell$) for fixed transverse momentum.

\bigskip

\vbox{
\begin{center}
\epsfig{file=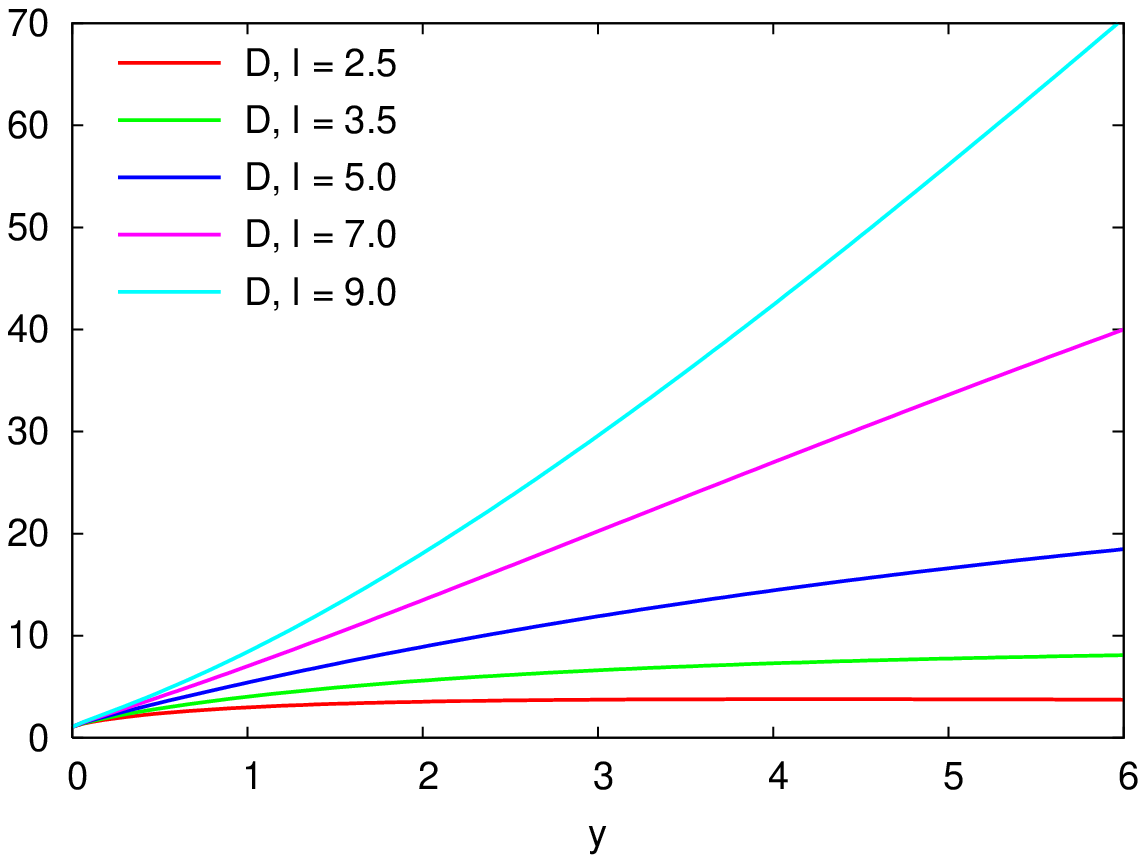, height=5truecm,width=7.5truecm}
\hfill
\epsfig{file=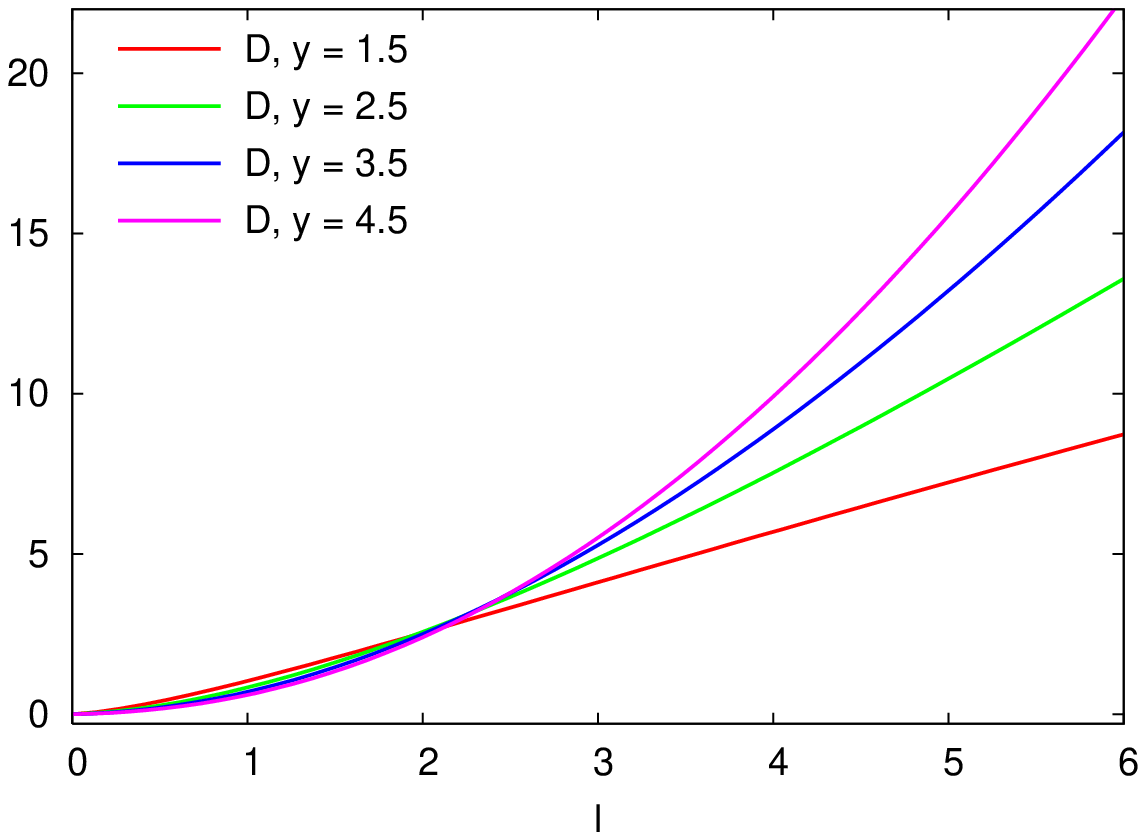, height=5truecm,width=7.5truecm}
\end{center}

\centerline{\em Fig.~12: spectrum $\tilde D(\ell,y)$ of emitted partons}

\centerline{\em as functions 
of transverse momentum (left) and energy (right)}

}

\bigskip
\bigskip

Fig.~13 shows enlargements of Fig.~12 for small values of $y$ and $\ell$
respectively; they ease the understanding of the curves for the derivatives
of the spectrum presented in subsection \ref{subsection:deriv}.

\bigskip
\bigskip

\vbox{
\begin{center}
\epsfig{file=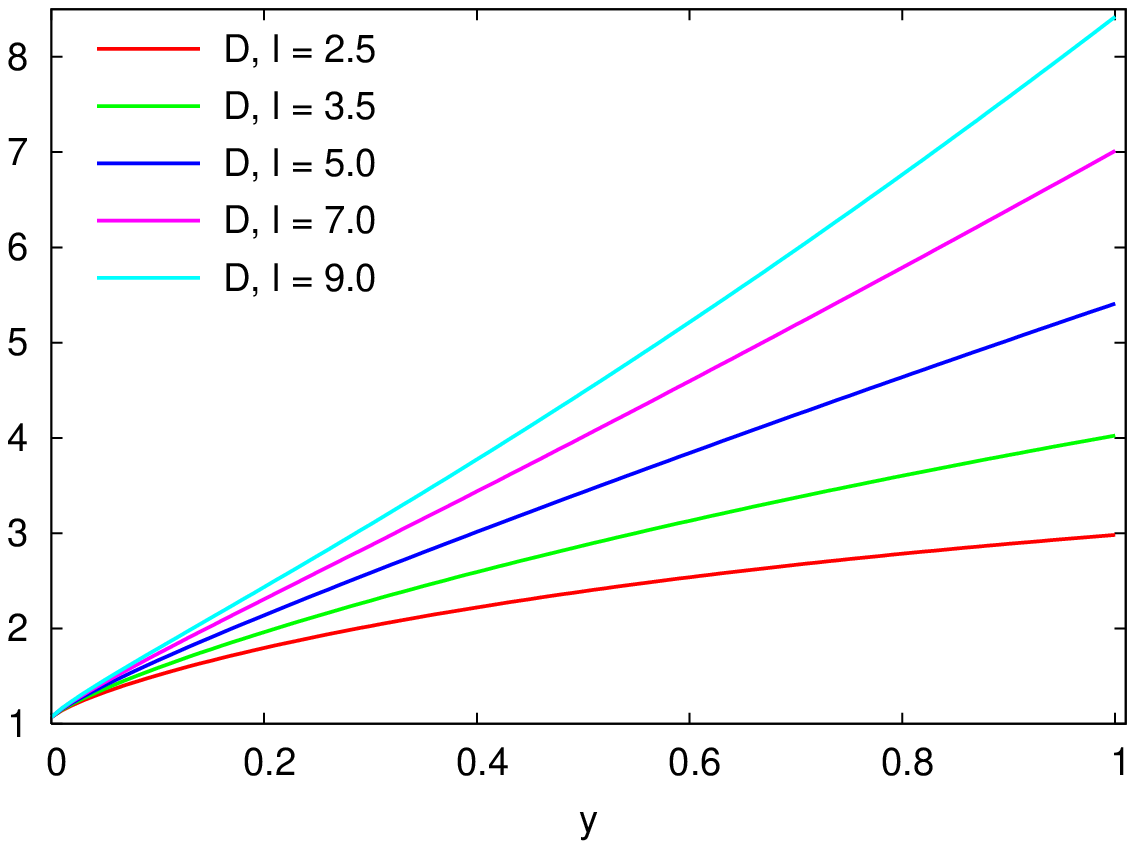, height=5truecm,width=7.5truecm}
\hfill
\epsfig{file=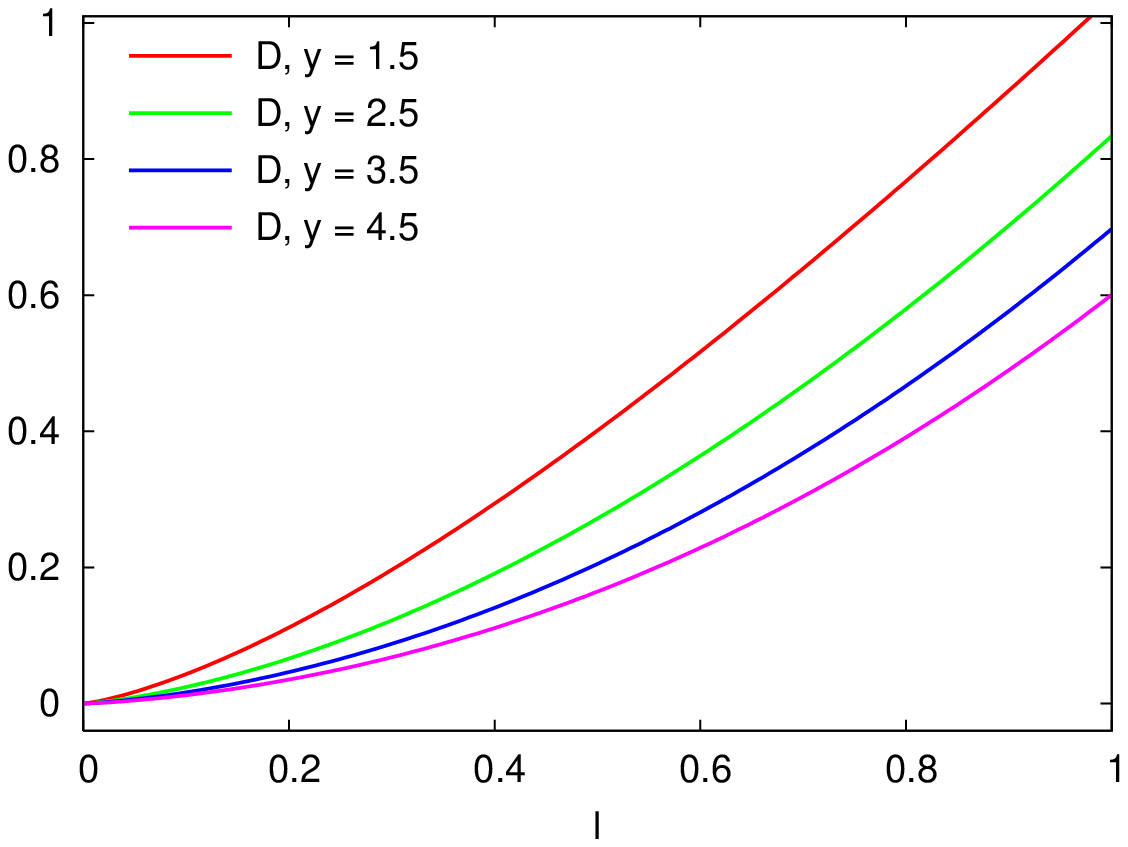, height=5truecm,width=7.5truecm}
\end{center}

\centerline{\em Fig.~13: spectrum $\tilde D(\ell,y)$ of emitted partons}

\centerline{\em as functions 
of transverse momentum (left) and energy (right): enlargement of Fig.~12}}

\bigskip

A comparison between MLLA and DLA calculations of the spectrum is done in
appendix \ref{subsection:DLAspec}.

\vskip .75 cm

\subsection{Derivatives of the spectrum}
\label{subsection:deriv}
%%%%%%%%%%%%%%%%%%%%%%%%%%%%%%%%%%%%%%%%%

\vskip .5cm

We evaluate below the derivatives of the spectrum w.r.t. $\ln k_{\perp}$
and $\ln(1/x)$.

We make use of the following property for the
confluent hypergeometric functions $\Phi$ ~\cite{SDP}: 
\begin{eqnarray}\label{eq:derivconf}
\frac{d}{d\ell}\Phi\left(A+1,B+2,\omega\left(\ell+y\right)\right)
&\equiv&\frac{d}{dy}\Phi\left(A+1,B+2,\omega\left(\ell+y\right)\right)\cr
&=&\omega\frac{A+1}{B+2}\Phi\left(A+2,B+3,\omega\left(\ell+y\right)\right).
\end{eqnarray} 

$\bullet$\quad We first determine the derivative w.r.t. $\ell\equiv\ln(1/x)$.
Differentiating  (\ref{eq:confrep}) w.r.t. $\ell$, and
expanding (\ref{eq:derivconf}), one gets
\footnote{(\ref{eq:derivl}) and (\ref{eq:derivy}) have also been
checked by numerically differentiating (\ref{eq:ifD}).}
 \cite{Perez} 
\begin{equation}
\frac{d}{d\ell}\tilde  D_{g}\left(\ell,y\right)
 = 2\frac{\Gamma(B)}{\beta} \int_0^{\frac{\pi}2}\frac{d\tau}{\pi}\,
 e^{-B\alpha}
 \left[\frac1{\ell + y}\left(1+2e^{\alpha}
\sinh{\alpha}\right){\cal{F}}_B
+\frac1{\beta}e^{\alpha}{\cal{F}}_{B+1}\right];
\label{eq:derivl}
\end{equation}

$\bullet$\quad Differentiating w.r.t.
$y\equiv\ln\displaystyle{\frac{k_{\perp}}{Q_0}}$ yields
\begin{equation}
\frac{d}{dy}\tilde D_{g}\left(\ell,y\right)
= 2 \frac{\Gamma(B)}{\beta} \int_0^{\frac{\pi}2}
\frac{d\tau}{\pi}\,  e^{-B\alpha}
 \left[\frac1{\ell + y}
\left(1+2e^{\alpha}\sinh{\alpha}\right)
 {\cal{F}}_B
 +\frac1{\beta}
 e^{\alpha}{\cal{F}}_{B+1}\right.
\left.-\frac{2\sinh\alpha}{\ell +
y}{\cal{F}}_{B-1}\right].
\label{eq:derivy}
\end{equation}
In Fig.~14,  Fig.~15,  Fig.~16 and Fig.~17  below, we draw the curves for:

\smallskip

$\ast$\ $\displaystyle\frac{d\tilde D_g(\ell,y)}{dy}$ as a function
of $y$, for different values of $\ell$ fixed;

$\ast$\ $\displaystyle\frac{d\tilde D_g(\ell,y)}{dy}$ as a function
of $\ell$, for different values of $y$ fixed;

$\ast$\ $\displaystyle\frac{ d\tilde D_g(\ell,y)}{d\ell}$ as a
function of $\ell$ for different values of $y$ fixed;

$\ast$\ $\displaystyle\frac{ d\tilde D_g(\ell, y)}{d\ell}$
as a function of $y$ for different values of $\ell$ fixed.

\bigskip

In each case the right figure is an enlargement,
close to the origin of axes, of the left figure.

\vskip .75 cm

%\vbox{
\vbox{
\begin{center}
\epsfig{file=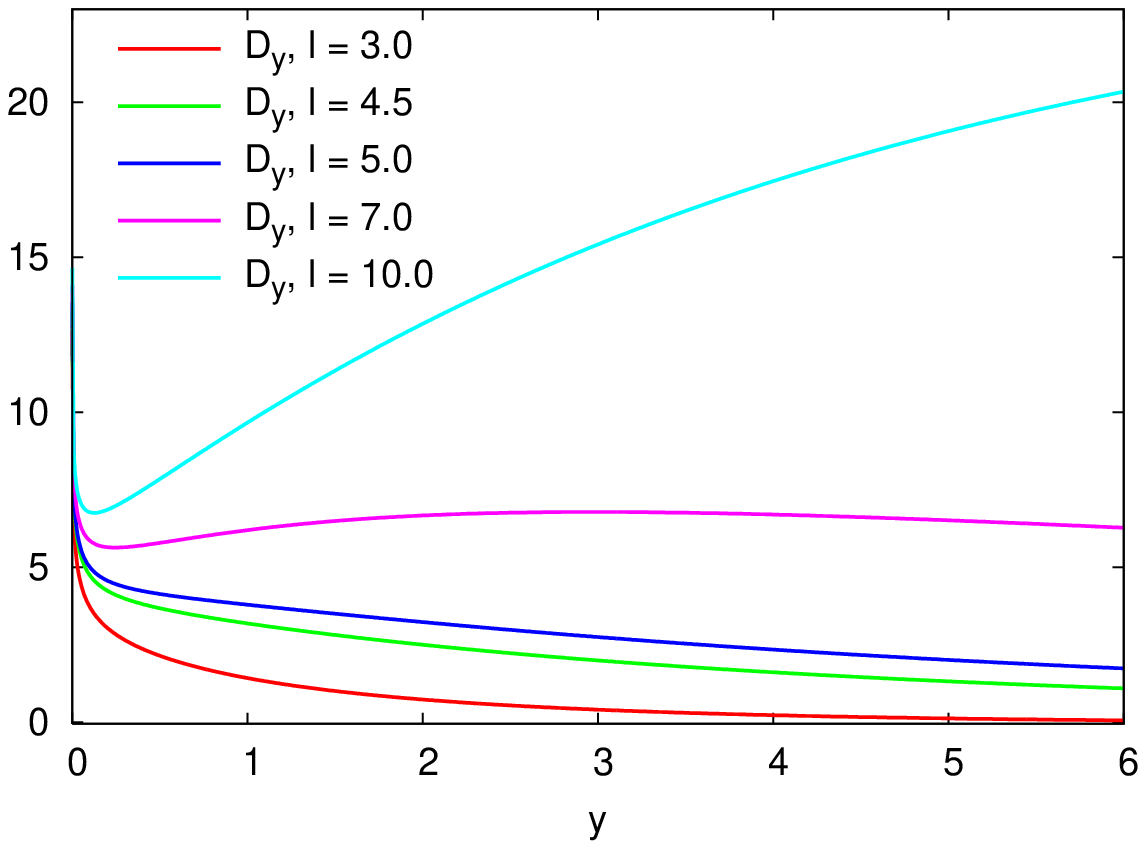, height=5truecm,width=7.5truecm}
\hfill
\epsfig{file=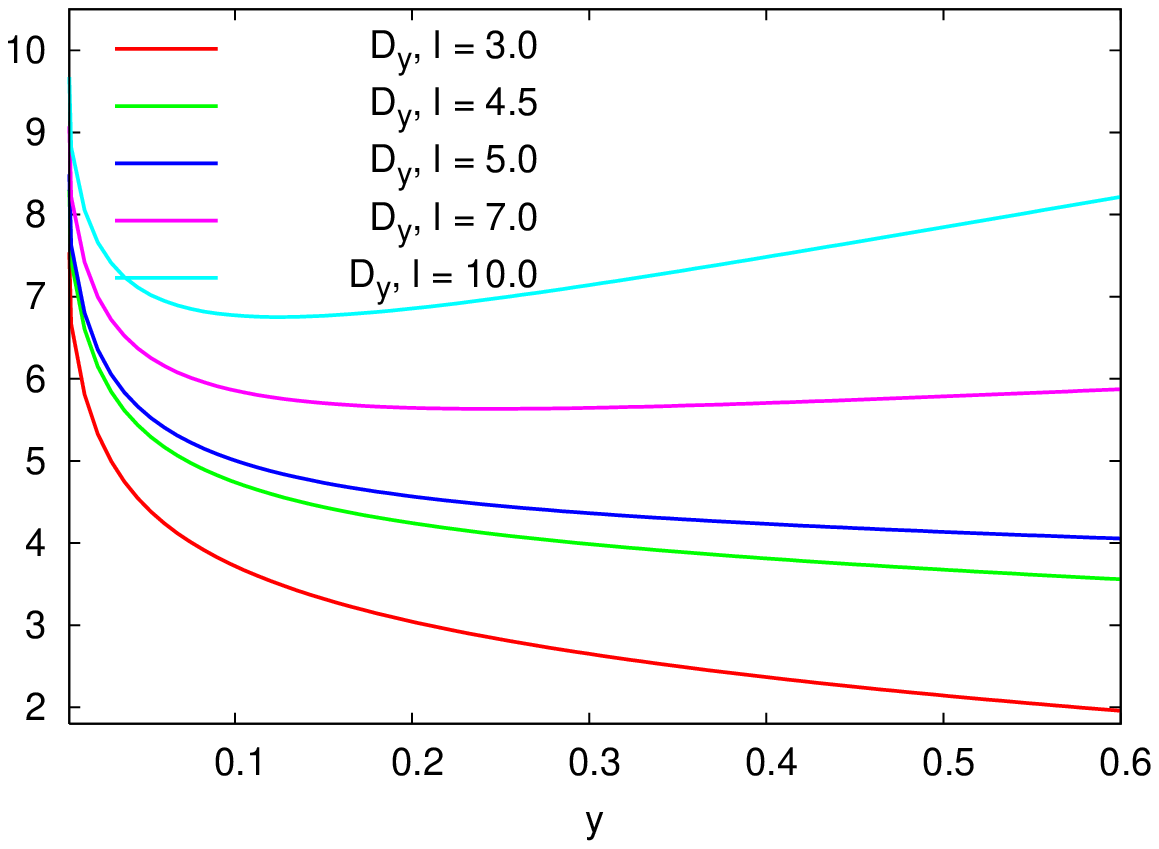, height=5truecm,width=7.5truecm}
\end{center}

\centerline{\em Fig.~14: $\frac{d\tilde D_g(\ell,y)}{dy}$ as a function of $y$
 for different values of $\ell$}

}

\medskip

\vbox{
\begin{center}
\epsfig{file=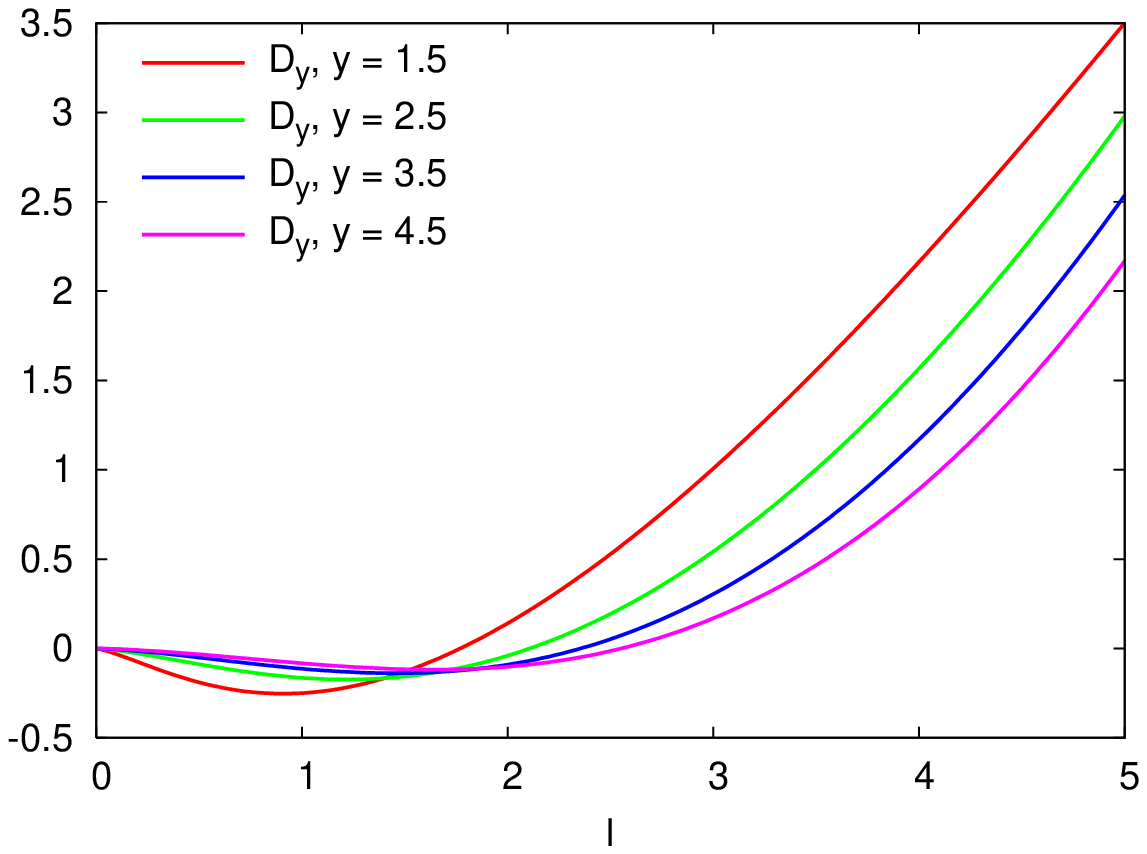,height=5truecm,width=7.5truecm}
\hfill
\epsfig{file=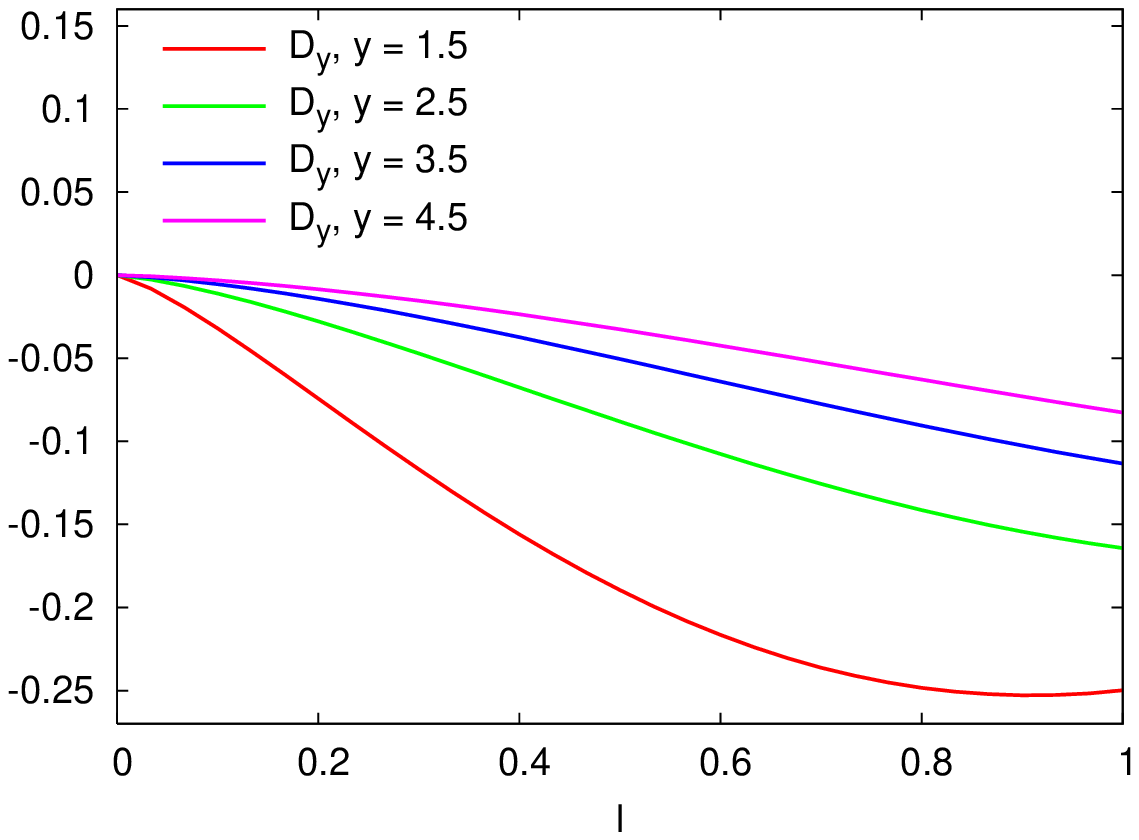, height=5truecm,width=7.5truecm}
\end{center}

\centerline{\em Fig.~15: $\frac{d\tilde D_g(\ell,y)}{dy}$ as a function of
$\ell$ for different values of $y$}

}

\medskip

\vbox{
\begin{center}
\epsfig{file=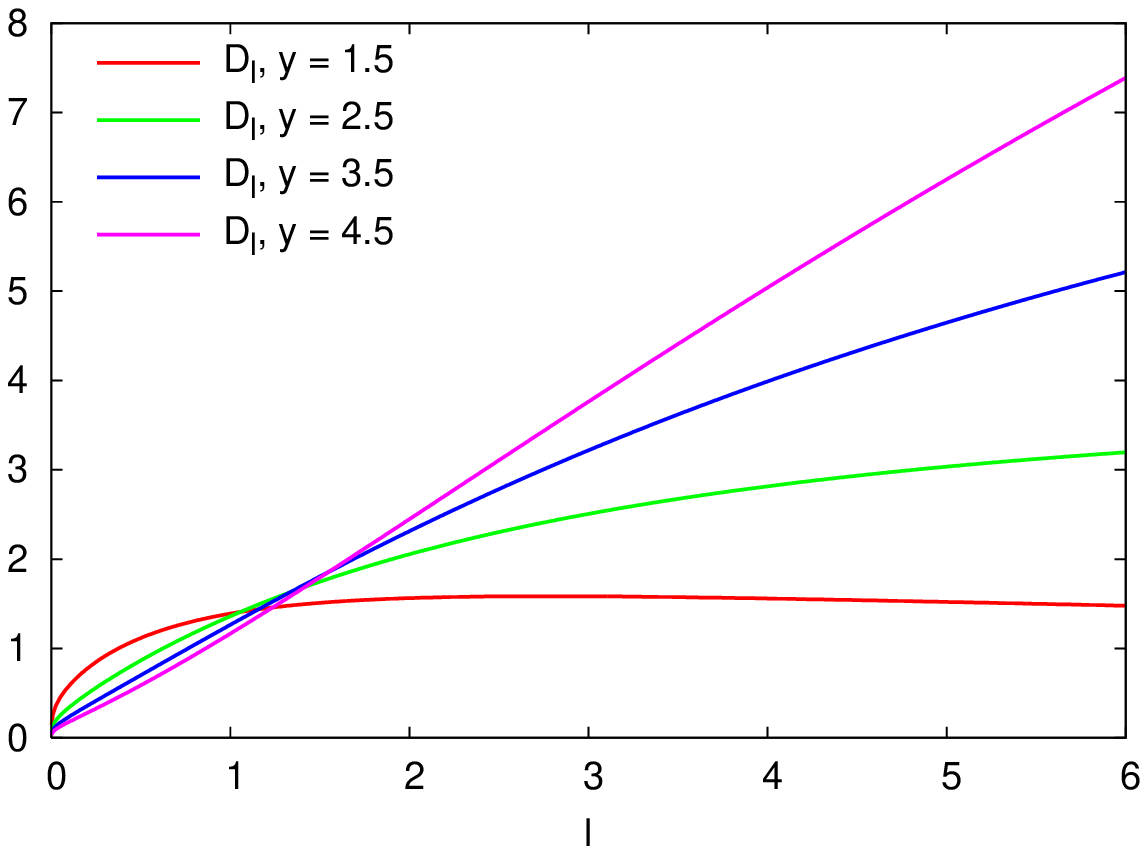, height=5truecm,width=7.5truecm}
\hfill
\epsfig{file=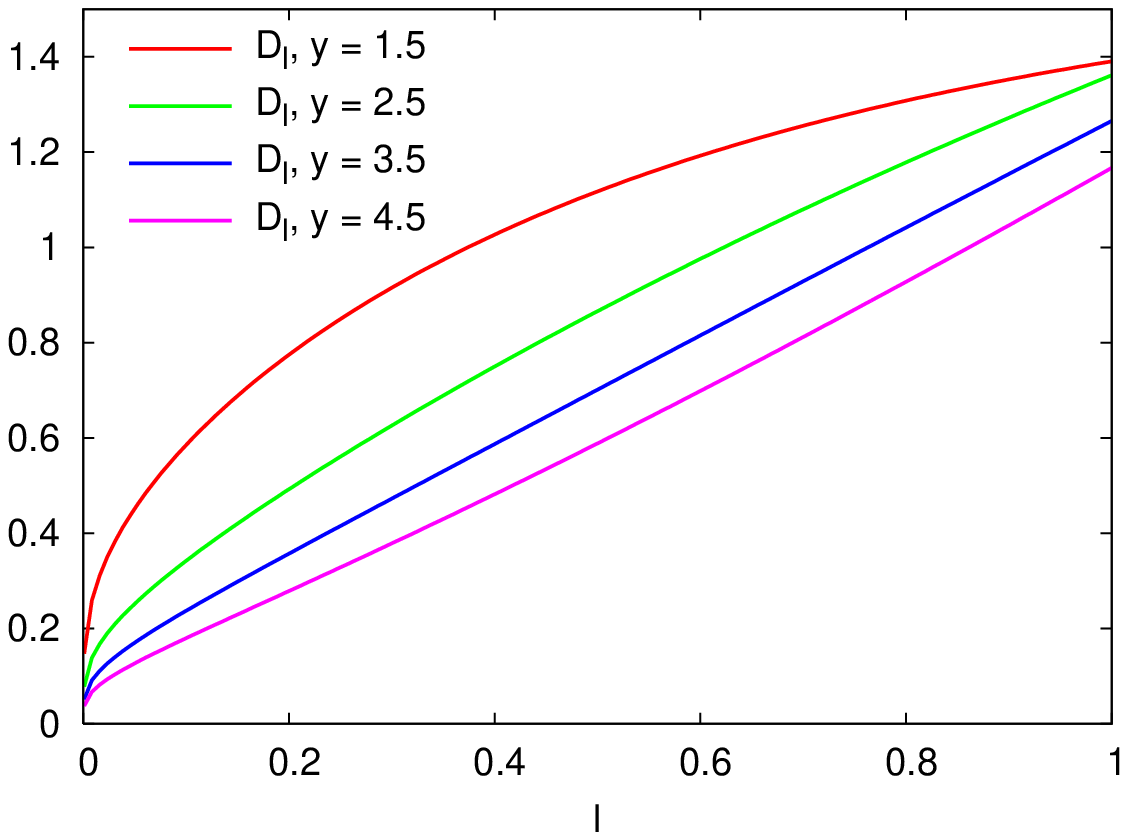, height=5truecm,width=7.5truecm}
\end{center}

\centerline{\em Fig.~16: $\frac{d\tilde D_g(\ell,y)}{d\ell}$ as a function of
$\ell$ for different values of $y$}
}

\medskip

\vbox{
\begin{center}
\epsfig{file=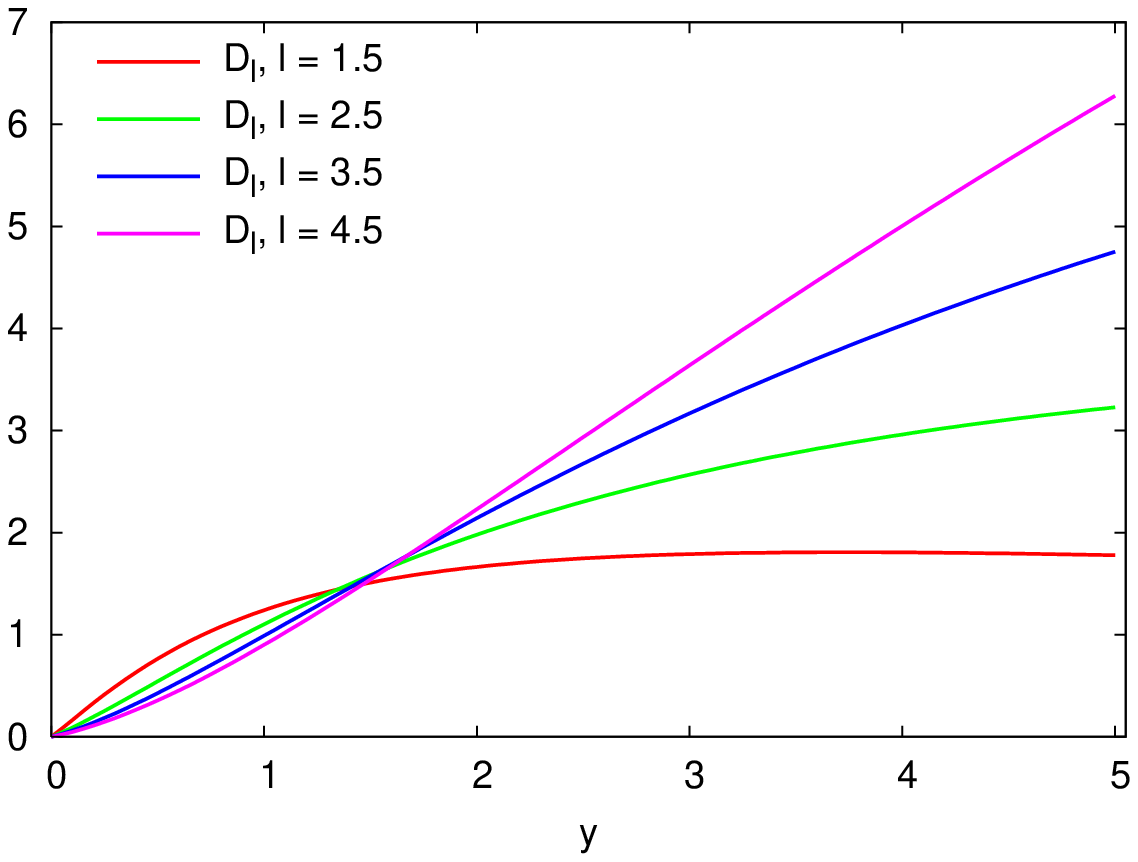, height=5truecm,width=7.5truecm}
\hfill
\epsfig{file=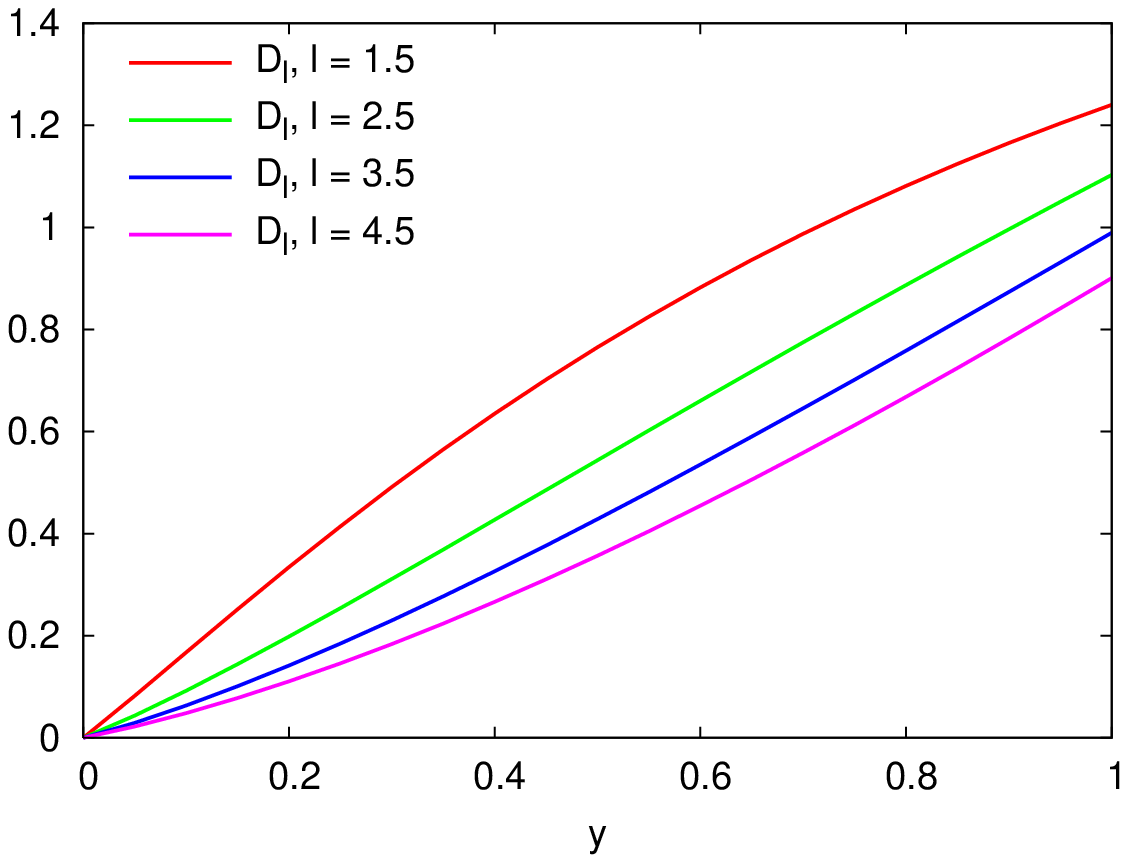, height=5truecm,width=7.5truecm}
\end{center}

\centerline{\em Fig.~17: $\frac{d\tilde D_g(\ell,y)}{d\ell}$ as a function of
$y$  for different values of $\ell$}
}
%}

\vskip .5cm

That $\displaystyle\frac{d\tilde D_g(\ell,y)}{dy}$ goes to infinity
when  $y \to 0$ is
in agreement with the analytic behavior in $\ln(\ell/y)$ of this
derivative.

\vskip .75 cm

%%%%%%%%%%%%%%%%%%%%%%%%%%%%%%%%%%%%%%%%%%%%%%%%%%%%%%%%%%%%%%%%%%%%%%%%%%%%
\section{LEADING CONTRIBUTIONS TO  $\boldsymbol{x_1 F_{A_0}^{h_1}(x_1, \Theta,
E, \Theta_0)}$ AT SMALL $\boldsymbol{x_1}$}
\label{section:leadingxF}
%%%%%%%%%%%%%%%%%%%%%%%%%%%%%%%%%%%%%%%%%%%%%%%%%%%%%%%%%%%%%%%%%%%%%%%%%%%%

\vskip .5cm

Using (\ref{eq:DgDq}), the leading terms of $x_1 F_{A_0}^{h_1}(x_1, \Theta,
E, \Theta_0)$ (\ref{eq:Fdev}) calculated at small $x_1$ read
\begin{eqnarray}
x_1 F_{g}^{h_1}\left(x_1,\Theta,E,\Theta_0\right)^0
&\approx&
\tilde D_g(\ell_1,y_1)\left(<u>^g_{g}
+ \frac{C_F}{N_c}<u>^q_{g}\right)
= \frac{<C>_g^0}{N_c}\;\tilde D_g(\ell_1,y_1),\cr
x_1 F_{q}^{h_1}\left(x_1,\Theta,E,\Theta_0\right)^0
&\approx&
\tilde D_g(\ell_1,y_1)\left(<u>^g_{q} 
+ \frac{C_F}{N_c}<u>^q_{q}\right)
=\frac{<C>_q^0}{N_c}\;\tilde D_g(\ell_1,y_1).\cr
&&
\label{eq:FA0}
\end{eqnarray}

The leading  $<C>_{g}^0$ and $<C>_{q}^0$ in (\ref{eq:C})
 for  a quark and a gluon jet are
 given respectively by (see \cite{EvEqI}, chapt. 9
\footnote{The coefficient $\beta$, omitted in the exponents of eqs.
(9.12a), (9.12b), (9.12c) of \cite{EvEqI} has been restored here. The factor
$4N_c$ is due to our normalization (see the beginning of section
\ref{section:descri}).}
)

\begin{eqnarray}
 <C>_{q}^0
&=&<C>_{\infty}-c_1\left(N_c-C_F\right)\left(
 \frac{\ln\left(E\Theta/\Lambda_{QCD}\right)}
 {\ln\left(E\Theta_0/\Lambda_{QCD}\right)}\right)^{(c_3/4N_c\beta)}\cr
&=&<C>_{\infty}-c_1\left(N_c-C_F\right)\left(
 \frac{Y_\Theta +\lambda}
 {Y_{\Theta_0} +\lambda}\right)^{(c_3/4N_c\beta)},
\label{eq:Cquark}
\end{eqnarray}
\begin{eqnarray}
 <C>_{g}^0
&=&<C>_{\infty}+c_2\left(N_c-C_F\right)\left(
 \frac{\ln\left(E\Theta/\Lambda_{QCD}\right)}
 {\ln\left(E\Theta_0/\Lambda_{QCD}\right)}\right)^{(c_3/4N_c\beta)}\cr
&=&<C>_{\infty}+c_2\left(N_c-C_F\right)\left(
 \frac{Y_{\Theta} +\lambda}
 {Y_{\Theta_0} +\lambda}\right)^{(c_3/4N_c\beta)},
\label{eq:Cgluon}
\end{eqnarray}
with
\begin{eqnarray}
<C>_{\infty} &=& c_1 N_c+c_2\, C_F,\cr
&& \cr
c_1=\displaystyle{\frac83}\displaystyle{\frac{C_F}{c_3}},\quad c_2 
&=&1-c_1=\displaystyle{\frac{2}{3}}\displaystyle{\frac{n_f}{c_3}},
\quad c_3=\displaystyle{\frac83}C_F+\displaystyle{\frac23}n_f;
\label{eq:c1c2c3}
\end{eqnarray}
in the r.h.s of (\ref{eq:Cquark}) (\ref{eq:Cgluon})
 we have used the definitions (\ref{eq:defY}) (\ref{eq:Y0}).
$<C>_\infty$ corresponds to the limit  $E\to \infty, \Theta \to 0$.
 
In practice, we take in this work
\begin{equation}
Q_0 \approx \Lambda_{QCD} \Leftrightarrow \lambda \approx 0,
\label{eq:lambdanul}
\end{equation}
which ensures in particular the consistency with the analytical
calculation of the MLLA spectrum (appendix \ref{section:exactsol}),
which can only be explicitly achieved in this limit.

\vskip .75 cm

%%%%%%%%%%%%%%%%%%%%%%%%%%%%%%%%%%%%%%%%%%%%%%%%%%%%%%%%%%%%%%%%%%%%%%%%%%%%
\section{ CALCULATION OF $\boldsymbol{\delta\!<C>_g}$ and $\boldsymbol{\delta\!<C>_q}$
OF SECTION \ref{section:lowEA}}
\label{section:udeltau}
%%%%%%%%%%%%%%%%%%%%%%%%%%%%%%%%%%%%%%%%%%%%%%%%%%%%%%%%%%%%%%%%%%%%%%%%%%%%

\vskip .5cm

\subsection{Explicit expressions for $\boldsymbol{<u>_{A_0}^A}$ and
$\boldsymbol{\delta\!<u>_{A_0}^A}$ defined in (\ref{eq:udef})}
\label{subsection:udu}
%%%%%%%%%%%%%%%%%%%%%%%%%%%%%%%%%%%%%%%%%%%%%%%%%%%%%%%%%%%%%%%%%%%%

\vskip .5cm

The expressions (\ref{eq:udef}) for $<u>^A_{A_0}$ and $\delta \!<u>^A_{A_0}$
are conveniently obtained from the Mellin-transformed DGLAP fragmentation
functions
\cite{EvEqI}
\begin{equation}
        {\cal D}(j,\xi) = \int_0^1 du\, u^{j-1}  D(u,\xi), 
\label{eq:mellin}
\end{equation}
which, if one deals with $D_A^B(u,r^2,s^2)$, depends in reality on
the difference $\xi(r^2) - \xi(s^2)$:
\begin{equation}
\xi(Q^2) = \int_{\mu^2}^{Q^2} \;
\frac{dk^2}{k^2}\frac{\alpha_s(k^2)}{4\pi}, \quad
\xi(r^2) - \xi(s^2) \approx \frac{1}{4N_c\beta}
\ln\left(\frac{\ln(r^2/\Lambda_{QCD}^2)}{\ln(s^2/\Lambda_{QCD}^2)}\right).
\end{equation}
One has accordingly
\begin{equation}
<u>^A_{A_0} = {\cal D}_{A_0}^A (2,\xi(E\Theta_0)-\xi(E\Theta)),\quad
\delta\!<u>^A_{A_0} = \frac{d}{dj}{\cal D}_{A_0}^A(j,\xi(E\Theta_0)-\xi(E\Theta))\Big|_{j=2}.  \label{eq:udu}
\end{equation}
The DGLAP functions ${\cal D}(j,\xi)$ are expressed \cite{EvEqI}
% chapt.~1)
in terms
of the anomalous dimensions $\nu_F(j)$, $\nu_G(j)$ and $\nu_\pm(j)$, the
$j$ dependence of which is  in particular known.

For the sake of completeness, we give below the expressions for the $<u>$'s
and $\delta<u>$'s.
\begin{eqnarray}
 <u>^q_g &=& {\frac {9}{25}}\, \left(  \left( {\frac {{
Y_{\Theta_0}}+\lambda}{Y_\Theta+\lambda}}
 \right) ^{{\frac {50}{81}}}-1 \right)  \left( {\frac {{ Y_{\Theta_0}}+
\lambda}{Y_\Theta+\lambda}} \right) ^{-{\frac {50}{81}}},\cr
&&\cr
&&\cr
<u>^g_g &=&
1/25\, \left( 16\, \left( {\frac {{ Y_{\Theta_0}}+\lambda}{Y_\Theta+\lambda}}
 \right) ^{{\frac {50}{81}}}+9 \right)  \left( {\frac {{ Y_{\Theta_0}}+
\lambda}{Y_\Theta+\lambda}} \right) ^{-{\frac {50}{81}}},\cr
&&\cr
&&\cr
<u>^g_q &=&
{\frac {16}{25}}\, \left(  \left( {\frac {{
Y_{\Theta_0}}+\lambda}{Y_\Theta+\lambda}
} \right) ^{{\frac {50}{81}}}-1 \right)  \left( {\frac {{ Y_{\Theta_0}}+
\lambda}{Y_\Theta+\lambda}} \right) ^{-{\frac {50}{81}}},\cr
&&\cr
&&\cr
&& \hskip-4cm <u>^{sea}_q =
-1/25\, \left( -9\, \left( {\frac {{ Y_{\Theta_0}}+\lambda}{Y_\Theta+\lambda}}
 \right) ^{{\frac {50}{81}}}-16+25\, \left( {\frac {{ Y_{\Theta_0}}+\lambda}{
Y_\Theta+\lambda}} \right) ^{2/9} \right)  \left( {\frac {{
Y_{\Theta_0}}+\lambda}{Y_\Theta
+\lambda}} \right) ^{-{\frac {50}{81}}},\cr
&&\cr
&&\cr
<u>^{val} &=&
\left( {\frac {{ Y_{\Theta_0}}+\lambda}{Y_\Theta+\lambda}} \right) ^{-{\frac {32}{
81}}},\cr
&&\cr
&&\cr
<u>^{sea}_q + <u>^{val} &=& 1/25\, \left( 9\, \left( {\frac {{
Y_{\Theta_0}}+\lambda}{Y_\Theta+\lambda}}
 \right) ^{{\frac {50}{81}}}+16 \right)  \left( {\frac {{ Y_{\Theta_0}}+
\lambda}{Y_\Theta+\lambda}} \right) ^{-{\frac {50}{81}}};\cr
&&\cr
&&\cr
\delta <u>^q_g &=&
-{\frac {1}{337500}}\, \left( -43011\, \left( {\frac {{ Y_{\Theta_0}}+\lambda
}{Y_\Theta+\lambda}} \right) ^{{\frac {50}{81}}}+43011
-6804\,{\pi }^{2}
 \left( {\frac {{ Y_{\Theta_0}}+\lambda}{Y_\Theta+\lambda}} \right) ^{{\frac {50}{81
}}}
\right. \cr && \left.
+6804\,{\pi }^{2}-48600\,\ln  \left( {\frac {{
Y_{\Theta_0}}+\lambda}{Y_\Theta+
\lambda}} \right)  \left( {\frac {{ Y_{\Theta_0}}+\lambda}{Y_\Theta+\lambda}}
 \right) ^{{\frac {50}{81}}}
\right. \cr && \left.
+21600\,\ln  \left( {\frac {{ Y_{\Theta_0}}+
\lambda}{Y_\Theta+\lambda}} \right)  \left( {\frac {{
Y_{\Theta_0}}+\lambda}{Y_\Theta+
\lambda}} \right) ^{{\frac {50}{81}}}{\pi }^{2}+109525\,\ln  \left( {
\frac {{ Y_{\Theta_0}}+\lambda}{Y_\Theta+\lambda}} \right)
\right. \cr && \left.
 -17400\,\ln  \left( {
\frac {{ Y_{\Theta_0}}+\lambda}{Y_\Theta+\lambda}} \right) {\pi }^{2} \right) 
 \left( {\frac {{ Y_{\Theta_0}}+\lambda}{Y_\Theta+\lambda}} \right) ^{-{\frac {50}{
81}}},\cr
&&\cr
&&\cr
\delta <u>^g_g &=&
-{\frac {1}{337500}}\, \left( -11664\, \left( {\frac {{ Y_{\Theta_0}}+\lambda
}{Y_\Theta+\lambda}} \right) ^{{\frac {50}{81}}}+31104\,{\pi }^{2} \left( {
\frac {{ Y_{\Theta_0}}+\lambda}{Y_\Theta+\lambda}} \right) ^{{\frac {50}{81}}}
\right. \cr && \left.
-86400
\,\ln  \left( {\frac {{ Y_{\Theta_0}}+\lambda}{Y_\Theta+\lambda}} \right)  \left( {
\frac {{ Y_{\Theta_0}}+\lambda}{Y_\Theta+\lambda}} \right) ^{{\frac {50}{81}}}
\right. \cr && \left.
+38400
\,\ln  \left( {\frac {{ Y_{\Theta_0}}+\lambda}{Y_\Theta+\lambda}} \right)  \left( {
\frac {{ Y_{\Theta_0}}+\lambda}{Y_\Theta+\lambda}} \right) ^{{\frac {50}{81}}}{\pi }
^{2}
\right. \cr && \left.
+11664-31104\,{\pi }^{2}-109525\,\ln  \left( {\frac {{ Y_{\Theta_0}}+
\lambda}{Y_\Theta+\lambda}} \right)
\right. \cr && \left.
 +17400\,\ln  \left( {\frac {{ Y_{\Theta_0}}+
\lambda}{Y_\Theta+\lambda}} \right) {\pi }^{2} \right)  \left( {\frac {{ 
Y_{\Theta_0}}+\lambda}{Y_\Theta+\lambda}} \right) ^{-{\frac {50}{81}}},\cr
&&\cr
&&\cr
\delta <u>^g_q &=&
-{\frac {4}{759375}}\, \left( 48114\, \left( {\frac {{ Y_{\Theta_0}}+\lambda}
{Y_\Theta+\lambda}} \right) ^{{\frac {50}{81}}}-48114-6804\,{\pi }^{2}
 \left( {\frac {{ Y_{\Theta_0}}+\lambda}{Y_\Theta+\lambda}} \right) ^{{\frac {50}{81
}}}
\right. \cr && \left.
+6804\,{\pi }^{2}-48600\,\ln  \left( {\frac {{
Y_{\Theta_0}}+\lambda}{Y_\Theta+
\lambda}} \right)  \left( {\frac {{ Y_{\Theta_0}}+\lambda}{Y_\Theta+\lambda}}
 \right) ^{{\frac {50}{81}}}
\right. \cr && \left.
+21600\,\ln  \left( {\frac {{ Y_{\Theta_0}}+
\lambda}{Y_\Theta+\lambda}} \right)  \left( {\frac {{
Y_{\Theta_0}}+\lambda}{Y_\Theta+
\lambda}} \right) ^{{\frac {50}{81}}}{\pi }^{2}+109525\,\ln  \left( {
\frac {{ Y_{\Theta_0}}+\lambda}{Y_\Theta+\lambda}} \right)
\right. \cr && \left.
 -17400\,\ln  \left( {
\frac {{ Y_{\Theta_0}}+\lambda}{Y_\Theta+\lambda}} \right) {\pi }^{2} \right) 
 \left( {\frac {{ Y_{\Theta_0}}+\lambda}{Y_\Theta+\lambda}} \right) ^{-{\frac {50}{
81}}},\cr
&&\cr
&&\cr
\delta <u>^{sea}_q &=&
{\frac {2}{759375}}\, \left( -13122\, \left( {\frac {{ Y_{\Theta_0}}+\lambda}
{Y_\Theta+\lambda}} \right) ^{{\frac {50}{81}}}+34992\,{\pi }^{2} \left( {
\frac {{ Y_{\Theta_0}}+\lambda}{Y_\Theta+\lambda}} \right) ^{{\frac {50}{81}}}
\right. \cr && \left.
+54675
\,\ln  \left( {\frac {{ Y_{\Theta_0}}+\lambda}{Y_\Theta+\lambda}} \right)  \left( {
\frac {{ Y_{\Theta_0}}+\lambda}{Y_\Theta+\lambda}} \right) ^{{\frac {50}{81}}}
\right. \cr && \left.
-24300
\,\ln  \left( {\frac {{ Y_{\Theta_0}}+\lambda}{Y_\Theta+\lambda}} \right)  \left( {
\frac {{ Y_{\Theta_0}}+\lambda}{Y_\Theta+\lambda}} \right) ^{{\frac {50}{81}}}{\pi }
^{2}
\right. \cr && \left.
+13122-34992\,{\pi }^{2}+219050\,\ln  \left( {\frac {{ Y_{\Theta_0}}+
\lambda}{Y_\Theta+\lambda}} \right)
-34800\,\ln  \left( {\frac {{ Y_{\Theta_0}}+
\lambda}{Y_\Theta+\lambda}} \right) {\pi }^{2}
\right. \cr && \left.
-265625\,\ln  \left( {\frac {{
 Y_{\Theta_0}}+\lambda}{Y_\Theta+\lambda}} \right)  \left( {\frac {{
Y_{\Theta_0}}+\lambda}
{Y_\Theta+\lambda}} \right) ^{2/9}
\right. \cr && \left.
+37500\,\ln  \left( {\frac {{ Y_{\Theta_0}}+
\lambda}{Y_\Theta+\lambda}} \right)  \left( {\frac {{
Y_{\Theta_0}}+\lambda}{Y_\Theta+
\lambda}} \right) ^{2/9}{\pi }^{2} \right)  \left( {\frac {{
Y_{\Theta_0}}+
\lambda}{Y_\Theta+\lambda}} \right) ^{-{\frac {50}{81}}},\cr
&&\cr
&&\cr
\delta <u>^{val} &=&
-{\frac {2}{243}}\, \left( -85+12\,{\pi }^{2} \right) \ln  \left( {
\frac {{ Y_{\Theta_0}}+\lambda}{Y_\Theta+\lambda}} \right)  \left( {\frac {{
Y_{\Theta_0}}+
\lambda}{Y_\Theta+\lambda}} \right) ^{-{\frac {32}{81}}},\cr
&&\cr
&&\cr
&& \hskip -6cm\delta <u>^{val} + \delta <u>^{sea}_q =
-{\frac {2}{759375}}\, \left( 13122\, \left( {\frac {{ Y_{\Theta_0}}+\lambda}
{Y_\Theta+\lambda}} \right) ^{{\frac {50}{81}}}-34992\,{\pi }^{2} \left( {
\frac {{ Y_{\Theta_0}}+\lambda}{Y_\Theta+\lambda}} \right) ^{{\frac {50}{81}}}
\right. \cr
&& \hskip -2.5cm \left.
-54675
\,\ln  \left( {\frac {{ Y_{\Theta_0}}+\lambda}{Y_\Theta+\lambda}} \right)  \left( {
\frac {{ Y_{\Theta_0}}+\lambda}{Y_\Theta+\lambda}} \right) ^{{\frac {50}{81}}}
\right. \cr
&& \hskip -2.5cm \left.
+24300
\,\ln  \left( {\frac {{ Y_{\Theta_0}}+\lambda}{Y_\Theta+\lambda}} \right)  \left( {
\frac {{ Y_{\Theta_0}}+\lambda}{Y_\Theta+\lambda}} \right) ^{{\frac {50}{81}}}{\pi }
^{2}-13122+34992\,{\pi }^{2}
\right. \cr
&& \hskip -2.5cm \left.
-219050\,\ln  \left( {\frac {{ Y_{\Theta_0}}+
\lambda}{Y_\Theta+\lambda}} \right) +34800\,\ln  \left( {\frac {{ Y_{\Theta_0}}+
\lambda}{Y_\Theta+\lambda}} \right) {\pi }^{2} \right)  \left( {\frac {{ 
Y_{\Theta_0}}+\lambda}{Y_\Theta+\lambda}} \right) ^{-{\frac {50}{81}}}.\cr
&& \hskip -2.5cm
\label{eq:deltaus}
\end{eqnarray}

\bigskip

When $\Theta \to \Theta_0$, all $\delta\! <u>$'s vanish, ensuring that the
limits $\xi(E\Theta_0) -\xi(E\Theta) \to 0$ of the $(<C>_{A_0}^0 +
\delta\!<C>_{A_0})$'s
are the same as the ones of the $<C>_{A_0}^0$'s.

\vskip .75 cm

\subsection{$\boldsymbol{\delta\!<C>_q}$ and $\boldsymbol{\delta\!<C>_g}$}
%%%%%%%%%%%%%%%%%%%%%%%%%%%%%%%%%%%%%%%%%%%%%%%%%%%%%%%%%%%%%%%%%%%%%%%%%%

\vskip .5cm

They are given in (\ref{eq:deltaC}), and one uses
(\ref{eq:DgDq}) such that only $\psi_{g,\ell_1}$ (see (\ref{eq:psidef}))
appears.
Their full analytical expressions for the $\delta\!<C>$'s are 
too complicated to be easily written and manipulated.

Using the formul{\ae} of \ref{subsection:udu}, one gets the
approximate results
\begin{eqnarray}
\delta\!<C>_q &\approx& \left(
1.4676
- 1.4676\left(\frac{Y_{\Theta_0}+\lambda}{Y_\Theta+\lambda}\right)^{-\frac{50}{81}}
- 3.2510 \ln \left(\frac{Y_{\Theta_0}+\lambda}{Y_\Theta+\lambda}\right)\right.\cr
&& \left.  \hskip 2cm
+ 0.5461
\left(\frac{Y_{\Theta_0}+\lambda}{Y_\Theta+\lambda}\right)^{-\frac{50}{81}}
\ln \left(\frac{Y_{\Theta_0}+\lambda}{Y_\Theta+\lambda}\right)
\right)
\psi_{g,\ell_1}(\ell_1,y_1),\cr
&&
\label{eq:deltaCq}
\end{eqnarray}
and

\vbox{
\begin{eqnarray}
\delta\!<C>_g &\approx& \left(
-2.1898 
+ 2.1898 \left(\frac{Y_{\Theta_0}+\lambda}{Y_\Theta+\lambda}\right)^{-\frac{50}{81}}
- 3.2510 \ln \left(\frac{Y_{\Theta_0}+\lambda}{Y_\Theta+\lambda}\right)\right.\cr
&& \left.  \hskip 2cm
- 0.3072
\left(\frac{Y_{\Theta_0}+\lambda}{Y_\Theta+\lambda}\right)^{-\frac{50}{81}}
\ln \left(\frac{Y_{\Theta_0}+\lambda}{Y_\Theta+\lambda}\right)
\right)
\psi_{g,\ell_1}(\ell_1,y_1).\cr
&&
\label{eq:deltaCg}
\end{eqnarray}
}

The logarithmic derivative $\psi_{g,\ell_1}(\ell_1,y_1)$
(\ref{eq:psidef}) of the MLLA
spectrum $\tilde D_g(\ell_1,y_1)$ is obtained from (\ref{eq:ifD})
of appendix \ref{section:exactsol}.

\vskip .75 cm

%%%%%%%%%%%%%%%%%%%%%%%%%%%%%%%%%%%%%%%%%%%%%%%%%%%%%%%%%%%%%%
\section{AT LEP AND TEVATRON}
\label{section:LEP}
%%%%%%%%%%%%%%%%%%%%%%%%%%%%%%%%%%%%%%%%%%%%%%%%%%%%%%%%%%%%%%

\vskip .5cm

At LEP energy, the working conditions correspond to $Y_{\Theta_0}\approx
5.2$; they are not very different at the Tevatron where $Y_{\Theta_0}
\approx 5.6$. We first present the curves for LEP, then, after the
discussion concerning the size of the corrections and the domain of
validity of our calculations, we give our predictions for the Tevatron.

\vskip .5cm

\subsection{The average color current}
\label{subsection:accLEP}
%%%%%%%%%%%%%%%%%%%%%%%%%%%%%%%%%%%%%%%%

\bigskip

\vbox{
\begin{center}
\epsfig{file=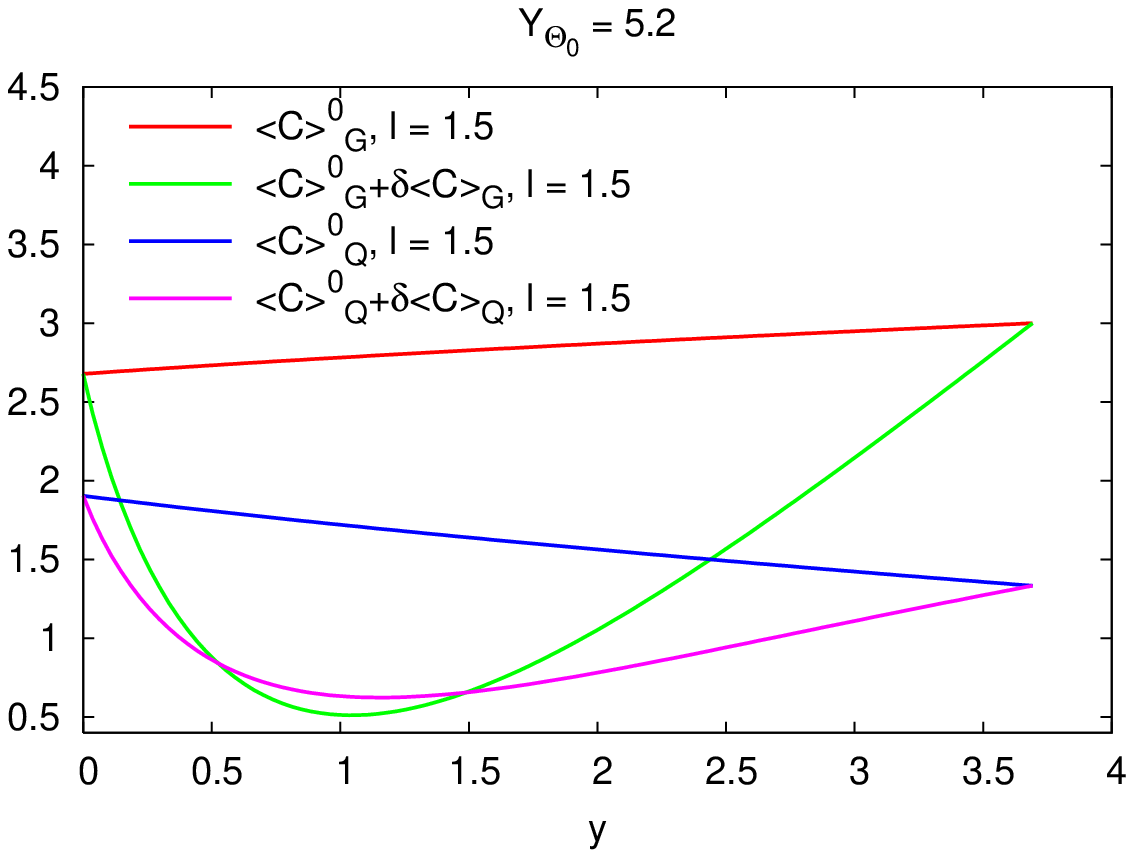, height=5truecm,width=7.5truecm}
\hfill
\epsfig{file=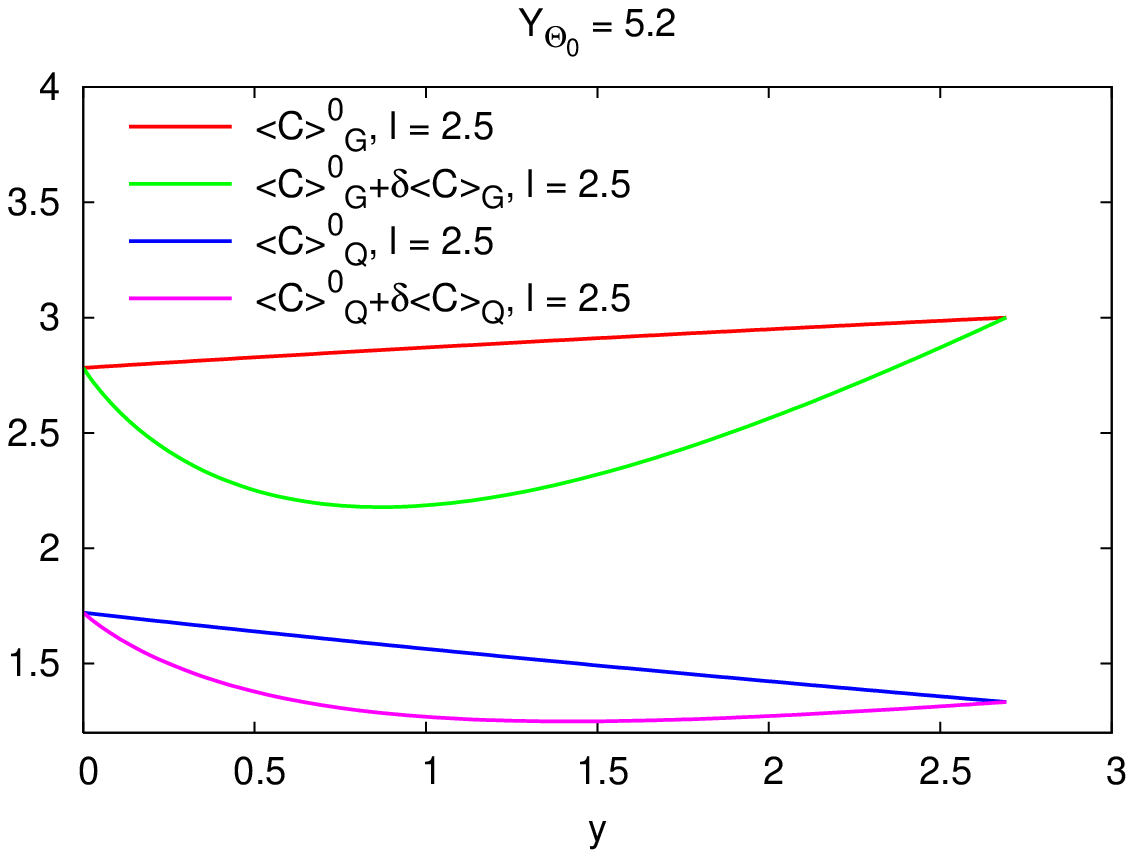, height=5truecm,width=7.5truecm}
\end{center}

\centerline{\em Fig.~18 $<C>_{A_0}^0$ and $<C>_{A_0}^0 + \delta\!<C>_{A_0}$
for quark and gluon jets, as functions of $y$,}

\centerline{\em for $Y_{\Theta_0}=5.2$, $\ell=1.5$ on the left and
$\ell=2.5$ on the right.}
}

\bigskip

Owing to the size of the (MLLA) corrections to the  $<C>$'s and their $y$
derivatives, we will keep to the lower bound $\ell_1 \geq 2.5$.

\vskip .5cm

\subsection{$\boldsymbol{\displaystyle\frac{d^2N}{d\ell_1\;d\ln k_\perp}}$
for a gluon jet}
\label{subsection:d2NgLEP}
%%%%%%%%%%%%%%%%%%%%%%%%%%%%%%%%%%%%%%%%%%%%%%%%%%%%%%%%%%%%%%%

\bigskip

We plot below $\frac{d^2N}{d\ell_1\;d\ln k_\perp}$ for the two values
$\ell=1.5$ and $\ell=2.5$.

\vbox{
\begin{center}
\epsfig{file=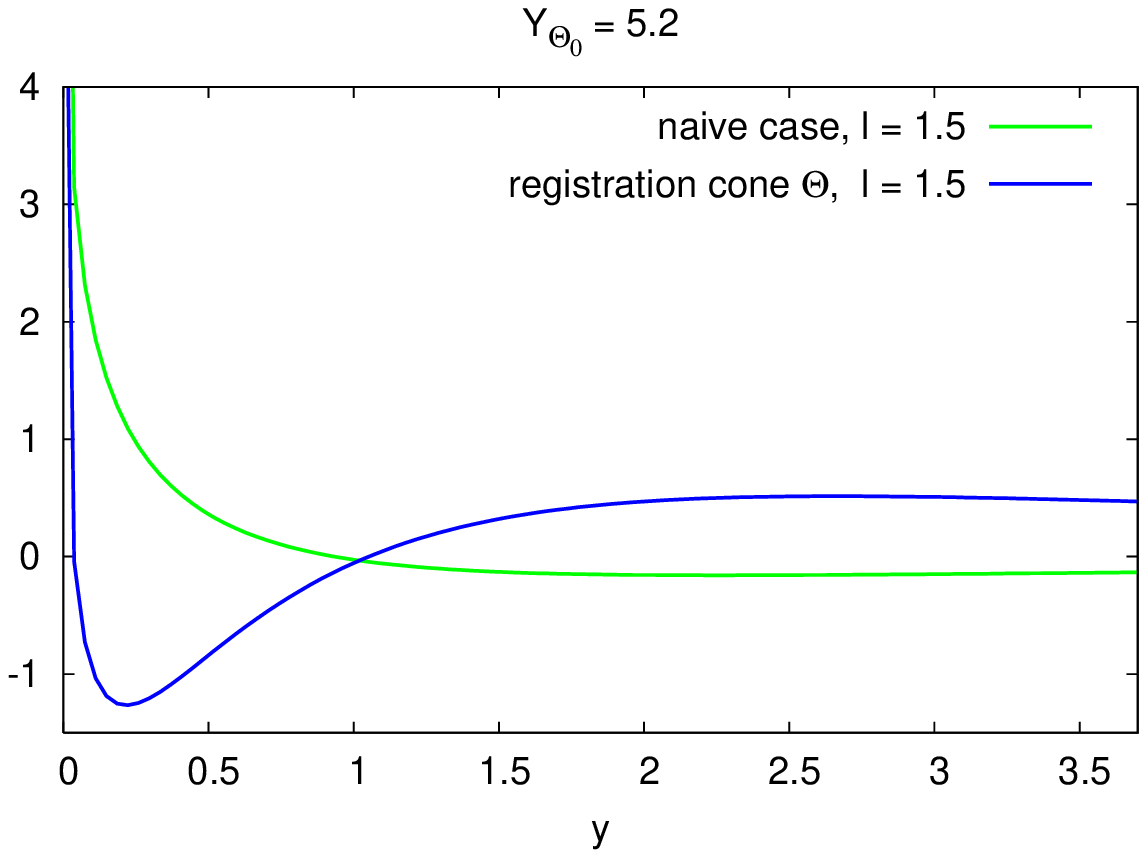, height=5truecm,width=7.5truecm}
\hfill
\epsfig{file=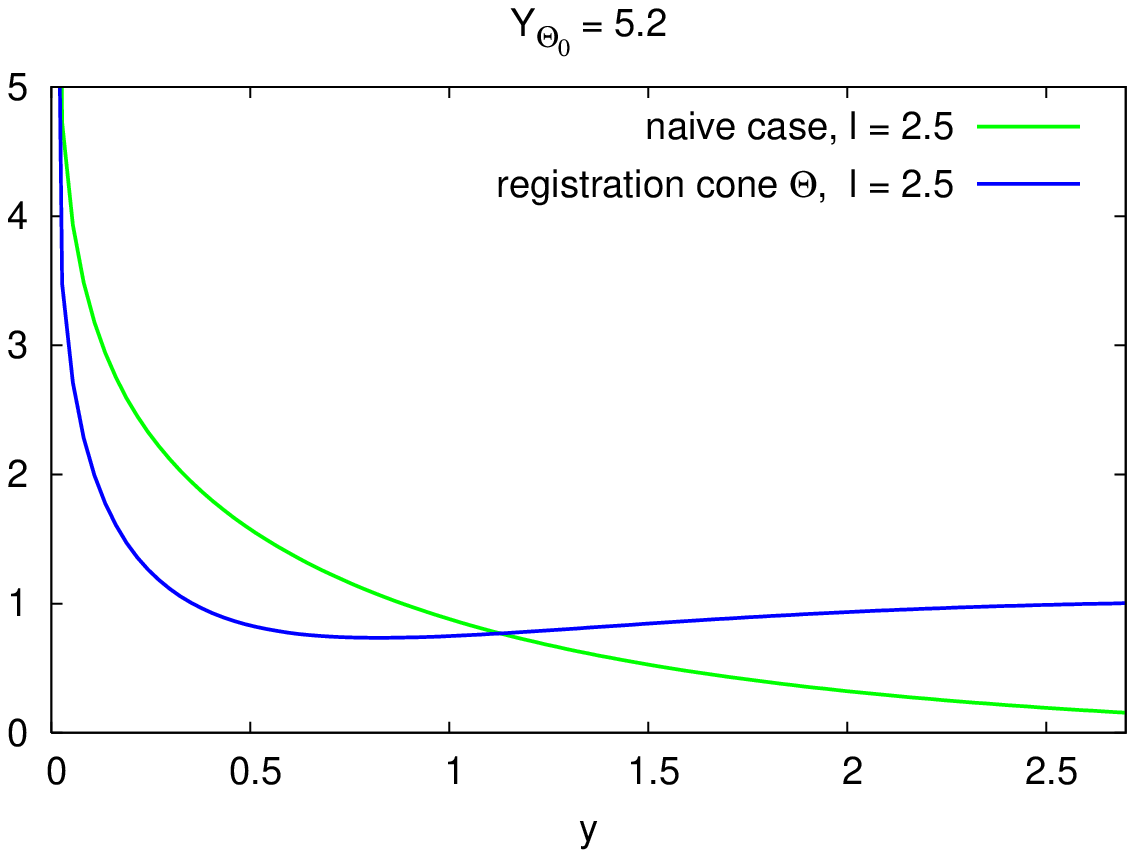, height=5truecm,width=7.5truecm}
\end{center}

\centerline{\em Fig.~19: $\frac{d^2N}{d\ell_1\;d\ln k_\perp}$ for a gluon jet
at fixed $\ell_1$,  MLLA and naive approach.}
}

\bigskip

The excessive size of the $\delta\!<C>$ corrections emphasized in
subsection \ref{subsection:accLEP} translates here into the loss  of
the positivity for $\frac{d^2N}{d\ell_1\;d\ln k_\perp}$ at $\ell=1.5$ for
$y<1$: our approximation is clearly not trustable there.

\vskip .5cm

\subsection{$\boldsymbol{\displaystyle\frac{d^2N}{d\ell_1\;d\ln k_\perp}}$
for a quark jet}
\label{subsection:d2NqLEP}
%%%%%%%%%%%%%%%%%%%%%%%%%%%%%%%%%%%%%%%%%%%%%%%%%%%%%%%%%%%%%%%

\bigskip

We consider the same two values of $\ell$ as above.

\vbox{
\begin{center}
\epsfig{file=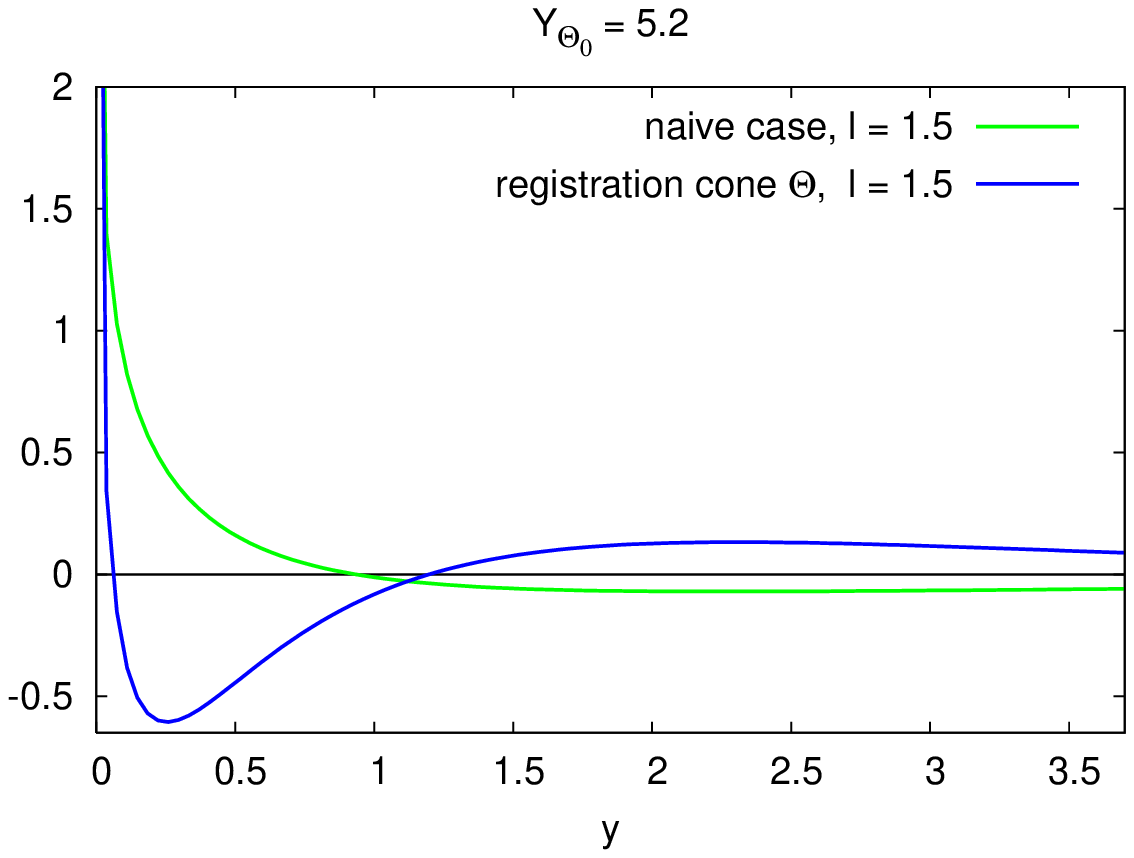, height=5truecm,width=7.5truecm}
\hfill
\epsfig{file=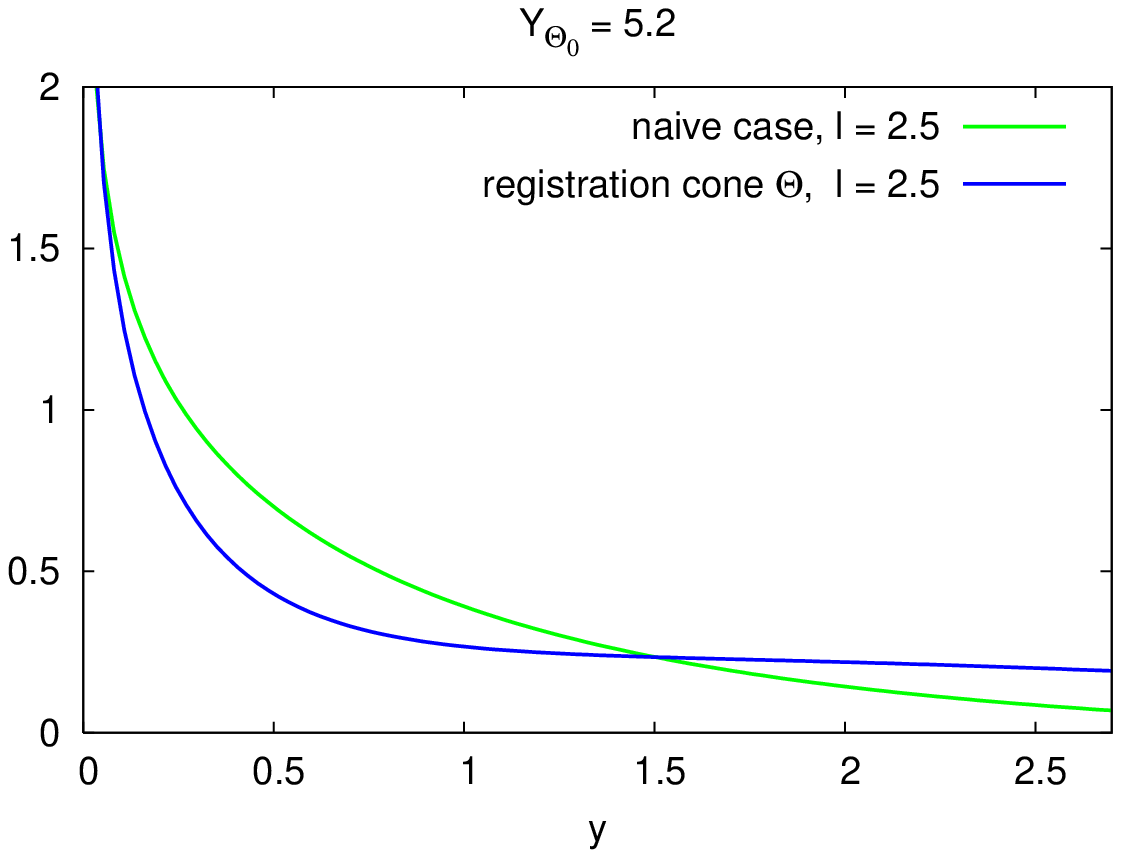, height=5truecm,width=7.5truecm}
\end{center}

\centerline{\em Fig.~20: $\frac{d^2N}{d\ell_1\;d\ln k_\perp}$ for a quark jet
at fixed $\ell_1$,  MLLA and naive approach.}
}

\bigskip

Like for the gluon jet, we encounter positivity problems at $\ell=1.5$ for
$y< 1.25$.

\vskip .5cm

\subsection{$\boldsymbol{\displaystyle\frac{dN}{d\ln k_\perp}}$ for a gluon jet}
\label{subsection:dNgLEP}
%%%%%%%%%%%%%%%%%%%%%%%%%%%%%%%%%%%%%%%%%%%%%%%%%%%%%%%%%%%%%%%%%%%%%%%%%%%%

\bigskip

We plot below $\frac{dN}{d\ln k_\perp}$ for a gluon jet obtained by the
``naive'' approach and including the jet evolution from $\Theta_0$ to
$\Theta$;  on the right is
an enlargement which shows how positivity is recovered when MLLA corrections
are included.

\vbox{
\begin{center}
\epsfig{file=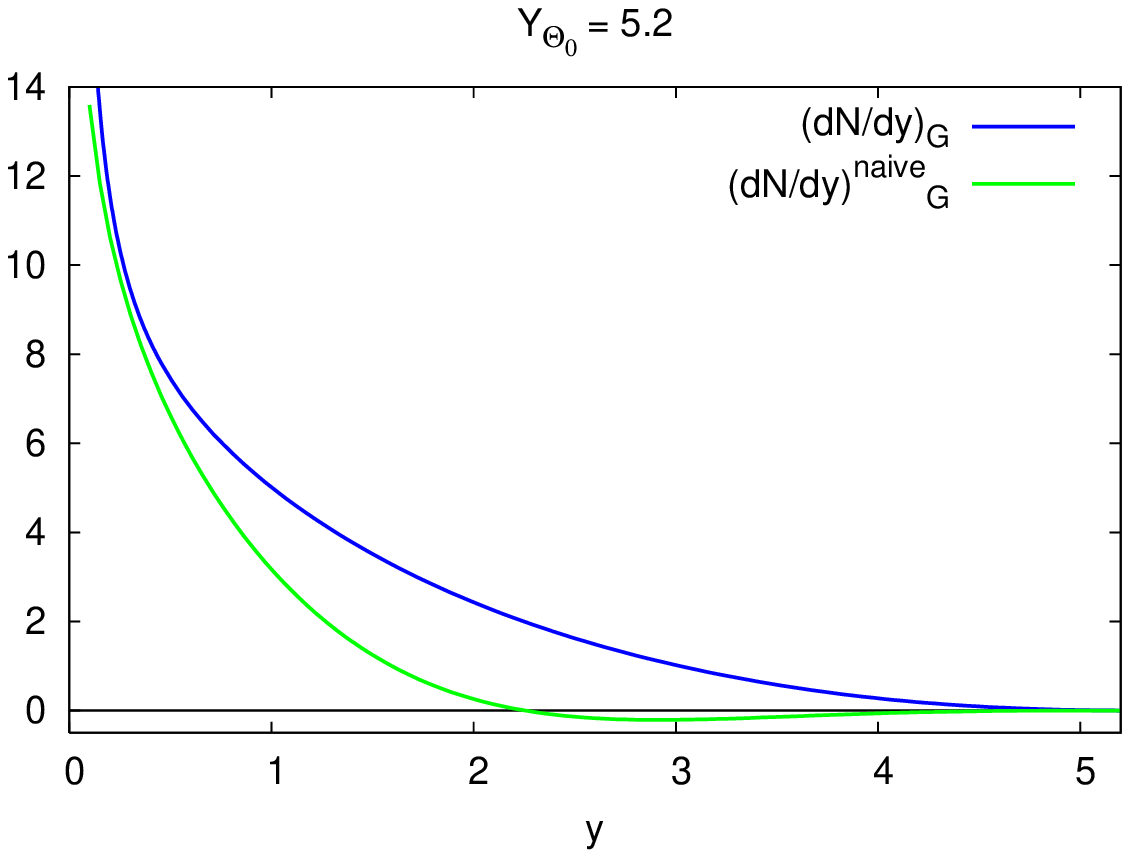, height=5truecm,width=7.5truecm}
\hfill
\epsfig{file=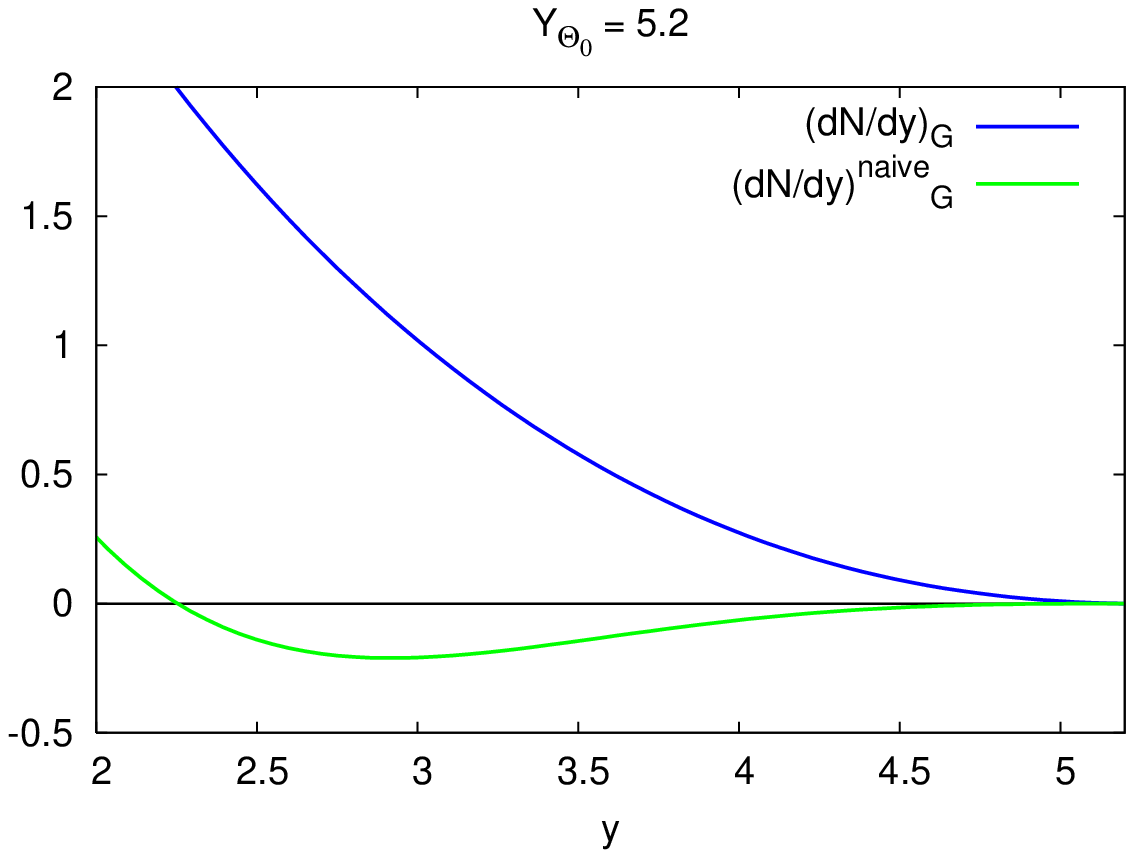, height=5truecm,width=7.5truecm}
\end{center}

\centerline{\em Fig.~21: $\frac{dN}{d\ln k_\perp}$ for a gluon jet,
MLLA and naive approach.}
}

\vskip .5cm

\subsection{$\boldsymbol{\displaystyle\frac{dN}{d\ln k_\perp}}$ for a quark jet}
\label{subsection:dNqLEP}
%%%%%%%%%%%%%%%%%%%%%%%%%%%%%%%%%%%%%%%%%%%%%%%%%%%%%%%%%%%%%%%%%%%%%%%%%%%%

\bigskip

We proceed like for a gluon jet. The curves below show the restoration of
positivity by MLLA corrections.

\vbox{
\begin{center}
\epsfig{file=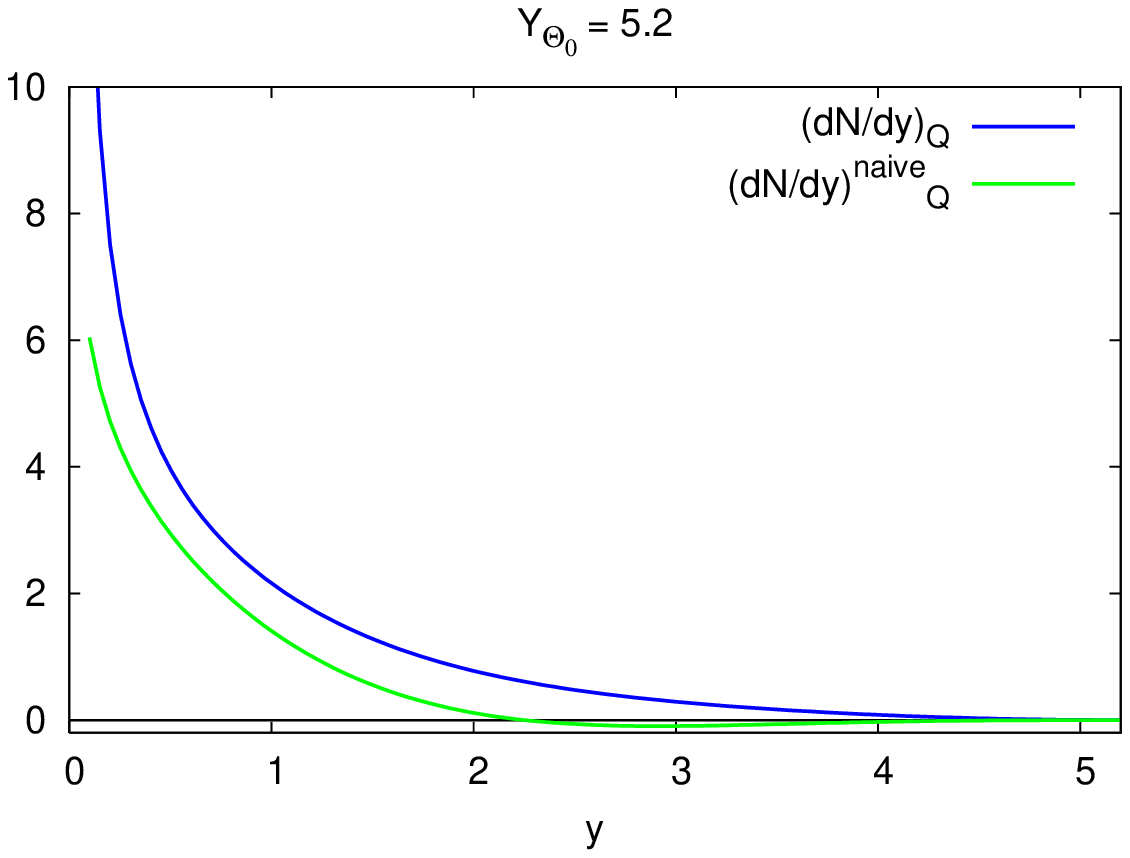, height=5truecm,width=7.5truecm}
\hfill
\epsfig{file=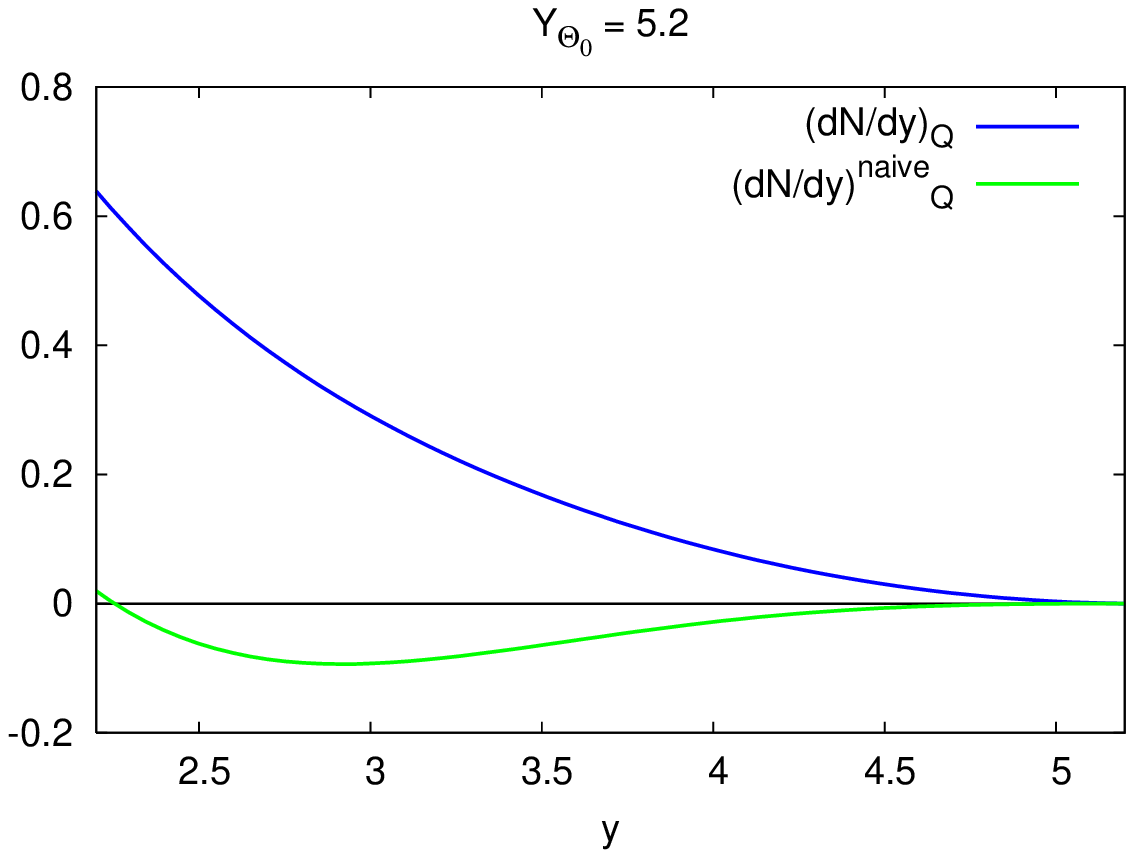, height=5truecm,width=7.5truecm}
\end{center}

\centerline{\em Fig.~22: $\frac{dN}{d\ln k_\perp}$ for a quark jet,
MLLA and naive approach.}
}

\vskip .5cm

That the upper bound of the $\ell_1$ domain of integration defining
$\frac{dN}{d\ln k_\perp}$ corresponds to a large enough $\ell_1 \geq 2.5$
requires that, for LEP, $y_1$ should be smaller that $5.2 - 2.5 = 2.7$;
combined with the necessity to stay in the perturbative regime, it yields
$1 \leq y_1 \leq 2.7$.

\subsection{Discussion and predictions for the Tevatron}
\label{subsection:discussion}
%%%%%%%%%%%%%%%%%%%%%%%%%%%%%%%%%%%%%%%%%%%%%%%%%%%%%%%%

The similar condition at Tevatron is $1 \leq y_1 \leq 5.6 - 2.5 = 3.1$;
like for LEP, it  does not extend to large values of $k_\perp$
because, there, the small $x$ approximation is no longer valid.
We give below the curves that we predict in this confidence
interval.

\vbox{
\begin{center}
\epsfig{file=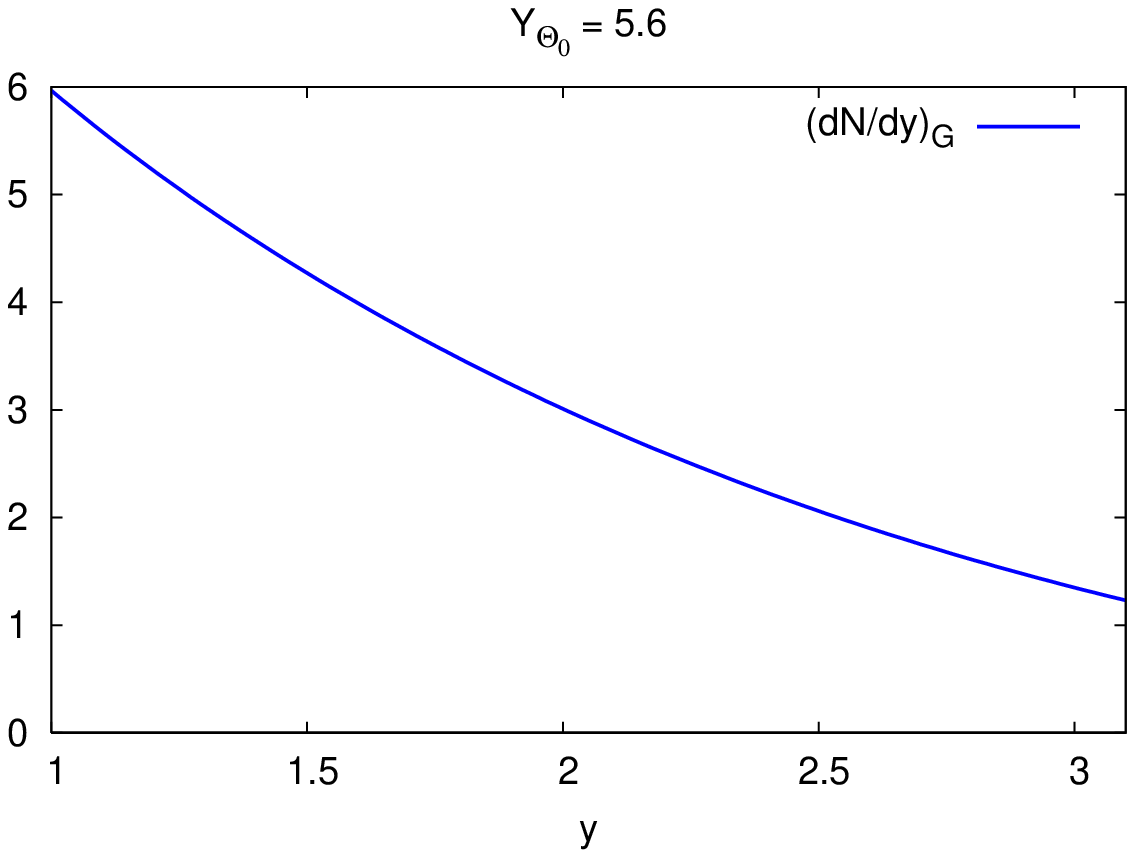, height=5truecm,width=7.5truecm}
\hfill
\epsfig{file=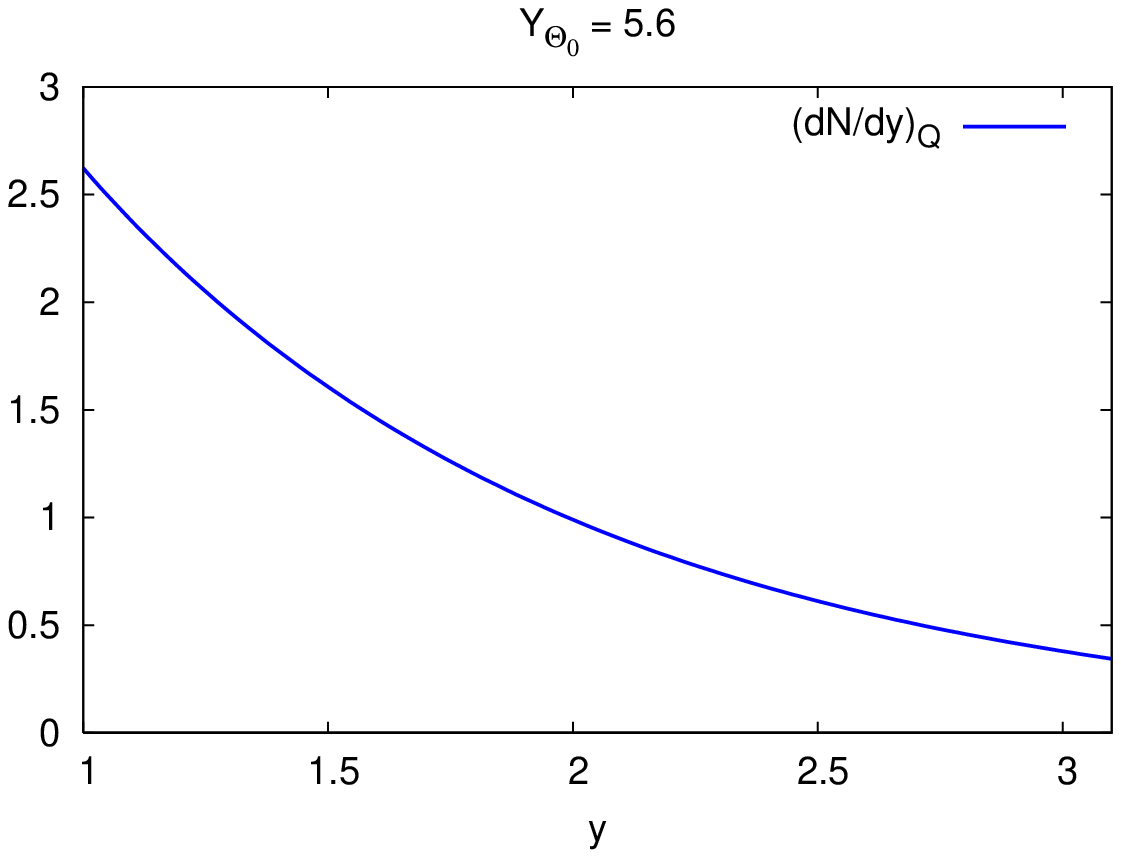, height=5truecm,width=7.5truecm}
\end{center}

\centerline{\em Fig.~23: $\frac{dN}{d\ln k_\perp}$ for a gluon (left) and
a quark (right) jets, MLLA predictions for the Tevatron.}
}

\bigskip

Since experimental results involve a mixture of gluon and quark jets, the
mixing parameter $\omega$ (subsection \ref{subsection:mixed}) has to be
introduced in the comparison with theoretical curves, together with the
phenomenological factor ${\cal K}^{ch}$ 
 normalizing partonic to charge hadrons distributions.

\vskip .75 cm

%%%%%%%%%%%%%%%%%%%%%%%%%%%%%%%%%%%%%%%%%%%%%%%%%%%%%%%%%%%%%%%%%%%%%%%%%%%%
\section{COMPARING DLA AND MLLA APPROXIMATIONS}
\label{section:DLAI}
%%%%%%%%%%%%%%%%%%%%%%%%%%%%%%%%%%%%%%%%%%%%%%%%%%%%%%%%%%%%%%%%%%%%%%%%%%%%

\vskip .5cm

DLA \cite{DLAI} \cite{DLA1} and MLLA approximations are very different
\cite{EvEqI}; in particular,
the exact balance of energy (recoil effects of partons) is not accounted
for in DLA.

We compare  DLA and MLLA results for the two distributions of concern
in this work.
 Studying first their difference for the spectrum itself
eases the rest of the comparison.

We choose the two values $Y_{\Theta_0}=7.5$ and $Y_{\Theta_0}=15$.
While the first corresponds to the LHC working
conditions (see footnote \ref{footnote:LHC}), the second is purely academic
since, taking for example
$\Theta_0 \approx .5$ and $Q_0 \approx 250\; MeV$, it corresponds to an
energy of  $1635\; TeV$; it is however suitable, as we shall see in
subsection \ref{subsection:ktDLA} to disentangle the effects of coherence and
the ones of the divergence of $\alpha_s$ at low energy in the calculation
of the inclusive $k_\perp$ distribution.

\vskip .75 cm

\subsection{The spectrum}
\label{subsection:DLAspec}
%%%%%%%%%%%%%%%%%%%%%%%%%

\vskip .5cm

Fixing  $\alpha_s$  in DLA at the largest scale of the process, the
collision energy, enormously damps the corresponding spectrum (it does not
take into account the growing of $\alpha_s$ accompanying parton cascading),
which gives an unrealistic aspect to the comparison.

This is why, as far as the spectra are concerned, we shall compare their
MLLA evaluation with that obtained from the latter by 
taking to zero the coefficient $a$ given in (\ref{eq:adef}),
which also entails $B=0$; ${\cal F}_0(\tau,y,\ell)$ in
(\ref{eq:calFdef}) becomes $I_0(2\sqrt{Z(\tau,y.\ell)}$.
The infinite normalization that occurs in (\ref{eq:ifD}) because of
$\Gamma(B=0)$ we replace by a constant such that the two calculations can
be easily compared. 
This realizes a DLA  approximation (no accounting for recoil effects)
 ``with running $\alpha_s$''.

On Fig.~24 below are plotted the spectrum $\tilde D_g(\ell,y\equiv
Y_{\Theta_0}-\ell)$ for gluon jets in the MLLA and DLA ``with running
$\alpha_s$'' approximations.

\bigskip

\vbox{
\begin{center}
\epsfig{file=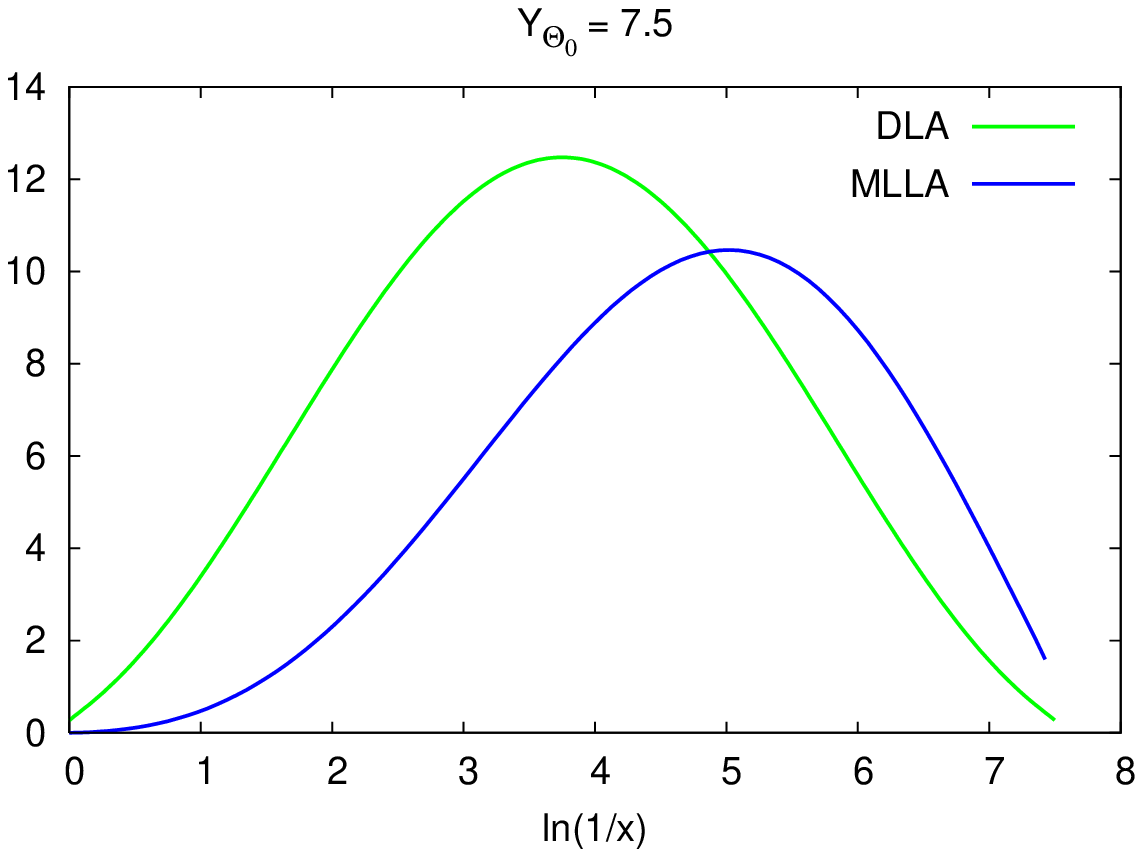, height=5truecm,width=7.5truecm}
\hfill
\epsfig{file=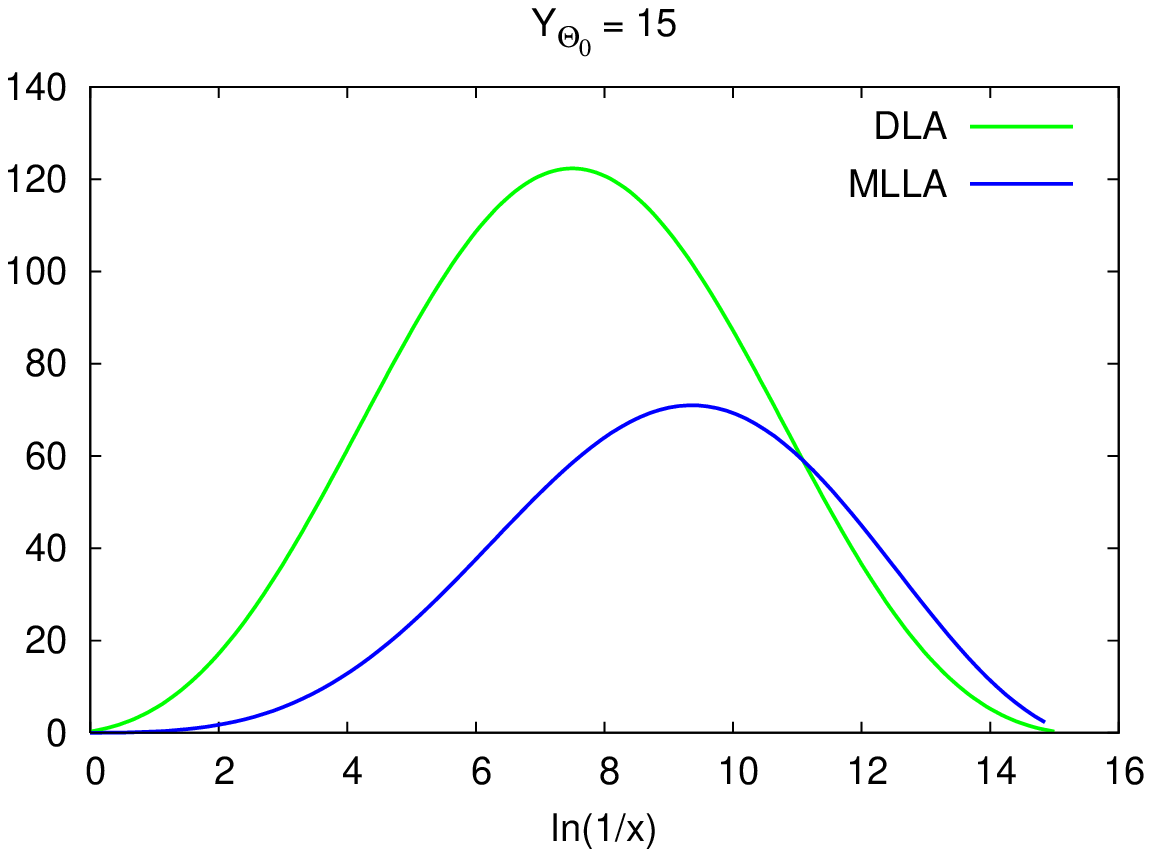, height=5truecm,width=7.5truecm}
\end{center}

\centerline{\em Fig.~24: the spectrum $\tilde D_g(\ell,Y_{\Theta_0}-\ell)$
for gluon jets;}
\centerline{\em  comparison between MLLA and DLA (``with running
$\alpha_s$'') calculations.}
}

\bigskip

The peak of the MLLA spectrum is seen, as expected,
 to occur at smaller values of the energy than that of DLA.

\vskip .5cm

\subsection{Double differential 1-particle inclusive distribution}
\label{subsection:doubleDLA}
%%%%%%%%%%%%%%%%%%%%%%%%%%%%%%%%%%%%%%%%%%%%%%%%%%%%%%%%%%%%%%%%%%%%%%%%%%%%

\vskip .5cm

The genuine MLLA calculations being already shown on Figs.~3 and 5, 
Fig.~25 displays, on the left, a ``modified''
MLLA calculation obtained by dividing by
$\alpha_s(k_\perp^2) \approx \frac{\pi}{2N_c \beta y}$
(see (\ref{eq:gamma0}) with $\lambda \to 0$);
subtracting in the MLLA calculations the dependence on $k_\perp$ due to the
running of $\alpha_s(k_\perp^2)$ allows a better comparison with DLA (with
fixed $\alpha_s$) by getting rid of the divergence when $k_\perp \to Q_0$.

 On the right are plotted the DLA results for gluon jets, in which
$\alpha_s$ has been fixed at the collision energy (it is thus very small).
Since their normalizations are now different, only the {\em shapes} of the two
types of curves must be compared;
we indeed observe that the DLA growing  of
$\frac{d^2N}{d\ell_1\,d\ln k_\perp}$ with $k_\perp$
(or $y_1$) also occurs in the ``modified'' MLLA curves.

\vskip .7cm 

\vbox{
\begin{center}
\epsfig{file=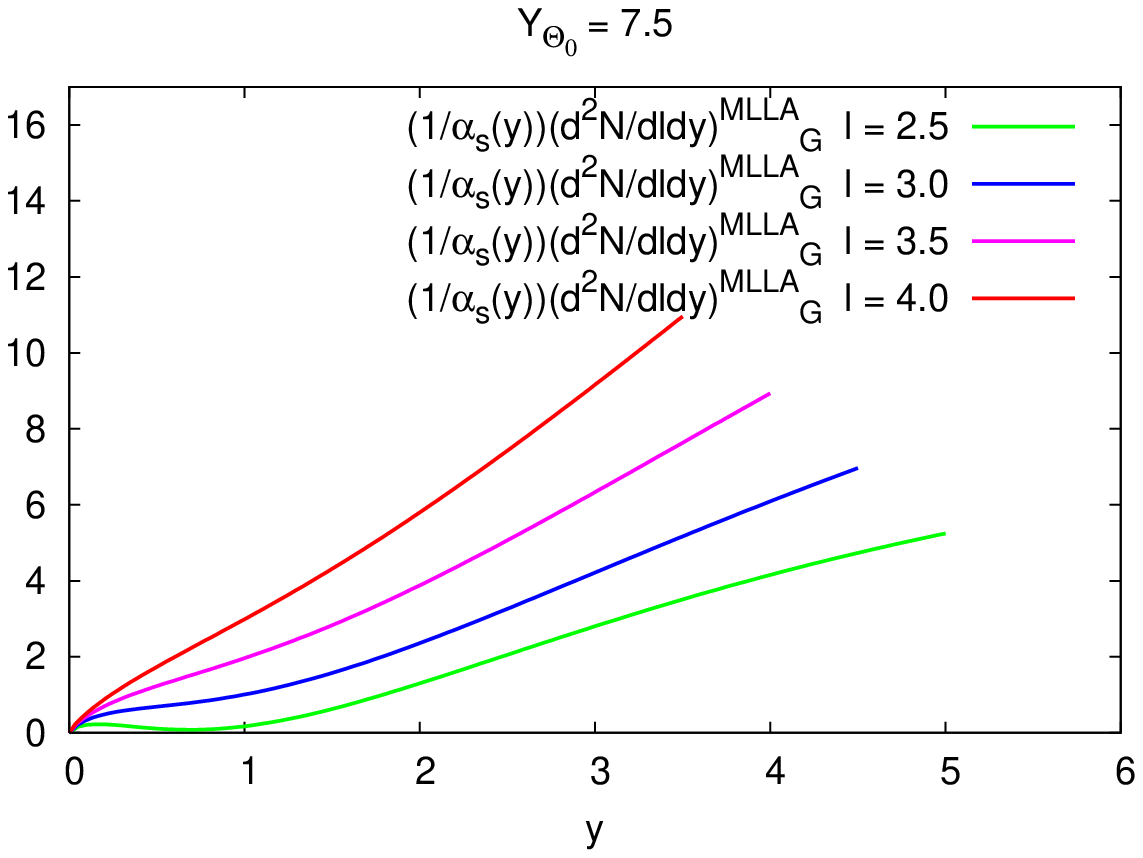, height=5truecm,width=7.5truecm}
\hfill
\epsfig{file=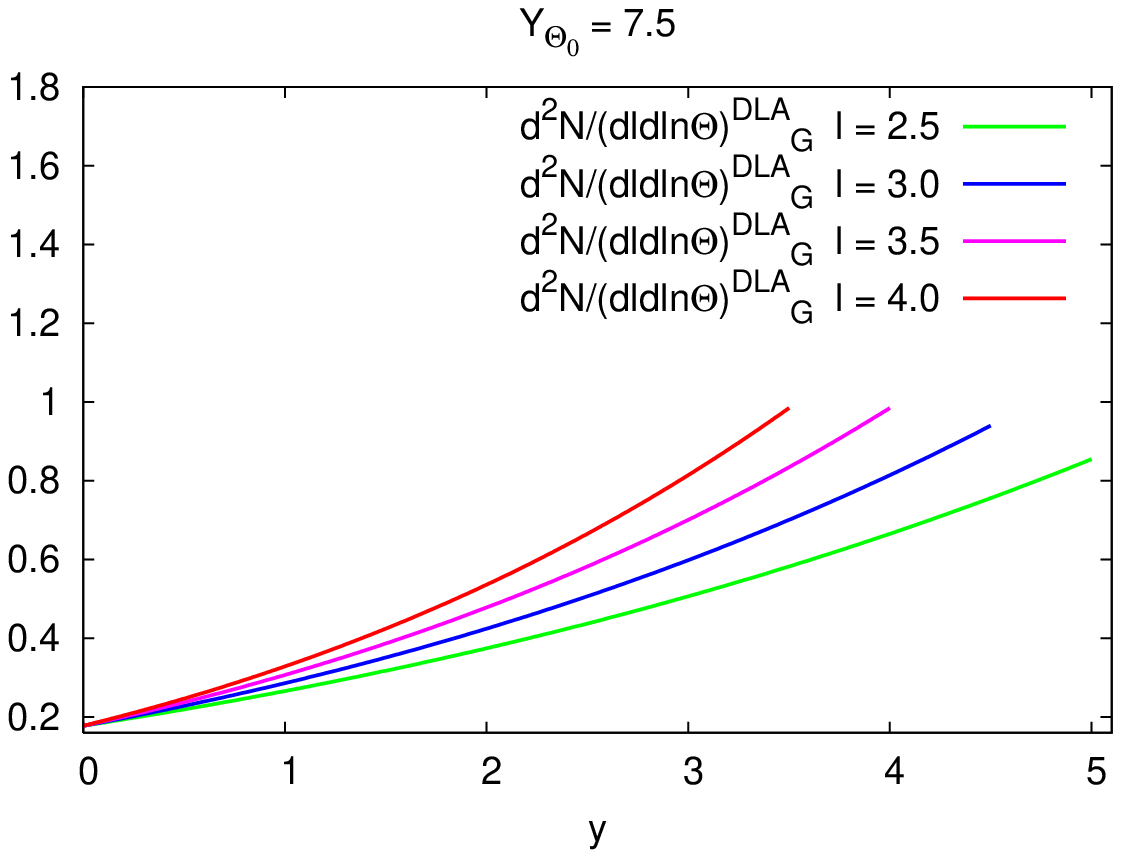, height=5truecm,width=7.5truecm}
\end{center}

\centerline{\em Fig.~25: comparison between MLLA (after dividing by
$\alpha_s(k_\perp^2)$, on the left)}
\centerline{\em  and DLA calculation with $\alpha_s$ fixed (on the right)
of $\frac{d^2N}{dy\,d\ln k_\perp}$ for gluon jets.}
}

\bigskip

The DLA distribution for quark jets is obtained from that of gluon jets
by multiplication by the factor $C_F/N_c$; it it thus also a
growing function of $y_1$.

The MLLA distribution for quark jets, which is, unlike that for gluon jets,
a decreasing function of $y_1$ (see Fig.~6), becomes, like the latter, 
growing, after the dependence on $\alpha_s(k_\perp^2)$ has been
factored out: one  finds the same behavior as in DLA.

\vskip .75 cm

\subsection{Inclusive $\boldsymbol{k_\perp}$ distribution}
\label{subsection:ktDLA}
%%%%%%%%%%%%%%%%%%%%%%%%%%%%%%%%%%%%%%%%%%%%%%%%%%%%%%%%%%%%%%%%%%%%%%%%%%%%

\vskip .5cm

On Fig.~26  we have plotted, at $Y_{\Theta_0}= 7.5$:

- the MLLA calculation of $\frac{dN}{d\ln k_\perp}$
divided by $\alpha_s(k_\perp^2)$,
such that the divergence due to the running of $\alpha_s$
has been factored out, leaving  unperturbed the damping  due to
coherence effects;

- the DLA calculation of  $\frac{dN}{d\ln k_\perp}$ with $\alpha_s$ fixed
  at the collision energy.

Like in \ref{subsection:doubleDLA}, because of the division by $\alpha_s$,
the two curves are  not normalized alike, such that only
their {\em shapes} should be compared.

The comparison of the DLA curve (at fixed $\alpha_s$)
 with the genuine MLLA calculation displayed in Fig.~7  (left)
 shows how different are the outputs of the two
approximations; while at large $k_\perp$ they are both decreasing, at small
$k_\perp$  the running of $\alpha_s$  makes the sole MLLA
distribution diverge when $k_\perp \to Q_0$ (non-perturbative
domain).

\vskip .5cm

\vbox{
\begin{center}
\epsfig{file=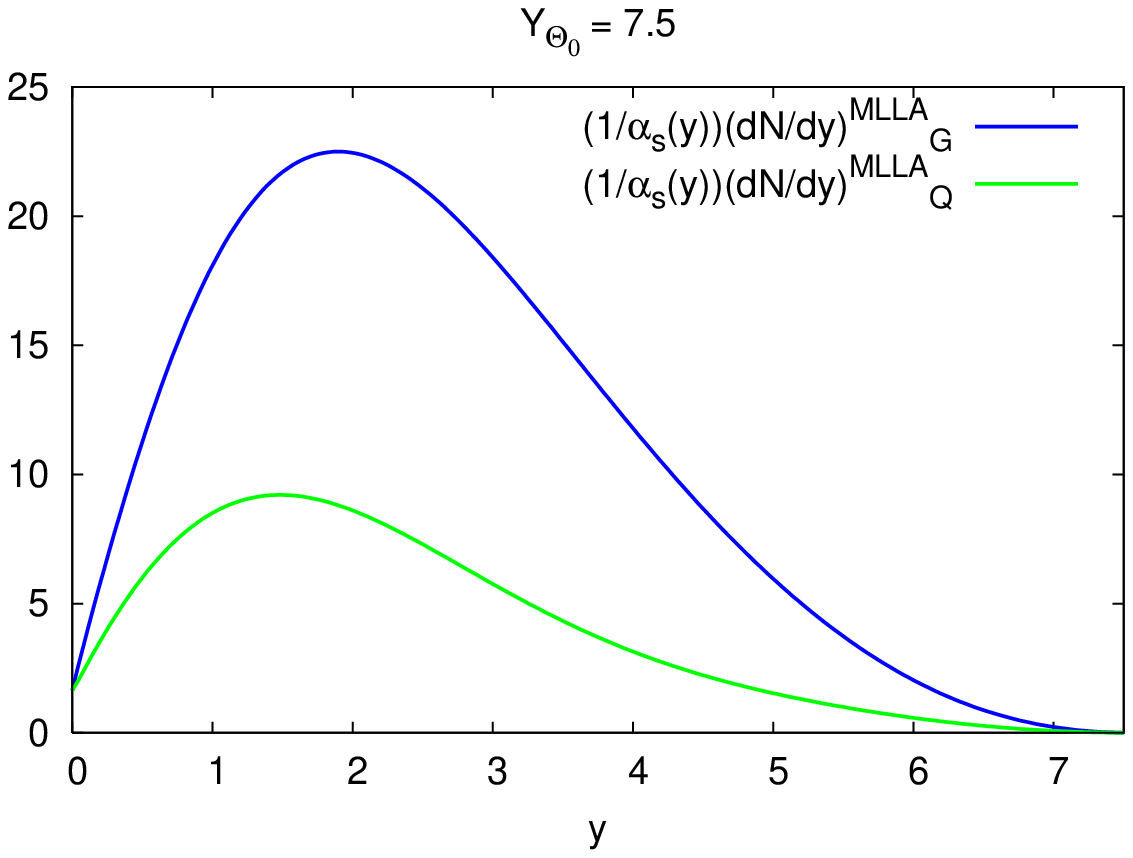, height=5truecm,width=7.5truecm}
\hfill
\epsfig{file=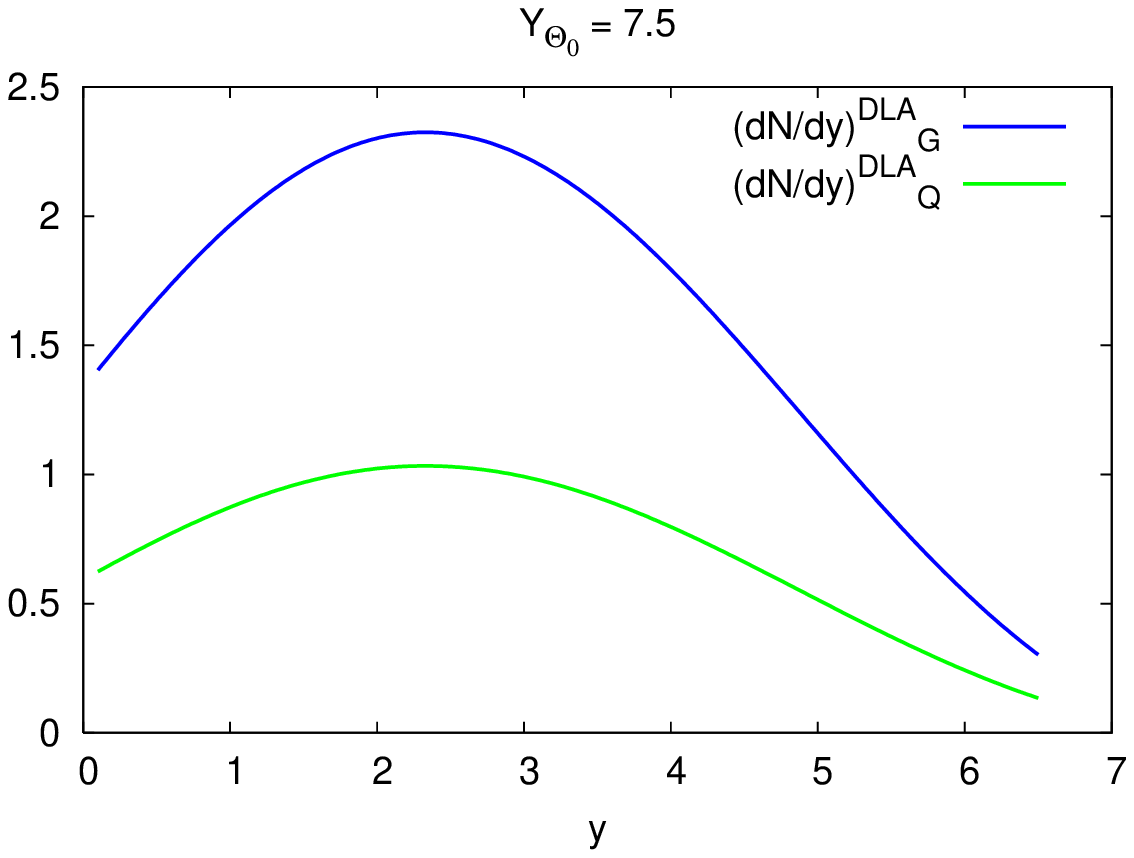, height=5truecm,width=7.5truecm}
\end{center}

\centerline{\em Fig.~26: $Y_{\Theta_0}=7.5$:
comparing  MLLA and DLA calculations
of $\frac{dN}{d\ln k_\perp}$ (see also Fig.~7);}
\centerline{\em from left to right: 
$\frac{1}{\alpha_s(k_T^2)}$MLLA
and DLA ($\alpha_s$ fixed).}
}

\vskip .5cm

In the extremely high domain of energy $Y_{\Theta_0}=15$ used for Fig.~27,
the two competing phenomena occurring
 at small $y_1$ can then be neatly distinguished.

The first plot, showing MLLA results,  cleanly separates
coherence effects from the running of  $\alpha_s$;
in the second figure we have plotted the  MLLA calculation
divided by $\alpha_s(k_\perp^2)$: damping at small $y_1$ due to coherence
effects appears now unspoiled;
finally, DLA calculations  clearly exhibit, too, the damping due to coherence
\footnote{The DLA points corresponding to $y_1=0$ can be analytically
determined to be $4N_c/n_f$ (gluon jet) and $4C_F/n_f$ (quark jet); they
are independent of the energy $Y_{\Theta_0}$.}
.

The large difference of magnitude  observed between the first
(genuine MLLA) and the last (DLA) plots occurs because DLA calculations
have been performed with $\alpha_s$ fixed at the very high collision energy.

Like in \ref{subsection:doubleDLA}, because of the division by $\alpha_s$,
the second curve is not normalized like the two others, such that only
its {\em shape} should be compared with theirs.

\begin{center}
\epsfig{file=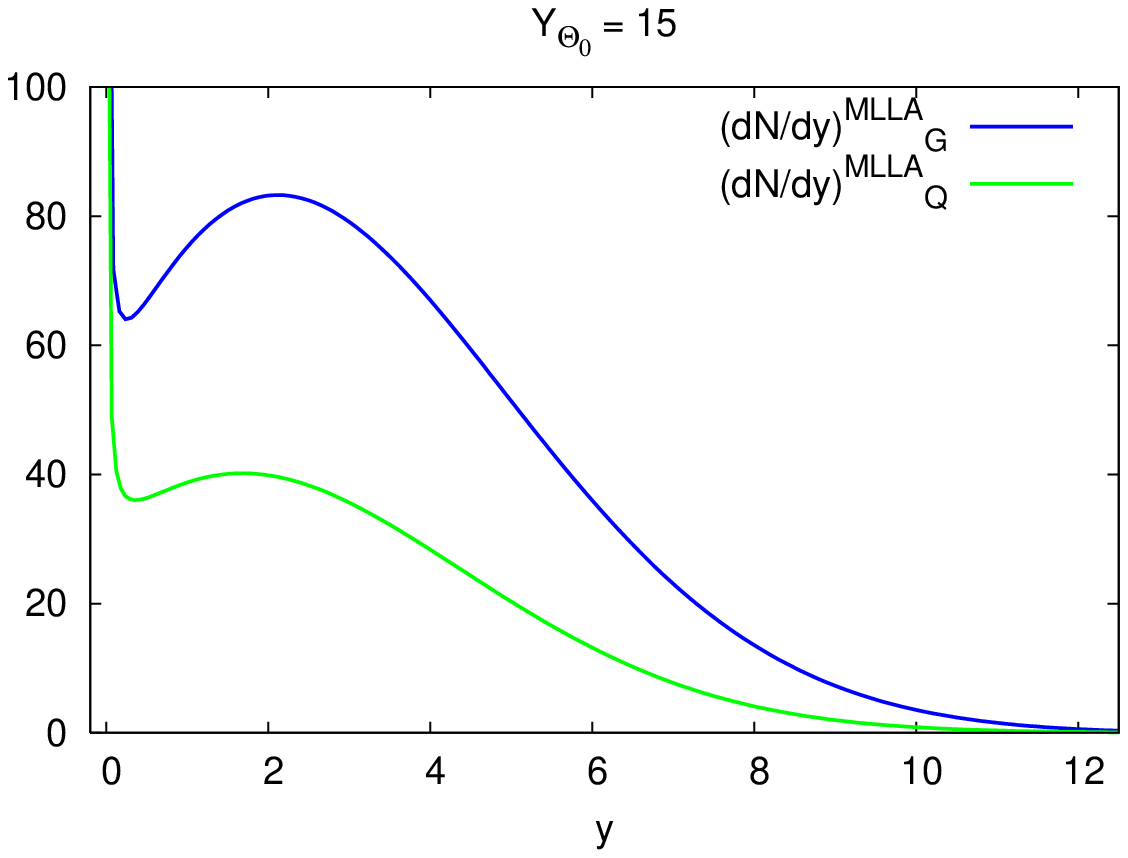, height=5truecm,width=7.5truecm}
\hfill
\epsfig{file=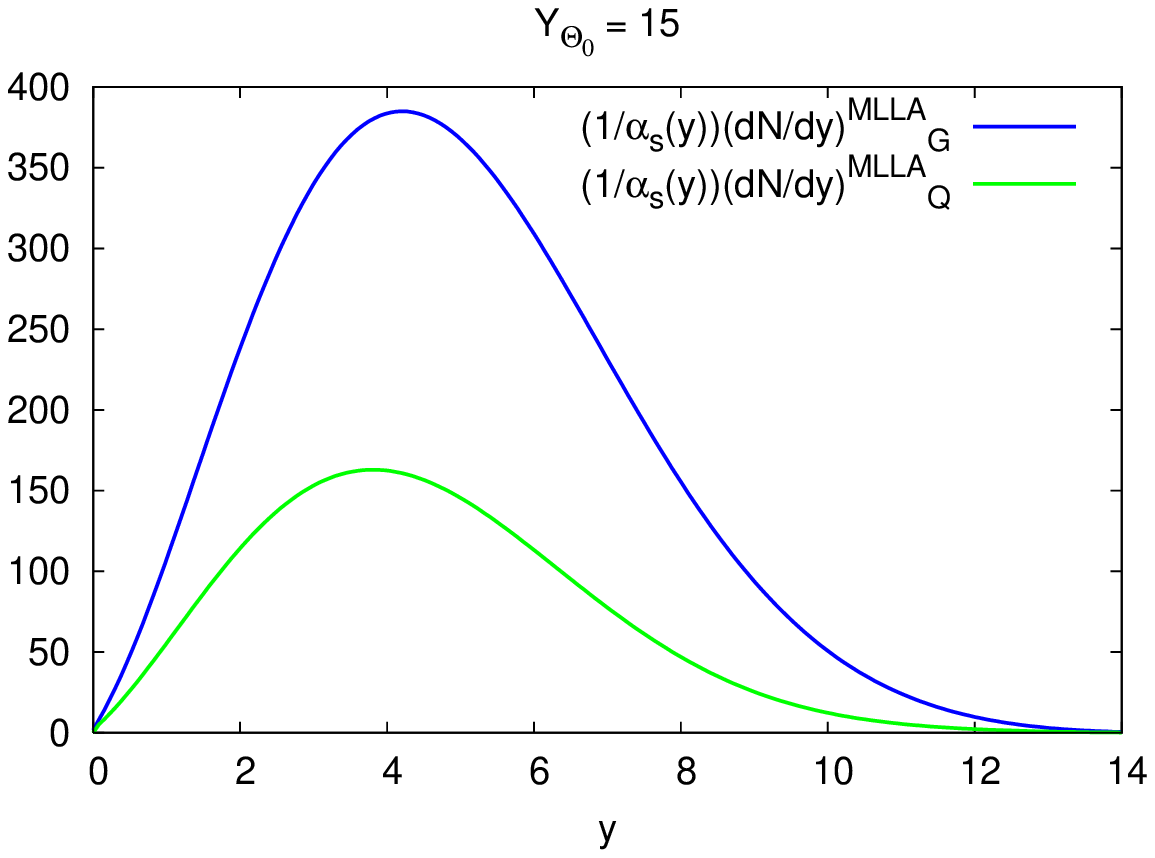, height=5truecm,width=7.5truecm}
\end{center}

\vbox{
\begin{center}
\epsfig{file=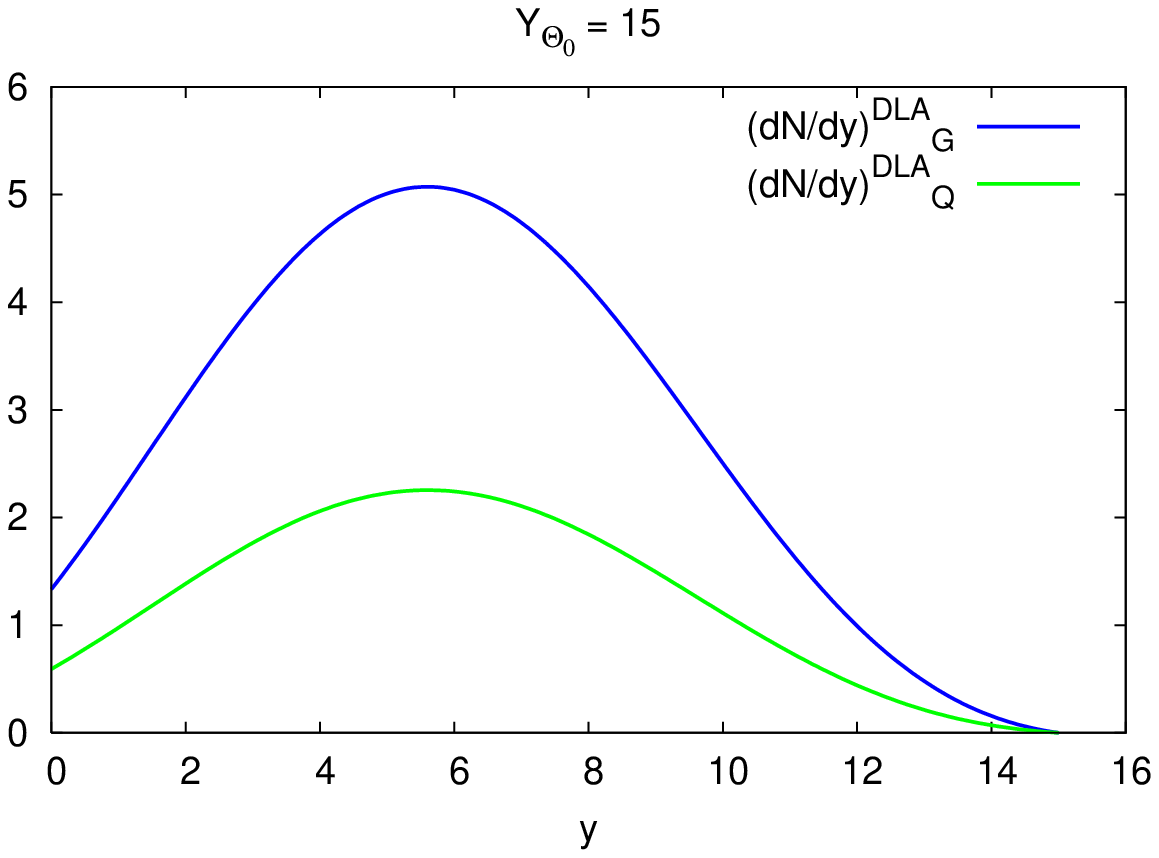, height=5truecm,width=7.5truecm}
\end{center}

\centerline{\em Fig.~27: $Y_{\Theta_0} = 15$: 
comparing  MLLA and DLA calculations
of $\frac{dN}{d\ln k_\perp}$;}
\centerline{\em from left to right: MLLA,
$\frac{1}{\alpha_s(k_T^2)}$MLLA and DLA ($\alpha_s$ fixed).}
}

\bigskip

%%%%%%%%%%%%%%%%%%%%%%%%%%%%%%%%%%%%%%%%%%%%%%%%%%%%%%%%%%%%%%%%%%%%%%%%%%%%
%%%%%%%%%%%%%%%%%%%%%%%%%%%%%%%%%%%%%%%%%%%%%%%%%%%%%%%%%%%%%%%%%%%%%%%%%%%%
\newpage

{\bf\Large Figure captions}

\vskip .75 cm

Fig.~1: the process under consideration: two hadrons $h_1$ and $h_2$ inside
one jet;

\smallskip

Fig.~2: $<C>_{A_0}^0$ and $<C>_{A_0}^0 + \delta<C>_{A_0}$ for quark and gluon
jets, as functions of $y$,
for $Y_{\Theta_0}=7.5$, $\ell=2.5$  and $\ell=3.5$;

\smallskip

Fig.~3: $\frac{d^2N}{d\ell_1\;d\ln k_\perp}$ for a gluon jet,
$Y_{\Theta_0} =7.5$ and $Y_{\Theta_0}=10$;

\smallskip

Fig.~4: $\frac{d^2N}{d\ell_1\;d\ln k_\perp}$ at fixed $\ell_1$ for a gluon jet,
 comparison between MLLA and the naive approach;

\smallskip

Fig.~5: $\frac{d^2N}{d\ell_1\;d\ln k_\perp}$ for a quark jet,
 $Y_{\Theta_0} =7.5$ and $Y_{\Theta_0}=10$;

\smallskip

Fig.~6: $\frac{d^2N}{d\ell_1\;d\ln k_\perp}$ at fixed $\ell_1$ for a quark jet,
 comparison between MLLA and the naive approach;

\smallskip

Fig.~7: inclusive $k_\perp$ distribution $\frac{d{N}} {d\ln k_\perp}$
for a gluon jet,  $Y_{\Theta_0} =7.5$ and $Y_{\Theta_0}=10$,
MLLA and naive approach, both for $\ell_{min}=0$;

\smallskip

Fig.~8: enlargements of Fig.~6 at large $k_\perp$;

\smallskip

Fig.~9: inclusive $k_\perp$ distribution  $\frac{d{N}} {d\ln k_\perp}$
for a quark jet,  $Y_{\Theta_0} =7.5$ and
$Y_{\Theta_0}=10$,
MLLA and naive approach, both for $\ell_{min}=0$;

\smallskip

Fig.~10: enlargements of Fig.~8 at large $k_\perp$;

\smallskip

Fig.~11: role of the upper limit of integration over $x_1$
in the inclusive $k_\perp$ distribution $\frac{dN}{d\ln\,k_\perp}$
for gluon (left) and quark (right) jet;

\smallskip

Fig.~12: spectrum $\tilde D_g(\ell,y)$ of emitted partons as functions 
of transverse momentum (left) and energy (right);

\smallskip

Fig.~13:  enlargements of Fig.~11 close to the origin;

\smallskip

Fig.~14:  $\frac{d\tilde D_g(\ell,y)}{dy}$ as a function of $y$
for different values of $\ell$;

\smallskip

Fig.~15: $\frac{d\tilde D_g(\ell,y)}{dy}$ as a function of $\ell$
for different values of $y$;

\smallskip

Fig.~16: $\frac{d\tilde D_g(\ell,y)}{d\ell}$ as a function of $\ell$
for different values of $y$;

\smallskip

Fig.~17: $\frac{d\tilde D_g(\ell,y)}{d\ell}$ as a function of $y$
for different values of $\ell$;

\smallskip

Fig.~18: $<C>_{A_0}^0$ and $<C>_{A_0}^0 + \delta\!<C>_{A_0}$
for quark and gluon jets, as functions of $y$,
for $Y_{\Theta_0}=5.2$, $\ell=1.5$ and $\ell=2.5$;

\smallskip

Fig.~19: $\frac{d^2N}{d\ell_1\,d\ln k_\perp}$ for a gluon jet
for $Y_{\Theta_0}=5.2$ at fixed $\ell_1$,  MLLA and naive approach;

\smallskip

Fig.~20: $\frac{d^2N}{d\ell_1\,d\ln k_\perp}$ for a quark jet
for $Y_{\Theta_0}=5.2$ at fixed $\ell_1$,  MLLA and naive approach;

\smallskip

Fig.~21: $\frac{dN}{d\ln k_\perp}$ for a gluon jet
for $Y_{\Theta_0}=5.2$,  MLLA and naive approach;

\smallskip

Fig.~22: $\frac{dN}{d\ln k_\perp}$ for a quark jet
for $Y_{\Theta_0}=5.2$,  MLLA and naive approach;

\smallskip

Fig.~23: $\frac{dN}{d\ln k_\perp}$ for a gluon and a quark
jets, MLLA predictions for the Tevatron.

\smallskip

Fig.~24: the spectrum $\tilde D_g(\ell,Y_{\Theta_0}-\ell)$
for gluon jets;  comparison between MLLA and DLA (``with running
$\alpha_s$'') calculations;

\smallskip

Fig.~25:  comparison between MLLA (after dividing by
$\alpha_s(k_\perp^2)$, on the left)
and DLA calculation with $\alpha_s$ fixed (on the right)
of $\frac{d^2N}{dy\,d\ln k_\perp}$ for gluon jets;

\smallskip

Fig.~26: $Y_{\Theta_0}=7.5$: comparing  MLLA and DLA calculations
of $\frac{dN}{d\ln k_\perp}$ (see also Fig.~6); from left to right: 
$\frac{1}{\alpha_s(k_T^2)}$MLLA and DLA ($\alpha_s$ fixed);

\smallskip

Fig.~27: $Y_{\Theta_0} = 15$: comparing  MLLA and DLA calculations
of $\frac{dN}{d\ln k_\perp}$; from left to right: MLLA,
$\frac{1}{\alpha_s(k_T^2)}$MLLA and DLA ($\alpha_s$ fixed).

%%%%%%%%%%%%%%%%%%%%%%%%%%%%%%%%%%%%%%%%%%%%%%%%%%%%%%%%%%%%%%%%%%%%%%%%%%%%%%
%%%%%%%%%%%%%%%%%%%%%%%%%%%%%%%%%%%%%%%%%%%%%%%%%%%%%%%%%%%%%%%%%%%%%%%%%%%%%%

\setcounter{page}{167}

%%%%%%%%%%%%%%%%%%%%%%%%%%%%%%%%%%%%%%%%%%%%%%%%%%%%%%%%%%%%%%%%%%%%%%%%%%%%

\null

\noindent

\vskip 10 cm

\chapter{Two-particle correlations inside one jet at 
``Modified Leading Logarithmic Approximation'' of Quantum Chromodynamics ;
\newline I :
Exact solution of the evolution equations at small $\boldsymbol x$}
\label{sub:article2}

\begin{titlepage}

\setcounter{page}{181}

%%%%%%%%%%%%%%%%%%%%%%%%%%%%%%%%%%%%%%%%%%%%%%%%%%%%%%%%%%%%%%%%%%%%%%%%%%%%%%
May 2006 \hfill hep-ph/0605083    , JHEP 06 (2006) 019

\vskip 4cm

\centerline{\bf TWO PARTICLE CORRELATIONS INSIDE ONE JET}
\medskip
\centerline{\bf  AT ``MODIFIED LEADING LOGARITHMIC APPROXIMATION''}
\medskip
\centerline{\bf  OF QUANTUM CHROMODYNAMICS}
\medskip
\centerline{\bf  I: EXACT SOLUTION OF THE EVOLUTION EQUATIONS AT SMALL
        $\boldsymbol{X}$}

\vskip 1cm

\centerline{Redamy Perez-Ramos
\footnote{E-mail: perez@lpthe.jussieu.fr}
}

\baselineskip=15pt

\smallskip
\centerline{\em Laboratoire de Physique Th\'eorique et Hautes Energies
\footnote{LPTHE, tour 24-25, 5\raise 3pt \hbox{\tiny \`eme} \'etage,
Universit\'e P. et M. Curie, BP 126, 4 place Jussieu,
F-75252 Paris Cedex 05 (France)}}
\centerline{\em Unit\'e Mixte de Recherche UMR 7589}
\centerline{\em Universit\'e Pierre et Marie Curie-Paris6; CNRS;
Universit\'e Denis Diderot-Paris7}

\vskip 2cm

{\bf Abstract}: We discuss correlations between two particles in jets 
at high energy colliders and exactly solve the MLLA evolution equations
in the small $x$ limit. We thus extend the Fong-Webber analysis 
to the region away from the hump of the single inclusive energy spectrum.
We give our results for LEP, Tevatron and LHC energies, and compare with 
existing experimental data.

\vskip 1 cm

{\em Keywords: Perturbative Quantum Chromodynamics, Particle Correlations
in jets, High Energy Colliders}

\vfill

%\null\hfil\epsfig{file=Figs/LOGOP7.eps, height=1.2cm,width=4.5cm}

\vskip .5cm

%%%%%%%%%%%%%%%%%%%%%%%%%%%%%%%%%%%%%%%%%%%%%%%%%%%%%%%%%%%%%%%%%%%%%%%%%%%%%

\end{titlepage}

%%%%%%%%%%%%%%%%%%%%%%%%%%%%%%%%%%%%%%%%%%%%%%%%%%%%%%%%%%%%%%%%%%%%%%%%%%%%%

%\tableofcontents

%\newpage
%%%%%%%%%%%%%%%%%%%%%%%%%%%%%%%%%%%%%%%%%%%%%%%%%%%%%%%%%%%%%%%%%%%%%%%%%%%%%

%%%%%%%%%%%%%%%%%%%%%%%%%%%%%%%%%%%%%%%%%%%%%%%%%%%%%%%%%%%%%%%%%%%%%%%%%%%%%
\section{INTRODUCTION}
\label{section:introC}
%%%%%%%%%%%%%%%%%%%%%%%%%%%%%%%%%%%%%%%%%%%%%%%%%%%%%%%%%%%%%%%%%%%%%%%%%%%%%

Perturbative QCD (pQCD) successfully predicts inclusive energy
spectra of particles in jets.  To this end it was enough to make one
step beyond the leading ``Double Logarithmic Approximation'' (DLA) which
is known to overestimate soft gluon multiplication, and to describe
parton cascades with account of first sub-leading single logarithmic
(SL) effects.
Essential SL corrections to DLA arise from:

$\ast$\  the running coupling $\alpha_s(k_\perp^2)$;

$\ast$\  decays of a parton into two with comparable
energies, $z\sim 1$ (the so called ``hard corrections'', taken care of
by employing exact DGLAP \cite{DGLAPC} splitting functions);

$\ast$\  kinematical regions of successive parton decay angles of the same
order of magnitude, $\Theta_{i+1} \sim \Theta_i$. The solution
to the latter problem turned
out to be extremely simple namely, the replacement of the {\em
strong} angular ordering (AO), $\Theta_{i+1} \ll \Theta_i$,
imposed by gluon coherence in  DLA , by the {\em
exact} AO condition $\Theta_{i+1} \le \Theta_i$ (see \cite{EvEqC}
and references therein).
The corresponding approximation is known as MLLA (Modified Leading
Logarithm Approximation) and embodies the next-to-leading correction, of 
order $\gamma_0^2$, to the parton evolution ``Hamiltonian'', 
$\gamma_0\propto\sqrt{\alpha_s}$ being the DLA multiplicity anomalous
dimension \cite{EvEqC}.

So doing, single inclusive charged hadron spectra (dominated by pions)
were found to be mathematically similar to that of the MLLA parton
spectrum, with an overall proportionality coefficient ${\cal K}^{ch}$
normalizing partonic distributions to the ones of charged hadrons; ${\cal K}^{ch}$ 
depends neither on the jet hardness nor on the particle energy. This finding was
interpreted as an experimental confirmation of the Local
Parton--Hadron Duality hypothesis (LPHD) (for a review
see  \cite{DKTMC}\cite{KOC} and references therein). However, in the ratio of two particle distribution and the product of two single particle distributions that determine the correlation, this non-perturbative parameter cancels. Therefore, one expects this 
observable to provide a more stringent test of parton dynamics.
At the same time, it constitutes much harder a problem for the naive perturbative
QCD (pQCD) approach.
 
The correlation between two soft gluons  was tackled in DLA 
in \cite{DLAC}.  The first realistic prediction with account
of next-to-leading (SL) effects was derived by Fong and Webber in
1990 \cite{FWC}. They obtained the expression for the two
particle correlator in the kinematical region where both particles
were close in energy to the maximum ("hump") of the single inclusive
distribution.  In \cite{OPALC} this pQCD result was compared with the
OPAL $e^+e^-$ annihilation data at the $Z^0$ peak: the analytical
calculations were found to have largely overestimated the measured
correlations.

In this paper we use the formalism of jet generating functionals \cite{KUV}
to derive the MLLA evolution equations for particle
correlators (two particle inclusive distributions). We then use the
soft approximation for the energies of the two particle by neglecting terms
proportional to powers of $x_1,x_2\ll1$ ($x$ is the fraction of the jet
energy carried away by the corresponding particle).  Thus simplified, the
evolution equations can be solved  iteratively and their  solutions 
are given explicitly in terms of logarithmic derivatives of single particle
distributions.

This allows us to achieve two goals. First, we generalize the
Fong--Webber result by extending its domain of application to the full
kinematical range of soft particle energies.  Secondly, by doing this, we
follow the same logic as was applied in describing inclusive spectra
namely, treating {\em exactly} {\em approximate} evolution
equations. Strictly speaking, such a solution, when formally expanded,
inevitably bears sub-sub-leading terms that exceed the accuracy with
which the equations themselves were derived.  This logic, however, was
proved successful in the case of single inclusive spectra \cite{OPALTASSO}, 
which demonstrated that MLLA equations, though approximate, fully take into
account essential physical ingredients of parton cascading:
energy conservation, coherence, running coupling constant.
Applying the same logic to double inclusive distributions
should help to elucidate the problem of particle correlations in QCD
jets.

The paper is organized as follows. 

$\bullet$\quad in section \ref{section:evol} we recall the formalism of jet
generating functionals and their evolution equations; we specialize first
 to inclusive energy spectrum, and then to 2-particle correlations; 

$\bullet$\quad in section \ref{section:soft}, we solve exactly the evolution
equations in the low energy (small $x$) limit; how various
corrections are estimated and controlled is specially emphasized;

$\bullet$\quad section \ref{section:corglu} is dedicated to correlations in a
gluon jet; the equation to be solved iteratively is exhibited, and an
estimate of the order of magnitudes of various contributions is given;

$\bullet$\quad section \ref{section:quark} is dedicated to correlations in a
quark jet, and follows the same lines as section \ref{section:corglu};

$\bullet$\quad in section \ref{section:numer} we give all  numerical
results, for LEP-I,  Tevatron and LHC. They are commented, compared with
Fong-Webber for OPAL, but all detailed numerical investigations concerning
 the size of various corrections is postponed, for the sake of clarity,
to appendix \ref{section:numcorr};

$\bullet$\quad a conclusion summarizes this work.

Six appendices provide all necessary theoretical demonstrations and
numerical investigations.

$\bullet$\quad in appendix \ref{section:Gcorr} and \ref{section:Qcorr} 
we derive the exact solution of the evolution equations for the gluon and 
quark jet correlators;

$\bullet$\quad appendix \ref{section:inspiredDLA} is a technical
complement to subsection \ref{subsection:estimate};

$\bullet$\quad in appendix \ref{section:ESEE} we demonstrate the exact
solution of the MLLA evolution equation for the inclusive spectrum and
give analytic expressions for its derivatives;

$\bullet$\quad appendix \ref{section:numcorr} is dedicated to a
numerical analysis of all corrections that occur in the
iterative solutions of the evolution equations;

$\bullet$\quad in appendix \ref{section:DLAcomp} we perform a comparison
between DLA and MLLA correlators.

%%%%%%%%%%%%%%%%%%%%%%%%%%%%%%%%%%%%%%%%%%%%%%%%%%%%%%%%%%%%%%%%%%%%%%%%%%%%%
\section{EVOLUTION EQUATIONS FOR JET GENERATING FUNCTIONALS}
\label{section:evol}
%%%%%%%%%%%%%%%%%%%%%%%%%%%%%%%%%%%%%%%%%%%%%%%%%%%%%%%%%%%%%%%%%%%%%%%%%%%%%

\begin{figure}
\vbox{
\begin{center}
\epsfig{file=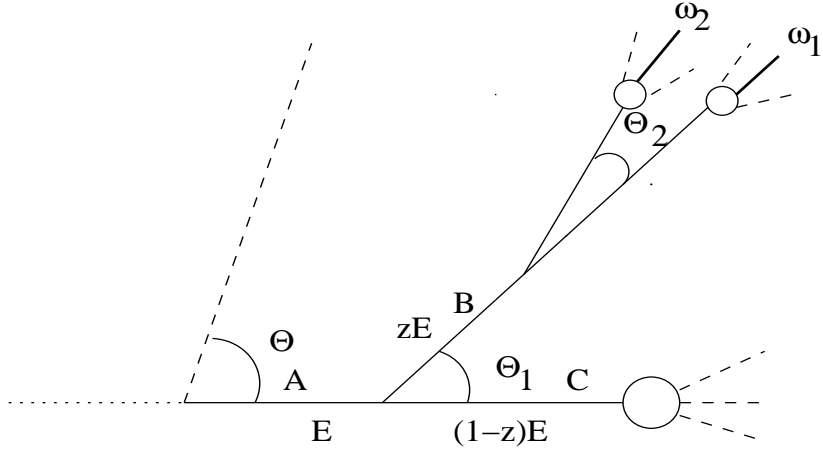, height=6truecm,width=11truecm}
\vskip .5cm
\caption{Two-particle correlations and Angular Ordering}
\label{fig:process}
\end{center}
}
\end{figure}

Consider (see Fig.~\ref{fig:process})
 a jet generated by a parton of type $A$ (quark or gluon)
with 4-momentum $p=(p_0\equiv E,\vec p)$.

 A generating functional $Z(E,\Theta;
\{u\})$ can be constructed \cite{KUV} that describes the azimuth averaged
 parton content of a jet of energy $E$ with a given opening half-angle $\Theta$;
by virtue of the exact angular ordering (MLLA),
it satisfies the following integro-differential evolution equation
\cite{EvEqC}

\medskip
\vbox{
\begin{eqnarray}
\frac{d}{d\ln\Theta}Z_A\left(p,\Theta;\{u\}\right)
\!\!&\!\!=\!\!&\!\!\frac{1}{2}\sum_{B,C}\int_{0}^{1}dz\>
\Phi_A^{B[C]}(z)\ \frac{\alpha_s\left(k^2_{\perp}\right)}{\pi} \cr
&&
\Big(Z_B\big(zp,\Theta;\{u\}\big)\ Z_C\big((1-z)p,\Theta;\{u\}\big)
-Z_A\big(p,\Theta;\{u\}\big)\Big); \cr
&&
\label{eq:red1}
\end{eqnarray}
}
\medskip

in (\ref{eq:red1}), $z$ and $(1-z)$ are the energy-momentum fractions
carried away by the two
offspring of the $A\to B C$ parton decay described by the standard
one loop splitting functions

\medskip

\begin{eqnarray}
\label{eq:split}
&& \Phi_q^{q[g]}(z) = C_F\, \frac{1+z^2}{1-z} , \quad
\Phi_q^{g[q]}(z)=C_F\, \frac{1+(1-z)^2}{z} , \\\notag\\
\label{eq:cst}
&&\hskip -1.5cm \Phi_g^{q[\bar{q}]}(z) = T_R\left( z^2+(1-z)^2\right) , \quad
\Phi_g^{g[g]}(z)= 2C_A\left(\frac{1-z}{z}+ \frac{z}{1-z} +
z(1-z)\right),\\\notag\\
&& C_A=N_c,\quad C_F=(N_c^2-1)/2N_c,\quad T_R=1/2,
\end{eqnarray}

\medskip

where $N_c$ is the number of colors;
$Z_A$ in the integral in the r.h.s. of (\ref{eq:red1})
 accounts for 1-loop virtual corrections, which exponentiate into
Sudakov form factors.

\vskip 0.5cm

$\alpha_s(q^2)$ is the running coupling constant of QCD
\begin{equation}
\alpha_s(q^2)= \frac{4\pi}{4N_c\beta
\ln\displaystyle\frac{q^2}{\Lambda^2_{QCD}}},
\label{eq:alphasC}
\end{equation}
where $\Lambda_{QCD} \approx$ a few hundred $MeV$'s is the intrinsic scale
of QCD, and
\begin{equation}
\beta = \frac{1}{4N_c}\Big(\frac{11}{3}N_c - \frac{4}{3} n_f T_R\Big)
\label{eq:betaC}
\end{equation}
is the first term in the perturbative expansion of the $\beta$ function,
$n_f$ the number of light quark flavors.

If the radiated parton with 4-momentum $k=(k_0,\vec k)$
 is emitted with an angle $\Theta$ with respect to
the direction of the jet, one has ($k_\perp$ is the modulus of the
transverse trivector $\vec k_\perp$ orthogonal to the direction of the jet)
$k_\perp\simeq  |\vec k| \Theta \approx k_0 \Theta \approx 
        z E\Theta$ when $z \ll 1$  or $(1-z)E\Theta$ when $z\to 1$,
and a collinear cutoff $k_\perp\geq Q_0$ is imposed. 

In (\ref{eq:red1}) the symbol $\{u\}$ denotes a set of {\em probing
functions} $u_a(k)$ with $k$ the 4-momentum of a secondary parton of
type $a$.
The jet functional is normalized to the total jet production cross
section such that 
\begin{equation}
Z_A(p,\Theta;u\equiv1)=1;
\label{eq:norm}
\end{equation}
for vanishingly small opening
angle it reduces to the probing function of the single initial parton
\begin{equation}
Z_A(p,\Theta\to 0;\{u\})= u_A(k\equiv p).
\end{equation}

To obtain {\em exclusive} $n$-particle distributions one takes $n$
variational derivatives of $Z_A$ over $u(k_i)$ with appropriate
particle momenta, $i=1 \ldots n$, and sets $u\equiv 0$ after wards;
{\em inclusive} distributions are generated by taking variational
derivatives around $u\equiv 1$.

\subsection{Inclusive particle energy spectrum}
\label{subsection:incluspec}
%%%%%%%%%%%%%%%%%%%%%%%%%%%%%%%%%%%%%%%%%%%%%%%

The probability of soft gluon radiation off a color charge
(moving in the $z$ direction) has the polar angle dependence 

\medskip

\begin{equation*}
\frac{\sin\theta\,d\theta}{2(1-\cos\theta)}
= \frac{d\sin(\theta/2)}{\sin(\theta/2)}
 \simeq\frac{d\theta}{\theta};
\end{equation*}

\medskip

therefore, we choose the angular evolution parameter to be

\medskip

\begin{equation}
Y = \ln \frac{2E\sin(\Theta/2)}{Q_0}
\Rightarrow dY = \frac{d\sin(\Theta/2)}{\sin(\Theta/2)};
\label{eq:Ydef}
\end{equation}

\medskip

this choice accounts for finite angles ${\cal O}(1)$ up to
the full opening half-angle $\Theta=\pi$, at which

\medskip

\begin{equation*}
 Y_{\Theta=\pi}=\ln\frac{2E}{Q_0}
,
\end{equation*}

\medskip

where $2E$ is the center-of-mass annihilation energy of the process
$e^+e^- \to q\bar{q}$. 
For small angles (\ref{eq:Ydef}) reduces to

\medskip

\begin{equation}
Y\simeq \ln\frac{E\Theta}{Q_0}, \quad \Theta\ll 1,
\quad \frac{d}{dY} = \frac{d}{d\ln\Theta},
\label{eq:Ydef2}
\end{equation}

\medskip

where $E\Theta$ is the maximal transverse momentum of a parton inside the jet 
with opening half-angle $\Theta$. 

To obtain the inclusive energy distribution of parton $a$ emitted at angles smaller 
than $\Theta$ with momentum $k_a$, energy $E_a = x E$  in a jet $A$, {\em i.e.}
the fragmentation function
${D}_A^{\,a}(x,Y)$, we take the variational derivative of 
(\ref{eq:red1}) over $u_a(k)$ and set $u\equiv1$ (which also corresponds to
 $Z=1$) according to

\medskip

\begin{equation}
x{D_A^a}(x,Y)
= E_a\frac{\delta}{\delta u(k_a)}
Z_A\left(k,\Theta;\left\{u\right\}\right) \Big\vert_{u=1},
\end{equation}

\medskip

where we have chosen the variables $x$ and $Y$ rather than $k_a$ and $\Theta$.

Two configurations must be accounted for:
$B$ carrying away the fraction $z$ and $C$
the fraction $(1-z)$ of the jet energy, and the symmetric one in which the
role of $B$ and $C$ is exchanged. Upon functional differentiation they
give the same result, which cancels the factor $1/2$. The
system of coupled linear integro-differential equations that comes out is

\medskip

\begin{equation}
\frac{d}{dY} \,
x{D}^a_A(x,Y) = \int_0^1 dz
\sum_B \Phi_A^B(z)\,\frac{\alpha_s}{\pi}
 \left[
\frac{x}{z}  {D}^a_B\left(\frac{x}{z},Y+\ln z\right)
- \frac1{2} x{D}^a_A(x,Y) \right].
\label{eq:SIeveq}
\end{equation}

\medskip

We will be interested in the region of small $x$ where fragmentation functions behave as 
\begin{equation}
x{D} (x) \stackrel{x\ll1}{\sim} \rho(\ln x),
\label{eq:rho}
\end{equation}
with $\rho$ a smooth function of $\ln x$.
Introducing logarithmic {\em parton densities} 
\begin{equation}
Q= x {D}^{\,a}_Q (x,Y),\quad G=x {D}^{\,a}_G (x,Y),
\label{eq:QGdef}
\end{equation}
respectively for quark and gluon jets, we obtain from (\ref{eq:SIeveq}) 

\medskip

\begin{eqnarray}
Q_{y}\equiv \frac{dQ}{dy}\!\!&\!\!=\!\!&\!\!\int_0^1 dz\>  \frac{\alpha_s}{\pi} \>\Phi_q^g(z)\>
\bigg[ \Big(Q(1-z)-Q\Big) + G(z) \bigg],
\label{eq:qpr}\\
G_{y}\equiv \frac{dG}{dy}\!\!&\!\!=\!\!&\!\!\int_0^1 dz\> \frac{\alpha_s}{\pi} \>
\bigg[\Phi_g^g(z) \Big(G(z)-zG\Big)
+n_f\; \Phi_g^q(z)\, \Big(2Q(z)-G\Big) \bigg],
\label{eq:gpr}
\end{eqnarray}

\medskip

where, for the sake of clarity,
  we have suppressed $x$ and $Y$ and only kept
the dependence on the integration variable $z$, e.g.,
\begin{equation}
G(z) \equiv  \frac{x}{z}  {D}^a_G\bigg(\frac{x}{z}, Y + \ln z\bigg),
\label{eq:shortnot}
\end{equation}
such that
\begin{equation}
G=G(1),\quad Q=Q(1).
\label{eq:GQ1}
\end{equation}
Some comments are in order concerning these equations. 
\begin{itemize}
\item
We chose to express the derivative with respect to the jet opening angle
$\Theta$ on the l.h.s.'s of equations (\ref{eq:qpr})(\ref{eq:gpr})
 in terms of 
\begin{equation}
y\equiv  Y-\ell=\ln\frac{xE\Theta}{Q_0}=\ln\frac{E_a\Theta}{Q_0}, 
\quad \ell\equiv\ln\frac1x=\ln\frac{E}{E_a},
\label{eq:yelldef}
\end{equation}
instead of $Y$ defined in (\ref{eq:Ydef}). The variable $y$
is convenient for imposing the collinear cutoff condition
$k_\perp\simeq xE\sin\theta \ge Q_0$ since, for small angles, 
it translates simply into $y\ge 0$;

\item
to obtain (\ref{eq:qpr}) one proceeds as follows. When $B$ 
is a quark in (\ref{eq:SIeveq}) , since $A$ is also a quark, 
one gets two contributions:
the real contribution ${D_{B=q}^a}$ and the virtual one 
$-\frac{1}{2}{D_{A=q}^a}$;

\begin{itemize}
\item

in the virtual contribution, since $\Phi_q^q(z) =
\Phi_q^g(1-z)$, the sum over $B$ cancels the factor $1/2$;

\item

in the real contribution,  when it is a
quark, it is associated with $\Phi_q^q(z)$ and, when it is a gluon, with
$\Phi_q^g(z)$; we use like above the symmetry $\Phi_q^q(z) =
\Phi_q^g(1-z)$  to only keep one of the two, namely $\Phi_q^q$, at the
price of changing  the  corresponding $ D(z)$ into
${D}(1-z)$;

\end{itemize}
\item
to obtain (\ref{eq:gpr}), one goes along the following steps;
now $A=g$ and $B=q$ or $g$;

\begin{itemize}

\item
like before, the subtraction term does not depend on $B$ and is summed
over $B=q$ and $B=g$, with the corresponding splitting functions $\Phi_g^q$
and $\Phi_g^g$. In the term $\Phi_g^g$, using the property
$\Phi_g^g(z)=\Phi_g^g(1-z)$ allows us to replace $\frac12\int_0^1dz\Phi_g^g (z)=
\int_0^1z\Phi_g^g(z)$. This yields upon functional differentiation the $-zG$ term in
(\ref{eq:gpr}).
For $B=q$, $2n_f$ flavors 
($n_f$ flavors of quarks and $n_f$ flavors of
anti-quarks) yield identical contributions, which, owing to
the initial factor $1/2$ finally yields $n_f$;

\item
concerning the real terms,
$\Phi_g^g G$ in (\ref{eq:gpr}) comes directly from
$\Phi_g^{g}\frac{x}{z}{D_{g}^a}$ in (\ref{eq:SIeveq}).
For $B=q$, $2n_f$ flavors  of quarks and antiquarks contribute equally since at 
$x\ll1$ sea quarks are produced via gluons
\footnote{accompanied by  a relatively small fraction
 ${\cal O}(\sqrt{\alpha_s})$ of (flavor singlet) sea quark pairs,
 while the valence (non-singlet) quark
distributions are suppressed as ${\cal O}(x)$.}.
This is why we have multiplied
$Q(z)$ by $2n_f$  in (\ref{eq:gpr}).

\end{itemize}
\end{itemize}
Now we recall that both splitting functions $\Phi_q^g(z)$ and $\Phi_g^g$ 
are singular at $z=0$; the symmetric gluon-gluon splitting $\Phi_g^g(z)$ 
is singular at $z=1$ as well.
The latter singularity in (\ref{eq:gpr}) gets regularized by the
factor $\big(G(z)-zG\big)$ which vanishes at $z\to1$.
This regularization can be made explicit as follows

$$
\int_0^1 dz \Phi_g^g(z)\> \Big(G(z)-zG\Big)
\equiv \int_0^1 dz \Phi_g^g(z)\> \bigg[(1-z)G(z) + z\Big(G(z)-G\Big)\bigg];
$$
since $\Phi_g^g(z) = \Phi_g^g(1-z)$, while leaving the first term
$\int_0^1 dz \Phi_g^g(z)(1-z)G(z)$ unchanged, we can rewrite the second
$$
\int_0^1 dz \Phi_g^g(z)z\Big(G(z)-G\Big) =
\int_0^1 dz \Phi_g^g(z)(1-z)\Big(G(1-z)-G\Big),
$$ 
such that, re-summing the two, $(1-z)$ gets factorized and one gets
\begin{equation}
 \int_0^1 dz \Phi_g^g(z)\> \Big(G(z)-zG\Big) =
\int_0^1 dz \Phi_g^g(z)(1-z)\bigg[ G(z) + \Big(G(1-z) -G\Big)\bigg].
\label{eq:regul1}
\end{equation}

Terms proportional to $G(z)$ on r.h.s.'s of equations 
(\ref{eq:qpr})(\ref{eq:gpr})  remain singular at $z\!\to\! 0$ and   
produce enhanced contributions due to the logarithmic integration
over the region $x\ll z\ll 1$. 

\medskip

Before discussing the MLLA evolution equations following from (\ref{eq:qpr})
and (\ref{eq:gpr}), let us derive similar equation for two particle correlations
inside one jet.

\subsection{Two parton correlations}
\label{subsection:2pc}
%%%%%%%%%%%%%%%%%%%%%%%%%%%%%%%%%%%%%%

We study correlation between two particles with fixed energies $x_1=\omega_1/E$, $x_2=\omega_2/E$ ($x_1>x_2$) emitted at arbitrary angles $\Theta_1$ and $\Theta_2$
smaller than the jet opening angle $\Theta$. If these partons are emitted in a 
cascading process, then $\Theta_1\geq\Theta_2$ by the AO property; see 
Fig.~\ref{fig:process}.

\subsubsection{Equations}
\label{subsub:eqs}
%%%%%%%%%%%%%%%%%%%%%%%%%

Taking the second variational derivative of (\ref{eq:red1})
with respect to $u(k_1)$ and $u(k_2)$, one gets a system of equations
for the two-particle distributions $G^{(2)}$ and $Q^{(2)}$
in gluon and quark jets, respectively:

\vbox{
\begin{eqnarray}
\label{eq:Q2pr}
&& \hskip -4cm Q^{(2)}_{y} =  \int dz\> \frac{\alpha_s}{\pi}\>
 \Phi_q^g(z)\bigg[ G^{(2)}(z)\!+\! \Big(Q^{(2)}(1\!-\!z)\!\!-Q^{(2)}\Big) 
  + G_1(z)Q_2(1\!-\!z)\! +\! G_2(z)Q_1(1\!-\!z) \bigg],\\\notag\\
\label{eq:G2pr}
 G^{(2)}_{y} \!\!\!&\!\!=\!\!&\!\!\!  \int dz\>  \frac{\alpha_s}{\pi}\> \Phi_g^g(z)
\bigg[ \Big(G^{(2)}(z)\!-\!zG^{(2)}\Big) 
  + G_1(z)G_2(1\!-\!z) \bigg] \cr
 && +  \int dz\> \frac{\alpha_s}{\pi}\> n_f\Phi^q_g(z)
\bigg[ \Big(2Q^{(2)}(z)\!-\!G^{(2)}\Big)\! +\! 2Q_1(z)Q_2(1\!-\!z) \bigg] .
\end{eqnarray}
}

Like before, the notations have been lightened to a maximum, such that
 $Q^{(2)} = Q^{(2)}(z=1),\ G^{(2)} = G^{(2)}(z=1)$.  More details
about the variables on which $Q^{(2)}$ depend are given in subsection
\ref{subsection:MLLAcor}. % after (\ref{eq:eveeqqC}) (\ref{eq:eveeqgluC}).
Now using (\ref{eq:qpr}) we construct the $y$-derivative of the
product of single inclusive spectra. Symbolically, 

\vbox{
\begin{eqnarray}
(Q_1Q_2)_{y}\!\! &\!\!=\!\!&\!\! Q_2\int_0^1 dz \frac{\alpha_s}{\pi}\Phi_q^g(x)
\Big[\big(Q_1(1-z)-Q_1\big)+G_1(z)\Big] \cr
         && \hskip 1cm +  Q_1 \int_0^1 dz \frac{\alpha_s}{\pi}\Phi_q^g(x)
\Big[\big(Q_2(1-z)-Q_2\big)+G_2(z)\Big].
\label{eq:Qprod}
\end{eqnarray}
}

Subtracting this expression from (\ref{eq:Q2pr}) we get

\medskip

\vbox{
\begin{eqnarray}
\label{eq:Q2prsub}
(Q^{(2)}-Q_1Q_2)_{y}  \!\!&\!\!=\!\!& \!\! \int \!dz\>\frac{\alpha_s}{\pi}\> 
 \Phi_q^g(z)\>\bigg[ G^{(2)}(z)+ \Big(Q^{(2)}(1-z)-Q^{(2)}\Big) \cr
&& \hskip -4cm + \Big(G_1(z)-Q_1\Big)\Big(Q_2(1-z) - Q_2\Big) +  \Big(G_2(z)-
Q_2\Big)\Big(Q_1(1-z)- Q_1\Big)\bigg].
\end{eqnarray}
}

\medskip

For the gluon jet,  making use of (\ref{eq:gpr})  we  analogously obtain
from (\ref{eq:G2pr})

\medskip

\vbox{
\begin{eqnarray}
\label{eq:G2prsub}
(G^{(2)}-G_1G_2)_{y} \!\!&\!\!=\!\!&\!\!
\int dz\> \frac{\alpha_s}{\pi}
\Phi_g^g(z)\>\bigg[ \Big(G^{(2)}(z)-zG^{(2)}\Big) 
  + \Big(G_1(z)-G_1\Big)\Big(G_2(1-z)-G_2\Big) \bigg] \cr
\!\!&\!\!+\!\!&\!\!  \int dz\>\frac{\alpha_s}{\pi}\, n_f\Phi_g^q(z)\>
\bigg[ 2\Big(Q^{(2)}(z)-Q_1(z)Q_2(z)\Big) - \Big(G^{(2)}-G_1G_2\Big)\cr
\!\!&\!\! +\!\!&\!\! \Big(2Q_1(z)-G_1\Big)\Big(2Q_2(1-z)-G_2\Big) \bigg].
\end{eqnarray}
}
The combinations on the l.h.s.'s of (\ref{eq:Q2prsub}) and (\ref{eq:G2prsub}) form {\em correlation functions} which vanish when particles 1 and 2 are produced independently.
They represent the combined probability of emitting particle 2 with $\ell_2, y_2,\ldots$ when particle 1 with $\ell_1,y_1,\ldots$ is emitted, too.
This way of representing the r.h.s.'s of the equations is convenient
for estimating the magnitude of the various terms.

%%%%%%%%%%%%%%%%%%%%%%%%%%%%%%%%%%%%%%%%%%%%%%%%%%%%%%%%%%%%%%%%%%%%%%%%%%%%%
\section{SOFT PARTICLE APPROXIMATION}
\label{section:soft}
%%%%%%%%%%%%%%%%%%%%%%%%%%%%%%%%%%%%%%%%%%%%%%%%%%%%%%%%%%%%%%%%%%%%%%%%%%%%%

In the standard DGLAP region $x={\cal O}(1)$
($\ell={\cal O}(0)$), the $x$ dependence of parton distributions is fast while scaling
violation is small

\begin{equation}
\frac{\partial_\ell {D}_{G,Q}(\ell,y)}{D_{G,Q}}
\equiv \psi_{\ell} = {\cal O}(1),\qquad
\frac{\partial_y
{D}_{G,Q}(\ell,y)}{D_{G,Q}} \equiv \psi_{y} =
{\cal O}(\alpha_s).
\end{equation}

\medskip

With $x$ decreasing, the running coupling gets enhanced while the
$x$-dependence slows down so that, in the kinematical region of the
{\em maximum} ("hump") of the inclusive spectrum the two logarithmic
derivatives become of the same order:
\begin{equation}
 \psi_{y} \sim \psi_{\ell} = {\cal O}(\sqrt{\alpha_s}), \quad
y\simeq \ell \simeq \textstyle {\frac12} Y.
\end{equation}
This allows to  significantly simplify  the equations for inclusive
spectra (\ref{eq:qpr})(\ref{eq:gpr}) and two particle correlations
(\ref{eq:Q2prsub})(\ref{eq:G2prsub}) for soft particles, $x_i\ll 1$,
which determine the bulk of parton multiplicity in jets. 
We shall estimate various contributions to evolution equations 
in order to single out the leading and first sub-leading terms in $\sqrt{\alpha_s}$
to construct the MLLA equations.

\subsection{MLLA spectrum}
\label{subsection:MLLAspec}
%%%%%%%%%%%%%%%%%%%%%%%%%%%

We start by recalling the logic of the MLLA analysis of the inclusive
spectrum. In fact (\ref{eq:qpr})(\ref{eq:gpr}) are
identical to the DGLAP evolution equations but for one detail: the shift
 $\ln z$ in the variable $Y$ characterizing the evolution of the jet
hardness $Q$.  Being the consequence of
exact angular ordering, this modification is negligible, within leading log
accuracy in $\alpha_s Y$, for energetic partons when 
 $|\ln z| < |\ln x| ={\cal O}(1)$. For soft particles, however,
ignoring this effect amounts to 
corrections of order ${\cal O}((\alpha_s\ln^2x)^n)$ that drastically
modify the character of the parton yield in time-like jets as compared
with space-like deep inelastic scattering (DIS) parton distributions. 

The MLLA logic consists of keeping the leading term  and the first
next-to-leading term in the right hand sides of  evolution equations
(\ref{eq:qpr})(\ref{eq:gpr}). 
Meanwhile, the combinations $\Big(Q(1-z)-Q\Big)$ in (\ref{eq:qpr}) and
$\Big(G(1-z)-G\Big)$ in (\ref{eq:regul1}) produce next-to-MLLA
corrections that can be omitted; 
indeed, in the small-$x$ region the parton densities $G(x)$ and $Q(x)$
are smooth functions (see \ref{eq:rho}) of $\ln x$ and we can estimate, say, $G(1-z)-G$,
using (\ref{eq:rho}), as
\begin{equation*}
 G(1-z)-G \equiv
 G\Big(\frac{x}{1-z}, Y+\ln(1-z) \Big) -  G\big(x,Y \big)
 \simeq   \psi_{\ell}\;  G\; \ln(1-z).
\end{equation*}
Since $\psi_{\ell} \sim \sqrt{\alpha_s}$ (see \ref{eq:psimodel}), combined with $\alpha_s$
 this gives a next-to-MLLA  correction 
${\cal O}(\alpha^{3/2})$ to the r.h.s.\ of (\ref{eq:gpr}).
Neglecting these corrections we arrive at 
\begin{eqnarray}
\label{eq:qappr}
Q_{y}  \!\!&\!\!=\!\!&\!\!  \int_x^1 dz\>  \frac{\alpha_s}{\pi} \>\Phi_q^g(z)  G(z), \\
\label{eq:gappr}
G_{y}\!\!&\!\!=\!\!&\!\! \int_x^1 dz\> \frac{\alpha_s}{\pi} \bigg[(1-z)\Phi_g^g(z) G(z)
+ n_f \Phi_g^q(z) \Big(2Q(z)-G\Big)\bigg].
\end{eqnarray}

To evaluate (\ref{eq:qappr}), we rewrite (see (\ref{eq:split}))
$$
\Phi_q^g(z) = C_F\left(\frac{2}{z} + z-2\right).
$$
The singularity in $1/z$
yields the leading (DLA) term; since $G(z)$ is a smoothly varying function
of $\ln z$ (see (\ref{eq:rho})(\ref{eq:QGdef})),
the main $z$ dependence of this non-singular
part of the integrand we only slightly alter by replacing
$(z-2) G(z)$ by $(z-2)G$, which yields
\footnote{ since $x\ll1$, the lower bound of integration is set to ``$0$'' in the 
sub-leading pieces of (\ref{eq:qappr}) and (\ref{eq:gappr})}
\begin{eqnarray}
Q_{y}=\int_x^1 dz\>  \frac{\alpha_s}{\pi} C_F
\bigg(\frac{2}{z} G(z) + (z-2)G\bigg)
&=& \frac{C_F}{N_c}\int_x^1 \frac{dz}z\> \frac{2N_c\alpha_s}{\pi}
G(z) -\frac{3}{4}\frac{C_F}{N_c}\frac{2N_c\alpha_s}{\pi} G\cr
&&
\label{eq:qap2}
\end{eqnarray}
where $\alpha_s=\alpha_s(\ln z)$ in the integral term while in the second, 
it is just a constant. To get the last term in (\ref{eq:qap2}) we used
\begin{equation}\label{eq:int32}
\int_0^1 dz(z-2) = -\frac{3}{2}.
\end{equation}

To evaluate (\ref{eq:gappr}) we go along similar steps. $\Phi_g^q$ being a
regular function of $z$, we replace $2Q(z)-G$ with $2Q-G$;
$\Phi_g^g(z)$ also reads (see (\ref{eq:split}))

$$
\Phi_g^g(z)=2C_A\bigg(\frac{1}{z(1-z)} -2 + z(1-z)\bigg).
$$

The singularity in $1/(1-z)$ disappears, the one in $1/z$ we leave unchanged,
and in the regular part we replace $G(z)$ with $G$. This yields

\begin{eqnarray}\nonumber
G_{y}\!\!&\!\!=\!\!&\!\! \int_x^1 dz\> \frac{\alpha_s}{\pi} \bigg[2C_A \bigg(
\frac1z G(z) +(1-z)\Big(-2 + z(1-z)\Big)G\bigg)
+ n_f T_R\Big(z^2 + (1-z)^2\Big)\Big(2Q-G\Big)\bigg]\\\nonumber\\
\!\!&\!\!=\!\!&\!\!2C_A \int_x^1 \frac{dz}z\> \frac{\alpha_s}{\pi}
G(z) - \bigg(\frac{11}{6}C_A + \frac{2}{3}n_f T_R\bigg)\frac{\alpha_s}{\pi}G
+\frac{4}{3}n_f T_R\frac{\alpha_s}{\pi}\; Q;\label{eq:gap2}
\end{eqnarray}
the comparison of the singular leading (DLA) terms of (\ref{eq:qap2}) and
(\ref{eq:gap2}) shows that

\begin{equation}
Q \stackrel{DLA}{=} \frac{C_F}{C_A}G,
\label{eq:DLAratio}
\end{equation}

which one uses to replace $Q$ accordingly, in the last (sub-leading)
term of (\ref{eq:gap2}) (the corrections would be next-to-MLLA (see \ref{eq:ratio}) 
and can be neglected).
This yields the MLLA equation for $G$ where we set $C_A=N_c$:
\begin{eqnarray}
\label{eq:gap3}
G_{y}\!\!&\!\!=\!\!&\!\!
\int_x^1 \frac{dz}z\> \;\frac{2N_c\alpha_s}{\pi}G(z) -a\frac{2N_c\alpha_s}{\pi} G
\end{eqnarray}
with
\begin{equation}
a = \frac{11}{12} + \frac{n_f T_R}{3N_c}\bigg(1-\frac{2C_F}{N_c}\bigg)
\>=\> \frac{1}{4N_c}\bigg[\frac{11}{3}N_c + \frac{4}{3}n_f T_R
 \bigg(1-\frac{2C_F}{N_c}\bigg)\bigg]\stackrel{n_f=3}{=}0.935.
\label{eq:adefC}
\end{equation}

$a$ parametrizes ``hard'' corrections to soft gluon multiplication and
sub-leading $g\to q\bar{q}$ splittings
\footnote{The present formula for $a$ differs from  (47) in
\cite{PerezMachetC} because, there, we defined $T_R=n_f/2$, instead of
$T_R=1/2$ here.}.

We define conveniently the integration variables $z$ and $\Theta'$ satisfying
$x\leq z\leq 1$ and $xE/Q_0\leq\Theta'\leq\Theta$
\footnote{the lower bound on $\Theta'$ follows from the kinematical condition
$k_{\perp}\approx xE\Theta'\geq Q_0$} through
\begin{equation}
\ell'=\ln\frac{z}{x}\quad \text{and}\quad y'=\ln\frac{xE\Theta'}{Q_0}
\end{equation}

The condition $x\leq z\leq 1$ is then equivalent to $0\leq\ell'\leq\ell$ and
$xE/Q_0\leq\Theta'\leq\Theta$ is $0\!\leq\! y'\!\leq\!y$. Therefore,

$$
\int_x^1\frac{dz}z=\int_0^{\ell}d\ell',\qquad\int_{Q_0/xE}^{\Theta}\frac{d\Theta'}
{\Theta'}=\int_0^{y}dy.
$$

\vskip 0.5cm

We end up  with the following system of integral equations of (\ref{eq:qap2})
and (\ref{eq:gap3}) for the
spectrum of one particle inside a quark and a gluon jet

\begin{equation}
Q(\ell,y)= \delta(\ell) + \frac{C_F}{N_c}\bigg[\int_0^\ell d\ell'\int_0^y dy'
\gamma_0^2(\ell'+y')\Big(1
-\frac34\delta(\ell'-\ell) \Big) G(\ell',y')\bigg],
\label{eq:solqC}
\end{equation}

\begin{equation}
G(\ell,y) = \delta(\ell)
+\int_0^{\ell} d\ell'\int_0^{y} dy' \gamma_0^2(\ell'+y')\Big(
 1  -a\delta(\ell'-\ell) \Big) G(\ell',y')
\label{eq:solgC}
\end{equation}

\vskip 0.5cm

that we write in terms of the anomalous dimension
\begin{equation}
\gamma_0 = \gamma_0(\alpha_s) = \sqrt{\frac{2N_c\alpha_s}{\pi}}
\label{eq:gammadef}
\end{equation}
which determines the rate of multiplicity growth with energy.
Indeed, using (\ref{eq:alphasC}), (\ref{eq:yelldef}) and (\ref{eq:gammadef}) one gets

$$
\gamma_0^2(zE\Theta')=\frac{1}{\beta\ln\left(\displaystyle{\frac{zE\Theta'}
{\Lambda_{QCD}}}\right)}=
\frac1{\beta\left(\ln\displaystyle{\frac{z}{x}}+\displaystyle{\frac{xE\Theta'}{Q_0}}+
\lambda\right)}\equiv\gamma_0^2(\ell'+y')=
\frac{1}{\beta(\ell'+y'+\lambda)}.
$$

with $\lambda=\ln(Q_0/\Lambda_{QCD})$.
In particular, for $z=1$ and $\Theta'=\Theta$ one has

\begin{equation}
\gamma_0^2 = \frac1{\beta(\ell+y+\lambda)}=\frac{1}{\beta(Y+\lambda)},\qquad \ell+y=Y.
\label{eq:gammabeta}
\end{equation}

\medskip

The DLA relation (\ref{eq:DLAratio}) can be refined to

\begin{eqnarray}
Q(\ell, y) =
\frac{C_F}{C_A}
\Big[1 +\left(a-{\textstyle{\frac34}}\right)\Big(\psi_{\ell}+
a\big(\psi_{\ell}^2+\psi_{\ell\,\ell}\big)\Big)
+ {\cal O}(\gamma_0^2)\Big]G(\ell, y),
\label{eq:ratio}
\end{eqnarray}
where
$$\psi_{\ell}=\frac{1}{G(\ell,y)}
\frac{dG(\ell,y)}{d\ell},\quad \psi_{\ell}^2+\psi_{\ell\,\ell}=
\frac1{G(\ell,y)}\frac{d^2G(\ell,y)}{d\ell^2}.
$$ 

Indeed subtracting (\ref{eq:solgC}) and (\ref{eq:solqC}) gives
\begin{equation}
\label{eq:ratiogq}
Q(\ell,y) - \frac{C_F}{N_c}G(\ell,y) = \frac{C_F}{N_c}\Big(a-\frac34\Big)
\int_0^y dy' \gamma_0^2 G(\ell,y');
\end{equation}

iterating twice (\ref{eq:solgC}) yields

$$
\int_0^y dy' \gamma_0^2 G(\ell,y')=G_{\ell} + aG_{\ell\,\ell}+{\cal O}(\gamma_0^2)=
G(\ell,y)\Big(\psi_{\ell}+a\big(\psi_{\ell}^2+\psi_{\ell\,\ell}\big)\Big)+ {\cal O}(\gamma_0^2)
$$
which is then plugged in (\ref{eq:ratiogq}) to get (\ref{eq:ratio}). 
$\psi_{\ell}^2+\psi_{\ell\,\ell}$ can be easily estimated from subsection \ref{subsection:estimate} to be ${\cal O}(\gamma_0^2)$.
In MLLA, (\ref{eq:ratio}) reduces to 

\begin{equation}
Q(\ell, y) =
\frac{C_F}{C_A}
\Big[1 +\left(a-{\textstyle{\frac34}}\right)\psi_{\ell}(\ell,y)+{\cal O}(\gamma_0^2)\Big]
G(\ell, y).
\label{eq:ratioMLLA}
\end{equation}

\subsection{MLLA correlation}
\label{subsection:MLLAcor}
%%%%%%%%%%%%%%%%%%%%%%%%%%%%%%

We estimate analogously the magnitude of various terms on the r.h.s.
of (\ref{eq:Q2prsub}) and (\ref{eq:G2prsub}).
Terms proportional to $Q_2(1-z)-Q_2$ and to  $Q_1(1-z)-Q_1$ in the second
line of (\ref{eq:Q2prsub}) will produce next-to-MLLA corrections that we
drop out.  In the first line, $Q^{(2)}(1-z)-Q^{(2)}$ 
($Q^{(2)}(z)$ is also a smooth function of $\ln z$) will also produce
higher order corrections that we neglect. We get

\begin{equation}
(Q^{(2)}-Q_1Q_2)_{y}=
 \int_{x_1}^1 dz\>\frac{\alpha_s}{\pi}\>  \Phi_q^g(z)\> G^{(2)}(z),
\label{eq:QMLLA}
\end{equation}

where we consider $z\!\geq\! x_1\! \geq\! x_2$.
In the first line of (\ref{eq:G2prsub}) we drop for identical reasons the
term proportional to $G_2(1-z)-G_2$, and the term $G^{(2)}(z)-zG^{(2)}$ is
regularized in the same way as we did for $G(z)-zG$ in (\ref{eq:gpr}).
In the second non-singular line, we use the smooth behavior
of $\phi_g^q(z)$ to neglect the $z$ dependence in all $G^{(2)}$, $Q^{(2)}$,
$G$ and $Q$ so that it factorizes and gives

\begin{eqnarray}\nonumber
(G^{(2)}-G_1G_2)_{y} &=&
\int_{x_1}^1 dz\> \frac{\alpha_s}{\pi}\, (1-z)\Phi_g^g(z)\> G^{(2)}(z)\\\nonumber\\
&&\hskip -3cm +  \int_0^1 dz\>\frac{\alpha_s}{\pi}\, n_f\Phi_g^q(z)\>
\bigg[ 2\big(Q^{(2)}-Q_1Q_2\big) - \big(G^{(2)}-G_1G_2\big) 
 +   (2Q_1-G_1)\,(2Q_2-G_2) \bigg].\cr
&&
\label{eq:GMLLA}
\end{eqnarray}

At the same level of approximation, we use the leading order relations

\begin{equation}
 Q_i = \frac{C_F}{N_c}\, G_i,  \qquad
 Q^{(2)}-Q_1Q_2 = \frac{C_F}{N_c} \,\left(G^{(2)}-G_1G_2\right);
\label{eq:app2}
\end{equation}
the last will be proved consistent in the following.
This makes the equation for the correlation in the gluon jet self
contained, we then get
\begin{eqnarray}\nonumber
(G^{(2)}-G_1G_2)_{y} \!\!&\!\!=\!\!&\!\!
\int_{x_1}^1 dz\> \frac{\alpha_s}{\pi}\, (1-z)\Phi_g^g(z)\> G^{(2)}(z) \\\nonumber\\
&&\hskip -3cm +  \int_0^1 dz\>\frac{\alpha_s}{\pi}\, n_f\Phi_g^q(z)\>
\bigg(2\frac{C_F}{N_c}-1\bigg)\bigg[ \big(G^{(2)}-G_1G_2\big) 
 + \bigg(2\frac{C_F}{N_c}-1\bigg)  G_1G_2) \bigg].
\label{eq:GMLLA2}\cr
&&
\end{eqnarray}

Like for the spectra, we isolate the singular terms $2C_F/z$ and $2C_A/z(1-z)$
of the splitting functions $\phi_q^g$ and $\phi_g^g$ respectively
(see(\ref{eq:split}) and (\ref{eq:cst})). We then write (\ref{eq:QMLLA}) and 
(\ref{eq:GMLLA2}) as follows
\begin{eqnarray}\label{eq:nonsing}
(Q^{(2)}-Q_1Q_2)_{y}=
 \int_{x_1}^1 dz\>\frac{\alpha_s}{\pi}\> 2C_F 
\bigg[\frac1zG^{(2)}(z) + \frac{1}{2}(z-2) G^{(2)}\bigg],
\end{eqnarray}
\begin{eqnarray}
(G^{(2)}-G_1G_2)_{y} \!\!&\!\!=\!\!&\!\!
\int_{x_1}^1 dz\> \frac{\alpha_s}{\pi}\, 
2C_A \bigg[\frac1z G^{(2)}(z) + (1-z)\Big(-2 +z(1-z)\Big)G^{(2)}\bigg]\notag\\\notag\\
&&\hskip -3.5cm +  \int_0^1 dz\>\frac{\alpha_s}{\pi}\, n_f
T_R\Big[z^2+(1-z)^2\Big]
\bigg(2\frac{C_F}{N_c}-1\bigg)\bigg[ \big(G^{(2)}-G_1G_2\big) 
 + \bigg(2\frac{C_F}{N_c}-1\bigg)  G_1G_2 \bigg]\label{eq:nonsingbis},\cr
&&
\end{eqnarray}
which already justifies {\em a posteriori} the last equation in
(\ref{eq:app2}).
One then proceeds with the $z$ integration of the polynomials that occur in
the non-singular terms (that of (\ref{eq:nonsing}) was already written in 
(\ref{eq:int32})). For the term $\propto G^{(2)}$ which we factorize by $2C_A$,
we find (see (\ref{eq:adefC}) for the expression of $a$) in (\ref{eq:nonsingbis})

\begin{eqnarray}
\int_0^1 dz\bigg[(1-z)\Big(\!-2 +z(1-z)\Big)
+ \frac{n_fT_R}{2C_A}\Big(z^2+(1-z)^2\Big)
\bigg(2\frac{C_F}{N_c}-1\bigg)\bigg] = -a,\cr
&&
\end{eqnarray}
while in the one $\propto G_1G_2$ we have simply

\begin{eqnarray}\label{eq:g1g2}
\frac{n_fT_R}{C_A}\bigg(1-2\frac{C_F}{N_c}\bigg)\bigg(1-\frac{C_F}{N_c}\bigg)
\int_0^1 dz\left[z^2+(1-z)^2\right]
= \frac{2n_fT_R}{3C_A}\bigg(1-2\frac{C_F}{N_c}
\bigg)\bigg(1-\frac{C_F}{N_c}\bigg).\cr
&&
\end{eqnarray}
Introducing

\begin{equation}
b = \frac{11}{12} - \frac{n_fT_R}{3N_c}
\left(1-\frac{2C_F}{N_c}\right)^2
= \frac{1}{4N_c}\bigg[\frac{11}{3}N_c -\frac{4}{3}n_f T_R
\bigg(1-2\frac{C_F}{N_c}\bigg)^2\bigg]\stackrel{n_f=3}{=}0.915
\label{eq:bdefC}
\end{equation}

allows us to express (\ref{eq:g1g2}) with $C_A=N_c$ as

\begin{equation}
a-b=
\frac{2n_fT_R}{3N_c}\bigg(1-\frac{2C_F}{N_c}\bigg)\bigg(1-\frac{C_F}{N_c}\bigg)
\stackrel{n_f=3}=0.02,
\label{eq:a-b}
\end{equation}

such that (\ref{eq:nonsing}) and (\ref{eq:nonsingbis}) can be easily rewritten 
in the form

\begin{equation}\label{eq:eqqq}
\left(Q^{(2)}-Q_1Q_2\right)_y=\frac{C_F}{N_c}\int_{x_1}^1\frac{dz}z
\frac{2N_c\alpha_s}{\pi}G^{(2)}(z)-\frac34\frac{C_F}{N_c}\frac{2N_c\alpha_s}
{\pi}G^{(2)},
\end{equation}

\begin{equation}\label{eq:eqgg}
\left(G^{(2)}-G_1G_2\right)_y=\int_{x_1}^1\frac{dz}z
\frac{2N_c\alpha_s}{\pi}G^{(2)}(z)-a\frac{2N_c\alpha_s}{\pi}G^{(2)}+
(a-b)\frac{2N_c\alpha_s}{\pi}G_1G_2.
\end{equation}

Again, $\alpha_s=\alpha_s(\ln z)$ in the leading contribution while in the
sub-leading ones it is a constant.
We now introduce the following convenient variables and notations
to rewrite correlation evolution equations

\begin{eqnarray}
\ell_{i} = \ln\frac{1}{x_{i}}= \ln\frac{E}{\omega_{i}},\quad i=1,2
\end{eqnarray}

\begin{equation}
y_{i}= \ln\frac{\omega_{i}\Theta}{Q_0} =
\ln\frac{x_{i}E\Theta}{Q_0}= Y - \ell_{i}\quad \text{and} \quad
\eta=\ln\frac{x_1}{x_2}=\ell_2-\ell_1=y_1-y_2>0.
\label{eq:vars}
\end{equation}

\medskip

The transverse momentum of parton with energy $zE$ is $k_{\perp}\approx zE\Theta_1$. We 
conveniently define the integration variables $z$ and $\Theta_1$ satisfying
$x_1\!\leq\! z\!\leq\!1$ and
$\Theta_2\!\leq\!\Theta_1\!\leq\!\Theta$ with $\Theta_2\!\geq\!(\Theta_2)_{min}\!=\!Q_0/\omega_2$
through

\begin{equation}\label{eq:intvar}
\ell=\ln\frac{z}{x_1},\qquad y=\ln\frac{x_2E\Theta_1}{Q_0},
\end{equation}

then we write

\begin{equation}\label{eq:alpgasbis}
\gamma_0^2(zE\Theta_1)=
\frac{1}{\beta\left(\ln\displaystyle{\frac{z}{x_1}}+
\ln\displaystyle{\frac{x_2E\Theta_1}{Q_0}}+
\ln\displaystyle{\frac{x_1}{x_2}}+\lambda\right)}
\equiv\gamma_0^2(\ell+y)=\frac1{\beta(\ell+y+\eta+\lambda)}.
\end{equation}

\bigskip

In particular, for $z=1$ and $\Theta_1=\Theta$ we have

$$
\gamma_0^2=\frac1{\beta(\ell_1+y_2+\eta+\lambda)}=\frac1{\beta(Y+\lambda)},\qquad
\ell_1+y_2+\eta=Y.
$$

The condition $x_1\!\leq\! z\!\leq\!1$ translates into $0\!\leq\!\ell\!\leq\!\ell_1$, while
$(\Theta_2)_{min}\!\leq\!\Theta_1\!\leq\!\Theta$ becomes $0\!\leq\! y\!\leq\!y_2$.
Therefore, 

$$
\int_{x_1}^1\frac{dz}z=\int_0^{\ell_1}d\ell\quad \text{and}\quad \int_{Q_0/\omega_2}^{\Theta}\frac{d\Theta_1}{\Theta_1}=\int_0^{y_2}dy.
$$

\bigskip

One gets finally the MLLA system of equations of (\ref{eq:eqqq})(\ref{eq:eqgg})
for quark and gluon jets correlations

\bigskip

\vbox{
\begin{eqnarray}
\label{eq:eveeqqC}
\hskip -3cm Q^{(2)}(\ell_1,y_2,\eta)\!-\! Q_1(\ell_1,y_1)Q_2(\ell_2,y_2)
\!\!\!&\!\!=\!\!&\!\!\! \frac{C_F}{N_c}\!\!
\int_0^{\ell_1}\!\!\! d\ell\!\int_0^{y_2}\!\!\! dy\,
\gamma_0^2(\ell+y) \Big[\!1\!-\!\frac34 \delta(\ell-\ell_1) \!\Big]
G^{(2)}(\ell,y,\eta),\\\notag\\\notag\\
\notag
\hskip -3cm G^{(2)}(\ell_1,y_2,\eta) - G_1(\ell_1,y_1)G_2(\ell_2,y_2)
\!\!\!&\!\!\!=\!\!\!&\!\!\!\! 
\int_0^{\ell_1}\!\! d\ell\!\int_0^{y_2}\!\!dy\, \gamma_0^2(\ell+y)
\Big[\!1 - a \delta(\ell-\ell_1) \!\Big] G^{(2)}(\ell,y,\eta)\\\notag\\
\!\!\!&\!\!\!+\!\!\!&\!\!\! (a-b) \int_0^{y_2}dy \> \gamma_0^2(\ell_1+y)
G(\ell_1,y+\eta)G(\ell_1+\eta,y).
 \label{eq:eveeqgluC} 
\end{eqnarray}
}

In the last line of (\ref{eq:eveeqgluC}) we have made used of (\ref{eq:vars})
to write

\begin{equation}
G_1\equiv G(\ell_1,y_1) = G(\ell_1, y_2+\eta),\quad
G_2\equiv G(\ell_2,y_2) = G(\ell_1+\eta, y_2).
\label{eq:G1G2}
\end{equation}

\bigskip

The first term in (\ref{eq:eveeqqC}) and (\ref{eq:eveeqgluC}) represents the
DLA contribution;  the terms proportional to $\delta$ functions or to
$a$, $b$, represent MLLA corrections.
$a-b$ appearing in (\ref{eq:eveeqgluC}) and defined in (\ref{eq:a-b}) 
is proportional to $n_f$, positive and color suppressed.
%

%%%%%%%%%%%%%%%%%%%%%%%%%%%%%%%%%%%%%%%%%%%%%%%%%%%%%%%%%%%%%%%%%%%%%%%%%%%%%
\section{TWO PARTICLE CORRELATION IN A GLUON JET} 
\label{section:corglu}
%%%%%%%%%%%%%%%%%%%%%%%%%%%%%%%%%%%%%%%%%%%%%%%%%%%%%%%%%%%%%%%%%%%%%%%%%%%%%

\subsection{Iterative solution}
\label{subsection:iterglue}
%%%%%%%%%%%%%%%%%%%%%%%%%%%%%%%

Since equation (\ref{eq:eveeqgluC}) for a gluon  jet is self contained, it
is our starting point.
We define the normalized correlator ${\cal C}_g$ by
\begin{equation}
 G^{(2)} = {\cal C}_g (\ell_1,y_2,\eta)\ G_1\,G_2,
\label{eq:Gnor}
\end{equation}
where $G_1$ and $G_2$ are expressed in (\ref{eq:G1G2}).
Substituting (\ref{eq:Gnor}) into (\ref{eq:eveeqgluC}) one gets
(see appendix \ref{section:Gcorr})  the following expression for
the correlator 

\begin{eqnarray}
 {\cal C}_g -1
=\frac{1 -\delta_1 -b\left(\psi_{1,\ell} +\psi_{2,\ell}-
  [\beta\gamma_0^2] \right) - \left[a\chi_{\ell} + \delta_2\right]}
{1+ \Delta
+\delta_1 + \Big[a\left(\chi_{\ell} +{[\beta\gamma_0^2]}\right)+\delta_2\Big]}
\label{eq:CGfull}
\end{eqnarray}
which is to be evaluated numerically. We have introduces
the following notations and variables
\begin{eqnarray}\label{eq:nota4bis}
&& \quad \chi =  \ln {\cal C}_g,\qquad
\chi_{\ell} = \frac{d\chi}{d\ell},\qquad \chi_{y}=\frac{d\chi}{dy};\\\notag\\
&&\label{eq:psi1}\psi_{1} = \ln G_{1},\qquad
\psi_{1,\ell}= \frac{1}{G_1}\frac{dG_1}{d\ell},\qquad
\psi_{1,y}= \frac{1}{G_1}\frac{dG_1}{dy};\\\notag\\
&&\label{eq:psi2}\psi_{2} = \ln G_{2},\qquad
\psi_{2,\ell}= \frac{1}{G_2}\frac{dG_2}{d\ell},\qquad
\psi_{2,y}= \frac{1}{G_2}\frac{dG_2}{dy};\\\notag\\
&&\label{eq:deltabisC}\Delta = \gamma_0^{-2}
\Big(\psi_{1,\ell}\psi_{2,y}+\psi_{1,y}\psi_{2,\ell}\Big);\\\notag\\
&&\delta_1 = \gamma_0^{-2}\Big[\chi_{\ell}(\psi_{1,y}+\psi_{2,y}) +
   \chi_{y}(\psi_{1,\ell}+\psi_{2,\ell})\Big];\label{eq:delta1C}\\\notag\\
&&\delta_2 = \gamma_0^{-2}\Big(\chi_{\ell}\chi_{y} + \chi_{\ell\,y}\Big).
\label{eq:nota4C}
\end{eqnarray}
As long as ${\cal C}_g$ is changing slowly with
$\ell$ and $y$, (\ref{eq:CGfull}) can be solved iteratively. The expressions
of $\psi_{\ell}$ and $\psi_{y}$, as well as the numerical analysis of 
the other quantities are explicitly given in appendices \ref{subsection:Logder} and \ref{section:numcorr} for $\lambda=0$ ($Q_0=\Lambda_{QCD}$), the so call 
``limiting spectrum''. Consequently, (\ref{eq:CGfull}) will be computed in the same limit.

\subsection{Estimate of magnitude of various contributions}
\label{subsection:estimate}
%%%%%%%%%%%%%%%%%%%%%%%%%%%%%%%%%%%%%%%%%%%%%%%%%%%%%%%%%%%

To estimate the relative r\^ole of various terms in (\ref{eq:CGfull}) we
can make use of a simplified model for the MLLA spectrum in which one neglects
the variation of $\alpha_s$, hence of $\gamma_0$ in (\ref{eq:gap3}).
It becomes, after differentiating with respect to $\ell$
\begin{equation}
G_{\ell\,y} = \gamma_0^2\big(G - a\, G_{\ell}\big).
\label{eq:Gsimp}
\end{equation}
The solution of this equation is the function for
$\gamma_0^2=const$ (see appendix \ref{section:inspiredDLA} for details)
\begin{eqnarray}
G(\ell,y) \stackrel{x\ll1}{\simeq} \exp{\Big( 2\gamma_0 \sqrt{\ell\,y} -a\gamma_0^2\,y\Big)}.
\label{eq:Gmod}
\end{eqnarray}

The subtraction term $\propto a$ in (\ref{eq:Gmod}) accounts for hard 
corrections (MLLA) that shifts the position of the maximum of the single inclusive 
distribution toward larger values of $\ell$ (smaller $x$) and partially 
guarantees the energy balance during soft gluons cascading (see
\cite{EvEqC}\cite{KOC}  and
references therein). The position of the maximum follows from (\ref{eq:Gmod})
$$
\ell_{max}=\frac{Y}2(1+a\gamma_0).
$$
From (\ref{eq:Gmod}) one gets
\begin{eqnarray}
\psi_{\ell} = \gamma_0 \sqrt{\frac{y}{\ell}},\qquad  \psi_{y} = \gamma_0
\sqrt{\frac{\ell}{y}} - a\gamma_0^2,\qquad
\psi_{\ell\,y}\sim\psi_{\ell\,\ell}\sim\psi_{y\,y} = {\cal O}(\gamma_0^3),\quad
\ell^{-1}\sim y^{-1}={\cal O}(\gamma_0^2)\cr
&&
\label{eq:psimodel} 
\end{eqnarray}
and the function $\Delta$ in (\ref{eq:deltabisC}) becomes
\begin{eqnarray}
\Delta \!\!&\!\!=\!\!&\!\! \left( \sqrt{\frac{y_1\ell_2}{\ell_1 y_2}} +
 \sqrt{\frac{\ell_1 y_2}{y_1 \ell_2}} \right)
-a\gamma_0\left(\sqrt{\frac{y_1}{\ell_1}}+\sqrt{\frac{y_2}{\ell_2}} \right)
\cr
\!\!&\!\!=\!\!&\!\! 2\cosh(\mu_1-\mu_2) - a\gamma_0(e^{\mu_1}+e^{\mu_2}); \qquad
\mu_{i}= \textstyle {\frac12} \displaystyle\ln\frac{y_{i}}{\ell_{i}}.  
\label{eq:Deltamodel} 
\end{eqnarray}
We see that $\Delta={\cal O}(1)$
and depends on the ratio of logarithmic variables $\ell$ and $y$. One step further is needed before we can estimate the order of magnitude of $\chi_{\ell}$, $\chi_{y}$ and $\chi_{\ell\,y}$. Indeed, the leading contribution to these quantities is obtained by taking
the leading (DLA) piece of (\ref{eq:CGfull}), that is

$$
\chi\stackrel{DLA}{\simeq}\ln\left(1+\frac1{1+\Delta}\right);
$$
then, it is easy to get

$$
\chi_{\ell}=-\frac{\Delta_{\ell}}{(1+\Delta)(2+\Delta)},\quad
\chi_{y}=-\frac{\Delta_{y}}{(1+\Delta)(2+\Delta)};
$$
we have roughly
$$
\chi_{\ell}\propto\mu_{\ell},\quad \chi_{y}\propto\mu_{y},\quad\chi_{\ell\,y}
\propto\mu_{\ell}\,\mu_{y};
$$
since $\mu_{i,\ell}=\mu_{i,y}={\cal O}(\gamma_0^2)$ one gets

\begin{equation}
\chi_{\ell} \sim \chi_{y} = {\cal O}(\gamma_0^2), \qquad
\chi_{\ell\,y}\sim\chi_{\ell}\chi_{y}={\cal O}(\gamma_0^4),
\label{eq:magn1}
\end{equation}
which entails for the corrections terms $\delta_1$ and
$\delta_2$ in (\ref{eq:delta1C}) (\ref{eq:nota4C})
\begin{equation}
\delta_1 ={\cal O}(\gamma_0), \qquad \delta_2={\cal O}(\gamma_0^2).
\label{eq:magn2}
\end{equation}
The term $\delta_1$ constitutes a MLLA correction
while $\delta_2$ as well as other terms that are displayed in square brackets 
in (\ref{eq:CGfull}) are of order $\gamma_0^2$ and are, formally 
speaking, beyond the MLLA accuracy.

\subsection{MLLA reduction of (\ref{eq:CGfull})}
%%%%%%%%%%%%%%%%%%%%%%%%%%%%%%%%%%%%%%%%%%%%%%%%%%%%

Dropping ${\cal O}(\gamma_0^2)$ terms , the expression for the correlator would simplify to
\begin{equation}
{\cal C}_g-1 \stackrel{MLLA}{\approx} \frac{1-b\left(\psi_{1,\ell}
    +\psi_{2,\ell} \right)-\delta_1} {1+ \Delta + \delta_1}.
\label{eq:CGMLLAC}
\end{equation}

\subsection{$\boldsymbol{{\cal C}_g\ge 0}$ in the soft approximation}
\label{subsection:Cgpos}
%%%%%%%%%%%%%%%%%%%%%%%%%%%%%%%%%%%%%%%%%%%%%%%%%%%%%%%%%%%%%%%%%%%

${\cal C}_g$ must obviously be positive. By looking at ${\cal C}_g\ge 0$
one determines the region of applicability of our soft approximation.
Using (\ref{eq:CGMLLAC}), the condition reads
\begin{equation}
   2+\Delta >  b(\psi_{1,\ell}+\psi_{2,\ell}).
\label{eq:Gsoft}
\end{equation}
For the sake of simplicity,  we employ the model
(\ref{eq:Gmod})(\ref{eq:psimodel})(\ref{eq:Deltamodel}), this gives
\begin{equation}
 2\big(1+\cosh(\mu_1-\mu_2)\big)>\gamma_0(a+b) \big(e^{\mu_1}+ e^{\mu_2}\big), 
\end{equation}
which translates into 
\begin{equation}
\label{eq:elliyi}
\sqrt{\frac{\ell_1}{y_1}} + \sqrt{\frac{\ell_2}{y_2}} > \gamma_0\,(a+b).
\end{equation}
For $\ell_1$, $\ell_2 \ll Y$ we can set $y_1\simeq y_2\simeq Y$ and, using  
$\gamma_0^2 \simeq 1/{\beta Y}$
\footnote{for $n_f=3$, $\beta=0.75$},  we get the condition
\begin{equation}
\sqrt{\ell_1} + \sqrt{\ell_2}  >  \frac{a+b}{\sqrt{\beta}} \simeq 2.1,
\end{equation}
which is satisfied as soon as $\ell_1>1$ ($\ell_2>\ell_1$);  
so, for $x_1\lesssim 0.4 , x_2 < x_1$, the correlation $\cal C$ is
positive.

\subsection{The sign of $\boldsymbol{({\cal C}_g-1)}$}
\label{subsection:signG}
%%%%%%%%%%%%%%%%%%%%%%%%%%%%%%%%%%%%%%%%%%%%%%%%%%%%

In the region of relatively hard particles $({\cal C}_g -1)$ becomes
negative. To find out at which value of $\ell$ it happens, we use the simplified
model and take, for simplicity, $\ell_1=\ell_2=\ell_\pm$.

The condition $1 = \delta_1 + b \big(\psi_{1,\ell} + \psi_{2,\ell}\big)$, using  (\ref{eq:yelldef})(\ref{eq:gammabeta})(\ref{eq:psimodel}) 
and neglecting $\delta_1$ which vanishes 
at $\ell_1\approx\ell_2$ reads
\begin{equation}
1- b\gamma_0\cdot 2 \sqrt{\frac{Y-\ell_\pm}{\ell_\pm}} = 0
\Leftrightarrow
\ell_\pm = \displaystyle\frac{M_g}{1+\frac{M_g}{Y}},\quad
M_g= \frac{4b^2}{\beta} \simeq 4.5. 
\label{eq:pmest} 
\end{equation}
Thus in the $Y\to \infty$ limit the correlation between two equal energy
partons in a gluon jet turns negative at a fixed value, $x> x_\pm
\simeq \exp(4.5)=1/90$.  For finite energies this energy is
essentially larger; in particular, for $Y=5.2$ (which corresponds to
LEP-I energy) (\ref{eq:pmest}) gives $\ell_\pm\simeq 2.4$ ($x_\pm
\simeq 1/11$).

For the Tevatron, let us for instance take the typical value $Y=6.0$, one has 
$\ell_{\pm}\simeq2.6$ and finally, for the LHC we take the typical one, 
$Y=7.5$, one gets the corresponding
$\ell_{\pm}\simeq2.8$. This is confirmed numerically in Figs.~\ref{fig:3gbandsLEP}, \ref{fig:3gbandsTeV} and \ref{fig:3gbandsLHC}.

%%%%%%%%%%%%%%%%%%%%%%%%%%%%%%%%%%%%%%%%%%%%%%%%%%%%%%%%%%%%%%%%%%%%%%%%%%%%%
\section{TWO PARTICLE CORRELATIONS IN A QUARK JET}
\label{section:quark}
%%%%%%%%%%%%%%%%%%%%%%%%%%%%%%%%%%%%%%%%%%%%%%%%%%%%%%%%%%%%%%%%%%%%%%%%%%%%%

\subsection{Iterative solution}
\label{subsection:iterQ}
%%%%%%%%%%%%%%%%%%%%%%%%%%%%%%%%

We define the normalized correlator ${\cal C}_q$ by
\begin{equation}
 Q^{(2)} = {\cal C}_q (\ell_1,y_2,\eta)\ Q_1\,Q_2,
\label{eq:Qnor}
\end{equation}
where $Q_1$ and $Q_2$ are expressed like in (\ref{eq:G1G2}) for $G_1$ and
$G_2$.
By differentiating (\ref{eq:eveeqqC}) with respect to $\ell_1$ and $y_2$,
one gets (see appendix \ref{section:Qcorr})

\begin{equation}
{\cal C}_q-1
=\displaystyle\frac{
     \frac {N_c}{C_F} {\cal C}_g
       \Big[ 1-\textstyle {\frac34}\Big(\psi_{1,\ell}
  +\psi_{2,\ell} +[\chi_{\ell}] - [\beta\gamma_0^2]\Big) \Big]
\frac{C_F}{N_c}\frac{G_1}{Q_1} \frac{C_F}{N_c}\frac{G_2}{Q_2}
-\tilde\delta_1 -[\tilde\delta_2]}
{\widetilde\Delta + \Big[1-\textstyle {\frac34}
  \big(\psi_{1,\ell}-[\beta\gamma_0^2]\big)\Big]\frac{C_F}{N_c}\frac{G_1}{Q_1}
+\Big[1-\textstyle {\frac34} \big(\psi_{2,\ell}-[\beta\gamma_0^2]\big)\Big]
\frac{C_F}{N_c}\frac{G_2}{Q_2} 
+ \tilde\delta_{1}+[\tilde\delta_{2}]},  
\label{eq:Qcorr}
\end{equation}

which is used for numerical analysis. $G_i/Q_i$ is computed using (\ref{eq:ratio}).
The terms ${\cal O}(\gamma_0^2)$ are the one that can be neglected
when staying at MLLA (see \ref{subsection:MLLAreduc}).
We have introduced, in addition to (\ref{eq:nota4bis})-(\ref{eq:nota4C}), the following notations
\begin{eqnarray}
\widetilde\Delta \!\!&\!\!=\!\!&\!\!
\gamma_0^{-2}\Big(\varphi_{1,\ell}\varphi_{2,y}+\varphi_{1,y}\varphi_{2,\ell}\Big),\\\notag\\
\tilde\delta_1\!\!&\!\!=\!\!&\!\!
\gamma_0^{-2}\Big[\sigma_{\ell}(\varphi_{1,y}+\varphi_{2,y}) +
\sigma_{y}(\varphi_{1,\ell}+\varphi_{2,\ell})\Big],\\\notag\\
\tilde\delta_2 \!\!&\!\!=\!\!&\!\!
\gamma_0^{-2}\Big(\sigma_{\ell}\sigma_{y}+\sigma_{\ell\,y}\Big),
\label{eq:nota5}
\end{eqnarray}
with
\begin{equation}
\varphi_k = \ln Q_k,\quad \sigma= \ln {\cal C}_q.
\label{eq:phisigma}
\end{equation}
Accordingly, (\ref{eq:Qcorr}) will be computed for $\lambda=0$, the analysis of the
previous functions is done in appendix \ref{section:numcorr}.
\subsection{MLLA reduction of (\ref{eq:Qcorr})}
\label{subsection:MLLAreduc}
%%%%%%%%%%%%%%%%%%%%%%%%%%%%%%%%%%%%%%%%%%%%%%%%%%

Using (\ref{eq:ratioMLLA}), which entails
$\frac{C_F}{N_c}\frac{G_i}{Q_i} \simeq 1 - \big(a-\frac34\big)\psi_{i,\ell}
+ {\cal O}(\gamma_0^2)$, reduces (\ref{eq:Qrel}) and (\ref{eq:Qcorr}) respectively to

\vbox{
\begin{eqnarray}\label{eq:Qrelbis}
{\cal C}_q-1 \!\!&\!\!=\!\!&\!\!\displaystyle\frac {
\frac{N_c}{C_F}{\cal C}_g\Big[1 -a\big(\psi_{1,\ell}+\psi_{2,\ell}\big)
-\frac34[\chi_{\ell} -\beta\gamma_0^2]\Big] -{\cal C}_q(\tilde\delta_1
+[\tilde\delta_2])}
{2 +\widetilde\Delta -a\big(\psi_{1,\ell}+\psi_{2,\ell}\big)
 +[\frac32 \beta\gamma_0^2]}\label{eq:Qcorr3}\\\notag\\
\!\!&\!\!=\!\!&\!\!\displaystyle\frac {
\frac{N_c}{C_F}{\cal C}_g\Big[1 -a\big(\psi_{1,\ell}+\psi_{2,\ell}\big)
-\frac34[\chi_{\ell} -\beta\gamma_0^2]\Big] -\tilde\delta_1
-[\tilde\delta_2]}
{2 +\widetilde\Delta -a\big(\psi_{1,\ell}+\psi_{2,\ell}\big)
 +[\frac32 \beta\gamma_0^2] + \tilde\delta_1 + \tilde\delta_2}.
\label{eq:Qcorr1}
\end{eqnarray}
}

As demonstrated in appendix \ref{subsection:corrections},
$\tilde\Delta = \Delta  + {\cal O}(\gamma_0^2)$ and
\begin{equation}
{\cal C}_q (\tilde\delta_1 + \tilde\delta_2) \simeq \frac{N_c}{C_F}{\cal
C}_g (\delta_1 + \delta_2);
\label{eq:numdel}
\end{equation}
such that (\ref{eq:Qrelbis}) becomes
\begin{equation}
{\cal C}_q-1 \approx\displaystyle\frac{N_c}{C_F}\,\frac {
{\cal C}_g\Big[1 -a\big(\psi_{1,\ell}+\psi_{2,\ell}\big)
-\frac34[\chi_{\ell} -\beta\gamma_0^2] -\delta_1
-[\delta_2]\Big]}
{2 + \Delta -a\big(\psi_{1,\ell}+\psi_{2,\ell}\big)
 +[\frac32 \beta\gamma_0^2]]}.
\label{eq:Qcorr4}
\end{equation}

\medskip

Would we neglect, according to (\ref{eq:magn1})(\ref{eq:magn2}),
next to MLLA terms, which
amounts to dropping
all ${\cal O}(\gamma_0^2)$ corrections,
(\ref{eq:Qcorr1}) would simply reduce to

\medskip

\begin{equation}
{\cal C}_q-1 \!\!\stackrel{MLLA}{\approx}\!\!\frac{N_c}{C_F}\,\frac {
\displaystyle{\cal C}_g\Big[1 -a\big(\psi_{1,\ell}+\psi_{2,\ell}\big)
\Big] -\delta_1 }
{2 + \Delta -a\big(\psi_{1,\ell}+\psi_{2,\ell}\big) +\delta_1}.
\label{eq:QMLLAap}
\end{equation}

\medskip

Furthermore, comparing (\ref{eq:Qcorr4}) and (\ref{eq:CGMLLAC}) and
using the magnitude estimates of subsection \ref{subsection:estimate}
allows to
make an expansion in the small ${\cal O}(\gamma_0)$ corrections $\delta_1$, $\psi_{1,\ell}$ and
$\psi_{2,\ell}$ to get

\medskip

\begin{eqnarray}
\frac{{\cal C}_q-1}{{\cal C}_g-1} &\stackrel{MLLA}{\simeq}&
\frac{N_c}{C_F}\bigg[1+(b-a)(\psi_{1,\ell} + \psi_{2,\ell})
\frac{1+\Delta}{2+\Delta}\bigg]\cr
&\approx&
\frac{N_c}{C_F}\Big[1+(b-a)(\psi_{1,\ell} + \psi_{2,\ell})\Big({\cal C}_g^{DLA}\Big)^{-1}
\Big],
\label{eq:rapMLLAC}
\end{eqnarray}

\medskip

where we have consistently used the DLA expression ${\cal C}_g^{DLA} =
\displaystyle\frac{2+\Delta}{1+\Delta}$.
$(a-b)$ is given in (\ref{eq:a-b}).
The deviation of the ratio from the DLA value $N_c/C_F$ is proportional to $n_f$, 
is color suppressed and numerical small.

\subsection{$\boldsymbol{{\cal C}_q\ge 0}$ in the soft approximation}
\label{subsection:Cqpos}
%%%%%%%%%%%%%%%%%%%%%%%%%%%%%%%%%%%%%%%%%%%%%%%%%%%%%%%%%%%%%%%%%%%

Since we neglect NMLLA corrections and the running of $\alpha_s$, we can make use 
of (\ref{eq:rapMLLAC}) in order to derive the positivity constrain for the quark 
correlator. In the r.h.s. of (\ref{eq:rapMLLAC}) we can indeed neglect the MLLA 
correction in the square brackets because it is numerically small
(for instance, for $\gamma_0\simeq0.5$ it is $\approx10^{-3}$). Therefore, ${\cal C}_q$
changes sign when 

$$
{\cal C}_g\ge1-\frac{C_F}{N_c}=\frac59\approx\frac12,
$$

(\ref{eq:elliyi}) gets therefore replaced by
$$
\sqrt{\frac{\ell_1}{y_1}}+\sqrt{\frac{\ell_2}{y_2}}>\frac45(a+2b)\gamma_0,
$$
which finally, following the same steps, gives
$$
\sqrt{\ell_1}+\sqrt{\ell_2}>\frac45\frac{a+2b}{\sqrt{\beta}}\simeq2.6.
$$
The last inequality is satisfied as soon as $\ell_1>1.6$ ($\ell_2>\ell_1$). This condition
slightly differs from that of the gluon correlator in \ref{subsection:signG}.

\subsection{The sign of $\boldsymbol{({\cal C}_q-1)}$}
\label{subsection:signQ}
%%%%%%%%%%%%%%%%%%%%%%%%%%%%%%%%%%%%%%%%%%%%%%%%%%%%

From (\ref{eq:QMLLAap}), ${\cal C}_q-1$ changes sign for

\begin{equation}
{\cal C}_q-1\approx\frac{N_c}{C_F}\,\frac {
\displaystyle{\cal C}_g\Big[1 -a\big(\psi_{1,\ell}+\psi_{2,\ell}\big)
\Big]}
{2 + \Delta -a\big(\psi_{1,\ell}+\psi_{2,\ell}\big)}>0
\end{equation}

which gives the condition

$$
1=a\big(\psi_{1,\ell}+\psi_{2,\ell}\big).
$$

This gives a formula identical to (\ref{eq:pmest}) with the exchange
$b\rightarrow a$; $a$ being slightly larger than $b$, we find now a parameter
$M_q = 4a^2/\beta \simeq 4.66$. The corresponding $\ell_\pm$ at
which $({\cal C}_q-1)$ will change sign is slightly higher than for gluons;
for example at $Y=5.2$, $\ell_\pm \simeq 2.5\; (x_\pm \simeq 1/12)$, $Y=6.0$, 
$\ell_\pm \simeq 2.7\;  (x_\pm \simeq 1/13)$, $Y=7.5$, $\ell_\pm \simeq 2.9\;  (x_\pm \simeq 1/16)$. This is confirmed numerically in figures \ref{fig:3qbandsLEP}, \ref{fig:3qbandsTeV} and \ref{fig:3qbandsLHC}.

%%%%%%%%%%%%%%%%%%%%%%%%%%%%%%%%%%%%%%%%%%%%%%%%%%%%%%%%%%%%%%%%%%%%%%%%%%%%%
\section{NUMERICAL RESULTS}
\label{section:numer}
%%%%%%%%%%%%%%%%%%%%%%%%%%%%%%%%%%%%%%%%%%%%%%%%%%%%%%%%%%%%%%%%%%%%%%%%%%%%%

In order to lighten the core of the paper, only the main lines and ideas of
the calculations, and the results, are given here; the numerical analysis of
(MLLA and NMLLA) corrections occurring in (\ref{eq:CGfull}) and 
(\ref{eq:Qcorr}) is the object of appendix \ref{section:numcorr}, that we
summarize in subsection \ref{subsection:commE} below. We present our 
results as functions of $(\ell_1+\ell_2)$ and  $(\ell_1-\ell_2)$.

\subsection{The gluon jet correlator}
\label{subsection:gcorr}
%%%%%%%%%%%%%%%%%%%%%%%%%%%%%%%%%%%%

In order to implement the iterative solution of the first line of
(\ref{eq:CGfull}), we define

\begin{equation}
\Upsilon_g = \ln\Bigg[1+\displaystyle\frac
{1-b(\psi_{1,\ell} + \psi_{2,\ell} -[\beta\gamma_0^2])}
{1+\Delta +[a\beta\gamma_0^2]}\Bigg]
\label{eq:upsg}
\end{equation}
as the starting point of the procedure. It represent the zeroth order of
the iteration for $\chi \equiv \ln{\cal C}_g$. The terms proportional to derivatives of 
$\chi$ in the numerator and denominator of (\ref{eq:CGfull}) are the objects
of the iteration and do not appear in (\ref{eq:upsg});  
the parameter $\Delta$ depends (see (\ref{eq:deltabisC}))
only on the logarithmic 
derivatives $\psi_{\ell},\psi_{y}$ of the inclusive spectrum $G$ which are
 determined at each step, by the exact solution (\ref{eq:ifDC})
(\ref{eq:calFdefC}) for $G$ demonstrated in appendix \ref{section:ESEE}. The leading 
piece (DLA) of (\ref{eq:upsg}) 

$$
\Upsilon_g \stackrel{DLA}{=} \ln\Bigg[1+\displaystyle\frac
{1}{1+\Delta}\Bigg]
\label{eq:upsgDLA}
$$

is the one that should be used when reducing (\ref{eq:CGfull}) to MLLA.
We have instead consistently kept sub-leading (MLLA and NMLLA) corrections in 
(\ref{eq:upsg}) in order to 
follow the same logic that proved successful for the single inclusive spectrum.

\subsection{The quark jet correlator}
\label{subsection:qcorr}
%%%%%%%%%%%%%%%%%%%%%%%%%%%%%%%%%%%%%%

We start now from (\ref{eq:Qcorr}) and define, like for
gluons

\begin{equation}
\Upsilon_q =\ln\left\{1+\displaystyle\frac{
     \frac {N_c}{C_F} {\cal C}_g
       \Big[ 1-\textstyle {\frac34}\Big(\psi_{1,\ell}
  +\psi_{2,\ell} +[\chi_{\ell} - \beta\gamma_0^2]\Big) \Big]
\frac{C_F}{N_c}\frac{G_1}{Q_1} \frac{C_F}{N_c}\frac{G_2}{Q_2}}
{\widetilde\Delta + \Big[1-\textstyle {\frac34}
  \big(\psi_{1,\ell}-[\beta\gamma_0^2]\big)\Big]\frac{C_F}{N_c}\frac{G_1}{Q_1}
+\Big[1-\textstyle {\frac34} \big(\psi_{2,\ell}-[\beta\gamma_0^2]\big)\Big]
\frac{C_F}{N_c}\frac{G_2}{Q_2}}\right\}
\label{eq:upsq}
\end{equation}

as the starting point of the iterative procedure, {\em i.e.} the zeroth
order of the iteration for $\sigma \equiv \ln{\cal C}_q$;
it again includes MLLA
(and some NMLLA) corrections. Since the iteration concerns ${\cal C}_q$,
the terms proportional to ${\cal C}_g$ and to its derivative $\chi_{\ell}$
must be present in (\ref{eq:upsq}).
All other functions are
determined, like above, by the exact solution of 
(\ref{eq:ifDC}) and (\ref{eq:calFdefC}) for $G$.

We have replaced in the denominator of (\ref{eq:upsq}) $\tilde\Delta$ with
$\Delta$, which amounts to neglecting ${\cal O}(\gamma_0^2)$ corrections,
 because the coefficient of $\gamma_0^{-2}(\tilde\Delta
-\Delta)$ is numerically very small; this occurs for two
combined reasons: it is proportional to $(a-3/4)$ which is small,
and the combination $(\psi_{1,\ell\,y}\psi_{2,\ell}
+\psi_{2,\ell,\ell}\psi_{1,y} +\psi_{2,\ell\,y}\psi_{1,\ell}
+\psi_{1,\ell\,\ell}\psi_{2,y} )$ that appears in (\ref{eq:Deltatilde})
is very small (see Fig.~\ref{fig:doublepsi}). Accordingly,

$$
\Upsilon_q \stackrel{DLA}{=} \ln\Bigg[1+\frac{N_c}{C_F}\displaystyle\frac
{1}{1+\Delta}\Bigg].
\label{eq:upsqDLA}
$$

We can use this simplified expression for the MLLA reduction of (\ref{eq:Qcorr}).

\subsection{The role of corrections; summary of appendix \ref{section:numcorr}}
\label{subsection:commE}
%%%%%%%%%%%%%%%%%%%%%%%%%%%%%%%%%%%%%%%%%%%%%%%%%%%%%%%%%%%%%%%%%%%%%%%%%%%%%%

Analysis have been done separately for a gluon and a quark jet; their
conclusions are very similar.

That $\psi_{\ell}$ and $\psi_{y}$, which are ${\cal O}(\gamma_0)$ should
not exceed reasonable values (fixed arbitrarily to $1$) provides an
interval of reliability of our calculations; for example, at LEP-I
\begin{equation}
2.5 \leq \ell \leq  4.5\ \text{or}\ 5 \leq \ell_1 + \ell_2 \leq 9,\quad Y=5.2.
\label{eq:confintLEP}
\end{equation}
This interval is shifted upwards and gets larger when $Y$ increases.

$\Upsilon_g$ and $\Upsilon_q$ defined in (\ref{eq:upsg}) and (\ref{eq:upsq})
and their derivatives are shown to behave smoothly in the confidence
interval (\ref{eq:confintLEP}).

The roles of all corrections $\delta_1, \delta_2, \Delta$ for a gluon jet,
$\tilde\delta_1, \tilde\delta_2, \tilde\Delta$ for a quark jet,
 have been investigated individually. They stay under control in
(\ref{eq:confintLEP}).
While, in its center, their relative values coincide
with what is expected from subsection \ref{subsection:estimate}, NMLLA
corrections can become larger than MLLA close to the bounds; this could
make our approximations questionable. Two cases may occur which depend on
NMLLA corrections  not included in the present frame of
calculation;  either they largely cancel with the included ones and the sum
of all NMLLA corrections is (much) smaller than those of MLLA: then pQCD is
trustable at $Y=5.2$; or they do not, the confidence in our results at
this energy is weak, despite the fast convergence of the iterative
procedure which occurs thanks to the ``accidental'' observed cancellation
between MLLA and those of NMLLA which are included.
The steepest descent method \cite{RPR3}\cite{these}, in which a better
control is obtained of MLLA corrections alone, will shed some more light
on this question.
The global role of all corrections in the iterative process 
does not exceed $30 \%$ for $Y=5.2$ (OPAL) at the bounds
of (\ref{eq:confintLEP}); it is generally much smaller, though never
negligible. In particular, $\delta_1 + \delta_2 + a\Upsilon_{g,\ell}$ for
gluons (or $\tilde\delta_1 + \tilde\delta_2$ for quarks) sum up to ${\cal
O}(10^{-2})$ at LEP energy scale (they reach their maximum
${\cal O}(10^{-1})$ at the bound of the
interval corresponding to the $30\%$ evoked above). 

The role of corrections decreases when the total energy $Y$ of the jet
increases, which makes our calculations all the more reliable.

\subsection{Results for LEP-I}
%%%%%%%%%%%%%%%%%%%%%%%%%%%%%%%%%%

\begin{figure}
\begin{center}
\epsfig{file=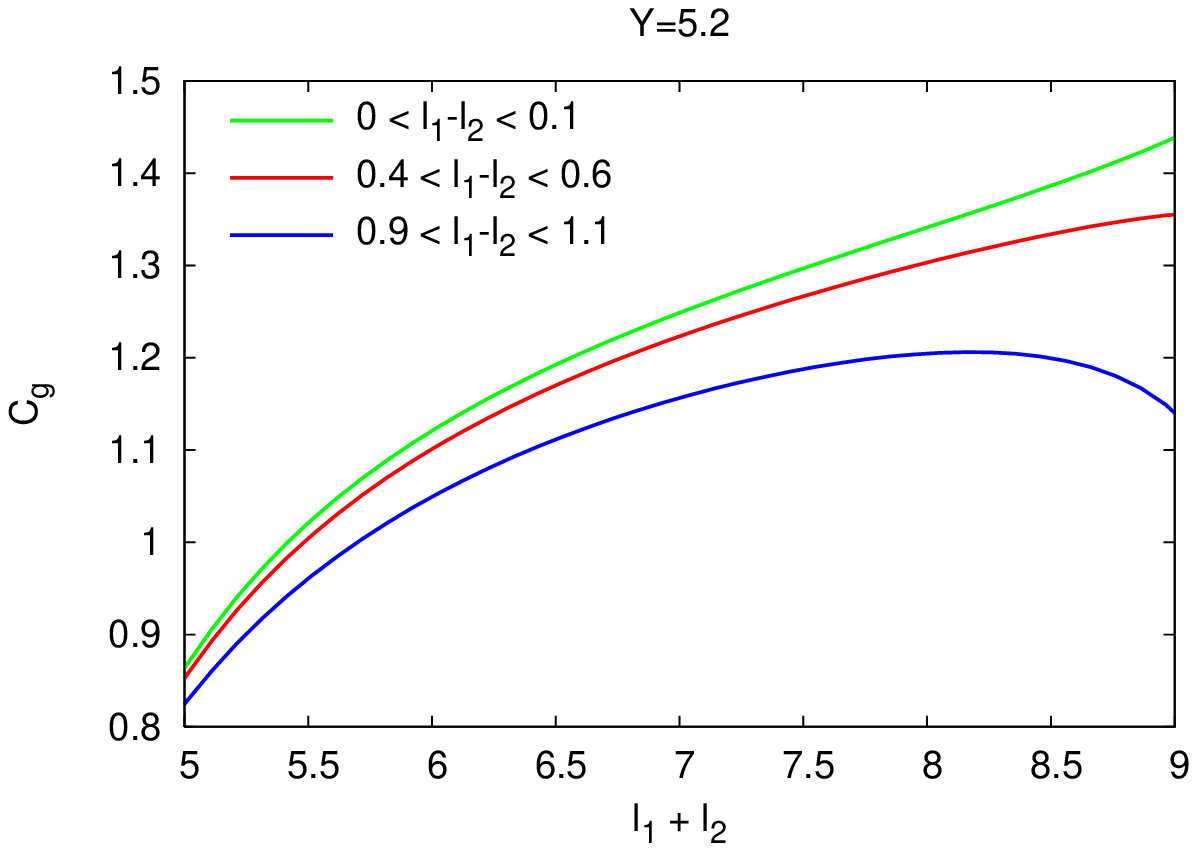, height=6truecm,width=0.47\tw}
\epsfig{file=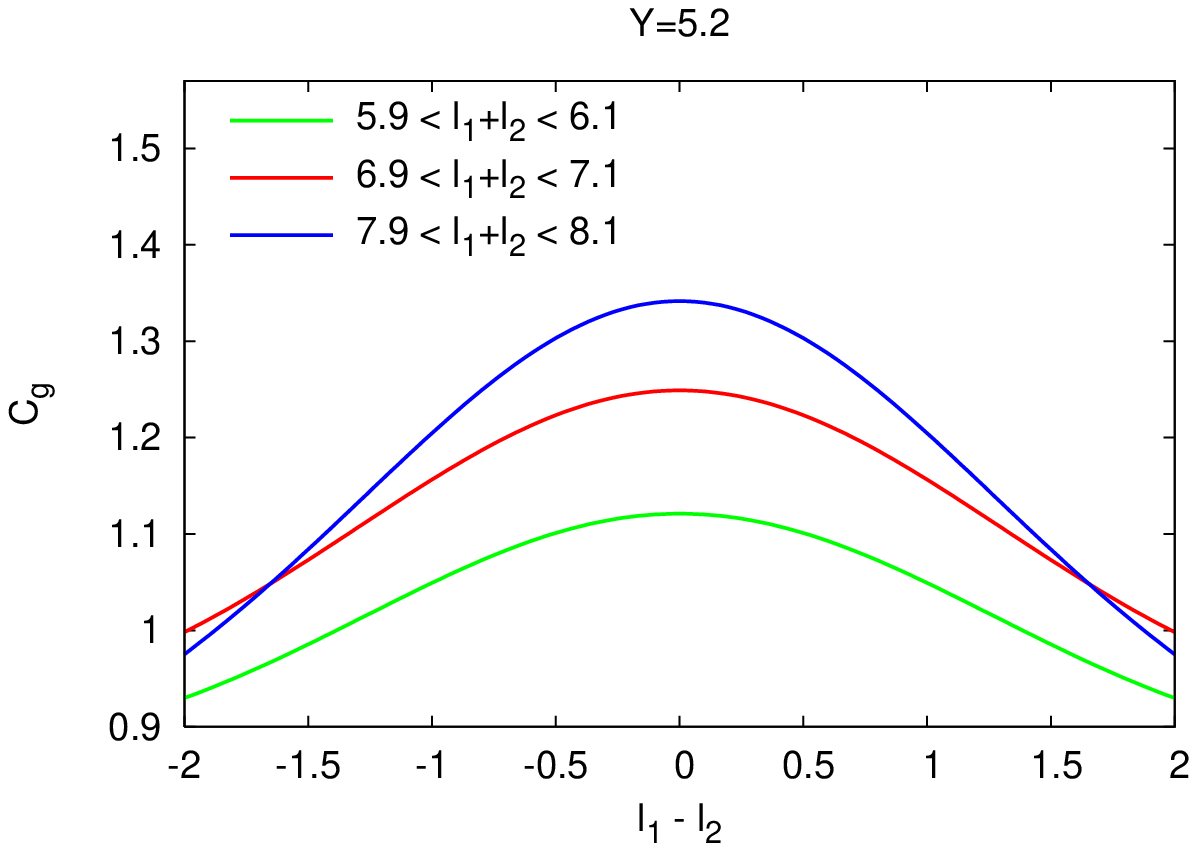, height=6truecm,width=0.47\tw}
\vskip .5cm
\caption{${\cal C}_g$ for the LEP-I ($Y=7.5$) inside a gluon jet as function of $\ell_1+\ell_2$ (left) and of $\ell_1-\ell_2$ (right)}
\label{fig:3gbandsLEP}
\end{center}
\end{figure}

\begin{figure}
\begin{center}
\epsfig{file=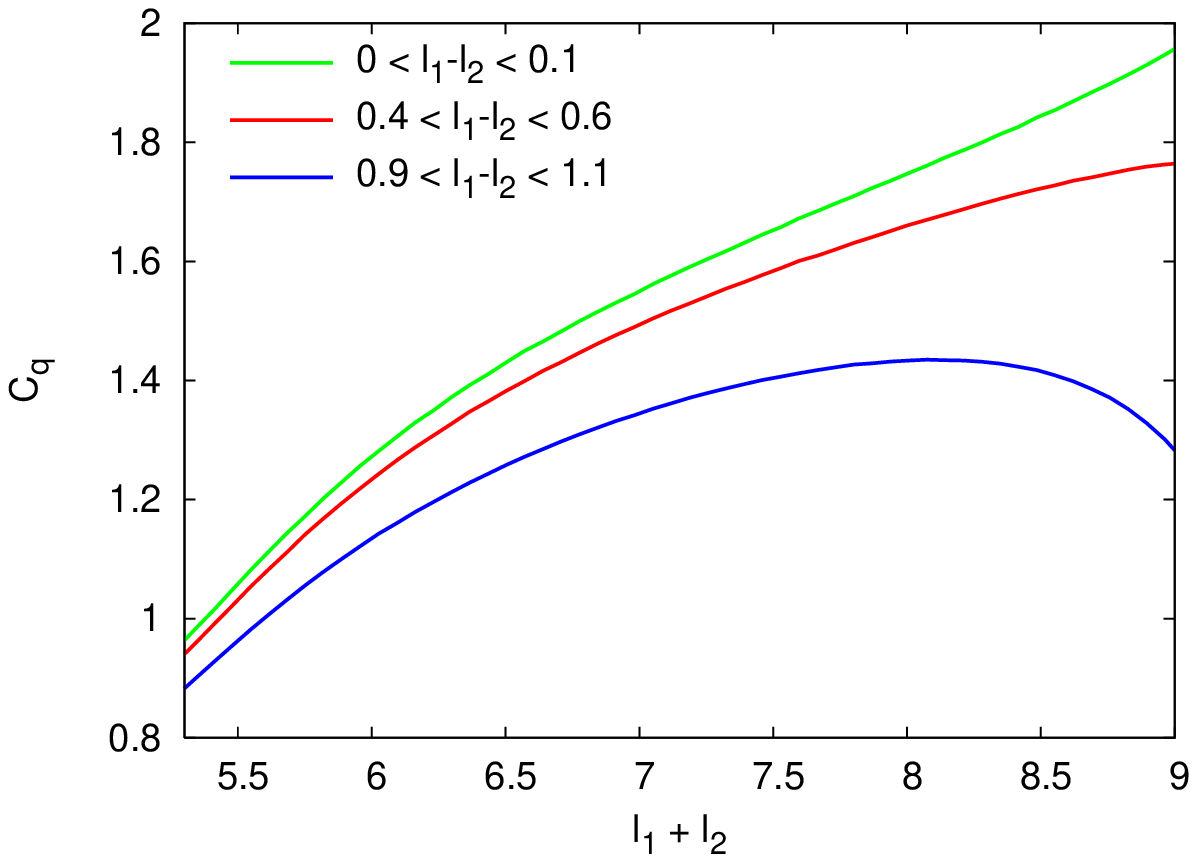, height=6truecm,width=0.47\tw}
\epsfig{file=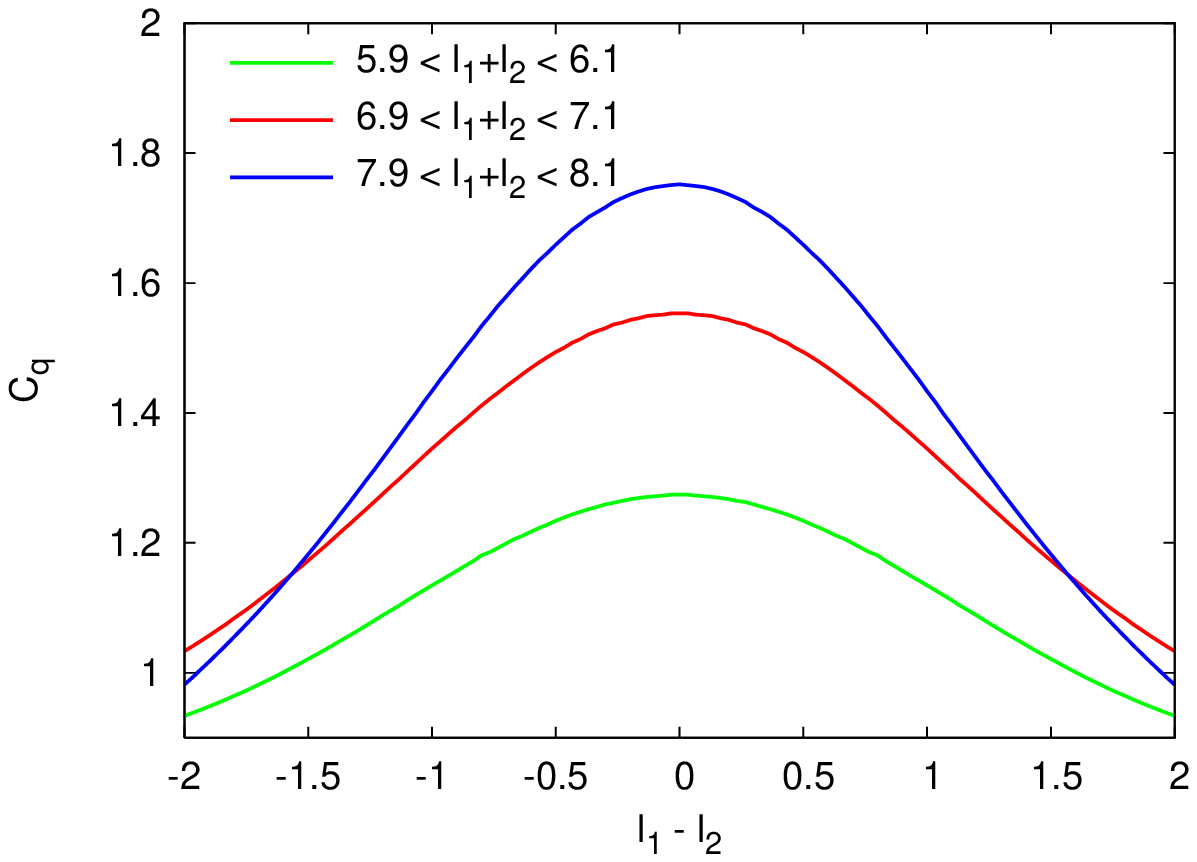, height=6truecm,width=0.47\tw}
\vskip .5cm
\caption{ ${\cal C}_q$ for the LEP-I ($Y=7.5$) inside  a quark jet as function of $\ell_1+\ell_2$ (left) and of $\ell_1-\ell_2$ (right)}
\label{fig:3qbandsLEP}
\end{center}
\end{figure}

In $e^+e^-\rightarrow q\bar{q}$ collisions 
at the $Z^0$ peak, $Q=91.2\,\text{GeV}$, $Y=5.2$, and $\gamma_0\simeq0.5$.
In Fig.~\ref{fig:3gbandsLEP} we give the results for  gluon jets  and
in Fig.~\ref{fig:3qbandsLEP} for quark jets.


\subsubsection{Comments}
\label{subsubsection:commentLEP}
%%%%%%%%%%%%%%%%%%%%%%%%%%%%%%%%

Near the maximum of the single inclusive distribution 
($\ell_1\approx\ell_2\approx \frac{Y}2(1+a\gamma_0)$) our curves are linear functions
of $(\ell_1+\ell_2)$ and quadratic functions of $(\ell_1-\ell_2)$, in
agreement with the Fong-Webber analysis \cite{FWC}.

$({\cal C}_q-1)$ is roughly twice $({\cal C}_g-1)$ since
gluons cascade twice more than quarks ($\frac{N_c}{C_F}\approx2$). The difference 
is clearly observed from Fig.~\ref{fig:3gbandsLEP} and Fig.~\ref{fig:3qbandsLEP} (left) 
near the hump of the single inclusive distribution ($\ell_1+\ell_2\simeq7.6$), 
that is where most of the partonic multiplication takes place.

In both cases, $\cal C$ reaches its largest value for $\ell_1\approx \ell_2$ 
and steadily increases as a function of $(\ell_1+\ell_2)$ 
(Fig.~\ref{fig:3gbandsLEP}, left);
for $\ell_1\ne \ell_2$, it increases with $(\ell_1+\ell_2)$, then flattens off 
and decreases.

Both ${\cal C}$'s decrease as $|\ell_1-\ell_2|$ becomes large 
(Fig.~\ref{fig:3gbandsLEP} and \ref{fig:3qbandsLEP}, right).
The quark's tail is steeper than the gluon's;
for $5.9<\ell_1+\ell_2<6.1$, $({\cal C}-1)$ becomes negative when $\ell_1-\ell_2$
increases; ${\cal C}\geq1$ as soon as $\ell_1,\ell_2\geq2.75\,(x_1,x_2\leq0.06)$;
this bounds is close to $\ell\geq2.4$ found in subsection \ref{subsection:signG} or 
$\ell\geq2.5$ of (\ref{eq:confint1}).

One finds the limit 

\begin{equation}
{\cal C}_{g\,or\,q}\stackrel{\ell_1+\ell_2\to 2Y}{\longrightarrow}1.
\end{equation}

\begin{figure}
\begin{center}
\epsfig{file=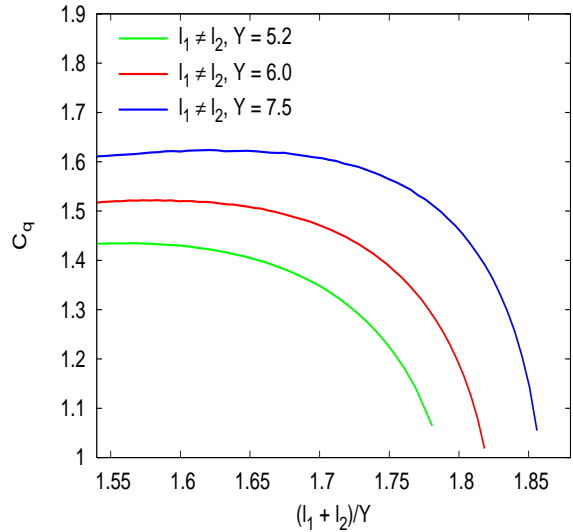, height=7truecm,width=0.48\tw}
\caption{Decrease of the correlation for $\ell_1\ne\ell_2$ at $Y=5.2$, $Y=6.0$ and $Y=7.5$}
\label{fig:Colliders}
\end{center}
\end{figure}

Actually, one observes on  Figs.~\ref{fig:3gbandsLEP}, \ref{fig:3qbandsLEP}
and \ref{fig:Colliders}
that a stronger statement holds. Namely, when 
 we take the limit $\ell_2\to Y$ for the softer particle,
the correlator goes to $1$.
This is the consequence of QCD coherence. The softer gluon is emitted at larger angles
by the total color charge of the jet and thus becomes de-correlated with the internal partonic structure of the jet.

The same phenomenon explains the flattening and the decrease of ${\cal C}$'s at $\ell_1\ne\ell_2$.

An interesting phenomenon is the seemingly continuous increase of ${\cal
C}_g$ and ${\cal C}_q$ at large $Y$ for $\ell_1 \approx \ell_2$ (green
curves in figs. \ref{fig:3gbandsLEP} and \ref{fig:3qbandsLEP} left). Like
we discussed in \cite{PerezMachetC} concerning inclusive distributions, here we
reach a domain where a perturbative analysis cannot be trusted {\em
because of the divergence of $\alpha_s$}. Indeed, when $(\ell_1+\ell_2)$ gets close to
its limiting kinematical value ($2Y$), both $y_1$ and
$y_2$ get close to $0$, such that the corresponding
$\alpha_s(k_{1\perp}^2)$ and $\alpha_s(k_{2\perp}^2)$ cannot but become out
of control. Away from the $\ell_1 \approx \ell_2$ diagonal, taking $\ell_2\to Y$ 
($y_2\to0$), we have $y_1\to\eta>0$ and the emission of the harder parton still stays under control.

The two limitations of our approach already pointed at in
\cite{PerezMachetC} are found again here:

$\ast$\ $x$ should be small enough such that our soft approximation stays valid;

$\ast$\ no running coupling constant should get too large such that pQCD stays reliable.

\subsection{Comparison with the data from LEP-I}
%%%%%%%%%%%%%%%%%%%%%%%%%%%%%%%%%%%%%%%%%%%%%%%%%

OPAL results are given in terms of

$$
 R\left(\ell_1, \ell_2, Y\right) \>=\>
 \frac1{2}+\frac1{2}{\cal C}_q\left(\ell_1, \ell_2, Y\right).
$$

\begin{figure}
\vbox{
\begin{center}
\epsfig{file=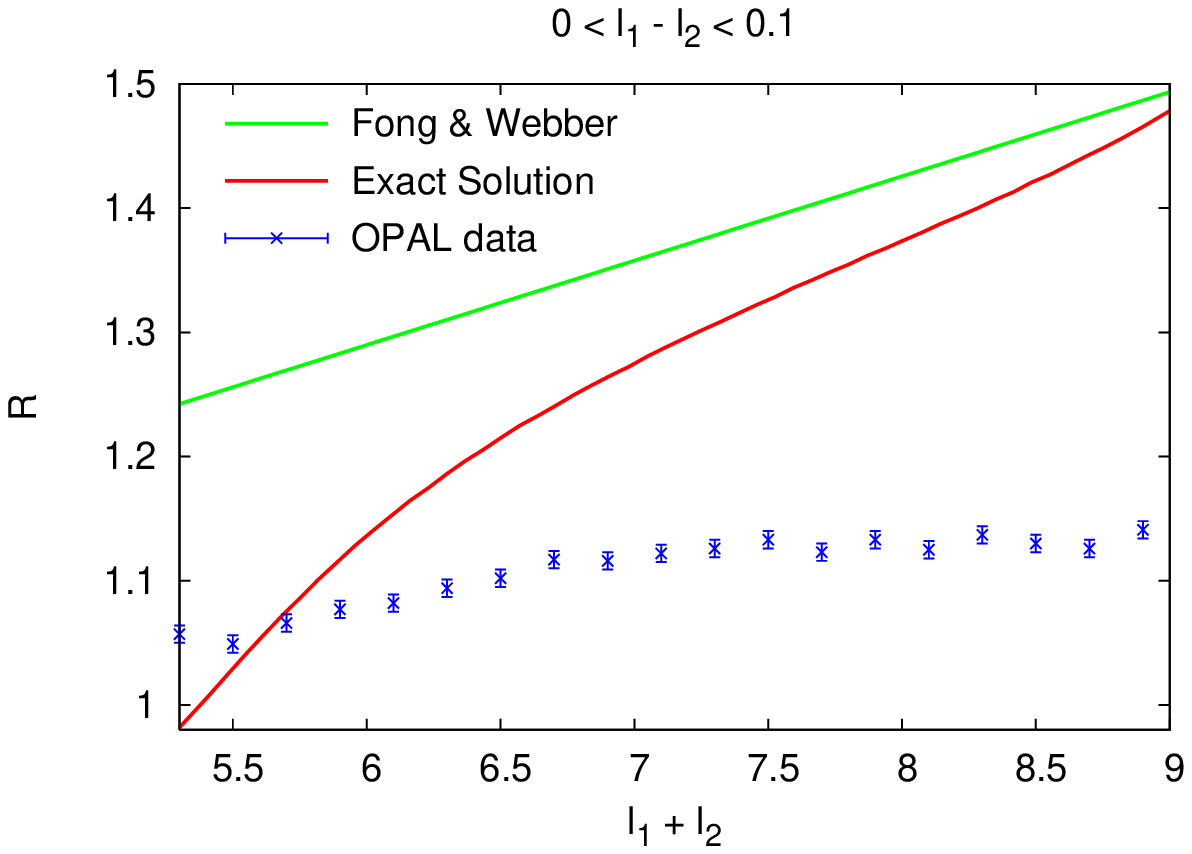, height=6truecm,width=0.48\tw}
\hfill
\epsfig{file=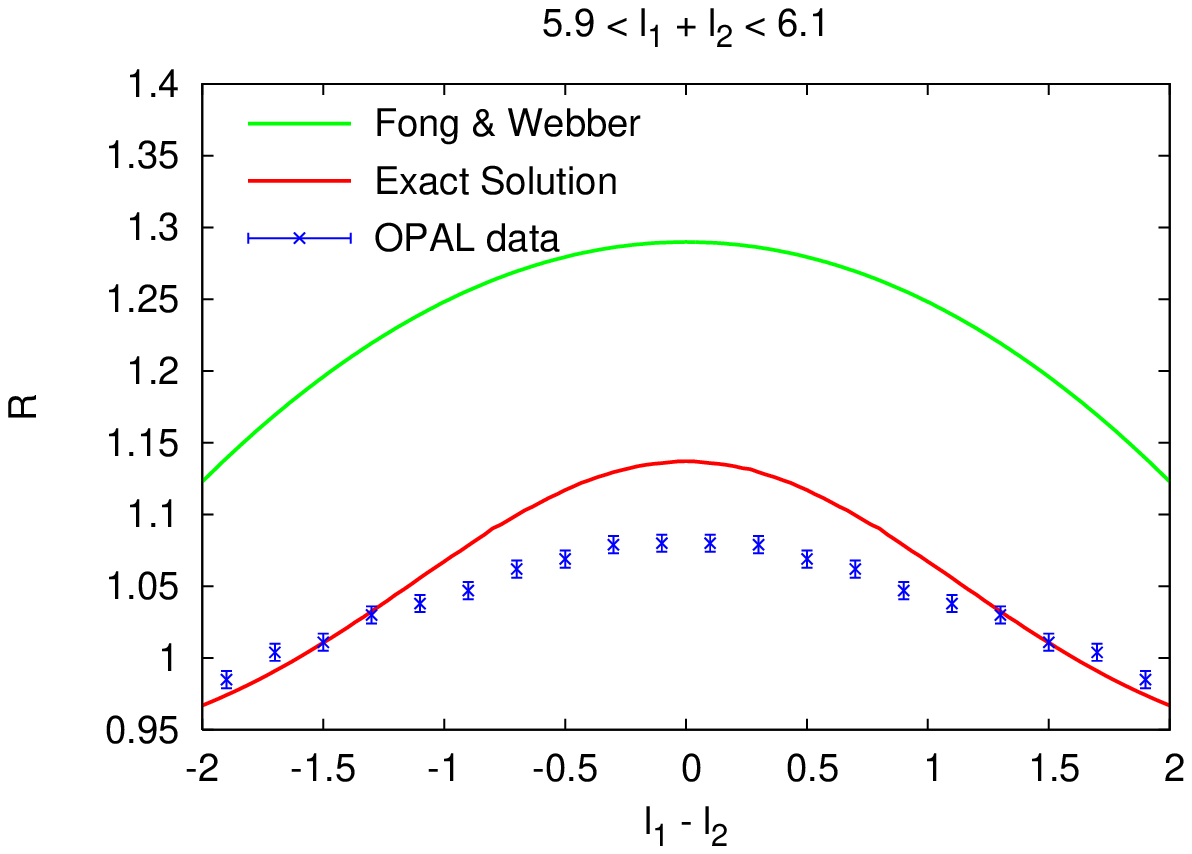, height=6truecm,width=0.48\tw}
\vskip .5cm
\epsfig{file=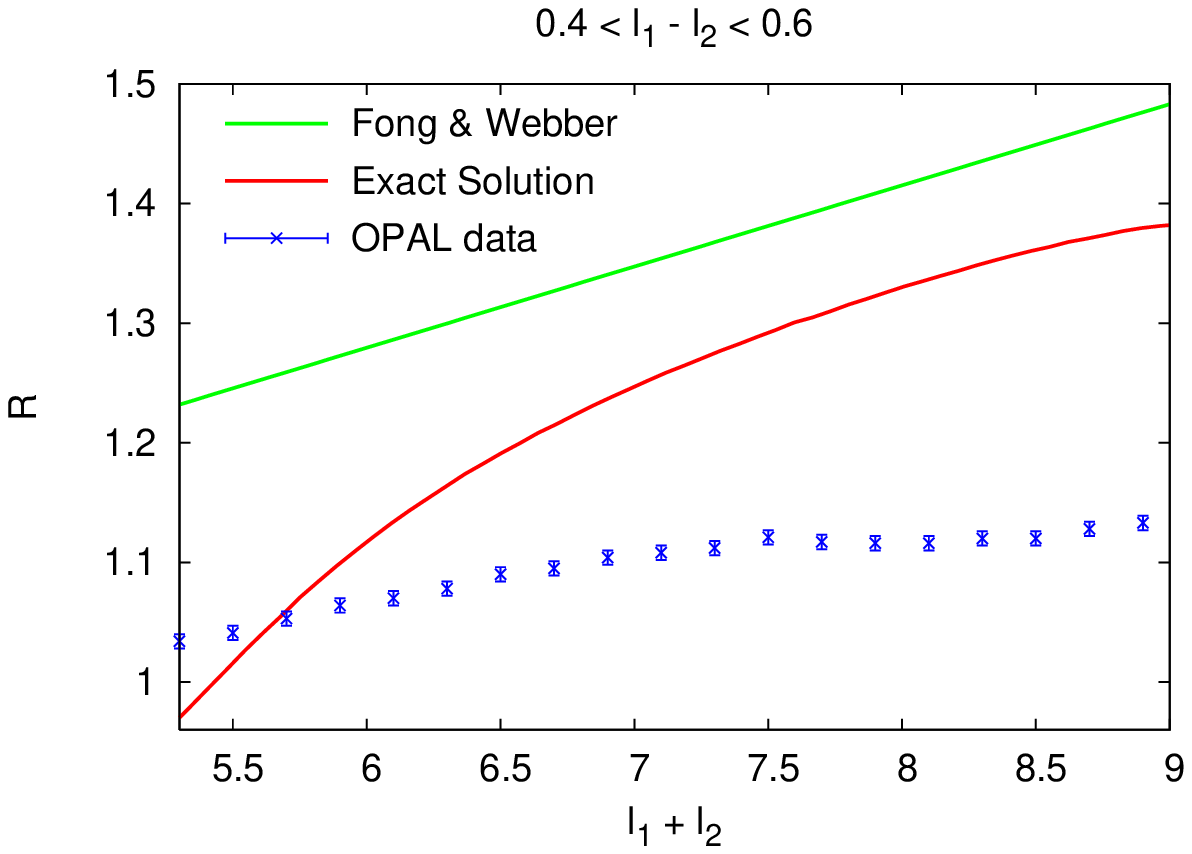, height=6truecm,width=0.48\tw}
\hfill
\epsfig{file=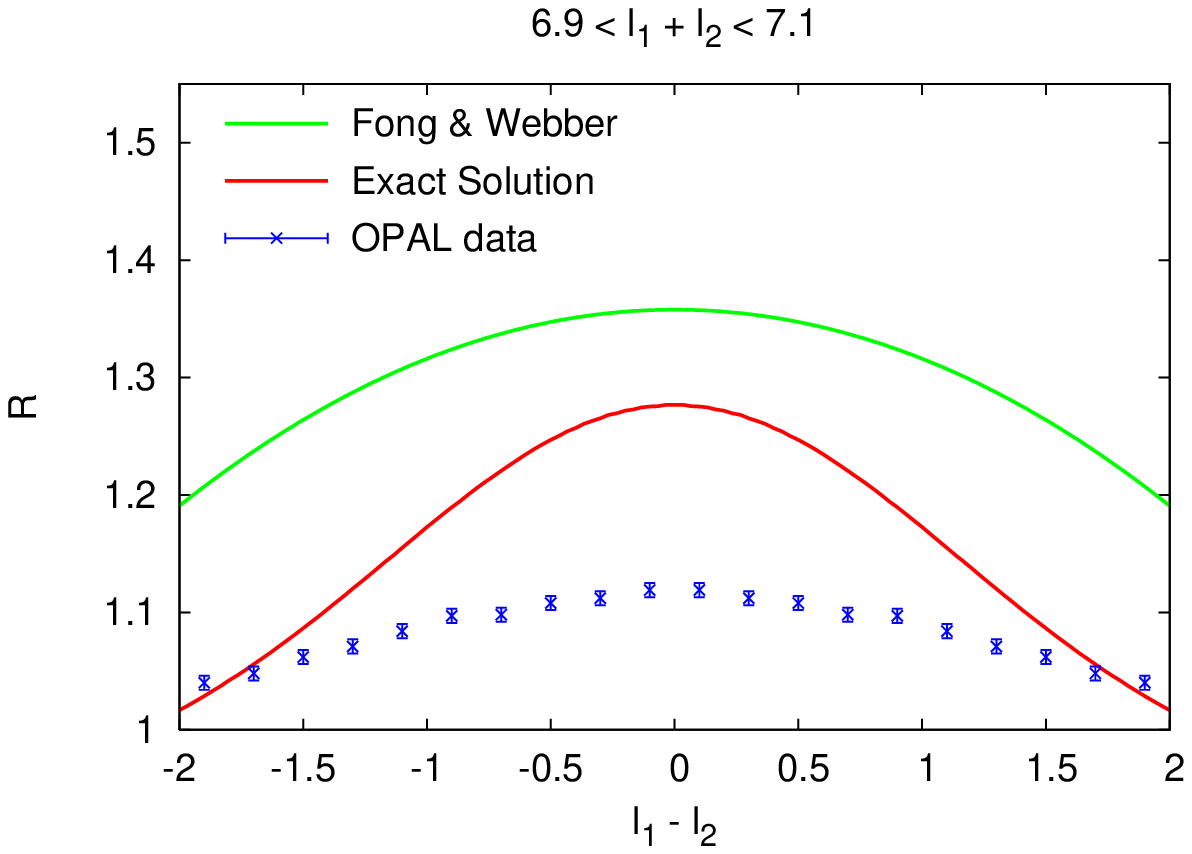, height=6truecm,width=0.48\tw}
\vskip .5cm
\epsfig{file=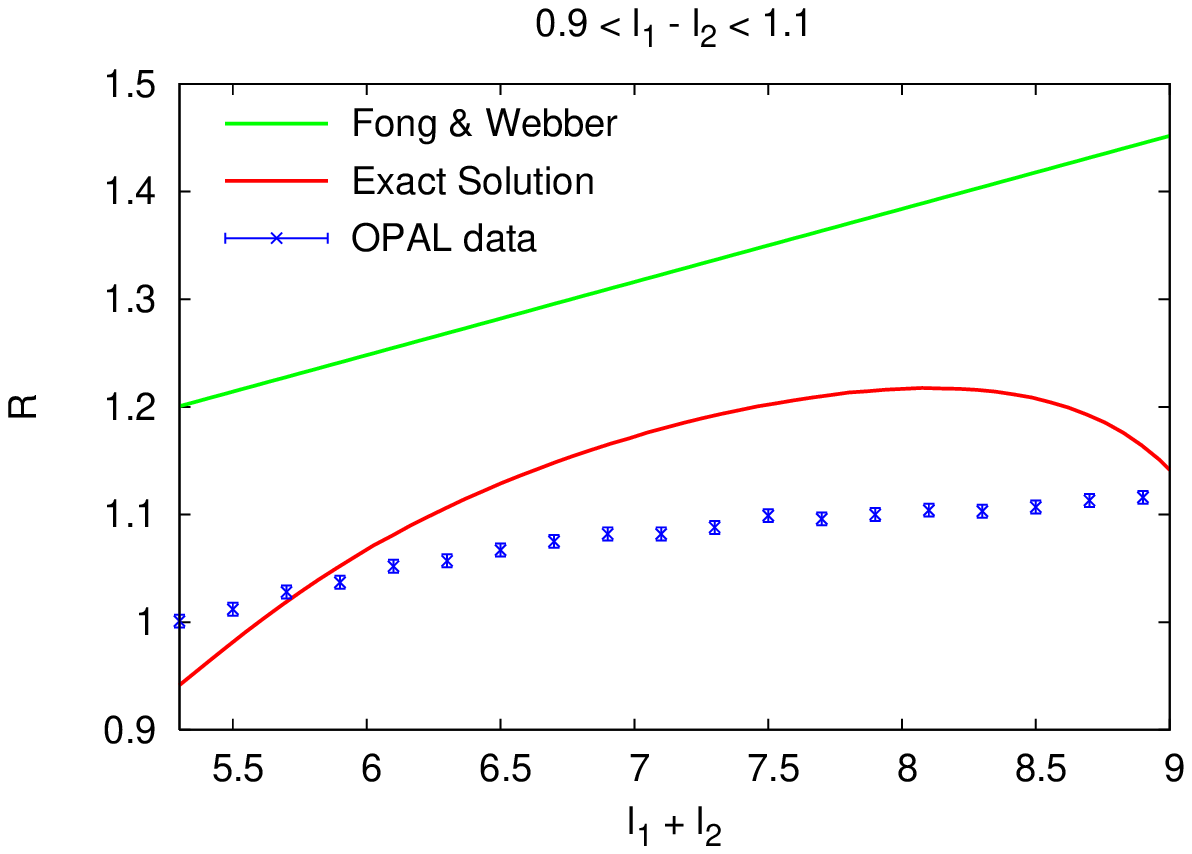, height=6truecm,width=0.48\tw}
\hfill
\epsfig{file=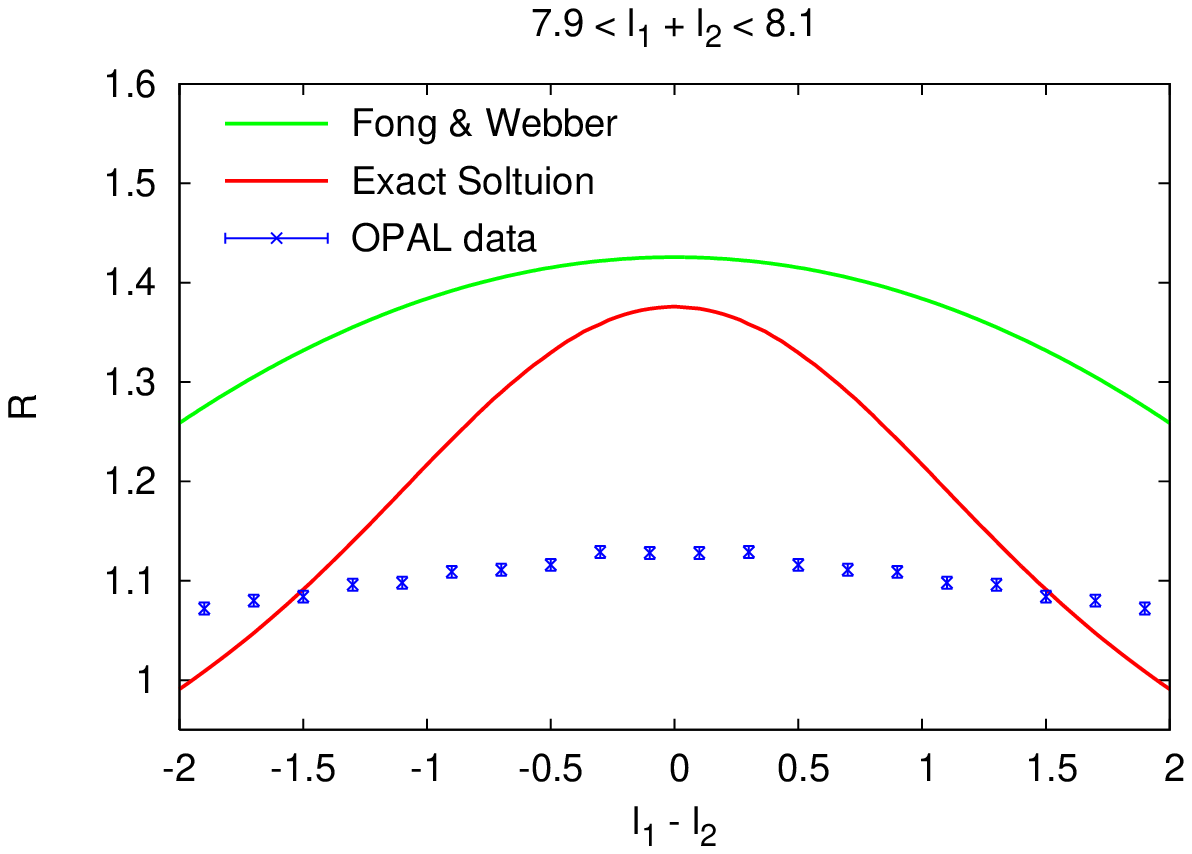, height=6truecm,width=0.48\tw}
\vskip .5cm
\caption{Correlations $R$ between two particles produced in $e^+e^-\rightarrow q\bar{q}$ 
compared with the OPAL data and the Fong-Webber approximation}
\label{fig:corrqqbar}
\end{center}
}
\end{figure}

In Fig.~\ref{fig:corrqqbar} we compare our prediction with the OPAL data 
\cite {OPALC} and the Fong-Webber curves (see subsection \ref{sub:FWC}
and \cite{FWC}).

\subsection{Comparing with the Fong-Webber approximation}
\label{sub:FWC}
%%%%%%%%%%%%%%%%%%%%%%%%%%%%%%%%%%%%%%%%%%%%%%%%%%%%%%%%%

The only pQCD analysis of two-particle correlations in jets
beyond DLA was performed by Fong and Webber in 1990. In~\cite{FWC}
the next-to-leading ${\cal O}({\gamma_0})$ correction,
${\cal C}_{g\, or\, q}=1+\sqrt{\alpha_s}+\cdots$, to the normalized two-particle
correlator was calculated. This expression was derived in the region 
$|\ell_1-\ell_2|/Y \ll 1$,
that is when the energies of the registered particles are 
close to each other (and to the maximum of the inclusive
distribution \cite{EvEqC}\cite{KOC}\cite{FW1C}).
In this approximation the correlation function is
quadratic in $(\ell_1-\ell_2)$ and increases linearly
with $(\ell_1+\ell_2)$, see (\ref{eq:FWS}). 
For example, if one replaces the expression of the 
single inclusive distribution distorted gaussian \cite{FW1C} 
(obtained in the region $\ell\approx \frac{Y}2(1+a\gamma_0)$) 
into (\ref{eq:CGMLLAC}) the MLLA result for a gluon jet reads

\begin{equation}
{\cal C}_g(\ell_1,\ell_2,Y)\approx1+ \frac{1- \bigg(5b-3b
\displaystyle{\frac{\ell_1+\ell_2}Y}\bigg)\gamma_0+{\cal O}(\gamma_0^2)}
{3+9\bigg(\displaystyle
{\frac{\ell_1-\ell_2}Y}\bigg)^2 
 -\left(2\beta + 5a-3a\displaystyle{\frac{\ell_1+\ell_2}Y}\right)\gamma_0+
{\cal O}(\gamma_0^2)},
\label{eq:CGMLLAFW}
\end{equation}

where we have neglected the MLLA correction $\delta_1\simeq(\ell_1-\ell_2)^2\sqrt{\alpha_s}\simeq0$ near
the hump of the single inclusive distribution ($\ell_1\approx\ell_2\approx\frac{Y}2(1+a\gamma_0)$). The Fong-Webber answer is obtained by
expanding (\ref{eq:CGMLLAFW}) in $\gamma_0$ to get 
\cite{FWC}

\begin{eqnarray}
{\cal C}_g^{(\mbox{\scriptsize FW})} \approx \frac43 - \left(\frac{\ell_1 - \ell_2}{Y}\right)^2 + \left[-\frac53\left(b-\frac13 a\right) +\frac{2}{9}\beta +\left(b-\frac13 a\right)
\left(\frac{\ell_1+\ell_2}{Y}\right)\right]\gamma_0 + {\cal O}(\gamma_0^2).\cr
&&
\label{eq:FWS}
\end{eqnarray}

\begin{figure}
\vbox{
\begin{center}
\epsfig{file=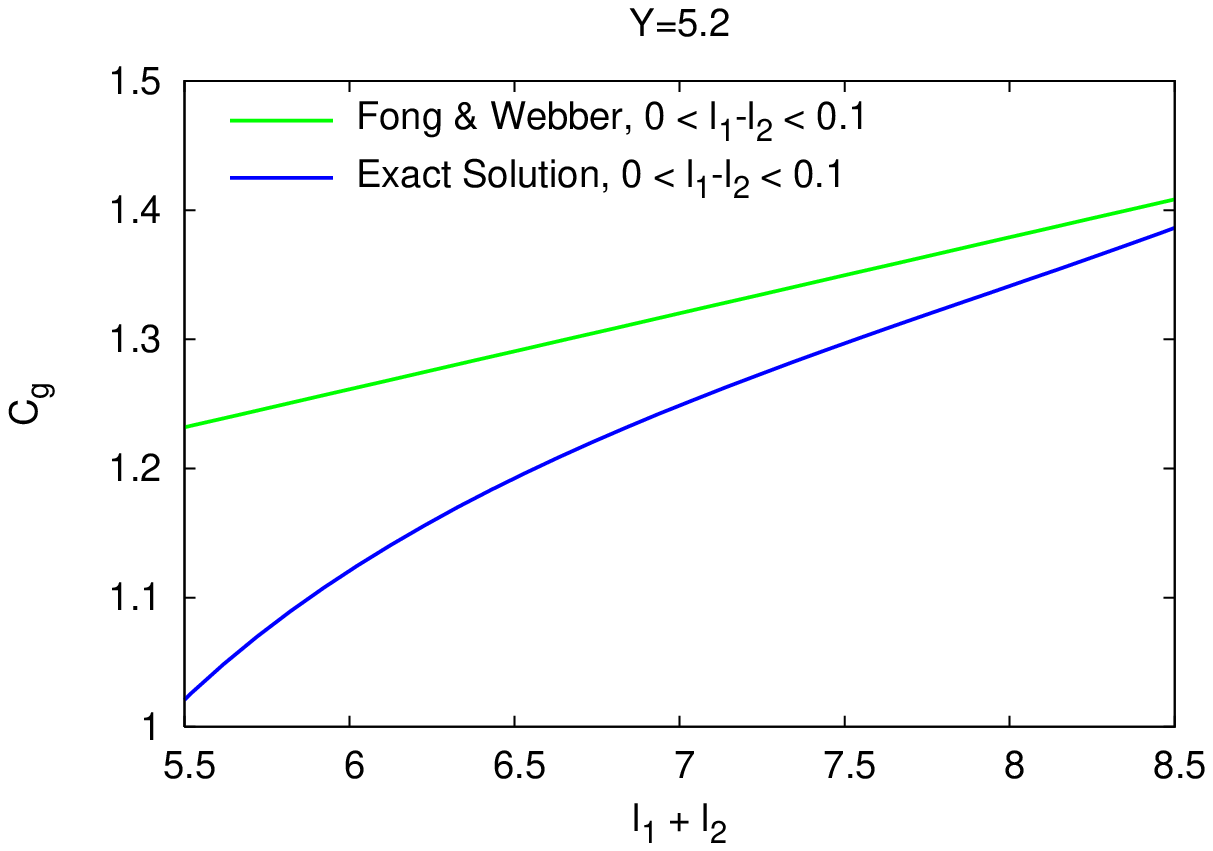, height=5truecm,width=0.48\tw}
\hfill
\epsfig{file=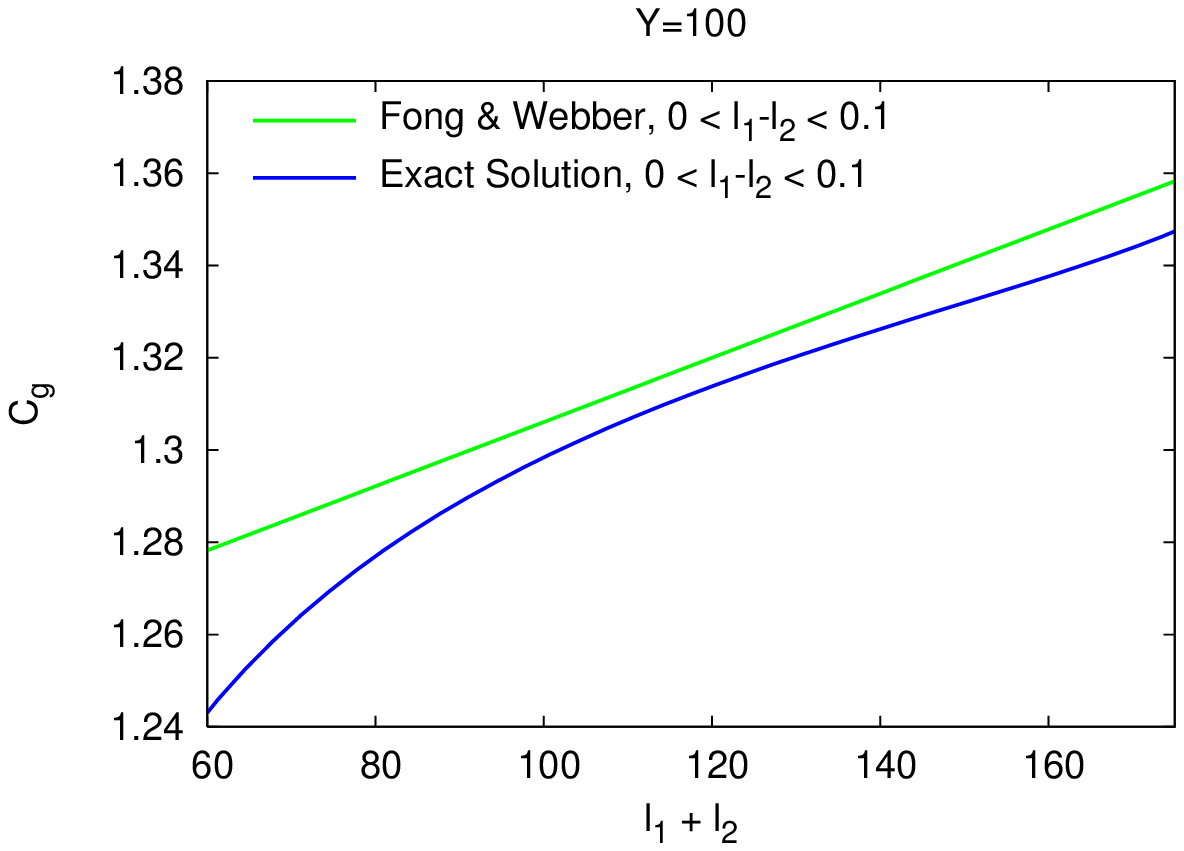, height=5truecm,width=0.48\tw}
\vskip .5cm
\caption{Exact ${\cal C}_g$ compared with Fong-Webber's at $Y=5.2$ (left) and 
$Y=100$ (right)}
\label{fig:FWES}
\end{center}
}
\end{figure}

In Fig.~\ref{fig:FWES} we compare, choosing for pedagogical reasons $Y=5.2$ and
$Y=100$, our exact solution of the evolution equation with
the Fong-Webber predictions \cite{FWC} for two particle correlations. 
The mismatch in both cases is, as seen on (\ref{eq:FWS}),  
${\cal O}(\gamma_0^2)$, and decreases for
smaller values of the perturbative expansion parameter $\gamma_0$. 
In particular, at $Y=100$, ($\gamma_0^2\simeq0.01$) 
the exact solution 
(\ref{eq:CGfull}) gets close to (\ref{eq:FWS}).
This comparison is analogous in the case of a quark jet.

We do not perform in the present work such an expansion but keep instead the ratios
(\ref{eq:CGfull}) and  (\ref{eq:Qcorr}) as exact solutions of the evolution equations.

\subsection{Predictions for Tevatron and LHC}
\label{subsection:tevatronlhc}
%%%%%%%%%%%%%%%%%%%%%%%%%%%%%%%%%%%%%%%%%%%%%%

In hadronic high energy colliders, the nature of the jet (quark or gluon) is not
determined, and one simply detects outgoing hadrons, which can originate
from  either type; one then introduces a ``mixing'' parameter $\omega$,
which is to be determined experimentally, such that, the expression for two particle
correlations can be written as a linear combination of ${\cal C}_g$ and ${\cal C}_q$

\begin{equation}
{\cal C}^{mixed}(\omega;\ell_1,\ell_2,Y)=A(\omega;\ell_1,\ell_2,Y)\,
{\cal C}_q(\ell_1,\ell_2,Y)+ B(\omega;\ell_1,\ell_2,Y)\,{\cal C}_g(\ell_1,\ell_2,Y),
\label{eq:corrmix}
\end{equation}

where

$$
A(\omega;\ell_1,\ell_2,Y)=
\frac{\omega\, \left[\displaystyle{\frac{Q(\ell_1,Y)}{G(\ell_1,Y)}}
\displaystyle{\frac{Q(\ell_2,Y)}{G(\ell_2,Y)}}\right]}
{\bigg[1+\omega\bigg(\displaystyle{\frac{Q(\ell_1,Y)}{G(\ell_1,Y)}}-1\bigg)\bigg]
\bigg[1+\omega\bigg(\displaystyle{\frac{Q(\ell_2,Y)}{G(\ell_2,Y)}}-1\bigg)\bigg]}
$$

and

$$ B(\omega;\ell_1,\ell_2,Y)=
\frac{(1-\omega)}{\bigg[1+\omega\bigg(\displaystyle{\frac{Q(\ell_1,Y)}{G(\ell_1,Y)}}
-1\bigg)\bigg]
\bigg[1+\omega\bigg(\displaystyle{\frac{Q(\ell_2,Y)}{G(\ell_2,Y)}}-1\bigg)\bigg]}.
$$

\begin{figure}
\begin{center}
\epsfig{file=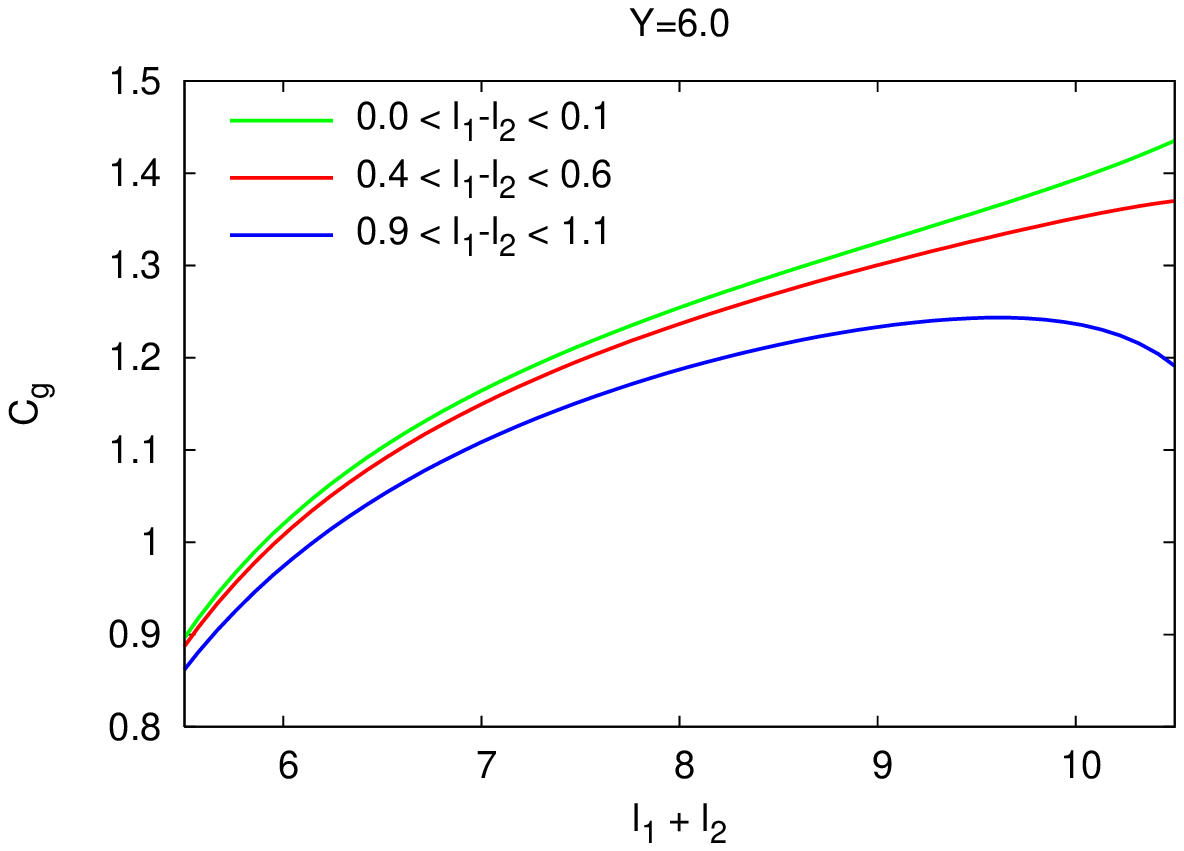, height=6truecm,width=0.47\tw}
\epsfig{file=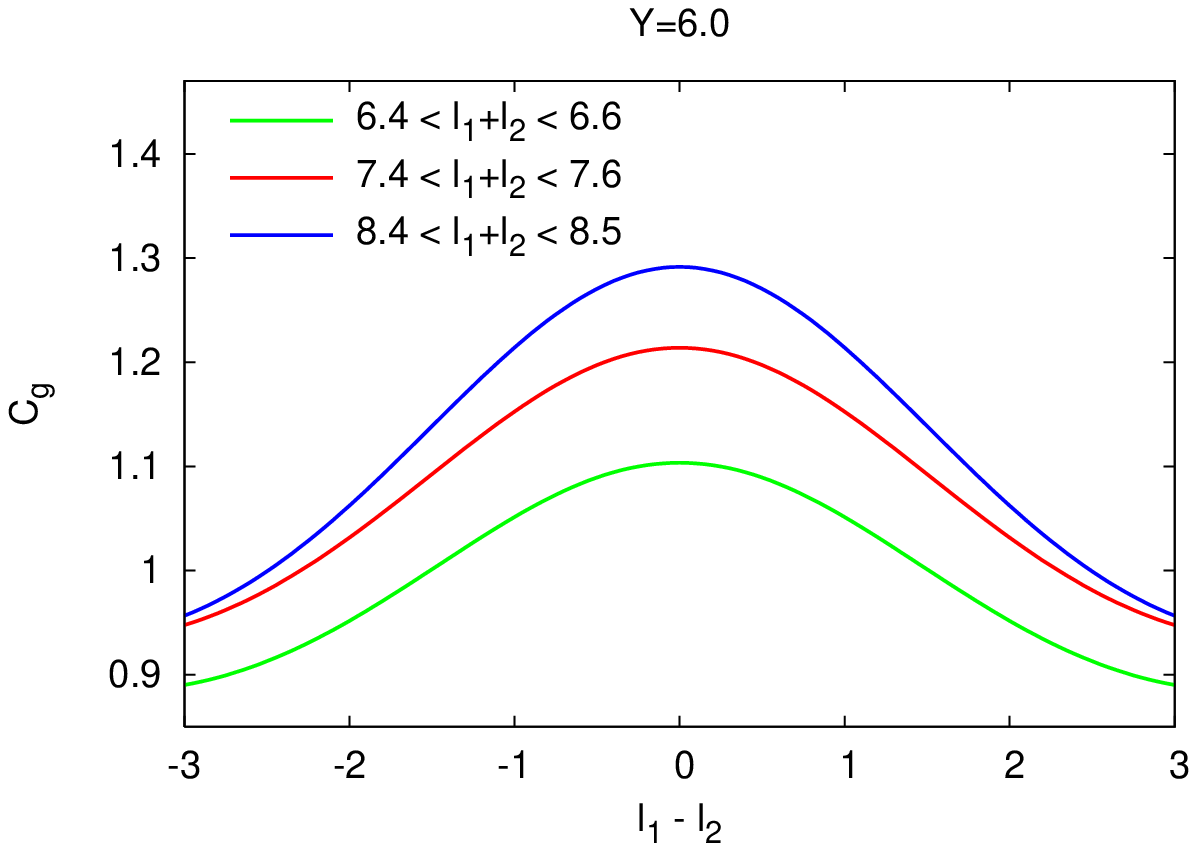, height=6truecm,width=0.47\tw}
\vskip .5cm
\caption{${\cal C}_g$ for the Tevatron ($Y=6.0$) as function of $\ell_1+\ell_2$ (left)
and of $\ell_1-\ell_2$ (right)}
\label{fig:3gbandsTeV}
\end{center}
\end{figure}

\begin{figure}
\begin{center}
\epsfig{file=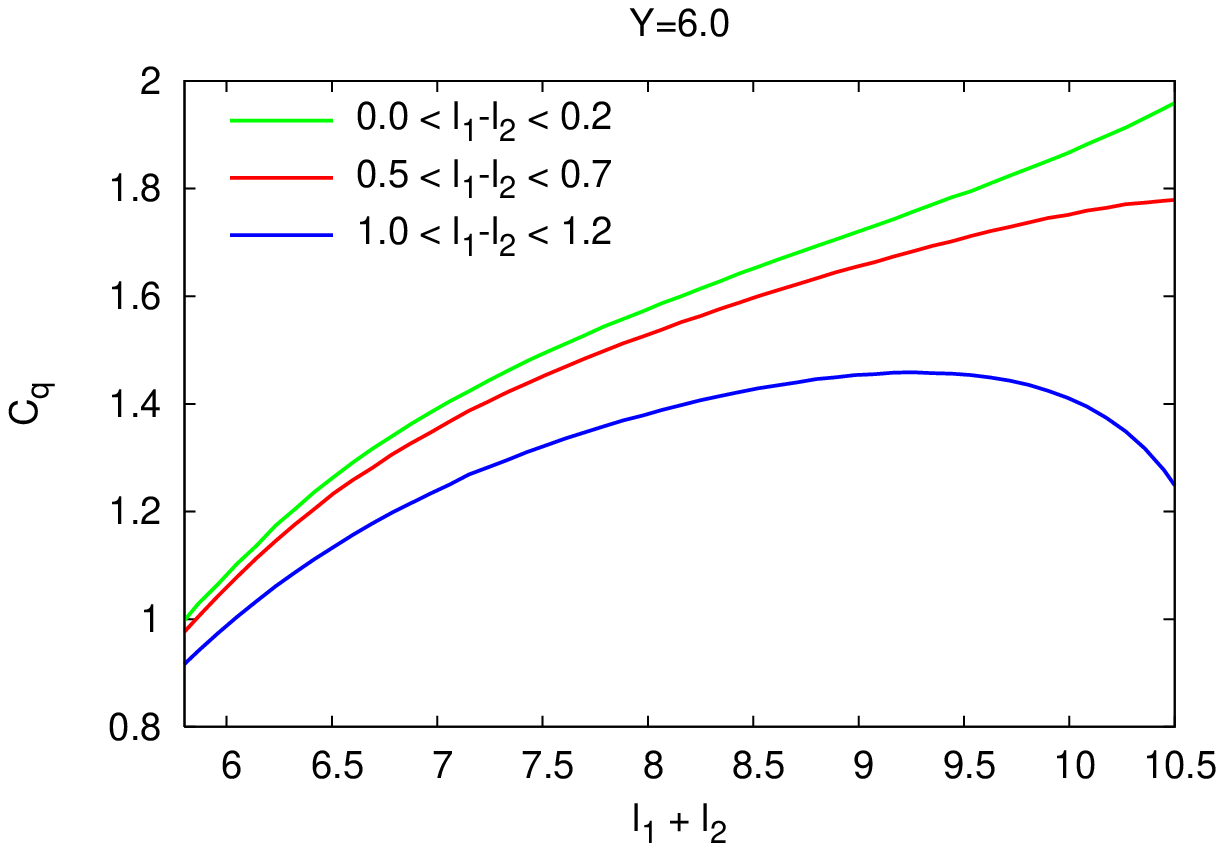, height=6truecm,width=0.47\tw}
\epsfig{file=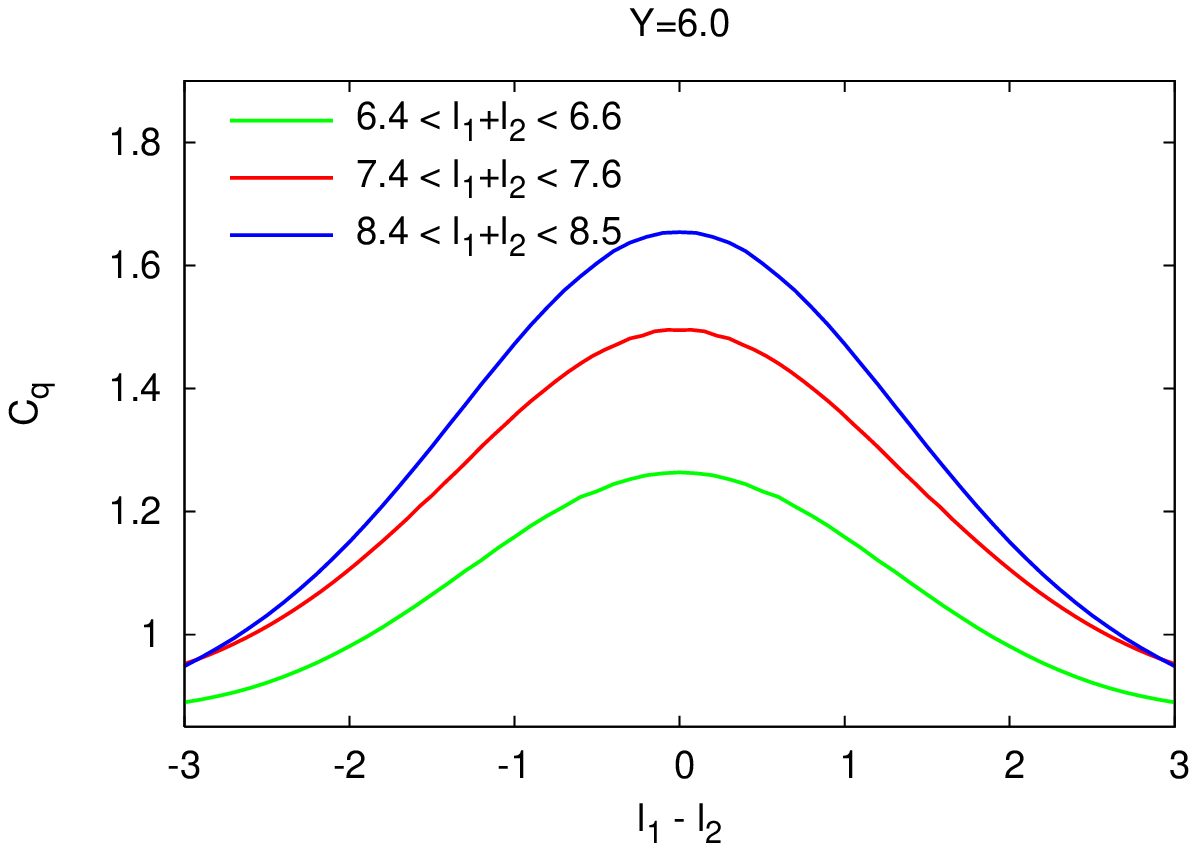, height=6truecm,width=0.47\tw}
\vskip .5cm
\caption{${\cal C}_q$ for the Tevatron ($Y=6.0$) as function of $\ell_1+\ell_2$ (left)
and of $\ell_1-\ell_2$ (right)}
\label{fig:3qbandsTeV}
\end{center}
\end{figure}

\begin{figure}
\begin{center}
\epsfig{file=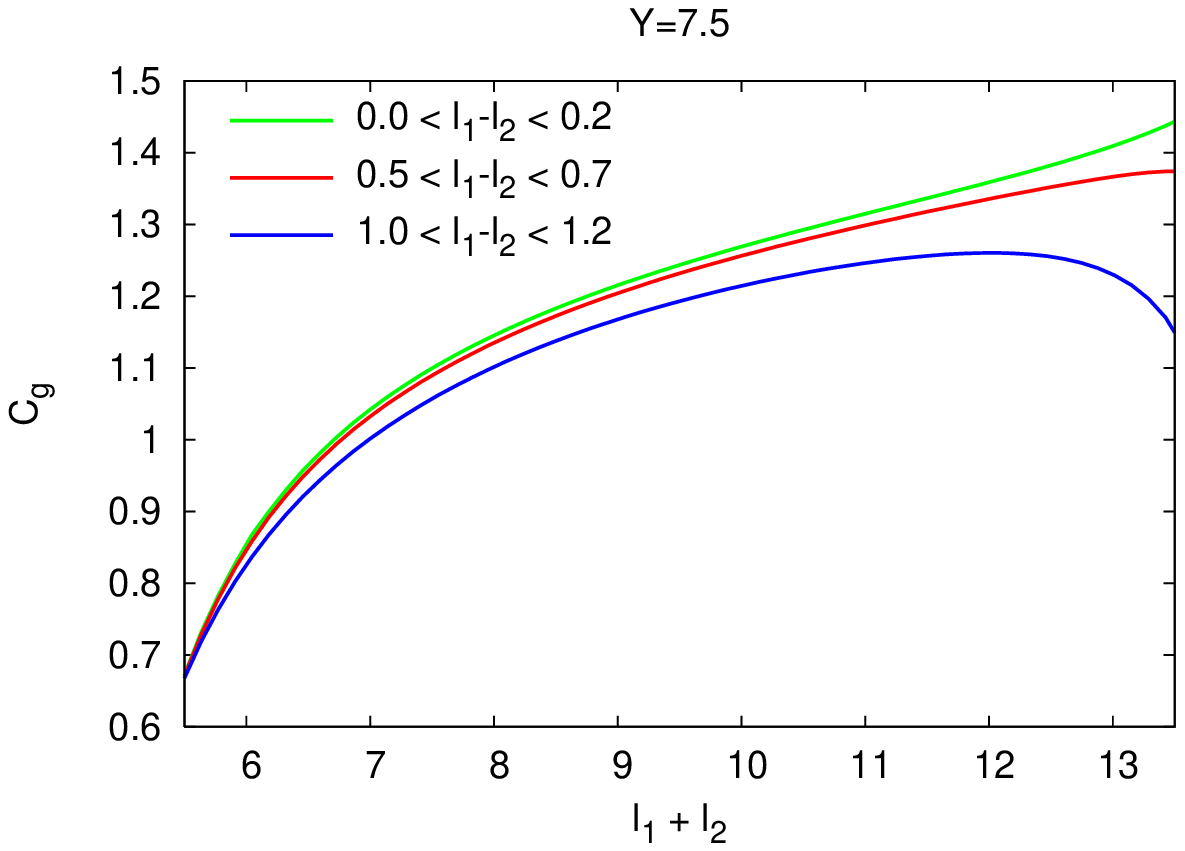, height=6truecm,width=0.47\tw}
\epsfig{file=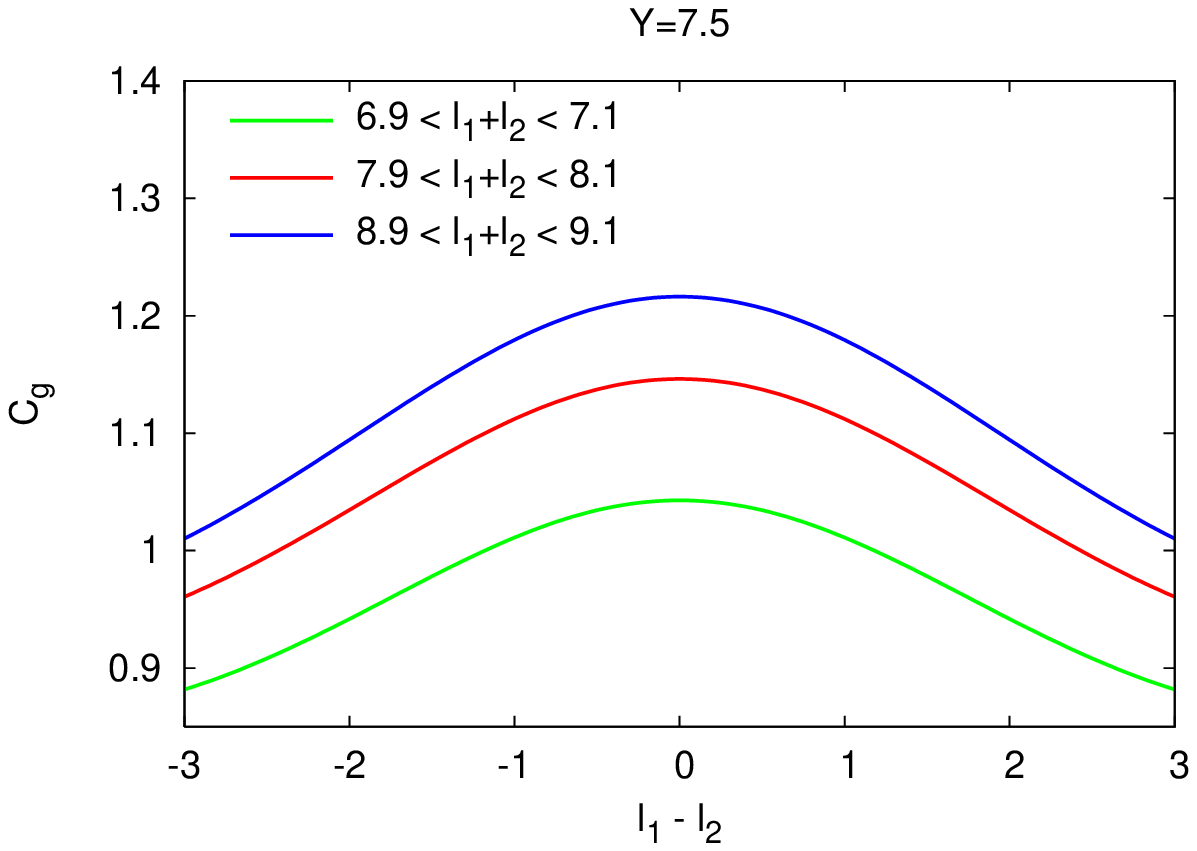, height=6truecm,width=0.47\tw}\vskip .5cm
\caption{ ${\cal C}_g$ for the LHC ($Y=7.5$) inside a gluon jet as function of $\ell_1+\ell_2$ (left) and of $\ell_1-\ell_2$ (right)}
\label{fig:3gbandsLHC}
\end{center}
\end{figure}

\begin{figure}
\begin{center}
\epsfig{file=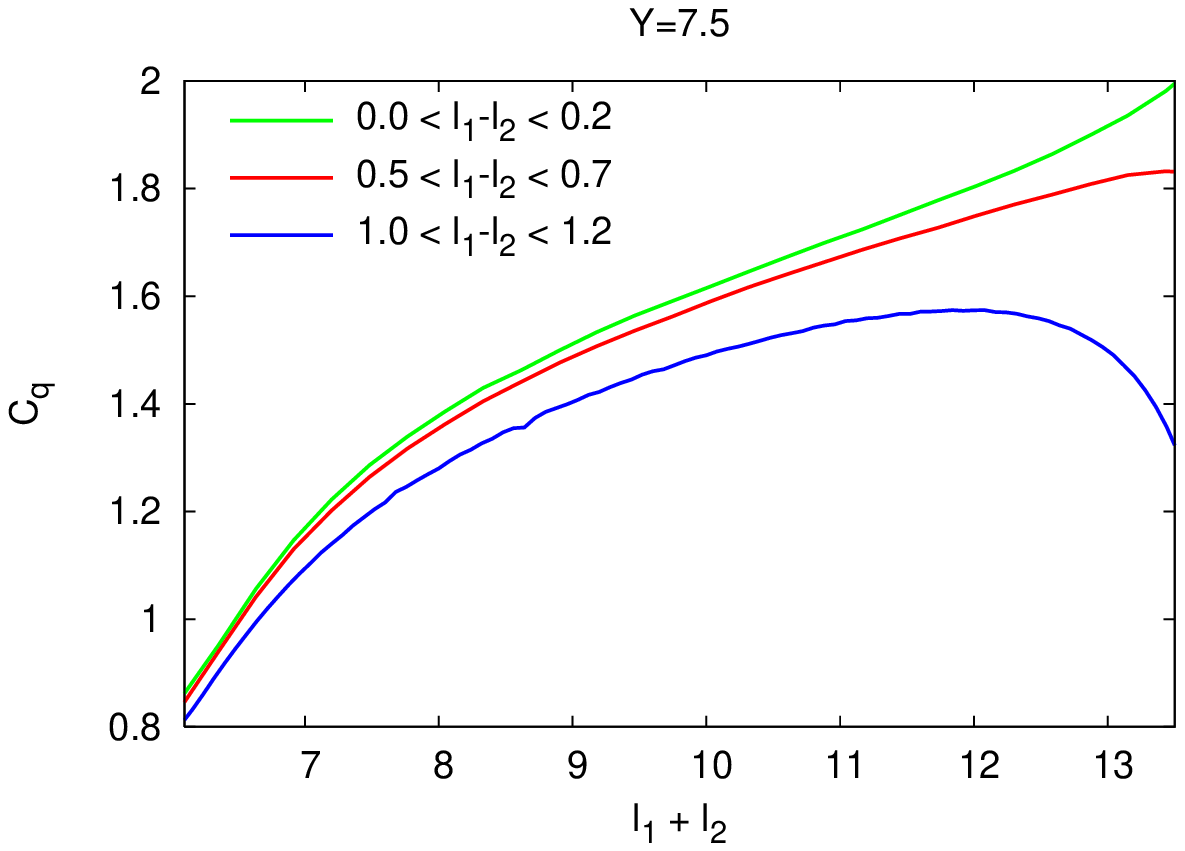, height=6truecm,width=0.47\tw}
\epsfig{file=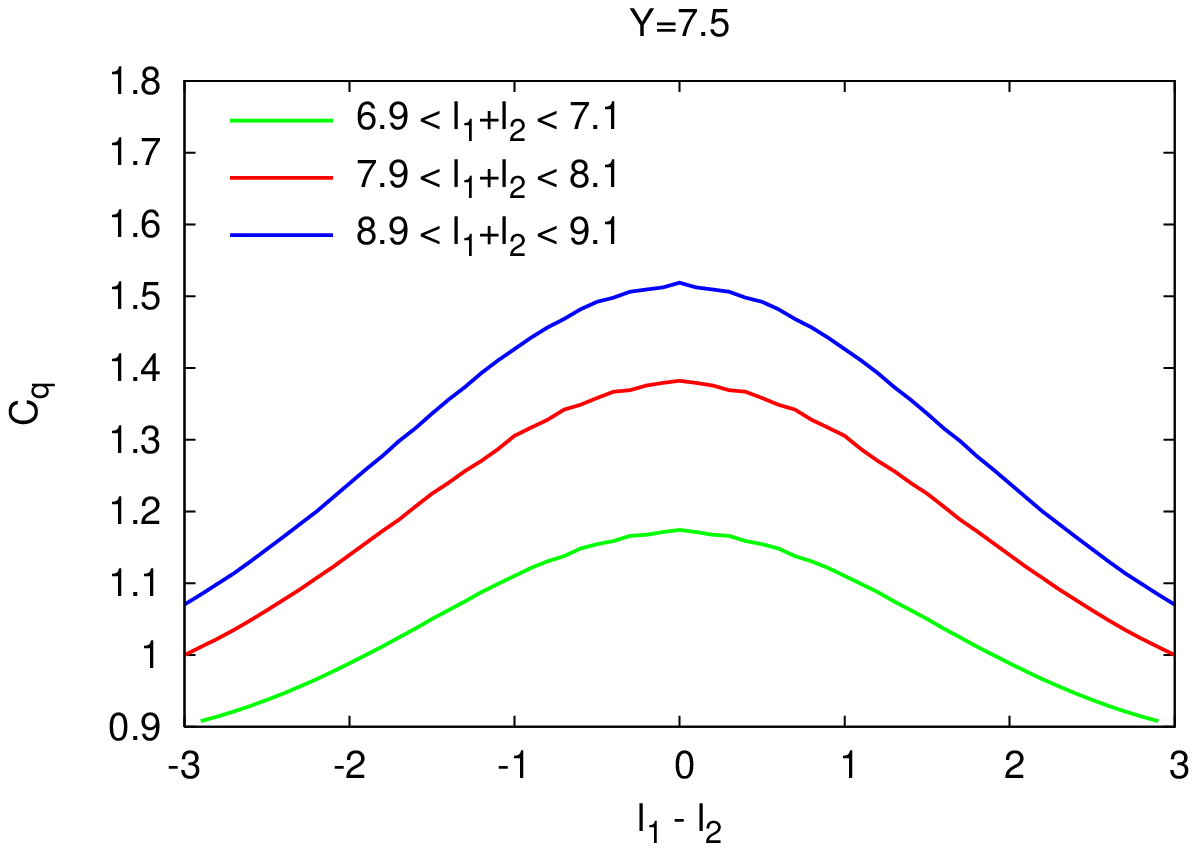, height=6truecm,width=0.47\tw}
\vskip .5cm
\caption{${\cal C}_q$ for the LHC ($Y=7.5$) inside a gluon jet as function of $\ell_1+\ell_2$ (left) and of $\ell_1-\ell_2$ (right)}
\label{fig:3qbandsLHC}
\end{center}
\end{figure}

We plug in respectively (\ref{eq:CGfull}) (\ref{eq:Qcorr}) for 
${\cal C}_g$ and ${\cal C}_q$; the predictions for the latter are given in
Figs.~\ref{fig:3gbandsTeV} and \ref{fig:3qbandsTeV} for the Tevatron,
Figs.~\ref{fig:3gbandsLHC} and \ref{fig:3qbandsLHC} for the LHC.



\subsubsection{Comments}
%%%%%%%%%%%%%%%%%%%%%%%%%

For both $Y=6.0$ (Tevatron) and $Y=7.5$ (LHC), the global behavior given in 
\ref{subsubsection:commentLEP} also holds. 
The interval corresponding to the condition ${\cal C}_{g\,or\,q}>1$ is shifted toward larger values of $\ell$ (smaller $x$) as compared with the $Y=5.2$ case, in agreement 
with the predictions of (\ref{subsection:signG}) and (\ref{subsection:signQ}).
Numerically, this is achieved for $\ell>2.9$ ($\ell>3.2$) at $Y=6.0$ ($Y=7.5$) in
a gluon jet at the Tevatron (LHC). For a quark jet, these values become
respectively $\ell>3.1$ ($\ell>3.3$) and one can check that they are close to
the approximated ones obtained in (\ref{subsection:signG}) and
(\ref{subsection:signQ}). 

One notices that correlations increase as the total energy (Y) increases (LHC $>$ TeV $>$ LEP-I).

\subsection{Asymptotic behavior of $\boldsymbol{{\cal C}_{g\,or\,q}}$}
%%%%%%%%%%%%%%%%%%%%%%%%%%%%%%%%%%%%%%%%%%%%%%%%%%%%%%%%%%%%%%%%%%%%%%%

\begin{figure}
\vbox{
\begin{center}
\epsfig{file=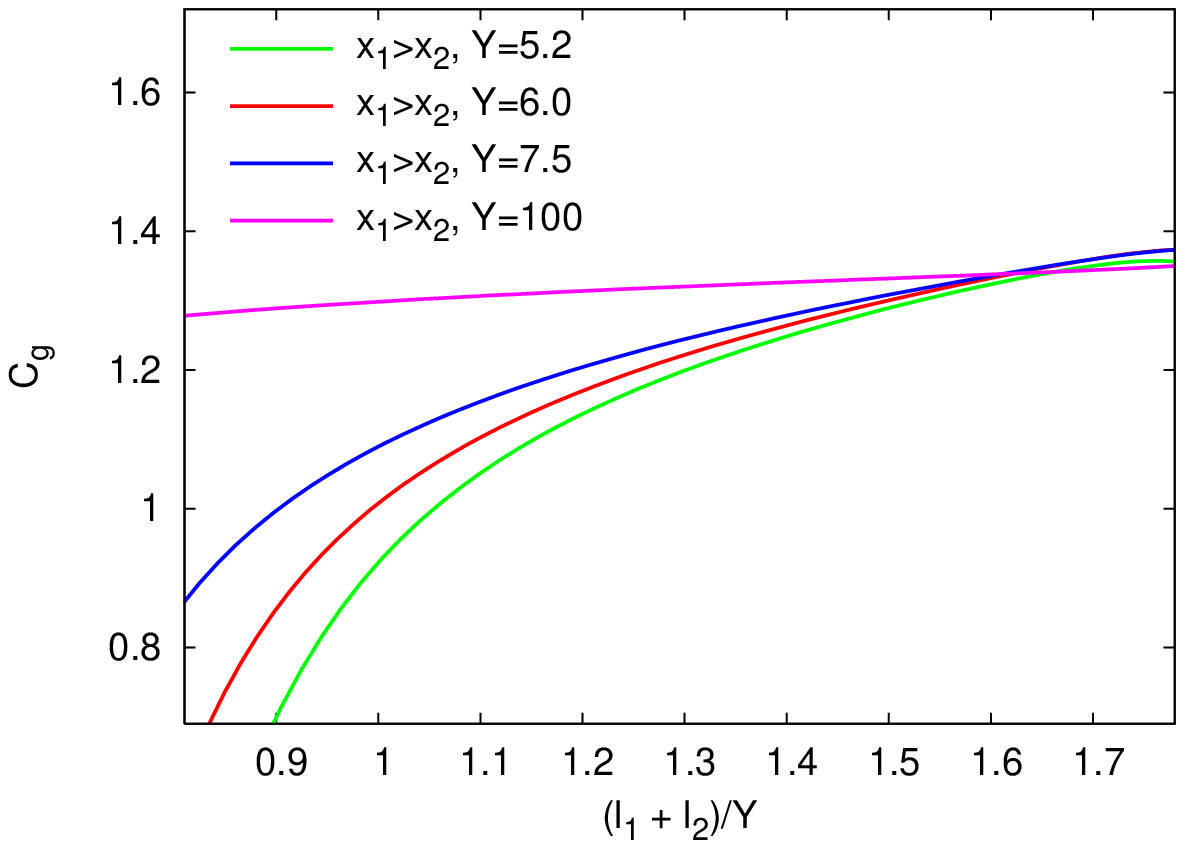, height=7truecm,width=0.48\tw}
\epsfig{file=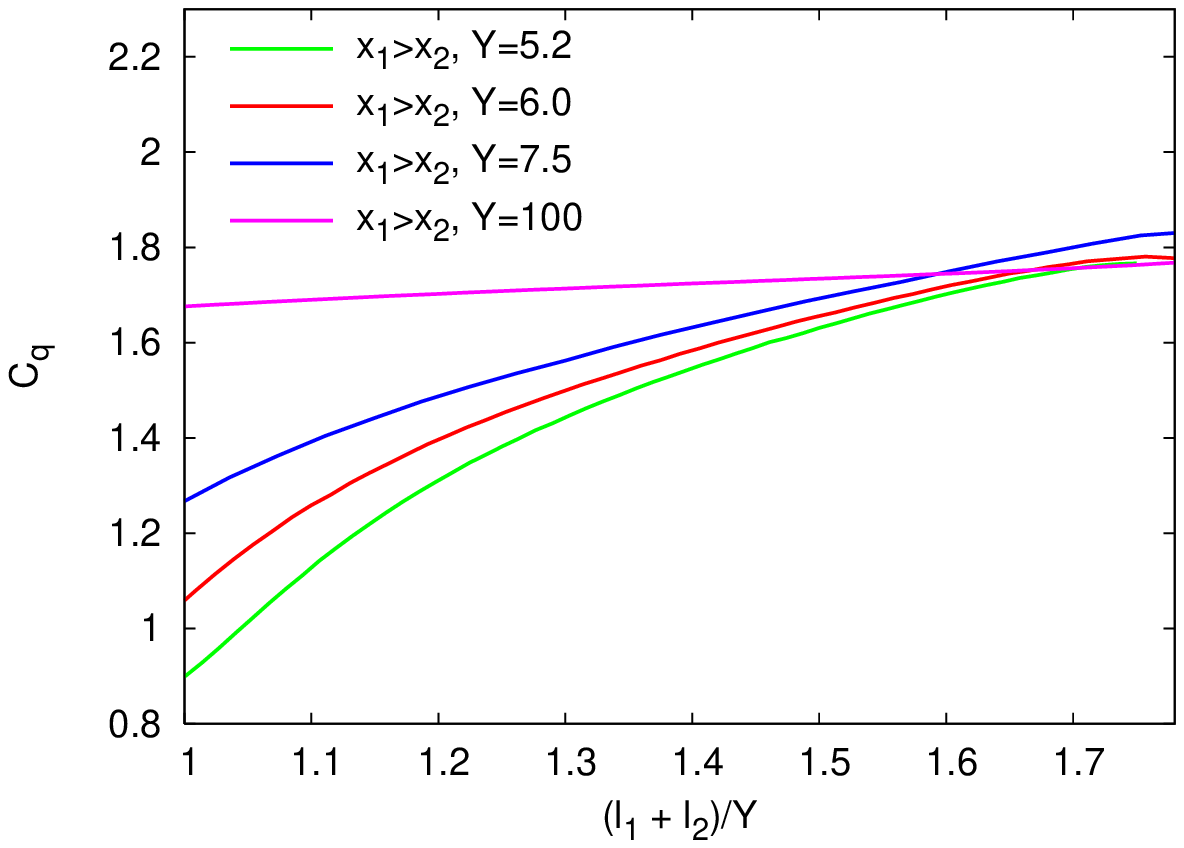, height=7truecm,width=0.48\tw}
\vskip .5cm
\caption{Asymptotic behavior of ${\cal C}_g$ and ${\cal C}_q$ when $Y$ increases}
\label{fig:corr0.1}
\end{center}
}
\end{figure}

We display in Fig.~\ref{fig:corr0.1}  the asymptotic behavior of
${\cal C}_g$ and ${\cal C}_q$ when $Y$ increases. 
$$
{\cal C}_g\stackrel{Y\to\infty}{\longrightarrow}\frac{<n(n-1)>_g}{<n>_g^2}\approx1+\frac13
\approx1.33,\quad 
{\cal C}_q\stackrel{Y\to\infty}{\longrightarrow}\frac{<n(n-1)>_q}{<n>_q^2}\approx1+\frac13
\frac{N_c}{C_F}=1.75,
$$
where $n$ is the multiplicity inside one jet. These limits coincide with
those of the DLA multiplicity correlator \cite{MWC}\cite{DFKC}.
It confirms the consistency of our approach.
%

%%%%%%%%%%%%%%%%%%%%%%%%%%%%%%%%%%%%%%%%%%%%%%%%%%%%%%%%%%%%%%%%%%%%%%%%%%%%%
\section{CONCLUSION}
\label{section:conclusion}
%%%%%%%%%%%%%%%%%%%%%%%%%%%%%%%%%%%%%%%%%%%%%%%%%%%%%%%%%%%%%%%%%%%%%%%%%%%%%
In this paper two particle correlations between soft partons in quark
and gluon jets were considered.

Corresponding evolution equations for parton correlators were derived in
the next to leading approximation of perturbative QCD, known as MLLA,
which accounts for QCD coherence (angular ordering) on soft gluon
multiplication, hard corrections to parton splittings and the running
coupling effects.

The MLLA equations for correlators were analyzed and solved
iteratively. This allowed us to generalize the result previously
obtained by Fong and Webber in \cite{FWC} that was valid in the vicinity
of the maximum of the single inclusive parton energy distribution
("hump").

In particular, we have analyzed the regions of moderately small $x$
above which the correlation becomes "negative" (${\cal C}-1<0$). This happens
when suppression because of the limitation of the phase space takes over
the positive correlation due to gluon cascading.

Also, the correlation vanishes (${\cal C}\to1$) when one of the partons
becomes very soft ($\ell = \ln1/x \to Y=\ln E\Theta/Q_0$). The reason
for that is dynamical rather than kinematical: radiation of a soft
gluon occurs at {\em large angles}\/ which makes the radiation
coherent and thus insensitive to the internal parton structure of the
jet ensemble.

Qualitatively, our MLLA result agrees better with available OPAL data
than the Fong--Webber prediction. There remains however a significant
discrepancy, markedly at very small $x$. In this region
non-perturbative effects are likely to be more pronounced. They may
undermine the applicability {\em to particle correlations}\/ of the local
parton--hadron duality considerations that were successful in
translating parton level predictions to hadronic observations in the
case of more inclusive {\em single particle energy spectra}.

Forthcoming data from Tevatron as well as future studies at LHC should
help to elucidate the problem.

\vskip 1cm

{\underline{\em Acknowledgments}}: It is a great pleasure to thank Yuri Dokshitzer and Bruno Machet for their guidance and encouragements. I thank Fran\c cois Arl\'eo, Bruno Durin for many discussions and Gavin Salam for his expert help in numerical calculations.

%%%%%%%%%%%%%%%%%%%%%%%%%%%%%%%%%%%%%%%%%%%%%%%%%%%%%%%%%%%%%%%%%%%%%%%%%%%%%

\newpage
%\appendix

{\Large\bf APPENDIX}

%%%%%%%%%%%%%%%%%%%%%%%%%%%%%%%%%%%%%%%%%%%%%%%%%%%%%%%%%%%%%%%%%%%%%%%%%%%%%
\section{DERIVATION OF THE GLUON CORRELATOR $\boldsymbol{{\cal C}_g}$ IN
(\ref{eq:CGfull})}
\label{section:Gcorr}
%%%%%%%%%%%%%%%%%%%%%%%%%%%%%%%%%%%%%%%%%%%%%%%%%%%%%%%%%%%%%%%%%%%%%%%%%%%%%

One differentiates $G^{(2)} -G_1G_2 \equiv G_1G_2\big({\cal C}_g-1\big)$ 
 with respect to $\ell_1$ and $y_2$  and use the
evolution equations (\ref{eq:solgC}) and (\ref{eq:eveeqgluC}).
 
By explicit differentiation and using the definitions
(ref{eq:nota4bis})-(\ref{eq:nota4C}) one gets
\begin{eqnarray}
\Big[ G_1G_2\big({\cal C}_g-1\big)\Big]_{\ell\,y}\!\!&\!\!=\!\!&\!\!
G_1G_2\Big[{\cal C}_{g,\ell\,y}+{\cal
C}_{g,\ell}\big(\psi_{1,y}+\psi_{2,y}\big) +{\cal
C}_{g,y}\big(\psi_{1,\ell}+\psi_{2,\ell}\big)\Big]\cr
 \!\!&\!\!+\!\!&\!\!({\cal
C}_g-1)\Big[G_1G_2\big(\psi_{1,\ell}\psi_{2,y}+\psi_{2,\ell}\psi_{1,y}\big)
+G_1G_{2,\ell\,y} +G_2G_{1,\ell\,y}\Big];\cr
&&
\label{eq:gc2}
\end{eqnarray}
the definition (\ref{eq:nota4bis}) of $\chi$ entails
${\cal C}_{g,\ell}=\chi_{\ell}{\cal C}_g$,
${\cal C}_{g,y}=\chi_{y}{\cal C}_g$,
${\cal C}_{g,\ell\,y} = {\cal C}_g\big(\chi_{\ell\,y} +
\chi_{\ell}\chi_{y}\big)$, such that (\ref{eq:gc2}) rewrites
\begin{eqnarray}
\Big[G^{(2)} -G_1G_2\Big]_{\ell\,y} \!\!&\!\!=\!\!&\!\!
 {\cal C}_g\;G_1G_2 \Big[
\big(\chi_{\ell\,y} + \chi_{\ell}\chi_{y}\big) 
+ \chi_{\ell}\big(\psi_{1,y}+\psi_{2,y}\big)
+  \chi_{y}\big(\psi_{1,\ell}+\psi_{2,\ell}\big)
\Big]\cr
\!\!&\!\!+\!\!&\!\!({\cal C}_g-1)\Big[
 G_1G_2 \big( \psi_{1,\ell}\psi_{2,y}+\psi_{1,y}\psi_{2,\ell} \big)
 + G_{1} G_{2,\ell\,y}+ G_2G_{1,\ell\,y}
 \Big].\cr
&&
\label{eq:gc3}
\end{eqnarray}
By differentiating the evolution equation for the inclusive spectra
(\ref{eq:solgC}) with respect to $y$ and $\ell$ one gets 
\begin{equation}
  G_{k,\ell\,y} =  \gamma_0^2
  \Big(1-a\big(\psi_{k,\ell}- {\beta}\gamma_0^2\big)\Big) G_k,
\label{eq:gc4}
\end{equation}
where one has used the definition (\ref{eq:psi1})(\ref{eq:psi2}) of
$\psi_{k,\ell}$ to replace 
$\frac{dG_k}{d\ell}$ with $G_k \psi_{k,\ell}$, and (\ref{eq:gammabeta})
to evaluate $\frac{d}{d\ell}\gamma_0^2 = -\beta \gamma_0^4$.
Substituting into (\ref{eq:gc3}) yields
\begin{eqnarray}
&&\hskip -1cm \frac{l.h.s. (\ref{eq:eveeqgluC})\big\vert_{\ell\,y}} {\gamma_0^2G_1G_2 }=
({\cal C}_g-1)\bigg( 2- a\left(\psi_{1,\ell}+\psi_{2,\ell}\right) +
\frac{ \psi_{1,\ell}\psi_{2,y}+\psi_{1,y}\psi_{2,\ell}}
{\gamma_0^{2}} + 2a {\beta}\gamma_0^2 \bigg)
+  {\cal C}_g\big(\delta_1+\delta_2\big), \cr
&&
\label{eq:gc5}
\end{eqnarray}
where $\delta_1$ and $\delta_2$ are defined in (\ref{eq:delta1C})
(\ref{eq:nota4C}).

Differentiating now the r.h.s. of (\ref{eq:eveeqgluC}) with respect to $y_2$
and $\ell_1$, one gets
\begin{eqnarray}
\frac{r.h.s. (\ref{eq:eveeqgluC})\Big\vert_{\ell\,y}}
{\gamma_0^2\,G_1G_2 }=
{\cal C}_g \big( 1-a\left(\psi_{1,\ell} +\psi_{2,\ell} 
    -{\beta}\gamma_0^2\right)\Big)
- {\cal C}_g a\chi_{\ell}
 + (a-b)\left(\psi_{1,\ell} +\psi_{2,\ell}- {\beta}\gamma_0^2 \right).  \cr
&&
\label{eq:rhsdiff2} 
\end{eqnarray}
Equating the expressions (\ref{eq:gc5}) and (\ref{eq:rhsdiff2})
for the correlation function we derive 
\begin{eqnarray} 
({\cal C}_g-1)
\Big(1+\Delta + \delta_1
+ a\left(\chi_{\ell} + {\beta\gamma_0^2}\right) + \delta_2\Big)
= 1-b\big(\psi_{1,\ell} +\psi_{2,\ell} - \beta\gamma_0^2 \big)
-\delta_1 - (a\chi_{\ell}+ \delta_2),\cr
&&
\end{eqnarray}
which gives (\ref{eq:CGfull}).

%%%%%%%%%%%%%%%%%%%%%%%%%%%%%%%%%%%%%%%%%%%%%%%%%%%%%%%%%%%%%%%%%%%%%%%%%%%%%
\section{DERIVATION OF THE QUARK CORRELATOR $\boldsymbol{{\cal C}_q}$ IN
(\ref{eq:Qcorr})}
\label{section:Qcorr}
%%%%%%%%%%%%%%%%%%%%%%%%%%%%%%%%%%%%%%%%%%%%%%%%%%%%%%%%%%%%%%%%%%%%%%%%%%%%%

\subsection{Derivation of (\ref{eq:Qcorr})}
\label{subsection:deriveq}
%%%%%%%%%%%%%%%%%%%%%%%%%%%%%%%%%%%%%%%

The method is the same as in appendix \ref{section:Gcorr}:
one evaluates now $\big[Q^{(2)} -{Q_1}{Q_2}\big]_{\ell\,y}
\equiv \Big[{\cal C}_q-1)Q_1Q_2\Big]_{\ell\,y}$.

First, by differentiating the evolution equation (\ref{eq:eveeqqC}), one gets
\begin{equation}
\big[Q^{(2)} -{Q_1}{Q_2}\big]_{\ell\,y} =
\frac{C_F}{N_c}\gamma_0^2 {\cal C}_g G_1G_2 
\Big(1-\frac34
\big(\psi_{1,\ell}+\psi_{2,\ell}+\chi_{\ell}-\beta\gamma_0^2\big)\Big);
\label{eq:rhsQ}
\end{equation}
then, one explicitly differentiates $\Big[({\cal C}_q-1)Q_1Q_2\Big]$ and
makes use of
\begin{equation}
Q_{k,\ell\,y}=\frac{C_F}{N_c}\gamma_0^2 G_k\Big(1-\frac34(\psi_{k,\ell}
-\beta\gamma_0^2)\Big),
\label{eq:relQ}
\end{equation}
which comes directly from differentiating the r.h.s of (\ref{eq:solqC})
with respect to $\ell$ and $y$; this yields
\begin{eqnarray}
\big[Q^{(2)} -{Q_1}{Q_2}\big]_{\ell\,y}\!\!&\!\!=\!\!&\!\!
{\cal C}_q Q_1Q_2\Big[\sigma_{\ell}\big(\varphi_{1,y} + \varphi_{2,y}\big)
+\sigma_{\ell}\big(\varphi_{1,\ell}+\varphi_{2,\ell}\big) +\sigma_{\ell\,y}
+\sigma_{\ell}\sigma_{y}\Big]\cr
&&+ \big({\cal C}_q-1\big)Q_1Q_2\gamma_0^2
\Big[\varphi_{1,\ell}\varphi_{2,y}+\varphi_{1,y}\varphi_{2,\ell}\Big]\cr
&&\hskip -2cm +\big({\cal C}_q-1\big)\gamma_0^2\frac{C_F}{N_c}\Big[
\big(G_1Q_2+Q_1G_2\big)
-\frac34 G_1Q_2\big(\psi_{1,\ell}-\beta\gamma_0^2\big)
-\frac34 Q_1G_2\big(\psi_{2,\ell}-\beta\gamma_0^2\big)\Big];\cr
&&
\label{eq:explidif}
\end{eqnarray}
equating (\ref{eq:rhsQ}) and (\ref{eq:explidif}) gives
\begin{eqnarray}
{\cal C}_q-1 = \frac{\frac{N_c}{C_F}{\cal C}_g
\bigg[ 1-\frac34\Big(
\psi_{1,\ell}+\psi_{2,\ell}+\chi_{\ell}-\beta\gamma_0^2 \Big)\bigg]
\frac{C_F}{N_c}\frac{G_1}{Q_1} \frac{C_F}{N_c}\frac{G_2}{Q_2}
-{\cal C}_q\big(\tilde\delta_1 +\tilde\delta_2\big)}
{ \widetilde\Delta +\Big[1-\frac34\big(\psi_{1,\ell}-\beta\gamma_0^2\big)\Big]
\frac{C_F}{N_c}\frac{G_1}{Q_1}
+\Big[1-\frac34\big(\psi_{2,\ell}-\beta\gamma_0^2\big)\Big]
\frac{C_F}{N_c}\frac{G_2}{Q_2}},
\label{eq:Qrel}
\end{eqnarray}
which leads (\ref{eq:Qcorr}).

\subsection{Expressing $\boldsymbol{\widetilde\Delta}$,
$\boldsymbol{\tilde\delta_1}$ and
$\boldsymbol{\tilde\delta_2}$ in terms of gluon-related quantities}
\label{subsection:corrections}
%%%%%%%%%%%%%%%%%%%%%%%%%%%%%%%%%%%%%%%%%%%%%%%%%%%%%%%%%%%%%%%%

All the intricacies of (\ref{eq:Qrel}) lie in $\widetilde\Delta$,
$\tilde\delta_1$ and $\tilde\delta_2$ defined in (\ref{eq:nota5}),
which involve
the quark related quantities $\sigma$ and $\varphi$ (\ref{eq:phisigma}).
In what follows, we will express them in terms of the gluon related
quantities $\chi$ and $\psi$
(\ref{eq:nota4bis})(\ref{eq:psi1})(\ref{eq:psi2}).

\subsubsection{Expression for $\boldsymbol{\tilde\Delta}$}
%%%%%%%%%%%%%%%%%%%%%%%%%%%%%%%%%%%%%%%%%%%%%%%%%%%%%%%%%

Differentiating (\ref{eq:ratio}) with respect to $\ell$
yields
\begin{equation}
Q_{k,\ell}=\frac{C_F}{N_c}G_{k,\ell}\bigg[1+\Big(a-\frac34\Big)\psi_{k,\ell}\bigg]+
\frac{C_F}{N_c}G_k\bigg(a-\frac34\bigg)\psi_{k,\ell\,\ell} + {\cal
O}(\gamma_0^4);
\end{equation}
then  
\begin{equation}
\varphi_{\ell}=\frac{Q_{k,\ell}}{Q_k}=\left\{\frac{C_F}{N_c}G_{k,\ell}
\Big[1+\Big(a-\frac34\Big)\psi_{k,\ell}\Big]+
\frac{C_F}{N_c}G_k\Big(a-\frac34\Big)\psi_{k,\ell\,\ell}\right\}\Big[G_{k}^{-1}-
\Big(a-\frac34\Big)\psi_{k,\ell}G_{k}^{-1}\Big]
\end{equation}
yields
\begin{equation}
\varphi_{k,\ell} = \psi_{k,\ell} + \Big(a-\frac34\Big)\psi_{k,\ell\,\ell}
+{\cal O}(\gamma_0^4).
\label{eq:varphil}
\end{equation}

Differentiating (\ref{eq:ratio}) with respect to $y$ yields
\begin{equation}
Q_{k,y}=\frac{C_F}{N_c}G_{k,y}\bigg[1+\Big(a-\frac34\Big)\psi_{k,\ell}\bigg]+
\frac{C_F}{N_c}G_k\bigg(a-\frac34\bigg)\psi_{k,\ell\,y} + {\cal
O}(\gamma_0^4),
\end{equation}
and, finally,
\begin{equation}
\varphi_{k,y} = \psi_{k,y} +\Big( a-\frac34\Big) \psi_{k,\ell\,y}
+ {\cal O}(\gamma_0^4).
\label{eq:varphiy2}
\end{equation}
Using (\ref{eq:varphil}) and (\ref{eq:varphiy2}) in
$\widetilde{\Delta}$ given by (\ref{eq:nota5}) gives

\begin{equation}
\widetilde\Delta \approx
\Delta +\Big(a-\frac34\Big)
\Big(\psi_{1,\ell\,y}\psi_{2,\ell}
+\psi_{2,\ell\,\ell}\psi_{1,y} +\psi_{2,\ell\,y}\psi_{1,\ell}
+\psi_{1,\ell\,\ell}\psi_{2,y} \Big)\gamma_0^{-2},
\label{eq:Deltatilde}
\end{equation}
which shows in particular, that
$\widetilde\Delta \approx \Delta +{\cal O}(\gamma_0^2)$.

\subsubsection{Expression for $\boldsymbol{\tilde\delta_1,
\tilde\delta_2}$}
%%%%%%%%%%%%%%%%%%%%%%%%%%%%%%%%%%%%%%%%%%%%%%%%%%%%%%%%%%%%%%%%%%%%%%%%%%%%

(\ref{eq:nota5}) entails
${\cal C}_q \gamma_0^2 \tilde\delta_1 = {\cal
C}_{q,\ell}\big(\varphi_{1,y}+\varphi_{2,y}\big) + {\cal
C}_{q,y}\big(\varphi_{1,\ell}+\varphi_{2,\ell}\big)$; since ${\cal
C}_{q,\ell}$ and ${\cal C}_{q,y}$ are ${\cal O}(\gamma_0^2)$ and
considering (\ref{eq:varphiy2}) and (\ref{eq:varphil}), we can
approximate
\begin{equation}
{\cal C}_q \gamma_0^2 \tilde\delta_1 = {\cal
C}_{q,\ell}\big(\psi_{1,y}+\psi_{2,y}\big) + {\cal
C}_{q,y}\big(\psi_{1,\ell}+\psi_{2,\ell}\big) +{\cal O}(\gamma_0^5),
\label{eq:delta1ap}
\end{equation}
which needs evaluating ${\cal C}_{q,\ell}$ and ${\cal C}_{q,y}$ in terms of
gluonic quantities.
Actually, since ${\cal C}_q\tilde\delta_1$ and ${\cal C}_q\tilde\delta_2$
occur as MLLA and NMLLA corrections in (\ref{eq:Qrel}),
it is enough to take the leading (DLA) term of ${\cal C}_q$ to estimate them
\begin{equation}
{\cal C}_q^{DLA}=1+\frac{N_c}{C_F}\frac1{1+\Delta}=1-\frac{N_c}{C_F}+\frac{N_c}{C_F}
\Big(1+\frac{1}{1+\Delta}\Big);
\end{equation}
differentiating then over $\ell$ and $y$ yields
\begin{eqnarray}\label{eq:cqell}
{\cal C}_{q,\ell}^{DLA}&=&-\frac{N_c}{C_F}
\frac{\Delta_{\ell}}{\Big(1+\Delta\Big)^2}=\frac{N_c}{C_F}{\cal
C}_{g,\ell}^{DLA},\\
\label{eq:cqy}
{\cal C}_{q,y}^{DLA}&=&-\frac{N_c}{C_F}
\frac{\Delta_{y}}{\Big(1+\Delta\Big)^2}=\frac{N_c}{C_F}{\cal C}_{g,y}^{DLA}.
\end{eqnarray}
Substituting (\ref{eq:cqell}), (\ref{eq:cqy}) into
(\ref{eq:delta1ap})
one gets
\begin{equation}
{\cal C}_q\tilde\delta_1={\cal C}_g\delta_1+{\cal {O}}(\gamma_0^3).
\end{equation}
Likewise, calculating $\gamma_0^2 {\cal C}_q \tilde\delta_2$ needs
evaluating ${\cal C}_{q,\ell\,y}^{DLA}$ in terms of gluonic quantities. Using
(\ref{eq:cqell}) one gets
\begin{equation}
{\cal C}_q\tilde\delta_2={\cal C}_g\delta_2+{\cal {O}}(\gamma_0^4).
\end{equation}
Accordingly, ${\cal C}_q(\tilde\delta_1+\tilde\delta_2)$ can be replaced by 
${\cal C}_g(\delta_1+\delta_2)$ to get the solution (\ref{eq:Qrel}). This 
approximation is used to get the MLLA solution (\ref{eq:QMLLAap}) of 
(\ref{eq:Qrel}).

\section{DLA INSPIRED  SOLUTION OF THE MLLA EVOLUTION EQUATIONS
FOR THE INCLUSIVE SPECTRUM}
\label{section:inspiredDLA}
%%%%%%%%%%%%%%%%%%%%%%%%%%%%%%%%%%%%%%%%%%%%%%%%%%%%%%%%%%%%%%%%

This appendix completes subsection \ref{subsection:estimate}.
For pedagogical reasons we will estimate the solution of 
(\ref{eq:solgC}) when neglecting the running of 
$\alpha_s$ (constant-$\gamma_0^2$) (see \cite{EvEqC}\cite{KOC} and references therein). 
We perform a Mellin's transformation of
$G(\ell,y)$

\begin{equation}
G\left(\ell,y\right)=\iint_{C}\frac{d\omega\, d\nu}
{\left(2\pi i\right)^2}\,e^{\omega \ell}
\,e^{\nu y}\,{\cal G}\left(\omega,\nu\right).
\label{eq:GG}
\end{equation}

The contour C lies to the right of all singularities. In (\ref{eq:solgC})
one set the lower
bounds for $\ell$ and $y$ to $-\infty$ since these integrals are
vanishing when one closes the C-contour to the right. Using the Mellin's representation
for $\delta(\ell)$

\begin{equation}
\delta(\ell)=\iint_C\frac{d\omega d\nu}
{\left(2\pi i\right)^2}\,e^{\omega \ell}\,e^{\nu y}\frac{1}{\nu},
\label{eq:deltaell}
\end{equation}

one gets

\begin{equation}\label{eq:GSMellin}
{\cal {G}}\left(\omega,\nu\right)=\displaystyle{\frac{1}
{\nu-\gamma_0^2\big(1/\omega-a\big)}}.
\end{equation}

Substituting (\ref{eq:GSMellin}) into (\ref{eq:GG}) and extracting the pole 
($\nu_0=\gamma_0^2\big(1/\omega-a\big)$)
from the denominator of (\ref{eq:GSMellin}) one gets rid of the integration over $\nu$
and obtains the following representation
\footnote{by making use of Cauchy's theorem.}

\begin{equation}
G\left(\ell,y\right)=\int_{C}\frac{d\omega}{2\pi i}\exp{\Big[\omega\ell+\gamma_0^2\big(1/\omega-a\big) y\Big]};
\label{eq:naiveG}
\end{equation}

finally treating $\ell$ as a large variable (soft approximation $x\ll1$)
allows us to have an estimate of (\ref{eq:naiveG}) by performing the steepest
descent method; one then has

\begin{equation}
G(\ell,y)\stackrel{x\ll1}{\approx}
\frac12\sqrt{\frac{\gamma_0\,y^{1/2}}{{\pi\,\ell^{3/2}}}}
\exp{\left(2\gamma_0\sqrt{\ell\,y}-a\gamma_0^2\,y\right)}.
\label{eq:naivesol}
\end{equation}

However, since we are interested in getting logarithmic derivatives;
in this approximation we can drop the normalization factor of (\ref{eq:naivesol})
which leads to sub-leading corrections that we do not take into account here;
we can use instead

\begin{equation}
G(\ell,y)\stackrel{x\ll1}{\simeq}
\exp{\left(2\gamma_0\sqrt{\ell\,y}-a\gamma_0^2\,y\right)},
\label{eq:naivesolbis}
\end{equation}

which is (\ref{eq:Gmod}).

%%%%%%%%%%%%%%%%%%%%%%%%%%%%%%%%%%%%%%%%%%%%%%%%%%%%%%%%%%%%%%%%%%%%%%%%%%%%%
\section{EXACT SOLUTION OF THE MLLA EVOLUTION EQUATION FOR THE
INCLUSIVE SPECTRUM}
\label{section:ESEE}
%%%%%%%%%%%%%%%%%%%%%%%%%%%%%%%%%%%%%%%%%%%%%%%%%%%%%%%%%%%%%%%%%%%%%%%%%%%%%

We  solve  (\ref{eq:solgC}) by performing a
Mellin's transformation of the following function ($\gamma_0^2$ 
, $\beta$ and $\lambda$ are defined in (\ref{eq:gammabeta}), (\ref{eq:betaC})):

\begin{equation*}
F\left(\ell,y\right)=\gamma_0^2(\ell+y)G\left(\ell,y\right),
\end{equation*}

that is,

\begin{equation}\label{eq:MellRep}
F\left(\ell,y\right)=\iint_{C}\frac{d\omega d\nu}{\left(2\pi
    i\right)^2}\,e^{\omega \ell}\,
 e^{\nu y}\,{\cal F}\left(\omega,\nu\right).
\end{equation}

Substituting (\ref{eq:MellRep}) into (\ref{eq:solgC}) we 
obtain:
\begin{eqnarray*}
{\beta}\left(\ell+y+\lambda\right)\iint\frac{d\omega d\nu}{\left(2\pi i\right)^2}\,e^{\omega \ell}\,e^{\nu y}\,{\cal F}\left(\omega,\nu\right)\!\!&\!\!=\!\!&\!\!\iint\frac{d\omega d\nu}{\left(2\pi i\right)^2}\,e^{\omega \ell}\,e^{\nu y}\left[\frac{1}{\nu}+\frac{{\cal F}\left(\omega,\nu\right)}{\omega\nu}\right]\\\nonumber\\
\!\!&\!\!-\!\!&\!\!a\!\!\iint\frac{d\omega d\nu}{\left(2\pi i\right)^2}\,e^{\omega \ell}\,e^{\nu y}\frac{{\cal F}(\omega,\nu)}{\nu},
\end{eqnarray*}

where we have again replaced $\delta(\ell)$ by its Mellin's representation (\ref{eq:deltaell}).
Then using the equivalence $\ell\leftrightarrow\frac{\partial}{\partial\omega},\, y\leftrightarrow\frac{\partial}{\partial\nu}$, we integrate the l.h.s. by parts and obtain:

\begin{eqnarray*}
&&\beta\!\!\iint\frac{d\omega d\nu}{\left(2\pi i\right)^2}\left[\left(\frac{\partial}{\partial\omega}\!+\!\frac{\partial}
{\partial\nu}\!+\!\lambda\right)e^{\omega\ell+\nu y}\right]{\cal F}\left(\omega,\nu\right)
=\beta\!\!\iint\frac{d\omega d\nu}{\left(2\pi i\right)^2}
\left(\lambda {\cal F}\!-\!\frac{\partial {\cal F}}
{\partial\omega}\!-\!\frac{\partial {\cal F}}
{\partial\nu}\right)\,e^{\omega\ell+\nu y}.
\end{eqnarray*}

We are finally left with the following inhomogeneous differential equation:

\begin{equation}\label{eq:diffeq}
\beta\left(\lambda {\cal F}\!-\!\frac{\partial {\cal F}}{\partial\omega}\!-\!\frac{\partial {\cal F}}{\partial\nu}\right)=\frac{1}{\nu}\!+\!\frac{{\cal F}}{\omega\nu}-a\frac{{\cal F}}{\nu}.
\end{equation}

The variables $\omega$ and $\nu$ can be changed conveniently to

$$
\omega'=\frac{\omega+\nu}{2},\qquad \nu'=\frac{\omega-\nu}{2},
$$

such that (\ref{eq:diffeq}) is now decoupled and can be 
easily solved:

\begin{equation*}
\beta\left(\lambda {\cal F}-\frac{d{\cal F}}{d\omega'}\right)=\frac{1}{\omega'-\nu'}+\frac{{\cal F}}{\omega'^2-\nu'^2}
-a\frac{{\cal F}}{\omega'-\nu'}.
\end{equation*}
The solution of the corresponding homogeneous equation, written as a function of
$\omega$ and $\nu$, is the following:

\begin{equation*}
{\cal F}^h\left(\omega,\nu\right)=\frac1{\beta}\int_{0}^{\infty}
\frac{ds}{\nu+s}\left(\frac{\omega\left(\nu+s\right)}
{\left(\omega+s\right)\nu}\right)^{1/\beta\left(\omega-\nu\right)}
\left(\frac{\nu}{\nu+s}\right)^{a/\beta}.
\end{equation*}

We finally obtain the exact solution of (\ref{eq:solgC}) given by the following 
Mellin's representation:

\begin{equation}\label{eq:MLLAalphasrunC}
G\left(\ell,y\right)=\left(\ell\!+\!y\!+\!\lambda\right)\!\!\iint\frac{d\omega\, d\nu}
{\left(2\pi i\right)^2}e^{\omega\ell+\nu y}
\!\!\int_{0}^{\infty}\frac{ds}{\nu+s}\!\!
\left(\frac{\omega\left(\nu+s\right)}
{\left(\omega+s\right)\nu}\right)^{1/\beta\left(\omega-\nu\right)}\!\!
\left(\frac{\nu}{\nu+s}\right)^{a/\beta}\,e^{-\lambda s}.
\end{equation}

(\ref{eq:MLLAalphasrunC}) will be estimated using 
the steepest descent method in a forthcoming work that will treat two particles
correlations at $Q_0\geq\Lambda_{QCD}$
($\lambda=\ln(Q_0/\Lambda_{QCD})\ne0$) \cite{RPR3}\cite{these}.
Substituting (\ref{eq:MLLAalphasrunC}) into (\ref{eq:eveeqgluC}) one has the Mellin's 
representation inside a quark jet

\begin{eqnarray*}
Q(\ell,y)\!\!=\!\!(\ell\!+\!y\!+\!\lambda)\!\!\iint\frac{d\omega\, d\nu}
{\left(2\pi i\right)^2}e^{\omega\ell+\nu y}
\left(\frac{\gamma_0^2}{\omega\nu}-\frac34\frac{\gamma_0^2}{\nu}\right)
\!\!\int_{0}^{\infty}\frac{ds}{\nu+s}\!\!
\left(\!\frac{\omega\left(\nu+s\right)}
{\left(\omega+s\right)\nu}\!\right)^{1/\beta\left(\omega-\nu\right)}\!\!
\left(\frac{\nu}{\nu+s}\right)^{a/\beta}\!\!e^{-\lambda s};
\end{eqnarray*}
where $\gamma_0^2/\omega\nu={\cal O}(1)$ and the second term is the MLLA 
correction $\gamma_0^2/\nu={\cal O}(\gamma_0)$.

\subsection{Limiting Spectrum, $\boldsymbol{\lambda=0}$}
%%%%%%%%%%%%%%%%%%%%%%%%%%%%%%%%%%%%%%%%%%%%%%%%%%%%%%%%

We set $\lambda=0$ (that is $Q_0=\Lambda_{QCD}$) in (\ref{eq:MLLAalphasrunC})
and change variables as follows

$$
\bar{\omega}=\omega-\nu,\quad 
s+\bar{\omega}t=\bar{\omega}/u,\quad A\equiv A(\bar{\omega})=\frac1{\beta\bar{\omega}},\quad
B=a/\beta
$$
to get ($\ell+y=Y$ is used as a variable)

\begin{equation}
    G\left(\ell,Y\right)=\int_{\epsilon_1-i\infty}^{\epsilon_1+i\infty}
    \frac{d\bar{\omega}}{2\pi i}
    x^{-\bar{\omega}}\bar{\omega}Y\int_{\epsilon_2-i\infty}^{\epsilon_2+i\infty}
    \frac{dt}{2\pi i}\,e^{\bar{\omega}Yt}\left(\frac{t}{1+t}\right)^{-A}t^{B}
    \int_{0}^{t^{-1}}\,du\,u^{B-1}\left(1+u\right)^{-A};
\label{eq:red12}
\end{equation}
the last integral of (\ref{eq:red12}) is the representation of the
hypergeometric functions of the second kind (see ~\cite{GRC})
\begin{equation*}
 \int_{0}^{t^{-1}}\,du\,u^{B-1}\left(1+u\right)^{-A}
 =\frac{t^{-B}}{B}\,_2F_1\left(A,B;B+1;-t^{-1}\right);
\end{equation*}

for $\Re B>0$, we also have

\begin{equation*}
_2F_1\left(a,b;c;x\right)=\sum_{n=0}^{\infty}
 \frac{\left(a\right)_n\left(b\right)_nx^n}{\left(c\right)_n n!},
\end{equation*}
where for example
$$
\left(a\right)_n=\frac{\Gamma\left(a+n\right)}{\Gamma\left(a\right)}=a\left(a+1\right)
...\left(a+n-1\right).
$$
Therefore (\ref{eq:red12}) can be rewritten in the form:

\begin{eqnarray}\label{eq:red13}
G\left(\ell,Y\right)=
\frac{Y}{B}\int_{\epsilon_1-i\infty}^{\epsilon_1+i\infty}
\frac{d\bar{\omega}}{2\pi  i}
x^{-\bar{\omega}}\bar{\omega}\int_{\epsilon_2-i\infty}^
{\epsilon_2+i\infty}\frac{dt}{2\pi i}\,
 e^{\bar{\omega}Yt}\left(\frac{t}{1+t}\right)^{-A}\,
 _2F_1\left(A,B;B+1;-t^{-1}\right).\cr
&&
\end{eqnarray}

\vskip 0.5cm

By making use of the identity \cite{SDPC}:
\begin{eqnarray}
 \left(1+t^{-1}\right)\,_2F_1\left(-A+B+1,1;B+1;-t^{-1}\right)
=\left(\frac{t}{1+t}\right)^{-A}\,_2F_1\left(A,B;B+1;-t^{-1}\right),\notag
\end{eqnarray}
we split (\ref{eq:red13}) into two integrals. The solution of the
second one is given by the hypergeometric function of the first kind
~\cite{SDPC}:

\begin{eqnarray}
 \int_{\epsilon_2-i\infty}^{\epsilon_2+i\infty}\frac{dt}{2\pi i}
 \,e^{\bar{\omega}Yt}\,
 t^{-1}\,_2F_1\left(-A+B+1,1;B+1;-t^{-1}\right)
\!=\!_1F_1\left(-A+B+1;B+1;-\bar{\omega}Y\right).\cr
&&
\label{eq:red14}
\end{eqnarray}

Taking the derivative of (\ref{eq:red14}) over $(\bar{\omega}Y)$ we
obtain:

\begin{eqnarray*}
\int_{\epsilon_2-i\infty}^{\epsilon_2+i\infty}
 \frac{dt}{2\pi i}\,e^{\bar{\omega}Yt}\,
 _2F_1\left(-A+B+1,1;B+1;-t^{-1}\right)
\!=\!-\frac{d}{d\left(-\bar{\omega}Y\right)}
 \,_1F_1\left(-A+B+1;B+1;-\bar{\omega}Y\right),
\end{eqnarray*}

where,

\begin{equation*}
 _1F_1\left(a;b;x\right)\equiv\Phi\left(a;b;x\right)=\sum_{n=0}^{\infty}
 \frac{\left(a\right)_nx^n}{\left(b\right)_n n!}.
\end{equation*}

We finally make use of the identity ~\cite{SDPC}:

\begin{eqnarray*}
_1F_1\left(-A+B+1;B+2;-\bar{\omega}Y\right)\!\!&\!\!=\!\!&\!\!\frac{B+1}{A}
 \Big[\,_1F_1\left(-A+B+1;B+1;-\bar{\omega}Y\right)\Big.\\
\!\!&\!\!-\!\!&\!\!\Big.\frac{d}{d\left(-\bar{\omega}Y\right)}\,
 _1F_1\left(-A+B+1;B+1;-\bar{\omega}Y\right)\Big]
\end{eqnarray*}

to get ($_1F_1\equiv\Phi$):

\begin{equation}\label{eq:hyprep}
G\left(\ell,Y\right)=\frac{Y}{\beta B\left(B+1\right)}
 \int_{\epsilon-i\infty}^{\epsilon+i\infty}\frac{d\bar{\omega}}{2\pi  i}
 x^{-\bar{\omega}}\Phi\left(-A+B+1,B+2,-\bar{\omega} Y\right);
\end{equation}

we can rename $\bar{\omega}\rightarrow\omega$ and set $Y=\ell+y$, 
which yields

\begin{eqnarray}\nonumber
G\left(\ell,y\right)\!\!&\!\!=\!\!&\!\!\frac{\ell+y}{\beta B\left(B+1\right)}
 \int_{\epsilon-i\infty}^{\epsilon+i\infty}\frac{d{\omega}}{2\pi  i}
 x^{-{\omega}}\Phi\left(-A+B+1,B+2,-{\omega} (\ell+y)\right)\\\nonumber\\
\!\!&\!\!=\!\!&\!\!\frac{\ell+y}{\beta B\left(B+1\right)}
 \int_{\epsilon-i\infty}^{\epsilon+i\infty}\frac{d{\omega}}{2\pi  i}
 e^{-{\omega}y}\Phi\left(A+1,B+2,{\omega} (\ell+y)\right).\label{eq:hyprep1}
\end{eqnarray}

We thus demonstrated that the integral representation
(\ref{eq:MLLAalphasrunC}) 
is equivalent to (\ref{eq:hyprep}) in the limit $\lambda=0$. In this problem
all functions are derived 
using  (\ref{eq:hyprep1}), and one fixes
the value of $Y=\ln(Q/Q_0)$ (that is fixing the hardness of the process under 
consideration), such that each result is presented as a function of the 
energy fraction in the logarithmic scale $\ell=\ln(1/x)$. As demonstrated
in \cite{EvEqC}
\cite{PerezMachetC}, the inclusive spectrum can be obtained using (\ref{eq:hyprep}) and
the result is

\begin{equation}
G(\ell,Y) = 2\frac{\Gamma(B)}{\beta}
\Re\left( \int_0^\frac{\pi}{2}
  \frac{d\tau}{\pi}\, e^{-B\alpha}\  {\cal F}_B(\tau,y,\ell)\right),
\label{eq:ifDC}
\end{equation}
where the integration is performed with respect to $\tau$ defined by
$\displaystyle \alpha = \frac{1}{2}\ln\left(\frac{Y}{\ell}-1\right)  + i\tau$,
\begin{eqnarray}
{\cal F}_B(\tau,\ell,Y) \!\!&\!\!=\!\!&\!\! \left[ \frac{\cosh\alpha
-\displaystyle{\left(1-\frac{2\ell}{Y}\right)}
\sinh\alpha} 
 {\displaystyle \frac{
Y}{\beta}\,\frac{\alpha}{\sinh\alpha}} \right]^{B/2}
  I_B\Big(2\sqrt{Z(\tau,\ell,Y)}\Big), \cr
&& \cr
&& \cr
 Z(\tau,\ell,Y) \!\!&\!\!=\!\!&\!\!
\frac{Y}{\beta}\,
\frac{\alpha}{\sinh\alpha}\,
 \left[\cosh\alpha
%+ (1-2\zeta)
-\left(1-\frac{2\ell}{Y}\right)
\sinh\alpha\right]; 
\label{eq:calFdefC}
\end{eqnarray}
$I_B$ is the modified Bessel function of the first kind.

\subsection{Logarithmic derivatives of the spectrum, $\boldsymbol{\lambda=0}$}
\label{subsection:Logder}
%%%%%%%%%%%%%%%%%%%%%%%%%%%%%%%%%%%%%%%%%%%%%%%%%%%

Using the expressions derived in \cite{PerezMachetC} and fixing the sum $\ell+y=Y$,
one gets
\begin{equation}
\frac{d}{d\ell}  G\left(\ell,Y\right)
 = 2\frac{\Gamma(B)}{\beta} \int_0^{\frac{\pi}2}\frac{d\tau}{\pi}\,
 e^{-B\alpha}
 \left[\frac1{Y}\left(1+2e^{\alpha}
\sinh{\alpha}\right){\cal{F}}_B
+\frac1{\beta}e^{\alpha}{\cal{F}}_{B+1}\right];
\label{eq:derivlC}
\end{equation}
and
\begin{equation}
\frac{d}{dy} G\left(\ell,y\right)
\!=\! 2 \frac{\Gamma(B)}{\beta}\! \int_0^{\frac{\pi}2}
\frac{d\tau}{\pi}\,  e^{-B\alpha}
 \left[\frac1{Y}
\left(1+2e^{\alpha}\sinh{\alpha}\right)
 {\cal{F}}_B
 +\frac1{\beta}
 e^{\alpha}{\cal{F}}_{B+1}\right.
\left.-\frac{2\sinh\alpha}{Y}{\cal{F}}_{B-1}\right].
\label{eq:derivyC}
\end{equation}

Logarithmic derivatives $\psi_{\ell}$ and $\psi_{y}$ are then constructed
according to their definition
 (\ref{eq:psi1})(\ref{eq:psi2}) by dividing (\ref{eq:derivlC}) and 
(\ref{eq:derivyC}) by the inclusive spectrum (\ref{eq:ifDC}).
 
Using the expression of Bessel's series, one gets

$\bullet$\quad for $\ell\rightarrow 0$;
\begin{eqnarray}
\psi_{\ell}&\stackrel{\ell \to 0}{\simeq}&
\frac{a}{\beta\ell} + c_1\ln\left(\frac{Y}{\ell}-1\right)
\to \infty,\cr
c_1 &=& \frac{2^{a/\beta+2}}{\pi(a+2\beta)}\int_{0}^{\pi/2}d\tau\,
(\cos\tau)^{a/\beta+2}
\left[\cos\left(\frac{a}{\beta}\tau\right)-\tan\tau\sin\left(\frac{a}
{\beta}\tau\right)\right]=0.4097>0,\cr
\psi_{y}&\stackrel{\ell\to 0}{\simeq}&-a\gamma_0^2+c_1\frac{\ell}{y}
\to -a\gamma_0^2.
\label{eq:psilzero}
\end{eqnarray}

$\bullet$\quad for $\ell\rightarrow Y \Leftrightarrow y\rightarrow0$;
\begin{eqnarray}
\psi_{\ell}&\stackrel{y\to 0}{\simeq}& c_2\left(\frac{Y}{\ell}-1\right)
\to 0,\cr
c_2 &=&\frac{2^{a/\beta+2}}{\pi(a+2\beta)}\int_{0}^{\pi/2}d\tau\,
(\cos\tau)^{a/\beta+2}
\left[\cos\left(\frac{a}{\beta}\tau\right)+\tan\tau\sin\left(\frac{a}
{\beta}\tau\right)\right]=0.9218>0;\cr
\psi_{y}&\stackrel{y\to 0}{\simeq}&-c_2\ln\left(\frac{Y}\ell-1\right)
\to \infty.
\label{eq:psiyzero}
\end{eqnarray}

\vskip 0.5cm

They are represented in Fig.~\ref{fig:psily} as functions of $\ell$
for two different values of $Y$ ($=5.2, 15$).

\subsection{Double derivatives}
\label{subsection:doublederiv}
%%%%%%%%%%%%%%%%%%%%%%%%%%%%%%%

In the core of this paper we also need the expression for $\psi_{,\ell,\ell}$

\begin{equation}
\psi_{\ell\,\ell}=\frac1G G_{\ell\,\ell}-(\psi_{\ell})^2.
\end{equation}

By differentiating twice (\ref{eq:hyprep1}) with respect to $\ell$, one gets

\begin{eqnarray}
G_{\ell\,\ell}(\ell,y)&\!\!=\!\!&\frac2{\ell+y}\left(G_{\ell}(\ell,y)-
\frac1{\ell+y}G(\ell,y)\right)\cr
&&\cr
&&+\frac{(\ell+y)\Gamma(B)}{\beta \Gamma(B+3)}
 \int_{\epsilon-i\infty}^{\epsilon+i\infty}\frac{d\omega}{2\pi  i}
 e^{-\omega y}\omega^2
 \left(A^2+3A+2\right)
\Phi\left(A+3,B+4;\omega (\ell+y)\right).\cr
&&
\end{eqnarray}
Using the procedure of \cite{PerezMachetC} (appendix A.2) and setting
$y=Y-\ell$, the result for $G_{\ell\,\ell}$ reads

\begin{eqnarray}
G_{\ell\,\ell}(\ell,Y)&\!\!=\!\!&\frac2{Y}\left(G_{\ell}(\ell,Y)-
\frac1{Y}G(\ell,Y)\right)\cr
&&\cr
&& + 2\frac{\Gamma(B)}{\beta} 
\int_{0}^{\frac{\pi}2} \frac{d\alpha}{\pi}\, e^{-(B-2)\alpha}\left[\frac1{\beta^2}\,{\cal{F}}_{B+2}+\frac6{\beta Y}\sinh\alpha\,
{\cal{F}}_{B+1}+\frac8{Y^2}\sinh^2\alpha\,{\cal{F}}_{B}\right].\cr
&&
\end{eqnarray}

Likewise, for
\begin{equation}
\psi_{y\,y}=\frac1G G_{y\,y}-(\psi_{y})^2,
\end{equation}
where

\begin{eqnarray}
G_{y\,y}(\ell,y)&\!\!=\!\!&\gamma_0^2 G(\ell,y) + \frac1{Y}\bigg(G_{y}(\ell,y)-G_{\ell}
(\ell,y)\bigg)\cr
&&\cr
&& +
\frac1{\beta}\frac{(\ell+y)\Gamma(B)}{\Gamma(B+2)}\int_{\epsilon-i\infty}^
{\epsilon+i\infty}\frac{d\omega}{2\pi  i} e^{-\omega y}
\left(\omega^2-\frac{\omega}{\beta}\right)
\Phi\left(A+1,B+3;\omega (\ell+y)\right),\cr
&&
\end{eqnarray}
one gets

\vbox{
\begin{eqnarray}
G_{y\,y}(\ell,Y)&\!\!=\!\!&\gamma_0^2 G(\ell,Y)
+ \frac1{Y}\bigg(G_{y}(\ell,Y)-G_{\ell}(\ell,Y)\bigg)\cr
&&\cr
&& + 4\frac{\Gamma(B)}{\beta}\int_{0}^{\frac{\pi}2} \frac{d\alpha}{\pi}\, 
e^{-(B+1)\alpha}
\left[2(B+1)\frac{\sinh^2\alpha}{Y^2}\,{\cal{F}}_{B-1}
-\frac1{\beta}\frac{\sinh\alpha}{Y}\,{\cal{F}}_{B}\right].\cr
&&
\end{eqnarray}
}

Finally, 
\begin{eqnarray}\nonumber
\psi_{\ell\, y} = \psi_{y\,\ell} = \gamma_0^2\left[1-a\left(\psi_{\ell}-\beta\gamma_0^2\right)\right]-\psi_{\ell}\psi_{y}. 
\end{eqnarray}
$\psi_{\ell\,\ell}$, $\psi_{y\,y}$ and $\psi_{\ell\, y}$ are drawn 
in Fig.~\ref{fig:doublepsi} of appendix \ref{subsub:psinum} as 
functions of $\ell$ for fixed $Y$. They are all ${\cal {O}}(\gamma_0^3)$.

%%%%%%%%%%%%%%%%%%%%%%%%%%%%%%%%%%%%%%%%%%%%%%%%%%%%%%%%%%%%%%%%%%%%%%%%%%%%%
\section{NUMERICAL ANALYSIS OF CORRECTIONS}
\label{section:numcorr}
%%%%%%%%%%%%%%%%%%%%%%%%%%%%%%%%%%%%%%%%%%%%%%%%%%%%%%%%%%%%%%%%%%%%%%%%%%%%%

In this section, we present plots for the derivatives of $\psi$,
and $\varphi$ (see (\ref{eq:psi1})(\ref{eq:psi2}) and (\ref{eq:phisigma})),
for $\Upsilon$ and its derivatives  (see (\ref{eq:upsg})(\ref{eq:upsq})),
for $\Delta$, $\delta_1$, $\delta_2$ (see
(\ref{eq:nota4bis})-(\ref{eq:nota4C}))
and the combination $\delta_c\equiv \delta_1 + \delta_2 + a\Upsilon_{\ell}$,
$\tilde\delta_c\equiv \tilde\delta_1 + \tilde\delta_2$.

\subsection{Gluon jet}
\label{subsection:glucorr}
%%%%%%%%%%%%%%%%%%%%%%%%%%%%

\subsubsection{$\boldsymbol\psi$ and its derivatives}
\label{subsub:psinum}
%%%%%%%%%%%%%%%%%%%%%%%%%%%%%%%%%%%%%%%%%%%%%%%%%%%%%

This subsection is associated with appendices \ref{subsection:Logder}
and \ref{subsection:doublederiv} . It enables in particular
to visualize the behaviors of $\psi_{\ell}$ and $\psi_{y}$ when $\ell
\to 0$ or $y\to 0$, as described in (\ref{eq:psilzero}) and
(\ref{eq:psiyzero}), and to set the $\ell$ interval within which our calculation
can be trusted.

\begin{figure}
\vbox{
\begin{center}
\epsfig{file=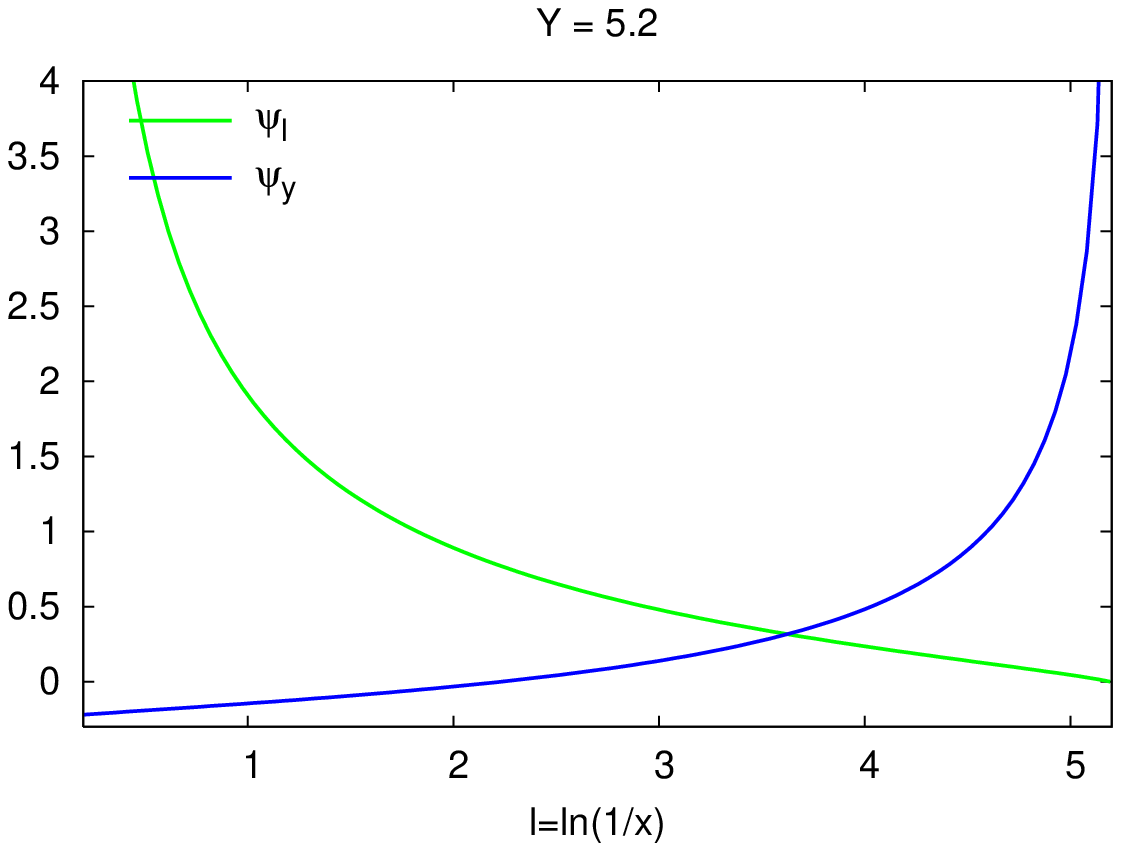, height=6truecm,width=0.45\tw}
\hfill
\epsfig{file=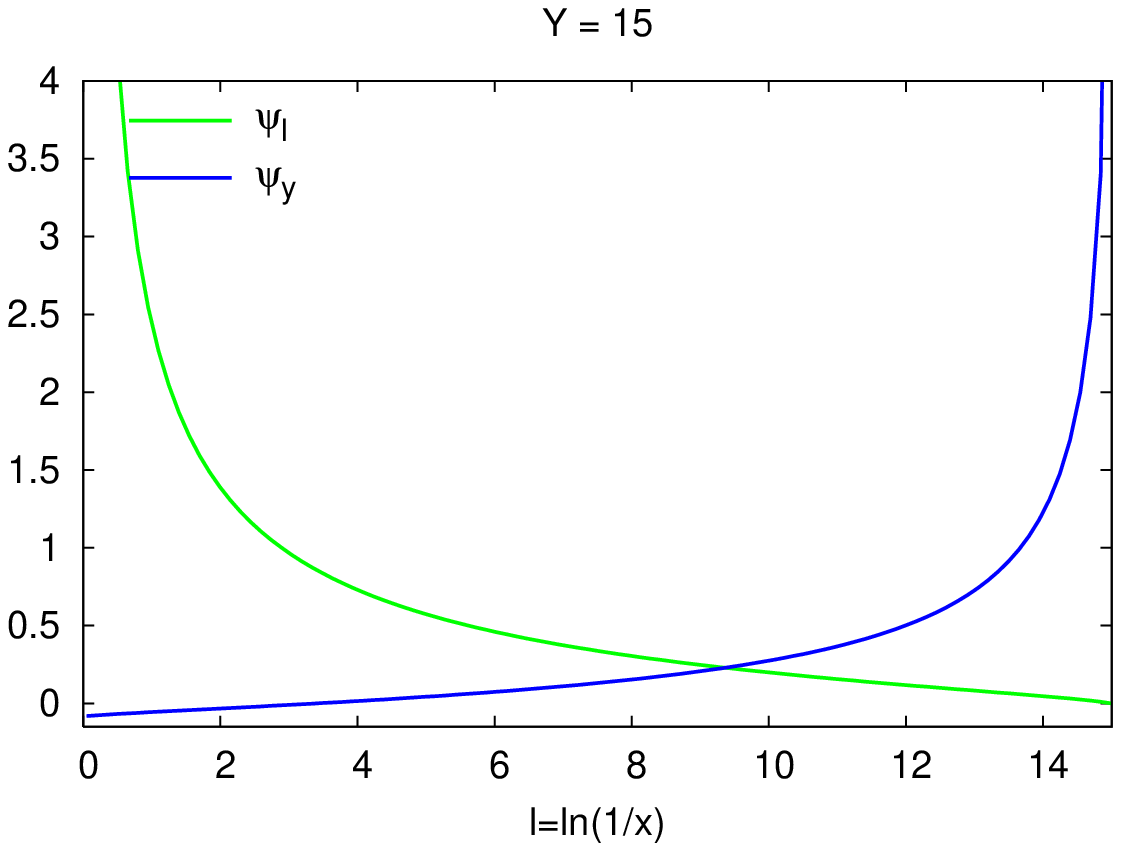, height=6truecm,width=0.45\tw}
\vskip .5cm
\caption{Derivatives $\psi_{\ell}$ and $\psi_{y}$ as functions of $\ell$ at fixed 
$Y=5.2$ (left) and $Y=15$ (right)}
\label{fig:psily}
\end{center}
}
\end{figure}

\begin{figure}
\vbox{
\begin{center}
\epsfig{file=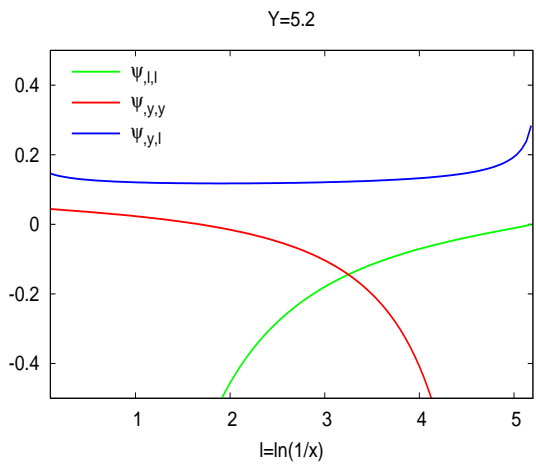, height=6truecm,width=0.45\tw}
\vskip .5cm
\caption{Double derivatives $\psi_{\ell\,\ell}$, $\psi_{\ell\,y}$ and
$\psi_{y\,y}$ as functions of $\ell$ at fixed $Y$}
\label{fig:doublepsi}
\end{center}
}
\end{figure}

In Fig.\ref{fig:psily} are drawn $\psi_{\ell}$ and $\psi_{y}$ as
functions of $\ell$ for two values $Y=5.2$ corresponding to LEP
working conditions, and $Y=15$ corresponding to an unrealistic ``high
energy limit''.

$\psi_{\ell}$ and ($\psi_{y}$) being both ${\cal {O}}(\gamma_0)$,
they should not exceed a ``reasonable value''; setting this value to
$1$, $|\psi_{\ell}|<1$ and  $|\psi_{y}|<1$ set, for $Y=5.2$, a confidence
interval 
\begin{equation}
2.5 \leq \ell \leq 4.5.
\label{eq:confint1}
\end{equation}

In the high energy limit $Y=15$, this interval becomes,  $4.5 \leq \ell
\leq 13$, in agreement with \ref{subsection:signG}.

\subsubsection{$\boldsymbol{\Delta(\ell_1,\ell_2,Y)}$}
\label{subsub:Deltanum}
%%%%%%%%%%%%%%%%%%%%%%%%%%%%%%%%%%%%%%%%%%%%%%%%%%%%%%

$\Delta$ has been defined in (\ref{eq:deltabisC}), in which $\psi_{1,\ell}$ and
$\psi_{1,y}$ are functions of $\ell_1$ and $Y$, $\psi_{2,\ell}$ and
$\psi_{2,y}$ are functions of $\ell_2$ and $Y$.

Studying the limits $\ell \to0$ and $\ell \to Y$ of subsection 
\ref{subsection:Logder}:

\begin{itemize}
\item for $\ell_1,\,\ell_2\rightarrow Y$ one gets (using the results of
\ref{subsection:Logder})
\begin{equation}
\Delta\simeq-c_2^{\,2}\left(\frac{Y-\ell_1}{\ell_1}\ln\frac{Y-\ell_2}{\ell_2}+
\frac{Y-\ell_2}{\ell_2}\ln\frac{Y-\ell_1}{\ell_1}\right),
\end{equation}
such that
\begin{equation}
\Delta \stackrel{\ell_1-\ell_2\to 0}{\longrightarrow}0,\quad \Delta
\stackrel{\ell_1-\ell_2\to\infty}{\longrightarrow}+\infty.
\end{equation}

\item for $\ell_1,\,\ell_2\rightarrow 0$ one gets (according to
\ref{subsection:Logder}):
\begin{equation}
\Delta\simeq-a\gamma_0^2\left[\frac{a}\beta\left( \frac1\ell_1+\frac1\ell_2 \right)+
c_1\left(\ln\frac{Y-\ell_1}{\ell_1}+\ln\frac{Y-\ell_2}{\ell_2}\right)\right]
\to -\infty.
\end{equation}

\end{itemize}

\begin{figure}
\vbox{
\begin{center}
\epsfig{file=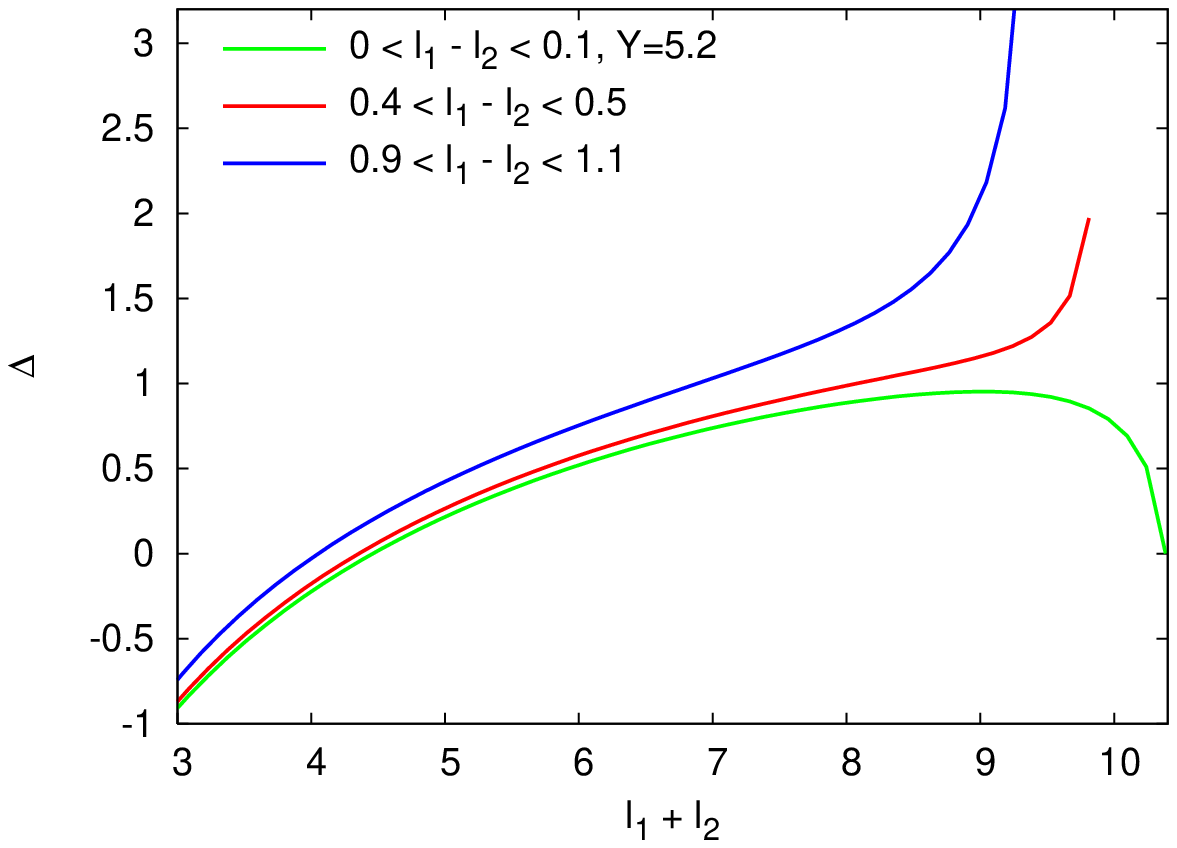, height=6truecm,width=0.47\tw}
\hfill
\epsfig{file=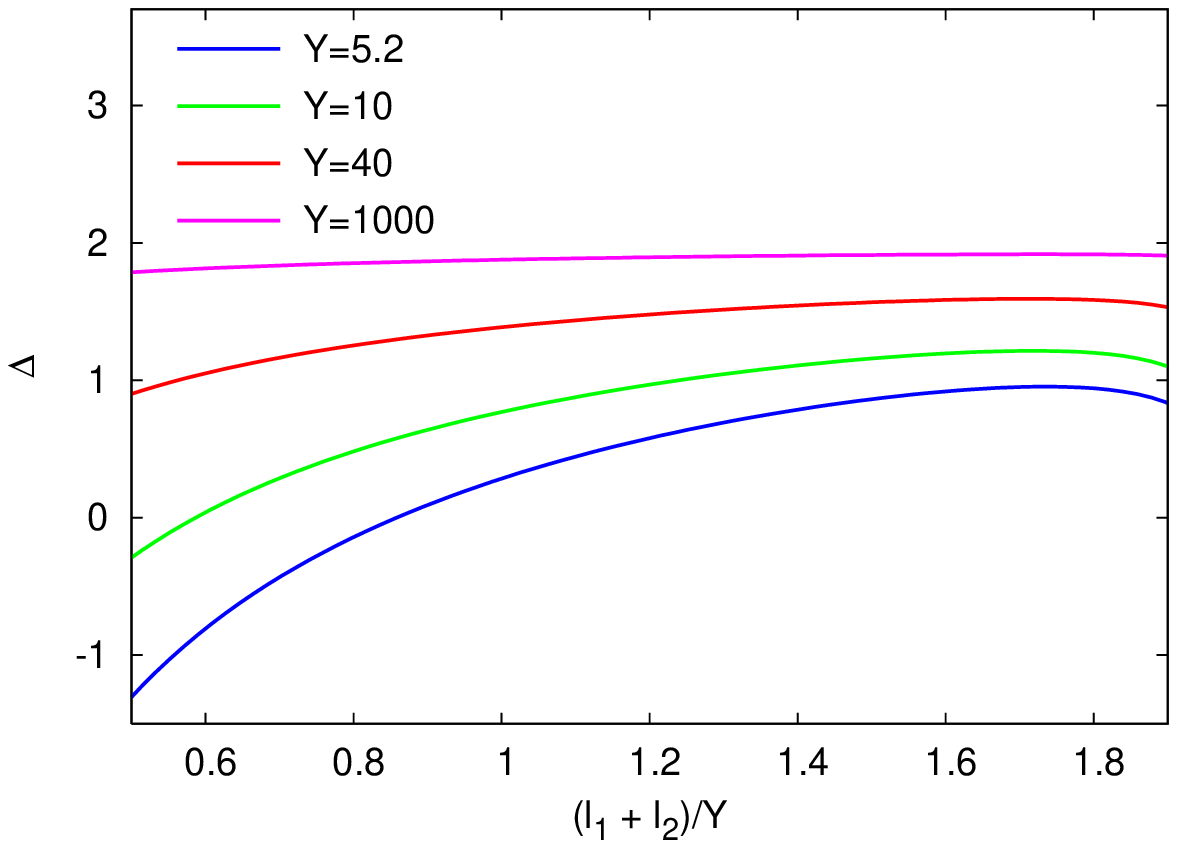, height=6truecm,width=0.47\tw}
\vskip .5cm
\caption{$\Delta$ as a function of $\ell_1+\ell_2$ for $Y=5.2$ (left) 
and its asymptotic behavior (right, $\ell_1-\ell_2=0.1$)}
\label{fig:Delta}
\end{center}
}
\end{figure}

In Fig.~\ref{fig:Delta} (left)  $\Delta$ is plotted as a function
of $\ell_1+\ell_2$ for three different values of $\ell_1-\ell_2$ $(0.1,\,0.5,\,1.0)$;
the condition (\ref{eq:confint1}) translates into
\begin{equation}
5.0\leq\ell_1+\ell_2\leq9.0;
\label{eq:confint2}
\end{equation}
on Fig.~\ref{fig:Delta} (right) the asymptotic limit $\Delta\rightarrow2$ 
for very large $Y$ clearly appears (we have taken $\ell_1-\ell_2=0.1$);
it is actually its DLA
value \cite{EvEqC}; this is not surprising since, in the high energy
limit $\gamma_0$ becomes very small and sub-leading
corrections (hard corrections and running coupling effects) get suppressed.

\subsubsection{$\boldsymbol{\Upsilon_g}$ and its derivatives}
\label{subsub:upsgnum}
%%%%%%%%%%%%%%%%%%%%%%%%%%%%%%%%%%%%%%%%%%%%%%%%%%%%%%%%%%%%%

\begin{figure}
\vbox{
\begin{center}
\epsfig{file=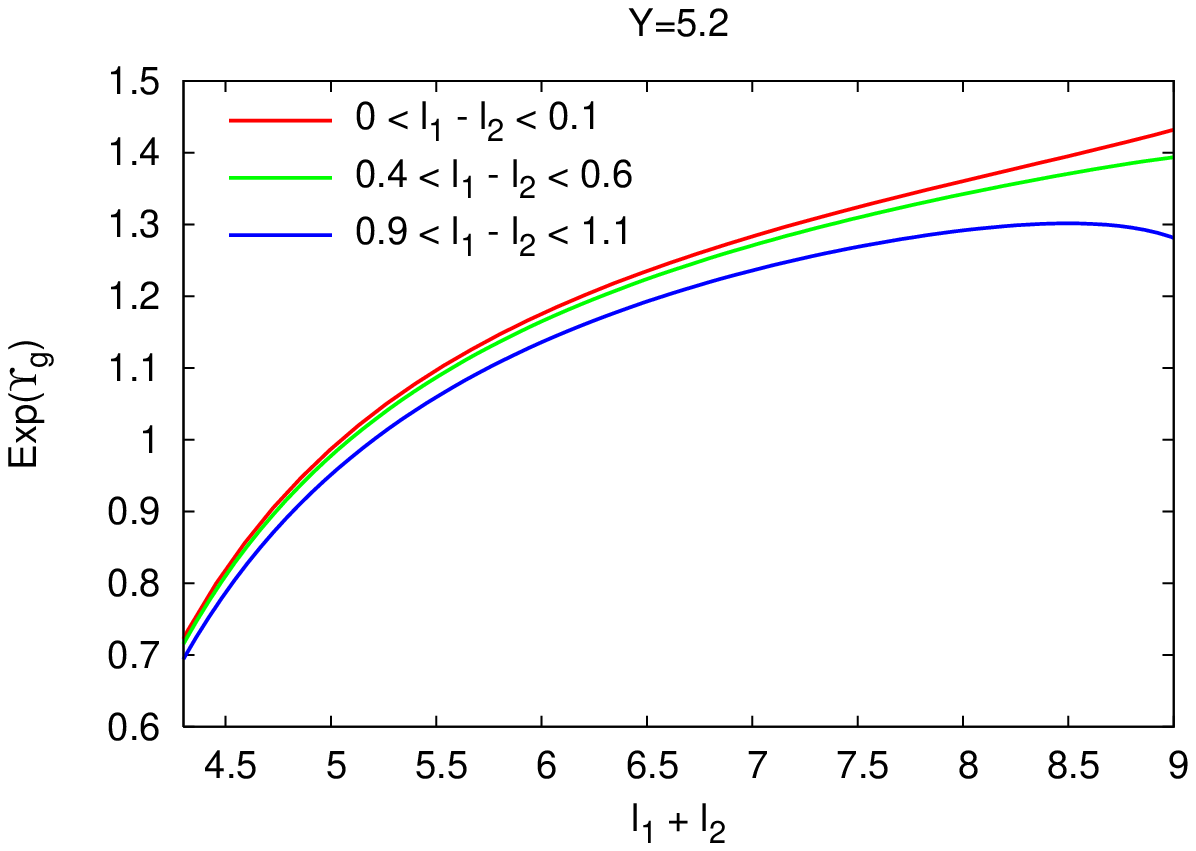, height=6truecm,width=0.47\tw}
\epsfig{file=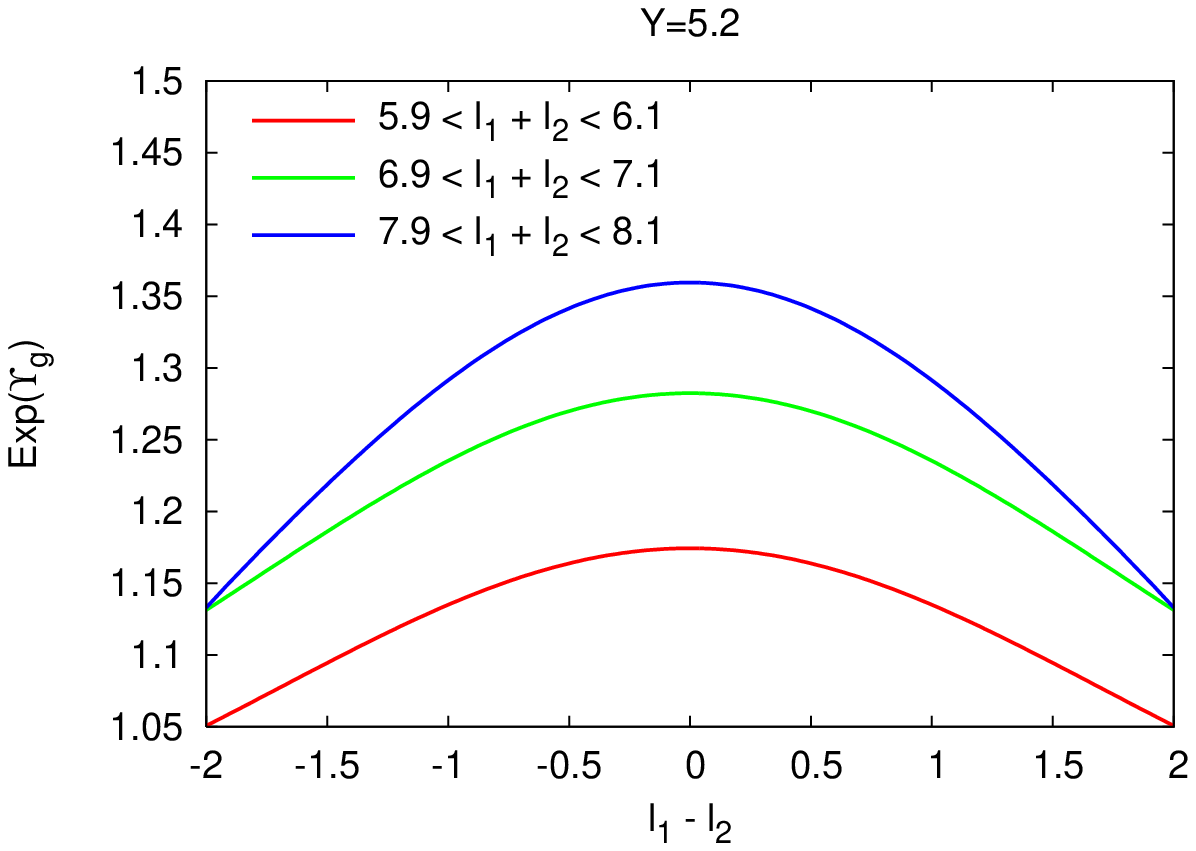, height=6truecm,width=0.47\tw}
\vskip .5cm
\caption{$\exp{(\Upsilon_g)}$ as a function of $\ell_1+\ell_2$ (left) and, 
$\ell_1-\ell_2$ (right) for $Y=5.2$}
\label{fig:Upsilong}
\end{center}
}
\end{figure}

Fig.~\ref{fig:Upsilong} exhibits the smooth behavior of
$\exp{(\Upsilon_g)}$ as a function of $(\ell_1+\ell_2)$ 
in the whole range of applicability of our approximation (we have chosen
the same values of $(\ell_1-\ell_2)$ as for Fig.~\ref{fig:Delta}),
and as a function of
$(\ell_1-\ell_2)$ for three values of $(\ell_1+\ell_2)$ 
($6.0, 7.0, 8.0$). So, the iterative procedure is 
safe and corrections stay under control.

\begin{figure}
\vbox{
\begin{center}
\epsfig{file=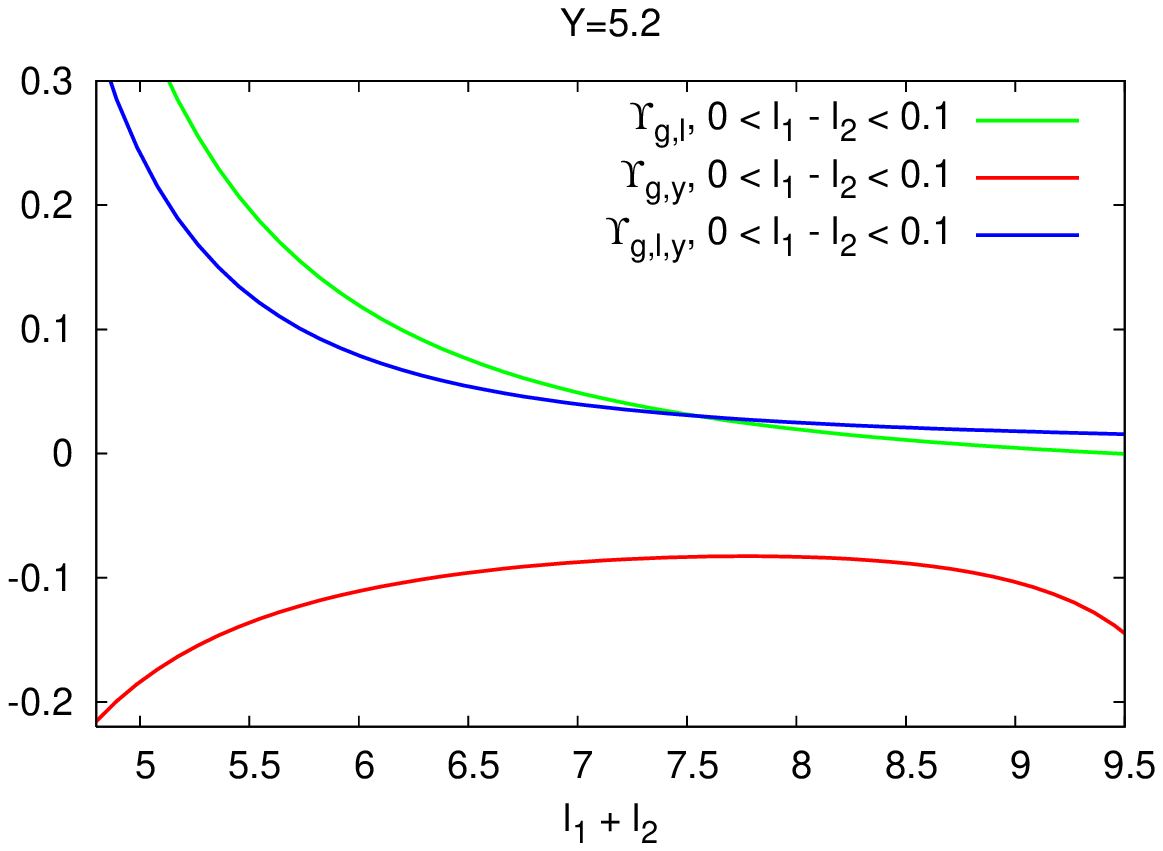, height=6truecm,width=0.47\tw}
\hfill
\epsfig{file=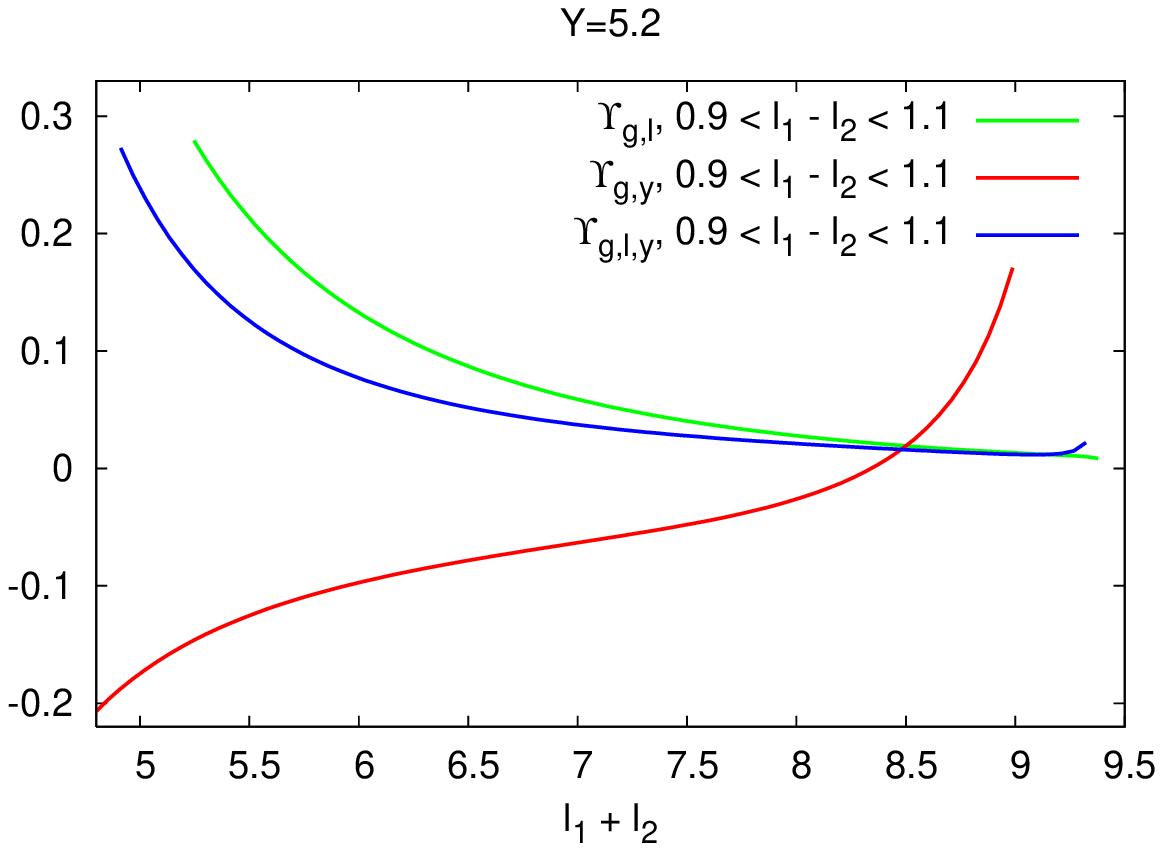, height=6truecm,width=0.47\tw}
\vskip .5cm
\caption{$\Upsilon_{g,\ell}$, $\Upsilon_{g,y}$ and $\Upsilon_{g,\ell\,y}$
as functions of $\ell_1+\ell_2$ for $Y=5.2$, $\ell_1-\ell_2=0.1$ (left) and 
$\ell_1-\ell_2=1.0$ (right)}
\label{fig:DUpsilon}
\end{center}
}
\end{figure}

Fig.~\ref{fig:DUpsilon} displays the derivatives of $\Upsilon_g$.
(\ref{eq:DUpsilonl}), (\ref{eq:DUpsilony}) and (\ref{eq:DUpsilonly}) 
have been plotted at $Y=5.2$, for  $(\ell_1-\ell_2)=0.1$ (left) and
$(\ell_1-\ell_2)=1.0$ (right).
The size and shape of these corrections agree with our expectations 
($\Upsilon_{g,\ell}=\Upsilon_{g,y}={\cal {O}}(\gamma_0^2)$, 
$\Upsilon_{g,\ell\,y}={\cal {O}}(\gamma_0^4)$).

For explicit calculations, we have used 
\begin{eqnarray}\label{eq:DUpsilonl}
&&\hskip -4cm \Upsilon_{g,\ell}=-\frac{\left[1\!-\!b\left(\psi_{1,\ell}
\!+\!\psi_{2,\ell}\!-\!
{\beta}\gamma_0^2 \right)\right]\left(\Delta_{\ell}\!-\!a\beta^2\gamma_0^4
\right)}{\left(1\!+\!\Delta\!+\!a\beta\gamma_0^2\right)\left[2\!+\!\Delta
  \!-\!b\left(\psi_{1,\ell}
\!+\!\psi_{2,\ell}\!-\!
  \,{\beta}\gamma_0^2 \right)\right]}
 \!-\!\frac{b\left(\psi_{1,\ell\,\ell}
\!+\!\psi_{2,\ell\,\ell}\!+\!
  {\beta^2}\gamma_0^4 \right)}
  {2\!+\!\Delta
  \!-\!b\left(\psi_{1,\ell}
\!+\!\psi_{2,\ell}\!-\!
  {\beta}\gamma_0^2 \right)},\\\nonumber\\
\label{eq:DUpsilony}
&& \hskip -4cm\Upsilon_{g,y}= -\frac{\left[1\!-\!b\left(\psi_{1,\ell}
\!+\!\psi_{2,\ell}\!-\!
 {\beta}\gamma_0^2\right)\right]\left(\Delta_{y}\!-\!a\beta^2\gamma_0^4
\right)}{\left(1\!+\!\Delta\!+\!a\beta\gamma_0^2\right)\left[2\!+\!\Delta
  \!-\!b\left(\psi_{1,\ell}
\!+\!\psi_{2,\ell}\!-\!
 {\beta}\gamma_0^2 \right)\right]}
 \!-\!\frac{b\left(\psi_{1,\ell\, y}
\!+\!\psi_{2,\ell\, y}\!+\!
 {\beta^2}\gamma_0^4 \right)}
  {2\!+\!\Delta
  \!-\!b\left(\psi_{1,\ell}
\!+\!\psi_{2,\ell}\!-\!
  {\beta}\gamma_0^2 \right)},\\\nonumber\\ 
\label{eq:DUpsilonly}
\Upsilon_{g,\ell\,y}\!\!\!&\!\!\!=\!\!\!&\!\!\!\frac{\partial\Upsilon_{g,y}}{\partial\ell},
\end{eqnarray}
where
\begin{eqnarray}
\Delta_{\ell}\!\!&\!\!=\!\!&\!\!\gamma_0^{-2}\left[\psi_{1,\ell\,\ell}\psi_{2,y}+
\psi_{1,\ell}\psi_{2,y\,\ell}+\psi_{2,\ell\,\ell}\psi_{1,y}+\psi_{2,\ell}
\psi_{1,y\,\ell}\right]+\beta\gamma_0^2\Delta,\cr
\Delta_{y}\!\!&\!\!=\!\!&\!\!\gamma_0^{-2}\left[\psi_{1,\ell\, y}\psi_{2,y}+
\psi_{1,\ell}\psi_{2,y\,y}+\psi_{2,\ell y}\psi_{1,y}+\psi_{2,\ell}
\psi_{1,y\,y}\right]+\beta\gamma_0^2\Delta.
\end{eqnarray}
For the expressions of $\psi_{\ell\,\ell}$, $\psi_{\ell\,y}=\psi_{y\,\ell}$ and 
$\psi_{y\,y}$, the reader is directed to  \ref{subsection:doublederiv}.
(\ref{eq:DUpsilonly})  has been computed numerically (its analytical
expression is too heavy to be easily manipulated).

\subsubsection{$\boldsymbol{\delta_1}$, $\boldsymbol{\delta_2}$,
$\boldsymbol{\delta_c}$}
\label{subsub:deltanum}
%%%%%%%%%%%%%%%%%%%%%%%%%%%%%%%%%%%%%%%%%%%%%%%%%%%%%%%%%%%%%%%%

$\delta_1$ and $\delta_2$ are defined in
(\ref{eq:nota4bis})(\ref{eq:nota4C}). We also define
\begin{equation}
\delta_c=\delta_1+\delta_2+a\Upsilon_{\ell},
\label{eq:deltac}
\end{equation}
which appears in the numerator of the first line of (\ref{eq:CGfull}).

\begin{figure}
\vbox{
\begin{center}
\epsfig{file=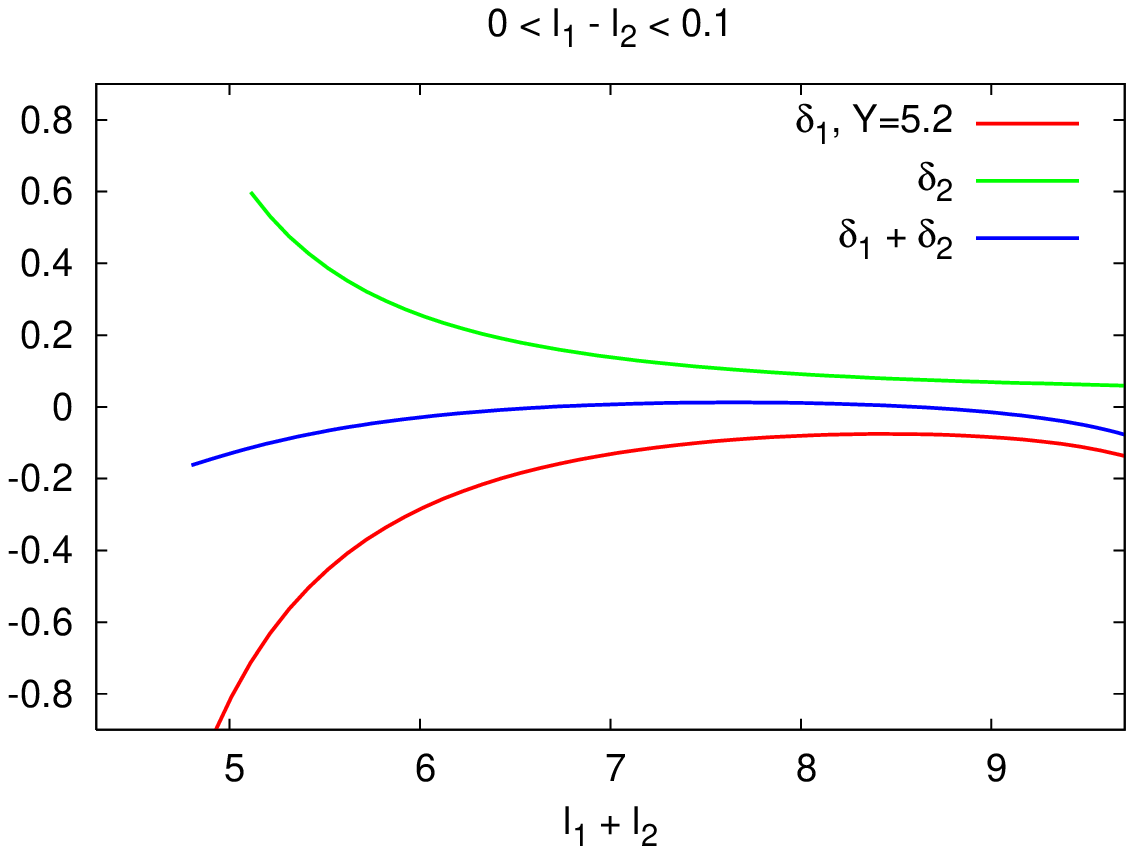, height=6truecm,width=0.48\tw}
\hfill
\epsfig{file=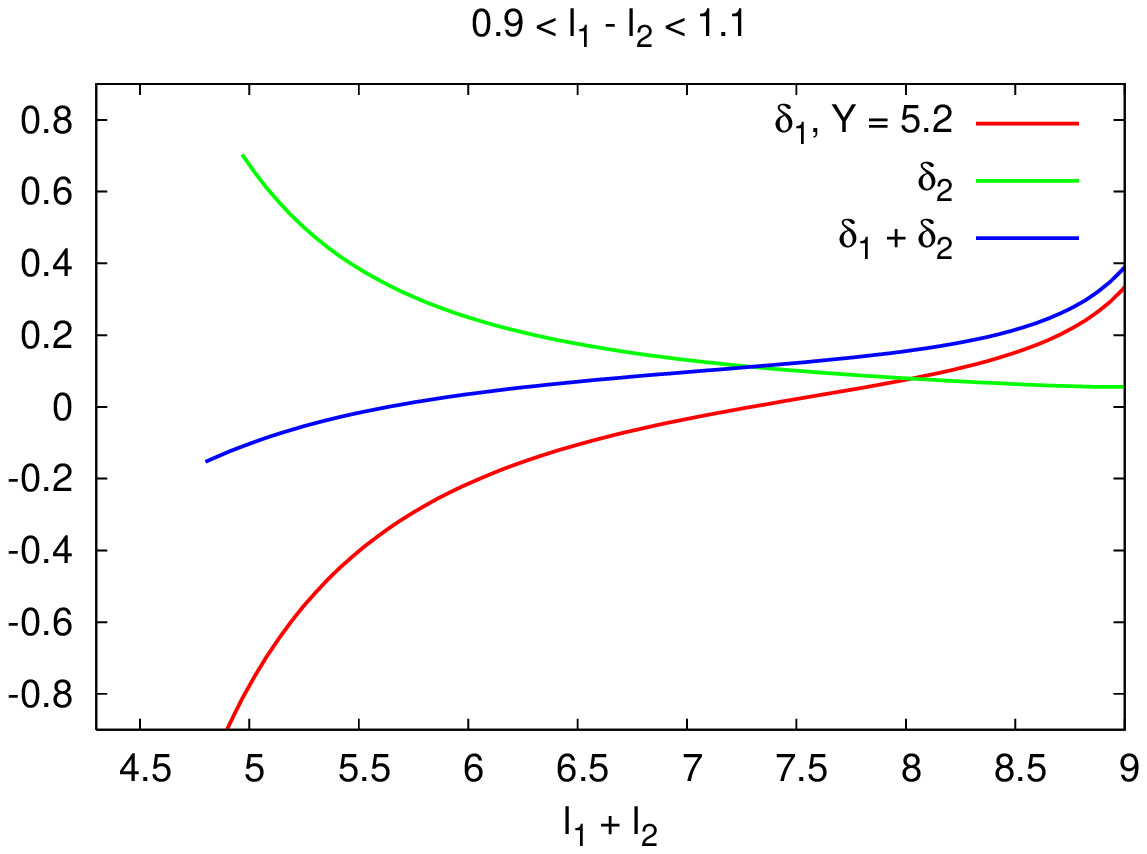, height=6truecm,width=0.48\tw}
\vskip .5cm
\caption{$\delta_1$, $\delta_2$ and $\delta_1+\delta_2$ as functions of
$\ell_1+\ell_2$ for $\ell_1-\ell_2=0.1$ (left) and $\ell_1-\ell_2=1.0$ (right)}
\label{fig:delta12}
\end{center}
}
\end{figure}

Fig.~\ref{fig:delta12} displays the behavior of $\delta_1$, $\delta_2$ and
$\delta_1+\delta_2$ at $Y=5.2$ for $\ell_1-\ell_2 =0.1$ and $\ell_1-\ell_2 = 1.0$.
We recall that these curves can only be reasonably trusted in the interval
(\ref{eq:confint2}).

Though $|\delta_1|={\cal {O}}(\gamma_0)$ (MLLA) should be numerically larger
than $|\delta_2|={\cal {O}}(\gamma_0^2)$ (NMLLA), it turns out that for 
relatively large $\gamma_0\sim0.5$ (Y=5.2), $|\delta_1|\sim|\delta_2|$, and 
 that strong
cancellations occur in their sum. As $\gamma_0$ decreases (or $Y$ increases) $|\delta_1|\gg|\delta_2|$,
in agreement with the perturbative expansion conditions.

\begin{figure}
\vbox{
\begin{center}
\epsfig{file=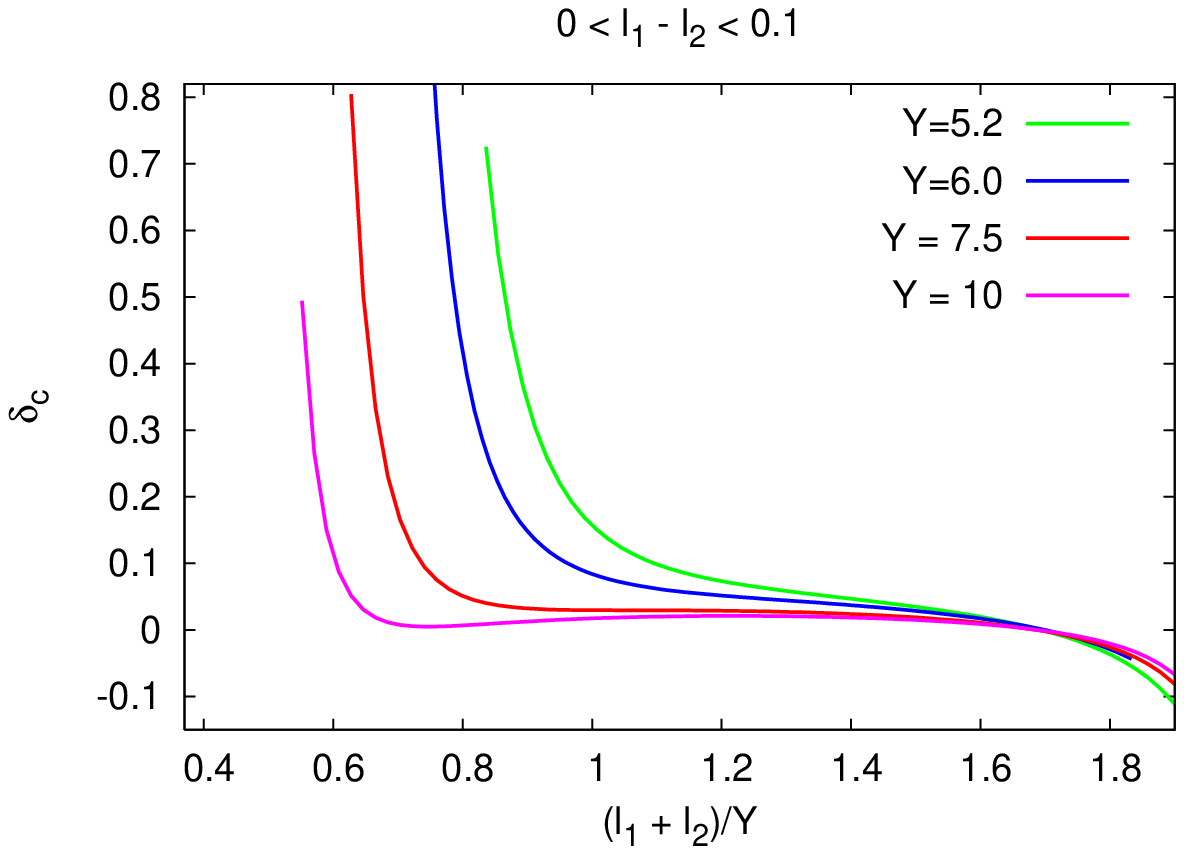, height=6truecm,width=0.48\tw}
\hfill
\epsfig{file=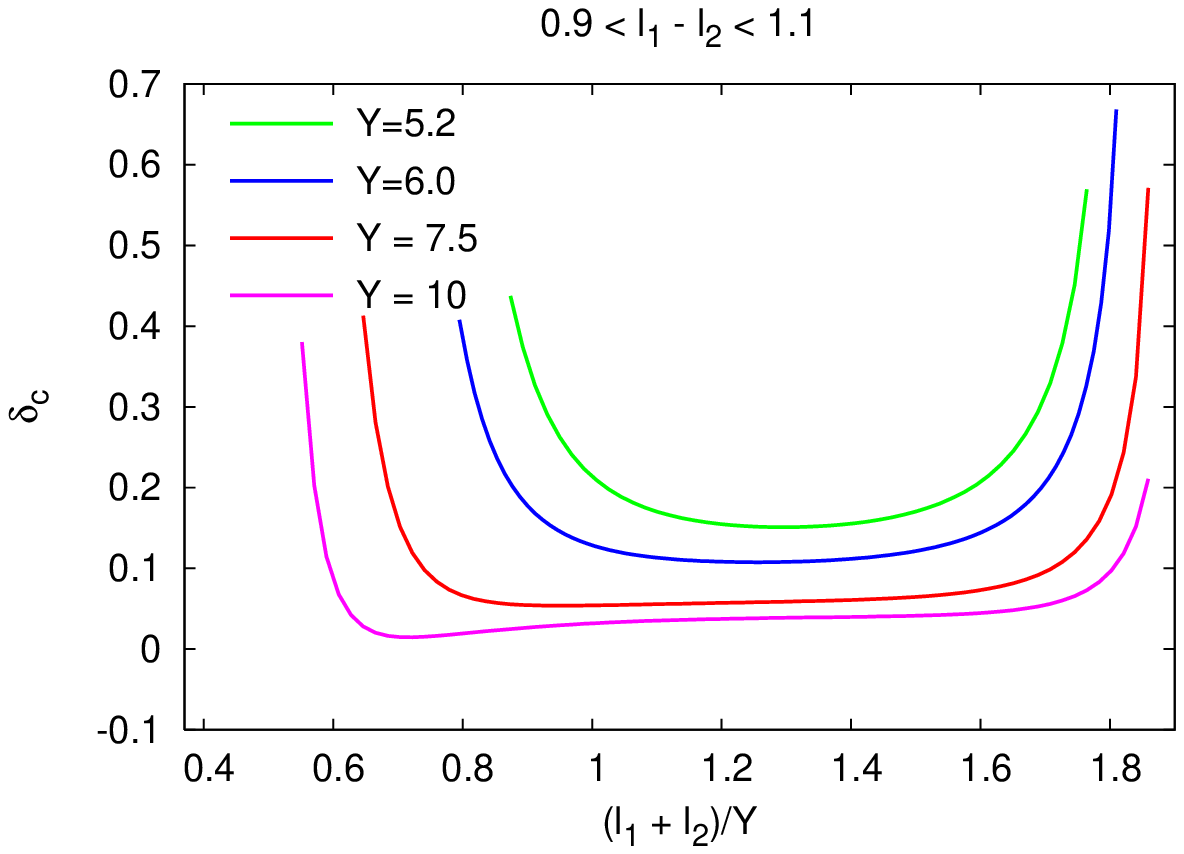, height=6truecm,width=0.48\tw}
\vskip .5cm
\caption{$\delta_c$ as a function of
$(\ell_1+\ell_2)/Y$ for $\ell_1-\ell_2=0.1$ (left) and $\ell_1-\ell_2=1.0$ (right)}
\label{fig:delta12chi}
\end{center}
}
\end{figure}

In Fig.~\ref{fig:delta12chi} we represent $\delta_c$ for different
values of $Y$;  it shows how the sum of corrections (MLLA and NMLLA) stay under
control in the confidence interval (\ref{eq:confint2}).
For $Y=5.2$ one reaches a regime where it
becomes slightly larger than $0.1$ away from the region $x_1\approx x_2$ 
(see upper curve on the right of Fig.~\ref{fig:delta12chi}) but still, since $1$ (which
is the leading term in the numerator of (\ref{eq:CGfull})) $\gg0.1$, our 
approximation can be trusted.

\subsubsection{The global role of corrections in the iterative procedure}
\label{subsub:allgcor}
%%%%%%%%%%%%%%%%%%%%%%%%%%%%%%%%%%%%%%%%%%%%%%%

\begin{figure}
\vbox{
\begin{center}
\epsfig{file=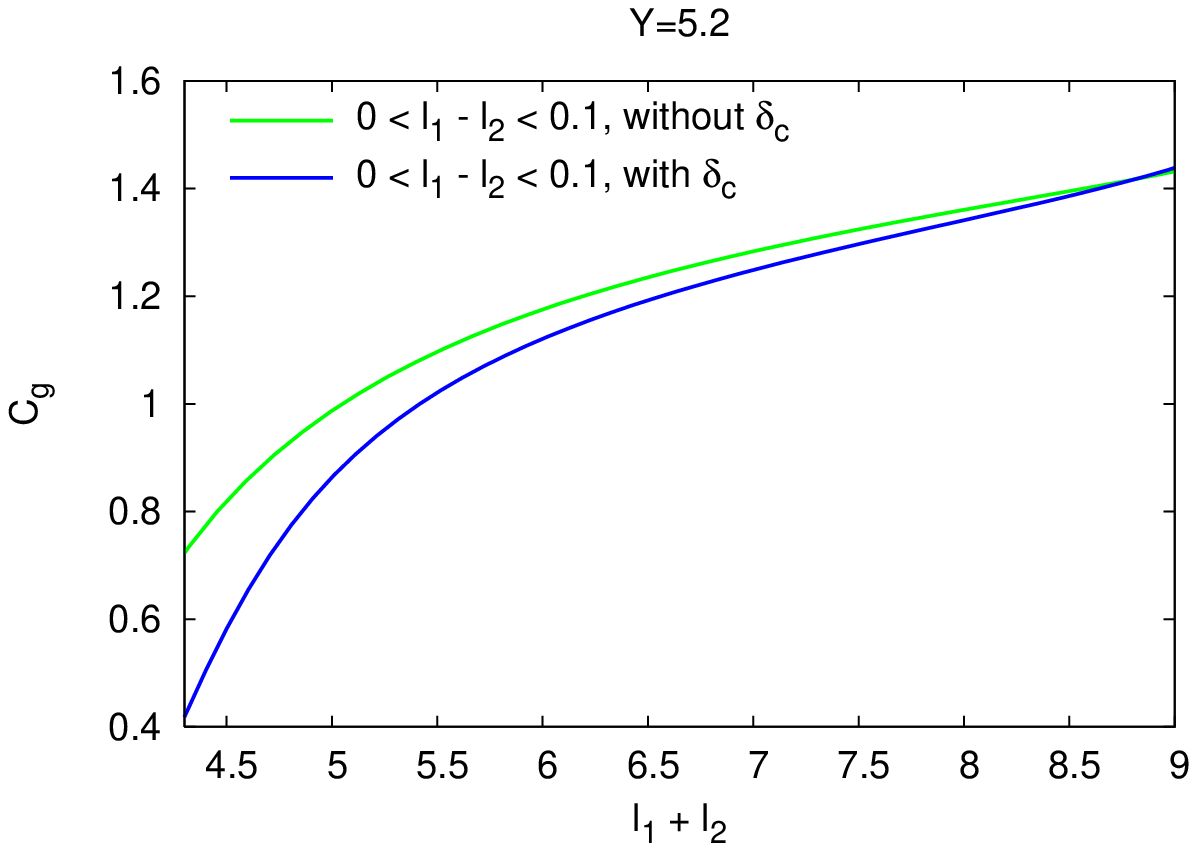, height=6truecm,width=0.48\tw}
\hfill
\epsfig{file=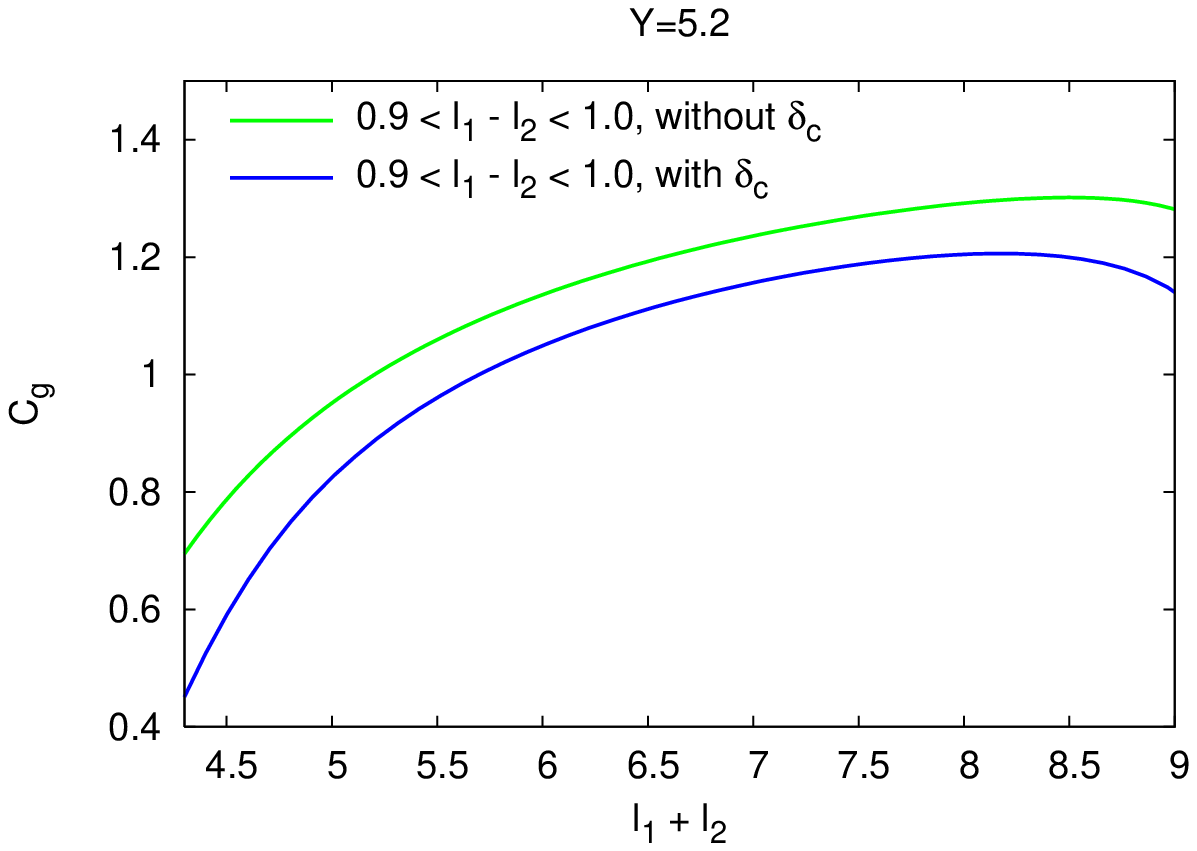, height=6truecm,width=0.48\tw}
\vskip 0.5cm
\epsfig{file=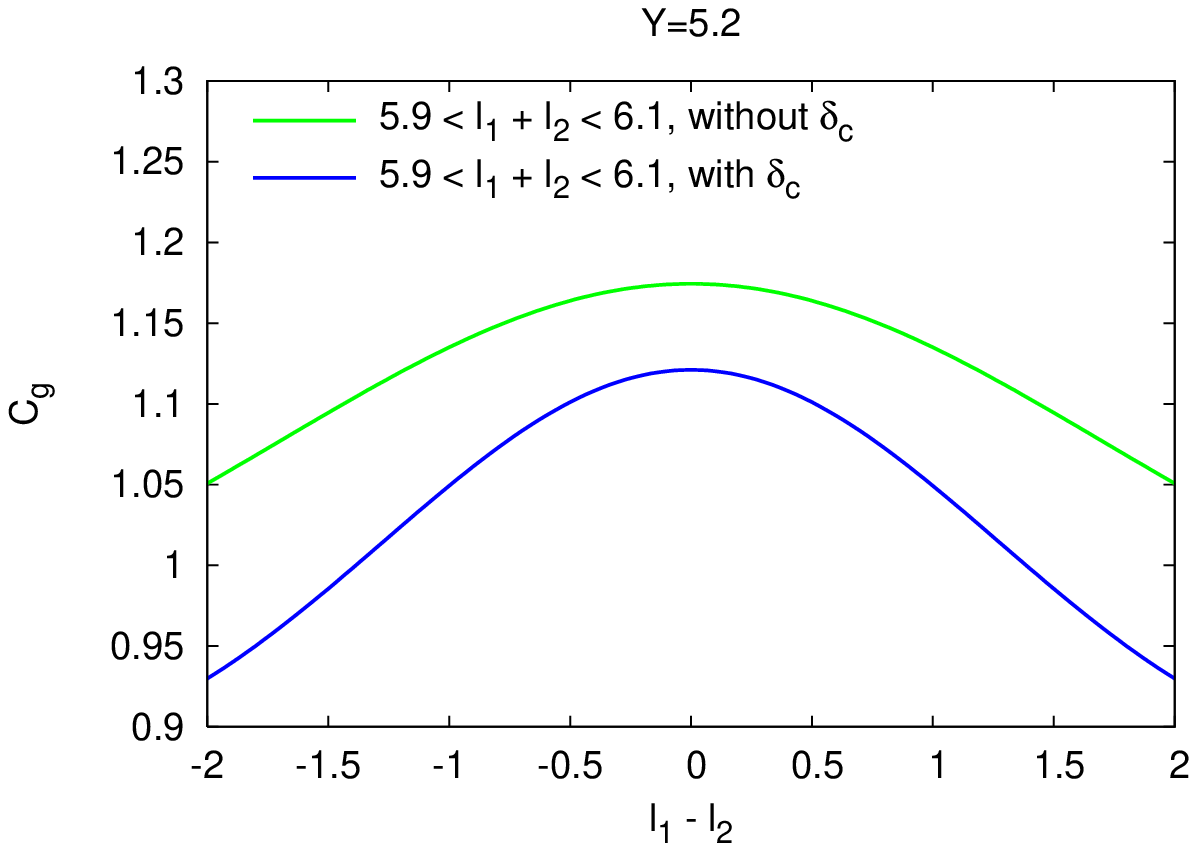, height=6truecm,width=0.48\tw}
\hfill
\epsfig{file=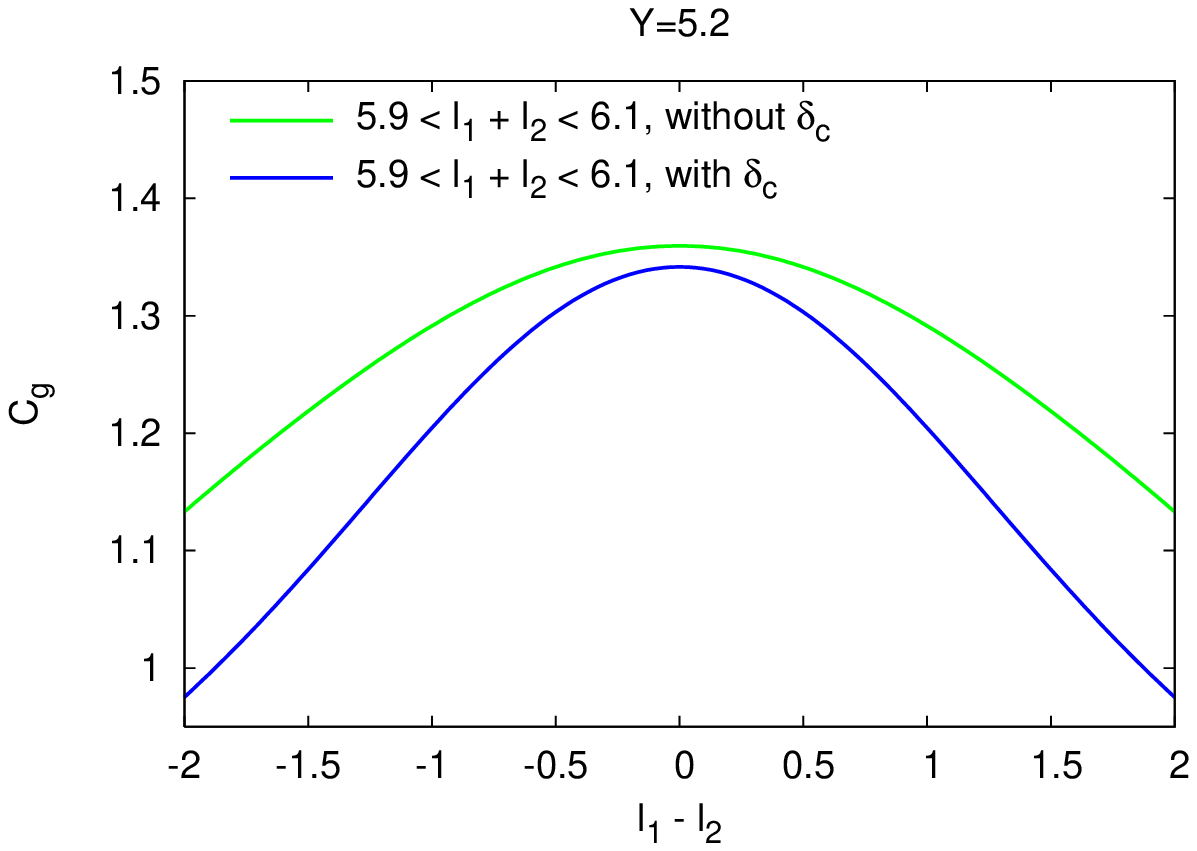, height=6truecm,width=0.48\tw}
\vskip .5cm
\caption{${\cal C}_g$ (blue) compared with $\exp\Upsilon_g$
(green)}
\label{fig:corrUpsilon}
\end{center}
}
\end{figure}

Fig.~\ref{fig:corrUpsilon} shows the role of $\delta_c$ on the 
correlation function: we represent the bare 
function $\exp{\Upsilon_g}$ (see \ref{eq:upsg}) as in Fig.~\ref{fig:Upsilong}, 
together with (\ref{eq:CGfull}). For $(\ell_1-\ell_2)=0.1$
($\ell_1\approx \ell_2$) and $(\ell_1-\ell_2)=1.0$, it is shown how
$\delta_c$ modifies the shape and size of $\exp{\Upsilon_g}$.
When $\ell_1\ne \ell_2$ ($(\ell_1-\ell_2)=1.0$),
$\delta_c$ decreases the correlations. 
They are also represented as a function of $(\ell_1-\ell_2)$
when $(\ell_1+\ell_2)$ is fixed ( to $6.0$ and $7.0$). The 
increase of $\delta_c$ 
as one goes away from the diagonal
$\ell_1\!\approx\! \ell_2$ (see Fig.~\ref{fig:delta12chi} for
$(\ell_1-\ell_2)=1.0$)
explain the difference between the green and blue curves; this substantially modifies
the tail of the correlations. 

When $Y$ gets larger, the role of $\delta_c$
decreases: at $Y=7.5$ (LHC conditions) the difference between the two curves becomes negligible.

\subsection{Quark jet}
\label{subsection:quacorr}
%%%%%%%%%%%%%%%%%%%%%%%%%%%

\subsubsection{$\boldsymbol\varphi$ and its derivatives}
\label{subsub:phinum}
%%%%%%%%%%%%%%%%%%%%%%%%%%%%%%%%%%%%%%%%%%%%%%%%%%%%%%%%

\begin{figure}
\vbox{
\begin{center}
\epsfig{file=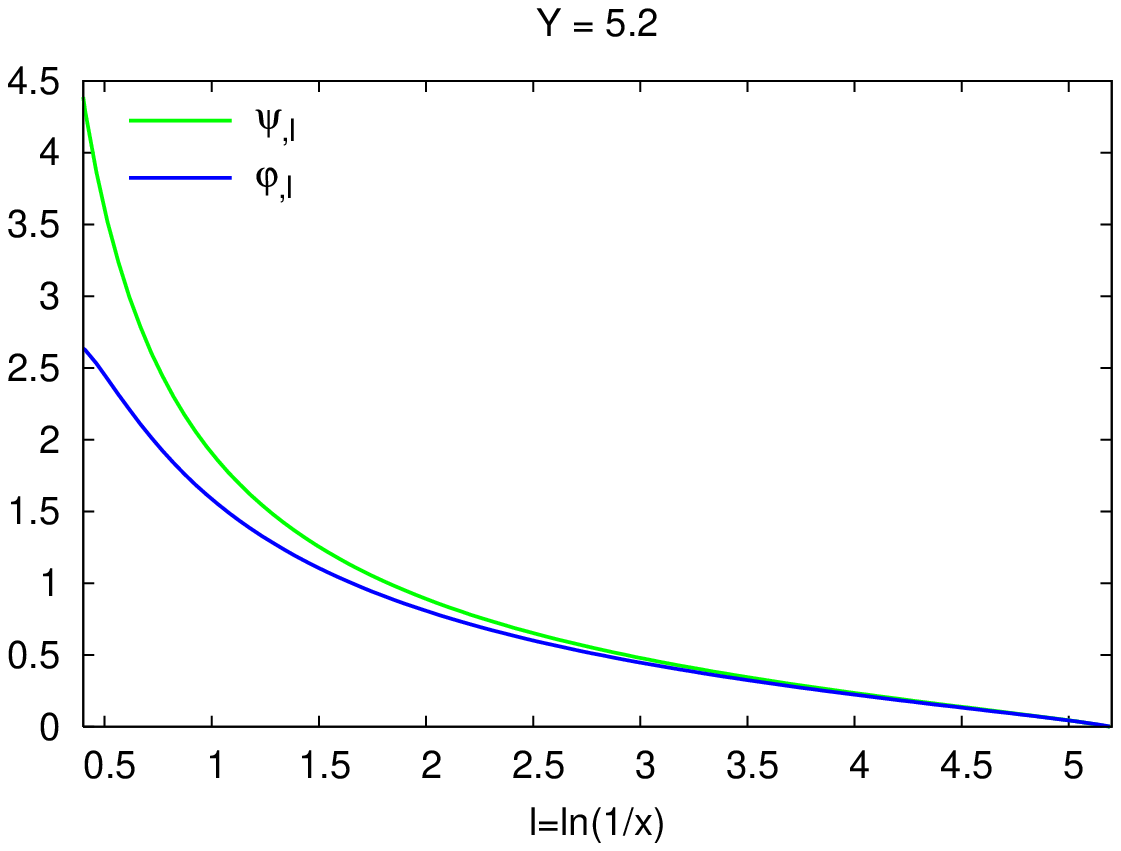, height=6truecm,width=0.45\tw}
\hfill
\epsfig{file=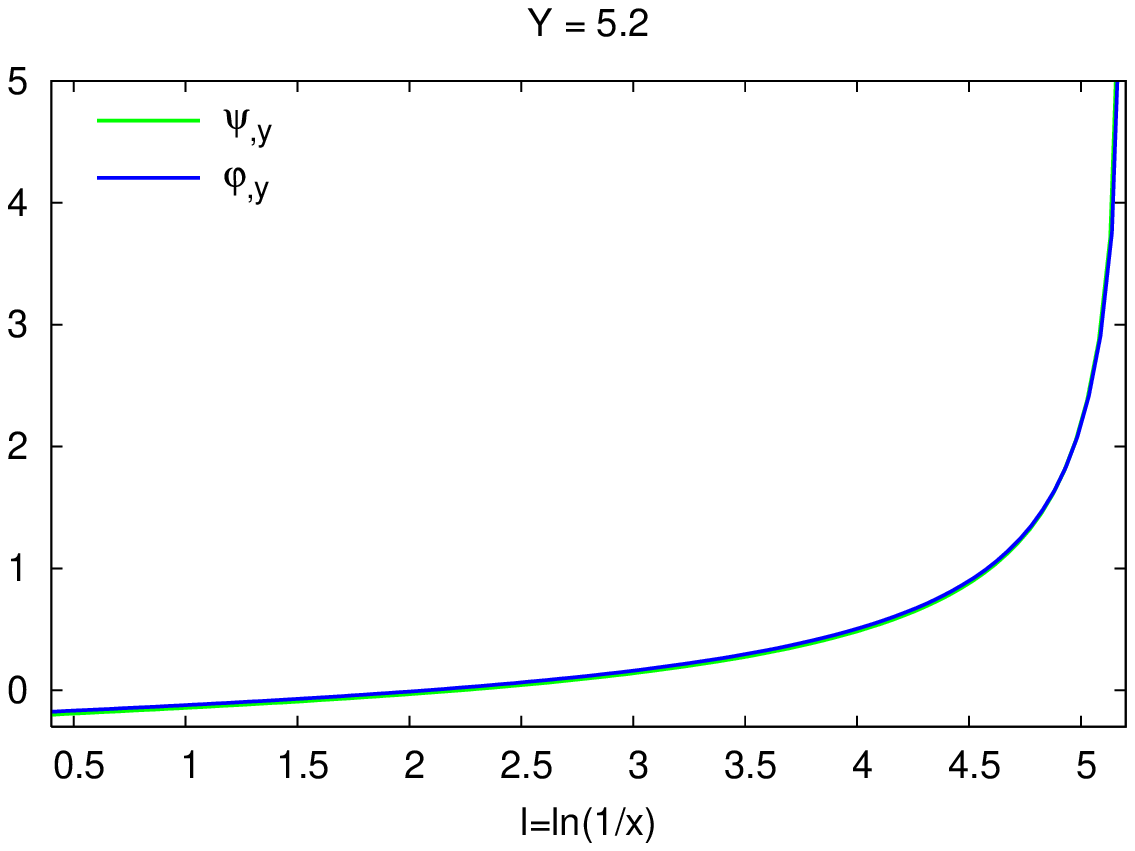, height=6truecm,width=0.45\tw}
\vskip .5cm
\caption{Derivatives $\varphi_{g,\ell}$ and $\varphi_{gy}$ as functions of $\ell$ at fixed 
$Y=5.2$ (left), compared with $\psi_{\ell}$ and $\psi_{y}$}
\label{fig:psiqly}
\end{center}
}
\end{figure}

Fig.~\ref{fig:psiqly} displays the derivatives $\varphi_{\ell}$ and $\varphi_{y}$
together with those $\psi_{\ell}$ and $\psi_{y}$ for the gluon jet, at $Y=5.2$. 
There sizes and shapes are the 
same since the logarithmic derivatives of the single inclusive 
distributions inside a gluon or a quark jet only depend on their shapes (the 
normalizations cancel in the ratio), 
which is the same in both cases. The mismatch at small $\ell$ between
$\varphi_{\ell}$ and $\psi_{\ell}$ stems from the behavior of 
$\psi_{\ell\,\ell}$ $\psi_{\ell\,\ell}\stackrel{\ell\to0}{\longrightarrow}-\infty$. 
Therefore, in the interval of applicability of the soft approximation (\ref{eq:varphil}) and (\ref{eq:varphiy2})
can be approximated by $\psi_{\ell}$ and $\psi_{y}$ respectively.

\subsubsection{$\boldsymbol{\tilde\Delta(\ell_1,\ell_2,Y)}$}
\label{subsub:tildeDelta}
%%%%%%%%%%%%%%%%%%%%%%%%%%%%%%%%%%%%%%%%%%%%%%%%%%%%%%%%

The last statement in \ref{subsub:phinum} numerically supports
the approximation (\ref{eq:Deltatilde}), that is

$$
\tilde\Delta\approx\Delta+{\cal O}(\gamma_0^2).
$$

We get rid of the heavy ${\cal O}(\gamma_0^2)$ factor in (\ref{eq:Deltatilde})
to ease our numerical calculations. Hence, the behavior of $\tilde\Delta$ is
already given in Fig.~\ref{fig:Delta}.

\subsubsection{$\boldsymbol{\Upsilon_q}$ and its derivatives}
\label{subsub:upsqnum}
%%%%%%%%%%%%%%%%%%%%%%%%%%%%%%%%%%%%%%%%%%%%%%%%%%%%%%%%

\begin{figure}
\vbox{
\begin{center}
\epsfig{file=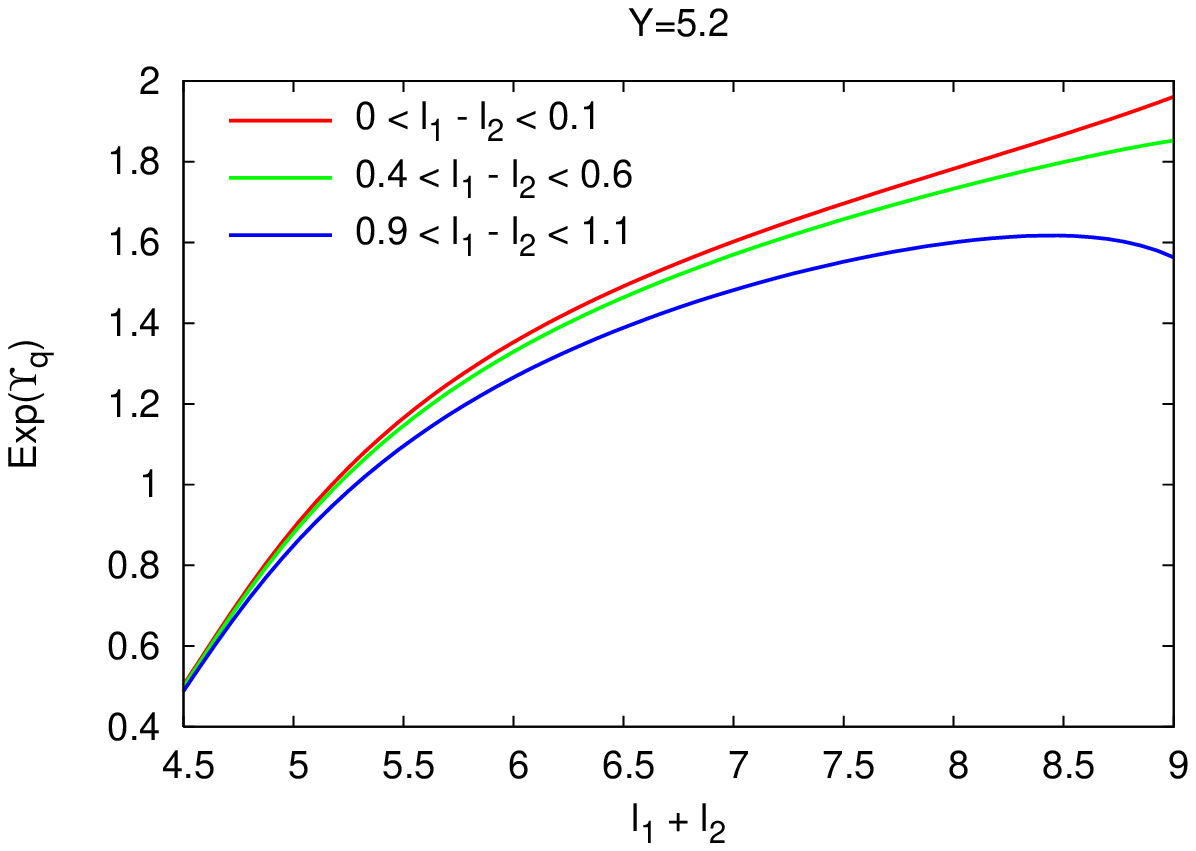, height=6truecm,width=0.47\tw}
\epsfig{file=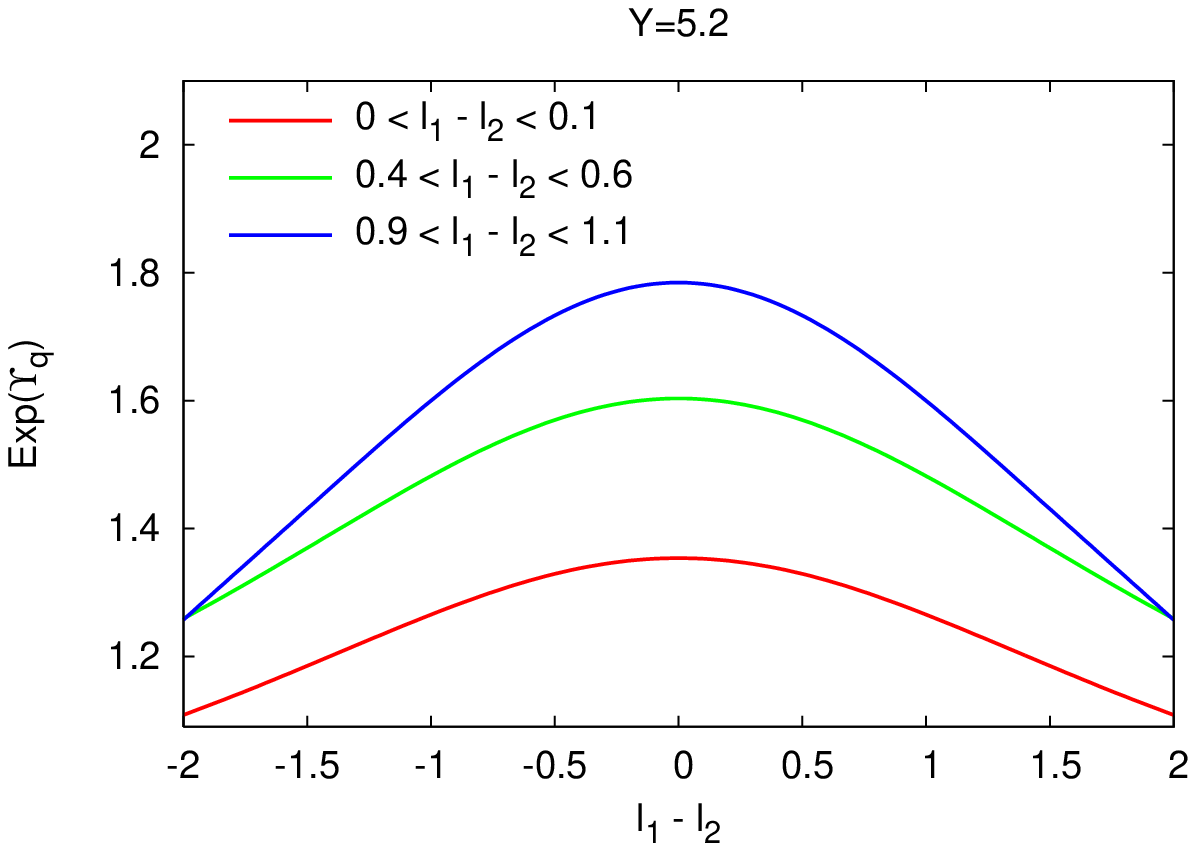, height=6truecm,width=0.47\tw}
\vskip .5cm
\caption{$\exp{(\Upsilon_q)}$ as a function of $\ell_1+\ell_2$ (left) and, 
$\ell_1-\ell_2$ (right) for $Y=5.2$}
\label{fig:Upsilonq}
\end{center}
}
\end{figure}

\begin{figure}
\vbox{
\begin{center}
\epsfig{file=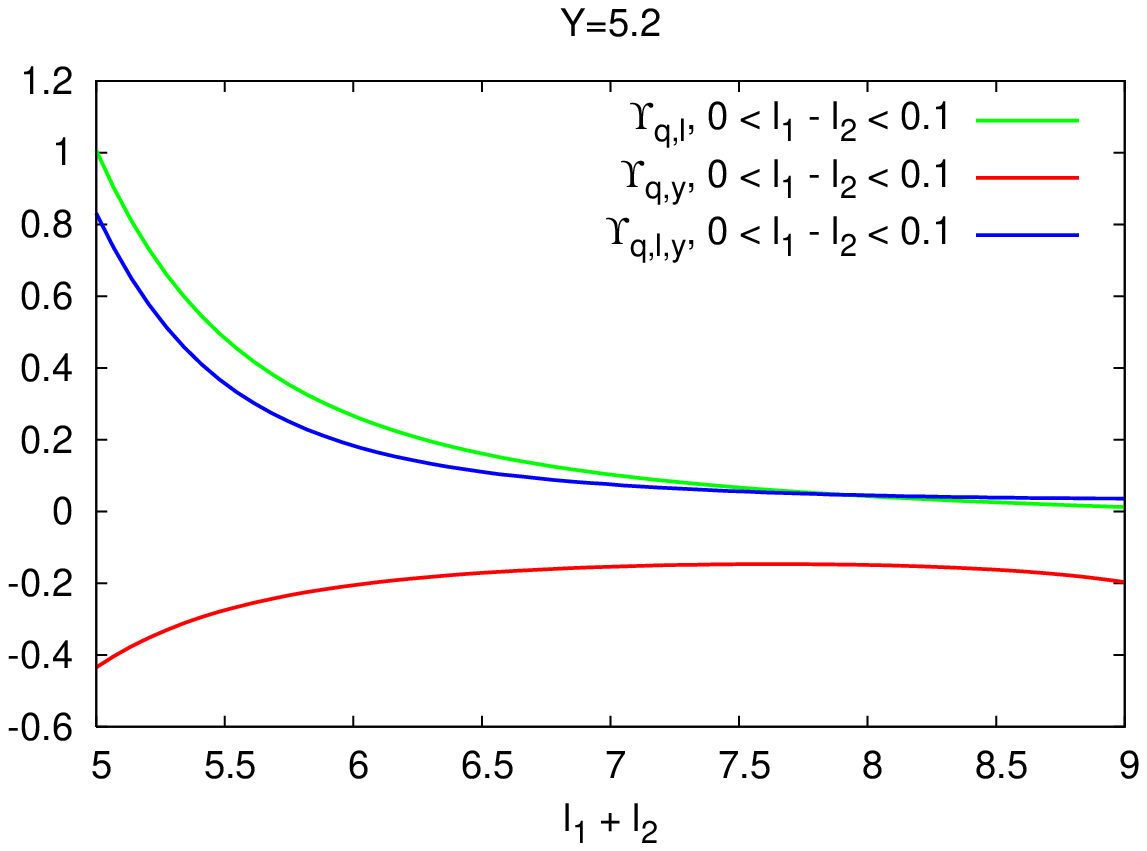, height=6truecm,width=0.48\tw}
\hfill
\epsfig{file=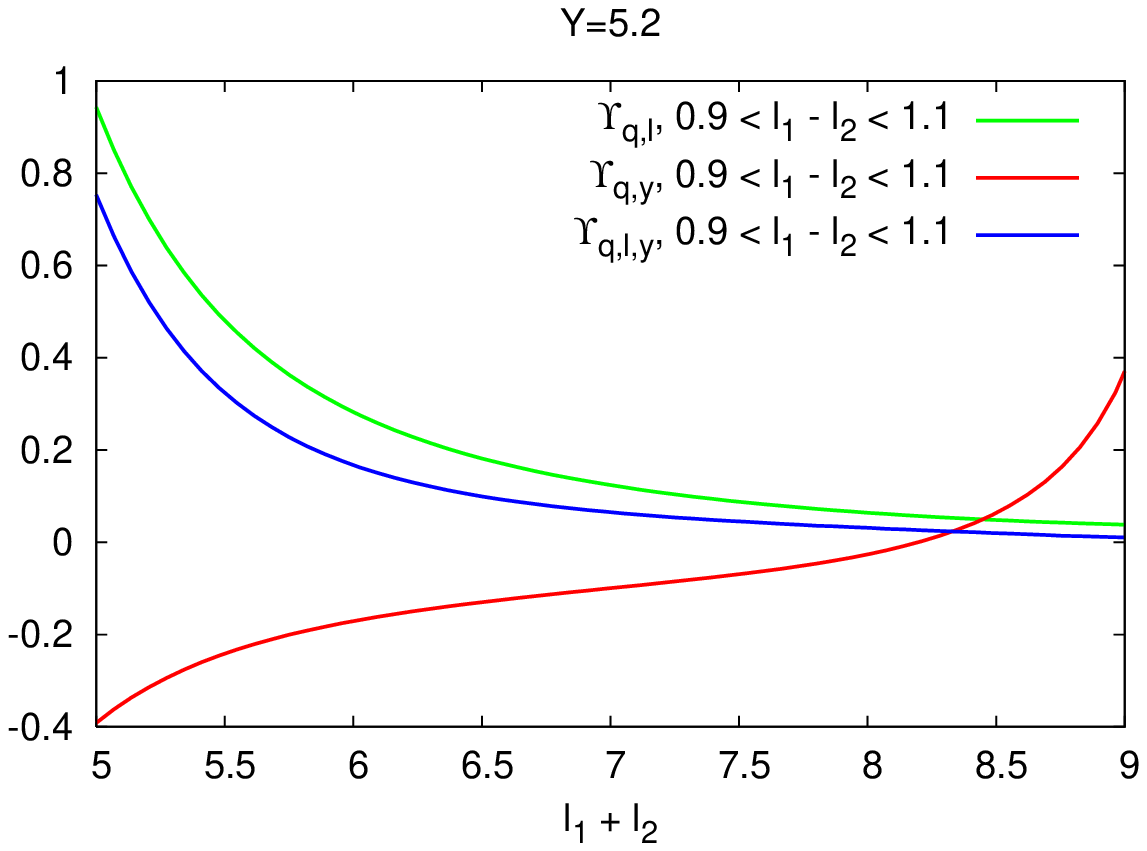, height=6truecm,width=0.48\tw}
\vskip .5cm
\caption{Corrections $\Upsilon_{q,\ell}$, $\Upsilon_{q,y}$ and 
$\Upsilon_{q,\ell,y}$ as functions of $\ell_1+\ell_2$ for $\ell_1-\ell_2=0.1$ (left) 
and $\ell_1-\ell_2=1.0$ (right) at $Y=5.2$} 
\label{fig:derUpsilonq}
\end{center}
}
\end{figure}

The smooth behavior of $\exp\Upsilon_q$ is displayed in Fig.~\ref{fig:Upsilonq}
as a function of the sum $(\ell_1+\ell_2)$ for fixed $(\ell_1-\ell_2)$ and vice versa.
The normalization of $(\exp\Upsilon_q-1)$ is roughly twice larger
($\times\frac{Nc}{CF}\approx2$) than that of $(\exp\Upsilon_g-1)$.
We  then consider derivatives of
this expression to get the corresponding iterative corrections shown in 
Fig.~\ref{fig:derUpsilonq}. The behavior of $\Upsilon_{q,\ell} ({\cal O}(\gamma_0^2))$, $\Upsilon_{q,y} ({\cal O}(\gamma_0^2))$
and $\Upsilon_{q,\ell\,y} ({\cal O}(\gamma_0^4))$ is in good agreement with 
our expectations as far as the order of magnitude and the normalization are concerned
(see also Fig.~\ref{fig:DUpsilon})
\footnote{it is also important to remark that $\Upsilon_{q,\ell},\,
\Upsilon_{q,\ell},\,\Upsilon_{q,\ell\,y}$ are $\times\frac{N_c}{C_F}\Upsilon_{g,\ell},\,
\Upsilon_{g,\ell},\,\Upsilon_{g,\ell\,y}$ .}.

\subsubsection{$\boldsymbol{\tilde{\delta}_1}$,
$\boldsymbol{\tilde{\delta}_2}$ and $\boldsymbol{\tilde{\delta}_c}$}
\label{subsub:tidelq}
%%%%%%%%%%%%%%%%%%%%%%%%%%%%%%%%%%%%%%%%%%%%%%%%%%%%%%%%%%%%%%%%%%%%%%%%%%%

\begin{figure}
\vbox{
\begin{center}
\epsfig{file=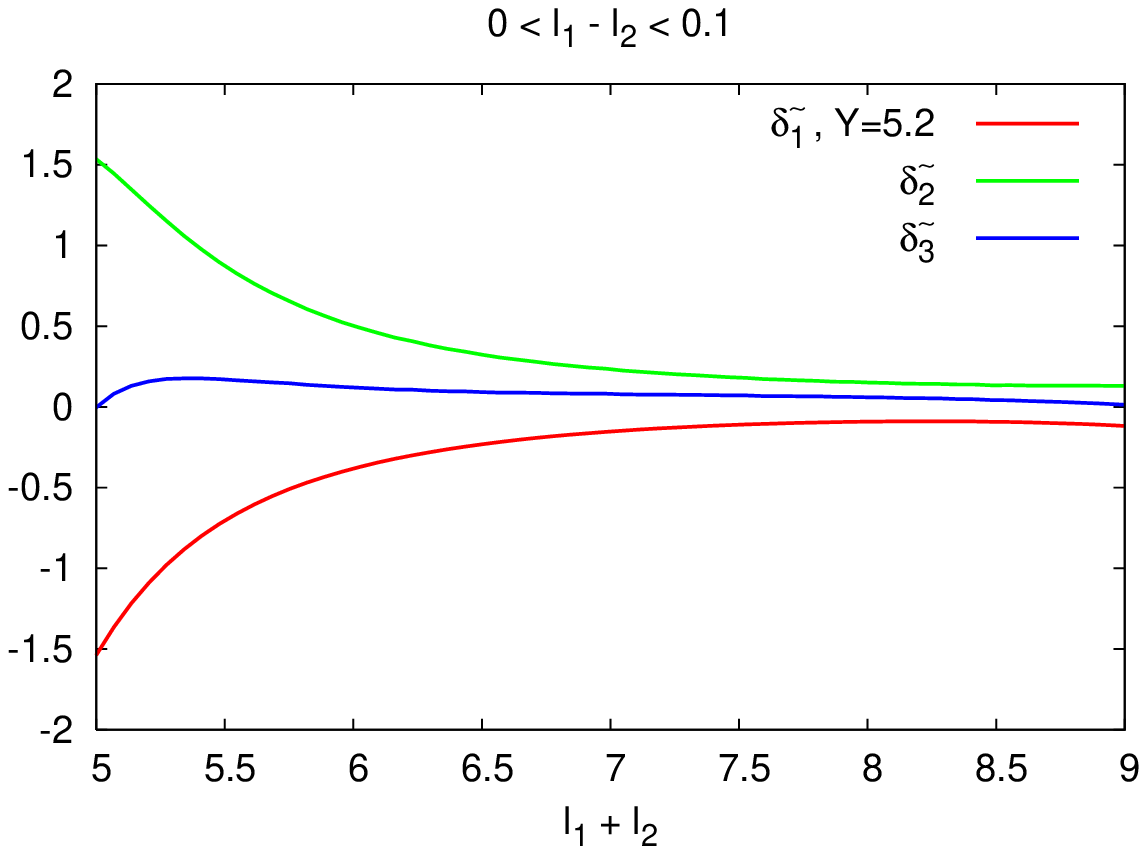, height=6truecm,width=0.48\tw}
\hfill
\epsfig{file=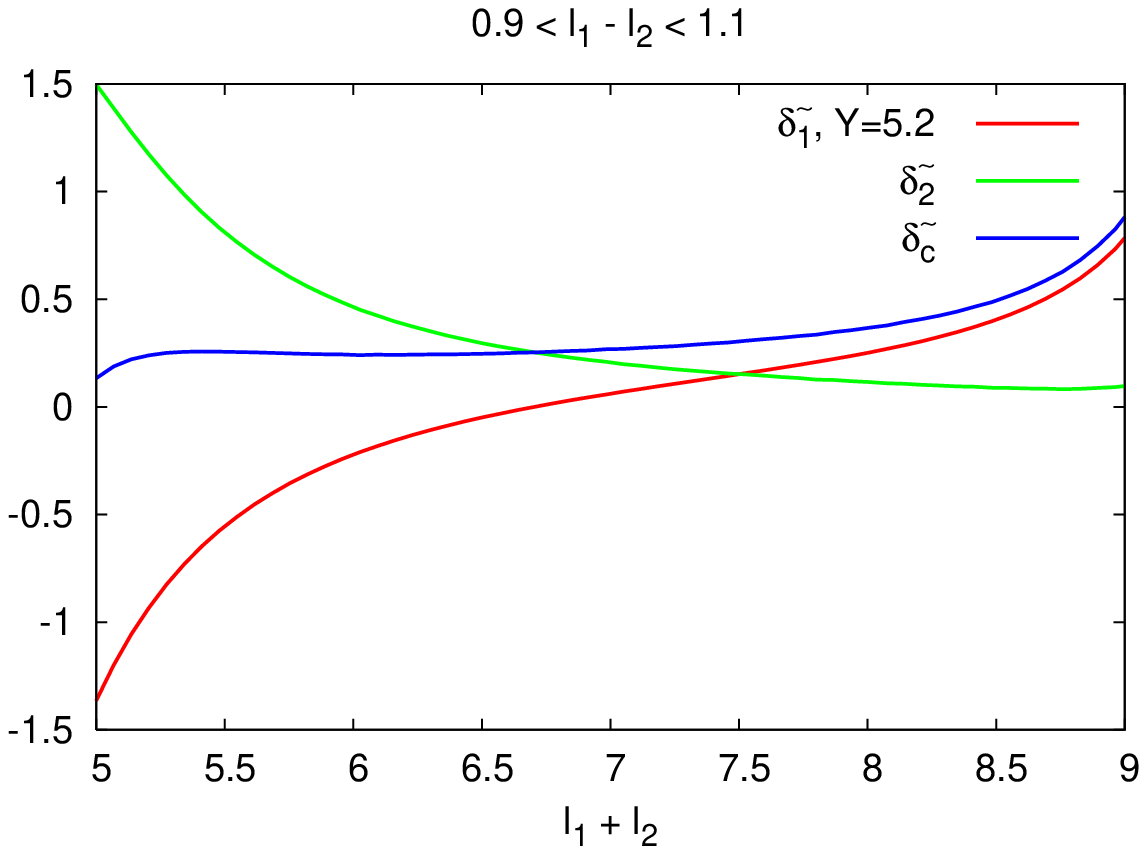, height=6truecm,width=0.48\tw}
\vskip .5cm
\caption{$\tilde\delta_1$, $\tilde\delta_2$ and $\tilde\delta_1+\tilde\delta_2$
as functions of $\ell_1+\ell_2$ for $\ell_1-\ell_2=0.1$ (left) and 
$\ell_1-\ell_2=1.0$ (right) at $Y=5.2$} 
\label{fig:tildedelta}
\end{center}
}
\end{figure}

We define
$$
\tilde{\delta}_{c}=\tilde{\delta}_1 + \tilde{\delta}_2
$$
as it appears in both the numerator and denominator of (\ref{eq:Qcorr}). 
In Fig.~\ref{fig:tildedelta} are displayed $\tilde\delta_1$, $\tilde\delta_2$
and their sum $\tilde\delta_c$ as functions of the sum
$(\ell_1+\ell_2)$ at fixed $(\ell_1-\ell_2)$
($\ell_1-\ell_2=0.1$, left) ($\ell_1-\ell_2=1.0$, right).

At $Y=5.2$, which corresponds to $\gamma_0 \approx 0.5$, the relative
magnitude of $\tilde\delta_1$ and $\tilde\delta_2$ is inverted
\footnote{it has been numerically investigated that the expected relative order of magnitude
of $\tilde\delta_1$ and $\tilde\delta_2$ is recovered for $Y\geq8.0$ (this value can be
eventually reached at LHC).} with respect to what is expected from respectively MLLA and NMLLA corrections
(see subsection \ref{subsection:estimate}). This is the only hint that, at
this energy, the expansion should be pushed to include all NMLLA
corrections to be reliable.

\begin{figure}
\vbox{
\begin{center}
\epsfig{file=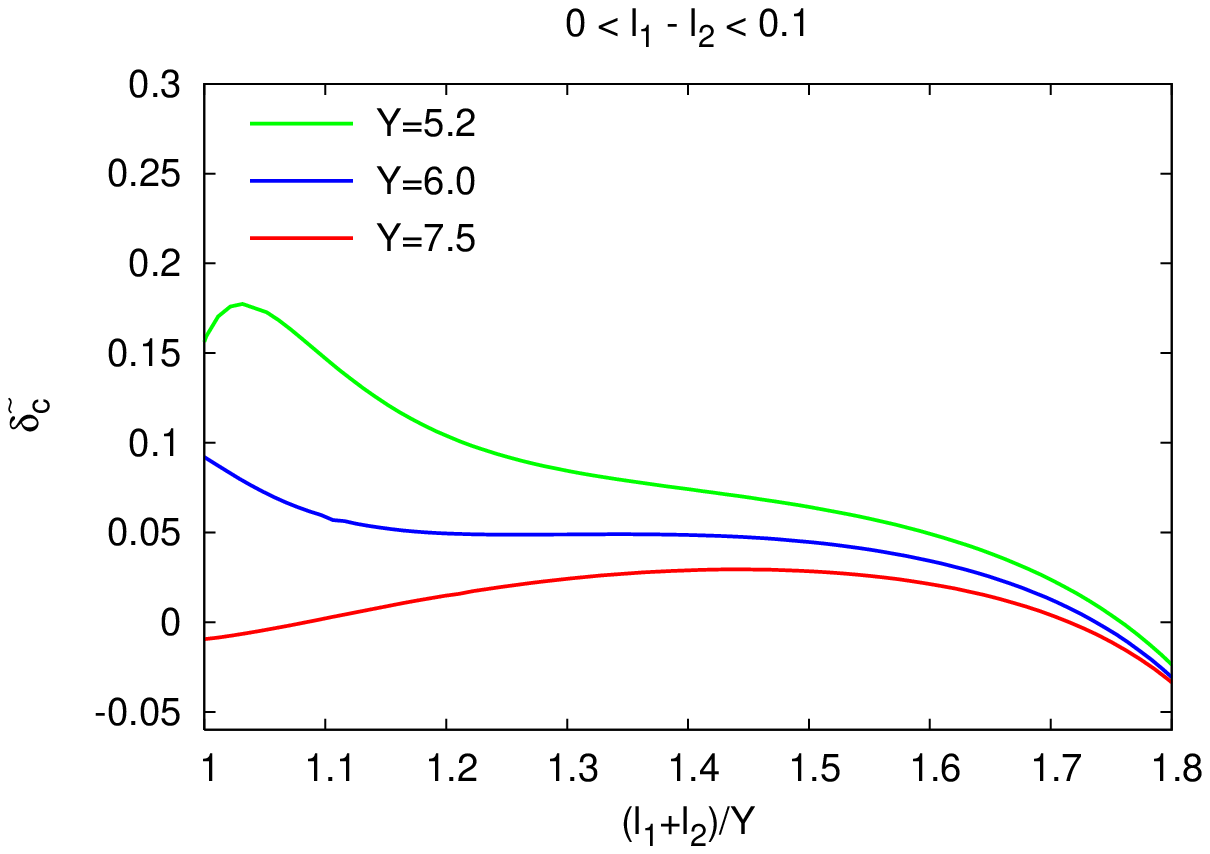, height=6truecm,width=0.48\tw}
\hfill
\epsfig{file=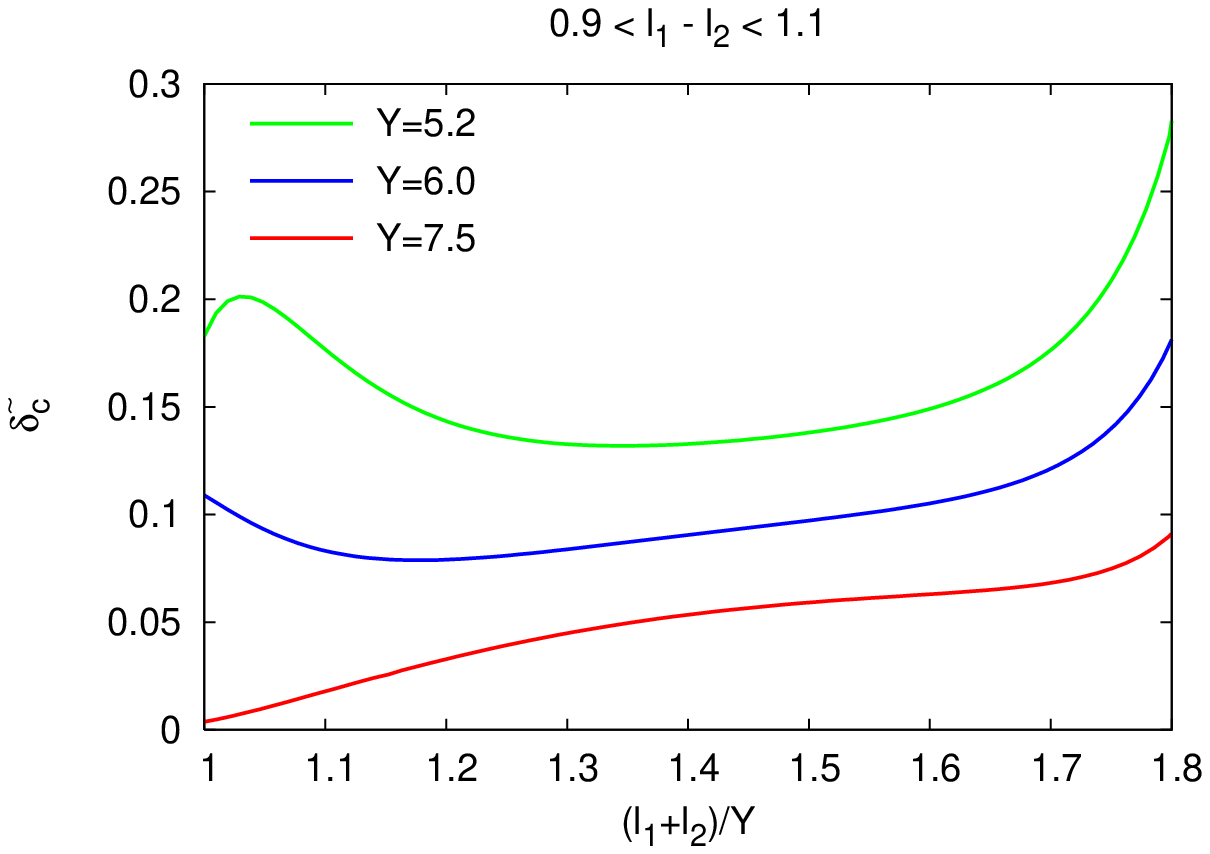, height=6truecm,width=0.48\tw}
\vskip .5cm
\caption{$\tilde\delta_c$ as a function of
$(\ell_1+\ell_2)/Y$ for $\ell_1-\ell_2=0.1$ (left) and $\ell_1-\ell_2=1.0$ (right)
at $Y=5.2,\,6.0,\,7.5$}
\label{fig:delta12chiQ}
\end{center}
}
\end{figure}

Large cancellations are, like for gluons, seen to occur in
$\tilde\delta_c$, making the sum of corrections quite small.
In order to study the behavior of $\tilde\delta_c$ as $Y$
increases, it is enough to look at Fig.~\ref{fig:delta12chiQ} where
we compare $\tilde\delta_c$ at $Y=5.2,\,6.0,\,7.5$.

\subsubsection{Global role of corrections in the iterative procedure}
%%%%%%%%%%%%%%%%%%%%%%%%%%%%%%%%%%%%%%%%%%%%%%%%%%%%%%%

\begin{figure}
\vbox{
\begin{center}
\epsfig{file=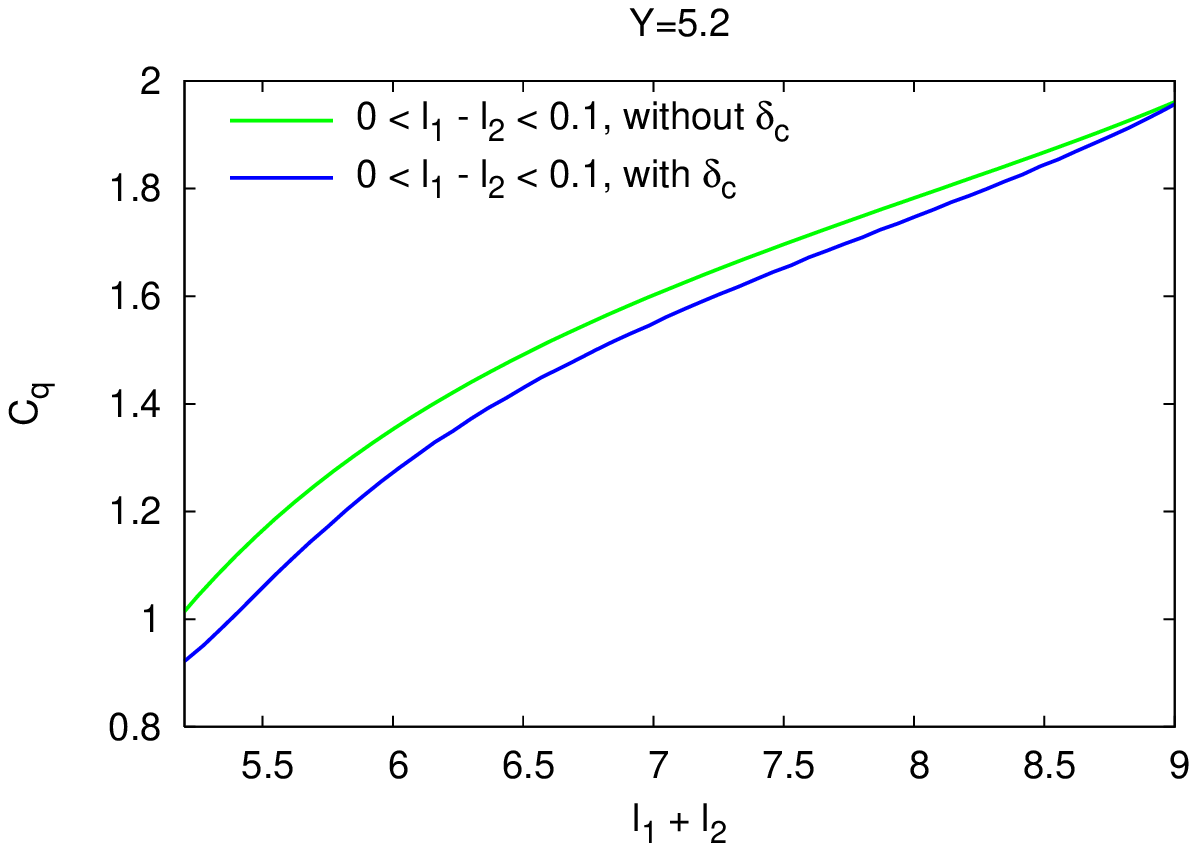, height=6truecm,width=0.48\tw}
\hfill
\epsfig{file=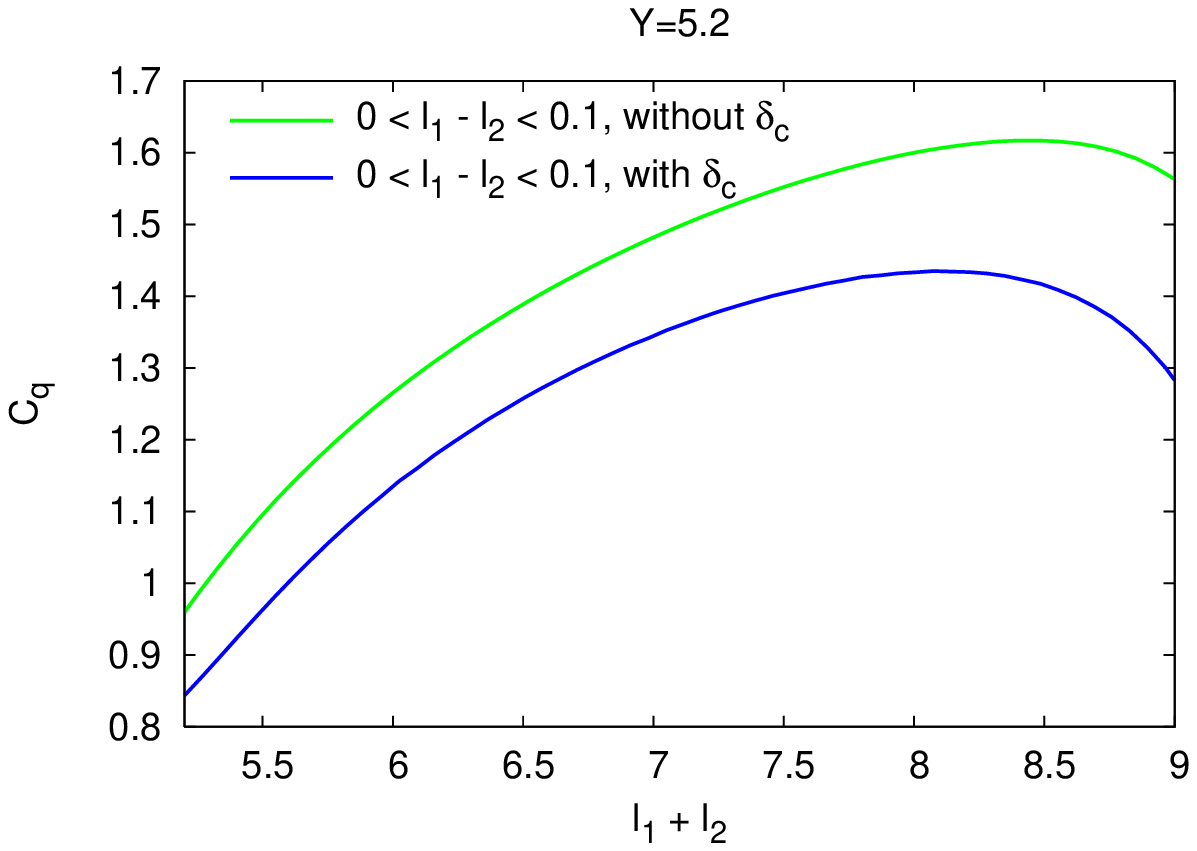, height=6truecm,width=0.48\tw}
\vskip 0.5cm
\epsfig{file=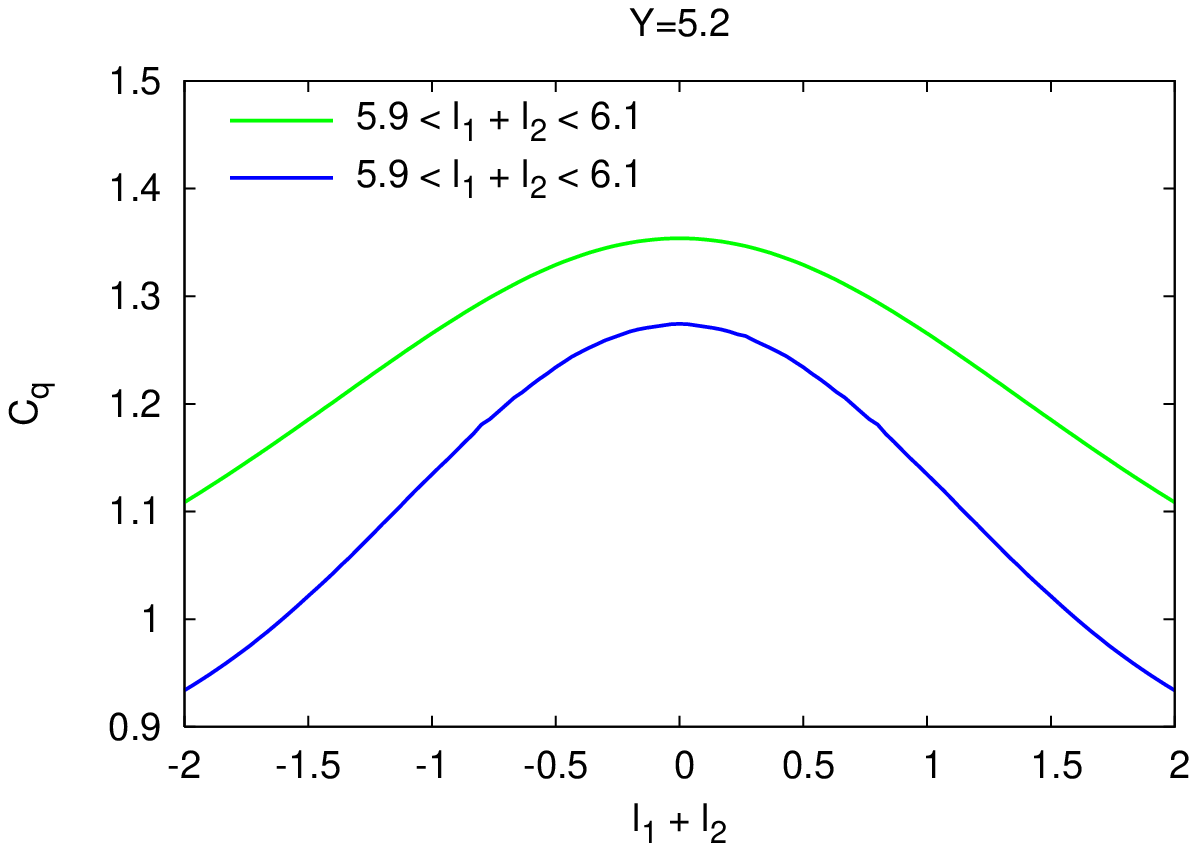, height=6truecm,width=0.48\tw}
\hfill
\epsfig{file=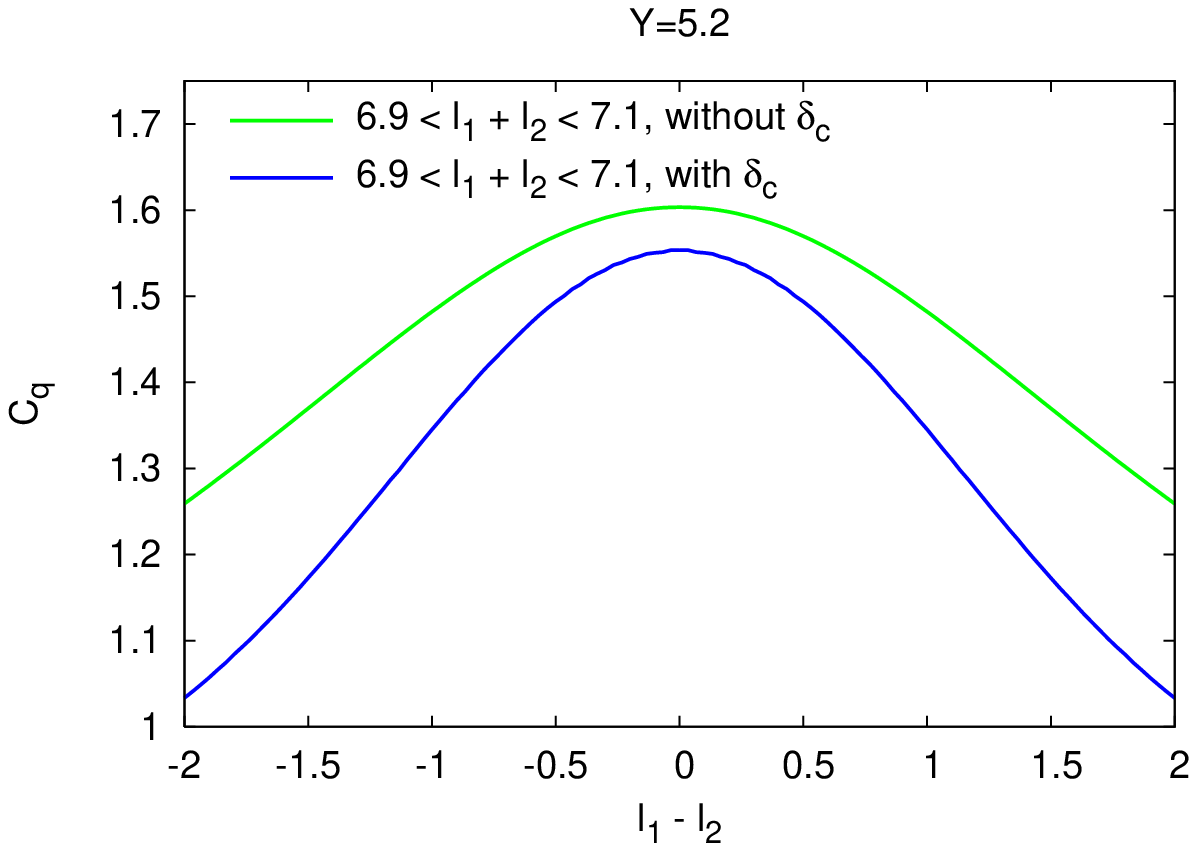, height=6truecm,width=0.48\tw}
\vskip .5cm
\caption{${\cal C}_q$ (blue) compared with $\exp\Upsilon_q$ 
(green)}
\label{fig:corrqUpsilon}
\end{center}
}
\end{figure}

It is displayed in Fig.~\ref{fig:corrqUpsilon}.
${\tilde\delta_c}$ does not affect $\exp\Upsilon_q$ near the main diagonal ($\ell_1=\ell_2$), but it 
does far from it.  We find the same behavior as in the case of a gluon jet.

\section{COMPARING DLA AND MLLA CORRELATIONS}
\label{section:DLAcomp}
%%%%%%%%%%%%%%%%%%%%%%%%%%%%%%%%%%%%%%%%%%%%%%

\begin{figure}
\vbox{
\begin{center}
\epsfig{file=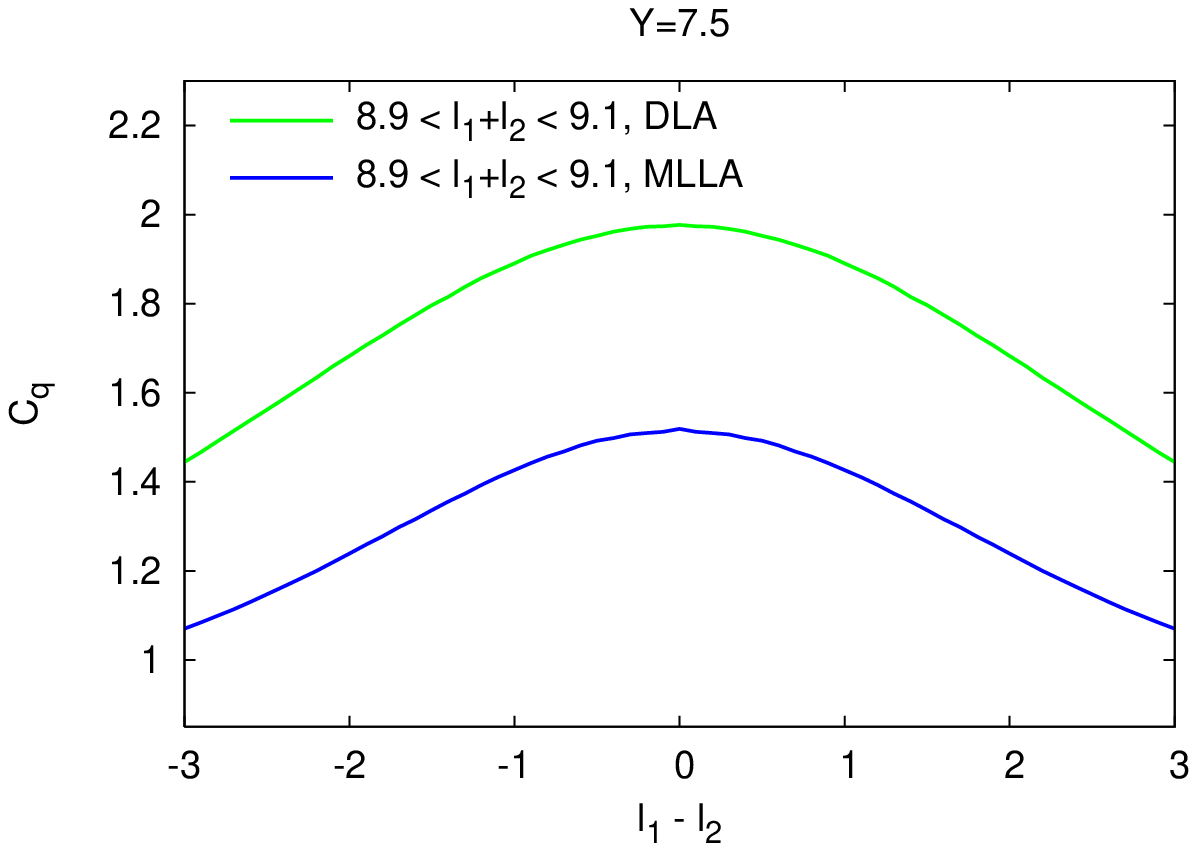, height=6truecm,width=0.48\tw}
\hfill
\epsfig{file=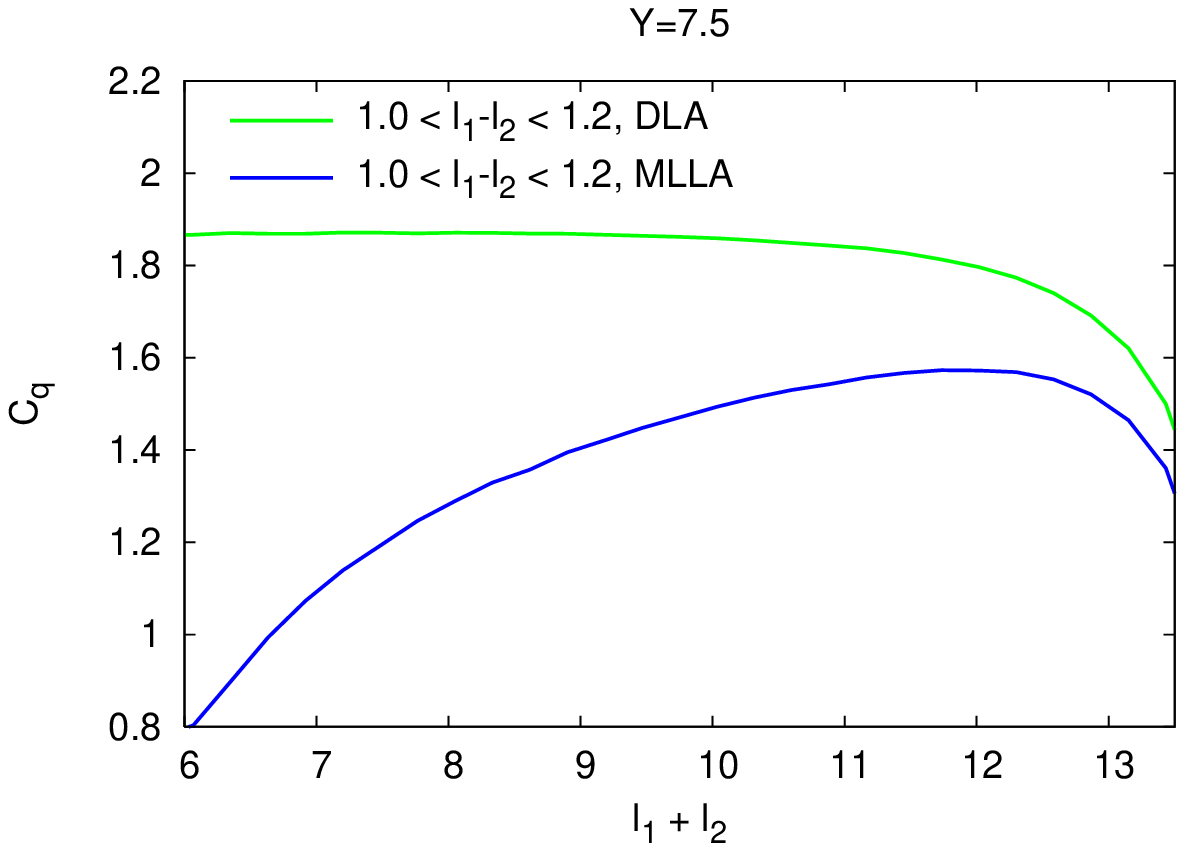, height=6truecm,width=0.48\tw}
\vskip .5cm
\caption{Comparing DLA and MLLA correlations}
\label{fig:DLAMLLA}
\end{center}
}
\end{figure}

In Fig.~\ref{fig:DLAMLLA} we compare the quark correlator at DLA and MLLA.
The large gap between the two curves accounts for the energy balance
that is partially restored in MLLA
by introducing hard corrections in the partonic evolution equations
%(\ref{eq:CGfull}) and (\ref{eq:Qcorr})
(terms $\propto$ $a$, $b$ and $\frac34$);
the DLA curve is obtained by setting $a$, $b$ and $\frac34$
to zero in (\ref{eq:CGfull}) and (\ref{eq:Qcorr});
${\cal C}_q$ is a practically constant function of
$\ell_1+\ell_2$ in almost the whole range, and decreases
when $\ell_1+\ell_2\rightarrow 2Y$ by the running of $\alpha_s$.
The MLLA increase of ${\cal C}_q$ with $\ell_1+\ell_2$ follows from energy
conservation.
Similar results are obtained for ${\cal C}_g$.

%\null\newpage

%%%%%%%%%%%%%%%%%%%%%%%%%%%%%%%%%%%%%%%%%%%%%%%%%%%%%%%%%%%%%%%%%%%%%%%%%%%%%%
%\null

%\listoffigures

%%%%%%%%%%%%%%%%%%%%%%%%%%%%%%%%%%%%%%%%%%%%%%%%%%%%%%%%%%%%%%%%%%%%%%%%%%%%%%
\newpage
%%%%%%%%%%%%%%%%%%%%%%%%%%%%%%%%%%%%%%%%%%%%%%%%%%%%%%%%%%%%%%%%%%%%%%%%%%%%%

%%%%%%%%%%%%%%%%%%%%%%%%%%%%%%%%%%%%%%%%%%%%%%%%%%%%%%%%%%%%%%%%%%%%%%%%%%%%%

% \include{Correl}

%%%%%%%%%%%%%%%%%%%%%%%%%%%%%%%%%%%%%%%%%%%%%%%%%%%%%%%%%%%%%%%%%%%%%%%%%%%%%

%\newpage

\null

\noindent

\vskip 10 cm

\chapter{Single inclusive distribution and 2-particle correlations
inside one jet at ``Modified Leading Logarithmic Approximation''
of Quantum Chromodynamics ;\newline
 II : Steepest descent evaluation at small $\boldsymbol x$}
\label{sub:article3}

%%%%%%%%%%%%%%%%%%%%%%%%%%%%%%%%%%%%%%%%%%%%%%%%%%%%%%%%%%%%%%%%%%%%%%%%%%

%\newpage\null

% \include{SteepDes}

\begin{titlepage}

\setcounter{page}{235}

%%%%%%%%%%%%%%%%%%%%%%%%%%%%%%%%%%%%%%%%%%%%%%%%%%%%%%%%%%%%%%%%%%%%%%%%%%%%%%
July 2006 \hfill hep-ph/0607223   , JHEP 09 (2006) 014

\vskip 5.5cm

\centerline{\bf SINGLE INCLUSIVE DISTRIBUTION AND
TWO-PARTICLE CORRELATIONS INSIDE ONE JET}
\medskip
\centerline{\bf  AT ``MODIFIED LEADING LOGARITHMIC APPROXIMATION''
 OF QUANTUM CHROMODYNAMICS}
\medskip
\centerline{\bf  II : STEEPEST DESCENT EVALUATION AT SMALL $\boldsymbol{X}$}

\vskip 1cm

\centerline{Redamy Perez-Ramos
\footnote{E-mail: perez@lpthe.jussieu.fr}
}

\baselineskip=15pt

\smallskip
\centerline{\em Laboratoire de Physique Th\'eorique et Hautes Energies
\footnote{LPTHE, tour 24-25, 5\raise 3pt \hbox{\tiny \`eme} \'etage,
Universit\'e P. et M. Curie, BP 126, 4 place Jussieu,
F-75252 Paris Cedex 05 (France)}}
\centerline{\em Unit\'e Mixte de Recherche UMR 7589}
\centerline{\em Universit\'e Pierre et Marie Curie-Paris6; CNRS;
Universit\'e Denis Diderot-Paris7}

\vskip 2cm

{\bf Abstract}: The MLLA single inclusive distribution inside one
high energy (gluon) jet at small $x$ is estimated by the steepest
descent method.
Its analytical expression is obtained outside the ``limiting spectrum''.
It is then used to evaluate 2-particle correlations at the same level of
generality. The dependence of both observables on the ratio between the
infrared cutoff $Q_0$ and $\Lambda_{QCD}$ is studied.
Fong \& Webber's results for correlations are recovered at
the limits when this ratio goes to $1$ and when one stays close to
the peak of the single inclusive distribution.

\vskip 1 cm

{\em Keywords: Perturbative Quantum Chromodynamics, Particle Correlations
in jets, High Energy Colliders}

\vfill

%%%%%%%%%%%%%%%%%%%%%%%%%%%%%%%%%%%%%%%%%%%%%%%%%%%%%%%%%%%%%%%%%%%%%%%%%%%%%

\end{titlepage}

%%%%%%%%%%%%%%%%%%%%%%%%%%%%%%%%%%%%%%%%%%%%%%%%%%%%%%%%%%%%%%%%%%%%%%%%%%%%%

%\tableofcontents

%\newpage
%%%%%%%%%%%%%%%%%%%%%%%%%%%%%%%%%%%%%%%%%%%%%%%%%%%%%%%%%%%%%%%%%%%%%%%%%%%%%

%%%%%%%%%%%%%%%%%%%%%%%%%%%%%%%%%%%%%%%%%%%%%%%%%%%%%%%%%%%%%%%%%%%%%%%%%%%%%
\section{INTRODUCTION}
\label{section:introSD}
%%%%%%%%%%%%%%%%%%%%%%%%%%%%%%%%%%%%%%%%%%%%%%%%%%%%%%%%%%%%%%%%%%%%%%%%%%%%%

Exactly solving the MLLA evolution equations
for the quark and gluon
inclusive spectra and for 2-particle correlations inside one jet
provided, at small $x$, in \cite{RPR2}, analytical
expressions  for these observables, which were unfortunately
limited, for technical reasons to the ``limiting spectrum''
$\lambda \equiv \ln(Q_0/\Lambda_{QCD} =0$. The goal of this second work is
to go beyond this limit in an approximate scheme which proves
very economical and powerful: the steepest descent (SD) method.
It offers sizable technical progress in the calculation of both
observables.

First, we perform a SD evaluation of the (quark and) gluon
single inclusive distributions. Their full dependence on $\lambda$ is
given, including the normalization. The well known  shift to smaller
values of $x$ of the maximum of the distribution, as compared with
DLA calculations is checked, as well as its Gaussian shape
around the maximum. Comparison with the results obtained numerically in
\cite{DKTSD} is done.

As shown in \cite{RPR2}, knowing the logarithmic derivatives of
the inclusive spectra immediately gives access to 2-particle correlations.
This is accordingly our next step. Since, in particular, the former prove
to be infra-red stable in the limit $\lambda\to 0$, the result can be
safely compared with the exact one obtained in \cite{RPR2}. The agreement
turns out to be excellent, and increases with the energy scale of the
process.

Last, we evaluate 2-particle correlations inside one high energy jet
and study their behavior at $Q_0\ne\Lambda_{QCD}$.
That one recovers the results of Fong \& Webber \cite{FWSD} close to the peak
of the single inclusive distribution and when $\lambda
\to 0$ is  an important test of the validity and efficiency
of the SD method. The  quantitative predictions do not substantially
differ from the ones of \cite{RPR2} for the ``limiting spectrum'', which
stays  the best candidate to reproduce experimental results.

A conclusion summarizes the achievements, limitations and expectations
of \cite{RPR2} and of the present work. It is completed with two technical
appendices.

%%%%%%%%%%%%%%%%%%%%%%%%%%%%%%%%%%%%%%%%%%%%%%%%%%%%%%%%%%%%%%%%%%%%%%%%%%%%%
\section{STEEPEST DESCENT EVALUATION OF THE SINGLE INCLUSIVE DISTRIBUTION}
\label{sec:SD}
%%%%%%%%%%%%%%%%%%%%%%%%%%%%%%%%%%%%%%%%%%%%%%%%%%%%%%%%%%%%%%%%%%%%%%%%%%%%%

We consider  the production of one hadron inside a quark or a 
gluon jet in a hard process. It carries the fraction $x$ of the total energy
$E$ of the jet. $\Theta_0$ is the half opening angle of the jet while
$\Theta$ is the angle corresponding to the first splitting with
 energy fraction $x\ll z\ll1$.

\subsection{Variables and kinematics}
%%%%%%%%%%%%%%%%%%%%%%%%%%%%%%%%%%%%%

The variables
and  kinematics of the process under consideration are the same as in section
3.1 of \cite{RPR2}.

\subsection{Evolution equations for particle spectra at MLLA}
%%%%%%%%%%%%%%%%%%%%%%%%%%%%%%%%%%%%%%%%%%%%%%%%%%%%%%%%%%%%%%

We define like in \cite{RPR2} the logarithmic parton densities 

$$
Q(\ell)\equiv xD_Q(x),\qquad G(\ell)=xD_G(x)
$$

for quark and gluon jets in terms of which
the system of evolution equations for particle
spectra at small $x$ (see eqs.~(42) and (43) of \cite{RPR2}) read

\begin{equation}
Q(\ell,y)= \delta(\ell) + \frac{C_F}{N_c}\!\!\int_0^\ell d\ell'\int_0^y dy'
\gamma_0^2(\ell'+y')\Big(1
-\frac34\delta(\ell'-\ell) \Big) G(\ell',y'),
\label{eq:solqSD}
\end{equation}

\begin{equation}
G(\ell,y) = \delta(\ell)
+\int_0^{\ell} d\ell'\int_0^{y} dy' \gamma_0^2(\ell'+y')\Big(
 1  -a\delta(\ell'-\ell) \Big) G(\ell',y'),
\label{eq:solgSD}
\end{equation}
where
\begin{equation}
a = \frac{1}{4N_c}\bigg[\frac{11}{3}N_c + \frac{4}{3}n_f T_R
 \bigg(\!1-\frac{2C_F}{N_c}\!\bigg)\bigg]\stackrel{n_f=3}{=}0.935.
\label{eq:adefSD}
\end{equation}
The terms $\propto\frac34$ in (\ref{eq:solqSD}) and $\propto a$ in
(\ref{eq:solgSD}) 
account for hard
corrections to soft gluon multiplication, sub-leading $g\!\to\!q\bar q$ splittings,
strict angular ordering and energy conservation.

\subsection{Evolution equations; steepest descent evaluation}
\label{subsection:SDeval}
%%%%%%%%%%%%%%%%%%%%%%%%%%%%%%%%%%%%%%%%%%%%%%%%%%%%%%%%%%

The exact solution of (\ref{eq:solgSD}) is demonstrated in \cite{RPR2} to be given
by the Mellin's integral representation

\begin{eqnarray}\label{eq:MLLAalphasrunSD}
G\left(\ell,y\right) &=& \left(\ell\!+\!y\!+\!\lambda\right)\!\!\iint
\frac{d\omega\, d\nu}{\left(2\pi i\right)^2}e^{\omega\ell+\nu y}
\!\!\int_{0}^{\infty}\frac{ds}{\nu+s}\!\!\left(\!\frac{\omega
\left(\nu+s\right)}{\left(\omega+s\right)\nu}\!\right)^{1/\beta
\left(\omega-\nu\right)}\!\!\left(\!\frac{\nu}{\nu+s}\!\right)^
{a/\beta}\,e^{-\lambda s}\cr
&=&\left(\ell\!+\!y\!+\!\lambda\right)\!\!\iint
\frac{d\omega\, d\nu}{\left(2\pi i\right)^2}e^{\omega\ell+\nu y}
\!\!\int_{0}^{\infty}\frac{ds}{\nu+s}\left(\!\frac{\nu}{\nu+s}\!\right)^
{a/\beta}e^{\sigma(s)},
\end{eqnarray}

where we have exponentiated the kernel (symmetrical in $(\omega,\nu)$)
\begin{equation}\label{eq:sigma}
\sigma(s)=\frac1{\beta(\omega-\nu)}\ln\left(\!\frac{\omega(\nu+s)}
{\nu(\omega+s)}\!\right)-\lambda s.
\end{equation}

(\ref{eq:MLLAalphasrunSD}) will be estimated by the SD method.
The value $s_0$ of the saddle point, satisfying
$\frac{d\sigma(s)}{ds}\Big\vert_{s=s_0}=0$, reads (see \cite{DLASD})

\begin{equation}\label{eq:saddlepoint}
s_0(\omega,\nu)=\frac12\left[\sqrt{\frac4{\beta\lambda}+(\omega-\nu)^2}-(\omega+\nu)
\right].
\end{equation}
One makes a Taylor expansion of $\sigma(s)$ nearby $s_0$:

\begin{equation}\label{eq:sigmas0}
\sigma(s)=\sigma(s_0)+\frac12\sigma''(s_0)(s-s_0)^2+{\cal O}\left((s-s_0)^2\right),
\quad
\sigma''(s_0)=-\beta\lambda^2\sqrt{\frac{4}{\beta\lambda}+(\omega-\nu)^2}<0,
\end{equation}

such that

\begin{eqnarray}\label{eq:sigmavalue}
\int_{0}^{\infty}\frac{ds}{\nu+s}\!\!\left(\!\frac{\omega
\left(\nu+s\right)}{\left(\omega+s\right)\nu}\!\right)^{1/\beta
\left(\omega-\nu\right)}\!\!\left(\!\frac{\nu}{\nu+s}\!\right)^
{a/\beta}\,e^{-\lambda s}
\!\!&\!\!\stackrel{\lambda\gg1}{\approx}\!\!&\!\!
2\sqrt{\frac{\pi}2}\displaystyle{\frac{e^{\sigma(s_0)}}{(\nu+s_0)
\sqrt{\mid\!\sigma''(s_0)\!\mid}}}\left(\!\frac{\nu}{\nu+s_0}\!\right)^{a/\beta}\!\!.\cr
&&
\end{eqnarray}

The condition $\lambda\!\gg\!1$$\Rightarrow$$\alpha_s/\pi\!\ll\!1$ 
\footnote{in (\ref{eq:sigmas0}), $\lambda$ appears to the power $3/2>1$,
which guarantees the fast convergence of the SD as $\lambda$ increases.}
guarantees, in particular, the convergence of the perturbative approach. Substituting (\ref{eq:sigmavalue}) in 
(\ref{eq:MLLAalphasrunSD}) yields

\begin{equation}\label{eq:SpecDD}
G\left(\ell,y\right)\approx2\sqrt{\frac{\pi}2}(\ell+y+\lambda)
\iint\frac{d\omega\, d\nu}{\left(2\pi i\right)^2}\,
\displaystyle{\frac{e^{\phi\left(\omega,\nu,\ell,y\right)}}{(\nu+s_0)
\sqrt{\mid\!\sigma''(s_0)\!\mid}}}\left(\!\frac{\nu}{\nu+s_0}\!\right)^{a/\beta},
\end{equation}

where the argument of the exponential is

\begin{equation}\label{eq:phiexp}
\phi\left(\omega,\nu,\ell,y\right)=\omega\ell+\nu y+
\frac1{\beta\left(\omega-\nu\right)}
\ln{\frac{\omega\left(\nu+s_0\right)}{\left(\omega+s_0\right)\nu}}-
\lambda s_0.
\end{equation}

Once again, we perform the SD method to evaluate 
(\ref{eq:SpecDD}). The saddle point $(\omega_0,\nu_0)$ satisfies
the equations

\vbox{
\begin{subequations}
\begin{equation}\label{eq:deromega}
\frac{\partial\phi}{\partial\omega}=\ell-\frac1{\beta\left(\omega-\nu\right)^2}
\ln{\frac{\omega\left(\nu+s_0\right)}{\left(\omega+s_0\right)\nu}}+\frac1
{\beta\omega\left(\omega-\nu\right)}-\lambda\frac{\left(\nu+s_0\right)}
{\left(\omega-\nu\right)}=0,
\end{equation}

\begin{equation}\label{eq:dernu}
\frac{\partial\phi}{\partial\nu}=y+\frac1{\beta\left(\omega-\nu\right)^2}
\ln{\frac{\omega\left(\nu+s_0\right)}{\left(\omega+s_0\right)\nu}}-
\frac1{\beta\nu\left(\omega-\nu\right)}+\lambda
\frac{\left(\omega+s_0\right)}{\left(\omega-\nu\right)}=0.
\end{equation}
\end{subequations}
}

Adding and subtracting (\ref{eq:deromega}) and (\ref{eq:dernu}) gives respectively

\begin{subequations}
\begin{equation}
\omega_0\nu_0=\frac1{\beta\left(\ell+y+\lambda\right)},
\end{equation}

\begin{equation}\label{eq:yml}
y-\ell=\frac1{\beta\left(\omega_0-\nu_0\right)}
\left(\frac1{\omega_0}+\frac1{\nu_0}\right)-\frac2
{\beta\left(\omega_0-\nu_0\right)^2}\ln{\frac{\omega_0
\left(\nu_0+s_0\right)}{\left(\omega_0+s_0\right)\nu_0}}-
\lambda\frac{\omega_0+\nu_0+2s_0}{\omega_0-\nu_0};
\end{equation}
\end{subequations}

$(\omega_0,\nu_0)$ also satisfies (from (\ref{eq:saddlepoint}))

\begin{equation}
\left(\omega_0+s_0\right)\left(\nu_0+s_0\right)=\frac1{\beta\lambda}.
\end{equation}

One can substitute the expressions
(\ref{eq:deromega}) and (\ref{eq:dernu}) of  $\ell$ and $y$ 
into (\ref{eq:phiexp}), which yields

\begin{equation}\label{eq:phiexpbis}\
\varphi\equiv\phi(\omega_0,\nu_0,\ell,y)=\frac{2}
{\beta\left(\omega_0-\nu_0\right)}\ln{\frac{\omega_0\left(\nu_0+s_0\right)}
{\left(\omega_0+s_0\right)\nu_0}}.
\end{equation}

Introducing the variables $(\mu,\upsilon)$ \cite{DLASD} to parametrize
$(\omega_0,\nu_0)$ through
\begin{equation}\label{eq:muupsilon}
\omega_0\left(\nu_0\right)=\frac1{\sqrt{\beta(\ell\!+\!y\!+\!\lambda)}}
e^{\pm\mu(\ell,y)},\qquad \left(\omega_0+s_0\right)
\left(\nu_0+s_0\right)=\frac1{\sqrt{\beta\lambda}}e^{\pm\upsilon(\ell,y)},
\end{equation}
one rewrites (\ref{eq:phiexpbis}) and (\ref{eq:yml}) respectively
in the form
\begin{equation}\label{eq:phi}
\varphi(\mu,\upsilon)=\frac2{\sqrt{\beta}}\left(\sqrt{\ell+y+\lambda}-\sqrt{\lambda}\right)
\,\frac{\mu-\upsilon}{\sinh\mu-\sinh\upsilon},
\end{equation}

\begin{subequations}
\begin{equation}\label{eq:ratiomunu}
\frac{y-\ell}{y+\ell}=\frac{\left(\sinh 2\mu-2\mu\right)-\left(\sinh 2\upsilon-2\upsilon\right)}{2\left(\sinh^2\mu-\sinh^2\upsilon\right)};
\end{equation}

moreover, since $\omega_0-\nu_0=(\omega_0-s_0)-(\nu_0-s_0)$, $(\mu,\upsilon)$ also
satisfy

\begin{equation}\label{eq:relmunu}
\frac{\sinh\upsilon}{\sqrt{\lambda}}=\frac{\sinh\mu}{\sqrt{\ell+y+\lambda}}.
\end{equation}
\end{subequations}

Performing a Taylor expansion of $\phi(\omega,\nu,\ell,y)$ around $(\omega_0,\nu_0)$,
which needs evaluating $\frac{\partial^2\phi}{\partial\omega^2}$, $\frac{\partial^2\phi}{\partial\nu^2}$ and 
$\frac{\partial^2\phi}{\partial\omega\partial\nu}$ (see appendix \ref{sec:DDDet}),
treating $(Y+\lambda)$ as a large parameter and making use of 
(\ref{eq:muupsilon}) provides the SD result

\begin{eqnarray}
G(\ell,y)\approx{\cal N}(\mu,\upsilon,\lambda)\exp\Big[
\frac2{\sqrt{\beta}}\left(\sqrt{\ell+y+\lambda}-\sqrt{\lambda}\right)
\frac{\mu-\upsilon}{\sinh\mu-\sinh\upsilon}+\upsilon
-\frac{a}{\beta}(\mu-\upsilon)\Big],\cr
&&
\label{eq:Specalphasrunmlla}
\end{eqnarray}

where
$$
{\cal N}(\mu,\upsilon,\lambda)=\frac12(\ell\!+\!y\!+\!\lambda)
\frac{\left(\frac{\beta}{\lambda}\right)^{1/4}}{\sqrt{\pi\cosh\upsilon\,
DetA(\mu,\upsilon)}}\left(\frac{\lambda}{\ell+y+\lambda}\right)^{a/2\beta}
$$

with (see details in appendix \ref{sec:DDDet})

\begin{equation}\label{eq:determinant}
DetA(\mu,\nu)=\beta\,(\ell\!+\!y\!+\!\lambda)^3
\left[\frac{(\mu\!-\!\upsilon)\cosh\mu\cosh\upsilon\!+\!\cosh\mu\sinh\upsilon
\!-\!\sinh\mu\cosh\upsilon}
{\sinh^3\mu\cosh\upsilon}\right].
\end{equation}

\subsubsection{Shape of the spectrum given in eq.~(\ref{eq:Specalphasrunmlla})}
\label{subsub:shape}
%%%%%%%%%%%%%%%%%%%%%%%%%%%%%%%%%%%%%%%%%%%%%%%%%%%%%%%%%%%%%%%%%%

We normalize (\ref{eq:Specalphasrunmlla}) by
the MLLA mean multiplicity inside one jet \cite{EvEqSD}
$$
\bar{n}(Y)\stackrel{\lambda\gg1}{\approx}\frac12\left(\frac{Y+\lambda}{\lambda}\right)^
{\displaystyle{-\frac12\frac{a}{\beta}+\frac14}}
\exp\left[\frac2{\sqrt{\beta}}\left(\sqrt{Y+\lambda}-\sqrt{\lambda}\right)\right].
$$
The normalized expression for the single inclusive distribution
as a function of $\ell=\ln(1/x)$ is accordingly
 obtained by setting $y=Y-\ell$ in (\ref{eq:Specalphasrunmlla})

\begin{eqnarray}
\frac{G(\ell,Y)}{\bar{n}(Y)}\approx\sqrt{\frac{\beta^{1/2}(Y\!+\!\lambda)^{3/2}}
{\pi\cosh\upsilon DetA(\mu,\upsilon)}}\!\exp\!\Big[
\frac2{\sqrt{\beta}}\left(\sqrt{Y\!+\!\lambda}\!-\!\sqrt{\lambda}\right)
\!\left(\frac{\mu\!-\!\upsilon}{\sinh\mu\!-\!\sinh\upsilon}\!-\!1\right)\!+\!\upsilon
\!-\!\frac{a}{\beta}
(\mu\!-\!\upsilon)\Big].\cr
&&
\label{eq:SpecNormMLLA}
\end{eqnarray}

One can explicitly verify that (\ref{eq:SpecNormMLLA}) preserves the position 
of the maximum \cite{EvEqSD}\cite{KOSD}\cite{FW1SD} at

\begin{equation}\label{eq:ellmaxmlla}
\ell_{max}=\frac{Y}2+\frac12\frac{a}{\beta}
\left(\sqrt{Y+\lambda}-\sqrt{\lambda}\right)>\frac{Y}2,
\end{equation}
as well as the gaussian shape of the distribution around (\ref{eq:ellmaxmlla}) 
(see appendix \ref{subsec:DeTA})
\begin{equation}\label{eq:gaussian}
\frac{G(\ell,Y)}{\bar{n}(Y)}\approx\left(\frac{3}{\pi\sqrt{\beta}
\left[(Y+\lambda)^{3/2}
-\lambda^{3/2}\right]}\right)^{1/2}\!\!\exp
\left(-\frac2{\sqrt{\beta}}\,\frac3{(Y+\lambda)^{3/2}
-\lambda^{3/2}}\frac{\left(\ell-\ell_{max}\right)^2}2\right).
\end{equation}

\begin{figure}[h]
\begin{center}
\epsfig{file=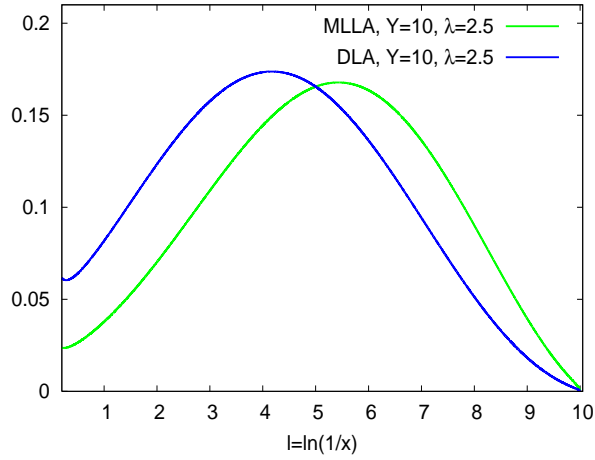, height=6truecm,width=0.5\tw}
\caption{\label{fig:MLLAlambda}SD normalized spectrum: DLA (blue),  MLLA 
 (green);  $Y=10.0$, $\lambda=2.5$.}
\end{center}
\end{figure}

In Fig.\ref{fig:MLLAlambda} we compare for $Y=10$ and $\lambda=2.5$ 
the MLLA curve with DLA (by setting $a=0$ in (\ref{eq:SpecNormMLLA})). 
The general features of the MLLA curve (\ref{eq:SpecNormMLLA}) 
at $\lambda\ne0$ are in good agreement with those of \cite{DKTSD}.

The shape of the single inclusive spectrum given by (\ref{eq:SpecNormMLLA}) 
can easily be proved to be ``infrared stable'' (it has indeed a final limit
when $\lambda\to0$).

\subsection{Logarithmic derivatives}
%%%%%%%%%%%%%%%%%%%%%%%%%%%%%%%%%%%%

Their calculation is important since they appear in the expressions of
2-particle correlations.

Exponentiating the $(\ell,y)$ dependence of the factor ${\cal N}$ in (\ref{eq:Specalphasrunmlla}), we decompose the whole expression in two pieces
\begin{equation}
\psi=\varphi+\delta\psi,
\label{eq:SP}
\end{equation}

where $\varphi$, given in (\ref{eq:phi}), is the DLA term for the shape of 
the distribution \cite{DLASD}, and
\begin{equation}\label{eq:phiprime}
\delta\psi=-\frac12\left(1+\frac{a}{\beta}\right)\ln(\ell+y+\lambda)-\frac{a}{\beta}\mu
+\left(1+\frac{a}{\beta}\right)\upsilon+\frac12\ln[Q(\mu,\upsilon)]
\end{equation}

is the sub-leading contribution (in the sense that its derivative gives
the MLLA correction), where

$$
Q(\mu,\upsilon)\equiv
\frac{\beta(\ell+y+\lambda)^3}{\cosh\upsilon Det\,A(\mu,\upsilon)}=\frac{\sinh^3\mu}
{(\mu\!-\!\upsilon)\cosh\mu\cosh\upsilon\!+\!\cosh\mu\sinh\upsilon\!-\!\sinh\mu
\cosh\upsilon}.
$$

By the definition of the saddle point, the derivatives of
(\ref{eq:phi}) over $\ell$ and $y$ respectively read:

\begin{equation}\label{eq:SPbis}
\varphi_{\ell}=\omega_0=\gamma_0e^{\mu},\qquad \qquad
\varphi_{y}=\nu_0=\gamma_0e^{-\mu}.
\end{equation}

\medskip

We introduce (see appendix \ref{subsec:LK})
\begin{eqnarray}
&&{\cal {L}}(\mu,\upsilon)=-\frac{a}{\beta}+L(\mu,\upsilon),\qquad\quad
L(\mu,\upsilon)=\frac12\frac{\partial}{\partial\mu}
\ln[Q(\mu,\upsilon)],\cr\cr\cr
&&{\cal {K}}(\mu,\upsilon)=1+\frac{a}{\beta}+K(\mu,\upsilon),\qquad\quad
K(\mu,\upsilon)=\frac12\frac{\partial}{\partial\upsilon}
\ln[Q(\mu,\upsilon)]\label{eq:LK}
\end{eqnarray}
and make use of

$$
\frac{\partial\upsilon}{\partial\ell}=\tanh\upsilon
\left(\coth\mu\frac{\partial\mu}{\partial\ell}-
\frac12\beta\gamma_0^2\right),\qquad\quad
\frac{\partial\upsilon}{\partial y}=\tanh\upsilon
\left(\coth\mu\frac{\partial\mu}{\partial y}-
\frac12\beta\gamma_0^2\right),
$$

that follows from (\ref{eq:relmunu}), to write 
$\delta\psi_{\ell},\,\,\delta\psi_y$ in terms of $\frac{\partial\mu}{\partial\ell}$, $\frac{\partial\mu}{\partial y}$

\begin{subequations}
\begin{eqnarray}\label{eq:derpsiprime1}
&&\hskip -3.5cm\delta\psi_{\ell}=-\frac12\left(1+\frac{a}{\beta}
+\tanh\upsilon\,{\cal {K}}(\mu,\upsilon)\right)\beta\gamma_0^2+
\bigg({\cal {L}}(\mu,\upsilon)+\tanh{\upsilon}
\coth{\mu}\,{\cal {K}}(\mu,\upsilon)\bigg)\frac{\partial\mu}{\partial\ell},
\\\notag\\
&&\hskip -3.5cm\delta\psi_{y}=-\frac12\left(1+\frac{a}{\beta}
+\tanh\upsilon\,{\cal {K}}(\mu,\upsilon)\right)\beta\gamma_0^2+
\bigg({\cal {L}}(\mu,\upsilon)+\tanh{\upsilon}
\coth{\mu}\,{\cal {K}}(\mu,\upsilon)\bigg)\frac{\partial\mu}{\partial y}
\label{eq:derpsiprime1bis}.
\end{eqnarray}
\end{subequations}

\begin{figure}[h]
\begin{center}
\epsfig{file=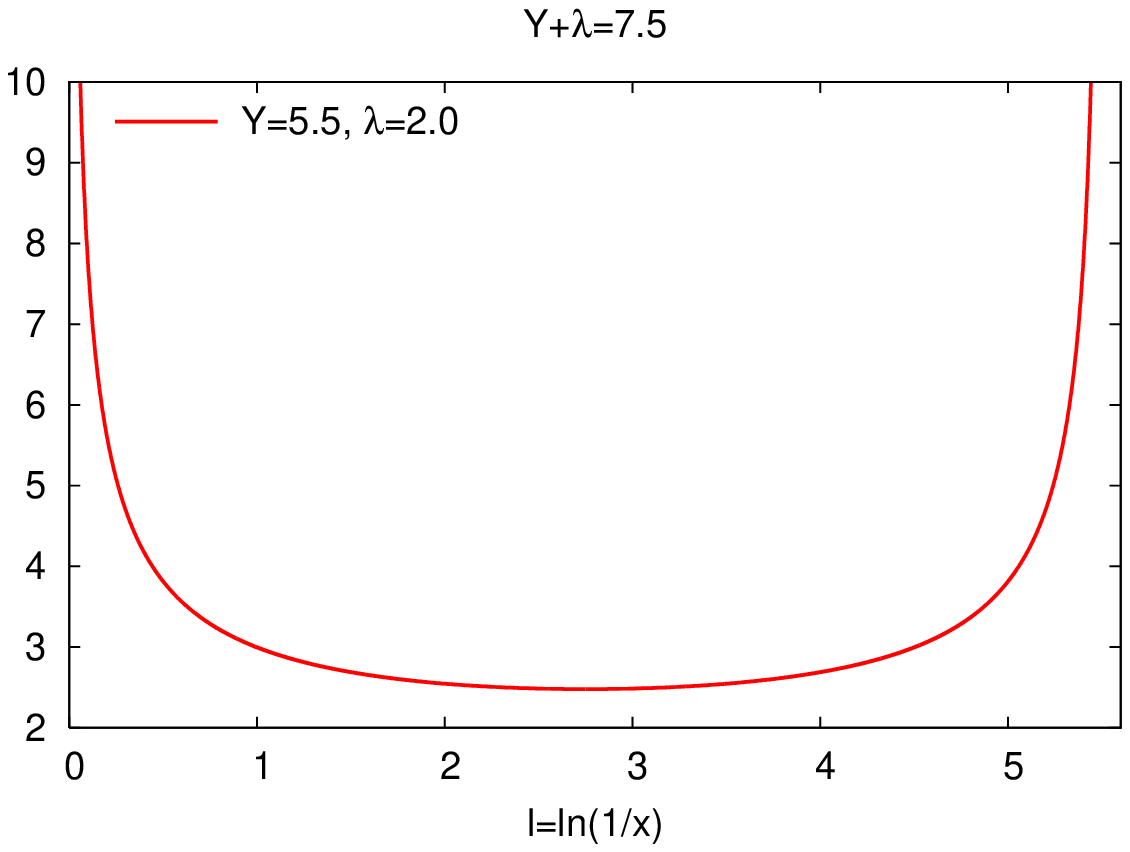, height=5truecm,width=0.45\tw}
\caption{\label{fig:tildeQ} Behavior of $\widetilde Q(\mu,\upsilon)$ as a function
of $\ell=\ln(1/x)$.}
\end{center}
\end{figure}

Using (\ref{eq:ratiomunu}) and (\ref{eq:relmunu}) we obtain

\begin{equation}\label{eq:dermul1}
\frac{\partial\mu}{\partial\ell}=-\frac12\beta\gamma_0^2
\left[1+e^{\mu}\widetilde{Q}(\mu,\upsilon)\right],\qquad\quad
\frac{\partial\mu}{\partial y}=\frac12\beta\gamma_0^2
\left[1+e^{-\mu}\widetilde{Q}(\mu,\upsilon)\right]
\end{equation}

where

\begin{equation}\label{eq:tildeQ}
\widetilde{Q}(\mu,\upsilon)=\frac{\cosh\mu\sinh\mu\cosh\upsilon-
(\mu-\upsilon)\cosh\upsilon-\sinh\upsilon}
{(\mu-\upsilon)\cosh\mu\cosh\upsilon+\cosh\mu\sinh\upsilon-\sinh\mu\cosh\upsilon},
\end{equation}

which we have displayed in Fig.\ref{fig:tildeQ} (useful for correlations).

\begin{figure}[h]
\begin{center}
\epsfig{file=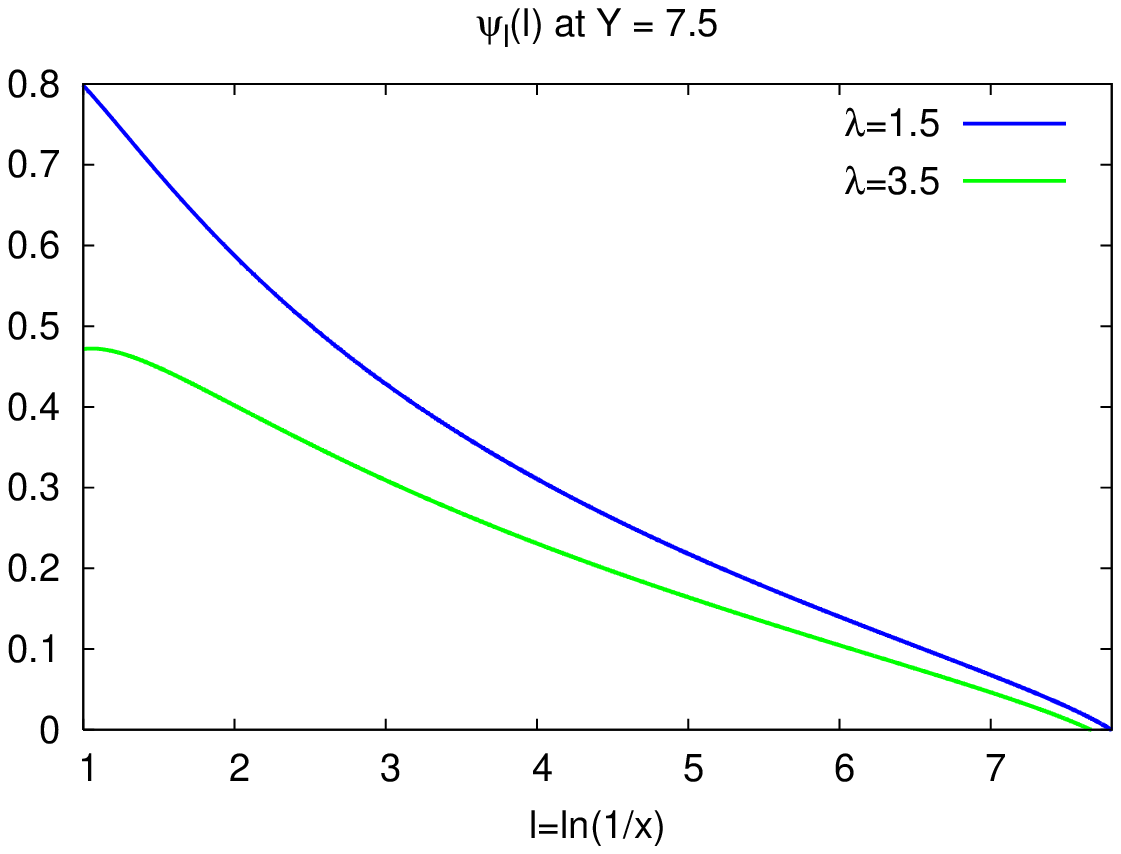, height=5truecm,width=0.48\tw}
\hfill
\epsfig{file=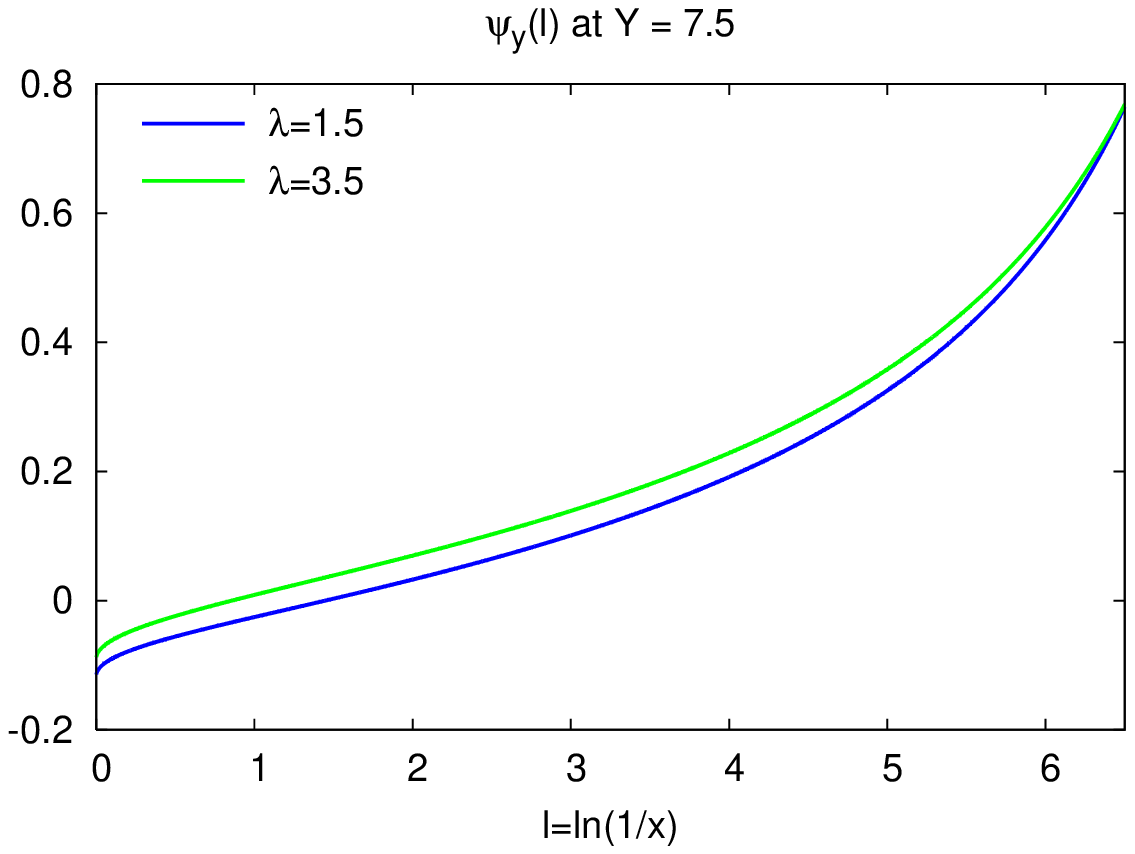, height=5truecm,width=0.48\tw}
\caption{\label{fig:psiellylambda} SD logarithmic derivatives $\psi_\ell$
and $\psi_y$ of the inclusive spectrum at $Y=7.5$,
for $\lambda=1.5$ and $\lambda=3.5$.}
\end{center}
\end{figure}

Inserting (\ref{eq:derpsiprime1bis}) and (\ref{eq:dermul1}) into (\ref{eq:derpsiprime1})\label{eq:derpsiprime2} 
gives the SD logarithmic derivatives of the single inclusive distribution

\begin{subequations}
\begin{eqnarray}
\psi_{\ell}(\mu,\upsilon)\!\!&\!\!=\!\!&\!\!\gamma_0e^{\mu}\!+\!
\frac12a\gamma_0^2\Big[e^{\mu}\widetilde{Q}(\mu,\upsilon)\!-\!\tanh\upsilon
\!-\!\tanh\upsilon\coth\mu\Big(1\!+\!e^{\mu}\widetilde{Q}(\mu,\upsilon)\Big)\Big]\cr\cr
&&\hskip -2.3cm-\frac12\beta\gamma_0^2\Big[1\!+\!\tanh\upsilon
\Big(1\!+\!K(\mu,\upsilon)\Big)+C(\mu,\upsilon)
\Big(1\!+\!e^{\mu}\widetilde{Q}(\mu,\upsilon)\Big)\Big]+{\cal O}(\gamma_0^2)
\label{eq:derpsi'l},
\end{eqnarray}
\begin{eqnarray}
\psi_{y}(\mu,\upsilon)\!\!&\!\!=\!\!&\!\!\gamma_0e^{-\mu}\!-\!\frac12a\gamma_0^2
\Big[2\!+\!e^{-\mu}\widetilde{Q}(\mu,\upsilon)\!+\!\tanh\upsilon
\!-\!\tanh\upsilon\coth\mu\Big(1\!+\!e^{-\mu}\widetilde{Q}(\mu,\upsilon)\Big)\Big]\cr\cr
&&\hskip -2.6cm-\frac12\beta\gamma_0^2\Big[1\!+\!\tanh\upsilon
\Big(1\!+\!K(\mu,\upsilon)\Big)-C(\mu,\upsilon)
\Big(1\!+\!e^{-\mu}\widetilde{Q}(\mu,\upsilon)\Big)\Big]+{\cal O}(\gamma_0^2)
\label{eq:derpsi'y}
\end{eqnarray}
\end{subequations}

where we have introduced ($L$ and $K$ have been written in (\ref{eq:LL})
and (\ref{eq:KK}))

\begin{equation}\label{eq:CC}
C(\mu,\upsilon)=L(\mu,\upsilon)+
\tanh\upsilon\coth\mu\Big(1 + K(\mu,\upsilon)\Big).
\end{equation}
$C$ does not diverge  when $\mu\sim\upsilon\to0$. One has indeed
$$
\lim_{\mu,\upsilon\rightarrow0}\left[L(\mu,\upsilon)+
\tanh\upsilon\coth\mu K(\mu,\upsilon)\right]=\lim_{\mu,\upsilon\rightarrow0}
\frac{2-3\frac{\upsilon^2}{\mu^2}-\frac{\upsilon^3}{\mu^3}}{4\left(1-\frac{\upsilon^3}
{\mu^3}\right)}\mu=0
$$
as well as
\begin{equation*}
\lim_{\mu,\upsilon\rightarrow0}
\tanh\upsilon\coth\mu\left(1+e^{\pm\mu}\widetilde{Q}(\mu,\upsilon)\right)=
\lim_{\mu,\upsilon\rightarrow0}=\frac{3\frac{\upsilon}{\mu}}{1-\frac{\upsilon^3}
{\mu^3}}=\frac{3\sqrt{\frac{\lambda}{Y+\lambda}}}{1-
\left(\frac{\lambda}{Y+\lambda}\right)^{3/2}}.
\end{equation*}
In (\ref{eq:derpsi'l}) and (\ref{eq:derpsi'y}) it is easy to keep trace of
leading and sub-leading contributions.
The first ${\cal O}(\gamma_0)$ term is DLA \cite{DLASD}
while the second ($\propto a\to$ ``hard corrections'') and third 
($\propto\beta\to$ ``running coupling effects'') terms
are MLLA corrections (${\cal O}(\gamma_0^2)$), of relative order
${\cal O}(\gamma_0)$ with respect to the leading one. In Fig.\ref{fig:psiellylambda}
we plot (\ref{eq:derpsi'l}) (left) and (\ref{eq:derpsi'y}) (right) for two 
different values of $\lambda$; one observes that $\psi_\ell$ 
($\psi_y$) decreases (increases) when $\lambda$ increases.

For further use in correlations, the  logarithmic derivatives
have the important property that they do not depend on the normalization
but only on the  shape of the single inclusive distribution.

\subsubsection{``Limiting spectrum'': $\boldsymbol{\lambda\to0\ (Q_0
=\Lambda_{QCD})}$}
%%%%%%%%%%%%%%%%%%%%%%%%%%%%%%%%%%%%%%%%%%%%%%%%%%%%%%%%%%%%%%%%%%

\begin{figure}[h]
\begin{center}
\epsfig{file=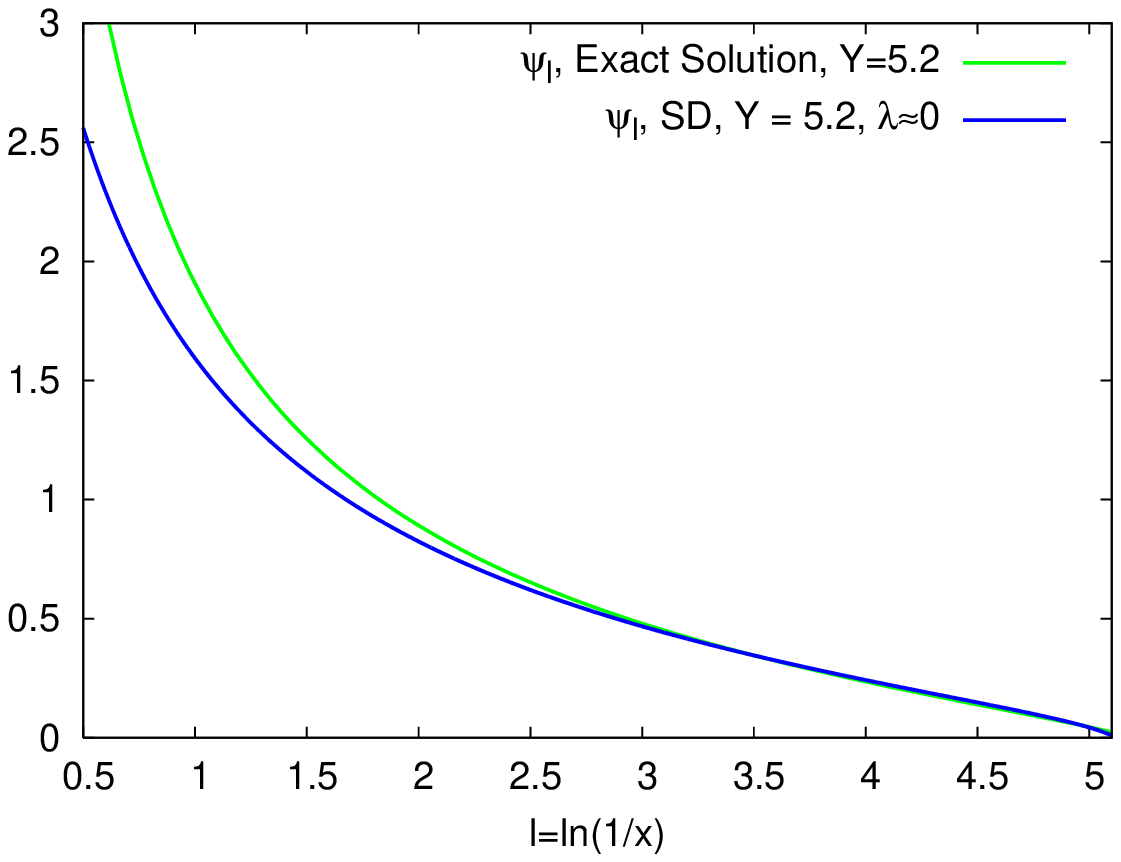, height=5truecm,width=0.48\tw}
\hfill
\epsfig{file=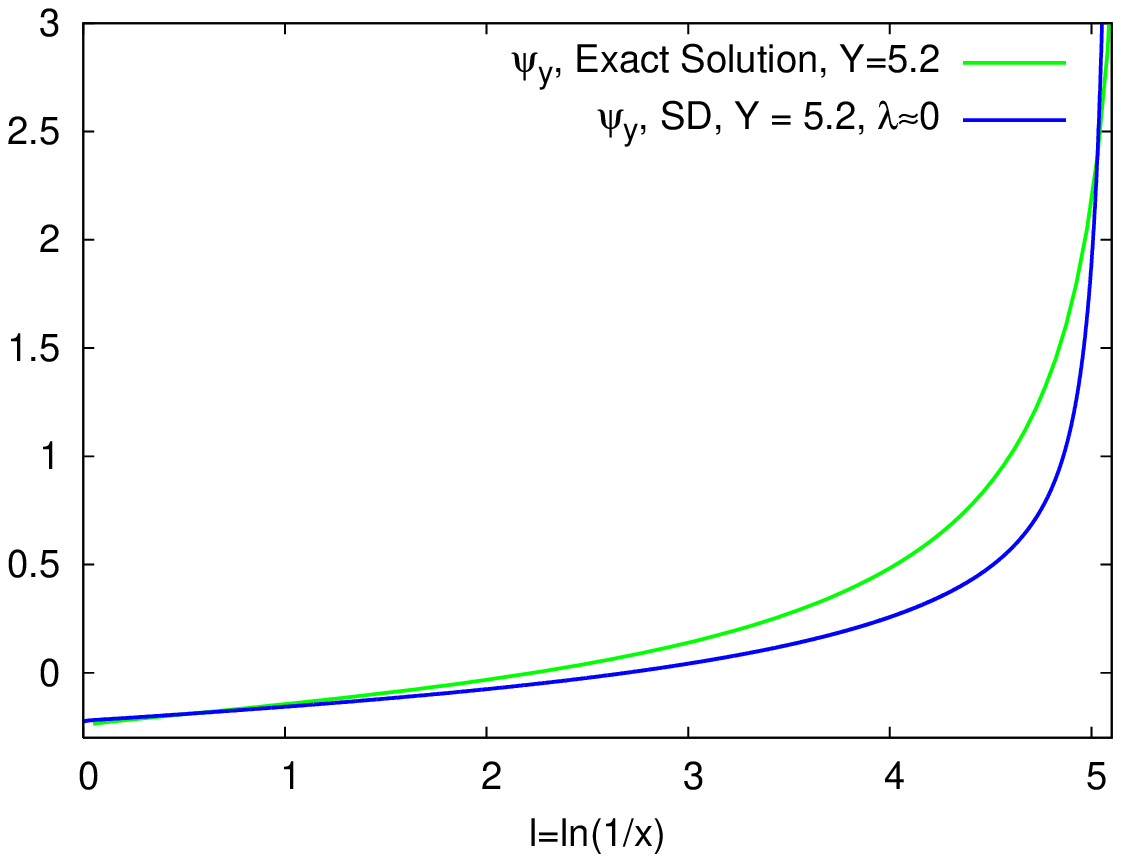, height=5truecm,width=0.48\tw}
\caption{\label{fig:MLLASTESpec} SD logarithmic derivatives $\psi_\ell$
(left) and $\psi_y$ (right) compared
with the ones of \cite{RPR2} at $Y=5.2$.}
\end{center}
\end{figure}

\begin{figure}[h]
\begin{center}
\epsfig{file=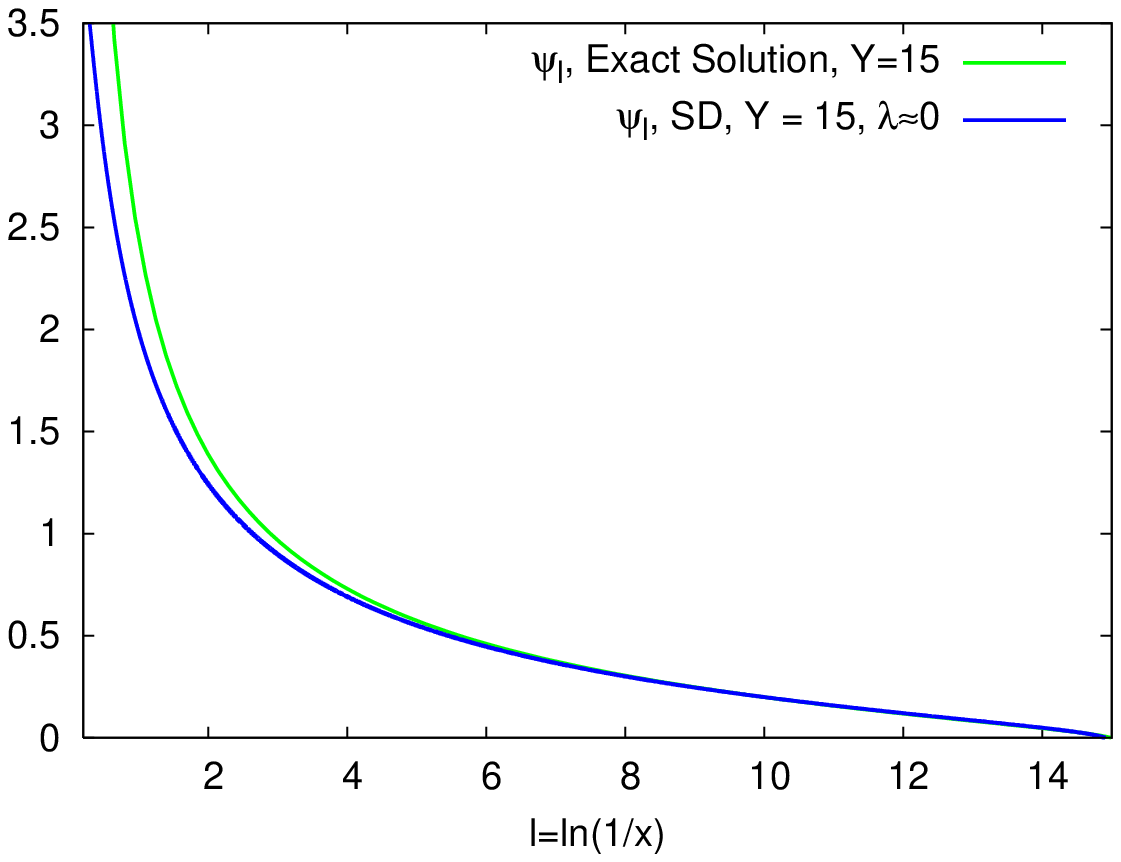, height=5truecm,width=0.48\tw}
\hfill
\epsfig{file=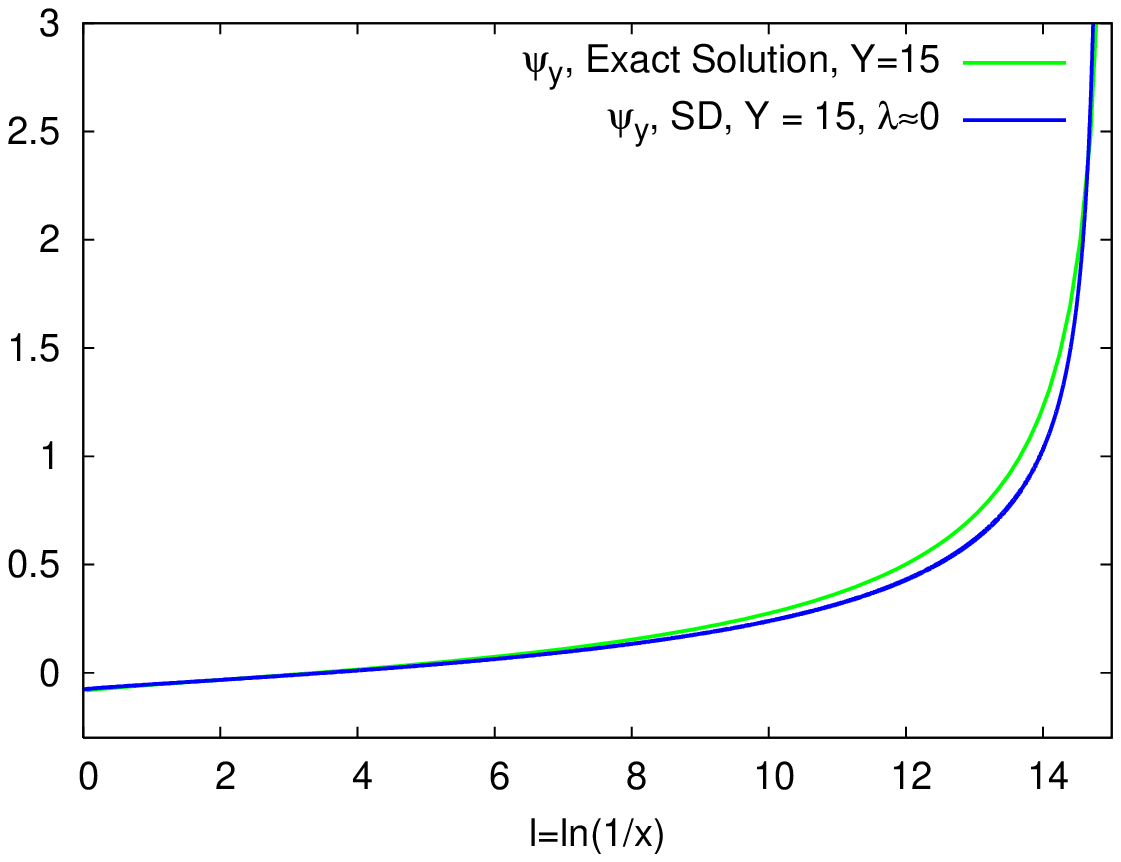, height=5truecm,width=0.48\tw}
\caption{\label{fig:MLLASTESpecbis} SD logarithmic derivatives $\psi_\ell$
(left) and $\psi_y$ (right) compared
with the ones of \cite{RPR2} at $Y=15$.}
\end{center}
\end{figure}

Since the logarithmic derivatives are ``infrared stable'' (see above),
we can take 
the limit $\lambda\to0$ in (\ref{eq:derpsi'l})(\ref{eq:derpsi'y})
\footnote{For this purpose, (\ref{eq:ratiomunu}) has been numerically
inverted.}, 
and compare their shapes with the
ones obtained in \cite{PerezMachetSD}; this is done
in  Figs.~\ref{fig:MLLASTESpec} and \ref{fig:MLLASTESpecbis}, at LEP-I energy
($E\Theta_0=91.2\,\text{GeV}$, $Y=5.2$) and at the unrealistic value $Y=15$.

The agreement between the SD and the exact logarithmic derivatives
is seen to be quite good.
The small deviations ($\leq 20\%$) that can be observed at large $\ell$
(the domain we deal with) arise from NMLLA
corrections that one does not control in the exact solution.
The agreement gets better and better as the energy increases.

It is checked in appendix (\ref{subsec:check}) that (\ref{eq:Specalphasrunmlla})
satisfies the evolution equation (\ref{eq:solgSD});
the SD logarithmic derivatives (\ref{eq:derpsi'l}) and (\ref{eq:derpsi'y}) can
therefore be used in the approximate calculation of
2-particle correlations at $\lambda\ne0$. This is what is done in the 
next section.

%%%%%%%%%%%%%%%%%%%%%%%%%%%%%%%%%%%%%%%%%%%%%%%%%%%%%
\section{2-PARTICLE CORRELATIONS INSIDE ONE JET AT 
$\boldsymbol{\lambda\!\not=\!0\ (Q_0\!\ne\!\Lambda_{QCD})}$}
\label{sec:TPC}
%%%%%%%%%%%%%%%%%%%%%%%%%%%%%%%%%%%%%%%%%%%%%%%%%%%%%

We study the correlation between 2-particles inside one jet of 
half opening angle $\Theta$ within the MLLA accuracy.
They have fixed energies $x_1=\omega_1/E$, $x_2=\omega_2/E$ ($\omega_1>\omega_2$)
and are emitted at arbitrary angles $\Theta_1$, $\Theta_2$. The constrain 
$\Theta_1\geq\Theta_2$ follows from the angular ordering in the cascading process. 
One has $\Theta\geq\Theta_1$ (see Fig.~1 of \cite{RPR2}).

\subsection{Variables and kinematics}
%%%%%%%%%%%%%%%%%%%%%%%%%%%%%%%%%%%%%

The variables and
kinematics of the cascading process are defined like in section 3.2 of \cite{RPR2}.

\subsection{MLLA evolution equations for correlations}
%%%%%%%%%%%%%%%%%%%%%%%%%%%%%%%%%%%%%%%%%%%%%%%%%%%%%%

The system of integral evolution equations 
for the quark and gluon jets two-particle 
correlation reads (see eqs.~(65) and (66) of \cite{RPR2})

\vbox{
\begin{eqnarray}
\label{eq:eveeqqSD}
\hskip -2.9cm Q^{(2)}(\ell_1,y_2,\eta)\!-\! Q_1(\ell_1,y_1)Q_2(\ell_2,y_2)
\!\!\!&\!\!=\!\!&\!\!\! \frac{C_F}{N_c}\!\!
\int_0^{\ell_1}\!\!\! d\ell\!\int_0^{y_2}\!\!\! dy\,
\gamma_0^2(\ell+y) \Big[\!1\!-\!\frac34 \delta(\ell-\ell_1) \!\Big]
G^{(2)}(\ell,y,\eta),\\\notag\\\notag\\
\notag
\hskip -2.9cm G^{(2)}(\ell_1,y_2,\eta) - G_1(\ell_1,y_1)G_2(\ell_2,y_2)
\!\!\!&\!\!\!=\!\!\!&\!\!\!\! 
\int_0^{\ell_1}\!\! d\ell\!\int_0^{y_2}\!\!dy\, \gamma_0^2(\ell+y)
\Big[\!1 - a \delta(\ell-\ell_1) \!\Big] G^{(2)}(\ell,y,\eta)\\\notag\\
\!\!\!&\!\!\!+\!\!\!&\!\!\! (a-b) \int_0^{y_2}dy \> \gamma_0^2(\ell_1+y)
G(\ell_1,y+\eta)G(\ell_1+\eta,y).
 \label{eq:eveeqgluSD} 
\end{eqnarray}
}

a is defined in (\ref{eq:adefSD}) while

\begin{equation}
b = \frac{1}{4N_c}\bigg[\frac{11}{3}N_c -\frac{4}{3}n_f T_R
\bigg(1-2\frac{C_F}{N_c}\bigg)^2\bigg]\stackrel{n_f=3}{=}0.915.
\label{eq:bdefSD}
\end{equation}

\subsection{MLLA solution at $\boldsymbol{\lambda \ne 0}$}
%%%%%%%%%%%%%%%%%%%%%%%%%%%%%%%%%%%%%%%%%%%%%%%%%%%%%%%%%%%%%%%%%%%%%

The quark and gluon jet correlators ${\cal C}_q$ and ${\cal C}_g$
have been exactly determined for any $\lambda$ in \cite{RPR2}
by respectively setting $Q^{(2)}={\cal C}_qQ_1Q_2$ and $G^{(2)}={\cal C}_gG_1G_2$
into (\ref{eq:eveeqqSD}) and (\ref{eq:eveeqgluSD}).
In the present work we limit ourselves to the exact MLLA solution which
consists in neglecting all ${\cal O}(\gamma_0^2)$ corrections in equations
(64) and (84) of \cite{RPR2}.

\subsubsection{Gluon jet}
%%%%%%%%%%%%%%%%%%%%%%%%%

At MLLA, the logarithmic derivatives of $\psi$ (\ref{eq:SP}) can be truncated to
the saddle point derivatives $\varphi_\ell,\,\varphi_y$ of (\ref{eq:phi}).
The MLLA solution of (\ref{eq:eveeqgluSD}) then reads (see (77) in \cite{RPR2})

\begin{equation}
{\cal C}_g-1 \stackrel{MLLA}{\approx} \frac{1-b\left(\varphi_{1,\ell}
    +\varphi_{2,\ell} \right)-\delta_1} {1+ \bar\Delta + \Delta' + \delta_1}
\label{eq:CGMLLASD}
\end{equation}

where we introduce
\begin{eqnarray}
&&\label{eq:deltabisSD}\bar\Delta = \gamma_0^{-2}
\Big(\varphi_{1,\ell}\varphi_{2,y}+\varphi_{1,y}\varphi_{2,\ell}\Big),
\\\notag\\
&&\label{eq:deltater}\Delta' = \gamma_0^{-2}
\Big(\varphi_{1,\ell}\delta\psi_{2,y}+\delta\psi_{1,y}\varphi_{2,\ell}+
\delta\psi_{1,\ell}\varphi_{2,y}+\varphi_{1,y}\delta\psi_{2,\ell}\Big);
\\\notag\\
&&\label{eq:chi} \chi=\ln\left(1+\frac1{1+\bar\Delta}\right),\quad 
\chi_{\ell}=\frac1{\chi}\frac{\partial\chi}{\partial\ell},\quad 
\chi_y=\frac1{\chi}\frac{\partial\chi}{\partial y};\\\notag\\
&&\delta_1 = \gamma_0^{-2}\Big[\chi_{\ell}(\varphi_{1,y}+\varphi_{2,y}) +
   \chi_{y}(\varphi_{1,\ell}+\varphi_{2,\ell})\Big].\label{eq:delta1SD}
\label{eq:nota4SD}
\end{eqnarray}

(\ref{eq:deltabisSD}) is obtained by using (\ref{eq:SPbis}):

\begin{equation}\label{eq:Delta}
\bar\Delta(\mu_1,\mu_2)=2\cosh(\mu_1-\mu_2)={\cal O}(1),
\end{equation}

which is the DLA contribution \cite{DLASD}, while (\ref{eq:deltater}) 
(see appendix \ref{subsec:Deltaprime})
is obtained by using (\ref{eq:SPbis}), (\ref{eq:derpsiprime1}) and 
(\ref{eq:derpsiprime1bis})

\begin{equation}\label{eq:Deltaprime}
\Delta'(\mu_1,\mu_2)=\frac{ e^{-\mu_1}\delta\psi_{2,\ell} + e^{-\mu_2}\delta\psi_{1,\ell} 
        + e^{\mu_1}\delta\psi_{2,y} + e^{\mu_2}\delta\psi_{1,y}}
{\gamma_0}={\cal O}(\gamma_0);
\end{equation}

it is a next-to-leading (MLLA) correction.
To get (\ref{eq:chi}), we first use (\ref{eq:Delta}), which gives
\begin{eqnarray}\label{eq:chielly}
\chi_\ell =  -\frac{\tanh{\frac{\mu_1-\mu_2}{2}}}{1+2\cosh(\mu_1\!-\!\mu_2)}
 \left(\frac{\partial\mu_1}{\partial\ell} -\frac{\partial\mu_2}{\partial\ell}
 \right),\quad
 \chi_y = -\frac{\tanh{\frac{\mu_1-\mu_2}{2}}}{1+2\cosh(\mu_1\!-\!\mu_2)}
 \left(\frac{\partial\mu_1}{\partial y} -\frac{\partial\mu_2}{\partial y} \right),\cr
&&
\end{eqnarray}

and then (\ref{eq:dermul1}) to get
\begin{eqnarray*}
  \chi_\ell\!=\! \beta\gamma_0^2\,
  \frac{\tanh{\frac{\mu_1-\mu_2}{2}}}{1\!+\!2\cosh(\mu_1\!-\!\mu_2)} \>
  \frac{e^{\mu_1}\widetilde{Q}_1-e^{\mu_2}\widetilde{Q}_2}{2},\quad
 \chi_y \!=\! -\beta\gamma_0^2
  \frac{\tanh{\frac{\mu_1-\mu_2}{2}}}{1\!+\!2\cosh(\mu_1\!-\!\mu_2)} \>
  \frac{e^{-\mu_1}\widetilde{Q}_1-e^{-\mu_2}\widetilde{Q}_2}{2}
\end{eqnarray*}

which are ${\cal O}(\gamma_0^2)$. They  should be plugged into
(\ref{eq:nota4SD}) together with (\ref{eq:SPbis}), which gives

\begin{equation}\label{eq:delta2}
 \delta_1 = \beta\gamma_0
\frac{2\sinh^2{\left(\frac{\mu_1-\mu_2}2\right)}}
{3+4\sinh^2{\left(\frac{\mu_1-\mu_2}2\right)}}
\Big(\widetilde Q(\mu_1,\upsilon_1)+
\widetilde Q(\mu_2,\upsilon_2)\Big)={\cal O}(\gamma_0);
\end{equation}

it is also a MLLA term. For $Q\gg Q_0\geq\Lambda_{QCD}$ we finally get,
\begin{equation}
{\cal C}_g(\ell_1,\ell_2,Y,\lambda)\stackrel{MLLA}{\approx}
1 + \frac{1 - b\gamma_0\left(e^{\mu_1} + e^{\mu_2} \right) - \delta_1} 
{1+ 2\cosh(\mu_1-\mu_2) + \Delta'(\mu_1,\mu_2) + \delta_1}
\label{eq:CGMLLAbis}
\end{equation}

where the expression for $\Delta'$ (\ref{eq:dDelta}) 
is written in appendix \ref{subsec:Deltaprime}.
It is important to notice that $\delta_1\simeq0$ near $\ell_1\approx\ell_2$ ($\mu_1\approx\mu_2$) while it 
is positive and increases as $\eta$ 
gets larger (see (\ref{eq:tildeQ}) and Fig.\ref{fig:tildeQ});
it makes the correlation function narrower in $|\ell_1-\ell_2|$.

\subsubsection{Quark jet}
%%%%%%%%%%%%%%%%%%%%%%%%%

The MLLA solution of (\ref{eq:eveeqqSD}) reads (see (93) in \cite{RPR2})
\begin{eqnarray}
\frac{{\cal C}_q-1}{{\cal C}_g-1} \stackrel{MLLA}{\approx}
\frac{N_c}{C_F}\bigg[1+(b-a)(\phi_{1,\ell} + \phi_{2,\ell})
\frac{1+\bar\Delta}{2+\bar\Delta}\bigg]
\label{eq:rapMLLASD}
\end{eqnarray}

Inserting (\ref{eq:deltabisSD})-(\ref{eq:nota4SD}) into (\ref{eq:rapMLLASD})  we get
\begin{equation*}
{\cal C}_q(\ell_1, \ell_2, Y, \lambda)\stackrel{MLLA}{\approx}1+
\frac{N_c}{C_F}\Big({\cal C}_g(\ell_1, \ell_2, Y, \lambda)-1\Big)
\bigg[1+(b-a)\gamma_0(e^{\mu_1} + e^{\mu_2})
\frac{1+2\cosh(\mu_1-\mu_2)}{2+2\cosh(\mu_1-\mu_2)}\bigg].
\end{equation*}

which finally reduces (for $Q\gg Q_0\geq\Lambda_{QCD}$) to
\begin{equation}\label{eq:rapMLLAbis}
{\cal C}_q(\ell_1,\ell_2, Y, \lambda)\stackrel{MLLA}{\approx}1+
\frac{N_c}{C_F}
\bigg[\Big({\cal C}_g(\ell_1, \ell_2, Y, \lambda)-1\Big)+\frac12(b-a)\gamma_0\frac{e^{\mu_1} + e^{\mu_2}}
{1+\cosh(\mu_1-\mu_2)}\bigg].
\end{equation}

\subsection{Sensitivity of the quark and gluon jets correlators to the value
of $\boldsymbol{\lambda}$}
%%%%%%%%%%%%%%%%%%%%%%%%%%%%%%%%%%%%%%%%%%%%%%%%%%%%%%%%%%%%%%%%%%%%%%%%%%%%

Increasing $\lambda$ translates into taking the limits $\beta,\,\Lambda_{QCD}\to0$
($Y=\ell+y\ll\lambda,\, Q\gg Q_0\gg\Lambda_{QCD}$) in the definition of the anomalous 
dimension via the running coupling constant ($\gamma_0=\gamma_0(\alpha_s)$, see (44) in 
\cite{RPR2}). It allows to neglect $\ell$, $y$ with respect to $\lambda$ as follows
\begin{equation}\label{eq:gammaconst}
\gamma_0^2(\ell+y)=\frac1{\beta(\ell+y+\lambda)}
\stackrel{\ell+y\ll\lambda}{\approx}\gamma_0^2=\frac1{\beta\lambda},
\end{equation}

such that $\gamma_0$ can be taken as a constant. Estimating
(\ref{eq:MLLAalphasrunSD}) 
in the region $\lambda\gg1\Leftrightarrow s\ll1$ needs evaluating the kernel
\begin{eqnarray}
\frac1{\nu+s}\left(\frac{\omega\left(\nu+s\right)}
{\left(\omega+s\right)\nu}\right)^{1/\beta\left(\omega-\nu\right)}\!\!
\left(\frac{\nu}{\nu+s}\right)^{a/\beta}
\!\!&\!\!\stackrel{s\ll1}{\approx}\!\!&\!\!
\frac1{\nu}\left(1+\frac{\omega-\nu}
{\omega\nu}s\right)^{1/\beta\left(\omega-\nu\right)}\left(1-\frac{s}{\nu}\right)^{a/\beta}
\cr\cr
&&\hskip -7cm
\approx\frac1{\nu}\left[1+\frac1{\nu}\left(\frac1\omega-a\right)\frac{s}\beta+\frac1{2!}
\frac1{\nu^2}\left(\frac1\omega-a\right)^2\frac{s^2}{\beta^2}+\frac1{3!}\frac1{\nu^3}
\left(\frac1\omega-a\right)^3\frac{s^3}{\beta^3}+\dots\right].\label{eq:MLLApropag}\cr
&&
\end{eqnarray}

Integrating (\ref{eq:MLLApropag}) over $s$, using (\ref{eq:gammaconst}) and
$\int_0^{\infty}s^n\,e^{-\lambda s}=\frac{n!}{\lambda^n}$, we get
\begin{eqnarray*}
{\cal {G}}(\omega,\nu)
\!\!&\!\!\approx\!\!&\!\!\frac1{\nu}
\left[1+\frac1{\nu}\left(\frac1\omega-a\right)\frac1{\beta\lambda}+
\frac1{\nu^2}\left(\frac1\omega-a\right)^2\left(\frac1{\beta\lambda}\right)^2+
\frac1{\nu^3}\left(\frac1\omega-a\right)^3
\left(\frac1{\beta\lambda}\right)^3+\dots\right]\cr\cr
\!\!&\!\!=\!\!&\!\!\frac1{\nu-\gamma_0^2\left(1/\omega-a\right)},
\end{eqnarray*}

which, after inverting the Mellin's representation (132) of \cite{RPR2}, gives

\begin{equation}\label{eq:Gexpres}
G(\ell,y)\stackrel{x\ll1}{\simeq}\exp(2\gamma_0\sqrt{\ell\,y}-a\gamma_0^2y).
\end{equation}

Taking the same limit in (\ref{eq:ratiomunu}) and (\ref{eq:relmunu}) gives
respectively

\begin{equation}\label{eq:tanhmu}
\frac{y-\ell}{y+\ell}\stackrel{\ell+y\ll\lambda}\approx\tanh\mu\Rightarrow
\mu=\frac12\ln\frac{y}{\ell},\qquad
\mu-\upsilon\stackrel{\ell+y\ll\lambda}{\approx}\frac12\frac{y-\ell}{\lambda}
\Rightarrow\mu\sim\upsilon.
\end{equation}

Furthermore, we use (\ref{eq:tanhmu}) to show how (\ref{eq:SP}) reduces to
the exponent in (\ref{eq:Gexpres})
\footnote{we set $\beta=0$ in (\ref{eq:derpsi'l}), (\ref{eq:derpsi'y}) 
and only consider terms $\propto a$}
\begin{eqnarray}
\phi\!\!&\!\!=\!\!&\!\!\frac2{\sqrt{\beta}}\frac{\mu-\upsilon}{\sinh\mu-\sinh\upsilon}
\stackrel{\ell+y\ll\lambda}{\approx}
2\gamma_0\sqrt{\ell\, y},\cr\cr\cr
\left(\frac{\nu_0}{\nu_0+s_0}\right)^{a/\beta}\!\!&\!\!=\!\!&\!\!
-\frac12\frac{a}\beta\ln\left(1+\frac{\ell+y}\lambda\right)-
\frac{a}{\beta}(\mu-\upsilon)\approx-\frac12\frac{a}\beta\frac{\ell+y}\lambda-
\frac{a}{\beta}(\mu-\upsilon)\cr\cr
\!\!&\!\!\stackrel{\ell+y\ll\lambda}{\approx}\!\!&\!\!-a\gamma_0^2\,y.
\end{eqnarray}

Thus, since $\mu=\frac12\ln\frac{y}{\ell}$ (\ref{eq:tanhmu}),
(\ref{eq:derpsi'l}) and (\ref{eq:derpsi'y}) simplify to

\begin{equation}
\psi_\ell\stackrel{\ell+y\ll\lambda}{\approx}\gamma_0e^{\mu}=\gamma_0\sqrt{\frac{y}{\ell}},
\qquad\psi_y\stackrel{\ell+y\ll\lambda}{\approx}\gamma_0e^{-\mu}-a\gamma_0^2=
\gamma_0\sqrt{\frac{\ell}{y}}-a\gamma_0^2.
\end{equation}

Therefore, taking the limit $\beta,\,\Lambda_{QCD}\to0$ ($\lambda\to\infty$)
leads to the simplified model
described in section 4.2 of \cite{RPR2}. Setting, for the sake of simplicity,
$\ell_1\approx\ell_2$ in (\ref{eq:CGMLLAbis})(\ref{eq:rapMLLASD}),
where $\delta_1$ vanishes, we obtain, in the high energy limit
\begin{equation}\label{eq:cgqmodel}
{\cal C}_g(\ell,y)\simeq1\!+\!\frac13\left[1\!-\!2\left(b\!-\!\frac13a\right)
\psi_{\ell}(\ell,y)\right],\quad
{\cal C}_q(\ell,y)\simeq1\!+\!\frac{N_c}{C_F}\left[\frac13\!-\!\frac12
\left(\frac53a\!+\!b\right)\psi_{\ell}(\ell,y)\right],
\end{equation}
where
\begin{eqnarray}
&&b-\frac13 a = \frac1{18}\left(11-8\frac{T_R}{N_c}
 +28\frac{T_R}{N_c}\frac{C_F}{N_c}-24\frac{T_R}{N_c}
\frac{C_F^2}{N_c^2}\right)\stackrel{n_f=3}{\approx}0.6,\cr\cr\cr
&&\frac53 a + b = \frac29\left(11 + \frac{T_R}{N_c}
 +\frac{T_R}{N_c}\frac{C_F}{N_c} -6\frac{T_R}{N_c}
\frac{C_F^2}{N_c^2}\right)\stackrel{n_f=3}{\approx}2.5.
\end{eqnarray}

\begin{figure}[h]
\begin{center}
\epsfig{file=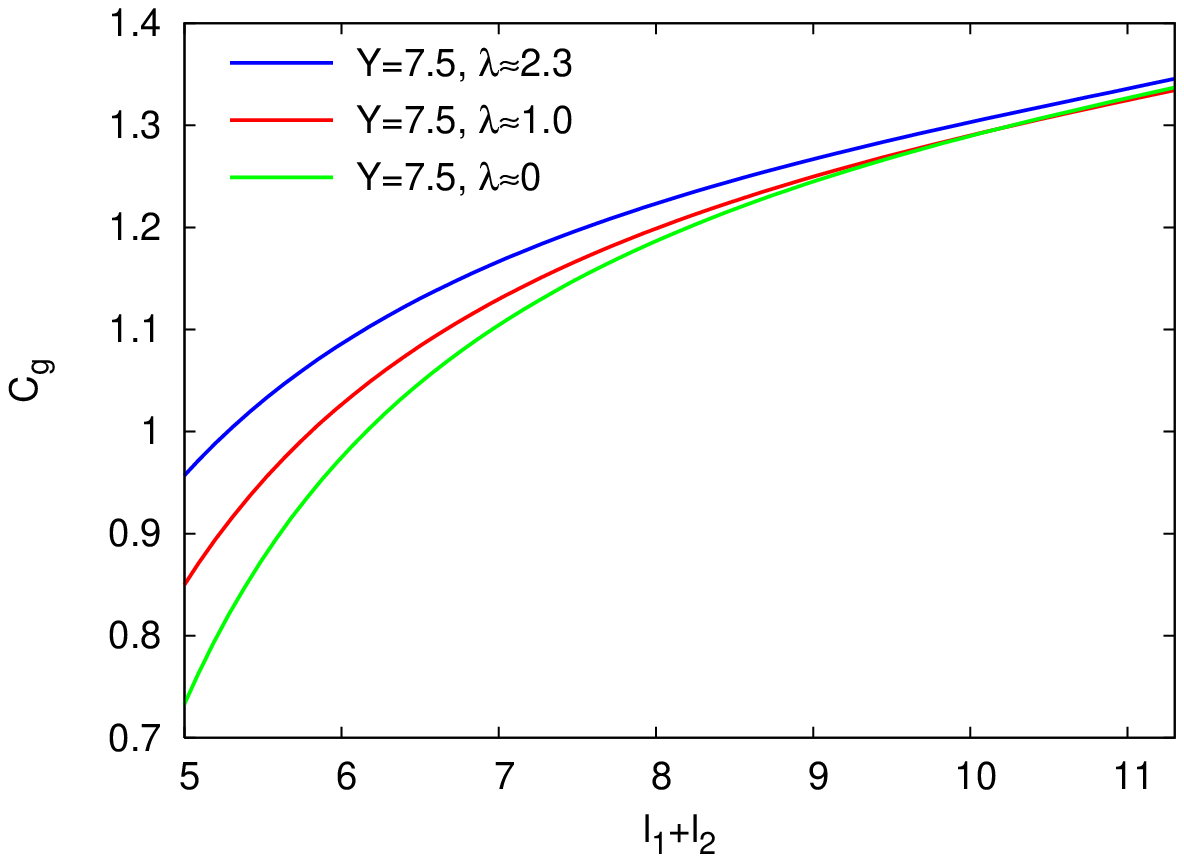, height=5truecm,width=0.48\tw}
\hfill
\epsfig{file=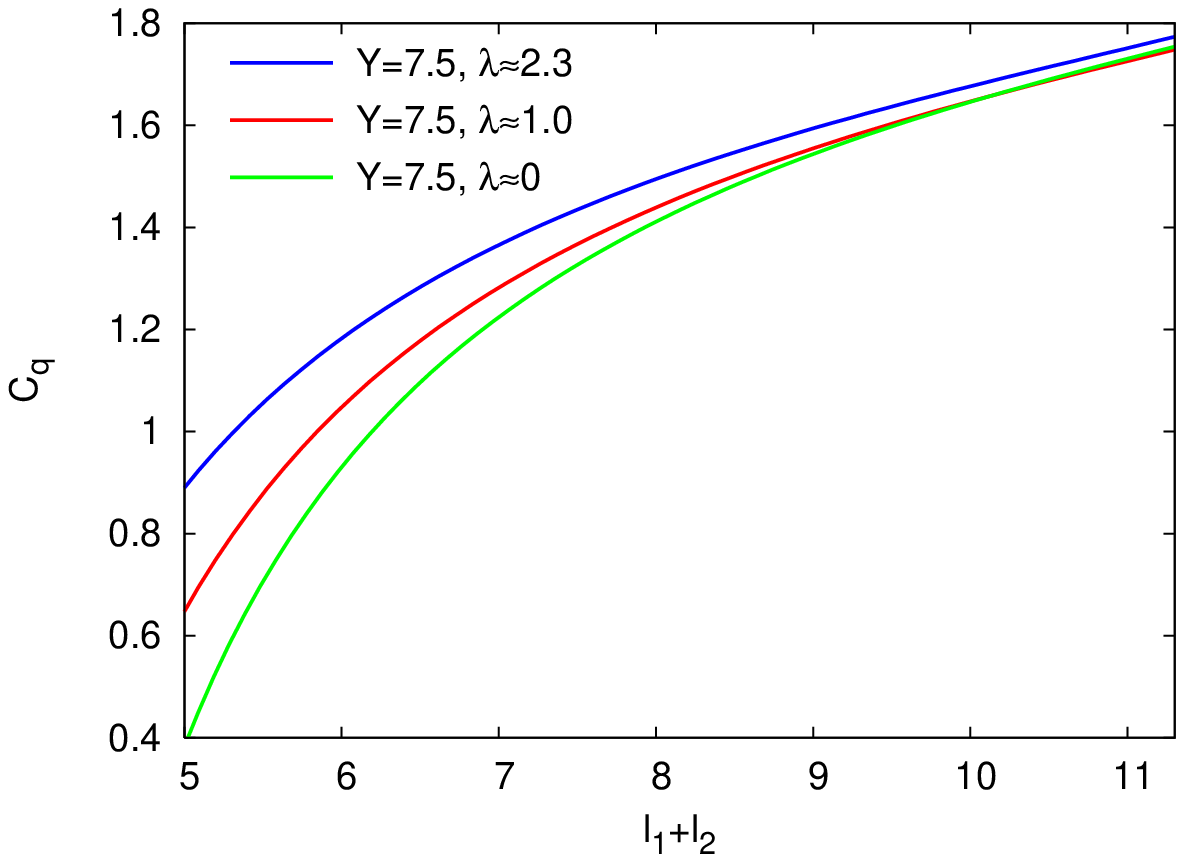, height=5truecm,width=0.48\tw}
\caption{\label{fig:LHCcorrg} Varying $\lambda$ at fixed $Q_0$;
$\Lambda_{QCD}$ dependence of  ${\cal C}_g$ (left) and ${\cal C}_q$ (right)}
\end{center}
\end{figure}

Thus, when $\lambda$ increases by decreasing $\Lambda_{QCD}$, 
$\psi_\ell\!\propto\!\gamma_0$ decreases and
the correlators (\ref{eq:cgqmodel}) increase.
For LHC, a typical value is $Y=7.5$ and we compare
in Fig.~\ref{fig:LHCcorrg}, at fixed $Q_0$,
 the limiting case $\lambda\approx0$
($Q_0\approx\Lambda_{QCD}\approx 253\,\text{MeV}$) with
$\lambda\approx1.0$ ($\Lambda_{QCD}=100\,\text{MeV}$) and $\lambda \approx
2.3$ ($\Lambda_{QCD}=25\,\text{MeV}$). As predicted by (\ref{eq:cgqmodel}),
the correlation increases when $\Lambda_{QCD}\to 0$ at fixed $Q_0$.

\begin{figure}
\vbox{
\begin{center}
\epsfig{file=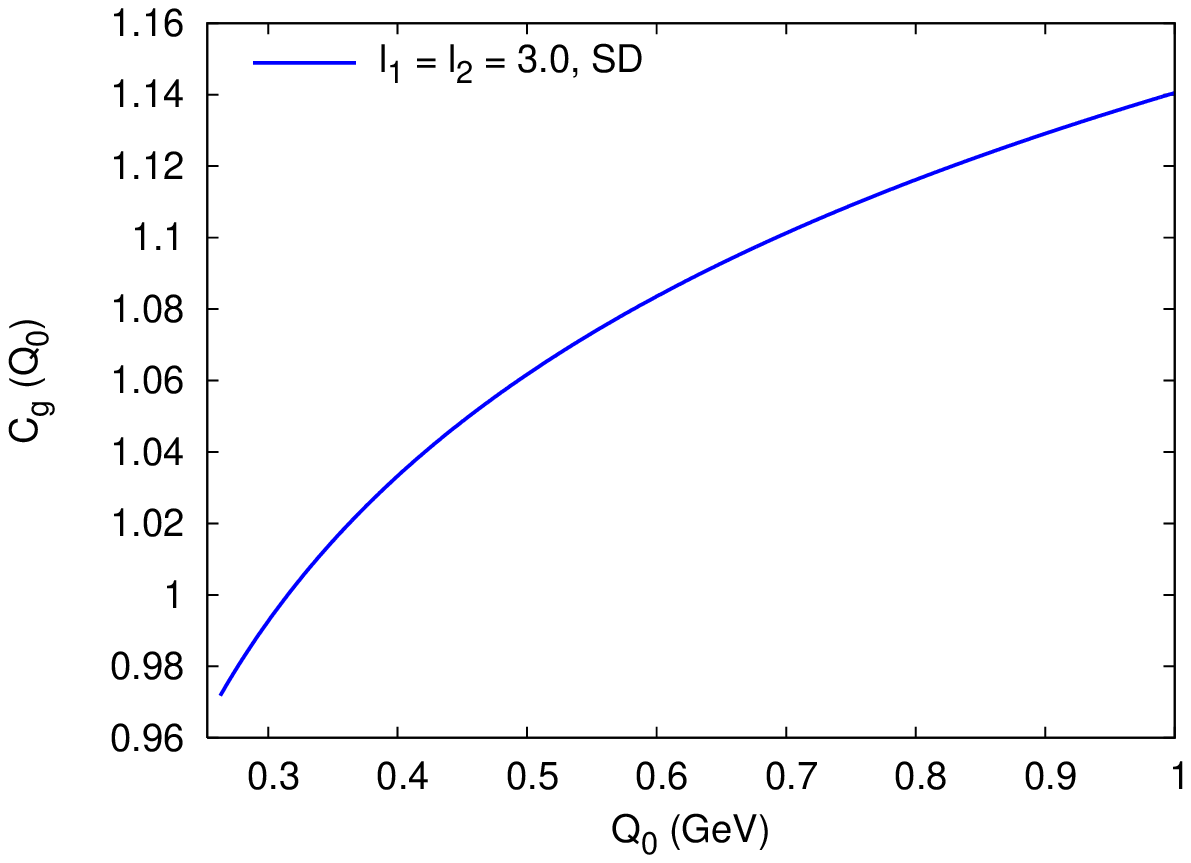, height=5truecm,width=0.48\tw}
\hfill
\epsfig{file=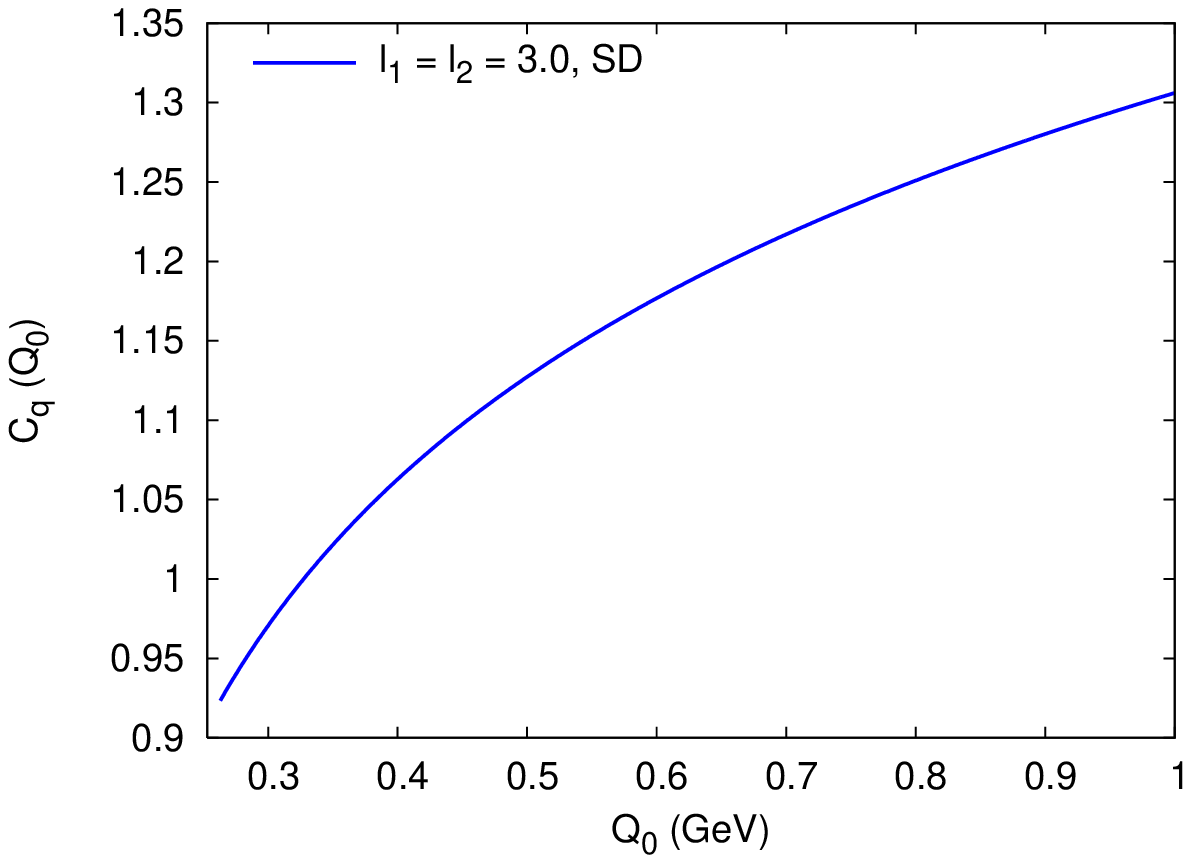, height=5truecm,width=0.48\tw}
\caption{Varying $\lambda$ at fixed $\Lambda_{QCD}=253\,MeV$; $Q_0$-dependence
of ${\cal C}_g$ (left) and ${\cal C}_q$ (right) 
at $\ell_1=\ell_2=3.0$}
\label{fig:Qeff}
\end{center}
}
\end{figure}

It is also sensitive to the value of $Q_0$. As
seen in (\ref{eq:cgqmodel}), since $y=\ln\frac{Q}{Q_0}-\ell$, if one increases
$Q_0$ (since $\Lambda_{QCD}$ is fixed, $\gamma_0$ does not change), 
thereby reducing the available phase space, the correlators increase.
This dependence of the correlators at fixed $\Lambda_{QCD}$
is displayed in Fig.\ref{fig:Qeff} for $0.3\,\text{GeV}\leq Q_0\leq1.0\,\text{GeV}$
at $\ell_1=\ell_2=3.0$ (soft parton).

In the simplified model which leads to (\ref{eq:cgqmodel}),
 ${\cal C}_g$ and ${\cal C}_q$
respectively go to the asymptotic values $4/3$ and $1 + N_c/3C_F$. This is
however not the case in the general situation  $\beta \not=0$, as can
be easily checked by using (\ref{eq:derpsi'l}) and (\ref{eq:derpsi'y}); 
for example, near the maximum
of the distribution ($\mu \sim v \to 0$), a contribution 
$\propto\lambda^{3/2}/[(Y+\lambda)^{3/2}-\lambda^{3/2}]$ occurs in
the term proportional to $\beta$ in (\ref{eq:cgqmodel}) that
yields negative values of $\psi_\ell$ when $\lambda$ increases.

\subsection{Extension of the Fong and Webber expansion;
its limit $\boldsymbol{\lambda=0}$}
%%%%%%%%%%%%%%%%%%%%%%%%%%%%%%%%%%%%%%%%%%%%%%%%%%%%%%%%

In the Fong-Webber regime, the energies of the two registered particles
stay very close to the peak of the inclusive hump-backed distribution that
is, $|\ell_i-\ell_{max}|\ll \sigma \propto [(Y+\lambda)^{3/2}-\lambda^{3/2}]^{1/2}$ 
(see (\ref{eq:gaussian})).

Near the maximum of the single inclusive distribution $\ell_1\sim\ell_2\simeq Y/2$ 
($\mu,\upsilon\to0$, see appendix \ref{subsec:DeTA})

$$
\lim_{\mu,\upsilon\rightarrow0}C=\left(\frac{\lambda}{Y+\lambda}\right)^{1/2}
,\quad\lim_{\mu,\upsilon\rightarrow0}K_i=\frac32\frac{\upsilon^2_i}{\mu^3_i-\upsilon^3_i},
\quad\lim_{\mu,\upsilon\rightarrow0}\widetilde{Q}=
\frac23+\frac13\left(\frac{\lambda}{Y+\lambda}\right)^{3/2},
$$

where $C$, $K_i$ and $\tilde Q$ are defined in (\ref{eq:CC}), (\ref{eq:KK}) and (\ref{eq:tildeQ}).
Keeping only the  terms linear in $\mu$ and the term quadratic in the difference 
$(\mu_1-\mu_2)$, one has
\begin{eqnarray}
\bar\Delta+\Delta'\stackrel{\ell_1\sim\ell_2\simeq Y/2}{\simeq}
2+(\mu_1-\mu_2)^2-a\gamma_0\left(2+\mu_1+\mu_2\right)-\beta\gamma_0\left[2
+3\frac{\lambda^{3/2}}{(Y+\lambda)^{3/2}-\lambda^{3/2}}\right]\cr
&&
\end{eqnarray}

and
\begin{equation}
\delta_1\stackrel{\ell_1\sim\ell_2\simeq Y/2}{\simeq}\frac19\beta\gamma_0
(\mu_1-\mu_2)^2\left[2+\left(\frac{\lambda}{Y+\lambda}\right)^{3/2}\right];
\end{equation}

$\delta_1$ can be neglected, since $\gamma_0(\mu_1-\mu_2)^2\ll(\mu_1-\mu_2)^2\ll1$.
Then, in the same limit, (\ref{eq:CGMLLAbis}), (\ref{eq:rapMLLAbis}) become
\begin{equation}\label{eq:FWgluon}
{\cal C}_g^0(\ell_1,\ell_2,Y,\lambda)\!\stackrel{\ell_1\sim\ell_2\simeq Y/2}{\simeq}\!1\!+\!\frac{1-b\gamma_0(2+\mu_1+\mu_2)}
{3+(\mu_1-\mu_2)^2-a\gamma_0\left(2+\mu_1+\mu_2\right)-\beta\gamma_0\left[2
+3\displaystyle\frac{\lambda^{3/2}}{(Y+\lambda)^{3/2}-\lambda^{3/2}}\right]},
\end{equation}

\begin{equation}\label{eq:FWquark}
{\cal C}_q^0(\ell_1, \ell_2, Y, \lambda)\stackrel{\ell_1\sim\ell_2\simeq Y/2}{\simeq}
1+\frac{N_c}{C_F}
\bigg[\Big({\cal C}_g^0(\ell_1, \ell_2, Y, \lambda)-1\Big)+
\frac14(b-a)\gamma_0\left(2+\mu_1+\mu_2\right)\bigg].
\end{equation}

Using (\ref{eq:ellmu}) one has
$$
(\mu_1\!-\!\mu_2)^2\simeq9\frac{Y+\lambda}
{\left[(Y+\lambda)^{3/2}-\lambda^{3/2}\right]^2}(\ell_1-\ell_2)^2,
\quad\mu_1\!+\!\mu_2\simeq3\frac{(Y+\lambda)^{1/2}}{(Y+\lambda)^{3/2}-\lambda^{3/2}}
\left[Y\!-\!(\ell_1+\ell_2)\right]
$$
such that the expansion of (\ref{eq:FWgluon}), (\ref{eq:FWquark}) 
in $\gamma_0\propto\sqrt{\alpha_s}$ reads

\vbox{
\begin{eqnarray}
{\cal C}_g^0(\ell_1,\ell_2,Y,\lambda)\!\!&\!\!\simeq\!\!&\!\!\frac43-
\left(\!\frac{(Y+\lambda)^{1/2}(\ell_1-\ell_2)}
{(Y+\lambda)^{3/2}-\lambda^{3/2}}\!\right)^2+
\left(\!\frac23+\frac{\left(Y+\lambda\!\right)^{1/2}Y}
{(Y+\lambda)^{3/2}-\lambda^{3/2}}\right)\!\!\left(\!\frac13a-b\!\right)\gamma_0\cr\cr
&&\hskip -2.5cm+\frac13\left(\frac23+\frac{\lambda^{3/2}}{(Y+\lambda)^{3/2}-\lambda^{3/2}}\right)
\beta\gamma_0+\left(\!b-\frac13a\!\right)\left(\frac{\left(Y+\lambda\right)^{1/2}(\ell_1+\ell_2)}
{(Y+\lambda)^{3/2}-\lambda^{3/2}}\right)\gamma_0+{\cal O}(\gamma_0^2),\cr
&&
\end{eqnarray}
}
\vbox{
\begin{eqnarray}
{\cal C}_q^0(\ell_1,\ell_2,Y,\lambda)\!\!&\!\!\simeq\!\!&\!\!1\!+\!\frac{N_c}{3C_F}\!+\!
\frac{N_c}{C_F}\Bigg[\!\!-\!
\left(\frac{(Y+\lambda)^{1/2}(\ell_1-\ell_2)}
{(Y+\lambda)^{3/2}-\lambda^{3/2}}\right)^2\!\!\!\!-\frac14\!\!
\left(\!\frac23\!+\!\frac{\left(Y+\lambda\right)^{1/2}Y}
{(Y+\lambda)^{3/2}-\lambda^{3/2}}\!\right)\!\!
\left(\!\frac53a\!+\!b\!\right)\gamma_0\bigg.\cr\cr&&\hskip -2.5cm\bigg.+\frac13\left(\frac23+\frac{\lambda^{3/2}}{(Y+\lambda)^{3/2}-\lambda^{3/2}}\right)
\beta\gamma_0+\frac14\left(\!\frac53a+b\!\right)
\left(\frac{\left(Y+\lambda\right)^{1/2}(\ell_1+\ell_2)}
{(Y+\lambda)^{3/2}-\lambda^{3/2}}\right)\gamma_0\Bigg]+{\cal O}(\gamma_0^2).\cr
&&
\end{eqnarray}
}
Therefore, near the hump of the single inclusive distribution,
(\ref{eq:CGMLLAbis}),(\ref{eq:rapMLLAbis}) behave as a linear functions
of the sum $(\ell_1+\ell_2)$ and as a quadratic functions of the difference
$(\ell_1-\ell_2)$. At the limit $\lambda=0$, one recovers the Fong-Webber expression 
\cite{FWSD}.

\subsection{Comparison with the exact solution of the evolution equations:
$\boldsymbol{\lambda=0}$}
%%%%%%%%%%%%%%%%%%%%%%%%%%%%%%%%%%%%%%%%%%%%%%%%%%%%%%%%%%%%%%%%%%%%%%%%%%

\begin{figure}[h]
\begin{center}
\epsfig{file=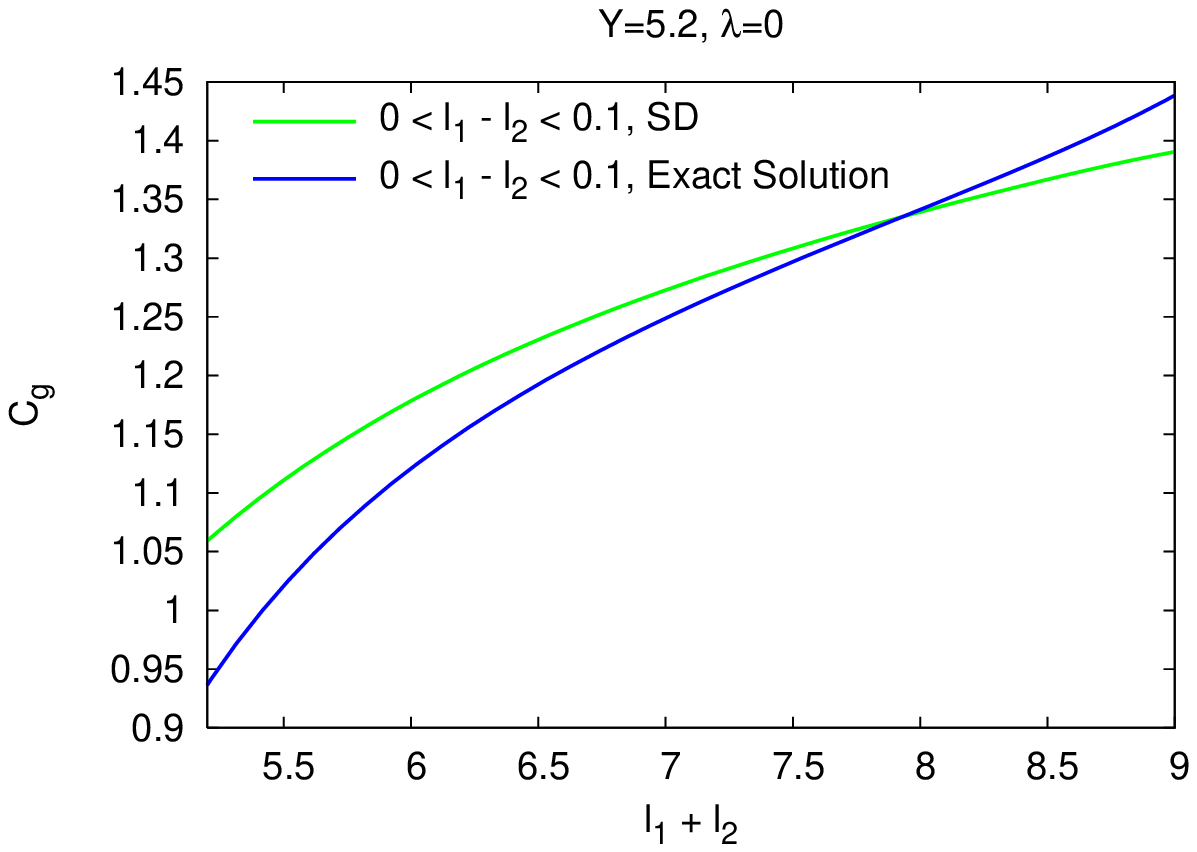, height=5truecm,width=0.48\tw}
\hfill
\epsfig{file=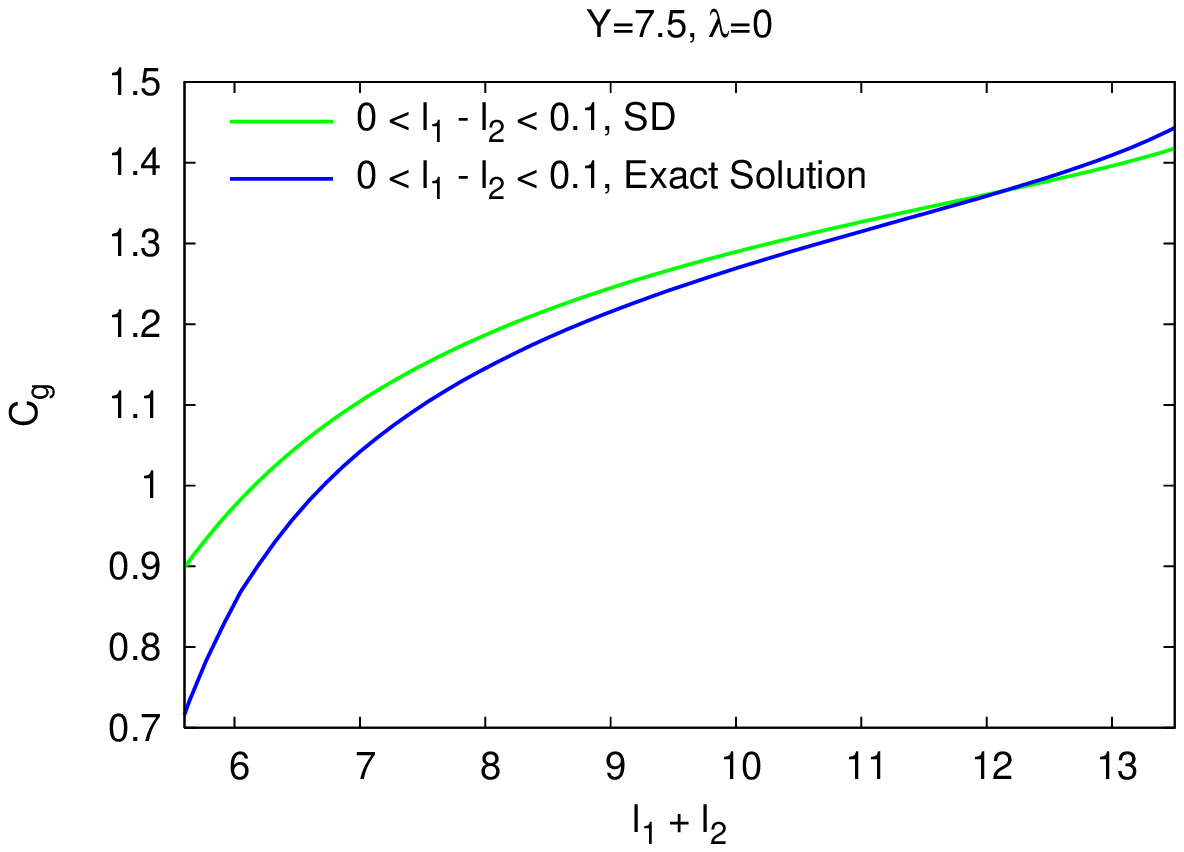, height=5truecm,width=0.48\tw}
\vskip 0.5cm
\epsfig{file=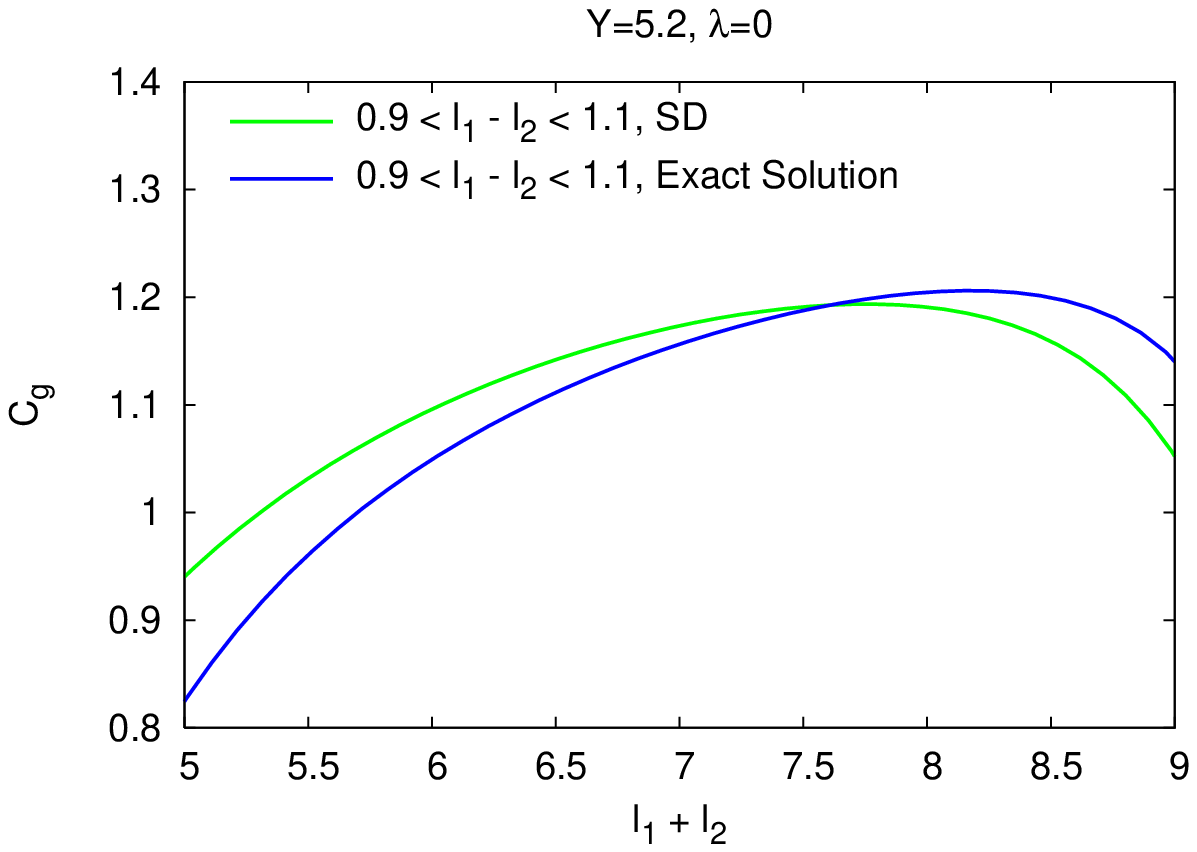, height=5truecm,width=0.48\tw}
\hfill
\epsfig{file=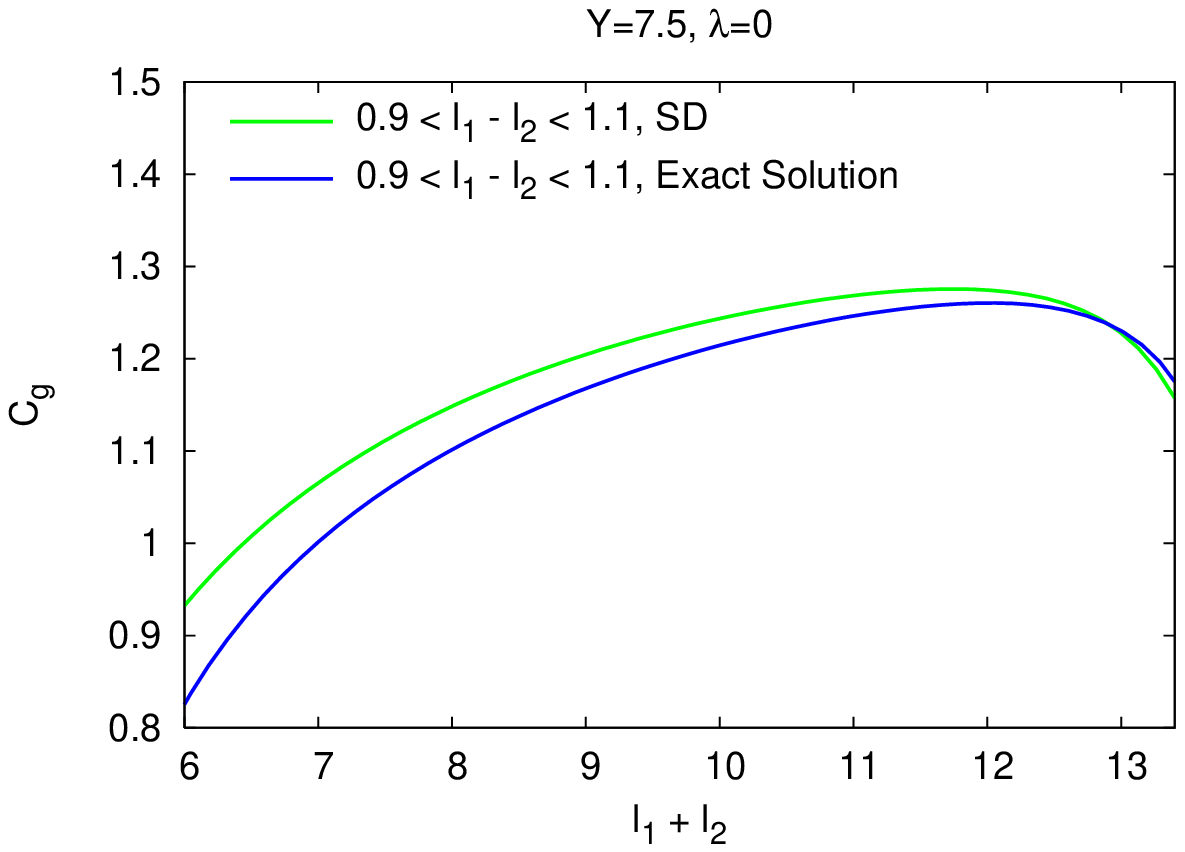, height=5truecm,width=0.48\tw}
\caption{\label{fig:MLLASTESpecter} Comparison between correlators
given by SD and in \cite{RPR2},  at $\lambda=0$.}
\end{center}
\end{figure}

\begin{figure}[h]
\begin{center}
\epsfig{file=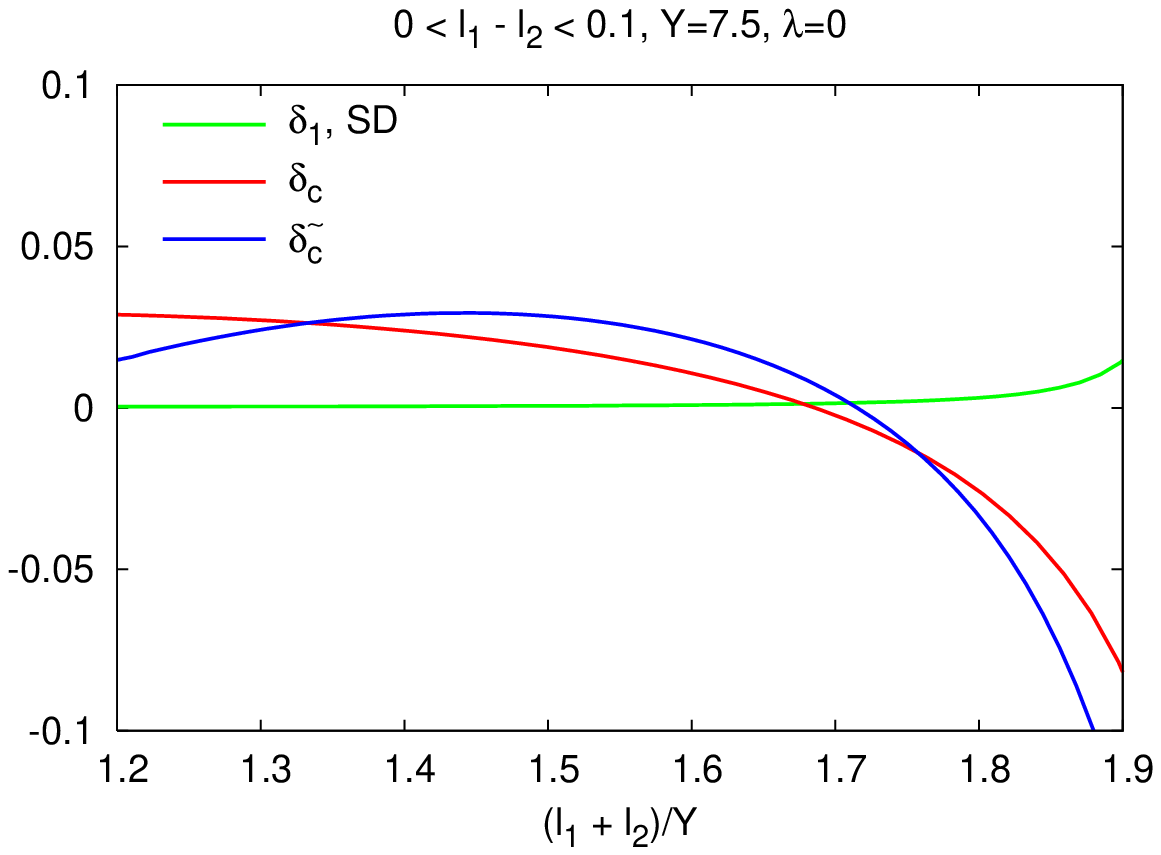, height=5truecm,width=0.48\tw}
\hfill
\epsfig{file=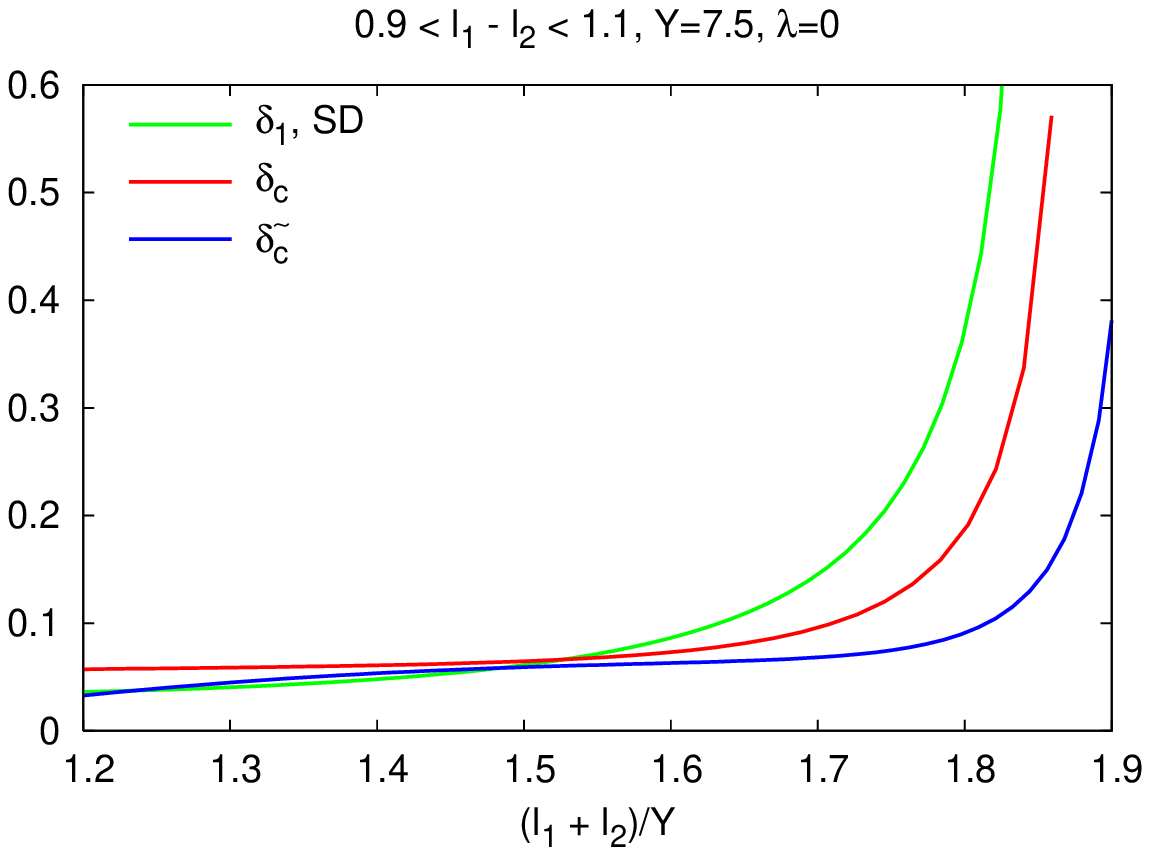, height=5truecm,width=0.48\tw}
\caption{\label{fig:deltas} Comparison between the SD $\delta_1$ and 
$\delta_c,\,\tilde\delta_c$ of \cite{RPR2} at $Y=7.5$, $\lambda=0$.}
\end{center}
\end{figure}

In Figs.\ref{fig:MLLASTESpecter} we compare the SD evaluation of the gluon correlator
with the exact solution of \cite{RPR2} at $\lambda=0$. The difference comes from sub-leading
corrections of order $\gamma_0^2$ that are not present in (\ref{eq:CGMLLAbis}). 
For example, $-\beta\gamma_0^2\approx-0.2$
at $Y=5.2$ occurring in the exact solution (69) of \cite{RPR2} 
is not negligible but is absent
in (\ref{eq:CGMLLAbis}) and (\ref{eq:rapMLLAbis}).
That is why, the SD MLLA curve lies slightly above 
the one of \cite{RPR2} at small $\ell_1+\ell_2$. The mismatch becomes smaller 
at $Y=7.5$, since $-\beta\gamma_0^2\approx-0.13$. However, when $\ell_1+\ell_2$
increases, the solution of \cite{RPR2} takes over, which can be explained by comparing
the behavior of the SD MLLA $\delta_1$ obtained in (\ref{eq:delta2}) and 
$\delta_c,\,\tilde\delta_c$ in \cite{RPR2}. Namely, while $\delta_1$ remains positive
and negligible for $\ell_1\approx\ell_2$, $\delta_c,\,\tilde\delta_c$ 
decrease and get negative when $\ell_1+\ell_2\to2Y$, see Fig.\ref{fig:deltas} (left),
which makes the correlations slightly bigger
in this region. As $|\ell_1-\ell_2|$ increases, $\delta_1$ is seen in
Fig.\ref{fig:deltas} (right) to play the same role as $\delta_c,\tilde\delta_c$ do
in the solution \cite{RPR2} and therefore, to decrease the correlation.
The agreement between both methods improves as the energy scale increases.
A similar behavior holds for the quark correlator.

In \cite{RPR2}, strong cancellations between the MLLA $\delta_1$ and the
NMLLA $\delta_2$  were seen to take place, giving very small
$\delta_c$ and $\tilde\delta_c$; this eased the
convergence of the iterative method but raised questions concerning the
relative size of MLLA and NMLLA corrections and the validity of the
perturbative expansion.
However, since $\delta_1$ is itself, there, entangled with {\em some} NMLLA
corrections, no definitive conclusions could be drawn.
The present work and Fig.~\ref{fig:deltas}, by showing that, below, 
$\delta_c$ and $\tilde \delta_c$ of \cite{RPR2} play the same role as
the {\em pure MLLA} $\delta_1$ which is now calculated, suggests (though it
is not a demonstration) that the perturbative series is safe.
It is indeed compatible with the following scheme: in \cite{RPR2},
the pure MLLA part of $\delta_1$ is the same as that in the present work;
the cancellations in \cite{RPR2} occur between NMLLA corrections
included in $\delta_1$ and $\delta_2$;  these are eventually of the same
order of magnitude as MLLA terms, but they are only parts of all
NMLLA corrections; this leaves the possibility that  the sum of all NMLLA
corrections to $\delta_1$ and all NMLLA terms of $\delta_2$ are separately
smaller than the pure MLLA terms of $\delta_1$, that is that strong
cancellations occur {\em between NMLLA corrections}, the ones
 included, because of the logic of the calculation, in \cite{RPR2}, and
those which were not be taken into account.

\subsection{Comparison with Fong-Webber and LEP-I data; how
$\boldsymbol{\lambda=0}$ is favored}
%%%%%%%%%%%%%%%%%%%%%%%%%%%%%%%%%%%%%%%%%%%%%%%%%%%%%%%%%%%%%%%%%%%%

Let us consider, at the $Z^0$ peak $Y=5.2$ ($E\Theta=91.2\,\text{GeV}\, \text{at}\, $LEP-I energy), the process $e^+e^-\to q\bar q$. As can be induced 
from Fig.\ref{fig:MLLASTESpecter},
the results obtained in the present work by the (approximate) SD method
are very close to the ones obtained in subsection 6.5 of \cite{RPR2}
by the exact solution of the evolution equations.
Accordingly, the same comparison as in \cite{RPR2} holds with respect to
both Fong \& Webber's results \cite{FWSD} and  OPAL data \cite{OPALSD}.

It is also noticeable that, since, at $\lambda=0$, correlations already lye
above (present) experimental curves, and since an increase of
$\lambda$  tends to increase the predictions, the limiting spectrum stays the
best candidate to bring agreement with experiments.

%%%%%%%%%%%%%%%%%%%%%
\section{CONCLUSION}
\label{sec:CONCL}
%%%%%%%%%%%%%%%%%%%%%

Let us, in a few words, summarize the achievements,  but also the
limitations of the two methods that have been used respectively
in \cite{RPR2} (exact solution of MLLA evolution equations) and in the
present work (steepest descent approximate evaluation of their solutions).

Achievements are threefold:\newline
- in \cite{RPR2}, MLLA evolution equations for 2-particle
correlations have been deduced at small $x$ and at any $\lambda$; their
(iterative) solution can unfortunately only be expressed analytically at
the limit $\lambda \to 0$; 
\newline
- by the steepest descent method, which is an approximate method,
analytical expressions for the spectrum could instead be obtained
for $\lambda \ne 0$, which enabled to calculate the correlation at the same
level of generality; \newline
- one could move away from the peak of the inclusive distribution.

So doing, the limitations of the work of Fong \& Webber have vanished.
Their results have been recovered at the appropriate limits.

The two methods numerically agree remarkably well, despite an unavoidable
entanglement of MLLA + some NMLLA corrections in the first one.

The limitations are the following:\newline
- the uncontrollable increase of  $\alpha_s$ when one goes to smaller
and smaller transverse momenta: improvements in this directions mainly
concern the inclusion of non-perturbative contributions;\newline
- departure from the limiting spectrum: it cannot of course appear as a
limitation, but we have seen that increasing the value of $\lambda$, by
increasing the correlations, does not bring better agreement with present
data; it confirms thus, at present, that
the limiting spectrum is the best possibility;\newline
- the LPHD  hypothesis: it works surprisingly well for
inclusive distributions; only forthcoming data will assert
whether its validity decreases  when one studies less inclusive
processes (like correlations);\newline
- last, the limitation to small $x$: it is still quite drastic;
departing from this limit most probably lye in the art of numerical
calculations, which makes part of forthcoming projects.

Expectations rest on  experimental data, which are being collected at the
Tevatron, and which will be at LHC. The higher the energy, the safer
perturbative QCD is, and the better the agreement should be with our
predictions. The remaining disagreement (but much smaller than
Fong-Webber's) between predictions and LEP-1 results for 2-particle
correlations stands as an open
question concerning the validity of the LPHD hypothesis for these
observables which are not ``so'' inclusive as the distributions studied in
\cite{PerezMachetSD}. The eventual necessity to include NMLLA corrections can
only be decided when new data appear.

\vskip 1cm
\underline{\em Acknowledgments:}

It is a pleasure to thank Yuri Dokshitzer for enticing me towards the steepest
descent method and for showing me its efficiency with simple examples. I am
grateful to Bruno Machet for his help and advice and to Gavin Salam for helping
me in numerically inverting formula (\ref{eq:ratiomunu}).

%%%%%%%%%%%%%%%%%%%%%%%%%%%%%%%%%%%%%%%%%%%%%%%%%%%%%%%%%%%%%%%%%%%%%%%%%%%%%

\newpage
%\appendix

{\large\bf APPENDIX}

%%%%%%%%%%%%%%%%%%%%%%%%%%%%%%%%%%%%%%%%%%%%%%%%%%%%%%%%%%%%%%%%%%%%%%%%%%%%%
\section{DOUBLE DERIVATIVES AND DETERMINANT}
\label{sec:SDdetails}
%%%%%%%%%%%%%%%%%%%%%%%%%%%%%%%%%%%%%%%%%%%%%%%%%%%%%%%%%%%%%%%%%%%%%%%%%%%%%

\subsection{Demonstration of eq.~(\ref{eq:determinant})}
\label{sec:DDDet}
%%%%%%%%%%%%%%%%%%%%%%%%%%%%%%%%%%%%%%%%%%%%%%%%%%%%

We conveniently rewrite (\ref{eq:deromega}) and (\ref{eq:dernu}) in the form

\begin{equation}\label{eq:derphiellbis}
\frac{\partial\phi}{\partial\omega}=\frac{2\omega-\nu}{\omega-\nu}\ell+
\frac{\nu}{\omega-\nu}-\frac{\phi}{\omega-\nu}-\lambda\frac{\nu+2s_0}{\omega-\nu}+
\frac1{\beta\omega(\omega-\nu)},
\end{equation}

\begin{equation}\label{eq:derphiybis}
\frac{\partial\phi}{\partial\nu}=\frac{\omega-2\nu}{\omega-\nu}y-
\frac{\omega}{\omega-\nu}+\frac{\phi}{\omega-\nu}+\lambda\frac{\omega+2s_0}{\omega-\nu}-
\frac1{\beta\nu(\omega-\nu)}.
\end{equation}
The Taylor expansion of (\ref{eq:phiexp}) in (\ref{eq:SpecDD}) reads
\begin{eqnarray}\nonumber
\phi(\omega,\nu,\ell,y)\!\!&\!\!\approx\!\!&\!\!\phi(\omega_0,\nu_0,\ell,y)+
\frac12\,\frac{\partial^2\phi}{\partial\omega^2}\,(\omega_0,\nu_0)(\omega-\omega_0)^2+
\frac12\,\frac{\partial^2\phi}{\partial\nu^2}\,(\omega_0,\nu_0)(\nu-\nu_0)^2\\\nonumber\\
&&+\frac{\partial^2\phi}{\partial\omega\partial\nu}
(\omega_0,\nu_0)\,(\omega-\omega_0)(\nu-\nu_0).
\end{eqnarray}

The expressions of the second derivatives follow directly from (\ref{eq:derphiellbis}) and (\ref{eq:derphiybis})
\begin{eqnarray}\nonumber
\frac{\partial^2\phi}{\partial\omega^2}=-\frac{\nu}{(\omega-\nu)^2}(\ell\!+\!y\!+\!\lambda)
+\frac{\phi}{(\omega-\nu)^2}-\frac{2\omega-\nu}{\beta\omega^2(\omega-\nu)^2}
+\frac4{\beta(\omega-\nu)^2(2s_0+\omega+\nu)},
\end{eqnarray}
\begin{eqnarray}\nonumber
\frac{\partial^2\phi}{\partial\nu^2}=-\frac{\omega}{(\omega-\nu)^2}(\ell\!+\!y\!+\!\lambda)
+\frac{\phi}{(\omega-\nu)^2}+\frac{\omega-2\nu}{\beta\nu^2(\omega-\nu)^2}
+\frac4{\beta(\omega-\nu)^2(2s_0+\omega+\nu)},
\end{eqnarray}
\begin{eqnarray}\nonumber
\frac{\partial^2\phi}{\partial\omega\partial\nu}=\frac{\omega}{(\omega-\nu)^2}(\ell\!+\!y\!+\!\lambda)
-\frac{\phi}{(\omega-\nu)^2}+\frac1{\beta\omega(\omega-\nu)^2}
-\frac4{\beta(\omega-\nu)^2(2s_0+\omega+\nu)}.
\end{eqnarray}

(\ref{eq:SpecDD}) and its solution can be written in the form

$$
G\simeq\iint d^2v\, e^{-\frac12 v^TAv}=\frac{2\pi}{\sqrt{Det\,A}}
$$
where 

$$
v=(\omega,\nu),\quad v^T=\left(\!\!
\begin{array}{c}
        \omega \cr
\cr
        \nu
 \end{array}
\!\!\right),\quad\begin{array}{c}
 DetA =\end{array} Det\left(
\begin{array}{c}
        \frac{\partial^2\phi}{\partial\omega^2}\qquad
 \frac{\partial^2\phi}{\partial\omega\partial\nu} \cr
\cr
        \frac{\partial^2\phi}{\partial\nu\partial\omega}\qquad
 \frac{\partial^2\phi}{\partial\nu^2}
 \end{array}
\right)=\frac{\partial^2\phi}{\partial\omega^2}\frac{\partial^2\phi}{\partial\nu^2}-
\left(\frac{\partial^2\phi}{\partial\omega\partial\nu}\right)^2. 
$$

An explicit calculation gives

$$
DetA=(\ell+y+\lambda)^2\left[\frac{\beta(\omega+\nu)\phi-4}{(\omega-\nu)^2}+\frac{4(\omega+\nu)}
{(\omega-\nu)^2(2s_0+\omega+\nu)}\right]
$$

which, by using (\ref{eq:muupsilon}) leads to (\ref{eq:determinant}).

\subsection{$\boldsymbol{DetA}$ (see eq.~(\ref{eq:determinant}))
around the maximum}
\label{subsec:DeTA}
%%%%%%%%%%%%%%%%%%%%%%%%%%%%%%%%%%%%%%%%%%%%%%%%%%%%%%%%%%%%%%%%%%%%%

This is an addendum to subsection \ref{subsection:SDeval} 

$\ell_{max}$ written in (\ref{eq:ellmaxmlla}) is close to the
DLA value $Y/2$ \cite{DLASD}\cite{EvEqSD}\cite{KOSD}. We then have
$\mu\sim\upsilon\to0$ for $\ell\approx y\simeq Y/2$. In this limit,
(\ref{eq:ratiomunu})  and (\ref{eq:relmunu}) respectively translate into

\begin{equation}\label{eq:ellmu}
Y-2\ell\stackrel{\mu,\upsilon\to0}{\approx}\frac23\,
\frac{\left(Y+\lambda\right)^{3/2}-\lambda^{3/2}}
{\left(Y+\lambda\right)^{1/2}}\,\mu,\quad 
\upsilon\stackrel{\mu,\upsilon\to0}{\approx}
\sqrt{\frac{\lambda}{Y+\lambda}}\mu,
\end{equation}

while

\begin{equation}\label{eq:dermull}
\frac{\partial\mu}{\partial\ell}\simeq-3\frac{(Y+\lambda)^{1/2}}
{(Y+\lambda)^{3/2}-\lambda^{3/2}}
\end{equation}

should be used to get (\ref{eq:gaussian}). An explicit calculation gives

$$
\lim_{\mu,\upsilon\rightarrow0}
\sqrt{\frac{\beta^{1/2}(Y+\lambda)^{3/2}}{\pi DetA(\mu,\upsilon)}}\!=\!
\left(\frac{3}{\pi\sqrt{\beta}\left[(Y+\lambda)^{3/2}
\!-\!\lambda^{3/2}\right]}\right)^{1/2},
$$

where

\begin{eqnarray}\nonumber
DetA\!&\!\stackrel{\mu,\upsilon\to0}\approx
\!&\!\beta(Y\!+\!\lambda)^3\frac{\left(\mu\!-\!\upsilon\right)
\left(1\!+\!\frac12
\mu^2\right)\left(1\!+\!\frac12\upsilon^2\right)\!+\!(1\!+\!\frac12
\mu^2)\left(\upsilon\!+\!\frac16\upsilon^3\right)\!-\!\left(\mu\!+\!\frac16\mu^3\right)
\left(1\!+\!\frac12\upsilon^2\right)}{\mu^3}\\\notag\\
\!&\!\simeq\!&\!\frac13\beta(Y\!+\!\lambda)^3\left(1\!-\!\frac{\upsilon^3}{\mu^3}\right)=
\frac13\beta(Y+\lambda)^3\left[1-\left(\frac{\lambda}{Y+\lambda}\right)^{3/2}\right].
\end{eqnarray}

\subsection{The functions $\boldsymbol{L(\mu,\upsilon)}$, 
$\boldsymbol{K(\mu,\upsilon)}$ in eq.~(\ref{eq:LK})}
\label{subsec:LK}
%%%%%%%%%%%%%%%%%%%%%%%%%%%%%%%%%%%%%%%%%%%%%%%%%%%%%%%%%

An explicit calculation gives
\begin{equation}\label{eq:LL}
L(\mu,\upsilon)=\frac32\frac{\cosh\mu}{\sinh\mu}-
\frac12\frac{(\mu-\upsilon)\cosh\upsilon\sinh\mu+\sinh\upsilon\sinh\mu}
{(\mu-\upsilon)\cosh\mu\cosh\upsilon+\cosh\mu\sinh\upsilon-\sinh\mu\cosh\upsilon},
\end{equation}
and
\begin{equation}\label{eq:KK}
K(\mu,\upsilon)=-\frac12\sinh\upsilon\frac{(\mu-\upsilon)\cosh\mu-\sinh\mu}
{(\mu-\upsilon)\cosh\mu\cosh\upsilon+
\cosh\mu\sinh\upsilon-\sinh\mu\cosh\upsilon}.
\end{equation}

\subsection{A consistency check}
\label{subsec:check}
%%%%%%%%%%%%%%%%%%%%%%%%%%%%%%%%

Let us verify that the evolution equation (\ref{eq:solgSD}) is satisfied
by (\ref{eq:SpecNormMLLA}) within the MLLA accuracy.
Differentiating (\ref{eq:solgSD}) with respect to $\ell$, $y$
yields the equivalent differential equation

$$
G_{\ell y} = \gamma_0^2 \left(G - a G_\ell\right) \!+\! {\cal {O}}{\big(\gamma_0^4G\big)}
$$

that can be rewritten in the form

\begin{equation}\label{eq:MLLApsi}
 \psi_\ell\psi_y +\psi_{\ell y} \>=\>
 \gamma_0^2\left(1-a\psi_\ell\right)  \>+\> {\cal {O}}\big({\gamma_0^4}\big);
\end{equation}

we have neglected next-to-MLLA corrections ${\cal O}(\gamma_0^4)$
(of relative order $\gamma_0^2$) coming from differentiating the coupling
$\gamma_0^2$ in the sub-leading (``hard correction'') term $\propto a$.

We have to make sure that (\ref{eq:MLLApsi}) holds including the terms
${\cal O}(\gamma_0^3)$. In the sub-leading terms we can set $\psi\to \varphi$
(see (\ref{eq:SPbis})):

\begin{equation}
  (\varphi_\ell + \delta\psi_\ell)(\varphi_y + \delta\psi_y) + \varphi_{\ell
  y} = \gamma_0^2(1-a\varphi_\ell). 
\end{equation}

Isolating correction terms and casting them all on the l.h.s. of the
equation we get

\begin{equation}\label{eq:collect}
  a\gamma_0^2\varphi_\ell 
+ \left[\,\varphi_\ell \delta\psi_y + \varphi_y\delta\psi_\ell\,\right] + \varphi_{\ell
  y} \>=\>  \gamma_0^2 -\varphi_\ell\varphi_y.
\end{equation}

By the definition (\ref{eq:SPbis}) of the saddle point
we conclude that the r.h.s. of (\ref{eq:collect}) is zero such that we have

\begin{equation}\label{eq:hastobe}
   \omega_0 a\gamma_0^2 + \left[\,\omega_0 \delta\psi_y +
     \nu_0\delta\psi_\ell\,\right] + \frac{d\omega_0}{dy} \>=\> 0\,,
\end{equation}

that is,

\begin{equation}\label{eq:omega0}
   \omega_0 \left(a\gamma_0^2 + \delta\psi_y\right) +
     \nu_0\delta\psi_\ell + \frac{d\omega_0}{dy} \>=\> 0\,.
\end{equation}

First, we select the terms $\propto a$:

\begin{eqnarray*}
&&a\gamma_0^3\left[-\frac12\widetilde{Q}-\frac12\tanh\upsilon\, e^{\mu}
+\frac12\tanh\upsilon\coth\mu\, e^{\mu} + \frac12\tanh\upsilon\coth\mu \,
\widetilde{Q}\right.\\
&&\left.+\frac12\widetilde{Q}-\frac12\tanh\upsilon\,e^{-\mu}-\frac12
\tanh\upsilon\coth\mu\, e^{-\mu}-\frac12\tanh\upsilon\coth\mu \,
\widetilde{Q}\right]\\
&&=a\gamma_0^3\left[-\tanh\upsilon\cosh\mu+\tanh\upsilon\coth\mu\sinh\mu\right]
\equiv0.
\end{eqnarray*}

From (\ref{eq:muupsilon}) one deduces

$$
\frac{d\omega_0}{dy}=\frac12\beta\gamma_0^3\widetilde{Q},
$$

that is inserted in (\ref{eq:omega0}) such that,
for terms $\propto\beta$, we have

\begin{eqnarray*}
&&-\beta\gamma_0^3\left[\frac12e^{\mu}+\frac12\tanh\upsilon
\Big(1\!+\!K\Big)e^{\mu}-\frac12C\,e^{\mu}-\frac12C\widetilde{Q}+
\frac12e^{-\mu}+\frac12\tanh\upsilon
\Big(1\!+\!K\Big)e^{-\mu}\right.\\
&&\left.+\frac12C\,e^{-\mu}+\frac12C\widetilde{Q}\right]
=-\beta\gamma_0^3\left[\cosh\mu+\tanh\upsilon\cosh\mu\Big(1\!+\!K\Big)
-C\sinh\mu-\frac12\widetilde{Q}\right],
\end{eqnarray*}

which gives

\begin{eqnarray*}
&&-\beta\gamma_0^3\left[\cosh\mu-\sinh\mu\,L-\frac12\widetilde{Q} \right].
\end{eqnarray*}

 Constructing (see (\ref{eq:tildeQ}) and appendix \ref{subsec:LK})

\begin{eqnarray*}
\widetilde{Q}(\mu,\upsilon)-2\cosh\mu\!\!&\!\!=\!\!&\!\!
-3\cosh\mu+\sinh\mu\frac{(\mu-\upsilon)\cosh\upsilon\sinh\mu+\sinh\upsilon\sinh\mu}
{(\mu-\upsilon)\cosh\mu\cosh\upsilon+\cosh\mu\sinh\upsilon-\sinh\mu\cosh\upsilon}\\
\!\!&\!\!=\!\!&\!\!-2\sinh\mu L(\mu,\upsilon)
\end{eqnarray*}

we have

$$
-\beta\gamma_0^3\left[\cosh\mu-\sinh\mu\,L-\frac12\widetilde{Q}\right]\equiv0.
$$

%%%%%%%%%%%%%%%%%%%%%%%%%%%%%%%%%%%%%%%%%%%%%%%%%%%%%%%%%%%%%%%%%%%%%%%
\section{ANALYTICAL EXPRESSION  OF $\boldsymbol{\Delta'(\mu_1,\mu_2)}$
OBTAINED FROM  EQ.~(\ref{eq:Deltaprime})}
\label{subsec:Deltaprime}
%%%%%%%%%%%%%%%%%%%%%%%%%%%%%%%%%%%%%%%%%%%%%%%%%%%%%%%%%%%%%%%%%%%%%%%

Replacing (\ref{eq:derpsi'l})(\ref{eq:derpsi'y}) in
(\ref{eq:Deltaprime}) and neglecting terms of relative order ${\cal O}(\gamma_0^3)$
which are beyond the MLLA accuracy, we obtain

\begin{eqnarray}\label{eq:dDelta}
\Delta'\!\!&\!\! = \!\!&\!\!
\frac{ e^{-\mu_1}\delta\psi_{2,\ell} + e^{-\mu_2}\delta\psi_{1,\ell} 
       + e^{\mu_1}\delta\psi_{2,y} + e^{\mu_2}\delta\psi_{1,y}} {\gamma_0}\cr\cr
\!\!&\!\!=\!\!&\!\!-a\gamma_0\left[e^{\mu_1}+e^{\mu_2}-\sinh(\mu_1-\mu_2)(\widetilde{Q}_1-
\widetilde{Q}_2)+\cosh\mu_1\tanh\upsilon_2+\cosh\mu_2\tanh\upsilon_1\right.\cr\cr
&&\hskip 1cm\left.-\sinh\mu_1\tanh\upsilon_2\coth\mu_2-
\sinh\mu_2\tanh\upsilon_1\coth\mu_1\right.\cr\cr
&&\hskip 1cm\left.+\sinh(\mu_1-\mu_2)\Big(\tanh\upsilon_1
\coth\mu_1\widetilde{Q}_1-\tanh\upsilon_2
\coth\mu_2\widetilde{Q}_2\Big)\right]\cr\cr
&&-\beta\gamma_0\left[\Big(\cosh\mu_1-\sinh\mu_1C_2\Big)+\Big(\cosh\mu_2-\sinh\mu_2C_1\Big)
+\sinh(\mu_1-\mu_2)(C_1\widetilde{Q}_1-C_2\widetilde{Q}_2)\right.\cr\cr
&&\hskip 1cm\left.+\cosh\mu_1\tanh\upsilon_2\Big(1+K_2\Big)+\cosh\mu_2\tanh\upsilon_1
\Big(1+K_1\Big)\right].
\end{eqnarray}

%%%%%%%%%%%%%%%%%%%%%%%%%%%%%%%%%%%%%%%%%%%%%%%%%%%%%%%%%%%%%%%%%%%%%%%%%%%%%%
\null

%\vskip 2cm

%\listoffigures

%%%%%%%%%%%%%%%%%%%%%%%%%%%%%%%%%%%%%%%%%%%%%%%%%%%%%%%%%%%%%%%%%%%%%%%%%%%%%%

%%%%%%%%%%%%%%%%%%%%%%%%%%%%%%%%%%%%%%%%%%%%%%%%%%%%%%%%%%%%%%%%%%%%%%%%%%%%
%%%%%%%%%%%%%%%%%%%%%%%%%%%%%%%%%%%%%%%%%%%%%%%%%%%%%%%%%%%%%%%%%%%%%%%%%%%%

%\newpage\null
%
%

%%%%%%%%%%%%%%%%%%%%%%%%%%%%%%%%%%%%%%%%%%%%%%%%%%%%%%%%%%%%%%%%%%%%%%%%%%%%%%%%%%%%%%%%%%%%%%%%

\end{document}